\documentclass{ucthesis}%
\usepackage{amsmath}
\usepackage{amsfonts}
\usepackage{amssymb}
\usepackage{graphicx}
\newtheorem{theorem}{Theorem}

\newtheorem{algorithm}[theorem]{Algorithm}

\newtheorem{conjecture}[theorem]{Conjecture}
\newtheorem{corollary}[theorem]{Corollary}

\newtheorem{definition}[theorem]{Definition}

\newtheorem{lemma}[theorem]{Lemma}

\newtheorem{proposition}[theorem]{Proposition}

\newenvironment{proof}[1][Proof]{\textbf{#1.} }{\ \rule{0.5em}{0.5em}}

\def\dsp{\def\baselinestretch{1.0}\large\normalsize}
\dsp 
\begin{document}

\title{Limits on Efficient Computation in the Physical World}
\author{Scott Joel Aaronson}
\degreeyear{2004} \degreesemester{Fall} \degree{Doctor of
Philosophy} \numberofmembers{3} \chair{Professor Umesh Vazirani}
\othermembers{Professor Luca Trevisan\\Professor K. Birgitta Whaley}
\prevdegrees{Bachelor of Science (Cornell University) 2000}
\field{Computer Science} \campus{Berkeley} \maketitle \approvalpage{

}
\copyrightpage{

}

\begin{abstract}
More than a speculative technology, quantum computing seems to
challenge our most basic intuitions about how the physical world
should behave. \ In this thesis I show that, while some intuitions
from classical computer science must be jettisoned in the light of
modern physics, many others emerge nearly unscathed; and I use
powerful tools from computational complexity theory to help
determine which are which.

In the first part of the thesis, I attack the common belief that
quantum computing resembles classical exponential parallelism, by
showing that quantum computers would face serious limitations on a
wider range of problems than was previously known. \ In particular,
any quantum algorithm that solves the \textit{collision
problem}---that of deciding whether a sequence of $n$ integers is
one-to-one or two-to-one---must query the sequence $\Omega\left(
n^{1/5}\right) $\ times. \ This resolves a question that was open
for years; previously no lower bound better than constant was known.
\ A corollary is that there is no \textquotedblleft
black-box\textquotedblright\ quantum algorithm to break
cryptographic hash functions or solve the Graph Isomorphism problem
in polynomial time. \ I also show that relative to an oracle,
quantum computers could not solve $\mathsf{NP}$-complete problems in
polynomial time, even with the help of nonuniform \textquotedblleft
quantum advice states\textquotedblright; and that any quantum
algorithm needs $\Omega\left( 2^{n/4}/n\right)  $\ queries to find a
local minimum of a black-box function on the $n$-dimensional
hypercube. \ Surprisingly, the latter result also leads to new
\textit{classical} lower bounds for the local search problem. \
Finally, I give new lower bounds on quantum one-way communication
complexity, and on the quantum query complexity of total Boolean
functions and recursive Fourier sampling.

The second part of the thesis studies the relationship of the quantum
computing model to physical reality. \ I first examine the arguments of Leonid
Levin, Stephen Wolfram, and others who believe quantum computing to be
fundamentally impossible. \ I find their arguments unconvincing without a
\textquotedblleft Sure/Shor separator\textquotedblright---a criterion that
separates the already-verified quantum states from those that appear in Shor's
factoring algorithm. \ I argue that such a separator should be based on a
complexity classification of quantum states, and go on to create such a
classification. \ Next I ask what happens to the quantum computing model if we
take into account that the speed of light is finite---and in particular,
whether Grover's algorithm still yields a quadratic speedup for searching a
database. \ Refuting a claim by Benioff, I show that the surprising answer is
yes. \ Finally, I analyze hypothetical models of computation that go even
beyond quantum computing. \ I show that many such models would be as powerful
as the complexity class $\mathsf{PP}$, and use this fact to give a simple,
quantum computing based proof that $\mathsf{PP}$\ is closed under
intersection. \ On the other hand, I also present one model---wherein we could
sample the entire history of a hidden variable---that appears to be more
powerful than standard quantum computing, but only \textit{slightly} so.

\abstractsignature

\end{abstract}

%\approvalpage
%\copyrightpage

\begin{frontmatter}
\tableofcontents \listoffigures \listoftables
\begin{acknowledgements}%
My adviser, Umesh Vazirani, once said that he admires the quantum
adiabatic algorithm because, like a great squash player, it achieves
its goal while moving as little as it can get away with. \
Throughout my four years at Berkeley, I saw Umesh inculcate by
example his \textquotedblleft adiabatic\textquotedblright\
philosophy of life: a philosophy about which papers are worth
reading, which deadlines worth meeting, and which research problems
worth a fight to the finish. \ Above all, the concept of
\textquotedblleft beyond hope\textquotedblright\ does not exist in
this philosophy, except possibly in regard to computational
problems. \ My debt to Umesh for his expert scientific guidance,
wise professional counsel, and generous support is obvious and
beyond my ability to embellish. \ My hope is that I graduate from
Berkeley a more adiabatic person than when I came.

Admittedly, if the push to finish this thesis could be called
adiabatic, then the spectral gap was exponentially small. \ As I
struggled to make the deadline, I relied on the help of David
Molnar, who generously agreed to file the thesis in Berkeley while I
remained in Princeton; and my committee---consisting of Umesh, Luca
Trevisan, and Birgitta Whaley---which met procrastination with
flexibility.

Silly as it sounds, a principal reason I came to Berkeley was to
breathe the same air that led Andris Ambainis to write his epochal
paper \textquotedblleft Quantum lower bounds by quantum
arguments.\textquotedblright\ \ Whether or not the air in 587 Soda
did me any good, Part \ref{LQC}\ of the thesis is essentially a
150-page tribute to Andris---a colleague whose unique combination of
genius and humility fills everyone who knows him with awe.

The direction of my research owes a great deal as well to Ronald de
Wolf, who periodically emerges from his hermit cave to challenge
non-rigorous statements, eat dubbel zout, or lament American
ignorance. \ While I can see eye-to-eye with Ronald about (say) the
$\operatorname*{D}\left( f\right)  $\ versus
$\operatorname*{bs}\left( f\right)  ^{2}$\ problem, I still feel
that Andrei Tarkovsky's \textit{Solaris} would benefit immensely
from a car chase.

For better or worse, my conception of what a thesis should be was
influenced by Dave Bacon, quantum computing's elder clown, who
entitled the first chapter of his own 451-page behemoth
\textquotedblleft Philosonomicon.\textquotedblright \ I'm also
indebted to Chris Fuchs and his \textit{samizdat}, for the idea that
a document about quantum mechanics more than 400 pages long can be
worth reading most of the way through.

I began working on the best-known result in this thesis, the quantum
lower bound for the collision problem, during an unforgettable
summer at Caltech. \ Leonard Schulman and Ashwin Nayak listened
patiently to one farfetched idea after another, while John
Preskill's weekly group meetings helped to ensure that the mysteries
of quantum mechanics, which inspired me to tackle the problem in the
first place, were never far from my mind. \ Besides Leonard, Ashwin,
and John, I'm grateful to Ann Harvey\ for putting up with the
growing mess in my office. \ For the record, I never once slept in
the office; the bedsheet was strictly for doing math on the floor.

I created the infamous Complexity Zoo web site during a summer at
CWI in Amsterdam,\ a visit enlivened by the presence of Harry
Buhrman, Hein R\"{o}hrig, Volker Nannen, Hartmut Klauck, and Troy
Lee. \ That summer I also had memorable conversations with David
Deutsch and Stephen Wolfram. \ Chapters \ref{PLS}, \ref{MLIN}, and
\ref{QCHV} partly came into being during a semester at the Hebrew
University in Jerusalem, a city where \textquotedblleft Aaron's
sons\textquotedblright\ were already obsessing about cubits three
thousand years ago. \ I thank Avi Wigderson, Dorit Aharonov, Michael
Ben-Or, Amnon Ta-Shma, and Michael Mallin for making that semester a
fruitful and enjoyable one. \ I also thank Avi for pointing me to
the then-unpublished results of Ran Raz on which Chapter \ref{MLIN}
is based, and Ran for sharing those results.

A significant chunk of the thesis was written or revised over two
summers at the Perimeter Institute for Theoretical Physics in
Waterloo. \ I thank Daniel Gottesman, Lee Smolin, and Ray Laflamme
for welcoming a physics doofus to their institute, someone who
thinks the string theory versus loop quantum gravity debate should
be resolved by looping over all possible strings. \ From Marie
Ericsson, Rob Spekkens, and Anthony Valentini I learned that
theoretical physicists have a better social life than theoretical
computer scientists, while from Dan Christensen I learned that
complexity and quantum gravity had better wait before going steady.

Several ideas were hatched or incubated during the yearly QIP
conferences; workshops in Toronto, Banff, and Leiden; and visits to
MIT, Los Alamos, and IBM Almaden. \ I'm grateful to Howard Barnum,
Andrew Childs, Elham Kashefi, Barbara Terhal, John Watrous, and many
others for productive exchanges on those occasions.

Back in Berkeley, people who enriched my grad-school experience
include Neha Dave, Julia Kempe, Simone Severini, Lawrence Ip,
Allison Coates, David Molnar, Kris Hildrum, Miriam Walker, and
Shelly Rosenfeld. \ Alex Fabrikant and Boriska Toth are forgiven for
the cruel caricature that they attached to my dissertation talk
announcement, provided they don't try anything like that ever again.
\ The results on one-way communication in Chapter \ref{ADV}\
benefited greatly from conversations with Oded Regev and Iordanis
Kerenidis, while Andrej Bogdanov kindly supplied the explicit
erasure code for Chapter \ref{MLIN}. \ I wrote Chapter \ref{PLS}\ to
answer a question of Christos Papadimitriou.

I did take some actual \ldots\ \textit{courses} at Berkeley, and I'm
grateful to John Kubiatowicz, Stuart Russell, Guido Bacciagaluppi,
Richard Karp, and Satish Rao for not failing me in theirs. \
Ironically, the course that most directly influenced this thesis was
Tom Farber's magnificent short fiction workshop. \ A story I wrote
for that workshop dealt with the problem of transtemporal identity,
which got me thinking about hidden-variable interpretations of
quantum mechanics, which led eventually to the collision lower
bound. \ No one seems to believe me, but it's true.

The students who took my \textquotedblleft Physics, Philosophy,
Pizza\textquotedblright\ course remain one of my greatest
inspirations. \ Though they were mainly undergraduates with liberal
arts backgrounds, they took nothing I said about special relativity
or G\"{o}del's Theorem on faith. \ If I have any confidence today in
my teaching abilities; if I think it possible for students to show
up to class, and to participate eagerly, without the usual
carrot-and-stick of grades and exams; or if I find certain
questions, such as how a superposition over exponentially many
`could-have-beens' can collapse to an `is,' too vertiginous to be
pondered only by nerds like me, then those pizza-eating students are
the reason.

Now comes the part devoted to the mist-enshrouded pre-Berkeley
years. \ My initiation into the wild world of quantum computing
research took place over three summer internships at Bell Labs: the
first with Eric Grosse, the second with Lov Grover, and the third
with Rob Pike. \ I thank all three of them for encouraging me to
pursue my interests, even if the payoff was remote and, in Eric's
case, not even related to why I was hired. \ Needless to say, I take
no responsibility for the subsequent crash of Lucent's stock.

As an undergraduate at Cornell, I was younger than my classmates,
invisible to many of the researchers I admired, and profoundly
unsure of whether I belonged there or had any future in science. \
What made the difference was the unwavering support of one
professor, Bart Selman. \ Busy as he was, Bart listened to my
harebrained ideas about genetic algorithms for SAT or quantum
chess-playing, invited me to give talks, guided me to the right
graduate programs, and generally treated me like a future colleague.
\ As a result, his conviction that I could succeed at research
gradually became my conviction too. \ Outside of research, Christine
Chung, Fion Luo, and my Telluride roommate Jason Stockmann helped to
warm the Ithaca winters, Lydia Fakundiny taught me what an essay is,
and Jerry Abrams provided a much-needed boost.

Turning the clock back further, my earliest research foray was a
paper on hypertext organization, written when I was fifteen and
spending the year at Clarkson University's unique Clarkson School
program. \ Christopher Lynch generously agreed to advise the
project, and offered invaluable help as I clumsily learned how to
write a C program, prove a problem $\mathsf{NP}$-hard, and conduct a
user experiment (one skill I've never needed again!). \ I was elated
to be trading ideas with a wise and experienced researcher, only
months after I'd escaped from the prison-house of high school. \
Later, the same week the rejection letters were arriving from
colleges, I learned that my first paper had been accepted to SIGIR,
the main information retrieval conference. \ I was filled with
boundless gratitude toward the entire scientific community---for
struggling, against the warp of human nature, to judge ideas rather
than the personal backgrounds of their authors. \ Eight years later,
my gratitude and amazement are undiminished.

Above all, I thank Alex Halderman for a friendship that's spanned
twelve years and thousands of miles, remaining as strong today as it
was amidst the \textit{Intellectualis minimi} of Newtown Junior High
School; my brother David for believing in me, and for making me
prouder than he realizes by doing all the things I didn't; and my
parents for twenty-three years of harping, kvelling, chicken noodle
soup, and never doubting for a Planck time that I'd live up to my
potential---even when I couldn't, and can't, share their certainty.
\end{acknowledgements}
\end{frontmatter}

\chapter{\textquotedblleft Aren't You Worried That Quantum Computing Won't Pan
Out?\textquotedblright\label{PROLOGUE}}

\begin{figure}
[ptb]
\begin{center}
\includegraphics[
trim=2.025710in 4.052040in 2.026746in 0.000000in, height=1.7883in,
width=3.0245in
]%
{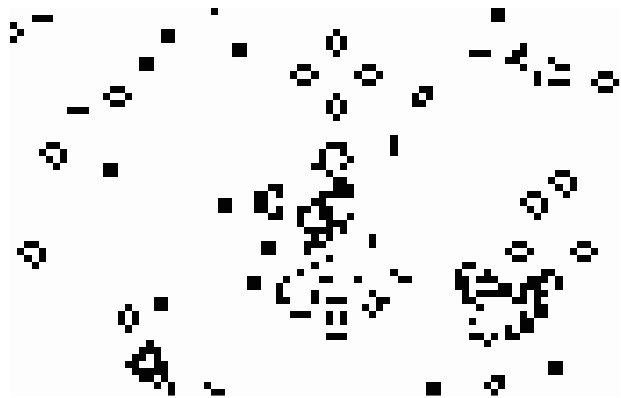}%
\caption[Conway's Game of Life]{In Conway's Game of Life, each cell of a 2D
square grid becomes `dead' or `alive' based on how many of its eight neighbors
were alive in the previous time step. \ A simple rule applied iteratively
leads to complex, unpredictable behavior. \ In what ways is our physical world
similar to Conway's, and in what ways is it different?}%
\label{conwayfig}%
\end{center}
\end{figure}

For a century now, physicists have been telling us strange things:
about twins who age at different rates, particles that look
different when rotated 360$^{\circ}$, a force that is transmitted by
gravitons but is also the curvature of spacetime, a negative-energy
electron sea that pervades empty space, and strangest of all,
\textquotedblleft probability waves\textquotedblright\ that produce
fringes on a screen when you don't look and don't when you do. \ Yet
ever since I learned to program, I suspected that such things were
all \textquotedblleft implementation details\textquotedblright\ in
the source code of Nature, their study only marginally relevant to
forming an accurate picture of reality. \ Physicists, I thought,
would eventually realize that the state of the universe can be
represented by a finite string of bits. \ These bits would be the
\textquotedblleft pixels\textquotedblright\ of space, creating the
illusion of continuity on a large scale much as a computer screen
does. \ As time passed, the bits would be updated according to
simple rules. \ The specific form of these rules was of no great
consequence---since according to the Extended Church-Turing Thesis,
any sufficiently complicated rules could simulate any other rules
with reasonable efficiency.\footnote{Here \textquotedblleft
extended\textquotedblright\ refers to the efficiency requirement,
which was not mentioned in the original Church-Turing Thesis. \
Also, I am simply using the standard terminology, sidestepping the
issue of whether Church and Turing themselves intended to make a
claim about physical reality.} \ So apart from practical
considerations, why worry about Maxwell's equations,\ or Lorentz
invariance, or even mass and energy, if the most fundamental aspects
of our universe already occur in Conway's Game of Life (see Figure
\ref{conwayfig})?%
%TCIMACRO{\FRAME{ftbpFU}{3.0245in}{1.7883in}{0pt}{\Qcb[Conway's Game of
%Life]{In Conway's Game of Life, each cell of a 2D square grid becomes `dead'
%or `alive' based on how many of its eight neighbors were alive in the previous
%time step. \ A simple rule applied iteratively leads to complex, unpredictable
%behavior. \ In what ways is our physical world similar to Conway's, and in
%what ways is it different?}}{\Qlb{conwayfig}}{conwayfig.eps}%
%width 3.0245in;  height 1.7883in;  depth 0pt;  original-width 10.3511in;
%original-height 7.7551in;  cropleft "0.1957";  croptop "1";
%cropright "0.8042";  cropbottom "0.5225";
%filename 'conwayfig.eps';file-properties "XNPEU";}}}%
%BeginExpansion
%EndExpansion

Then I heard about Shor's algorithm \cite{shor} for factoring
integers in polynomial time on a quantum computer. \ Then as now,
many people saw quantum computing as at best a speculative diversion
from the \textquotedblleft real\ work\textquotedblright\ of computer
science. \ Why devote one's research career to a type of computer
that might never see application within one's lifetime, that faces
daunting practical obstacles such as decoherence, and whose most
publicized success to date\ has been the confirmation that, with
high probability, $15=3\times5$ \cite{vsbysc}? \ Ironically, I might
have agreed with this view, had I not taken the Extended
Church-Turing Thesis so seriously as a claim about reality. \ For
Shor's algorithm forces us to accept that, under widely-believed
assumptions, that Thesis conflicts with the experimentally-tested
rules of quantum mechanics as we currently understand them. \ Either
the Extended Church-Turing Thesis is false, or quantum mechanics
must be modified, or the factoring problem is solvable in classical
polynomial time. \ All three possibilities seem like wild, crackpot
speculations---but at least one of them is true!

The above conundrum is what underlies my interest in quantum
computing, far more than any possible application. \ Part of the
reason is that I am neither greedy, nefarious, nor
number-theoretically curious enough ever to have hungered for the
factors of a $600$-digit integer. \ I \textit{do} think that quantum
computers would have benign uses, the most important one being the
simulation of quantum physics and chemistry.\footnote{Followed
closely by Recursive Fourier Sampling, parity in $n/2$\ queries, and
efficiently deciding whether a graph is a scorpion.} \ Also, as
transistors approach the atomic scale, ideas from quantum computing
are likely to become pertinent even for \textit{classical} computer
design. \ But none of this quickens my pulse.

For me, quantum computing matters because it combines two of the great
mysteries bequeathed to us by the twentieth century: the nature of quantum
mechanics, and the ultimate limits of computation. \ It would be astonishing
if such an elemental connection between these mysteries shed no new light on
either of them. \ And indeed, there is already a growing list of examples
\cite{aar:pls,ar,kw}---we will see several of them in this thesis---in which
ideas from quantum computing have led to new results about \textit{classical}
computation. \ This should not be surprising: after all, many celebrated
results in computer science involve only \textit{deterministic} computation,
yet it is hard to imagine how anyone could have proved them had computer
scientists not long ago \textquotedblleft taken randomness
aboard.\textquotedblright\footnote{A few examples are primality testing in
$\mathsf{P}$ \cite{aks}, undirected connectivity in $\mathsf{L}$
\cite{reingold}, and inapproximability of 3-SAT unless $\mathsf{P}%
=\mathsf{NP}$\ \cite{hastad}.}\ \ Likewise, taking quantum mechanics aboard
could lead to a new, more general perspective from which to revisit the
central questions of computational complexity theory.

The other direction, though, is the one that intrigues me even more.
\ In my view, quantum computing has brought us \textit{slightly}
closer to the elusive Beast that devours Bohmians for breakfast,
Copenhagenists for lunch, and a linear combination of many-worlders
and consistent historians for dinner---the Beast that tramples
popularizers, brushes off arXiv preprints like fleas, and snorts at
the word \textquotedblleft decoherence\textquotedblright---the Beast
so fearsome that physicists since Bohr and Heisenberg have tried to
argue it away, as if semantics could banish its unitary jaws and
complex-valued tusks. \ But no, the Beast is there whenever you
aren't paying attention, following all possible paths in
superposition. \ Look, and suddenly the Beast is gone. \ But what
does it\ even mean to look? \ If you're governed by the same
physical laws as everything else, then why don't \textit{you} evolve
in superposition too, perhaps until someone else looks at you and
thereby `collapses' you? \ But then who collapses whom first? \ Or
if you never collapse, then what determines what \textit{you}-you,
rather than the superposition of you's, experience? \ Such is the
riddle of the Beast,\footnote{Philosophers call the riddle of the
Beast the \textquotedblleft measurement problem,\textquotedblright\
which sounds less like something that should cause insomnia and
delirious raving in all who have understood it. \ Basically, the
problem is to reconcile a picture of the world in which ``everything
happens simultaneously'' with the fact that you (or at least I!)
have a sequence of definite experiences.} and it has filled many
with terror and awe.

The contribution of quantum computing, I think, has been to show
that the real nature of the Beast lies in its
\textit{exponentiality}.\ \ It is not just two, three, or a thousand
states held in ghostly superposition that quantum mechanics is
talking about, but an astronomical multitude, and these states could
in principle reveal their presence to us by factoring a
five-thousand-digit number. \ Much more than even Schr\"{o}dinger's
cat\ or the Bell inequalities, this particular discovery ups the
ante---forcing us either to swallow the full quantum brew, or to
stop saying that we believe in it. \ Of course, this is part of the
reason why Richard Feynman \cite{feynman:qc} and David Deutsch
\cite{deutsch:qc} introduced quantum computing in the first place,
and why Deutsch, in his defense of the many-worlds interpretation,\
issues a famous challenge to skeptics \cite[p. 217]{deutsch}: if
parallel universes are not physically real, then \textit{explain how
Shor's algorithm works}.

Unlike Deutsch, here I will not use quantum computing to defend the
many-worlds interpretation, or any of its competitors for that
matter. \ Roughly speaking, I agree with \textit{every}
interpretation of quantum mechanics to the extent that it
acknowledges the Beast's existence, and disagree to the extent that
it claims to have caged the Beast. \ I would adopt the same attitude
in computer science, if instead of freely admitting (for example)\
that $\mathsf{P}$ versus $\mathsf{NP}$ is an open problem,
researchers had split into \textquotedblleft equalist,\textquotedblright%
\ \textquotedblleft unequalist,\textquotedblright\ and \textquotedblleft
undecidabilist\textquotedblright\ schools of interpretation, with others
arguing that the whole problem is meaningless and should therefore be abandoned.

Instead, in this thesis I will show how adopting a computer science
perspective can lead us to \textit{ask better questions}---nontrivial but
answerable questions, which put old mysteries in a new light even when they
fall short of solving them. \ Let me give an example. \ One of the most
contentious questions about quantum mechanics is whether the individual
components of a wavefunction should be thought of as \textquotedblleft really
there\textquotedblright\ or as \textquotedblleft mere
potentialities.\textquotedblright\ \ When we don our computer scientist
goggles, this question morphs into a different one: \textit{what resources are
needed to make a particular component of the wavefunction manifest?}
\ Arguably the two questions are related, since something \textquotedblleft
real\textquotedblright\ ought to take less work to manifest than something
\textquotedblleft potential.\textquotedblright\ \ For example, this thesis
gradually became\ more real as less of it remained to be written.

Concretely, suppose our wavefunction has $2^{n}$\ components, all with equal
amplitude. \ Suppose also that we have a procedure to recognize a particular
component $x$ (i.e., a function $f$ such that $f\left(  x\right)  =1$\ and
$f\left(  y\right)  =0$\ for all $y\neq x$). \ Then how often must we apply
this procedure before we make $x$ manifest; that is, observable with
probability close to $1$? \ Bennett, Bernstein, Brassard, and Vazirani
\cite{bbbv}\ showed that $\sim2^{n/2}$\ applications are necessary, even if
$f$ can be applied to all $2^{n}$ components in superposition. \ Later Grover
\cite{grover}\ showed that $\sim2^{n/2}$ applications are also sufficient.
\ So if we imagine a spectrum with \textquotedblleft really
there\textquotedblright\ ($1$ application)\ on one end, and \textquotedblleft
mere potentiality\textquotedblright\ ($\sim2^{n}$ applications) on the other,
then we have landed somewhere in between: closer to the \textquotedblleft
real\textquotedblright\ end on an absolute scale, but closer to the
\textquotedblleft potential\textquotedblright\ end\ on the polynomial versus
exponential scale that is more natural for computer science.

Of course, we should be wary of drawing grand conclusions from a
single data point. \ So in this thesis, I will imagine a
hypothetical resident of Conway's Game of Life, who arrives in our
physical universe on a computational complexity safari---wanting to
know exactly which intuitions to keep and which to discard\
regarding the limits of efficient computation. \ Many popular
science writers would tell our visitor to throw \textit{all}
classical intuitions out the window, while quantum computing
skeptics would urge retaining them all. \ These positions are
actually two sides of the same coin, since the belief that a quantum
computer would necessitate the first is what generally leads to the
second. \ I will show, however, that neither position is justified.
\ Based on what we know today, there really is a Beast, but it
usually conceals its exponential underbelly.

I'll provide only one example from the thesis here; the rest are
summarized in Chapter \ref{OVER}. \ Suppose we are given a procedure
that computes a two-to-one function $f$,\ and want to find distinct
inputs $x$ and $y$ such that $f\left(  x\right)  =f\left(  y\right)
$. \ In this case, by simply preparing a uniform superposition over
all inputs to $f$, applying the procedure, and then measuring its
result, we can produce a state of the form $\left(  \left\vert
x\right\rangle +\left\vert y\right\rangle \right) /\sqrt{2}$, for
some $x$ and $y$ such that $f\left(  x\right)  =f\left( y\right)  $.
\ The only problem is that if we measure this state, then we see
either $x$ or $y$, but not both. \ The task, in other words, is no
longer to find a needle in a haystack, but just to find two needles
in an otherwise empty barn! \ Nevertheless, the \textit{collision
lower bound} in Chapter \ref{COL}\ will show that, if there are
$2^{n}$ inputs to $f$, then any quantum algorithm for this problem
must apply the procedure\ for $f$ at least $\sim2^{n/5}$\ times. \
Omitting technical details, this lower bound can be interpreted in
at least seven ways:

\begin{enumerate}
\item[(1)] Quantum computers need exponential time even to compute certain
\textit{global} properties of a function, not just local properties such as
whether there is an $x$ with $f\left(  x\right)  =1$.

\item[(2)] Simon's algorithm \cite{simon}, and the period-finding core of
Shor's algorithm \cite{shor}, cannot be generalized to functions with no
periodicity or other special structure.

\item[(3)] Any \textquotedblleft brute-force\textquotedblright\ quantum
algorithm needs exponential time, not just for $\mathsf{NP}$-complete
problems, but for many structured problems such as Graph Isomorphism,
approximating the shortest vector in a lattice, and finding collisions in
cryptographic hash functions.

\item[(4)] It is unlikely that all problems having \textquotedblleft
statistical zero-knowledge proofs\textquotedblright\ can be efficiently solved
on a quantum computer.

\item[(5)] Within the setting of a collision algorithm, the components
$\left\vert x\right\rangle $\ and $\left\vert y\right\rangle $\ in the state
$\left(  \left\vert x\right\rangle +\left\vert y\right\rangle \right)
/\sqrt{2}$\ should be thought of as more \textquotedblleft
potentially\textquotedblright\ than \textquotedblleft
actually\textquotedblright\ there, it being impossible to extract information
about both of them in a reasonable amount of time.

\item[(6)] The ability to map $\left\vert x\right\rangle $\ to $\left\vert
f\left(  x\right)  \right\rangle $, \textquotedblleft
uncomputing\textquotedblright\ $x$ in the process, can be exponentially more
powerful than the ability to map $\left\vert x\right\rangle $\ to $\left\vert
x\right\rangle \left\vert f\left(  x\right)  \right\rangle $.

\item[(7)] In hidden-variable interpretations of quantum mechanics, the
ability to sample the entire history of a hidden variable would yield even
more power than standard quantum computing.
\end{enumerate}

Interpretations (5), (6), and (7) are examples of what I\ mean by putting old
mysteries in a new light. \ We are not brought face-to-face with the Beast,
but at least we have fresh footprints and droppings.

Well then. \ \textit{Am} I worried that quantum computing won't pan out? \ My
usual answer is that I'd be \textit{thrilled} to know it will \textit{never}
pan out, since this would entail the discovery of a lifetime, that quantum
mechanics is false. \ But this is not what the questioner has in mind. \ What
if quantum mechanics holds up, but building a useful quantum computer turns
out to be so difficult and expensive that the world ends before anyone
succeeds? \ The questioner is usually a classical theoretical computer
scientist, someone who is not known to worry excessively that the world will
end before $\log\log n$\ exceeds $10$. \ Still, it would be nice to see
nontrivial quantum computers in my lifetime, and while I'm cautiously
optimistic, I'll admit to being \textit{slightly} worried that I won't. \ But
when faced with the evidence that one was born into a universe profoundly
unlike Conway's---indeed, that one is living one's life on the back of a
mysterious, exponential Beast comprising everything that ever could have
happened---what is one to do? \ \textquotedblleft Move right along\ldots
\ nothing to see here\ldots\textquotedblright

\chapter{Overview\label{OVER}}

\begin{quotation}
\textquotedblleft Let a computer smear---with the right kind of quantum
randomness---and you create, in effect, a `parallel' machine with an
astronomical number of processors \ldots\ All you have to do is be sure that
when you collapse the system, you choose the version that happened to find the
needle in the mathematical haystack.\textquotedblright

---From \textit{Quarantine} \cite{egan}, a 1992 science-fiction novel by Greg Egan
\end{quotation}

Many of the deepest discoveries of science are \textit{limitations}: for
example, no superluminal signalling, no perpetual-motion machines, and no
complete axiomatization for arithmetic. \ This thesis is broadly concerned
with limitations on what can efficiently be computed in the physical world.
\ The word \textquotedblleft quantum\textquotedblright\ is absent from the
title, in order to emphasize that the focus on quantum computing is not an
arbitrary choice, but rather an inevitable result of taking our current
physical theories seriously. \ The technical contributions of the thesis are
divided into two parts, according to whether they accept the quantum computing
model as given and study its fundamental limitations; or question, defend, or
go beyond that model in some way. \ Before launching into a detailed overview
of the contributions, let me make some preliminary remarks.

Since the early twentieth century, two
communities---physicists\footnote{As in Saul Steinberg's famous
\textit{New Yorker} world map, in which 9$^{th}$ Avenue and the
Hudson River take up more space than Japan and China, from my
perspective chemists, engineers, and even mathematicians who know
what a gauge field is are all \textquotedblleft
physicists.\textquotedblright} and computer scientists---have been
asking some of the deepest questions ever asked in almost total
intellectual isolation from each other. \ The great joy of quantum
computing research is that it brings these communities together. \
The trouble was initially that, although each community would nod
politely during the other's talks, eventually it would come out that
the physicists thought $\mathsf{NP}$\ stood for \textquotedblleft
Non Polynomial,\textquotedblright\ and the computer scientists had
no idea what a Hamiltonian was. \ Thankfully, the situation has
improved a lot---but my hope is that it improves further still, to
the point where computer scientists have internalized the problems
faced by physics and vice versa. \ For this reason, I have worked
hard to make the thesis as accessible as possible to both
communities. \ Thus, Chapter \ref{COMPLEXITY}\ provides a
\textquotedblleft complexity theory cheat sheet\textquotedblright\
that defines $\mathsf{NP}$, $\mathsf{P/poly}$, $\mathsf{AM}$, and
other computational complexity classes that appear in the thesis;
and that explains oracles and other important concepts. \ Then
Chapter \ref{QUANTUM}\ presents the quantum model of computation
with no reference to the underlying physics, before moving on to
fancier\ notions such as density matrices, trace distance, and
separability. \ Neither chapter is a rigorous introduction to its
subject;\ for that there are fine textbooks---such as
Papadimitriou's \textit{Computational Complexity} \cite{papa:book}\
and Nielsen and Chuang's \textit{Quantum Computation and Quantum
Information} \cite{nc}---as well as course lecture notes available
on the web. \ Depending on your background, you might want to skip
to Chapters \ref{COMPLEXITY}\ or \ref{QUANTUM}\ before continuing
any further, or you might want to skip past these chapters entirely.

Even the most irredeemably classical reader should\ take heart: of
the $103$ proofs in the thesis, $66$ do not contain a single ket\
symbol.\footnote{To be honest, a few of those do contain density
matrices---or the \textit{theorem} contains ket symbols, but not the
\textit{proof}.} \ Many of the proofs can be understood by simply
accepting certain facts about quantum computing on faith, such as
Ambainis's\footnote{Style manuals disagree about whether
\textit{Ambainis'} or \textit{Ambainis's} is preferable, but one
referee asked me to follow the latter rule with the following
deadpan remark: \textquotedblleft Exceptions to the rule generally
involve religiously
significant individuals, e.g., `Jesus' lower-bound method.' \textquotedblright%
} adversary theorem \cite{ambainis}\ or Beals et al.'s polynomial lemma
\cite{bbcmw}. \ On the other hand, one does run the risk that after one
understands the proofs, ket symbols will seem less frightening than before.

The results in the thesis have all previously appeared in published papers or
preprints
\cite{aar:rev,aar:col,aar:cer,aar:rfs,aar:isl,aar:adv,aar:pls,aar:mlin,aar:qchv,aa}%
, with the exception of the quantum computing based proof that
$\mathsf{PP}$ is closed under intersection in Chapter \ref{POST}. \
I thank Andris Ambainis for allowing me to include our joint results
from \cite{aa} on quantum search of spatial regions. \ Results of
mine that do \textit{not} appear in the
thesis include those on Boolean function query properties \cite{aar:bf}%
,\ stabilizer circuits \cite{ag}\ (joint work with Daniel Gottesman), and
agreement complexity \cite{aar:agr}.

In writing the thesis, one of the toughest choices I faced was whether to
refer to myself as `I' or `we.' \ Sometimes a personal voice seemed more
appropriate, and sometimes the Voice of Scientific Truth, but I wanted to be
consistent. \ Readers can decide whether I chose humbly or arrogantly.

\section{Limitations of Quantum Computers\label{OVERLQC}}

Part \ref{LQC} studies the fundamental limitations of quantum
computers within the usual model for them. \ With the exception of
Chapter \ref{ADV}\ on quantum advice, the contributions of Part
\ref{LQC} all deal with \textit{black-box} or \textit{query}
complexity, meaning that one counts only the number of queries to an
\textquotedblleft oracle,\textquotedblright\ not the number of
computational steps. \ Of course, the queries can be made in quantum
superposition. \ In Chapter \ref{INTROLQC}, I explain the quantum
black-box model, then offer a detailed justification for its
relevance to understanding the limits of quantum computers. \ Some
computer scientists say that black-box results should not be taken
too seriously; but I argue that, within quantum computing, they are
not taken seriously enough.

What follows is a (relatively) nontechnical overview of Chapters \ref{COL}\ to
\ref{ADV}, which contain the results of Part \ref{LQC}. \ Afterwards, Chapter
\ref{SUMLQC}\ summarizes the conceptual lessons that I\ believe can be drawn
from those results.

\subsection{The Collision Problem\label{OVERCOL}}

Chapter \ref{COL}\ presents my lower bound on the quantum query complexity of
the collision problem.\ \ Given a function $X$ from $\left\{  1,\ldots
,n\right\}  $ to $\left\{  1,\ldots,n\right\}  $ (where $n$ is even), the
collision problem is to decide whether $X$ is one-to-one or two-to-one,
promised that one of these is the case. \ Here the only way to learn about $X$
is to call a procedure that computes $X\left(  i\right)  $\ given $i$.
\ Clearly, any deterministic classical algorithm needs to call the procedure
$n/2+1$\ times to solve the problem. \ On the other hand, a randomized
algorithm can exploit the \textquotedblleft birthday paradox\textquotedblright%
: only $23$\ people have to enter a room before there's a $50\%$\
chance that two of them share the same birthday, since what matters
is the number of \textit{pairs} of people. \ Similarly, if $X$ is
two-to-one, and an algorithm queries $X$\ at $\sqrt{n}$ uniform
random locations,\ then with constant probability it will find two
locations $i\neq j$\ such that $X\left( i\right)  =X\left(  j\right)
$, thereby establishing that $X$ is two-to-one. \ This bound is
easily seen to be tight,\ meaning that the bounded-error randomized
query complexity of the collision problem is $\Theta\left(
\sqrt{n}\right)  $.

What about the \textit{quantum} complexity? \ In 1997, Brassard, H\o yer, and
Tapp \cite{bht} gave a quantum algorithm that uses only $O\left(
n^{1/3}\right)  $\ queries. \ The algorithm is simple to describe: in the
first phase, query $X$ classically at $n^{1/3}$ randomly chosen locations.
\ In the second phase, choose $n^{2/3}$ random locations, and run Grover's
algorithm on those locations, considering each location $i$\ as
\textquotedblleft marked\textquotedblright\ if $X\left(  i\right)  =X\left(
j\right)  $\ for some $j$ that was queried in the first phase. \ Notice that
both phases use order $n^{1/3}=\sqrt{n^{2/3}}$\ queries, and that the total
number of comparisons is $n^{2/3}n^{1/3}=n$. \ So, like its randomized
counterpart, the quantum algorithm finds a collision with constant probability
if $X$ is two-to-one.

What I show in Chapter \ref{COL}\ is that \textit{any} quantum algorithm for
the collision problem needs $\Omega\left(  n^{1/5}\right)  $\ queries.
\ Previously, no lower bound better than the trivial $\Omega\left(  1\right)
$ was known. \ I also show a lower bound of $\Omega\left(  n^{1/7}\right)
$\ for the following \textit{set comparison problem}: given oracle access to
injective functions $X:\left\{  1,\ldots,n\right\}  \rightarrow\left\{
1,\ldots,2n\right\}  $\ and $Y:\left\{  1,\ldots,n\right\}  \rightarrow
\left\{  1,\ldots,2n\right\}  $, decide whether%
\[
\left\{  X\left(  1\right)  ,\ldots,X\left(  n\right)  ,Y\left(  1\right)
,\ldots,Y\left(  n\right)  \right\}
\]
has at least $1.1n$\ elements or exactly $n$\ elements, promised that one of
these is the case. \ The set comparison problem is similar to the collision
problem, except that it lacks permutation symmetry, making it harder to prove
a lower bound. \ My results for these problems have been improved, simplified,
and generalized by Shi \cite{shi}, Kutin \cite{kutin}, Ambainis
\cite{ambainis}, and Midrijanis \cite{midrijanis}.

The implications of these results were already discussed in Chapter
\ref{PROLOGUE}: for example, they demonstrate that a \textquotedblleft
brute-force\textquotedblright\ approach will never yield efficient quantum
algorithms for the Graph Isomorphism, Approximate Shortest Vector, or
Nonabelian Hidden Subgroup problems; suggest that there could be cryptographic
hash functions secure against quantum attack; and imply that there exists\ an
oracle relative to which $\mathsf{SZK}\not \subset \mathsf{BQP}$, where
$\mathsf{SZK}$ is the class of problems having statistical zero-knowledge
proof protocols, and $\mathsf{BQP}$ is quantum polynomial time.

Both the original lower bounds and the subsequent improvements are based on
the \textit{polynomial method}, which was introduced by Nisan and Szegedy
\cite{ns},\ and first used to prove quantum lower bounds by Beals, Buhrman,
Cleve, Mosca, and de Wolf \cite{bbcmw}. \ In that method, given a quantum
algorithm that makes $T$ queries to an oracle $X$, we first represent the
algorithm's acceptance probability by a multilinear polynomial $p\left(
X\right)  $\ of degree at most $2T$. \ We then use results from a
well-developed area of mathematics called \textit{approximation theory} to
show a lower bound on the degree of $p$. \ This in turn implies a lower bound
on $T$.

In order to apply the polynomial method to the collision problem, first I
extend the collision problem's domain from one-to-one and two-to-one functions
to $g$-to-one functions for larger values of $g$. \ Next I replace the
multivariate polynomial $p\left(  X\right)  $ by a related \textit{univariate}
polynomial $q\left(  g\right)  $ whose degree is easier to lower-bound. \ The
latter step is the real \textquotedblleft magic\textquotedblright\ of the
proof; I\ still have no good intuitive explanation for why it works.

The polynomial method is one of two principal methods that we have for proving
lower bounds on quantum query complexity. \ The other is Ambainis's
\textit{quantum adversary method} \cite{ambainis}, which can be seen as a
far-reaching generalization of the \textquotedblleft hybrid
argument\textquotedblright\ that Bennett, Bernstein, Brassard, and Vazirani
\cite{bbbv} introduced in 1994 to show that a quantum computer needs
$\Omega\left(  \sqrt{n}\right)  $\ queries to search an unordered database of
size $n$\ for a marked item. \ In the adversary method, we consider a
bipartite quantum state, in which one part consists of a superposition over
possible inputs, and the other part consists of a quantum algorithm's work
space. \ We then upper-bound how much the \textit{entanglement} between the
two parts can increase as the result of a single query. \ This in turn implies
a lower bound on the number of queries, since the two parts must be highly
entangled by the end. \ The adversary method is more intrinsically
\textquotedblleft quantum\textquotedblright\ than the polynomial method; and
as Ambainis \cite{ambainis} showed, it is also applicable to a wider range of
problems, including those (such as game-tree search) that lack permutation
symmetry. \ Ambainis even gave problems for which the adversary method
\textit{provably} yields a better lower bound than the polynomial method
\cite{ambainis:deg}. \ It is ironic, then, that Ambainis's original goal in
developing the adversary method was to prove a lower bound for the collision
problem; and in this one instance, the polynomial method succeeded while the
adversary method failed.

\subsection{Local Search\label{OVERPLS}}

In Chapters \ref{PLS}, \ref{CER}, and \ref{RFS}, however, the adversary method
gets its revenge. \ Chapter \ref{PLS} deals with the \textit{local search}
problem: given an undirected graph $G=\left(  V,E\right)  $ and a black-box
function $f:V\rightarrow\mathbb{Z}$, find a local minimum of $f$---that is, a
vertex $v$\ such that $f\left(  v\right)  \leq f\left(  w\right)  $\ for all
neighbors $w$ of $v$. \ The graph $G$ is known in advance, so the complexity
measure is just the number of queries to $f$. \ This problem is central for
understanding the performance of the quantum adiabatic algorithm, as well as
classical algorithms such as simulated annealing. \ If $G$\ is the Boolean
hypercube $\left\{  0,1\right\}  ^{n}$, then previously Llewellyn, Tovey, and
Trick \cite{ltt} had shown that any deterministic algorithm needs
$\Omega\left(  2^{n}/\sqrt{n}\right)  $ queries to find a local minimum; and
Aldous \cite{aldous}\ had shown that any randomized algorithm needs
$2^{n/2-o\left(  n\right)  }$\ queries. \ What I show is that any quantum
algorithm needs $\Omega\left(  2^{n/4}/n\right)  $ queries. \ This is the
first nontrivial quantum lower bound for any local search problem; and it
implies that the complexity class $\mathsf{PLS}$\ (or \textquotedblleft
Polynomial Local Search\textquotedblright), defined by Johnson, Papadimitriou,
and Yannakakis \cite{jpy}, is not in quantum polynomial time relative to an oracle.

What will be more surprising to classical computer scientists is
that my proof technique, based on the quantum adversary method, also
yields new \textit{classical} lower bounds for local search. \ In
particular, I prove a classical analogue of Ambainis's quantum
adversary theorem, and show that it implies randomized lower bounds
up to quadratically better than the corresponding quantum lower
bounds. \ I then apply my theorem to show that any randomized
algorithm needs $\Omega\left( 2^{n/2}/n^{2}\right)  $\ queries to
find a local minimum of a function $f:\left\{  0,1\right\}  ^{n}%
\rightarrow\mathbb{Z}$. \ Not only does this improve on Aldous's
$2^{n/2-o\left(  n\right)  }$\ lower bound, bringing us closer to the known
upper bound of $O\left(  2^{n/2}\sqrt{n}\right)  $; but it does so in a
simpler way that does not depend on random walk analysis. \ In addition, I
show the first randomized \textit{or} quantum lower bounds for finding a local
minimum on a cube of constant dimension $3$ or greater. \ Along with recent
work by Bar-Yossef, Jayram, and Kerenidis \cite{bjk}\ and by Aharonov and
Regev \cite{ar}, these results provide one of the earliest examples of how
quantum ideas can help to resolve classical open problems. \ As I\ will
discuss in Chapter \ref{PLS}, my results on local search have subsequently
been improved by Santha and Szegedy \cite{ss}\ and by Ambainis
\cite{ambainis:pls}.

\subsection{Quantum Certificate Complexity\label{OVERCER}}

Chapters \ref{CER} and \ref{RFS} continue to explore the power of
Ambainis's lower bound method and the limitations of quantum
computers. \ Chapter \ref{CER}\ is inspired by the following theorem
of Beals et al.\ \cite{bbcmw}: if $f:\left\{  0,1\right\}
^{n}\rightarrow\left\{  0,1\right\}  $ is a total Boolean function,
then $\operatorname*{D}\left(  f\right)  =O\left(
\operatorname*{Q}_{2}\left(  f\right)  ^{6}\right)  $, where
$\operatorname*{D}\left(  f\right)  $\ is the deterministic
classical query complexity of $f$, and $\operatorname*{Q}_{2}\left(
f\right)  $\ is the bounded-error quantum query
complexity.\footnote{The subscript `$2$' means that the error is
two-sided.} \ This theorem is noteworthy for two reasons: first,
because it gives a case where quantum computers provide only a
polynomial speedup, in contrast to the exponential speedup of Shor's
algorithm; and second, because the exponent of $6$ seems so
arbitrary. \ The largest separation we know of is quadratic, and is
achieved by the $\operatorname*{OR}$\ function on $n$ bits:
$\operatorname*{D}\left( \operatorname*{OR}\right)  =n$, but
$\operatorname*{Q}_{2}\left( \operatorname*{OR}\right)  =O\left(
\sqrt{n}\right)  $\ because of Grover's search algorithm. \ It is a
longstanding open question whether this separation is optimal. \ In
Chapter \ref{CER}, I make the best progress so far toward
showing that it is. \ In particular I prove that%
\[
\operatorname*{R}\nolimits_{2}\left(  f\right)  =O\left(  \operatorname*{Q}%
\nolimits_{2}\left(  f\right)  ^{2}\operatorname*{Q}\nolimits_{0}\left(
f\right)  \log n\right)
\]
for all total Boolean functions $f:\left\{  0,1\right\}  ^{n}\rightarrow
\left\{  0,1\right\}  $. \ Here $\operatorname*{R}\nolimits_{2}\left(
f\right)  $\ is the bounded-error randomized query complexity of $f$, and
$\operatorname*{Q}\nolimits_{0}\left(  f\right)  $\ is the zero-error quantum
query complexity. \ To prove this result, I introduce two new query complexity
measures of independent interest: the \textit{randomized certificate
complexity} $\operatorname*{RC}\left(  f\right)  $ and the \textit{quantum
certificate complexity} $\operatorname*{QC}\left(  f\right)  $. \ Using
Ambainis's adversary method together with the minimax theorem, I relate these
measures \textit{exactly} to one another, showing that $\operatorname*{RC}%
\left(  f\right)  =\Theta\left(  \operatorname*{QC}\left(  f\right)
^{2}\right)  $.\ \ Then, using the polynomial method, I show that
$\operatorname*{R}\nolimits_{2}\left(  f\right)  =O\left(  \operatorname*{RC}%
\left(  f\right)  \operatorname*{Q}\nolimits_{0}\left(  f\right)  \log
n\right)  $ for all total Boolean $f$, which implies the above result since
$\operatorname*{QC}\left(  f\right)  \leq\operatorname*{Q}\nolimits_{2}\left(
f\right)  $. \ Chapter \ref{CER}\ contains several other results of interest
to researchers studying query complexity, such as a superquadratic gap between
$\operatorname*{QC}\left(  f\right)  $\ and the \textquotedblleft
ordinary\textquotedblright\ certificate complexity $\operatorname*{C}\left(
f\right)  $. \ But the main message is the unexpected versatility of our
quantum lower bound methods: we see the first use of the adversary method to
prove something about \textit{all} total functions, not just a specific
function; the first use of both the adversary and the polynomial methods at
different points in a proof; and the first combination of the adversary method
with a linear programming duality argument.

\subsection{The Need to Uncompute\label{OVERRFS}}

Next, Chapter \ref{RFS} illustrates how \textquotedblleft the need
to uncompute\textquotedblright\ imposes a fundamental limit on
efficient quantum computation. \ Like a classical algorithm, a
quantum algorithm can solve a problem recursively by calling itself
as a subroutine. \ When this is done, though, the quantum algorithm
typically needs to call itself \textit{twice} for each subproblem to
be solved. \ The second call's purpose is to \textquotedblleft
uncompute\textquotedblright\ garbage left over by the first call,
and thereby enable interference between different branches of the
computation. \ In a seminal paper, Bennett \cite{bennett}\
argued\footnote{Bennett's paper dealt with \textit{classical}
reversible computation, but this comment applies equally well to
quantum computation.} that uncomputation increases an algorithm's
running time by only a factor of $2$. \ Yet in the recursive
setting, the increase is by a factor of $2^{d}$, where $d$ is the
depth of recursion. \ Is there any way to avoid this exponential
blowup?

To make the question more concrete, Chapter \ref{RFS} focuses on the recursive
Fourier sampling problem of Bernstein and Vazirani \cite{bv}.\ \ This is a
problem that involves $d$ levels of recursion, and that takes a Boolean
function $g$ as a parameter. \ What Bernstein and Vazirani showed is that for
some choices of $g$, any classical randomized algorithm needs $n^{\Omega
\left(  d\right)  }$\ queries to solve the problem. \ By contrast, $2^{d}%
$\ queries always suffice for a quantum algorithm. \ The question I
ask is whether a quantum algorithm could get by with \textit{fewer}
than $2^{\Omega\left(  d\right)  }$\ queries, even while the
classical complexity remains large. \ I show that the answer is no:
for every $g$, either Ambainis's adversary method yields a
$2^{\Omega\left(  d\right)  }$\ lower bound\ on the quantum query
complexity, or else the classical and quantum query complexities\
are both $1$. \ The lower bound proof introduces a new parameter of
Boolean functions called the \textquotedblleft nonparity
coefficient,\textquotedblright\ which might be of independent
interest.

\subsection{Limitations of Quantum Advice\label{OVERADV}}

Chapter \ref{ADV} broadens the scope of Part \ref{LQC}, to include the
limitations of quantum computers equipped with \textquotedblleft quantum
advice states.\textquotedblright\ \ Ordinarily, we assume that a quantum
computer starts out in the standard \textquotedblleft all-$0$%
\textquotedblright\ state, $\left\vert 0\cdots0\right\rangle $. \ But it is
perfectly sensible to drop that assumption, and consider the effects of other
initial states. \ Most of the work doing so has concentrated on whether
universal quantum computing is still possible with highly mixed initial states
(see \cite{asv,sv:mixed} for example). \ But an equally interesting question
is whether there are states that could take exponential time to prepare, but
that would carry us far beyond the complexity-theoretic confines of
$\mathsf{BQP}$ were they given to us by a wizard. \ For even if quantum
mechanics is universally valid, we do not \textit{really} know whether such
states exist in Nature!

Let $\mathsf{BQP/qpoly}$\ be the class of problems solvable in
quantum polynomial time, with the help of a polynomial-size
\textquotedblleft quantum advice state\textquotedblright\
$\left\vert \psi_{n}\right\rangle $ that depends only on the input
length $n$ but that can otherwise be arbitrary. \ Then the question
is whether $\mathsf{BQP/poly}=\mathsf{BQP/qpoly}$, where
$\mathsf{BQP/poly}$\ is the class of the problems solvable in
quantum polynomial time using a polynomial-size \textit{classical}
advice string.\footnote{For clearly $\mathsf{BQP/poly}$\ and
$\mathsf{BQP/qpoly}$\ both contain uncomputable problems not in $\mathsf{BQP}%
$, such as whether the $n^{th}$\ Turing machine halts.} \ As usual, we could
try to prove an oracle separation. \ But why can't we show that quantum advice
is more powerful than classical advice, with \textit{no} oracle? \ Also, could
quantum advice be used (for example) to solve $\mathsf{NP}$-complete problems
in polynomial time?

The results in Chapter \ref{ADV}\ place strong limitations on the
power of quantum advice. \ First, I show that $\mathsf{BQP/qpoly}$
is contained in a classical complexity class called
$\mathsf{PP/poly}$. \ This means (roughly) that quantum advice can
always be replaced by classical advice, provided we're willing to
use exponentially more computation time. \ It also means that we
could not prove $\mathsf{BQP/poly}\neq\mathsf{BQP/qpoly}$ without
showing that $\mathsf{PP}$\ does not have polynomial-size circuits,
which is believed to be an extraordinarily hard problem. \ To prove
this result, I imagine that the advice state $\left\vert
\psi_{n}\right\rangle $\ is sent to the $\mathsf{BQP/qpoly}$\
machine by a benevolent \textquotedblleft
advisor,\textquotedblright\ through a one-way quantum communication
channel. \ I then give a novel protocol for simulating that quantum
channel using a classical channel. \ Besides showing that
$\mathsf{BQP/qpoly}\subseteq \mathsf{PP/poly}$, the simulation
protocol also implies that for all Boolean functions $f:\left\{
0,1\right\}  ^{n}\times\left\{  0,1\right\} ^{m}\rightarrow\left\{
0,1\right\}  $\ (partial or total), we have
$\operatorname*{D}^{1}\left(  f\right)  =O\left(  m\operatorname*{Q}_{2}%
^{1}\left(  f\right)  \log\operatorname*{Q}_{2}^{1}\left(  f\right)  \right)
$,\ where $\operatorname*{D}^{1}\left(  f\right)  $\ is the deterministic
one-way communication complexity of $f$, and $\operatorname*{Q}_{2}^{1}\left(
f\right)  $\ is the bounded-error quantum one-way communication complexity.
\ This can be considered a generalization of the \textquotedblleft dense
quantum coding\textquotedblright\ lower bound due to Ambainis, Nayak, Ta-Shma,
and Vazirani \cite{antv}.

The second result in Chapter \ref{ADV}\ is that there exists an
oracle relative to which $\mathsf{NP}\not \subset
\mathsf{BQP/qpoly}$. \ This extends the result of Bennett et al.\
\cite{bbbv}\ that there exists an oracle relative to which
$\mathsf{NP}\not \subset \mathsf{BQP}$, to handle quantum advice. \
Intuitively, even though the quantum state $\left\vert
\psi_{n}\right\rangle
$ could in some sense encode the solutions to exponentially many $\mathsf{NP}%
$\ search problems, only a miniscule fraction of that information could be
extracted by measuring the advice, at least in the black-box setting that we
understand today.

The proof of the oracle separation relies on another result of
independent interest: a \textit{direct product theorem} for quantum
search. \ This theorem says that given an unordered database with
$n$ items, $k$ of which are marked, any quantum algorithm that makes
$o\left(  \sqrt{n}\right)  $ queries\footnote{Subsequently Klauck,
\v{S}palek, and de Wolf \cite{ksw} improved this to $o\left(
\sqrt{nk}\right) $\ queries, which is tight.} has probability at
most $2^{-\Omega\left(  k\right)  }$\ of finding all $k$ of the
marked items.
\ In other words, there are no \textquotedblleft magical\textquotedblright%
\ correlations by which success in finding one marked item leads to success in
finding the others. \ This might seem intuitively obvious, but it does not
follow from the $\sqrt{n}$\ lower bound for Grover search,\ or any other
previous quantum lower bound for that matter. \ Previously, Klauck
\cite{klauck:ts} had given an incorrect proof of a direct product theorem,
based on Bennett et al.'s hybrid method. \ I give the first correct proof by
using the polynomial method, together with an inequality dealing with higher
derivatives of polynomials due to V. A. Markov, the younger brother of A. A. Markov.

The third result in Chapter \ref{ADV} is a new \textit{trace distance method}
for proving lower bounds on quantum one-way communication complexity. \ Using
this method, I obtain optimal quantum lower bounds for two problems of
Ambainis, for which no nontrivial lower bounds were previously known even for
classical randomized protocols.

\section{Models and Reality\label{OVERMAR}}

This thesis is concerned with the limits of efficient computation in
Nature. \ It is not obvious that these coincide with the limits of
the quantum computing model. \ Thus, Part \ref{MAR}\ studies the
relationship of the quantum computing model to physical reality. \
Of course, this is too grand a topic for any thesis,\ even a thesis
as long as this one. \ I therefore focus on three questions that
particularly interest me. \ First, how should we understand the
arguments of \textquotedblleft extreme\textquotedblright\ skeptics,
that quantum computing is impossible not only in practice but also
in principle? \ Second, what are the implications for quantum
computing if we recognize that the speed of light is finite, and
that according to widely-accepted principles, a bounded region of
space can store only a finite amount of information? \ And third,
are there reasonable changes to the quantum computing model that
make it even more powerful, and if so, how much more powerful do
they make it? \ Chapters \ref{SKEP} to\ \ref{QCHV}\ address these
questions from various angles; then Chapter \ref{SUMMAR}\
summarizes.

\subsection{Skepticism of Quantum Computing\label{OVERSKEP}}

Chapter \ref{SKEP} examines the arguments of skeptics who think that
large-scale quantum computing is impossible for a fundamental
physical reason. \ I first briefly consider the arguments of Leonid
Levin\ and other computer scientists, that quantum computing is
analogous to \textquotedblleft extravagant\textquotedblright\ models
of computation such as unit-cost arithmetic, and should be rejected
on essentially the same grounds. \ My response emphasizes the need
to grapple with the actual evidence for quantum mechanics, and to
propose an alternative picture of the world that is compatible with
that evidence but in which quantum computing is impossible. \ The
bulk of the chapter, though, deals with Stephen Wolfram's \textit{A
New Kind of Science} \cite{wolfram}, and in particular with one of
that book's most surprising claims: that a deterministic
cellular-automaton picture of the world is compatible with the
so-called \textit{Bell inequality violations} demonstrating the
effects of quantum entanglement. \ To achieve compatibility, Wolfram
posits \textquotedblleft long-range threads\textquotedblright\
between spacelike-separated points. \ I explain in detail why this
thread proposal violates Wolfram's own desiderata of relativistic
and causal invariance. \ Nothing in Chapter \ref{SKEP}\ is very
original technically, but it seems worthwhile to spell out what a
scientific argument against quantum computing would have to
accomplish, and why the existing arguments fail.

\subsection{Complexity Theory of Quantum States\label{OVERMLIN}}

Chapter \ref{MLIN} continues the train of thought begun in Chapter\ \ref{SKEP}%
, except that now the focus is more technical. \ I search for a natural
\textit{Sure/Shor separator}: a set of quantum states that can account for all
experiments performed to date, but that does not contain the states arising in
Shor's factoring algorithm. \ In my view, quantum computing skeptics would
strengthen their case by proposing specific examples of Sure/Shor separators,
since they could then offer testable hypotheses about \textit{where} the
assumptions of the quantum computing model break down (if not \textit{how}
they break down). \ So why am I doing the skeptics' work for them? \ Several
people have wrongly inferred from this that I too am a skeptic! \ My goal,
rather, is to illustrate what a scientific debate about the possibility of
quantum computing might look like.

Most of Chapter \ref{MLIN}\ deals with a candidate Sure/Shor separator that I
call \textit{tree states}. \ Any $n$-qubit pure state $\left\vert \psi
_{n}\right\rangle $ can be represented by a tree, in which each leaf is
labeled by $\left\vert 0\right\rangle $\ or $\left\vert 1\right\rangle $, and
each non-leaf vertex is labeled by either a linear combination or a tensor
product of its subtrees. \ Then the \textit{tree size} of $\left\vert \psi
_{n}\right\rangle $\ is just the minimum number of vertices in such a tree,
and a \textquotedblleft tree state\textquotedblright\ is an infinite family of
states whose tree size is bounded by a polynomial in $n$. \ The idea is to
keep a central axiom of quantum mechanics---that if $\left\vert \psi
\right\rangle $\ and $\left\vert \varphi\right\rangle $ are possible states,
so are $\left\vert \psi\right\rangle \otimes\left\vert \varphi\right\rangle $
and $\alpha\left\vert \psi\right\rangle +\beta\left\vert \varphi\right\rangle
$---but to limit oneself to polynomially many applications of the axiom.

The main results are \textit{superpolynomial lower bounds} on tree size for
explicit families of quantum states. \ Using a recent lower bound on
multilinear formula size due to Raz \cite{raz,raz:nc2}, I show that many
states arising in quantum error correction (for example, states based on
binary linear erasure codes) have tree size $n^{\Omega\left(  \log n\right)
}$. \ I\ show the same for the states arising in Shor's algorithm, assuming a
number-theoretic conjecture. \ Therefore, I argue, by demonstrating such
states in the lab on a large number of qubits, experimentalists could
weaken\footnote{Since tree size is an asymptotic notion (and for other reasons
discussed in Chapter \ref{MLIN}), strictly speaking experimentalists could
never refute the hypothesis---just push it beyond all bounds of plausibility.}
the hypothesis that all states in Nature are tree states.

Unfortunately, while I conjecture that the actual tree sizes are exponential,
Raz's method is currently only able to show lower bounds of the form
$n^{\Omega\left(  \log n\right)  }$. \ On the other hand, I do show
exponential lower bounds under a restriction, called \textquotedblleft
manifest orthogonality,\textquotedblright\ on the allowed linear combinations
of states.

More broadly, Chapter \ref{MLIN} develops a complexity classification of
quantum states, and---treating that classification as a subject in its own
right---proves many basic results about it. \ To give a few examples: if a
quantum computer is restricted to being in a tree state at every time step,
then it can be simulated in the third level of polynomial hierarchy
$\mathsf{PH}$. \ A random state cannot even be approximated by a state with
subexponential tree size. \ Any \textquotedblleft orthogonal tree
state\textquotedblright\ can be prepared by a polynomial-size quantum circuit.
\ Collapses of quantum state classes would imply collapses of ordinary
complexity classes, and vice versa. \ Many of these results involve unexpected
connections between quantum computing and classical circuit complexity. \ For
this reason, I think that the \textquotedblleft complexity theory of quantum
states\textquotedblright\ has an intrinsic computer-science motivation,
besides its possible role in making debates about quantum mechanics' range of
validity less philosophical and more scientific.

\subsection{Quantum Search of Spatial Regions}

A basic result in classical computer science says that Turing machines are
polynomially equivalent to random-access machines. \ In other words, we can
ignore the fact that the speed of light is finite for complexity purposes, so
long as we only care about polynomial equivalence. \ It is easy to see that
the same is true for quantum computing. \ Yet one of the two main quantum
algorithms, Grover's algorithm, provides only a polynomial
speedup.\footnote{If Grover's algorithm is applied to a combinatorial search
space of size $2^{n}$, then the speedup is by a factor of $2^{n/2}$---but in
this case the speedup is only conjectured, not proven.} \ So, does this
speedup disappear if we consider relativity as well as quantum mechanics?

More concretely, suppose a \textquotedblleft quantum robot\textquotedblright%
\ is searching a 2-D grid of size $\sqrt{n}\times\sqrt{n}$\ for a
single marked item. \ The robot can enter a superposition of grid
locations, but moving from one location to an adjacent one takes one
time step.\ \ How many steps are needed to find the marked item? \
If Grover's algorithm is implemented na\"{\i}vely, the answer is
order $n$---since each of the\ $\sqrt{n}$ Grover iterations takes
$\sqrt{n}$\ steps, just to move the robot across the grid and back.
\ This yields no improvement over classical search. \ Benioff
\cite{benioff:robot} noticed this defect of Grover's algorithm as
applied to a physical database, but failed to raise the question of
whether or not a faster algorithm exists.

Sadly, I was unable to prove a lower bound showing that the
na\"{\i}ve\ algorithm is optimal. \ But in joint work with Andris
Ambainis, we did the next best thing: we proved the
\textit{impossibility} of proving a lower bound, or to put it
crudely, gave an algorithm. \ In particular, Chapter \ref{GG}\ shows
how to search a $\sqrt{n}\times\sqrt{n}$\ grid for a unique marked
vertex in only $O\left(  \sqrt{n}\log^{3/2}n\right)  $ steps, by
using a carefully-optimized recursive Grover search. \ It also shows
how to search a $d$-dimensional hypercube in $O\left(
\sqrt{n}\right)  $\ steps for $d\geq3$. \ The latter result has an
unexpected implication: namely, that the quantum communication
complexity of the disjointness function is $O\left(  \sqrt
{n}\right)  $. \ This matches a lower bound of Razborov
\cite{razborov:cc}, and improves previous upper bounds due to
Buhrman, Cleve, and Wigderson \cite{bcw}\ and H\o yer and de Wolf
\cite{hoyerdewolf}.

Chapter \ref{GG} also generalizes our search algorithm to handle
multiple marked items, as well as graphs that are not hypercubes but
have sufficiently good expansion properties. \ More broadly, the
chapter\ develops a new model of \textit{quantum query complexity on
graphs}, and proves basic facts about that model, such as lower
bounds for search on \textquotedblleft starfish\textquotedblright\
graphs. \ Of particular interest to physicists will be Section
\ref{PHYS}, which relates our results to fundamental limits on
information processing imposed by the holographic principle. \ For
example, we can give an approximate answer to the following
question: assuming a positive cosmological constant $\Lambda>0$, and
assuming the only constraints (besides quantum mechanics) are the
speed of light and the holographic principle, how large a database
could ever be searched for a specific entry, before most of the
database receded past one's cosmological horizon?

\subsection{Quantum Computing and Postselection\label{OVERISL}}

There is at least one foolproof way to solve $\mathsf{NP}$-complete problems
in polynomial time: guess a random solution,\ then kill yourself if the
solution is incorrect. \ Conditioned on looking at anything at all, you will
be looking at a correct solution! \ It's a wonder that this approach is not
tried more often.

The general idea, of throwing out all runs of a computation except those that
yield a particular result, is called \textit{postselection}. \ Chapter
\ref{POST} explores the general power of postselection when combined with
quantum computing. \ I define a new complexity class called $\mathsf{PostBQP}%
$: the class of problems solvable in polynomial time on a quantum
computer, given the ability to measure a qubit and \textit{assume}
the outcome will be $\left\vert 1\right\rangle $ (or equivalently,
discard all runs in which the outcome is $\left\vert 0\right\rangle
$). \ I then show that $\mathsf{PostBQP}$\ coincides with the
classical complexity class $\mathsf{PP}$.

Surprisingly, this new characterization of $\mathsf{PP}$\ yields an
extremely simple, quantum computing based proof that $\mathsf{PP}$\
is closed under intersection. \ This had been an open problem for
two decades, and the previous proof, due to Beigel, Reingold, and
Spielman \cite{brs}, used highly nontrivial ideas about rational
approximations of the sign function. \ I also reestablish an
extension of the Beigel-Reingold-Spielman result due to Fortnow and
Reingold \cite{fr:pp}, that $\mathsf{PP}$\ is closed under
polynomial-time truth-table reductions. \ Indeed, I show that
$\mathsf{PP}$\ is closed under $\mathsf{BQP}$\ truth-table
reductions, which seems to be a new result.

The rest of Chapter \ref{POST}\ studies the computational effects of
simple changes to the axioms of quantum mechanics. \ In particular,
what if we allow linear but nonunitary transformations,\ or change
the measurement probabilities from $\left\vert \alpha\right\vert
^{2}$\ to $\left\vert \alpha\right\vert ^{p}$\ (suitably normalized)
for some $p\neq2$? \ I show that the first change would yield
exactly the power of $\mathsf{PostBQP}$, and
therefore of $\mathsf{PP}$; while the second change would yield $\mathsf{PP}%
$\ if $p\in\left\{  4,6,8,\ldots\right\}  $, and some class between
$\mathsf{PP}$\ and $\mathsf{PSPACE}$\ otherwise.

My results complement those of Abrams and Lloyd \cite{al}, who showed that
nonlinear quantum mechanics would let us solve $\mathsf{NP}$- and even
$\mathsf{\#P}$-complete problems in polynomial time; and Brun \cite{brun}\ and
Bacon \cite{bacon}, who showed the same for quantum computers involving closed
timelike curves. \ Taken together, these results lend credence to an
observation of Weinberg \cite{weinberg:dreams}: that quantum mechanics is a
\textquotedblleft brittle\textquotedblright\ theory, in the sense that even a
tiny change to it would have dramatic consequences.

\subsection{The Power of History\label{OVERQCHV}}

Contrary to widespread belief, what makes quantum mechanics so hard to swallow
is \textit{not} indeterminism about the\ future trajectory of a particle.
\ That is no more bizarre than a coin flip in a randomized algorithm. \ The
difficulty is that quantum mechanics also seems to require indeterminism about
a particle's \textit{past} trajectory. \ Or rather, the very notion of a
\textquotedblleft trajectory\textquotedblright\ is undefined---for until the
particle is measured, there is just an evolving wavefunction.

In spite of this, Schr\"{o}dinger \cite{schrodinger}, Bohm \cite{bohm}, Bell
\cite{bell}, and others proposed \textit{hidden-variable theories}, in which a
quantum state is supplemented by \textquotedblleft actual\textquotedblright%
\ values of certain observables. \ These actual values evolve in
time by a dynamical rule, in such a way that the predictions of
quantum mechanics are recovered at any individual time. \ On the
other hand, it now makes sense to ask questions like the following:
\textquotedblleft Given that a particle was at location $x_{1}$\ at
time $t_{1}$\ (even though it was not measured at $t_{1}$), what is
the probability of it being at location $x_{2}$\ at time
$t_{2}$?\textquotedblright\ \ The answers to such questions yield a
probability distribution over possible trajectories.

Chapter \ref{QCHV} initiates the study of hidden variables from the discrete,
abstract perspective of quantum computing. \ For me, a hidden-variable theory
is simply a way to convert a unitary matrix that maps one quantum state to
another, into a stochastic matrix that maps the initial probability
distribution to the final one in some fixed basis. \ I list five axioms that
we might want such a theory to satisfy, and investigate previous
hidden-variable theories of Dieks \cite{dieks}\ and Schr\"{o}dinger
\cite{schrodinger} in terms of these axioms. \ I also propose a new
hidden-variable theory based on \textit{network flows}, which are classic
objects of study in computer science, and prove that this theory satisfies two
axioms called \textquotedblleft indifference\textquotedblright\ and
\textquotedblleft robustness.\textquotedblright\ \ \textit{A priori}, it was
not at all obvious that these two key axioms could be satisfied simultaneously.

Next I turn to a new question: the computational complexity of
simulating hidden-variable theories. \ I show that, if we could
examine the entire history of a hidden variable, then we could
efficiently solve problems that are believed to be intractable even
for quantum computers. \ In particular, under any hidden-variable
theory satisfying the indifference axiom,\ we could solve the Graph
Isomorphism and Approximate Shortest Vector problems in
polynomial time, and indeed could simulate the entire class $\mathsf{SZK}%
$\ (Statistical Zero Knowledge). \ Combining this result with the collision
lower bound of Chapter \ref{COL}, we get an\ oracle relative to which
$\mathsf{BQP}$ is strictly contained in $\mathsf{DQP}$, where $\mathsf{DQP}$
(Dynamical Quantum Polynomial-Time) is the class of problems efficiently
solvable by sampling histories.

Using the histories model, I also show that one could search an $N$-item
database using $O\left(  N^{1/3}\right)  $ queries, as opposed to $O\left(
\sqrt{N}\right)  $ with Grover's algorithm. \ On the other hand, the $N^{1/3}$
bound is tight, meaning that one could probably \textit{not} solve
$\mathsf{NP}$-complete problems in polynomial time. \ We thus obtain the first
good example of a model of computation that appears \textit{slightly} more
powerful than the quantum computing model.

In summary, Chapter \ref{QCHV} ties together many of the themes of
this thesis: the black-box limitations of quantum computers; the
application of nontrivial computer science techniques; the obsession
with the computational resources needed to simulate our universe;
and finally, the use of quantum computing to shine light on the
mysteries of quantum mechanics itself.

\chapter{Complexity Theory Cheat Sheet\label{COMPLEXITY}}

\begin{quote}
\textquotedblleft If pigs can whistle, then donkeys can fly.\textquotedblright

(Summary of complexity theory, attributed to Richard Karp)
\end{quote}

To most people who are not theoretical computer scientists, the theory of
computational complexity---one of the great intellectual achievements of the
twentieth century---is simply a meaningless jumble of capital letters. \ The
goal of this chapter is to turn it into a meaningful jumble.

In computer science, a \textit{problem} is ordinarily an infinite set of
yes-or-no questions: for example, \textquotedblleft Given a graph, is it
connected?\textquotedblright\ \ Each particular graph is an \textit{instance}
of the general problem. \ An algorithm for the problem is
\textit{polynomial-time} if, given any instance as input, it outputs the
correct answer after at most $kn^{c}$\ steps, where $k$\ and $c$ are
constants, and $n$ is the \textit{length} of the instance, or the number of
bits needed to specify it. \ For example, in the case of a directed graph, $n$
is just the number of vertices squared. \ Then $\mathsf{P}$\ is the class of
all problems for which there exists a deterministic classical polynomial-time
algorithm. \ Examples of problems in $\mathsf{P}$\ include graph connectivity,
and (as was discovered two years ago \cite{aks}) deciding whether a positive
integer written in binary is prime or composite.

Now, $\mathsf{NP}$ (Nondeterministic Polynomial-Time)\ is the class of
problems for which, if the answer to a given instance is `yes', then an
omniscient wizard could provide a polynomial-size \textit{proof} of that fact,
which would enable us to verify it in deterministic polynomial time. \ As an
example, consider the Satisfiability problem: \textquotedblleft given a
formula involving the Boolean variables $x_{1},\ldots,x_{n}$\ and the logical
connectives $\wedge,\vee,\urcorner$\ (and, or, not), is there an assignment to
the variables that makes the formula true?\textquotedblright\ \ If there is
such an assignment, then a short, easily-verified proof is just the assignment
itself. \ On the other hand, it might be extremely difficult to \textit{find}
a satisfying assignment without the wizard's help---or for that matter, to
verify the \textit{absence} of a satisfying assignment, even given a purported
proof of its absence from the wizard. \ The question of whether there exist
polynomial-size proofs of \textit{un}satisfiability that can be verified in
polynomial time is called the $\mathsf{NP}$\ versus $\mathsf{coNP}$\ question.
\ Here $\mathsf{coNP}$\ is the class containing the \textit{complement}\ of
every $\mathsf{NP}$\ problem---for example, \textquotedblleft given a Boolean
formula, is it \textit{not} satisfiable?\textquotedblright

The Satisfiability problem turns out to be $\mathsf{NP}$-complete, which means
it is among the \textquotedblleft hardest\textquotedblright\ problems in
$\mathsf{NP}$: any instance of any $\mathsf{NP}$\ problem can be efficiently
converted into an instance of Satisfiability. \ The central question, of
course, is whether $\mathsf{NP}$-complete\ problems are solvable in polynomial
time, or equivalently whether $\mathsf{P}=\mathsf{NP}$ (it being clear that
$\mathsf{P}\subseteq\mathsf{NP}$). \ By definition, if any $\mathsf{NP}%
$-complete\ problem is solvable in polynomial time, then all of them are.
\ One thing we know is that if $\mathsf{P}\neq\mathsf{NP}$, as is almost
universally assumed, then there are problems in $\mathsf{NP}$\ that are
neither in $\mathsf{P}$\ nor $\mathsf{NP}$-complete \cite{ladner}.
\ Candidates for such \textquotedblleft intermediate\textquotedblright%
\ problems include deciding whether or not two graphs are isomorphic, and
integer factoring (e.g. given integers $N,M$\ written in binary, does $N$ have
a prime factor greater than $M$?). \ The $\mathsf{NP}$-intermediate problems
have been a major focus of quantum algorithms research.

\section{The Complexity Zoo Junior\label{ZOO}}

\begin{figure}
[ptb]
\begin{center}
\includegraphics[
trim=0.510309in 1.438571in 0.715261in 0.964735in,
height=3.2428in,
width=3.105in
]%
{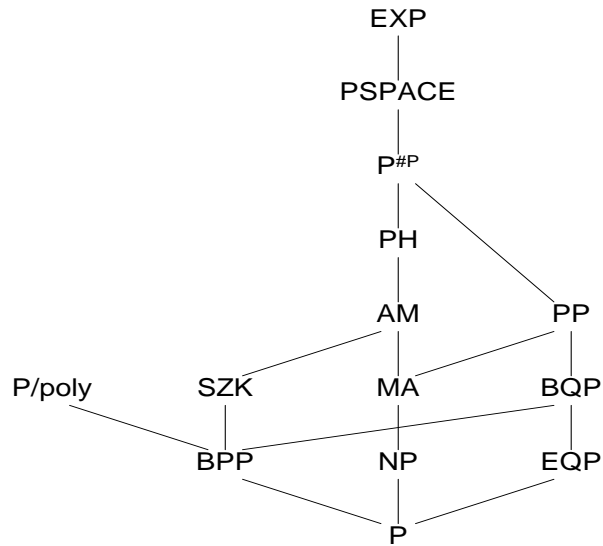}%
\caption[Known relations among $14$ complexity classes]{Known relations among
$14$ complexity classes.}%
\label{ccfig}%
\end{center}
\end{figure}

I now present a glossary of $12$ complexity classes besides $\mathsf{P}$ and
$\mathsf{NP}$\ that appear in this thesis; non-complexity-theorist readers
might wish to refer back to it as needed. \ The known relationships among
these classes are diagrammed in Figure \ref{ccfig}. \ These classes represent
a tiny sample of the more than $400$ classes described on my Complexity Zoo
web page (www.complexityzoo.com).

$\mathsf{PSPACE}$\textbf{ (Polynomial Space)} is the class of problems
solvable by a deterministic classical algorithm that uses a
polynomially-bounded amount of memory. \ Thus $\mathsf{NP}\subseteq
\mathsf{PSPACE}$, since a $\mathsf{PSPACE}$\ machine can loop through all
possible proofs.

$\mathsf{EXP}$ \textbf{(Exponential-Time)} is the class of problems solvable
by a deterministic classical algorithm that uses at most $2^{q\left(
n\right)  }$\ time steps, for some polynomial $q$. \ Thus $\mathsf{PSPACE}%
\subseteq\mathsf{EXP}$.

$\mathsf{BPP}$ \textbf{(Bounded-Error Probabilistic Polynomial-Time)} is the
class of problems solvable by a \textit{probabilistic} classical
polynomial-time algorithm, which given any instance, must output the correct
answer for that instance with probability at least $2/3$. \ Thus
$\mathsf{P}\subseteq\mathsf{BPP}\subseteq\mathsf{PSPACE}$. \ It is widely
conjectured that $\mathsf{BPP}=\mathsf{P}$\ \cite{iw}, but not even known that
$\mathsf{BPP}\subseteq\mathsf{NP}$.

$\mathsf{PP}$ \textbf{(Probabilistic Polynomial-Time)} is the class of
problems solvable by a probabilistic classical polynomial-time algorithm,
which given any instance, need only output the correct answer for that
instance with probability greater than $1/2$. \ The following problem is
$\mathsf{PP}$-complete: given a Boolean formula $\varphi$, decide whether at
least half of the possible truth assignments satisfy $\varphi$. \ We have
$\mathsf{NP}\subseteq\mathsf{PP}\subseteq\mathsf{PSPACE}$\ and also
$\mathsf{BPP}\subseteq\mathsf{PP}$.

$\mathsf{P}^{\mathsf{\#P}}$ \textbf{(pronounced \textquotedblleft P to the
sharp-P\textquotedblright)} is the class of problems solvable by a
$\mathsf{P}$\ machine that can access a \textquotedblleft
counting\ oracle.\textquotedblright\ \ Given a Boolean formula $\varphi$, this
oracle returns the number of truth assignments that satisfy $\varphi$. \ We
have $\mathsf{PP}\subseteq\mathsf{P}^{\mathsf{\#P}}\subseteq\mathsf{PSPACE}$.

$\mathsf{BQP}$ \textbf{(Bounded-Error Quantum Polynomial-Time)} is the class
of problems solvable by a quantum polynomial-time algorithm, which given any
instance, must output the correct answer for that instance with probability at
least $2/3$. \ More information is in Chapter \ref{QUANTUM}. \ We have
$\mathsf{BPP}\subseteq\mathsf{BQP}\subseteq\mathsf{PP}$ \cite{bv,adh}.

$\mathsf{EQP}$ \textbf{(Exact Quantum Polynomial-Time)} is similar to
$\mathsf{BQP}$, except that the probability of correctness must be $1$ instead
of $2/3$. \ This class is extremely artificial; it is not even clear how to
define it independently of the choice of gate set. \ But for any reasonable
choice, $\mathsf{P}\subseteq\mathsf{EQP}\subseteq\mathsf{BQP}$.

$\mathsf{P/poly}$ \textbf{(}$\mathsf{P}$\textbf{ with polynomial-size advice)}
is the class of problems solvable by a $\mathsf{P}$\ algorithm that, along
with a problem instance of length $n$, is also given an \textquotedblleft
advice string\textquotedblright\ $z_{n}$ of length bounded by a polynomial in
$n$. \ The only constraint is that $z_{n}$ can depend only on $n$, and not on
any other information about the instance. \ Otherwise the $z_{n}$'s\ can be
chosen arbitrarily to help the algorithm. \ It is not hard to show that
$\mathsf{BPP}\subseteq\mathsf{P/poly}$. \ Since the $z_{n}$'s can encode
noncomputable problems (for example, does the $n^{th}$\ Turing machine halt?),
$\mathsf{P/poly}$\ is not contained in any uniform complexity class,\ where
\textquotedblleft uniform\textquotedblright\ means that the same information
is available to an algorithm regardless of $n$. \ We can also add
polynomial-size advice to other complexity classes, obtaining
$\mathsf{EXP/poly}$, $\mathsf{PP/poly}$, and so on.

$\mathsf{PH}$ \textbf{(Polynomial-Time Hierarchy)} is the union of
$\mathsf{NP}$, $\mathsf{NP}^{\mathsf{NP}}$, $\mathsf{NP}^{\mathsf{NP}%
^{\mathsf{NP}}}$, etc. \ Equivalently, $\mathsf{PH}$\ is the class of problems
that are polynomial-time reducible to the following form: for all truth
assignments $x$, does there exist an assignment $y$ such that for all
assignments $z$, \ldots, $\varphi\left(  x,y,z,\ldots\right)  $ is satisfied,
where $\varphi$\ is a Boolean formula? \ Here the number of alternations
between \textquotedblleft for all\textquotedblright\ and \textquotedblleft
there exists\textquotedblright\ quantifiers is a constant independent of $n$.
\ Sipser \cite{sipser:bpp}\ and Lautemann \cite{lautemann}\ showed that
$\mathsf{BPP}\subseteq\mathsf{PH}$, while Toda \cite{toda}\ showed that
$\mathsf{PH}\subseteq\mathsf{P}^{\mathsf{\#P}}$.

$\mathsf{MA}$ \textbf{(Merlin Arthur)} is the class of problems for which, if
the answer to a given instance is `yes,' then an omniscient wizard could
provide a polynomial-size proof of that fact, which would enable us to verify
it in $\mathsf{BPP}$ (classical probabilistic polynomial-time,
with\ probability at most $1/3$ of accepting an invalid proof or rejecting a
valid one). \ We have $\mathsf{NP}\subseteq\mathsf{MA}\subseteq\mathsf{PP}$.

$\mathsf{AM}$ \textbf{(Arthur Merlin)} is the class of problems for which, if
the answer to a given instance is `yes,' then a $\mathsf{BPP}$ algorithm could
become convinced of that fact after a constant number of rounds of interaction
with an omniscient wizard. \ We have $\mathsf{MA}\subseteq\mathsf{AM}%
\subseteq\mathsf{PH}$. \ There is evidence that $\mathsf{AM}=\mathsf{MA}%
=\mathsf{NP}$\ \cite{kvm}.

$\mathsf{SZK}$ \textbf{(Statistical Zero Knowledge)} is the class of problems
that possess \textquotedblleft statistical zero-knowledge proof
protocols.\textquotedblright\ \ We have $\mathsf{BPP}\subseteq\mathsf{SZK}%
\subseteq\mathsf{AM}$. \ Although $\mathsf{SZK}$\ contains nontrivial problems
such as graph isomorphism \cite{gmw}, there are strong indications that it
does \textit{not} contain all of $\mathsf{NP}$\ \cite{bhz}.%
%TCIMACRO{\FRAME{ftbpFU}{3.105in}{3.2428in}{0pt}{\Qcb[Known relations among
%$14$ complexity classes]{Known relations among $14$ complexity classes.}%
%}{\Qlb{ccfig}}{cc.eps}{\special{ language "Scientific Word";  type "GRAPHIC";
%display "USEDEF";  valid_file "F";  width 3.105in;  height 3.2428in;
%depth 0pt;  original-width 10.3511in;  original-height 7.7551in;
%cropleft "0.0493";  croptop "0.8756";  cropright "0.9309";
%cropbottom "0.1855";  filename 'cc.eps';file-properties "XNPEU";}}}%
%BeginExpansion
%EndExpansion

Other complexity classes, such as $\mathsf{PLS}$,\ $\mathsf{TFNP}$,
$\mathsf{BQP/qpoly}$, and $\mathsf{BPP}_{\mathsf{path}}$, will be introduced
throughout the thesis as they are needed.

\section{Notation\label{NOTATION}}

In computer science, the following symbols are used to describe asymptotic
growth rates:

\begin{itemize}
\item $F\left(  n\right)  =O\left(  G\left(  n\right)  \right)  $\ means that
$F\left(  n\right)  $ is at most order $G\left(  n\right)  $; that is,
$F\left(  n\right)  \leq a+bG\left(  n\right)  $\ for all $n\geq0$\ and\ some
nonnegative constants $a,b$.

\item $F\left(  n\right)  =\Omega\left(  G\left(  n\right)  \right)  $ means
that $F\left(  n\right)  $\ is at least order $G\left(  n\right)  $; that is,
$G\left(  n\right)  =O\left(  F\left(  n\right)  \right)  $.

\item $F\left(  n\right)  =\Theta\left(  G\left(  n\right)  \right)  $ means
that $F\left(  n\right)  $\ is exactly order $G\left(  n\right)  $; that is,
$F\left(  n\right)  =O\left(  G\left(  n\right)  \right)  $\ and $F\left(
n\right)  =\Omega\left(  G\left(  n\right)  \right)  $.

\item $F\left(  n\right)  =o\left(  G\left(  n\right)  \right)  $ means that
$F\left(  n\right)  $\ is less than order $G\left(  n\right)  $; that is,
$F\left(  n\right)  =O\left(  G\left(  n\right)  \right)  $\ but not $F\left(
n\right)  =\Omega\left(  G\left(  n\right)  \right)  $.
\end{itemize}

The set of all $n$-bit strings is denoted $\left\{  0,1\right\}  ^{n}$. \ The
set of all binary strings, $%
%TCIMACRO{\tbigcup \nolimits_{n\geq0}}%
%BeginExpansion
{\textstyle\bigcup\nolimits_{n\geq0}}
%EndExpansion
\left\{  0,1\right\}  ^{n}$, is denoted $\left\{  0,1\right\}  ^{\ast}$.

\section{Oracles\label{ORACLES}}

One complexity-theoretic concept that will be needed again and again
in this thesis is that of an \textit{oracle}. \ An oracle is a
subroutine available to an algorithm, that is guaranteed to compute
some function even if we have no idea how. \ Oracles are denoted
using superscripts. \ For example, $\mathsf{P}^{\mathsf{NP}}$\ is
the class of problems solvable by a $\mathsf{P}$\ algorithm that,
given any instance of an $\mathsf{NP}$-complete problem such as
Satisfiability, can instantly find the solution for that instance by
calling the $\mathsf{NP}$\ oracle. \ The algorithm can make multiple
calls to the\ oracle, and these calls can be \textit{adaptive} (that
is, can depend on the outcomes of previous calls). \ If a quantum
algorithm makes oracle calls, then unless otherwise specified we
assume that the calls can be made in superposition. \ Further
details about the quantum oracle model are provided in Chapter
\ref{INTROLQC}.

We identify an oracle with the function that it computes, usually a Boolean
function $f:\left\{  0,1\right\}  ^{\ast}\rightarrow\left\{  0,1\right\}  $.
\ Often we think of $f$ as defining a problem instance, or rather an infinite
sequence of problem instances, one for each positive integer $n$. \ For
example, \textquotedblleft does there exist an $x\in\left\{  0,1\right\}
^{n}$ such that $f\left(  x\right)  =1$?\textquotedblright\ \ In these cases
the \textit{oracle string}, which consists of $f\left(  x\right)  $\ for every
$x\in\left\{  0,1\right\}  ^{n}$, can be thought of as an input that is
$2^{n}$\ bits long instead of $n$ bits. \ Of course, a classical algorithm
running in polynomial time could examine only a tiny fraction of such an
input, but maybe a quantum algorithm could do better. \ When discussing such
questions, we need to be careful to distinguish between two functions: $f$
itself, and the function of the oracle string that an algorithm is trying is
to compute.

\chapter{Quantum Computing Cheat Sheet\label{QUANTUM}}

\begin{quote}
\textquotedblleft Somebody says \ldots\ `You know those quantum mechanical
amplitudes you told me about, they're so complicated and absurd, what makes
you think those are right? \ Maybe they aren't right.' \ Such remarks are
obvious and are perfectly clear to anybody who is working on this problem.
\ It does not do any good to point this out.\textquotedblright

---Richard Feynman, \textit{The Character of Physical Law} \cite{feynman:cpl}
\end{quote}

Non-physicists often have the mistaken idea that quantum mechanics
is hard. \ Unfortunately, many physicists have done nothing to
correct that idea. \ But in newer textbooks, courses, and survey
articles
\cite{aharonov:review,fortnow:qc,mermin,nc,vazirani:course}, the
truth is starting to come out: if you wish to understand the central
`paradoxes' of quantum mechanics, together with almost the entire
body of research on quantum information and computing, then you do
not need to know anything about wave-particle duality, ultraviolet
catastrophes, Planck's constant, atomic spectra, boson-fermion
statistics, or even Schr\"{o}dinger's equation. \ All you need to
know is how to manipulate vectors whose entries are complex numbers.
\ If that is too difficult, then positive and negative \textit{real}
numbers turn out to suffice for most purposes as well. \ After you
have mastered these vectors, you will then have some context if you
wish to learn more about the underlying physics. \ But the
\textit{historical} order in which the ideas were discovered is
almost the reverse of the \textit{logical} order in which they are
easiest to learn!

What quantum mechanics says is that, if an object can be in either of two
perfectly distinguishable states, which we denote $\left\vert 0\right\rangle $
and $\left\vert 1\right\rangle $, then it can also be in a linear
\textquotedblleft superposition\textquotedblright\ of those states, denoted
$\alpha\left\vert 0\right\rangle +\beta\left\vert 1\right\rangle $. \ Here
$\alpha$\ and $\beta$\ are complex numbers called \textquotedblleft
amplitudes,\textquotedblright\ which satisfy $\left\vert \alpha\right\vert
^{2}+\left\vert \beta\right\vert ^{2}=1$. \ The asymmetric brackets
$\left\vert ~\right\rangle $\ are called \textquotedblleft Dirac ket
notation\textquotedblright; one gets used to them with time.

If we \textit{measure} the state $\alpha\left\vert 0\right\rangle
+\beta\left\vert 1\right\rangle $ in a standard way, then we see the
\textquotedblleft basis state\textquotedblright\ $\left\vert 0\right\rangle
$\ with probability $\left\vert \alpha\right\vert ^{2}$, and $\left\vert
1\right\rangle $\ with probability $\left\vert \beta\right\vert ^{2}$. \ Also,
the state \textit{changes} to whichever outcome we see---so if we see
$\left\vert 0\right\rangle $ and then measure again,\ nothing having happened
in the interim, we will still see $\left\vert 0\right\rangle $. \ The two
probabilities $\left\vert \alpha\right\vert ^{2}$ and $\left\vert
\beta\right\vert ^{2}$\ sum to $1$, as they ought to. \ So far, we might as
well have described the object using classical probabilities---for example,
\textquotedblleft this cat is alive with probability $1/2$\ and dead with
probability $1/2$; we simply don't know which.\textquotedblright

\begin{figure}
[ptb]
\begin{center}
\includegraphics[
trim=2.928326in 3.520815in 3.325809in 0.196979in,
height=2.0166in,
width=2.0465in
]%
{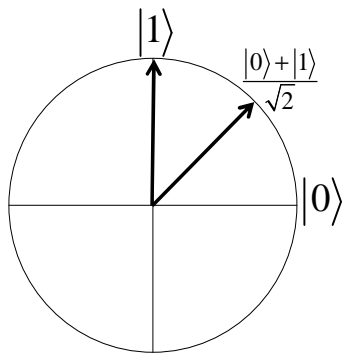}%
\caption[Quantum states of one qubit]{Quantum states of the form
$\alpha\left\vert 0\right\rangle +\beta\left\vert 1\right\rangle $, with
$\alpha$\ and $\beta$\ real, can be represented by unit vectors in the plane.
\ Then the operation $U$\ corresponds to a $45^{\circ}$\ counterclockwise
rotation.}%
\label{ufig}%
\end{center}
\end{figure}

The difference between classical probabilities and quantum amplitudes arises
in how the object's state changes when we perform an operation on it.
\ Classically, we can multiply a vector of probabilities by a
\textit{stochastic matrix}, which is a matrix of nonnegative real numbers each
of whose columns sums to $1$. \ Quantum-mechanically, we multiply the vector
of amplitudes by a \textit{unitary matrix}, which is a matrix of complex
numbers that maps any unit vector to another unit vector. \ (Equivalently, $U$
is unitary if and only if its inverse $U^{-1}$\ equals its conjugate transpose
$U^{\ast}$.) \ As an example, suppose we start with the state $\left\vert
0\right\rangle $, which corresponds to the vector of amplitudes%
\[
\left[
\begin{array}
[c]{c}%
1\\
0
\end{array}
\right]  .
\]
We then left-multiply this vector by the unitary matrix%
\[
U=\left[
\begin{array}
[c]{cc}%
\frac{1}{\sqrt{2}} & -\frac{1}{\sqrt{2}}\\
\frac{1}{\sqrt{2}} & \frac{1}{\sqrt{2}}%
\end{array}
\right]  ,
\]
which maps the vector to%
\[
\left[
\begin{array}
[c]{c}%
\frac{1}{\sqrt{2}}\\
\frac{1}{\sqrt{2}}%
\end{array}
\right]  ,
\]
and therefore the state $\left\vert 0\right\rangle $\ to%
\[
U\left\vert 0\right\rangle =\frac{1}{\sqrt{2}}\left\vert 0\right\rangle
+\frac{1}{\sqrt{2}}\left\vert 1\right\rangle .
\]
If we now measured, we would see $\left\vert 0\right\rangle $%
\ with\ probability $1/2$ and $\left\vert 1\right\rangle $\ with probability
$1/2$. \ The interesting part is what happens if we apply the same operation
$U$ a second time, without measuring. \ We get%
\[
\left[
\begin{array}
[c]{cc}%
\frac{1}{\sqrt{2}} & -\frac{1}{\sqrt{2}}\\
\frac{1}{\sqrt{2}} & \frac{1}{\sqrt{2}}%
\end{array}
\right]  \left[
\begin{array}
[c]{c}%
\frac{1}{\sqrt{2}}\\
\frac{1}{\sqrt{2}}%
\end{array}
\right]  =\left[
\begin{array}
[c]{c}%
0\\
1
\end{array}
\right]
\]
which is $\left\vert 1\right\rangle $\ with certainty (see Figure \ref{ufig}).%
%TCIMACRO{\FRAME{ftbpFU}{2.0465in}{2.0166in}{0pt}{\Qcb[Quantum states of one
%qubit]{Quantum states of the form $\alpha\left\vert 0\right\rangle
%+\beta\left\vert 1\right\rangle $, with $\alpha$\ and $\beta$\ real, can be
%represented by unit vectors in the plane. \ Then the operation $U$%
%\ corresponds to a $45^{\circ}$\ counterclockwise rotation.}}{\Qlb{ufig}%
%}{ufig.eps}{\special{ language "Scientific Word";  type "GRAPHIC";
%width 2.0465in;  height 2.0166in;  depth 0pt;  original-width 10.3511in;
%original-height 7.7551in;  cropleft "0.2829";  croptop "0.9746";
%cropright "0.6787";  cropbottom "0.4540";
%filename 'ufig.eps';file-properties "XNPEU";}}}%
%BeginExpansion
%EndExpansion
Applying a \textquotedblleft randomizing\textquotedblright\ operation to a
\textquotedblleft random\textquotedblright\ state produces a deterministic
outcome! \ The reason is that, whereas probabilities are always nonnegative,
amplitudes can be positive, negative, or even complex, and can therefore
cancel each other out. \ This interference of amplitudes can be considered the
source of all \textquotedblleft quantum weirdness.\textquotedblright

\section{Quantum Computers: $N$ Qubits\label{QCNQ}}

The above description applied to \textquotedblleft qubits,\textquotedblright%
\ or objects with only two distinguishable states. \ But it generalizes to
objects with a larger number of distinguishable states. \ Indeed, in quantum
computing we consider a system of $N$ qubits, each of which can be $\left\vert
0\right\rangle $\ or $\left\vert 1\right\rangle $. \ We then need to assign
amplitudes to all $2^{N}$\ possible outcomes of measuring the qubits in order
from first to last. \ So the computer's state has the form%
\[
\left\vert \psi\right\rangle =\sum_{z\in\left\{  0,1\right\}  ^{N}}\alpha
_{z}\left\vert z\right\rangle
\]
where%
\[
\sum_{z\in\left\{  0,1\right\}  ^{N}}\left\vert \alpha_{z}\right\vert ^{2}=1.
\]
What was just said is remarkable---for it suggests that Nature\ needs to keep
track of $2^{N}$\ complex numbers just to describe a state of $N$ interacting
particles. \ If $N=300$, then this is already more complex numbers than there
are particles in the known universe. \ The goal of quantum computing is to
exploit this strange sort of parallelism that is inherent in the laws of
physics as we currently understand them.

The difficulty is that, when the computer's state is measured, we only see one
of the \textquotedblleft basis states\textquotedblright\ $\left\vert
x\right\rangle $, not the entire collection of amplitudes. \ However, for a
few specific problems, we might be able to arrange things so that basis states
corresponding to wrong answers all have amplitudes close to $0$, because of
interference between positive and negative contributions. \ If we can do that,
then basis states corresponding to right answers will be measured with high probability.

More explicitly, a quantum computer applies a sequence of unitary matrices
called\ \textit{gates}, each of which acts on only one or two of the $N$
qubits (meaning that is a tensor product of the identity operation on $N-1$ or
$N-2$ qubits, and the operation of interest on the remaining qubits). \ As an
example, the \textit{controlled-NOT} or CNOT gate is a two-qubit gate that
flips a \textquotedblleft target\textquotedblright\ qubit if a
\textquotedblleft control\textquotedblright\ qubit is $1$, and otherwise does
nothing:%
\[
\left\vert 00\right\rangle \rightarrow\left\vert 00\right\rangle
,~~~~\left\vert 01\right\rangle \rightarrow\left\vert 01\right\rangle
,~~~~\left\vert 10\right\rangle \rightarrow\left\vert 11\right\rangle
,~~~~\left\vert 11\right\rangle \rightarrow\left\vert 10\right\rangle .
\]
The unitary matrix corresponding to the CNOT gate is%
\[
\left[
\begin{array}
[c]{cccc}%
1 & 0 & 0 & 0\\
0 & 1 & 0 & 0\\
0 & 0 & 0 & 1\\
0 & 0 & 1 & 0
\end{array}
\right]  .
\]
Adleman, DeMarrais, and Huang \cite{adh}\ showed that the CNOT gate, together
with the one-qubit gate%
\[
\left[
\begin{array}
[c]{cc}%
\frac{3}{5} & -\frac{4}{5}\\
\frac{4}{5} & \frac{3}{5}%
\end{array}
\right]  ,
\]
constitute a \textit{universal} set of quantum gates, in that they can be used
to approximate any other gate to any desired accuracy. \ Indeed, almost any
set of one- and two-qubit gates is universal in this sense \cite{dbe}.

A \textit{quantum circuit} is just a sequence of gates drawn from a finite
universal set. \ Without loss of generality, we can take the circuit's output
to be the result of a single measurement after all gates have been applied;
that is, $z\in\left\{  0,1\right\}  ^{N}$\ with probability $\left\vert
\alpha_{z}\right\vert ^{2}$. \ (If a binary output is needed, we simply throw
away the last $N-1$\ bits of $z$.) \ It is known that allowing intermediate
measurements does not yield any extra computational power \cite{bv}. \ The
circuit is \textit{polynomial-size} if both $N$ and the number of gates are
upper-bounded by a polynomial in the length $n$ of the input.

We can now define the important complexity class $\mathsf{BQP}$, or
Bounded-Error Quantum Polynomial-Time. \ Given an input $x\in\left\{
0,1\right\}  ^{n}$, first a polynomial-time classical algorithm $A$ prepares a
polynomial-size quantum circuit $U_{x}$. \ (The requirement that the circuit
itself be efficiently preparable is called \textit{uniformity}.) \ Then
$U_{x}$ is applied to the \textquotedblleft all-$0$\textquotedblright\ initial
state $\left\vert 0\right\rangle ^{\otimes N}$. \ We say a language
$L\subseteq\left\{  0,1\right\}  ^{n}$\ is in $\mathsf{BQP}$\ if there exists
an $A$ such that for all $x$,

\begin{enumerate}
\item[(i)] If $x\in L$\ then $U_{x}$\ outputs `1'\ with probability at least
$2/3$.

\item[(ii)] If $x\notin L$\ then $U_{x}$\ outputs `0'\ with probability at
least $2/3$.
\end{enumerate}

By running $U_{x}$\ multiple times and taking the majority answer, we can
boost the probability of success from $2/3$\ to\ $1-2^{-p\left(  n\right)  }%
$\ for any polynomial $p$.

$\mathsf{BQP}$\ was first defined in a 1993 paper by Bernstein and Vazirani
\cite{bv}.\footnote{As a historical note, Bernstein and Vazirani \cite{bv}
defined $\mathsf{BQP}$\ in terms of \textquotedblleft quantum Turing
machines.\textquotedblright\ \ However, Yao \cite{yao:bqp}\ showed that
Bernstein and Vazirani's definition is equivalent to the much simpler one
given here. \ Also, Berthiaume and Brassard \cite{bb:qc} had implicitly
defined $\mathsf{EQP}$\ (Exact Quantum Polynomial-Time) a year earlier, and
had shown that it lies outside $\mathsf{P}$\ and even $\mathsf{NP}$\ relative
to an oracle.}\ \ That paper marked a turning point. \ Before, quantum
computing had been an \textit{idea}, explored in pioneering work by Deutsch
\cite{deutsch:qc}, Feynman \cite{feynman:qc}, and others. \ Afterward, quantum
computing was a full-fledged \textit{model} in the sense of computational
complexity theory, which could be meaningfully compared against other models.
\ For example, Bernstein and Vazirani showed that\ $\mathsf{BPP}%
\subseteq\mathsf{BQP}\subseteq\mathsf{P}^{\mathsf{\#P}}$: informally, quantum
computers are \textit{at least} as powerful as classical probabilistic
computers, and at most exponentially \textit{more} powerful. \ (The
containment $\mathsf{BQP}\subseteq\mathsf{P}^{\mathsf{\#P}}$\ was later
improved to $\mathsf{BQP}\subseteq\mathsf{PP}$\ by Adleman, DeMarrais, and
Huang \cite{adh}.)

Bernstein and Vazirani\ also gave an oracle problem called Recursive Fourier
Sampling ($\operatorname*{RFS}$), and showed that it requires $n^{\Omega
\left(  \log n\right)  }$\ classical probabilistic queries but only $n$
quantum queries. \ This provided the first evidence that quantum computers are
strictly more powerful than classical probabilistic computers, i.e. that
$\mathsf{BPP}\neq\mathsf{BQP}$. \ Soon afterward, Simon \cite{simon}\ widened
the gap to polynomial versus exponential, by giving an oracle problem that
requires $\Omega\left(  2^{n/2}\right)  $\ classical probabilistic queries but
only $O\left(  n\right)  $\ quantum queries. \ However, these results
attracted limited attention because the problems seemed artificial.

People finally paid attention when Shor \cite{shor}\ showed that quantum
computers could factor integers and compute discrete logarithms in polynomial
time. \ The security of almost all modern cryptography rests on the presumed
intractability of those two problems. \ It had long been known \cite{miller}
that factoring is classically reducible to the following problem: given oracle
access to a periodic function $f:\left\{  1,\ldots,R\right\}  \rightarrow
\left\{  1,\ldots,R\right\}  $, where $R$ is exponentially large, find the
period of $f$. \ Shor gave an efficient quantum algorithm for this oracle
problem, by exploiting the \textit{quantum Fourier transform}, a tool that had
earlier been used by Simon. \ (The algorithm for the discrete logarithm
problem is more complicated but conceptually similar.)

Other results in the \textquotedblleft quantum
canon,\textquotedblright\ such as Grover's algorithm \cite{grover}\
and methods for quantum error-correction and fault-tolerance
\cite{ab,cs,gottesman:hamm,klz,steane}, will be discussed in this
thesis as the need arises.

\section{Further Concepts\label{FURTHERDEF}}

This section summarizes \textquotedblleft fancier\textquotedblright\ quantum
mechanics concepts, which are needed for Part \ref{MAR} and for Chapter
\ref{ADV} of Part \ref{LQC} (which deals with quantum advice). \ They are not
needed for the other chapters in Part \ref{LQC}.

\textbf{Tensor Product.} \ If $\left\vert \psi\right\rangle $\ and $\left\vert
\varphi\right\rangle $\ are two quantum states, then their \textit{tensor
product}, denoted $\left\vert \psi\right\rangle \otimes\left\vert
\varphi\right\rangle $\ or $\left\vert \psi\right\rangle \left\vert
\varphi\right\rangle $, is just a state that consists of $\left\vert
\psi\right\rangle $ and $\left\vert \varphi\right\rangle $\ next to each
other. \ For example, if $\left\vert \psi\right\rangle =\alpha\left\vert
0\right\rangle +\beta\left\vert 1\right\rangle $\ and $\left\vert
\varphi\right\rangle =\gamma\left\vert 0\right\rangle +\delta\left\vert
1\right\rangle $, then%
\[
\left\vert \psi\right\rangle \left\vert \varphi\right\rangle =\left(
\alpha\left\vert 0\right\rangle +\beta\left\vert 1\right\rangle \right)
\left(  \gamma\left\vert 0\right\rangle +\delta\left\vert 1\right\rangle
\right)  =\alpha\gamma\left\vert 00\right\rangle +\alpha\delta\left\vert
01\right\rangle +\beta\gamma\left\vert 10\right\rangle +\beta\delta\left\vert
11\right\rangle .
\]

\textbf{Inner Product.} \ The \textit{inner product} between two states
$\left\vert \psi\right\rangle =\alpha_{1}\left\vert 1\right\rangle
+\cdots+\alpha_{N}\left\vert N\right\rangle $\ and $\left\vert \varphi
\right\rangle =\beta_{1}\left\vert 1\right\rangle +\cdots+\beta_{N}\left\vert
N\right\rangle $\ is defined as%
\[
\left\langle \psi|\varphi\right\rangle =\alpha_{1}^{\ast}\beta_{1}%
+\cdots+\alpha_{N}^{\ast}\beta_{N}%
\]
where $\ast$\ denotes complex conjugate. \ The inner product satisfies all the
expected properties, such as $\left\langle \psi|\psi\right\rangle =1$ and%
\[
\left\langle \psi\right\vert \left(  \left\vert \varphi\right\rangle
+\left\vert \phi\right\rangle \right)  =\left\langle \psi|\varphi\right\rangle
+\left\langle \psi|\phi\right\rangle .
\]
If $\left\langle \psi|\varphi\right\rangle =0$\ then we say $\left\vert
\psi\right\rangle $\ and $\left\vert \varphi\right\rangle $\ are
\textit{orthogonal}.

\textbf{General Measurements.} \ In principle, we can choose any orthogonal
basis of states $\left\{  \left\vert \varphi_{1}\right\rangle ,\ldots
,\left\vert \varphi_{N}\right\rangle \right\}  $\ in which to measure a state
$\left\vert \psi\right\rangle $. \ (Whether that measurement can actually be
performed \textit{efficiently} is another matter.) \ Then the probability of
obtaining outcome $\left\vert \varphi_{j}\right\rangle $\ is $\left\vert
\left\langle \psi|\varphi_{j}\right\rangle \right\vert ^{2}$. \ We can even
measure in a \textit{non}-orthogonal basis, a concept called Positive Operator
Valued Measurements (POVM's) that\ I will not explain here. \ None of these
more general measurements increase the power of the quantum computing model,
since we can always produce the same effect by first applying a unitary matrix
(possibly using additional qubits called \textit{ancillas}), and then
measuring in a \textquotedblleft standard\textquotedblright\ basis such as
$\left\{  \left\vert 1\right\rangle ,\ldots,\left\vert N\right\rangle
\right\}  $.

\textbf{Mixed States.} \ Superposition states, such as $\alpha\left\vert
0\right\rangle +\beta\left\vert 1\right\rangle $, are also called \textit{pure
states}. \ This is to distinguish them from \textit{mixed states}, which are
the most general kind of state in quantum mechanics. \ Mixed states are just
classical probability distributions over pure states. \ There is a catch,
though: any mixed state can be decomposed into a probability distribution over
pure states in infinitely many nonequivalent ways. \ For example, if we have a
state that is $\left\vert 0\right\rangle $\ with probability $1/2$\ and
$\left\vert 1\right\rangle $\ with probability $1/2$, then no experiment could
ever distinguish it from a state that is $\left(  \left\vert 0\right\rangle
+\left\vert 1\right\rangle \right)  /\sqrt{2}$\ with probability $1/2$\ and
$\left(  \left\vert 0\right\rangle -\left\vert 1\right\rangle \right)
/\sqrt{2}$ with probability $1/2$. \ For regardless of what orthogonal basis
we measured in, the two possible outcomes of measuring would both occur with
probability $1/2$. \ Therefore, this state is called the one-qubit
\textit{maximally mixed state}.

\textbf{Density Matrices.} \ We can represent mixed states using a formalism
called \textit{density matrices}. \ The \textit{outer product} of $\left\vert
\psi\right\rangle =\alpha_{1}\left\vert 1\right\rangle +\cdots+\alpha
_{N}\left\vert N\right\rangle $\ with itself, denoted $\left\vert
\psi\right\rangle \left\langle \psi\right\vert $, is an $N\times N$\ complex
matrix whose $\left(  i,j\right)  $\ entry is $\alpha_{i}\alpha_{j}^{\ast}$.
\ Now suppose we have a state that is $\left\vert \varphi\right\rangle
=\alpha\left\vert 0\right\rangle +\beta\left\vert 1\right\rangle $\ with
probability $p$, and $\left\vert \phi\right\rangle =\gamma\left\vert
0\right\rangle +\delta\left\vert 1\right\rangle $\ with probability $1-p$.
\ We represent the state by a Hermitian positive definite matrix $\rho$\ with
trace $1$, as follows:%
\[
\rho=p\left\vert \varphi\right\rangle \left\langle \varphi\right\vert +\left(
1-p\right)  \left\vert \phi\right\rangle \left\langle \phi\right\vert
=p\left[
\begin{array}
[c]{ll}%
\alpha\alpha^{\ast} & \alpha\beta^{\ast}\\
\beta\alpha^{\ast} & \beta\beta^{\ast}%
\end{array}
\right]  +\left(  1-p\right)  \left[
\begin{array}
[c]{ll}%
\gamma\gamma^{\ast} & \gamma\delta^{\ast}\\
\delta\gamma^{\ast} & \delta\delta^{\ast}%
\end{array}
\right]  .
\]
When we apply a unitary operation $U$, the density matrix $\rho$\ changes to
$U\rho U^{-1}$. \ When we measure in the standard basis, the probability of
outcome $\left\vert j\right\rangle $\ is the $j^{th}$\ diagonal entry of
$\rho$. \ Proving that these rules are the correct ones, and that a density
matrix really is a unique description of a mixed state, are \textquotedblleft
exercises for the reader\textquotedblright\ (which as always means the author
was too lazy). \ Density matrices will mainly be used in Chapter \ref{ADV}.

\textbf{Trace Distance.} \ Suppose you are given a system that was prepared in
state $\rho$\ with probability $1/2$, and $\sigma$\ with probability $1/2$.
\ After making a measurement, you must guess which state the system was
prepared in. \ What is the maximum probability that you will be correct? \ The
answer turns out to be%
\[
\frac{1+\left\Vert \rho-\sigma\right\Vert _{\operatorname*{tr}}}{2}%
\]
where $\left\Vert \rho-\sigma\right\Vert _{\operatorname*{tr}}$\ is the
\textit{trace distance} between $\rho$\ and $\sigma$, defined as $\frac{1}%
{2}\sum_{i}\left\vert \lambda_{i}\right\vert $ where $\lambda_{1}%
,\ldots,\lambda_{N}$\ are the eigenvalues of $\rho-\sigma$.

\textbf{Entanglement.} \ Suppose $\rho$\ is a joint state of two systems. \ If
$\rho$\ can be written as a probability distribution over pure states of the
form $\left\vert \psi\right\rangle \otimes\left\vert \varphi\right\rangle $,
then we say $\rho$\ is \textit{separable}; otherwise $\rho$\ is
\textit{entangled}.

\textbf{Hamiltonians.} \ Instead of discrete unitary operations, we can
imagine that a quantum state evolves in time by a continuous rotation called a
\textit{Hamiltonian}. \ A Hamiltonian is an $N\times N$\ Hermitian matrix $H$.
\ To find the unitary operation $U\left(  t\right)  $\ that is effected by
\textquotedblleft leaving $H$ on\textquotedblright\ for $t$ time steps, the
rule\footnote{Here Planck's constant is set equal to $1$ as always.} is
$U\left(  t\right)  =e^{-iHt}$. \ The only place I use Hamiltonians is in
Chapter \ref{GG}, and even there the use is incidental.

\part{Limitations of Quantum Computers\label{LQC}}

\chapter{Introduction\label{INTROLQC}}

\begin{quotation}
\textquotedblleft A quantum possibility is less real than a classical reality,
but more real than a classical possibility.\textquotedblright

---Boris Tsirelson \cite{tsirelson}
\end{quotation}

Notwithstanding accounts in the popular press, a decade of research has made
it clear that quantum computers would not be a panacea. \ In particular, we
still do not have a quantum algorithm to solve $\mathsf{NP}$-complete problems
in polynomial time. \ But can we prove that no such algorithm exists, i.e.
that $\mathsf{NP}\not \subset \mathsf{BQP}$? \ The difficulty is that we can't
even prove no \textit{classical} algorithm exists; this is the $\mathsf{P}%
$\ versus $\mathsf{NP}$\ question.\ \ Of course, we could ask whether
$\mathsf{NP}\not \subset \mathsf{BQP}$\ \textit{assuming} that $\mathsf{P}%
\neq\mathsf{NP}$---but unfortunately, even this conditional question seems far
beyond our ability to answer. \ So we need to refine the question even
further: can quantum computers solve $\mathsf{NP}$-complete\ problems in
polynomial time, \textit{by brute force}?

What is meant by \textquotedblleft brute force\textquotedblright\ is the
following.\ \ In Shor's factoring algorithm \cite{shor}, we prepare a
superposition of the form%
\[
\frac{1}{\sqrt{R}}\sum_{r=1}^{R}\left\vert r\right\rangle \left\vert g\left(
r\right)  \right\rangle
\]
where $g\left(  r\right)  =x^{r}\operatorname{mod}N$ for some $x,N$. \ But as
far as the key step of the algorithm is concerned, the function $g$ is a
\textquotedblleft black box.\textquotedblright\ \ Given any superposition like
the one above, the algorithm will find the period of $g$ assuming $g$ is
periodic; it does not need further information about how $g$ was computed.
\ So in the language of Section \ref{ORACLES}, we might as well say that $g$
is computed by an oracle.

Now suppose we are given a Boolean formula $\varphi$\ over $n$ variables, and
are asked to decide whether $\varphi$ is satisfiable. \ One approach would be
to exploit the internal structure of $\varphi$: \textquotedblleft let's see,
if I set variable $x_{37}$\ to TRUE,\ then this clause here is satisfied, but
those other clauses aren't satisfied any longer \ldots
\ darn!\textquotedblright\ \ However, inspired by Shor's factoring algorithm,
we might hope for a cruder quantum algorithm that treats $\varphi$ merely as
an oracle, mapping an input string $x\in\left\{  0,1\right\}  ^{n}$\ to an
output bit $\varphi\left(  x\right)  $\ that is $1$ if and only if $x$
satisfies $\varphi$. \ The algorithm would have to decide whether there exists
an $x\in\left\{  0,1\right\}  ^{n}$\ such that $\varphi\left(  x\right)  =1$,
using as few calls to the $\varphi$\ oracle as possible, and not learning
about $\varphi$\ in any other way. \ This is what is meant by brute force.

A fundamental result of Bennett, Bernstein, Brassard, and Vazirani
\cite{bbbv}\ says that no brute-force quantum algorithm exists to solve
$\mathsf{NP}$-complete problems in polynomial time. \ In particular, for some
probability distribution over oracles, any quantum algorithm needs
$\Omega\left(  2^{n/2}\right)  $\ oracle calls\ to decide, with at least a
$2/3$\ chance of being correct, whether there exists an $x\in\left\{
0,1\right\}  ^{n}$\ such that $\varphi\left(  x\right)  =1$. \ On a classical
computer, of course, $\Theta\left(  2^{n}\right)  $\ oracle calls are
necessary and sufficient. \ But as it turns out, Bennett et al.'s quantum
lower bound is tight, since Grover's quantum search algorithm \cite{grover}%
\ can find a satisfying assignment (if it exists) quadratically faster than
any classical algorithm. \ Amusingly, Grover's algorithm was proven optimal
before it was discovered to exist!

A recurring theme in this thesis is the \textit{pervasiveness} of Bennett et
al.'s finding. \ I will show that, even if a problem has considerably more
structure than the basic Grover search problem, even if \textquotedblleft
quantum advice states\textquotedblright\ are available, or even if we could
examine the entire history of a hidden variable, still any brute-force quantum
algorithm would take exponential time.

\section{The Quantum Black-Box Model\label{BLACKBOX}}

The \textit{quantum black-box model} formalizes the idea of a brute-force
algorithm. \ For the time being, suppose that a quantum algorithm's goal is to
evaluate $f\left(  X\right)  $, where $f:\left\{  0,1\right\}  ^{n}%
\rightarrow\left\{  0,1\right\}  $\ is a Boolean function and $X=x_{1}\ldots
x_{n}$\ is an $n$-bit string. \ Then the algorithm's state at any time $t$ has
the form%
\[
\sum_{i,z}\alpha_{i,z}^{\left(  t\right)  }\left\vert i,z\right\rangle .
\]
Here $i\in\left\{  1,\ldots,N\right\}  $ is the index of an oracle bit $x_{i}%
$\ to query, and $z$ is an arbitrarily large string of bits called the
\textquotedblleft workspace,\textquotedblright\ containing whatever
information the algorithm wants to store there. \ The state evolves in time
via an alternating sequence of \textit{algorithm steps} and \textit{query
steps}. \ An algorithm step multiplies the vector of $\alpha_{i,z}$'s\ by an
arbitrary unitary matrix that does not depend on $X$. \ It does not matter how
many quantum gates would be needed to implement this matrix. \ A query step
maps each basis state $\left\vert i,z\right\rangle $\ to $\left\vert i,z\oplus
x_{i}\right\rangle $, effecting the transformation $\alpha_{i,z}^{\left(
t+1\right)  }=\alpha_{i,z\oplus x_{i}}^{\left(  t\right)  }$. \ Here $z\oplus
x_{i}$\ is the string $z$, with $x_{i}$\ exclusive-OR'ed into a particular
location in $z$ called the \textquotedblleft answer bit.\textquotedblright%
\ \ The reason exclusive-OR is used is that the query step has to be
reversible, or else it would not be unitary.

At the final step $T$, the state is measured in the standard basis, and the
output of the algorithm is taken to be (say) $z_{1}$, the first bit of $z$.
\ The algorithm succeeds if%
\[
\sum_{i,z~:~z_{1}=f\left(  X\right)  }\left\vert \alpha_{i,z}^{\left(
T\right)  }\right\vert ^{2}\geq\frac{2}{3}%
\]
for all $X\in\left\{  0,1\right\}  ^{n}$. \ Here the constant $2/3$\ is
arbitrary. \ Then the \textit{(bounded-error) quantum query complexity} of
$f$, denoted $\operatorname*{Q}_{2}\left(  f\right)  $, is the minimum over
all quantum algorithms $A$\ that succeed at evaluating $f$, of the number of
queries to $f$ made by $A$. \ Here the `$2$' represents the fact that the
error probability is two-sided. \ One can compare $\operatorname*{Q}%
_{2}\left(  f\right)  $\ with $\operatorname*{Q}_{0}\left(  f\right)  $, or
zero-error quantum query complexity; $\operatorname*{R}_{2}\left(  f\right)
$, or bounded-error classical randomized query complexity; and
$\operatorname*{D}\left(  f\right)  $,\ or deterministic query complexity,
among other complexity measures. \ Chapter \ref{CER} will define many such
measures and compare them in detail.

As a simple example of the black-box model, let
$\operatorname*{OR}\left( x_{1},\ldots,x_{n}\right)
=x_{1}\vee\cdots\vee x_{n}$. \ Then Grover's algorithm
\cite{grover}\ implies that $\operatorname*{Q}_{2}\left(
\operatorname*{OR}\right)  =O\left(  \sqrt{n}\right)  $, while the
lower bound of Bennett et al.\ \cite{bbbv} implies that
$\operatorname*{Q}_{2}\left( \operatorname*{OR}\right)
=\Omega\left(  \sqrt{n}\right)  $. \ By comparison, $D\left(
\operatorname*{OR}\right)  =R_{2}\left(  \operatorname*{OR}\right)
=\Theta\left(  n\right)  $.

The quantum black-box model has some simple generalizations, which I\ will use
when appropriate. \ First, $f$\ can be a partial function, defined only on a
subset of $\left\{  0,1\right\}  ^{n}$ (so we obtain what is called a
\textit{promise problem}).\ \ Second, the $x_{i}$'s do not need to be bits; in
Chapters \ref{COL}\ and \ref{PLS}\ they will take values from a larger range.
\ Third, in Chapter \ref{PLS}\ the output will not be Boolean, and there will
generally be more than one valid output (so we obtain what is called a
\textit{relation problem}).

\section{Oracle Separations\label{OSEP}}

\begin{quotation}
\textquotedblleft I do believe it

Against an oracle.\textquotedblright

---Shakespeare, \textit{The Tempest}
\end{quotation}

Several times in this thesis, I will use a lower bound on quantum query
complexity to show\ that a complexity class is not in $\mathsf{BQP}%
$\ \textquotedblleft relative to an oracle.\textquotedblright\ \ The
method for turning query complexity lower bounds into oracle
separations was invented by Baker, Gill, and Solovay \cite{bgs}\ to
show that there exists an oracle $A$\ relative to which
$\mathsf{P}^{A}\neq\mathsf{NP}^{A}$. \ Basically, they encoded into
$A$\ an infinite sequence of exponentially hard search problems, one
for each input length $n$, such that (i)\ a nondeterministic machine
can solve the $n^{th}$\ problem in time polynomial in $n$, but (ii)
any deterministic machine would need time exponential in $n$. \ They
guaranteed (ii) by \textquotedblleft
diagonalizing\textquotedblright\ against all possible deterministic
machines, similarly to how Turing created an uncomputable real
number by diagonalizing against all possible computable reals. \
Later, Bennett and Gill \cite{bg}\ showed that a simpler way to
guarantee (ii) is just to choose the search problems uniformly at
random. \ Throughout the thesis, I will cavalierly ignore such
issues, proceeding immediately from a query complexity lower bound
to the statement of the corresponding oracle separation.

The point of an oracle separation is to rule out certain approaches
to solving an open problem in complexity theory. \ For example, the
Baker-Gill-Solovay theorem implies that the standard techniques of
computability theory, which \textit{relativize} (that is, are
\textquotedblleft oblivious\textquotedblright\ to the presence of an
oracle), cannot be powerful enough to show that
$\mathsf{P}=\mathsf{NP}$. \ Similarly, the result of Bennett et al.\
\cite{bbbv}\ that $\operatorname*{Q}_{2}\left(
\operatorname*{OR}\right)  =\Omega\left(  \sqrt{n}\right)  $ implies
that there exists an oracle $A$ relative to which
$\mathsf{NP}^{A}\not \subset \mathsf{BQP}^{A}$. \ While this does
not show that $\mathsf{NP}\not \subset \mathsf{BQP}$, it does show
that any proof of $\mathsf{NP}\subseteq \mathsf{BQP}$\ would have to
use \textquotedblleft non-relativizing\textquotedblright\textit{
}techniques that are unlike anything we understand today.

However, many computer scientists are skeptical that anything can be learned
from oracles. \ The reason for their skepticism is that over the past $15$
years, they have seen several examples of non-relativizing results in
classical complexity theory. \ The most famous of these is Shamir's Theorem
\cite{shamir,lfkn}\ that $\mathsf{PSPACE}\subseteq\mathsf{IP}$,\ where
$\mathsf{IP}$\ is the class of problems that have \textit{interactive proof
systems}, meaning that if the answer for some instance is \textquotedblleft
yes,\textquotedblright\ then a polynomial-time verifier can become convinced
of that fact to any desired level of confidence by exchanging a sequence of
messages with an omniscient prover.\footnote{Arora, Impagliazzo, and Vazirani
\cite{aiv} claim the Cook-Levin Theorem, that Satisfiability is $\mathsf{NP}%
$-complete, as another non-relativizing result.\ \ But this hinges on what we
mean by \textquotedblleft non-relativizing,\textquotedblright\ far more than
the $\mathsf{PSPACE}\subseteq\mathsf{IP}$\ example.} \ By contrast, oracles
had been known relative to which not even $\mathsf{coNP}$, let alone
$\mathsf{PSPACE}$, is contained in $\mathsf{IP}$ \cite{fs}. \ So why should we
ever listen to oracles again, if they got interactive proofs so dramatically wrong?

My answer is threefold. \ First, essentially all quantum algorithms that we
know today---from Shor's algorithm, as discussed previously, to Grover's
algorithm, to the quantum adiabatic algorithm\footnote{Given an assignment $x$
to a 3SAT formula $\varphi$, the adiabatic algorithm actually queries an
oracle that returns the \textit{number of clauses} of $\varphi$\ that $x$
satisfies, not just whether $x$ satisfies $\varphi$\ or not. \ Furthermore,
van Dam, Mosca, and Vazirani \cite{dmv}\ have shown such an oracle is
sufficient to reconstruct $\varphi$. \ On the other hand, the adiabatic
algorithm itself would be just as happy with a fitness landscape that did not
correspond to any 3SAT instance, and that is what I mean by saying that it is
an oracle algorithm at the core.} \cite{fggllp}, to the algorithms of Hallgren
\cite{hallgren}\ and van Dam, Hallgren, and Ip \cite{dhi}---are oracle
algorithms at their core. \ We do not know of any non-relativizing quantum
algorithm technique analogous to the arithmetization technique that was used
to prove $\mathsf{PSPACE}\subseteq\mathsf{IP}$. \ If such a technique is ever
discovered, I will be one of the first to want to learn it.

The second response is that without oracle results, we do not have even the
beginnings of understanding. \ Once we know (for example) that $\mathsf{SZK}%
\not \subset \mathsf{BQP}$ relative to an oracle, we can \textit{then} ask the
far more difficult unrelativized question, knowing something about the hurdles
that any proof of $\mathsf{SZK}\subseteq\mathsf{BQP}$\ would have to overcome.

The third response is that \textquotedblleft the proof of the
pudding is in the proving.\textquotedblright\ \ In other words, the
real justification for the quantum black-box model is not the
\textit{a priori} plausibility of its assumptions, but the depth and
nontriviality of what can be (and has been) proved in it. \ For
example, the result that $\mathsf{coNP}\not \subset \mathsf{IP}$\
relative to an oracle \cite{fs}\ does not tell us much about
interactive proof systems. \ For given an exponentially long oracle
string $X$, it is intuitively obvious that nothing a prover could
say could convince a classical polynomial-time verifier that $X$ is
the all-$0$ string, even if the prover and verifier could interact.
\ The only issue is how to formalize that obvious fact by
diagonalizing against all possible proof systems. \ By contrast, the
quantum oracle separations that we have are \textit{not} intuitively
obvious in the same way; or rather, the act of understanding them
confers an intuition where none was previously present.

\chapter{The Collision Problem\label{COL}}

The collision problem of size $n$, or $\operatorname*{Col}_{n}$, is defined as
follows. \ Let $X=x_{1}\ldots x_{n}$\ be a sequence of $n$ integers drawn from
$\left\{  1,\ldots,n\right\}  $, with $n$ even. \ We are guaranteed that either

\begin{enumerate}
\item[(1)] $X$ is one-to-one (that is, a permutation of $\left\{
1,\ldots,n\right\}  $), or

\item[(2)] $X$ is two-to-one (that is, each element of $\left\{
1,\ldots,n\right\}  $ appears in $X$ twice or not at all).
\end{enumerate}

The problem is to decide whether (1) or (2)\ holds. \ (A variant asks us to
\textit{find} a collision in a given two-to-one function. \ Clearly a lower
bound for the collision problem as defined above implies an equivalent lower
bound for this variant.) \ Because of its simplicity, the collision problem
was widely considered a benchmark for our understanding of quantum query complexity.

I will show that $\operatorname*{Q}_{2}\left(  \operatorname*{Col}_{n}\right)
=\Omega\left(  n^{1/5}\right)  $, where $\operatorname*{Q}_{2}\left(
f\right)  $\ is the bounded-error quantum query complexity of function $f$.
\ The best known upper bound, due to Brassard, H\o yer, and Tapp
\cite{bht},\ is $O\left(  n^{1/3}\right)  $ (see Section \ref{OVERCOL}).
\ Previously, though, no lower bound better than the trivial $\Omega\left(
1\right)  $\ bound was known. \ How great a speedup quantum computers yield
for the problem was apparently first asked by Rains \cite{rains}.

Previous lower bound techniques failed for the problem because they
depended on a function's being sensitive to many disjoint changes to
the input. \ For example, Beals et al.\ \cite{bbcmw}\ showed that
for all total Boolean functions $f$, $\operatorname*{Q}_{2}\left(
f\right)  =\Omega\left( \sqrt{\operatorname*{bs}\left(  f\right)
}\right)  $,\ where $\operatorname*{bs}\left(  f\right)  $\ is the
block sensitivity, defined by Nisan \cite{nisan}\ to be, informally,
the maximum number of disjoint changes (to any particular input $X$)
to which $f$ is sensitive. \ In the case of the collision problem,
though, every one-to-one input differs from every two-to-one input
in at least $n/2$ places, so the block sensitivity is $O\left(
1\right)  $. \ Ambainis's adversary method \cite{ambainis} faces a
related obstacle. \ In that method we consider the algorithm and
input as a bipartite quantum state, and upper-bound how much the
entanglement of the state can increase via a single query. \ But
under the simplest measures of entanglement, it turns out that the
algorithm and input can become almost maximally entangled after
$O\left(  1\right)  $\ queries, again because every one-to-one input
is far from every two-to-one input.\footnote{More formally,
the adversary method cannot prove any lower bound on $\operatorname*{Q}%
_{2}\left(  f\right)  $\ better than $\operatorname*{RC}\left(  f\right)  $,
where $\operatorname*{RC}\left(  f\right)  $\ is the \textit{randomized
certificate complexity} of $f$ (to be defined in Chapter \ref{CER}). \ But for
the collision function, $\operatorname*{RC}\left(  \operatorname*{Col}%
_{n}\right)  =O\left(  1\right)  $.}

My proof is an adaptation of the polynomial method, introduced to
quantum computing by Beals et al.\ \cite{bbcmw}. \ Their idea was to
reduce questions about quantum algorithms to easier questions about
multivariate polynomials. \ In particular, if a quantum algorithm
makes $T$ queries, then its acceptance probability is a polynomial
over the input bits of degree at most $2T$. \ So by showing that any
polynomial approximating the desired output has high degree, one
obtains a lower bound on $T$.

To lower-bound the degree of a multivariate polynomial, a key
technical trick is to construct a related \textit{univariate}
polynomial. \ Beals et al.\ \cite{bbcmw}, using a lemma due to
Minsky and Papert \cite{mp}, replace a polynomial $p\left(  X\right)
$\ (where $X$ is a bit string) by $q\left( \left|  X\right|  \right)
$\ (where $\left|  X\right|  $\ denotes the Hamming weight of $X$),
satisfying
\[
q\left(  k\right)  =\operatorname*{EX}\limits_{\left|  X\right|  =k}p\left(
X\right)  \
\]
and $\deg\left(  q\right)  \leq\deg\left(  p\right)  $.

Here I construct the univariate polynomial in a different way. \ I consider a
uniform distribution over $g$-to-one inputs, where $g$ might be greater than
$2$. \ Even though the problem is to distinguish $g=1$\ from $g=2$, the
acceptance probability must lie in the interval $\left[  0,1\right]  $ for all
$g$, and that is a surprisingly strong constraint. \ I show that the
acceptance probability is \textit{close} to a univariate polynomial in $g$ of
degree at most $2T$. \ I then obtain a lower bound by generalizing a classical
approximation theory result of Ehlich and Zeller \cite{ez}\ and Rivlin and
Cheney \cite{rc}. \ Much of the proof deals with the complication that $g$
does not divide $n$ in general.

Shortly after this work was completed, Shi \cite{shi} improved it to
give a tight lower bound of $\Omega\left(  n^{1/3}\right)  $ for the
collision problem, when the $x_{i}$ range from $1$\ to $3n/2$\
rather than from $1$ to $n$. \ For a range of size $n$, his bound
was $\Omega\left(  n^{1/4}\right) $. \ Subsequently\ Kutin
\cite{kutin} and Ambainis \cite{ambainis:col}\ showed a lower bound
of $\Omega\left(  n^{1/3}\right)  $ for a range of size $n$ as well.
\ By a simple reduction, these results imply a lower bound of
$\Omega\left(  n^{2/3}\right)  $\ for the \textit{element
distinctness problem}---that of deciding whether there exist $i\neq
j$\ such that $x_{i}=x_{j}$. \ The previous best known lower bound
was $\Omega\left( n^{1/2}\right)  $, and at the time of Shi's work,
the best known upper bound was $O\left(  n^{3/4}\right)  $, due to
Buhrman et al.\ \cite{bdhhmsw}. \ Recently, however, Ambainis
\cite{ambainis:walk}\ gave a novel algorithm based on quantum walks
that matches the $n^{2/3}$\ lower bound.

The chapter is organized as follows. \ Section \ref{MOTIVATIONCOL}\ motivates
the collision lower bound within quantum computing, pointing out connections
to collision-resistant hash functions, the nonabelian hidden subgroup problem,
statistical zero-knowledge, and information erasure. \ Section \ref{PRELIMCOL}%
\ gives technical preliminaries, Section \ref{BIVAR}\ proves the crucial fact
that the acceptance probability is \textquotedblleft almost\textquotedblright%
\ a univariate polynomial, and Section \ref{LB}\ completes the lower bound
argument. \ I conclude in Section \ref{OPENCOL}\ with some open problems. \ In
Section \ref{SETCOMP}\ I show a lower bound of $\Omega\left(  n^{1/7}\right)
$\ for the \textit{set comparison problem}, a variant of the collision problem
needed for the application to information erasure.

\section{Motivation\label{MOTIVATIONCOL}}

In Chapter \ref{PROLOGUE} I listed seven implications of the
collision lower bound; this section discusses a few of those
implications in more detail. \ The implication that motivated me
personally---concerning the computational power of so-called
\textit{hidden-variable theories}---is deferred to Chapter
\ref{QCHV}.

\subsection{Oracle Hardness Results}

The original motivation for the collision problem was to model
\textit{(strongly) collision-resistant hash functions} in cryptography.
\ There is a large literature on collision-resistant hashing; see
\cite{damgard,bsp}\ for example. \ When building secure digital signature
schemes, it is useful to have a family of hash functions $\left\{
H_{i}\right\}  $, such that finding a distinct $\left(  x,y\right)  $ pair
with $H_{i}\left(  x\right)  =H_{i}\left(  y\right)  $\ is computationally
intractable. \ A quantum algorithm for finding collisions using $O\left(
\operatorname*{polylog}\left(  n\right)  \right)  $ queries\ would render
\textit{all} hash functions insecure against quantum attack in this sense.
\ (Shor's algorithm \cite{shor}\ already renders hash functions based on
modular arithmetic insecure.) \ My result indicates that collision-resistant
hashing might still be possible in a quantum setting.

The collision problem also models the \textit{nonabelian hidden subgroup
problem}, of which graph isomorphism is a special case. \ Given a group $G$
and subgroup $H\leq G$, suppose we have oracle access to a function
$f:G\rightarrow\mathbb{N}$\ such that for all $g_{1},g_{2}\in G$, $f\left(
g_{1}\right)  =f\left(  g_{2}\right)  $ if and only if $g_{1}$\ and $g_{2}%
$\ belong to the same coset of $H$. \ Is there then an efficient quantum
algorithm to determine $H$? \ If $G$ is abelian, the work of Simon
\cite{simon}, Shor \cite{shor}, and Kitaev \cite{kitaev:meas}\ implies an
affirmative answer. \ If $G$ is nonabelian, though, efficient quantum
algorithms are known only for special cases \cite{eh,gsvv}. \ An $O\left(
\operatorname*{polylog}\left(  n\right)  \right)  $-query algorithm for the
collision problem would yield a polynomial-time algorithm to distinguish
$\left\vert H\right\vert =1$\ from $\left\vert H\right\vert =2$, which does
not exploit the group structure at all. \ My result implies that no such
algorithm exists.

Finally, the collision lower bound implies that there exists an oracle
relative to which $\mathsf{SZK}\not \subset \mathsf{BQP}$,\ where
$\mathsf{SZK}$\ is the class of problems having statistical zero-knowledge
proof protocols. \ For suppose that a verifier $V$ and prover $P$ both have
oracle access to a sequence $X=x_{1}\ldots x_{2^{n}}$, which is either
one-to-one or two-to-one. \ To verify with zero knowledge that $X$ is
one-to-one, $V$ can repeatedly choose an $i\in_{R}\left\{  1,\ldots
,2^{n}\right\}  $ and send $x_{i}$\ to $P$, whereupon $P$ must send $i$ back
to $V$. \ Thus, using standard diagonalization techniques, one can produce an
oracle $A$ such that $\mathsf{SZK}^{A}\not \subset \mathsf{BQP}^{A}$.

\subsection{Information Erasure\label{ERASURE}}

Let $f:\left\{  0,1\right\}  ^{n}\rightarrow\left\{  0,1\right\}  ^{m}$ with
$m\geq n$\ be a one-to-one function. \ Then we can consider two kinds of
quantum oracle for $f$:

\begin{enumerate}
\item[(A)] a \textit{standard oracle}, one that maps $\left|  x\right\rangle
\left|  z\right\rangle $ to

$\left|  x\right\rangle \left|  z\oplus f\left(  x\right)  \right\rangle $, or

\item[(B)] an \textit{erasing oracle} (as proposed by Kashefi et
al.\
\cite{kashefi}), which maps\ $\left\vert x\right\rangle $ to $\left\vert
f\left(  x\right)  \right\rangle $, in effect \textquotedblleft
erasing\textquotedblright\ $\left\vert x\right\rangle $.
\end{enumerate}

Intuitively erasing oracles seem at least as strong as standard ones, though
it is not clear how to simulate the latter with the former without also having
access to an oracle that maps $\left|  y\right\rangle $ to $\left|
f^{-1}\left(  y\right)  \right\rangle $. \ The question that concerns us here
is whether erasing oracles are \textit{more} useful than standard ones for
some problems. \ One-way functions provide a clue: if $f$ is one-way, then (by
assumption) $\left|  x\right\rangle \left|  f\left(  x\right)  \right\rangle
$\ can be computed efficiently, but if $\left|  f\left(  x\right)
\right\rangle $\ could be computed efficiently given $\left|  x\right\rangle
$\ then so could $\left|  x\right\rangle $\ given $\left|  f\left(  x\right)
\right\rangle $, and hence $f$ could be inverted. \ But can we find, for some
problem, an exponential gap between query complexity given a standard oracle
and query complexity given an erasing oracle?

In Section \ref{SETCOMP}\ I extend the collision lower bound to show an
affirmative answer. \ Define the set comparison problem of size $n$, or
$\operatorname*{SetComp}_{n}$, as follows. \ We are given as input two
sequences, $X=x_{1}\ldots x_{n}$ and $Y=y_{1}\ldots y_{n}$, such that for each
$i$, $x_{i},y_{i}\in\left\{  1,\ldots,2n\right\}  $. \ A query has the form
$\left(  b,i\right)  $, where $b\in\left\{  0,1\right\}  \ $and $i\in\left\{
1,\ldots,n\right\}  $, and produces as output $\left(  0,x_{i}\right)  $\ if
$b=0$\ and $\left(  1,y_{i}\right)  $\ if $b=1$. $\ $Sequences $X$ and $Y$ are
both one-to-one; that is, $x_{i}\neq x_{j}$\ and $y_{i}\neq y_{j}$\ for all
$i\neq j$. \ We are furthermore guaranteed that either

\begin{enumerate}
\item[(1)] $X$ and $Y$ are equal as sets (that is, $\left\{  x_{1}%
,\ldots,x_{n}\right\}  =\left\{  y_{1},\ldots,y_{n}\right\}  $) or

\item[(2)] $X$ and $Y$ are far as sets (that is,

$\left|  \left\{  x_{1},\ldots,x_{n}\right\}  \cup\left\{  y_{1},\ldots
,y_{n}\right\}  \right|  \geq1.1n$).
\end{enumerate}

As before the problem is to decide whether (1) or (2) holds.

This problem can be solved with high probability in a constant number of
queries using an erasing oracle, by using a trick similar to that of Watrous
\cite{watrous}\ for verifying group non-membership. \ First, using the oracle,
we prepare the uniform superposition%
\[
\frac{1}{\sqrt{2n}}\sum_{i\in\left\{  1,\ldots,n\right\}  }\left(  \left|
0\right\rangle \left|  x_{i}\right\rangle +\left|  1\right\rangle \left|
y_{i}\right\rangle \right)  \text{.}%
\]
We then apply a Hadamard gate to the first register, and finally we measure
the first register. \ If $X$ and $Y$ are equal as sets, then interference
occurs between every $\left(  \left|  0\right\rangle \left|  z\right\rangle
,\left|  1\right\rangle \left|  z\right\rangle \right)  $\ pair and we observe
$\left|  0\right\rangle $\ with certainty. \ But if $X$ and $Y$ are far as
sets, then basis states $\left|  b\right\rangle \left|  z\right\rangle $\ with
no matching $\left|  1-b\right\rangle \left|  z\right\rangle $\ have
probability weight at least $1/10$, and hence we observe $\left|
1\right\rangle $\ with probability at least $1/20$.

In Section \ref{SETCOMP}\ I sketch a proof that $\operatorname*{Q}_{2}\left(
\operatorname*{SetComp}_{n}\right)  =\Omega\left(  n^{1/7}\right)  $; that is,
no efficient quantum algorithm using a standard oracle exists for this
problem. \ Recently, Midrijanis \cite{midrijanis}\ gave a lower bound of
$\Omega\left(  \left(  n/\log n\right)  ^{1/5}\right)  $\ not merely for the
set comparison problem, but for the \textit{set equality problem} (where we
are promised that $X$ and $Y$ are either equal or disjoint).

\section{Preliminaries\label{PRELIMCOL}}

Let $A$ be a quantum query algorithm as defined in Section \ref{BLACKBOX}. \ A
basis state of $A$ is written $\left\vert i,z\right\rangle $. \ Then a query
replaces each $\left\vert i,z\right\rangle $ by $\left\vert i,z\oplus
x_{i}\right\rangle $, where $x_{i}$\ is exclusive-OR'ed into some specified
location of $z$. \ Between queries, the algorithm can perform any unitary
operation that does not depend on the input. \ Let $T$ be the total number of
queries. \ Also, assume for simplicity that all amplitudes are real; this
restriction is without loss of generality \cite{bv}.

Let $\alpha_{i,z}^{\left(  t\right)  }\left(  X\right)  $\ be the
amplitude of basis state $\left\vert i,z\right\rangle $\ after $t$
queries when the input is $X$. \ Also, let $\Delta\left(
x_{i},h\right)  =1$\ if $x_{i}=h$, and $\Delta\left(  x_{i},h\right)
=0$ if $x_{i}\neq h$. \ Let $P\left(  X\right) $ be the probability
that $A$ returns \textquotedblleft two-to-one\textquotedblright\
when the input is $X$. \ Then we obtain a simple variant of a lemma
due to Beals et al.\ \cite{bbcmw}.

\begin{lemma}
\label{poly}$P\left(  X\right)  $ is a multilinear polynomial of degree at
most $2T$ over the $\Delta\left(  x_{i},h\right)  $.
\end{lemma}

\begin{proof}
We show, by induction on $t$, that for all basis states $\left\vert
i,z\right\rangle $, the amplitude\ $\alpha_{i,z}^{\left(  t\right)  }\left(
X\right)  $\ is a multilinear polynomial of degree at most $t$ over the
$\Delta\left(  x_{i},h\right)  $. \ Since $P\left(  X\right)  $\ is a sum of
squares of $\alpha_{i,z}^{\left(  t\right)  }$'s, the lemma follows.

The base case ($t=0$) holds since, before making any queries, each
$\alpha_{i,z}^{\left(  t\right)  }$\ is a degree-$0$ polynomial over the
$\Delta\left(  x_{i},h\right)  $. \ A unitary transformation on the algorithm
part replaces each $\alpha_{i,z}^{\left(  t\right)  }$\ by a linear
combination of $\alpha_{i,z}^{\left(  t\right)  }$'s, and hence cannot
increase the degree. \ Suppose the lemma holds prior to the $t^{th}$\ query.
\ Then%
\[
\alpha_{i,z}^{\left(  t+1\right)  }\left(  X\right)  =\sum_{1\leq h\leq
n}\alpha_{i,z\oplus h}^{\left(  t\right)  }\left(  X\right)  \Delta\left(
x_{i},h\right)  ,
\]
and we are done.
\end{proof}

\section{Reduction to Bivariate Polynomial\label{BIVAR}}

Call the point $\left(  g,N\right)  \in\Re^{2}$\ an $\left(  n,T\right)
$-\textit{quasilattice point} if and only if

\begin{enumerate}
\item[(1)] $g$ and $N$ are integers, with $g$ dividing $N$,

\item[(2)] $1\leq g\leq\sqrt{n}$,

\item[(3)] $n\leq N\leq n+n/\left(  10T\right)  $, and

\item[(4)] if $g=1$\ then $N=n$.
\end{enumerate}

For quasilattice point $\left(  g,N\right)  $,\ define $\mathcal{D}_{n}\left(
g,N\right)  $\ to be the uniform distribution over all size-$n$ subfunctions
of $g$-to-1\ functions having domain $\left\{  1,\ldots,N\right\}  $\ and
range a subset of $\left\{  1,\ldots,n\right\}  $. \ More precisely: to draw
an $X$ from $\mathcal{D}_{n}\left(  g,N\right)  $, we first choose a set
$S\subseteq\left\{  1,\ldots,n\right\}  $\ with $\left\vert S\right\vert
=N/g\leq n$\ uniformly at random. \ We then choose a $g$-to-1 function
$\widehat{X}=\widehat{x}_{1}\ldots\widehat{x}_{N}$ from $\left\{
1,\ldots,N\right\}  $\ to $S$ uniformly at random. \ Finally we let
$x_{i}=\widehat{x}_{i}$\ for each $1\leq i\leq n$.

Let $P\left(  g,N\right)  $\ be the probability that algorithm $A$\ returns
$z=2$\ when the input is chosen from $\mathcal{D}_{n}\left(  g,N\right)  $:%
\[
P\left(  g,N\right)  =\operatorname*{EX}\limits_{X\in\mathcal{D}_{n}\left(
g,N\right)  }\left[  P\left(  X\right)  \right]  .
\]
We then have the following surprising characterization:

\begin{lemma}
\label{univ}For all sufficiently large $n$ and if $T\leq\sqrt{n}/3$, there
exists a bivariate polynomial $q\left(  g,N\right)  $\ of degree at most $2T$
such that if $\left(  g,N\right)  $\ is a quasilattice point, then%
\[
\left|  P\left(  g,N\right)  -q\left(  g,N\right)  \right|  <0.182
\]
(where the constant $0.182$ can be made arbitrarily small by adjusting parameters).
\end{lemma}

\begin{proof}
Let $I$ be a product of $\Delta\left(  x_{i},h\right)  \ $variables, with
degree $r\left(  I\right)  $, and let $I\left(  X\right)  \in\left\{
0,1\right\}  $\ be $I$ evaluated on input $X$. \ Then define%
\[
\gamma\left(  I,g,N\right)  =\operatorname*{EX}\limits_{X\in\mathcal{D}%
_{n}\left(  g,N\right)  }\left[  I\left(  X\right)  \right]
\]
to be the probability that monomial $I$\ evaluates to $1$ when the input is
drawn from $\mathcal{D}_{n}\left(  g,N\right)  $. \ Then by Lemma \ref{poly},
$P\left(  X\right)  $\ is a polynomial of degree at most $2T$\ over $X$, so%
\begin{align*}
P\left(  g,N\right)   &  =\operatorname*{EX}\limits_{X\in\mathcal{D}%
_{n}\left(  g,N\right)  }\left[  P\left(  X\right)  \right] \\
&  =\operatorname*{EX}\limits_{X\in\mathcal{D}_{n}\left(  g,N\right)  }\left[
\sum_{I:r\left(  I\right)  \leq2t}\beta_{I}I\left(  X\right)  \right] \\
&  =\sum_{I:r\left(  I\right)  \leq2T}\beta_{I}\gamma\left(  I,g,N\right)
\end{align*}
for some coefficients $\beta_{I}$.

We now calculate $\gamma\left(  I,g,N\right)  $. \ Assume without loss of
generality that for all $\Delta\left(  x_{i},h_{1}\right)  ,\Delta\left(
x_{j},h_{2}\right)  \in I$, either $i\neq j$\ or $h_{1}=h_{2}$, since
otherwise $\gamma\left(  I,g,N\right)  =0$.

Define the ``range'' $Z\left(  I\right)  $\ of $I$ to be the set of all $h$
such that $\Delta\left(  x_{i},h\right)  \in I$. \ Let $w\left(  I\right)
=\left|  Z\left(  I\right)  \right|  $; then we write $Z\left(  I\right)
=\left\{  z_{1},\ldots,z_{w\left(  I\right)  }\right\}  $. $\ $Clearly
$\gamma\left(  I,g,N\right)  =0$\ unless $Z\left(  I\right)  \in S$, where $S$
is the range of $\widehat{X}$. \ By assumption,%
\[
\frac{N}{g}\geq\frac{n}{\sqrt{n}}\geq2T\geq r\left(  I\right)
\]
\ so the number of possible $S$ is $\dbinom{n}{N/g}$\ and, of these, the
number that contain $Z$\ is $\dbinom{n-w\left(  I\right)  }{N/g-w\left(
I\right)  }$.

Then, conditioned on $Z\in S$, what is the probability that $\gamma\left(
I,g,N\right)  =1$? \ The total number of $g$-to-1\ functions with domain size
$N$\ is $N!/\left(  g!\right)  ^{N/g},$ since we can permute the $N$\ function
values arbitrarily, but must not count permutations that act only within the
$N/g$\ constant-value blocks of size $g$.

Among these functions, how many satisfy $\gamma\left(  I,g,N\right)  =1$?
\ Suppose that, for each $1\leq j\leq w\left(  I\right)  $, there are
$r_{j}\left(  I\right)  $\ distinct $i$\ such that $\Delta\left(  x_{i}%
,z_{j}\right)  \in I$. \ Clearly%
\[
r_{1}\left(  I\right)  +\cdots+r_{w\left(  I\right)  }\left(  I\right)
=r\left(  I\right)  .
\]
Then we can permute the $\left(  N-r\left(  I\right)  \right)  !$ function
values outside of $I$ arbitrarily, but must not count permutations that act
only within the $N/g$ constant-value blocks, which have size either $g$ or
$g-r_{i}\left(  I\right)  $ for some $i$. \ So the number of functions for
which $\gamma\left(  I,g,N\right)  =1$\ is%
\[
\frac{\left(  N-r\left(  I\right)  \right)  !}{\left(  g!\right)
^{N/g-w\left(  I\right)  }%
%TCIMACRO{\dprod \nolimits_{i=1}^{w\left(  I\right)  }}%
%BeginExpansion
{\displaystyle\prod\nolimits_{i=1}^{w\left(  I\right)  }}
%EndExpansion
\left(  g-r_{i}\left(  I\right)  \right)  !}.
\]

Putting it all together, \
\begin{align*}
\gamma\left(  I,g,N\right)   &  =\frac{\dbinom{n-w\left(  I\right)
}{N/g-w\left(  I\right)  }}{\dbinom{n}{N/g}}\cdot\frac{\left(  N-r\left(
I\right)  \right)  !\left(  g!\right)  ^{N/g}}{\left(  g!\right)
^{N/g-w\left(  I\right)  }N!%
%TCIMACRO{\dprod \nolimits_{i=1}^{w\left(  I\right)  }}%
%BeginExpansion
{\displaystyle\prod\nolimits_{i=1}^{w\left(  I\right)  }}
%EndExpansion
\left(  g-r_{i}\left(  I\right)  \right)  !}\\
=  &  \frac{\left(  N-r\left(  I\right)  \right)  !\left(  n-w\left(
I\right)  \right)  !\left(  N/g\right)  !}{N!n!\left(  N/g-w\left(  I\right)
\right)  !}\cdot\frac{\left(  g!\right)  ^{w\left(  I\right)  }}{%
%TCIMACRO{\dprod \nolimits_{i=1}^{w\left(  I\right)  }}%
%BeginExpansion
{\displaystyle\prod\nolimits_{i=1}^{w\left(  I\right)  }}
%EndExpansion
\left(  g-r_{i}\left(  I\right)  \right)  !}\\
=  &  \frac{\left(  N-r\left(  I\right)  \right)  !}{N!}\frac{\left(
n-w\left(  I\right)  \right)  !}{n!}\cdot%
%TCIMACRO{\dprod \limits_{i=0}^{w\left(  I\right)  -1}}%
%BeginExpansion
{\displaystyle\prod\limits_{i=0}^{w\left(  I\right)  -1}}
%EndExpansion
\left(  \frac{N}{g}-i\right)
%TCIMACRO{\dprod \limits_{i=1}^{w\left(  I\right)  }}%
%BeginExpansion
{\displaystyle\prod\limits_{i=1}^{w\left(  I\right)  }}
%EndExpansion
\left[  g%
%TCIMACRO{\dprod \limits_{j=1}^{r_{i}\left(  I\right)  -1}}%
%BeginExpansion
{\displaystyle\prod\limits_{j=1}^{r_{i}\left(  I\right)  -1}}
%EndExpansion
\left(  g-j\right)  \right] \\
=  &  \frac{\left(  N-2T\right)  !n!}{N!\left(  n-2T\right)  !}\widetilde
{q}_{n,T,I}\left(  g,N\right)
\end{align*}
where%
\[
\widetilde{q}_{n,T,I}\left(  g,N\right)  =\frac{\left(  n-w\left(  I\right)
\right)  !\left(  n-2T\right)  !}{\left(  n!\right)  ^{2}}\cdot%
%TCIMACRO{\dprod \limits_{i=r\left(  I\right)  }^{2T-1}}%
%BeginExpansion
{\displaystyle\prod\limits_{i=r\left(  I\right)  }^{2T-1}}
%EndExpansion
\left(  N-i\right)
%TCIMACRO{\dprod \limits_{i=0}^{w\left(  I\right)  -1}}%
%BeginExpansion
{\displaystyle\prod\limits_{i=0}^{w\left(  I\right)  -1}}
%EndExpansion
\left(  N-gi\right)
%TCIMACRO{\dprod \limits_{i=1}^{w\left(  I\right)  }}%
%BeginExpansion
{\displaystyle\prod\limits_{i=1}^{w\left(  I\right)  }}
%EndExpansion%
%TCIMACRO{\dprod \limits_{j=1}^{r_{i}\left(  I\right)  -1}}%
%BeginExpansion
{\displaystyle\prod\limits_{j=1}^{r_{i}\left(  I\right)  -1}}
%EndExpansion
\left(  g-j\right)
\]
\ is a bivariate polynomial of total degree at most%
\[
\left(  2T-r\left(  I\right)  \right)  +w\left(  I\right)  +\left(  r\left(
I\right)  -w\left(  I\right)  \right)  =2T.
\]
(Note that in the case $r_{i}\left(  I\right)  >g$ for some $i$, this
polynomial evaluates to $0$, which is what it ought to do.) \ Hence%
\begin{align*}
P\left(  g,N\right)   &  =\sum_{I:r\left(  I\right)  \leq2T}\beta_{I}%
\gamma\left(  I,g,N\right) \\
&  =\frac{\left(  N-2T\right)  !n!}{N!\left(  n-2T\right)  !}q\left(
g,N\right)
\end{align*}
where%
\[
q\left(  g,N\right)  =\sum_{I:r\left(  I\right)  \leq2T}\beta_{I}\widetilde
{q}_{n,T,I}\left(  g,N\right)  .
\]

Clearly%
\[
\frac{\left(  N-2T\right)  !n!}{N!\left(  n-2T\right)  !}\leq1.
\]
Since $N\leq n+n/\left(  10T\right)  $ and $T\leq\sqrt{n}/3$, we also have%
\begin{align*}
\frac{\left(  N-2T\right)  !n!}{N!\left(  n-2T\right)  !}  &  \geq\left(
\frac{n-2T+1}{N-2T+1}\right)  ^{2T}\\
&  \geq\exp\left\{  -\frac{1}{5}\frac{n}{n-\left(  2T+1\right)  /n}\right\} \\
&  \geq0.818
\end{align*}
for all sufficiently large $n$.  Thus, since $0\leq P\left(  g,N\right)  \leq1$,%
\[
\left|  P\left(  g,N\right)  -q\left(  g,N\right)  \right|  <0.182
\]
and we are done.
\end{proof}

\section{Lower Bound\label{LB}}

We have seen that, if a quantum algorithm for the collision problem
makes few queries, then its acceptance probability can be
approximated by a low-degree bivariate polynomial. \ This section
completes the lower bound proof by showing that no such polynomial
exists. \ To do so, it generalizes an approximation theory result
due to Rivlin and Cheney \cite{rc}\ and (independently) Ehlich and
Zeller \cite{ez}. \ That result was applied to query complexity by
Nisan and Szegedy \cite{ns}\ and later by Beals et al.\
\cite{bbcmw}.

\begin{theorem}
\label{thetheorem}$\operatorname*{Q}_{2}\left(  \operatorname*{Col}%
_{n}\right)  =\Omega\left(  n^{1/5}\right)  .$
\end{theorem}

\begin{proof}
Let $g$\ have range $1\leq g\leq G$. \ Then the quasilattice points $\left(
g,N\right)  $\ all lie in the rectangular region $R=\left[  1,G\right]
\times\left[  n,n+n/\left(  10T\right)  \right]  $. \ Recalling the polynomial
$q\left(  g,N\right)  $\ from Lemma \ref{univ},\ define%
\[
d\left(  q\right)  =\max_{\left(  g,N\right)  \in R}\left(  \max\left\{
\left\vert \frac{\partial q}{\partial g}\right\vert ,\frac{n}{10T\left(
G-1\right)  }\cdot\left\vert \frac{\partial q}{\partial N}\right\vert
\right\}  \right)  .
\]
Suppose without loss of generality that we require%
\[
P\left(  1,n\right)  \leq\frac{1}{10}\ \ \text{and \ }P\left(  2,n\right)
\geq\frac{9}{10}%
\]
(that is, algorithm $A$ distinguishes 1-to-1 from 2-to-1 functions with error
probability at most $1/10$). \ Then, since%
\[
\left\vert P\left(  g,N\right)  -q\left(  g,N\right)  \right\vert <0.182
\]
by elementary calculus we have%
\[
d\left(  q\right)  \geq\max_{1\leq g\leq2}\frac{\partial q}{\partial
g}>0.8-2\left(  0.182\right)  =\allowbreak0.436.
\]
$\,$An inequality due to Markov (see \cite{cheney,ns})\ states that, for a
univariate polynomial $p$, if $b_{1}\leq p\left(  x\right)  \leq b_{2}$\ for
all\ $a_{1}\leq x\leq a_{2}$, then%
\[
\max_{a\left[  1\right]  \leq x\leq a\left[  2\right]  }\left\vert
\frac{dp\left(  x\right)  }{dx}\right\vert \leq\frac{b_{2}-b_{1}}{a_{2}-a_{1}%
}\deg\left(  p\right)  ^{2}.
\]
Clearly for every point $\left(  \widehat{g},\widehat{N}\right)  \in R$, there
exists a quasilattice point $\left(  g,N\right)  $\ for which%
\[
\left\vert g-\widehat{g}\right\vert \leq1\ \ \text{and \ }\left\vert
N-\widehat{N}\right\vert \leq G.
\]
For take $g=\left\lceil \widehat{g}\right\rceil $---or, in the
special case $\widehat{g}=1$, take $g=2$, since there is only one
quasilattice point with $g=1$.  Furthermore, since $P\left(
g,N\right)  $\ represents an acceptance
probability at such a point, we have%
\[
-0.182<q\left(  g,N\right)  <1.182.
\]
Observe that for all $\left(  \widehat{g},\widehat{N}\right)  \in R$, \
\[
-0.182-\left(  \frac{10TG\left(  G-1\right)  }{n}+1\right)  d\left(  q\right)
<q\left(  \widehat{g},\widehat{N}\right)  <1.182+\left(  \frac{10TG\left(
G-1\right)  }{n}+1\right)  d\left(  q\right)  .
\]

For consider a quasilattice point close to $\left(  \widehat{g},\widehat
{N}\right)  $, and note that the maximum-magnitude derivative is at most
$d\left(  q\right)  $\ in the $g$ direction and $10T\left(  G-1\right)
d\left(  q\right)  /n$ in the $N$ direction.

Let $\left(  g^{\ast},N^{\ast}\right)  $\ be a point in $R$ at which the
weighted maximum-magnitude derivative $d\left(  q\right)  $\ is attained.
\ Suppose first that the maximum is attained in the $g$\ direction. \ Then
$q\left(  g,N^{\ast}\right)  $\ (with $N^{\ast}$ constant) is a univariate
polynomial with%
\[
\left\vert \frac{dq\left(  g,N^{\ast}\right)  }{dg}\right\vert >0.436
\]
for some $1\leq g\leq G$. \ So%
\begin{align*}
2T  &  \geq\deg\left(  q\left(  g,N^{\ast}\right)  \right) \\
&  \geq\sqrt{\frac{d\left(  q\right)  \left(  G-1\right)  }{1.364+2d\left(
q\right)  \left(  1+10TG\left(  G-1\right)  /n\right)  }}\\
&  =\Omega\left(  \min\left\{  \sqrt{G},\sqrt{\frac{n}{TG}}\right\}  \right)
.
\end{align*}

Similarly, suppose the maximum $d\left(  q\right)  $\ is attained\ in the $N$
direction. \ Then $q\left(  g^{\ast},N\right)  $\ (with $g^{\ast}$ constant)
is a univariate polynomial with%
\[
\left|  \frac{dq\left(  g^{\ast},N\right)  }{dN}\right|  >\frac{0.436T\left(
G-1\right)  }{n}%
\]
for some $n\leq N\leq n+n/\left(  10T\right)  $. \ So%
\begin{align*}
2T  &  \geq\sqrt{\frac{\left(  10T\left(  G-1\right)  /n\right)  d\left(
q\right)  n/\left(  10T\right)  }{1.364+2d\left(  q\right)  \left(
1+10TG\left(  G-1\right)  /n\right)  }}\\
&  \geq\Omega\left(  \min\left\{  \sqrt{G},\sqrt{\frac{n}{TG}}\right\}
\right)  .
\end{align*}

One can show that the lower bound on $T$\ is optimized when we take
$G=n^{2/5}\leq\sqrt{n}$. \ Then%
\begin{align*}
T  &  =\Omega\left(  \min\left\{  n^{1/5},\frac{\sqrt{n}}{\sqrt{T}n^{1/5}%
}\right\}  \right)  ,\\
T  &  =\Omega\left(  n^{1/5}\right)
\end{align*}
and we are done.
\end{proof}

\section{Set Comparison\label{SETCOMP}}

Here I sketch a proof that $\operatorname*{Q}_{2}\left(
\operatorname*{SetComp}_{n}\right)  =\Omega\left(  n^{1/7}\right)  $, where
$\operatorname*{SetComp}_{n}$\ is the set comparison problem of size $n$ as
defined in Section \ref{ERASURE}. \

The idea is the following. \ We need a distribution of inputs with a parameter
$g$, such that the inputs are one-to-one when $g=1$ or $g=2$---since otherwise
the problem of distinguishing $g=1$\ from $g=2$\ would be ill-defined for
erasing oracles. \ On the other hand, the inputs must \textit{not} be
one-to-one for all $g>2$---since otherwise the lower bound for standard
oracles would apply also to erasing oracles, and we would not obtain a
separation between the two. \ Finally, the acceptance probability must be
close to a polynomial in $g$.

The solution is to consider $\kappa\left(  g\right)  $-to-one inputs, where%
\[
\kappa\left(  g\right)  =4g^{2}-12g+9.
\]
is a quadratic with $\kappa\left(  1\right)  =\kappa\left(  2\right)  =1$.
\ The total range of the inputs (on sequences $X$ and $Y$\ combined) has size
roughly $n/g$; thus, we can tell the $g=1$\ inputs apart from the
$g=2$\ inputs using an erasing oracle, even though $\kappa\left(  g\right)
$\ is the same for both. \ The disadvantage is that, because $\kappa\left(
g\right)  $\ increases quadratically rather than linearly in $g$, the
quasilattice points become sparse more quickly. \ That is what weakens the
lower bound from $\Omega\left(  n^{1/5}\right)  $\ to $\Omega\left(
n^{1/7}\right)  $. \ Note that, using the ideas of Shi \cite{shi}, one can
improve my lower bound on $\operatorname*{Q}_{2}\left(
\operatorname*{SetComp}_{n}\right)  $\ to $\Omega\left(  n^{1/6}\right)  $.

Call $\left(  g,N,M\right)  \in\Re^{3}$\ an $\left(  n,T\right)
$\textit{-super-quasilattice point} if and only if

\begin{enumerate}
\item[(1)] $g$ is an integer in $\left[  1,n^{1/3}\right]  $,

\item[(2)] $N$ and $M$ are integers in $\left[  n,n\left(  1+1/\left(
100T\right)  \right)  \right]  $,

\item[(3)] $g$ divides $N$,

\item[(4)] if $g=1$\ then $N=n$,

\item[(5)] $\kappa\left(  g\right)  $ divides $M$, and

\item[(6)] if $g=2$ then $M=n$.
\end{enumerate}

For super-quasilattice point $\left(  g,N,M\right)  $, we draw input $\left(
X,Y\right)  =\left(  x_{1}\ldots x_{n},y_{1}\ldots y_{n}\right)  $\ from
distribution $\mathcal{L}_{n}\left(  g,N,M\right)  $\ as follows. \ We first
choose a set $S\subseteq\left\{  1,\ldots,2n\right\}  $\ with $\left\vert
S\right\vert =2N/g\leq2n$ uniformly at random. \ We then choose two sets
$S_{X},S_{Y}\subseteq S$\ with $\left\vert S_{X}\right\vert =\left\vert
S_{X}\right\vert =M/\kappa\left(  g\right)  \leq\left\vert S\right\vert $,
uniformly at random and independently. \ Next we choose $\kappa\left(
g\right)  $-1 functions $\widehat{X}=\widehat{x}_{1}\ldots\widehat{x}_{N}$
$:\left\{  1,\ldots,M\right\}  \rightarrow S_{X}$ and $\widehat{Y}=\widehat
{y}_{1}\ldots\widehat{y}_{N}$ $:\left\{  1,\ldots,M\right\}  \rightarrow
S_{Y}$\ uniformly at random and independently. \ Finally we let $x_{i}%
=\widehat{x}_{i}$\ and $y_{i}=\widehat{y}_{i}$\ for each $1\leq i\leq n$.

Define sets $X_{S}=\left\{  x_{1},\ldots,x_{n}\right\}  $\ and $Y_{S}=\left\{
y_{1},\ldots,y_{n}\right\}  $. \ Suppose $g=1$ and $N=M=n$; then by Chernoff
bounds,%
\[
\Pr_{\left(  X,Y\right)  \in\mathcal{L}_{n}\left(  1,n,n\right)  }\left[
\left\vert X_{S}\cup Y_{S}\right\vert <1.1n\right]  \leq2e^{-n/10}.
\]
Thus, if algorithm $A$ can distinguish $\left\vert X_{S}\cup Y_{S}\right\vert
=n$\ from $\left\vert X_{S}\cup Y_{S}\right\vert \geq1.1n$\ with probability
at least $9/10$, then it can distinguish $\left(  X,Y\right)  \in
\mathcal{L}_{n}\left(  1,n,n\right)  $\ from $\left(  X,Y\right)
\in\mathcal{L}_{n}\left(  2,n,n\right)  $\ with probability at least
$9/10-2e^{-n/10}$. \ So a lower bound for the latter problem implies an
equivalent lower bound for the former.

Define $P\left(  X,Y\right)  $\ to be the probability that the algorithm
returns that $X$ and $Y$ are far on input $\left(  X,Y\right)  $, and let%
\[
P\left(  g,N,M\right)  =\operatorname*{EX}\limits_{\left(  X,Y\right)
\in\mathcal{L}_{n}\left(  g,N,M\right)  }\left[  P\left(  X,Y\right)  \right]
.
\]
We then have

\begin{lemma}
\label{univ2}For all sufficiently large $n$ and if $T\leq n^{1/3}/8$, there
exists a trivariate polynomial $q\left(  g,N,M\right)  $\ of degree at most
$8T$ such that if $\left(  g,N,M\right)  $\ is a super-quasilattice point,
then%
\[
\left\vert P\left(  g,N,M\right)  -q\left(  g,N,M\right)  \right\vert
<\varepsilon
\]
for some constant $0<\varepsilon<1/2$.
\end{lemma}

\begin{proof}
[Proof Sketch]By analogy to Lemma \ref{poly}, $P\left(  X,Y\right)  $\ is a
multilinear polynomial of degree at most $2T$\ over variables of the form
$\Delta\left(  x_{i},h\right)  $\ and $\Delta\left(  y_{i},h\right)  $. \ Let
$I\left(  X,Y\right)  =I_{X}\left(  X\right)  I_{Y}\left(  Y\right)  $ where
$I_{X}$\ is\ a product of $r_{X}\left(  I\right)  $\ distinct $\Delta\left(
x_{i},h\right)  \ $variables and $I_{Y}$ is a product of $r_{Y}\left(
I\right)  \ $distinct$\ \Delta\left(  y_{i},h\right)  $\ variables. \ Let
$r\left(  I\right)  =r_{X}\left(  I\right)  +r_{Y}\left(  I\right)  $.
\ Define%
\[
\gamma\left(  I,g,N,M\right)  =\operatorname*{EX}\limits_{\left(  X,Y\right)
\in\mathcal{L}_{n}\left(  g,N,M\right)  }\left[  I\left(  X,Y\right)  \right]
;
\]
then%
\[
P\left(  g,N,M\right)  =\sum_{I:r\left(  I\right)  \leq2T}\beta_{I}%
\gamma\left(  I,g,N,M\right)
\]
for some coefficients $\beta_{I}$. \ We now calculate $\gamma\left(
I,g,N,M\right)  $. \ As before we assume there are no pairs of variables
$\Delta\left(  x_{i},h_{1}\right)  ,\Delta\left(  x_{i},h_{2}\right)  \in I$
with $h_{1}\neq h_{2}$. \ Let $Z_{X}\left(  I\right)  $\ be the range of
$I_{X}$ and let $Z_{Y}\left(  I\right)  $\ be the range of $I_{Y}$.\ \ Then
let $Z\left(  I\right)  =Z_{X}\left(  I\right)  \cup Z_{Y}\left(  I\right)  $.
\ Let $w_{X}\left(  I\right)  =\left\vert Z_{X}\left(  I\right)  \right\vert
$, $w_{Y}\left(  I\right)  =\left\vert Z_{Y}\left(  I\right)  \right\vert $,
and $w\left(  I\right)  =\left\vert Z\left(  I\right)  \right\vert $. \ By
assumption%
\[
\frac{N}{g}\geq\frac{M}{\kappa\left(  g\right)  }\geq\frac{1}{4}n^{1/3}\geq2T
\]
so%
\[
\Pr\left[  Z\left(  I\right)  \subseteq S\right]  =\frac{\dbinom{2n-w\left(
I\right)  }{2N/g-w\left(  I\right)  }}{\dbinom{2n}{2N/g}}.
\]
The probabilities that $Z_{X}\left(  I\right)  \subseteq S_{X}$ given
$Z\left(  I\right)  \subseteq S$ and $Z_{Y}\left(  I\right)  \subseteq S_{Y}$
given $Z\left(  I\right)  \subseteq S$ can be calculated similarly.

Let $r_{X,1}\left(  I\right)  ,\ldots,r_{X,w\left[  X\right]  \left(
I\right)  }\left(  I\right)  $\ be the multiplicities of the range elements in
$Z_{X}\left(  I\right)  $, so that
\[
r_{X,1}\left(  I\right)  +\cdots+r_{X,w\left[  X\right]  \left(  I\right)
}\left(  I\right)  =r_{X}\left(  I\right)  .
\]
\ Then%
\[
\Pr\left[  I_{X}\left(  X\right)  \,\,|\,\,Z_{X}\left(  I\right)  \subseteq
S_{X}\right]  =\frac{\left(  M-r_{X}\left(  I\right)  \right)  !}{M!}%
%TCIMACRO{\dprod \limits_{i=1}^{w\left[  X\right]  \left(  I\right)  }}%
%BeginExpansion
{\displaystyle\prod\limits_{i=1}^{w\left[  X\right]  \left(  I\right)  }}
%EndExpansion%
%TCIMACRO{\dprod \limits_{j=0}^{r\left[  X,i\right]  \left(  I\right)  -1}}%
%BeginExpansion
{\displaystyle\prod\limits_{j=0}^{r\left[  X,i\right]  \left(  I\right)  -1}}
%EndExpansion
\left(  \kappa\left(  g\right)  -j\right)
\]
and similarly for $\Pr\left[  I_{Y}\left(  Y\right)
\,\,|\,\,Z_{Y}\left( I\right)  \subseteq S_{Y}\right]  $.

Putting it all together and manipulating, we obtain (analogously to Lemma
\ref{poly}) that%
\[
\gamma\left(  I,g,N,M\right)  \approx\widetilde{q}_{n,T,I}\left(
g,N,M\right)
\]
where $\widetilde{q}_{n,T,I}\left(  g,N,M\right)  $\ is a trivariate
polynomial in $\left(  g,N,M\right)  $ of total degree at most $8T$. \ Thus%
\[
P\left(  g,N,M\right)  \approx q\left(  g,N,M\right)
\]
where $q\left(  g,N,M\right)  $\ is a polynomial of total degree at most $8T$.
\ The argument that $q$\ approximates $P$\ to within a constant is analogous
to that of Lemma \ref{univ}.
\end{proof}

The remainder of the lower bound argument follows the lines of Theorem
\ref{thetheorem}.

\begin{theorem}
\label{thetheorem2}$\operatorname*{Q}_{2}\left(  \operatorname*{SetComp}%
_{n}\right)  =\Omega\left(  n^{1/7}\right)  $.
\end{theorem}

\begin{proof}
[Proof Sketch]Let $g\in\left[  1,G\right]  $ for some $G\leq n^{1/3}$. \ Then
the super-quasilattice points $\left(  g,N,M\right)  $\ all lie in $R=\left[
1,G\right]  \times\left[  n,n+n/\left(  100T\right)  \right]  ^{2}$. \ Define
$d(q)$ to be%
\[
\max_{\left(  g,N,M\right)  \in R}\left(  \operatorname*{max}\left\{
\left\vert \frac{\partial q}{\partial g}\right\vert ,\frac{n/100T}{\left(
G-1\right)  }\left\vert \frac{\partial q}{\partial N}\right\vert
,\frac{n/100T}{\left(  G-1\right)  }\left\vert \frac{\partial q}{\partial
M}\right\vert \right\}  \right)  .
\]
Then $d\left(  q\right)  \geq\delta$ for some constant $\delta>0$, by Lemma
\ref{univ2}.

For every point $\left(  \widehat{g},\widehat{N},\widehat{M}\right)  \in R$,
there exists a super-quasilattice point $\left(  g,N,M\right)  $\ such that
$\left\vert g-\widehat{g}\right\vert \leq1$, $\left\vert N-\widehat
{N}\right\vert \leq G$, and $\left\vert M-\widehat{M}\right\vert \leq
\kappa\left(  G\right)  .$ \ Hence, $q\left(  \widehat{g},\widehat{N}%
,\widehat{M}\right)  $\ can deviate from $\left[  0,1\right]  $\ by at most%
\[
O\left(  \left(  \frac{TG^{3}}{n}+1\right)  d\left(  q\right)  \right)  .
\]

Let $\left(  g^{\ast},N^{\ast},M^{\ast}\right)  $\ be a point in $R$ at which
$d\left(  q\right)  $\ is attained. \ Suppose $d\left(  q\right)  $\ is
attained in the $g$ direction; the cases of the $N$ and $M$ directions are
analogous. \ Then $q\left(  g,N^{\ast},M^{\ast}\right)  $\ is a univariate
polynomial in $g$, and%
\begin{align*}
8T  &  \geq\deg\left(  q\left(  g,N^{\ast},M^{\ast}\right)  \right) \\
&  =\Omega\left(  \min\left\{  \sqrt{G},\sqrt{\frac{n}{TG^{2}}}\right\}
\right)  .
\end{align*}
One can show that the bound is optimized when we take $G=n^{2/7}\leq n^{1/3}$.
\ Then%
\begin{align*}
T  &  =\Omega\left(  \min\left\{  n^{1/7},\frac{\sqrt{n}}{\sqrt{T}n^{2/7}%
}\right\}  \right)  ,\\
T  &  =\Omega\left(  n^{1/7}\right)  .
\end{align*}

\end{proof}

\section{Open Problems\label{OPENCOL}}

In my original paper on the collision problem, I listed four open problems:
improving the collision lower bound to $\Omega\left(  n^{1/3}\right)  $;
showing any nontrivial quantum lower bound for the set equality problem;
proving a time-space tradeoff lower bound for the collision problem; and
deciding whether quantum query complexity and degree as a real polynomial are
always asymptotically equal. \ Happily, three of these problems have since
been resolved \cite{kutin,midrijanis,ambainis:deg}, but the time-space
tradeoff remains wide open. \ We would like to say (for example) that if a
quantum computer is restricted to using $O\left(  \log n\right)  $\ qubits,
then it needs $\Theta\left(  \sqrt{n}\right)  $\ queries for the collision
problem, ordinary Grover search being the best possible algorithm.
\ Currently, we cannot show such a result for \textit{any} problem with
Boolean output, only for problems such as sorting with a large non-Boolean
output \cite{ksw}.

Another problem is to give an oracle relative to which $\mathsf{SZK}%
\not \subset \mathsf{QMA}$, where $\mathsf{QMA}$\ is Quantum Merlin Arthur as
defined in \cite{watrous}. \ In other words, show that if a function is
one-to-one rather than two-to-one, then this fact cannot be verified using a
small number of quantum queries, even with the help of a succinct quantum proof.

Finally, is it the case that for all (partial or total)\ functions $f$\ that
are invariant under permutation symmetry, $\operatorname*{R}_{2}\left(
f\right)  $ and $\operatorname*{Q}_{2}\left(  f\right)  $ are polynomially related?

\chapter{Local Search\label{PLS}}

This chapter deals with the following problem.

\textsc{Local Search.}\hfill\textit{Given an undirected graph
}$G=\left(V,E\right)$\textit{\ and function
}$f:V\rightarrow\mathbb{N}$\textit{,
find a local minimum of }$\mathit{f}$\textit{---that is, a vertex }%
$v$\textit{\ such that }$f\left(  v\right)  \leq f\left(  w\right)
$\textit{\ for all neighbors }$w$\textit{ of }$v$\textit{.}

We will be interested in the number of queries that an algorithm needs to
solve this problem, where a query just returns $f\left(  v\right)  $\ given
$v$. \ We will consider deterministic, randomized, and quantum algorithms.
\ Section \ref{MOTIVATIONPLS}\ motivates the problem theoretically and
practically; this section explains the results.

First, though, we need some simple observations. \ If $G$ is the
complete graph of size $N$, then clearly $\Omega\left(  N\right)  $\
queries are needed to find a local minimum (or $\Omega\left(
\sqrt{N}\right)  $\ with a quantum computer). \ At the other
extreme, if $G$\ is a line of length $N$, then even a deterministic
algorithm can find a local minimum in $O\left( \log N\right)  $\
queries, using binary search: query the middle two vertices, $v$ and
$w$. \ If $f\left(  v\right)  \leq f\left(  w\right)  $, then search
the line of length $\left(  N-2\right)  /2$\ connected to $v$;
otherwise search the line connected to $w$. \ Continue recursively
in this manner until a local minimum is found.

So the interesting case is when $G$ is a graph of `intermediate'
connectedness: for example, the Boolean hypercube $\left\{  0,1\right\}  ^{n}%
$, with two vertices adjacent if and only if they have Hamming distance
$1$.\ \ For this graph, Llewellyn, Tovey, and Trick \cite{ltt}\ showed a
$\Omega\left(  2^{n}/\sqrt{n}\right)  $ lower bound\ on the number of queries
needed by any deterministic algorithm, using a simple adversary argument.
\ Intuitively, until the set of vertices queried so far comprises a
\textit{vertex cut} (that is, splits the graph into two or more connected
components), an adversary is free to return a descending sequence of
$f$-values: $f\left(  v_{1}\right)  =2^{n}$\ for the first vertex $v_{1}%
$\ queried by the algorithm, $f\left(  v_{2}\right)  =2^{n}-1$\ for
the second vertex queried, and so on. \ Moreover, once the set of
queried vertices does comprise a cut, the adversary can choose the
largest connected component of unqueried vertices, and restrict the
problem recursively\ to that component. \ So to lower-bound the
deterministic query complexity, it suffices to lower-bound the size
of any cut that splits the graph into two reasonably large
components.\footnote{Llewellyn et al.\ actually give a tight
characterization of deterministic query complexity in terms of
vertex cuts.} \ For the Boolean hypercube, Llewellyn et al.\ showed
that the best one can do is essentially to query all $\Omega\left(
2^{n}/\sqrt{n}\right)  $\ vertices of Hamming weight $n/2$.

Llewellyn et al.'s argument fails completely in the case of randomized
algorithms. \ By Yao's minimax principle, what we want here is a fixed
\textit{distribution} $\mathcal{D}$ over functions $f:\left\{  0,1\right\}
^{n}\rightarrow\mathbb{N}$, such that any deterministic algorithm needs many
queries to find a local minimum of $f$, with high probability if $f$ is drawn
from $\mathcal{D}$. \ Taking $\mathcal{D}$\ to be uniform will not do, since a
local minimum of a uniform random function is easily found. \ However, Aldous
\cite{aldous}\ had the idea of defining $\mathcal{D}$\ via a \textit{random
walk}, as follows. \ Choose a vertex $v_{0}\in\left\{  0,1\right\}  ^{n}%
$\ uniformly at random; then perform an unbiased
walk\footnote{Actually, Aldous used a continuous-time random walk,
so the functions would be from $\left\{  0,1\right\}  ^{n}$\ to
$\mathbb{R}$.} $v_{0},v_{1},v_{2},\ldots $\ starting from $v_{0}$. \
For each vertex $v$, set $f\left(  v\right)  $ equal to the first
hitting time of the walk at $v$---that is, $f\left( v\right)
=\min\left\{  t:v_{t}=v\right\}  $. \ Clearly any $f$ produced in
this way has a unique local minimum at $v_{0}$, since for all $t>0$,
if vertex $v_{t}$\ is visited for the first time at step $t$ then
$f\left( v_{t}\right)  >f\left(  v_{t-1}\right)  $. \ Using
sophisticated random walk analysis, Aldous managed to show a lower
bound of $2^{n/2-o\left(  n\right) }$\ on the expected number of
queries needed by any randomized algorithm to find
$v_{0}$.\footnote{Independently and much later, Droste et al.\
\cite{djw}\ showed the weaker bound $2^{g\left(  n\right)  }$\ for
any $g\left(  n\right)  =o\left(  n\right)  $.} \ (As we will see in
Section \ref{PRELIMPLS}, this lower bound is close to tight.) \
Intuitively, since a random walk on the hypercube mixes in $O\left(
n\log n\right)  $\ steps, an algorithm that has not queried a $v$
with $f\left(  v\right)  <2^{n/2}$\ has almost no useful information
about where the unique minimum $v_{0}$\ is, so its next query will
just be a \textquotedblleft stab in the dark.\textquotedblright

However, Aldous's result leaves several questions about \textsc{Local
Search}\ unanswered. \ What if the graph $G$ is a $3$-D cube, on which a
random walk does \textit{not} mix very rapidly? \ Can we still lower-bound the
randomized query complexity of finding a local minimum? \ More generally, what
parameters of $G$ make the problem hard or easy? \ Also, what is the quantum
query complexity of \textsc{Local Search}?

This chapter presents a new approach to \textsc{Local Search},\
which I believe points the way to a complete understanding of its
complexity. \ The approach is based on Ambainis's quantum adversary
method \cite{ambainis}. \ Surprisingly, the approach yields new and
simpler lower bounds for the problem's \textit{classical} randomized
query complexity, in addition to quantum lower bounds. \ Thus, along
with recent work by Kerenidis and de Wolf \cite{kw}\ and by Aharonov
and Regev \cite{ar}, the results of this chapter illustrate how
quantum ideas can help to resolve classical open problems.

The results are as follows. \ For the Boolean hypercube $G=\left\{
0,1\right\}  ^{n}$, I show that any quantum algorithm needs $\Omega\left(
2^{n/4}/n\right)  $\ queries to find a local minimum on $G$, and any
randomized algorithm needs $\Omega\left(  2^{n/2}/n^{2}\right)  $\ queries
(improving the $2^{n/2-o\left(  n\right)  }$\ lower bound of Aldous
\cite{aldous}).\ \ The proofs are elementary and do not require random walk
analysis. \ By comparison, the best known upper bounds are $O\left(
2^{n/3}n^{1/6}\right)  $\ for a quantum algorithm and $O\left(  2^{n/2}%
\sqrt{n}\right)  $\ for a randomized algorithm. \ If $G$ is a $d$-dimensional
grid of size $N^{1/d}\times\cdots\times N^{1/d}$, where $d\geq3$\ is a
constant, then I show that any quantum algorithm needs $\Omega\left(
\sqrt{N^{1/2-1/d}/\log N}\right)  $\ queries to find a local minimum on $G$,
and any randomized algorithm needs $\Omega\left(  N^{1/2-1/d}/\log N\right)
$\ queries. \ No nontrivial lower bounds (randomized or quantum) were
previously known in this case.\footnote{A lower bound on deterministic query
complexity was known for such graphs \cite{lt}.}

In a preprint discussing these results, I raised as my \textquotedblleft most
ambitious\textquotedblright\ conjecture that the deterministic and quantum
query complexities of local search are polynomially related for \textit{every}
family of graphs. \ At the time, it was not even known whether deterministic
and \textit{randomized} query complexities were polynomially related, not even
for simple examples such as the $2$-dimensional square grid. \ Subsequently
Santha and Szegedy \cite{ss} spectacularly resolved the conjecture, by showing
that the quantum query complexity is always at least the $19^{th}$\ root (!)
of the deterministic complexity. \ On the other hand, in the specific case of
the hypercube, my lower bound is close to tight; Santha and Szegedy's is not.
\ Also, I give randomized lower bounds that are quadratically better than my
quantum lower bounds; Santha and Szegedy give only quantum lower bounds.

In another recent development, Ambainis \cite{ambainis:pls} has improved the
$\Omega\left(  2^{n/4}/n\right)  $\ quantum lower bound for local search on
the hypercube to $2^{n/3}/n^{O\left(  1\right)  }$, using a hybrid argument.
\ Note that Ambainis's lower bound matches the upper bound up to a polynomial factor.

The chapter is organized as follows. \ Section \ref{MOTIVATIONPLS}\ motivates
lower bounds on \textsc{Local Search}, pointing out connections to simulated
annealing, quantum adiabatic algorithms, and the complexity class
$\mathsf{TFNP}$\ of total function problems. \ Section \ref{PRELIMPLS}%
\ defines notation and reviews basic facts about \textsc{Local Search},
including upper bounds. \ In Section \ref{ADVERSARY}\ I give an intuitive
explanation of Ambainis's quantum adversary method, then state and prove a
classical analogue of Ambainis's main lower bound theorem. \ Section
\ref{SNAKE}\ introduces \textit{snakes}, a construction by which I apply the
two adversary methods to \textsc{Local Search}.\ \ I show there that to prove
lower bounds for any graph $G$, it suffices to upper-bound a combinatorial
parameter $\varepsilon$\ of a `snake distribution'\ on $G$. \ Section
\ref{GRAPHS} applies this framework to specific examples of graphs: the
Boolean hypercube in Section \ref{BOOLEAN}, and the $d$-dimensional grid in
Section \ref{DDIM}.

\section{Motivation\label{MOTIVATIONPLS}}

Local search is the most effective weapon ever devised against hard
optimization problems. \ For many real applications, neither
backtrack search, nor approximation algorithms, nor even Grover's
algorithm can compare. \ Furthermore, along with quantum computing,
local search (broadly defined) is one of the most interesting links
between computer science and Nature. \ It is related to evolutionary
biology via genetic algorithms, and to the physics of materials via
simulated annealing. \ Thus it is both practically and
scientifically important to understand its performance.

The conventional wisdom is that, although local search performs well in
practice, its central (indeed defining) flaw is a tendency to get stuck at
local optima. \ If this were correct, one corollary would be that the reason
local search performs so well is that the problem it really solves---finding a
local optimum---is intrinsically easy. \ It would thus be unnecessary to seek
further explanations for its performance. \ Another corollary would be that,
for \textit{unimodal} functions (which have no local optima besides the global
optimum), the global optimum would be easily found.

However, the conventional wisdom is false. \ The results of
Llewellyn et al.\ \cite{ltt} and Aldous \cite{aldous}\ show that
even if $f$ is unimodal, any classical algorithm that treats $f$ as
a black box needs exponential time to find the global minimum of $f$
in general.\ \ My results extend this conclusion to quantum
algorithms. \ In my view, the practical upshot of these results is
that they force us to confront the question: What is it about
`real-world'\ problems that makes it easy to find a local optimum? \
That is, why do exponentially long chains of descending values, such
as those used for lower bounds, almost never occur in practice, even
in functions with large range sizes?\ \ One possibility is that the
functions that occur in practice look ``globally'' like random
functions, but I do not know whether that is true in any meaningful
sense.

The results of this chapter are also relevant for physics. \ Many
physical systems, including folding proteins and networks of springs
and pulleys, can be understood as performing `local search' through
an energy landscape to reach a locally-minimal energy configuration.
\ A key question is, how long will the system take to reach its
ground state (that is, a globally-minimal configuration)? \ Of
course, if there are local optima, the system might \textit{never}
reach its ground state, just as a rock in a mountain crevice does
not roll to the bottom by going up first. \ But what if the energy
landscape is unimodal? \ And moreover, what if the physical system
is quantum? \ My results show that, for certain energy landscapes,
even a quantum system would take exponential time to reach its
ground state, regardless of what external Hamiltonian is applied to
``drive'' it. \ So in particular, the quantum adiabatic algorithm
proposed by Farhi et al.\ \cite{fggllp}, which can be seen as a
quantum analogue of simulated annealing, needs exponential time to
find a local minimum in the worst case.

Finally, this chapter's results have implications for so-called
\textit{total function problems} in complexity theory. \ Megiddo and
Papadimitriou \cite{megiddo}\ defined a complexity class
$\mathsf{TFNP}$, consisting (informally) of those $\mathsf{NP}$\
search problems for which a solution always exists. \ For example,
we might be given a function $f:\left\{  0,1\right\}
^{n}\rightarrow\left\{  0,1\right\} ^{n-1}$\ as a Boolean circuit,
and asked to find any distinct $x,y$ pair such that $f\left(
x\right)  =f\left(
y\right)  $. \ This particular problem belongs to a subclass of $\mathsf{TFNP}%
$ called $\mathsf{PPP}$ (Polynomial Pigeonhole Principle). \ Notice that no
promise is involved: the combinatorial nature of the problem itself forces a
solution to exist, even if we have no idea how to find it. \ In a recent talk,
Papadimitriou \cite{papa:talk} asked broadly whether such `nonconstructive
existence problems' might be good candidates for efficient quantum algorithms.
\ \ In the case of $\mathsf{PPP}$\ problems like the one above, the collision
lower bound of Chapter \ref{COL} implies a negative answer in the black-box
setting. \ For other subclasses of $\mathsf{TFNP}$, such as $\mathsf{PODN}%
$\ (Polynomial Odd-Degree Node), a quantum black-box lower bound follows
easily from the optimality of Grover's search algorithm.

However, there is one important subclass of $\mathsf{TFNP}$\ for
which no quantum lower bound was previously known. \ This is
$\mathsf{PLS}$ (Polynomial Local Search), defined by Johnson,
Papadimitriou, and Yannakakis \cite{jpy} as a class of optimization
problems whose cost function $f$ and neighborhood function $\eta$
(that is, the set of neighbors of a given point) are both computable
in polynomial time.\footnote{Some authors require only the
\textit{minimum} neighbor of a given point to be computable in
polynomial time, which does not seem like the ``right'' idealization
to me. \ In any case, for lower bound purposes we always assume the
algorithm knows the whole neighborhood structure in advance, and
does not need to make queries to learn about it.} \ Given such a
problem, the task is to output any local minimum of the cost
function: that is, a $v$ such that $f\left(  v\right)  \leq f\left(
w\right)  $\ for all $w\in\eta\left(  v\right)  $.\ \ The lower
bound of Llewellyn et al.\ \cite{ltt} yields an oracle $A$ relative
to which $\mathsf{FP}^{A}\neq\mathsf{PLS}^{A}$, by a standard
diagonalization argument along the lines of Baker, Gill, and Solovay
\cite{bgs}. \ Likewise, the lower bound of Aldous \cite{aldous}\
yields an oracle relative to which $\mathsf{PLS}\not \subset
\mathsf{FBPP}$, where $\mathsf{FBPP}$\ is simply the function
version of $\mathsf{BPP}$. \ The results of this chapter yield the
first oracle relative to which $\mathsf{PLS}\not \subset
\mathsf{FBQP}$. \ In light of this oracle separation, I raise an
admittedly vague question: is there a nontrivial \textquotedblleft
combinatorial\textquotedblright\ subclass of $\mathsf{TFNP}$\ that
we can show \textit{is} contained in $\mathsf{FBQP}$?

\section{Preliminaries\label{PRELIMPLS}}

In the \textsc{Local Search}\ problem, we are given an undirected graph
$G=\left(  V,E\right)  $ with $N=\left\vert V\right\vert $,\ and oracle access
to a function $f:V\rightarrow\mathbb{N}$. \ The goal is to find any
\textit{local minimum} of $f$, defined as a vertex $v\in V$ such that
$f\left(  v\right)  \leq f\left(  w\right)  $\ for all neighbors $w$\ of $v$.
\ Clearly such a local minimum exists. \ We want to find one using as few
queries as possible, where a query returns $f\left(  v\right)  $\ given $v$.
\ Queries can be adaptive; that is, can depend on the outcomes of previous
queries. \ We assume $G$ is known in advance, so that only $f$ needs to be
queried. \ Since we care only about query complexity, not computation time,
there is no difficulty in dealing with an infinite range for $f$---though for
lower bound purposes, it will turn out that a range of size $O\left(
\sqrt{\left\vert V\right\vert }\right)  $\ suffices. \ I do not know of any
case where a range larger than this makes the \textsc{Local Search}\ problem
harder, but I also do not know of a general reduction from large to small range.

The model of query algorithms is the standard one. \ Given a graph $G$, the
deterministic query complexity of \textsc{Local Search} on $G$, which we
denote $\operatorname*{DLS}\left(  G\right)  $, is $\min_{\Gamma}\max
_{f}T\left(  \Gamma,f,G\right)  $\ where the minimum ranges over all
deterministic algorithms $\Gamma$, the maximum ranges over all $f$, and
$T\left(  \Gamma,f,G\right)  $\ is the number of queries made to $f$ by
$\Gamma$\ before it halts and outputs a local minimum of $f$ (or $\infty$\ if
$\Gamma$ fails to do so). \ The randomized query complexity
$\operatorname*{RLS}\left(  G\right)  $\ is defined similarly, except that now
the algorithm has access to an infinite random string $R$, and must only
output a local minimum with probability at least $2/3$ over $R$. \ For
simplicity, one can assume that the number of queries $T$ is the same for all
$R$; clearly this assumption changes the complexity by at most a constant factor.

In the quantum model, an algorithm's state has the form $\sum_{v,z,s}%
\alpha_{v,z,s}\left|  v,z,s\right\rangle $, where $v$ is the label of a vertex
in $G$, and $z$ and $s$ are strings representing the answer register and
workspace respectively. \ The $\alpha_{v,z,s}$'s\ are complex amplitudes
satisfying $\sum_{v,z,s}\left|  \alpha_{v,z,s}\right|  ^{2}=1$. \ Starting
from an arbitrary (fixed) initial state, the algorithm proceeds by an
alternating sequence of \textit{queries} and \textit{algorithm steps}. \ A
query maps each $\left|  v,z,s\right\rangle $\ to $\left|  v,z\oplus f\left(
v\right)  ,s\right\rangle $, where $\oplus$\ denotes bitwise exclusive-OR.
\ An algorithm step multiplies the vector of $\alpha_{v,z,s}$'s\ by an
arbitrary unitary matrix that does not depend on $f$. \ Letting $\mathcal{M}%
_{f}$\ denote the set of local minima of $f$, the algorithm succeeds if at the
end $\sum_{v,z,s~:~v\in\mathcal{M}_{f}}\left|  \alpha_{v,z,s}\right|  ^{2}%
\geq\frac{2}{3}$. \ Then the bounded-error quantum query complexity, or
$\operatorname*{QLS}\left(  G\right)  $, is defined as the minimum number of
queries used by a quantum algorithm that succeeds on every $f$.

It is immediate that $\operatorname*{QLS}\left(  G\right)  \leq
\operatorname*{RLS}\left(  G\right)  \leq\operatorname*{DLS}\left(  G\right)
\leq N$. \ Also, letting $\delta$\ be the maximum degree of $G$, we have the
following trivial lower bound.

\begin{proposition}
\label{degree}$\operatorname*{RLS}\left(  G\right)  =\Omega\left(
\delta\right)  $\ and $\operatorname*{QLS}\left(  G\right)  =\Omega\left(
\sqrt{\delta}\right)  $.
\end{proposition}

\begin{proof}
Let $v$ be a vertex of $G$ with degree $\delta$. \ Choose a neighbor
$w$ of $v$ uniformly at random, and let $f\left(  w\right)  =1$. \
Let $f\left( v\right)  =2$, and $f\left(  u\right)  =3$ for all
neighbors $u$ of $v$ other than $w$. \ Let $S$\ be the neighbor set
of $v$ (including $v$ itself); then for all $x\notin S$, let
$f\left(  x\right)  =3+\Delta\left(  x,S\right) $\ where
$\Delta\left(  x,S\right)  $ is the minimum distance from $x$ to a
vertex in $S$. \ Clearly $f$ has a unique local minimum at $w$. \
However, finding $y$ requires exhaustive search among the $\delta$\
neighbors of $v$, which takes $\Omega\left(  \sqrt{\delta}\right)  $
quantum queries by Bennett et al.\ \cite{bbbv}.
\end{proof}

A corollary of Proposition \ref{degree}\ is that classically, zero-error
randomized query complexity is equivalent to bounded-error up to a constant
factor. \ For given a candidate local minimum $v$, one can check using
$O\left(  \delta\right)  $\ queries that $v$ is indeed a local minimum.
\ Since $\Omega\left(  \delta\right)  $\ queries are needed anyway, this
verification step does not affect the overall complexity.

As pointed out by Aldous \cite{aldous}, a classical randomized algorithm can
find a local minimum of $f$ with high probability in $O\left(  \sqrt{N\delta
}\right)  $\ queries. \ The algorithm just queries $\sqrt{N\delta}$\ vertices
uniformly at random, and lets $v_{0}$\ be a queried vertex for which $f\left(
v\right)  $\ is minimal. \ It then follows $v_{0}$\ to a local minimum by
steepest descent. \ That is, for $t=0,1,2,\ldots$, it queries all neighbors of
$v_{t}$, halts if $v_{t}$\ is a local minimum, and otherwise sets $v_{t+1}%
$\ to be the neighbor $w$ of $v_{t}$\ for which $f\left(  w\right)  $\ is
minimal (breaking ties by lexicographic ordering). \ A similar idea yields an
improved quantum upper bound.

\begin{proposition}
\label{upper}For any $G$, $\operatorname*{QLS}\left(  G\right)  =O\left(
N^{1/3}\delta^{1/6}\right)  $.
\end{proposition}

\begin{proof}
The algorithm first chooses $N^{2/3}\delta^{1/3}$\ vertices of $G$ uniformly
at random, then uses Grover search to find a chosen vertex $v_{0}$ for which
$f\left(  v\right)  $\ is minimal. \ By a result of D\"{u}rr and H\o yer
\cite{dh}, this can be done with high probability in $O\left(  N^{1/3}%
\delta^{1/6}\right)  $\ queries. \ Next, for $t=0,1,2,\ldots$, the algorithm
performs Grover search over all neighbors of $v_{t}$, looking for a neighbor
$w$ such that $f\left(  w\right)  <f\left(  v_{t}\right)  $. \ If it finds
such a $w$, then it sets $v_{t+1}:=w$ and continues to the next iteration.
\ Otherwise, it repeats the Grover search $\log\left(  N/\delta\right)
$\ times before finally giving up and returning $v_{t}$\ as a claimed local minimum.

The expected number of $u$ such that $f\left(  u\right)  <f\left(
v_{0}\right)  $ is at most $N/\left(  N^{2/3}\delta^{1/3}\right)  =\left(
N/\delta\right)  ^{1/3}$. \ Since $f\left(  v_{t+1}\right)  <f\left(
v_{t}\right)  $ for all $t$, clearly the number of such $u$ provides an upper
bound on $t$. \ Furthermore, assuming there exists a $w$ such that $f\left(
w\right)  <f\left(  v_{t}\right)  $, the expected number of repetitions of
Grover's algorithm until such a $w$ is found is $O\left(  1\right)  $. \ Since
each repetition takes $O\left(  \sqrt{\delta}\right)  $\ queries, by linearity
of expectation the total expected number of queries used by the algorithm is
therefore%
\[
O\left(  N^{1/3}\delta^{1/6}+\left(  N/\delta\right)  ^{1/3}\sqrt{\delta}%
+\log\left(  N/\delta\right)  \sqrt{\delta}\right)
\]
or $O\left(  N^{1/3}\delta^{1/6}\right)  $. \ To see that the algorithm finds
a local minimum with high probability, observe that for each $t$, the
probability of not finding a $w$ such that $f\left(  w\right)  <f\left(
v_{t}\right)  $, given that one exists, is at most $c^{-\log\left(
N/\delta\right)  }\leq\left(  \delta/N\right)  ^{1/3}/10$\ for a suitable
constant $c$. \ So by the union bound, the probability that the algorithm
returns a `false positive' is at most $\left(  N/\delta\right)  ^{1/3}%
\cdot\left(  \delta/N\right)  ^{1/3}/10=1/10$.
\end{proof}

\section{Relational Adversary Method\label{ADVERSARY}}

There are essentially two known methods for proving lower bounds on
quantum query complexity: the polynomial method of Beals et al.\
\cite{bbcmw}, and the quantum adversary method of Ambainis
\cite{ambainis}.\footnote{I am thinking here of the hybrid method
\cite{bbbv}\ as a cousin of the adversary method.} \ For a few
problems, such as the collision problem \cite{aar:col,shi}, the
polynomial method succeeded where the adversary method failed. \
However, for problems that lack permutation symmetry (such as
\textsc{Local Search}), the adversary method has proven more
effective.\footnote{Indeed, Ambainis \cite{ambainis:deg}\ has given
problems for which the adversary method provably yields a better
lower bound than the polynomial method.}

How could a quantum lower bound method possibly be applied classically? \ When
proving randomized lower bounds, the tendency is to attack ``bare-handed'':
fix a distribution over inputs, and let $x_{1},\ldots,x_{t}$\ be the locations
queried so far by the algorithm. \ Show that for small $t$, the posterior
distribution over inputs, \textit{conditioned} on $x_{1},\ldots,x_{t}$, is
still `hard' with high probability---so that the algorithm knows almost
nothing even about which location $x_{t+1}$\ to query next. \ This is
essentially the approach taken by Aldous \cite{aldous}\ to prove a
$2^{n/2-o\left(  n\right)  }$\ lower bound on $\operatorname*{RLS}\left(
\left\{  0,1\right\}  ^{n}\right)  $.

In the quantum case, however, it is unclear how to specify what an algorithm
`knows'\ after a given number of queries. \ So we are almost \textit{forced}
to step back, and identify general combinatorial properties of input sets that
make them hard to distinguish. \ Once we have such properties, we can then try
to exhibit them in functions of interest.

We will see, somewhat surprisingly, that this \textquotedblleft
gloved\textquotedblright\ approach is useful for classical lower bounds as
well as quantum ones. \ In the \textit{relational adversary method}, we assume
there exists a $T$-query randomized algorithm for function $F$. \ We consider
a set $\mathcal{A}$\ of $0$-inputs of $F$, a set $\mathcal{B}$\ of $1$-inputs,
and an arbitrary real-valued \textit{relation function} $R\left(  A,B\right)
\geq0$\ for $A\in\mathcal{A}$\ and $B\in\mathcal{B}$. \ Intuitively, $R\left(
A,B\right)  $\ should be large if $A$ and $B$ differ in only a few locations.
\ We then fix a probability distribution $\mathcal{D}$\ over inputs; by Yao's
minimax principle, there exists a $T$-query deterministic algorithm
$\Gamma^{\ast}$\ that succeeds with high probability on inputs drawn from
$\mathcal{D}$. \ Let $W_{A}$\ be the set of $0$-inputs and $W_{B}$\ the set of
$1$-inputs on which $\Gamma^{\ast}$\ succeeds. \ Using the relation function
$R$, we define a \textit{separation measure} $S$\ between $W_{A}$\ and $W_{B}%
$, and show that (1) initially $S=0$,\ (2) by the end of the computation
$S$\ must be large, and (3) $S$ increases by only a small amount as the result
of each query. \ It follows that $T$\ must be large.

The advantage of the relational method is that converts a \textquotedblleft
dynamic\textquotedblright\ opponent---an algorithm that queries
adaptively---into a relatively static one. \ It thereby makes it easier to
focus on what is unique about a problem, aspects of query complexity that are
common to all problems having been handled automatically. \ Furthermore, one
does not need to know anything about quantum computing to understand and apply
the method. \ On the other hand, I have no idea how one would come up with it
in the first place, without Ambainis's \textit{quantum} adversary method
\cite{ambainis} and the reasoning about entanglement that led to it.

The starting point is the ``most general''\ adversary theorem in Ambainis's
original paper (Theorem 6\ in \cite{ambainis}), which he introduced to prove a
quantum lower bound for the problem of inverting a permutation. \ Here the
input is a permutation $\sigma\left(  1\right)  ,\ldots,\sigma\left(
N\right)  $, and the task is to output $0$ if $\sigma^{-1}\left(  1\right)
\leq N/2$\ and $1$ otherwise. \ To lower-bound this problem's query
complexity, what we would like to say is this:

\textit{Given any }$0$\textit{-input }$\sigma$\textit{\ and any location }%
$x$\textit{, if we choose a random }$1$\textit{-input }$\tau$\textit{\ that is
`related' to }$\sigma$\textit{, then the probability }$\theta\left(
\sigma,x\right)  $\textit{ over }$\tau$\textit{\ that }$\sigma\left(
x\right)  $ \textit{does not equal }$\tau\left(  x\right)  $\textit{\ is
small. \ In other words, the algorithm is unlikely to distinguish }$\sigma
$\textit{\ from a random neighbor }$\tau$\textit{\ of }$\sigma$\textit{\ by
querying }$x$\textit{.}

Unfortunately, the above claim is false. \ Letting $x=\sigma^{-1}\left(
1\right)  $, we have that $\sigma\left(  x\right)  \neq\tau\left(  x\right)
$\ for \textit{every} $1$-input $\tau$, and thus $\theta\left(  \sigma
,x\right)  =1$. \ Ambainis resolves this difficulty by letting us take the
maximum, over all $0$-inputs $\sigma$\ and $1$-inputs $\tau$ that are related
and differ at $x$, of the \textit{geometric mean} $\allowbreak\sqrt
{\theta\left(  \sigma,x\right)  \theta\left(  \tau,x\right)  }$. \ Even if
$\theta\left(  \sigma,x\right)  =1$, the geometric mean is still small
provided that $\theta\left(  \tau,x\right)  $\ is small. \ More formally:

\begin{theorem}
[Ambainis]\label{ambthmls}Let $\mathcal{A}\subseteq F^{-1}\left(  0\right)  $
and $\mathcal{B}\subseteq F^{-1}\left(  1\right)  $\ be sets of inputs to
function $F$. \ Let $R\left(  A,B\right)  \geq0$ be a symmetric real-valued
function, and for $A\in\mathcal{A}$, $B\in\mathcal{B}$, and location $x$, let%
\begin{align*}
\theta\left(  A,x\right)   &  =\frac{\sum_{B^{\ast}\in\mathcal{B}~:~A\left(
x\right)  \neq B^{\ast}\left(  x\right)  }R\left(  A,B^{\ast}\right)  }%
{\sum_{B^{\ast}\in\mathcal{B}}R\left(  A,B^{\ast}\right)  },\\
\theta\left(  B,x\right)   &  =\frac{\sum_{A^{\ast}\in\mathcal{A}~:~A^{\ast
}\left(  x\right)  \neq B\left(  x\right)  }R\left(  A^{\ast},B\right)  }%
{\sum_{A^{\ast}\in\mathcal{A}}R\left(  A^{\ast},B\right)  },
\end{align*}
where the denominators are all nonzero. \ Then the number of quantum queries
needed to evaluate $F$ with at least $9/10$ probability is $\Omega\left(
1/\upsilon_{\operatorname*{geom}}\right)  $, where%
\[
\upsilon_{\operatorname*{geom}}=\max_{A\in\mathcal{A},~B\in\mathcal{B}%
,~x~:~R\left(  A,B\right)  >0,~A\left(  x\right)  \neq B\left(  x\right)
}\sqrt{\theta\left(  A,x\right)  \theta\left(  B,x\right)  }.
\]

\end{theorem}

The best way to understand Theorem \ref{ambthmls} is to see it used in an example.

\begin{proposition}
[Ambainis]The quantum query complexity of inverting a permutation is
$\Omega\left(  \sqrt{N}\right)  $.
\end{proposition}

\begin{proof}
Let $\mathcal{A}$\ be the set of all permutations $\sigma$\ such that
$\sigma^{-1}\left(  1\right)  $ $\leq N/2$, and $\mathcal{B}$\ be the set of
permutations $\tau$\ such that $\tau^{-1}\left(  1\right)  >N/2$. \ Given
$\sigma\in\mathcal{A}$\ and $\tau\in\mathcal{B}$, let $R\left(  \sigma
,\tau\right)  =1$\ if $\sigma$\ and $\tau$\ differ only at locations
$\sigma^{-1}\left(  1\right)  $\ and $\tau^{-1}\left(  1\right)  $, and
$R\left(  \sigma,\tau\right)  =0$\ otherwise. \ Then given $\sigma,\tau$\ with
$R\left(  \sigma,\tau\right)  =1$,\ if $x\neq\sigma^{-1}\left(  1\right)
$\ then $\theta\left(  \sigma,x\right)  =2/N$, and if $x\neq\tau^{-1}\left(
1\right)  $\ then $\theta\left(  \tau,x\right)  =2/N$. \ So $\max
_{x~:~\sigma\left(  x\right)  \neq\tau\left(  x\right)  }\sqrt{\theta\left(
\sigma,x\right)  \theta\left(  \tau,x\right)  }=\sqrt{2/N}$.
\end{proof}

The only difference between Theorem \ref{ambthmls}\ and my relational
adversary theorem is that in the latter, we take the \textit{minimum} of
$\theta\left(  A,x\right)  $ and $\theta\left(  B,x\right)  $\ instead of the
geometric mean. \ Taking the reciprocal then gives up to a quadratically
better lower bound: for example, we obtain that the randomized query
complexity of inverting a permutation is $\Omega\left(  N\right)  $.
\ However, the proofs of the two theorems are quite different.

\begin{theorem}
\label{classadv}Let $\mathcal{A},\mathcal{B},R,\theta$ be as in Theorem
\ref{ambthmls}. \ Then the number of randomized queries needed to evaluate
$F$\ with at least $9/10$ probability is $\Omega\left(  1/\upsilon_{\min
}\right)  $, where%
\[
\upsilon_{\min}=\max_{A\in\mathcal{A},~B\in\mathcal{B},~x~:\smallskip
\,~R\left(  A,B\right)  >0,~A\left(  x\right)  \neq B\left(  x\right)  }%
\min\left\{  \theta\left(  A,x\right)  ,\theta\left(  B,x\right)  \right\}  .
\]

\end{theorem}

\begin{proof}
Let $\Gamma$ be a randomized algorithm that, given an input $A$, returns
$F\left(  A\right)  $ with at least $9/10$ probability. \ Let $T$ be the
number of queries made by $\Gamma$. \ For all $A\in\mathcal{A}$,
$B\in\mathcal{B}$, define%
\begin{align*}
M\left(  A\right)   &  =\sum_{B^{\ast}\in\mathcal{B}}R\left(  A,B^{\ast
}\right)  ,\\
M\left(  B\right)   &  =\sum_{A^{\ast}\in\mathcal{A}}R\left(  A^{\ast
},B\right)  ,\\
M  &  =\sum_{A^{\ast}\in\mathcal{A}}M\left(  A^{\ast}\right)  =\sum_{B^{\ast
}\in\mathcal{B}}M\left(  B^{\ast}\right)  .
\end{align*}
Now let $\mathcal{D}_{A}$\ be the distribution over $A\in\mathcal{A}$\ in
which each $A$ is chosen with probability $M\left(  A\right)  /M$; and let
$\mathcal{D}_{B}$\ be the distribution over $B\in\mathcal{B}$\ in which each
$B$ is chosen with probability $M\left(  B\right)  /M$. \ Let $\mathcal{D}%
$\ be an equal mixture of $\mathcal{D}_{A}$\ and $\mathcal{D}_{B}$. \ By Yao's
minimax principle, there exists a deterministic algorithm $\Gamma^{\ast}%
$\ that makes $T$ queries, and succeeds with at least $9/10$ probability given
an input drawn from $\mathcal{D}$. \ Therefore $\Gamma^{\ast}$\ succeeds with
at least $4/5$\ probability given an input drawn from $\mathcal{D}_{A}$ alone,
or from $\mathcal{D}_{B}$ alone. \ In other words, letting $W_{A}$\ be the set
of $A\in\mathcal{A}$\ and $W_{B}$\ the set of $B\in\mathcal{B}$\ on which
$\Gamma^{\ast}$\ succeeds, we have%
\[
\sum_{A\in W_{A}}M\left(  A\right)  \geq\frac{4}{5}M,\,\,\,\,\,\,\sum_{B\in
W_{B}}M\left(  B\right)  \geq\frac{4}{5}M.
\]
Define a predicate $P^{\left(  t\right)  }\left(  A,B\right)  $, which is true
if $\Gamma^{\ast}$ has distinguished $A\in\mathcal{A}$ from $B\in\mathcal{B}%
$\ by the $t^{th}$\ query and false otherwise. \ (To distinguish $A$ from $B$
means to query an index $x$\ for which $A\left(  x\right)  \neq B\left(
x\right)  $, given either $A$ or $B$ as input.) \ Also, for all $A\in
\mathcal{A}$, define a score function%
\[
S^{\left(  t\right)  }\left(  A\right)  =\sum_{B^{\ast}\in\mathcal{B}%
~:~P^{\left(  t\right)  }\left(  A,B^{\ast}\right)  }R\left(  A,B^{\ast
}\right)  .
\]
This function measures how much \textquotedblleft progress\textquotedblright%
\ has been made so far in separating $A$ from $\mathcal{B}$-inputs, where the
$\mathcal{B}$-inputs\ are weighted by $R\left(  A,B\right)  $. \ Similarly,
for all $B\in\mathcal{B}$ define%
\[
S^{\left(  t\right)  }\left(  B\right)  =\sum_{A^{\ast}\in\mathcal{A}%
~:~P^{\left(  t\right)  }\left(  A^{\ast},B\right)  }R\left(  A^{\ast
},B\right)  .
\]
It is clear that for all $t$,%
\[
\sum_{A\in\mathcal{A}}S^{\left(  t\right)  }\left(  A\right)  =\sum
_{B\in\mathcal{B}}S^{\left(  t\right)  }\left(  B\right)  .
\]
So we can denote the above sum by $S^{\left(  t\right)  }$ and think of it as
a\ global progress measure. \ The proof relies on the following claims about
$S^{\left(  t\right)  }$:

\begin{enumerate}
\item[(i)] $S^{\left(  0\right)  }=0$ initially.

\item[(ii)] $S^{\left(  T\right)  }\geq3M/5$ by the end.

\item[(iii)] $\Delta S^{\left(  t\right)  }\leq3\upsilon_{\min}M$\ for all
$t$, where $\Delta S^{\left(  t\right)  }=S^{\left(  t\right)  }-S^{\left(
t-1\right)  }$ is the amount by which $S^{\left(  t\right)  }$\ increases as
the result of a single query.
\end{enumerate}

It follows from (i)-(iii) that%
\[
T\geq\frac{3M/5}{3\upsilon_{\min}M}=\frac{1}{5\upsilon_{\min}}%
\]
which establishes the theorem. \ Part (i) is obvious. \ For part (ii), observe
that for every pair $\left(  A,B\right)  $ with $A\in W_{A}$ and $B\in W_{B}$,
the algorithm $\Gamma^{\ast}$\ must query an $x$ such that $A\left(  x\right)
\neq B\left(  x\right)  $. \ Thus%
\begin{align*}
S^{\left(  T\right)  }  &  \geq\sum_{A\in W_{A},~B\in W_{B}}R\left(
A,B\right) \\
&  \geq\sum_{A\in W_{A}}M\left(  A\right)  -\sum_{B\notin W_{B}}M\left(
B\right) \\
&  \geq\frac{4}{5}M-\frac{1}{5}M.
\end{align*}
It remains only to show part (iii). \ Suppose $\Delta S^{\left(  t\right)
}>3\upsilon_{\min}M$\ for some $t$; we will obtain a contradiction.\ \ Let%
\[
\Delta S^{\left(  t\right)  }\left(  A\right)  =S^{\left(  t\right)  }\left(
A\right)  -S^{\left(  t-1\right)  }\left(  A\right)  ,
\]
and let $C_{A}$\ be the set of $A\in\mathcal{A}$ for which $\Delta S^{\left(
t\right)  }\left(  A\right)  >\upsilon_{\min}M\left(  A\right)  $. \ Since%
\[
\sum_{A\in\mathcal{A}}\Delta S^{\left(  t\right)  }\left(  A\right)  =\Delta
S^{\left(  t\right)  }>3\upsilon_{\min}M,
\]
it follows by Markov's inequality that%
\[
\sum_{A\in C_{A}}\Delta S^{\left(  t\right)  }\left(  A\right)  \geq\frac
{2}{3}\Delta S^{\left(  t\right)  }.
\]
Similarly, if we let $C_{B}$\ be the set of $B\in\mathcal{B}$ for which
$\Delta S^{\left(  t\right)  }\left(  B\right)  >\upsilon_{\min}M\left(
B\right)  $, we have%
\[
\sum_{B\in C_{B}}\Delta S^{\left(  t\right)  }\left(  B\right)  \geq\frac
{2}{3}\Delta S^{\left(  t\right)  }.
\]
In other words, at least $2/3$\ of the increase in $S^{\left(  t\right)  }%
$\ comes from $\left(  A,B\right)  $\ pairs such that $A\in C_{A}$,\ and at
least $2/3$\ comes from $\left(  A,B\right)  $\ pairs such that $B\in C_{B}$.
\ Hence, by a `pigeonhole' argument, there exists an $A\in C_{A}$ and $B\in
C_{B}$\ with $R\left(  A,B\right)  >0$\ that are distinguished by the $t^{th}%
$\ query. \ In other words, there exists an $x$ with $A\left(  x\right)  \neq
B\left(  x\right)  $, such that the $t^{th}$\ index queried by $\Gamma^{\ast}%
$\ is $x$ whether the input is $A$ or $B$. \ Then since $A\in C_{A}$, we have
$\upsilon_{\min}M\left(  A\right)  <\Delta S^{\left(  t\right)  }\left(
A\right)  $, and hence%
\begin{align*}
\upsilon_{\min}  &  <\frac{\Delta S^{\left(  t\right)  }\left(  A\right)
}{M\left(  A\right)  }\\
&  \leq\frac{\sum_{B^{\ast}\in\mathcal{B}~:~A\left(  x\right)  \neq B^{\ast
}\left(  x\right)  }R\left(  A,B^{\ast}\right)  }{\sum_{B^{\ast}\in
\mathcal{B}}R\left(  A,B^{\ast}\right)  }%
\end{align*}
which equals $\theta\left(  A,x\right)  $. \ Similarly $\upsilon_{\min}%
<\theta\left(  B,x\right)  $ since $B\in C_{B}$. \ This contradicts the
definition%
\[
\upsilon_{\min}=\max_{A\in\mathcal{A},~B\in\mathcal{B},~x~:\smallskip
\,~R\left(  A,B\right)  >0,~A\left(  x\right)  \neq B\left(  x\right)  }%
\min\left\{  \theta\left(  A,x\right)  ,\theta\left(  B,x\right)  \right\}  ,
\]
and we are done.
\end{proof}

\section{Snakes\label{SNAKE}}

For the lower bounds, it will be convenient to generalize random walks to
arbitrary distributions over paths, which we call \textit{snakes}.

\begin{definition}
\label{snake}Given a vertex $h$ in $G$ and a positive integer $L$, a
\textit{snake distribution} $\mathcal{D}_{h,L}$\ (parameterized by $h$ and
$L$) is a probability distribution over paths $\left(  x_{0},\ldots
,x_{L-1}\right)  $\ in $G$, such that each $x_{t}$\ is either equal or
adjacent to $x_{t+1}$, and $x_{L-1}=h$. \ Let $D_{h,L}$\ be the support of
$\mathcal{D}_{h,L}$. \ Then an element of $D_{h,L}$\ is called a
\textit{snake}; the part near $x_{0}$\ is the \textit{tail} and the part near
$x_{L-1}=h$\ is the \textit{head}.
\end{definition}

Given a snake $X$ and integer $t$, we use $X\left[  t\right]  $\ as shorthand
for $\left\{  x_{0},\ldots,x_{t}\right\}  $.

\begin{definition}
\label{elgood}We say a snake $X\in D_{h,L}$\ is $\varepsilon$\textit{-good}%
\ if the following holds. \ Choose $j$ uniformly at random from $\left\{
0,\ldots,L-1\right\}  $, and let $Y=\left(  y_{0},\ldots,y_{L-1}\right)  $\ be
a snake drawn from $\mathcal{D}_{h,L}$\ conditioned on $x_{t}=y_{t}$\ for all
$t>j$. \ Then

\begin{enumerate}
\item[(i)] Letting $S_{X,Y}$\ be the set of vertices $v$ in $X\cap Y$\ such
that $\min\left\{  t:x_{t}=v\right\}  =\min\left\{  t:y_{t}=v\right\}  $, we
have%
\[
\Pr_{j,Y}\left[  X\cap Y=S_{X,Y}\right]  \geq9/10.
\]

\item[(ii)] For all vertices $v$, $\Pr_{j,Y}\left[  v\in Y\left[  j\right]
\right]  \leq\varepsilon$.
\end{enumerate}
\end{definition}

The procedure above---wherein we choose a $j$ uniformly at random, then draw a
$Y$\ from $\mathcal{D}_{h,L}$\ consistent with $X$ on all steps later than
$j$---will be important in what follows.\ \ I call it \textit{the snake
}$\mathit{X}$\textit{ flicking its tail}. \ Intuitively, a snake is good if it
is spread out fairly evenly in $G$---so that when it flicks its tail, (1) with
high probability the old and new tails do not intersect, and (2) any
particular vertex is hit by the new tail with probability at most
$\varepsilon$.

I now explain the `snake method' for proving lower bounds for \textsc{Local
Search}. \ Given a snake $X$, we define an input $f_{X}$\ with a unique local
minimum at $x_{0}$, and $f$-values that decrease along $X$ from head to tail.
\ Then, given inputs $f_{X}$ and $f_{Y}$ with $X\cap Y=S_{X,Y}$, we let the
relation function $R\left(  f_{X},f_{Y}\right)  $ be proportional to the
probability that snake $Y$ is obtained by $X$ flicking its tail. \ (If $X\cap
Y\neq S_{X,Y}$ we let $R=0$.) \ Let $f_{X}$ and $g_{Y}$ be inputs with
$R\left(  f_{X},g_{Y}\right)  >0$, and let $v$ be a vertex such that
$f_{X}\left(  v\right)  \neq g_{Y}\left(  v\right)  $. \ Then if all snakes
were good, there would be two mutually exclusive cases: (1) $v$ belongs to the
tail of $X$, or (2) $v$ belongs to the tail of $Y$. \ In case (1), $v$ is hit
with small probability when $Y$\ flicks its tail, so $\theta\left(
f_{Y},v\right)  $\ is small. \ In case (2), $v$ is hit with small probability
when $X$ flicks its tail, so $\theta\left(  f_{X},v\right)  $\ is small. \ In
either case, then, the \textit{geometric mean} $\sqrt{\theta\left(
f_{X},v\right)  \theta\left(  f_{Y},v\right)  }$\ and \textit{minimum}
$\min\left\{  \theta\left(  f_{X},v\right)  ,\theta\left(  f_{Y},v\right)
\right\}  $\ are small. \ So even though $\theta\left(  f_{X},v\right)  $\ or
$\theta\left(  f_{Y},v\right)  $\ could be large individually, Theorems
\ref{ambthmls}\ and \ref{classadv} yield a good lower bound, as in the case of
inverting a permutation (see Figure 7.1).%
%TCIMACRO{\FRAME{ftbpFU}{3.4537in}{2.3163in}{0pt}{\Qcb[A snake of vertices
%flicks its tail]{For every vertex $v$ such that $f_{X}\left(  v\right)  \neq
%f_{Y}\left(  v\right)  $, either when snake $X$ flicks its tail $v$ is not hit
%with high probability, or when snake $Y$ flicks its tail $v$ is not hit with
%high probability.}}{\Qlb{snakefig}}{h:/public_html/snakefig.eps}%
%width 3.4537in;  height 2.3163in;  depth 0pt;  original-width 10.3511in;
%original-height 7.7551in;  cropleft "0.3627";  croptop "0.9493";
%cropright "0.6504";  cropbottom "0.6922";
%filename 'snakefig.eps';file-properties "XNPEU";}}}%
%BeginExpansion
\begin{figure}
[ptb]
\begin{center}
\includegraphics[
trim=3.754344in 5.368080in 3.618745in 0.393184in,
height=2.3163in,
width=3.4537in
]%
{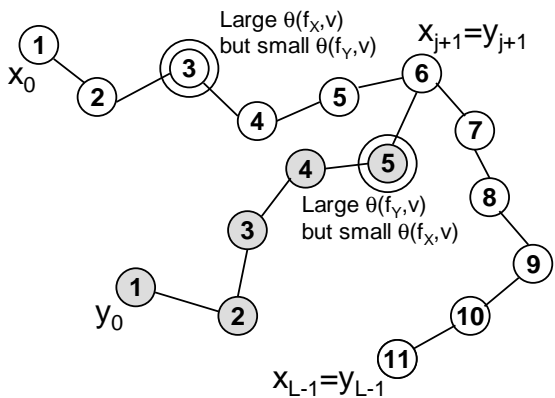}%
\caption[A snake of vertices flicks its tail]{For every vertex $v$ such that
$f_{X}\left(  v\right)  \neq f_{Y}\left(  v\right)  $, either when snake $X$
flicks its tail $v$ is not hit with high probability, or when snake $Y$ flicks
its tail $v$ is not hit with high probability.}%
\label{snakefig}%
\end{center}
\end{figure}
%EndExpansion

One difficulty is that not all snakes are good; at best, a large fraction of
them are. \ We could try deleting all inputs $f_{X}$\ such that $X$ is not
good, but that might ruin some remaining inputs, which would then have fewer
neighbors. \ So we would have to delete \textit{those} inputs as well, and so
on ad infinitum. \ What we need is basically a way to replace ``all
inputs''\ by ``most inputs''\ in Theorems \ref{ambthmls}\ and \ref{classadv}.

Fortunately, a simple graph-theoretic lemma can accomplish this. \ The lemma
(see Diestel \cite[p.6]{diestel} for example) says that any graph with average
degree at least $k$ contains an induced subgraph with \textit{minimum} degree
at least $k/2$. \ Below I prove a weighted analogue of the lemma.

\begin{lemma}
\label{subgraph}Let $p\left(  1\right)  ,\ldots,p\left(  m\right)  $\ be
positive reals summing to $1$. \ Also let $w\left(  i,j\right)  $ for
$i,j\in\left\{  1,\ldots,m\right\}  $\ be nonnegative reals satisfying
$w\left(  i,j\right)  =w\left(  j,i\right)  $\ and $\allowbreak\sum
_{i,j}w\left(  i,j\right)  \geq r$. \ Then there exists a nonempty subset
$U\subseteq\left\{  1,\ldots,m\right\}  $\ such that for all $i\in U$,
$\sum_{j\in U}w\left(  i,j\right)  \geq rp\left(  i\right)  /2.$
\end{lemma}

\begin{proof}
If $r=0$ then the lemma trivially holds, so assume $r>0$.\ \ We construct $U$
via an iterative procedure. \ Let $U\left(  0\right)  =\left\{  1,\ldots
,m\right\}  $. \ Then for all $t$, if there exists an\ $i^{\ast}\in U\left(
t\right)  $ for which%
\[
\sum_{j\in U\left(  t\right)  }w\left(  i^{\ast},j\right)  <\frac{r}%
{2}p\left(  i^{\ast}\right)  ,
\]
then set $U\left(  t+1\right)  =U\left(  t\right)  \setminus\left\{  i^{\ast
}\right\}  $. \ Otherwise halt and return $U=U\left(  t\right)  $. \ To see
that the $U$ so constructed is nonempty, observe that when we remove $i^{\ast
}$, the sum $\sum_{i\in U\left(  t\right)  }p\left(  i\right)  $\ decreases by
$p\left(  i^{\ast}\right)  $, while $\sum_{i,j\in U\left(  t\right)  }w\left(
i,j\right)  $\ decreases by at most%
\[
\sum_{j\in U\left(  t\right)  }w\left(  i^{\ast},j\right)  +\sum_{j\in
U\left(  t\right)  }w\left(  j,i^{\ast}\right)  <rp\left(  i^{\ast}\right)  .
\]
So since $\sum_{i,j\in U\left(  t\right)  }w\left(  i,j\right)  $\ was
positive to begin with, it must still be positive at the end of the procedure;
hence $U$\ must be nonempty.
\end{proof}

I can now prove the main result of the section.

\begin{theorem}
\label{kappathm}Suppose a snake drawn from $\mathcal{D}_{h,L}$\ is
$\varepsilon$\textit{-}good\ with probability at least $9/10$. \ Then%
\[
\operatorname*{RLS}\left(  G\right)  =\Omega\left(  1/\varepsilon\right)
,~~~~~~\operatorname*{QLS}\left(  G\right)  =\Omega\left(  \sqrt
{1/\varepsilon}\right)  .
\]

\end{theorem}

\begin{proof}
Given a snake $X\in D_{h,L}$, we construct an input function $f_{X}$ as
follows.\ \ For each $v\in X$, let $f_{X}\left(  v\right)  =\min\left\{
t:x_{t}=v\right\}  $; and for each $v\notin X$, let $f_{X}\left(  v\right)
=\Delta\left(  v,h\right)  +L$\ where $\Delta\left(  v,h\right)  $\ is the
distance from $v$ to $h$ in $G$. \ Clearly $f_{X}$ so defined has a unique
local minimum at $x_{0}$. \ To obtain a decision problem, we stipulate that
querying $x_{0}$\ reveals an answer bit ($0$ or $1$) in addition to
$f_{X}\left(  x_{1}\right)  $; the algorithm's goal is then to return the
answer bit. \ Obviously a lower bound for the decision problem implies a
corresponding lower bound for the search problem. Let us first prove the
theorem in the case that all snakes in $D_{h,L}$\ are $\varepsilon$%
\textit{-}good. \ Let $p\left(  X\right)  $\ be the probability of drawing
snake $X$ from $\mathcal{D}_{h,L}$. \ Also, given snakes $X,Y$\ and
$j\in\left\{  0,\ldots,L-1\right\}  $, let $q_{j}\left(  X,Y\right)  $\ be the
probability that $X^{\ast}=Y$, if $X^{\ast}$ is drawn from $\mathcal{D}_{h,L}$
conditioned on agreeing with $X$\ on all steps later than $j$. \ Then define%
\[
w\left(  X,Y\right)  =\frac{p\left(  X\right)  }{L}\sum_{j=0}^{L-1}%
q_{j}\left(  X,Y\right)  .
\]
The first claim is that $w$\ is symmetric; that is, $w\left(  X,Y\right)
=w\left(  Y,X\right)  $. \ It suffices to show that
\[
p\left(  X\right)  q_{j}\left(  X,Y\right)  =p\left(  Y\right)  q_{j}\left(
Y,X\right)
\]
for all $j$. \ We can assume $X$\ agrees with $Y$ on all steps later than $j$,
since otherwise $q_{j}\left(  X,Y\right)  =q_{j}\left(  Y,X\right)  =0$.
\ Given an $X^{\ast}\in D_{h,L}$, let $A$ denote the event that $X^{\ast}$
agrees with $X$ (or equivalently $Y$) on all steps later than $j$, and let
$B_{X}$ (resp. $B_{Y}$) denote the event that $X^{\ast}$ agrees with
$X$\ (resp. $Y$) on steps $1$ to $j$. \ Then%
\begin{align*}
p\left(  X\right)  q_{j}\left(  X,Y\right)   &  =\Pr\left[  A\right]
\Pr\left[  B_{X}|A\right]  \cdot\Pr\left[  B_{Y}|A\right] \\
&  =p\left(  Y\right)  q_{j}\left(  Y,X\right)  .
\end{align*}
Now let $E\left(  X,Y\right)  $\ denote the event that $X\cap Y=S_{X,Y}%
$,\ where $S_{X,Y}$\ is as in Definition \ref{elgood}. \ Also, let $f_{X}$\ be
the input obtained from $X$ that has answer bit $0$, and $g_{X}$\ be the input
that has answer bit $1$. \ To apply Theorems \ref{ambthmls}\ and
\ref{classadv}, take $\mathcal{A}=\left\{  f_{X}:X\in D_{h,L}\right\}  $ and
$\mathcal{B}=\left\{  g_{X}:X\in D_{h,L}\right\}  $. \ Then take $R\left(
f_{X},g_{Y}\right)  =w\left(  X,Y\right)  $\ if $E\left(  X,Y\right)
$\ holds, and $R\left(  f_{X},g_{Y}\right)  =0$\ otherwise. \ Given $f_{X}%
\in\mathcal{A}$\ and $g_{Y}\in\mathcal{B}$\ with $R\left(  f_{X},g_{Y}\right)
>0$, and letting $v$ be a vertex such that $f_{X}\left(  v\right)  \neq
g_{Y}\left(  v\right)  $, we must then have either $v\notin X$\ or $v\notin
Y$. \ Suppose the former case; then%
\[
\sum_{f_{X^{\ast}}\in\mathcal{A}~:~f_{X^{\ast}}\left(  v\right)  \neq
g_{Y}\left(  v\right)  }R\left(  f_{X^{\ast}},g_{Y}\right)  \leq
\sum_{f_{X^{\ast}}\in\mathcal{A}~:~f_{X^{\ast}}\left(  v\right)  \neq
g_{Y}\left(  v\right)  }\frac{p\left(  Y\right)  }{L}\sum_{j=0}^{L-1}%
q_{j}\left(  Y,X^{\ast}\right)  \leq\varepsilon p\left(  Y\right)  ,
\]
since $Y$\ is $\varepsilon$-good. \ Thus $\theta\left(  g_{Y},v\right)  $
equals
\[
\frac{\sum_{f_{X^{\ast}}\in\mathcal{A}~:~f_{X^{\ast}}\left(  v\right)  \neq
g_{Y}\left(  v\right)  }R\left(  f_{X^{\ast}},g_{Y}\right)  }{\sum
_{f_{X^{\ast}}\in\mathcal{A}}R\left(  f_{X^{\ast}},g_{Y}\right)  }\leq
\frac{\varepsilon p\left(  Y\right)  }{9p\left(  Y\right)  /10}.
\]
Similarly, if $v\notin Y$\ then $\theta\left(  f_{X},v\right)  \leq
10\varepsilon/9$\ by symmetry. \ Hence%
\begin{align*}
\upsilon_{\min}  &  =\max_{f_{X}\in\mathcal{A},~g_{Y}\in\mathcal{B}%
,~v~:~R\left(  f_{X},g_{Y}\right)  >0,~f_{X}\left(  v\right)  \neq
g_{Y}\left(  v\right)  }\min\left\{  \theta\left(  f_{X},v\right)
,\theta\left(  g_{Y},v\right)  \right\}  \leq\frac{\varepsilon}{9/10},\\
\upsilon_{\operatorname*{geom}}  &  =\max_{f_{X}\in\mathcal{A},~g_{Y}%
\in\mathcal{B},~v~:~R\left(  f_{X},g_{Y}\right)  >0,~f_{X}\left(  v\right)
\neq g_{Y}\left(  v\right)  }\sqrt{\theta\left(  f_{X},v\right)  \theta\left(
g_{Y},v\right)  }\leq\sqrt{\frac{\varepsilon}{9/10}},
\end{align*}
the latter since $\theta\left(  f_{X},v\right)  \leq1$\ and $\theta\left(
g_{Y},v\right)  \leq1$\ for all $f_{X},g_{Y}$ and $v$.

In the general case, all we know is that a snake drawn from $\mathcal{D}%
_{h,L}$\ is $\varepsilon$\textit{-}good\ with probability at least $9/10$.
\ Let $G\left(  X\right)  $\ denote the event that $X$ is $\varepsilon$-good.
\ Take $\mathcal{A}^{\ast}=\left\{  f_{X}\in\mathcal{A}:G\left(  X\right)
\right\}  $ and $\mathcal{B}^{\ast}=\left\{  g_{Y}\in\mathcal{B}:G\left(
Y\right)  \right\}  $, and\ take $\allowbreak R\left(  f_{X},g_{Y}\right)
$\ as before. \ Then since%
\[
\sum_{X,Y~:~E\left(  X,Y\right)  }w\left(  X,Y\right)  \geq\sum_{X}\frac
{9}{10}p\left(  X\right)  \geq\frac{9}{10},
\]
by the union bound we have%
\begin{align*}
\sum_{f_{X}\in\mathcal{A}^{\ast},~g_{Y}\in\mathcal{B}^{\ast}}R\left(
f_{X},g_{Y}\right)   &  \geq\sum_{X,Y~:~G\left(  X\right)  \wedge G\left(
Y\right)  \wedge E\left(  X,Y\right)  }w\left(  X,Y\right)  -\sum
_{X~:~\urcorner G\left(  X\right)  }p\left(  X\right)  -\sum_{Y~:~\urcorner
G\left(  Y\right)  }p\left(  Y\right) \\
&  \geq\frac{9}{10}-\frac{1}{10}-\frac{1}{10}\\
&  =\frac{7}{10}.
\end{align*}
So by Lemma \ref{subgraph}, there exist subsets $\widetilde{\mathcal{A}%
}\subseteq\mathcal{A}^{\ast}$\ and $\widetilde{\mathcal{B}}\subseteq
\mathcal{B}^{\ast}$\ such that for all $f_{X}\in\widetilde{\mathcal{A}}$ and
$g_{Y}\in\widetilde{\mathcal{B}}$,%
\begin{align*}
\sum_{g_{Y^{\ast}}\in\widetilde{\mathcal{B}}}R\left(  f_{X},g_{Y^{\ast}%
}\right)   &  \geq\frac{7p\left(  X\right)  }{20},\\
\sum_{f_{X^{\ast}}\in\widetilde{\mathcal{A}}}R\left(  f_{X^{\ast}}%
,g_{Y}\right)   &  \geq\frac{7p\left(  Y\right)  }{20}.
\end{align*}
So for all $f_{X},g_{Y}$\ with $R\left(  f_{X},g_{Y}\right)  >0$, and all $v$
such that $f_{X}\left(  v\right)  \neq g_{Y}\left(  v\right)  $, either
$\theta\left(  f_{X},v\right)  \leq20\varepsilon/7$\ or $\theta\left(
g_{Y},v\right)  \leq20\varepsilon/7$. \ Hence $\upsilon_{\min}\leq
20\varepsilon/7$\ and $\upsilon_{\operatorname*{geom}}\leq\sqrt{20\varepsilon
/7}$.
\end{proof}

\section{Specific Graphs\label{GRAPHS}}

In this section I apply the `snake method' developed in Section \ref{SNAKE} to
specific examples of graphs: the Boolean hypercube in Section \ref{BOOLEAN},
and the $d$-dimensional cubic grid (for $d\geq3$)\ in Section \ref{DDIM}.

\subsection{Boolean Hypercube\label{BOOLEAN}}

Abusing notation, let $\left\{  0,1\right\}  ^{n}$\ denote the $n$-dimensional
Boolean hypercube---that is, the graph whose vertices are $n$-bit
strings,\ with two vertices adjacent if and only if they have Hamming distance
$1$. \ Given a vertex $v\in\left\{  0,1\right\}  ^{n}$, let $v\left[
0\right]  ,\ldots,v\left[  n-1\right]  $\ denote the $n$ bits of $v$, and let
$v^{\left(  i\right)  }$ denote the neighbor obtained by flipping bit
$v\left[  i\right]  $. \ In this section I lower-bound $\operatorname*{RLS}%
\left(  \left\{  0,1\right\}  ^{n}\right)  $\ and $\operatorname*{QLS}\left(
\left\{  0,1\right\}  ^{n}\right)  $.

Fix a `snake head' $h\in\left\{  0,1\right\}  ^{n}$ and take $L=2^{n/2}/100$.
\ I define the snake distribution $\mathcal{D}_{h,L}$\ via what I call a
\textit{coordinate loop}, as follows. \ Starting from $x_{0}=h$,\ for each $t$
take $x_{t+1}=x_{t}$\ with $1/2$ probability, and $x_{t+1}=x_{t}^{\left(
t\operatorname{mod}n\right)  }$\ with $1/2$ probability. \ The following is a
basic fact about this distribution.

\begin{proposition}
\label{mixtime}The coordinate loop mixes completely in $n$ steps, in the sense
that if $t^{\ast}\geq t+n$, then $x_{t^{\ast}}$\ is a uniform random
vertex\ conditioned on $x_{t}$.
\end{proposition}

One could also use the random walk distribution, following Aldous
\cite{aldous}. \ However, not only is the coordinate loop distribution easier
to work with (since it produces fewer self-intersections), it also yields a
better lower bound (since it mixes completely in $n$ steps, as opposed to
approximately in $n\log n$ steps).

I first upper-bound the probability, over $X$, $j$, and $Y\left[  j\right]  $,
that $X\cap Y\neq S_{X,Y}$ (where $S_{X,Y}$\ is as in Definition \ref{elgood}).

\begin{lemma}
\label{intersect}Suppose $X$ is drawn from $\mathcal{D}_{h,L}$, $j$ is drawn
uniformly from $\left\{  0,\ldots,L-1\right\}  $, and $Y\left[  j\right]
$\ is drawn from $\mathcal{D}_{x_{j},j}$. \ Then $\Pr_{X,j,Y\left[  j\right]
}\left[  X\cap Y=S_{X,Y}\right]  \geq0.9999$.
\end{lemma}

\begin{proof}
Call a \textit{disagreement} a vertex $v$\ such that%
\[
\min\left\{  t:x_{t}=v\right\}  \neq\min\left\{  t^{\ast}:y_{t^{\ast}%
}=v\right\}  .
\]
Clearly if there are no disagreements then $X\cap Y=S_{X,Y}$. \ If $v$\ is a
disagreement, then by the definition of $\mathcal{D}_{h,L}$\ we cannot have
both $t>j-n$\ and $t^{\ast}>j-n$. \ So by Proposition \ref{mixtime}, either
$y_{t^{\ast}}$\ is uniformly random conditioned on $X$, or $x_{t}$\ is
uniformly random conditioned on $Y\left[  j\right]  $. \ Hence $\Pr
_{X,j,Y\left[  j\right]  }\left[  x_{t}=y_{t^{\ast}}\right]  =1/2^{n}$. \ So
by the union bound,%
\[
\Pr_{X,j,Y\left[  j\right]  }\left[  X\cap Y\neq S_{X,Y}\right]  \leq
\frac{L^{2}}{2^{n}}=0.0001.
\]

\end{proof}

I now argue that, unless $X$ spends a `pathological' amount of time in one
part of the hypercube, the probability of any vertex $v$ being hit when $X$
flicks its tail is small. \ To prove this, I define a notion of
\textit{sparseness}, and then show that (1) almost all snakes drawn from
$\mathcal{D}_{h,L}$\ are sparse (Lemma \ref{hammingball}), and (2) sparse
snakes are unlikely to hit any given vertex $v$ (Lemma \ref{sparsegood}).

\begin{definition}
\label{sparse}Given vertices $v,w$ and $i\in\left\{  0,\ldots,n-1\right\}  $,
let $\Delta\left(  x,v,i\right)  $\ be the number of steps needed to reach $v$
from $x$ by first setting $x\left[  i\right]  :=v\left[  i\right]  $, then
setting $x\left[  i-1\right]  :=v\left[  i-1\right]  $, and so on. \ (After we
set $x\left[  0\right]  $\ we wrap around to $x\left[  n-1\right]  $.) \ Then
$X$ is \textit{sparse} if there exists a constant $c$ such that for all
$v\in\left\{  0,1\right\}  ^{n}$ and all $k$,%
\[
\left\vert \left\{  t:\Delta\left(  x_{t},v,t\operatorname{mod}n\right)
=k\right\}  \right\vert \leq cn\left(  n+\frac{L}{2^{n-k}}\right)  .
\]

\end{definition}

\begin{lemma}
\label{hammingball}If $X$ is drawn from $\mathcal{D}_{h,L}$, then $X$ is
sparse with probability $1-o\left(  1\right)  $.
\end{lemma}

\begin{proof}
For each $i\in\left\{  0,\ldots,n-1\right\}  $, the number of $t\in\left\{
0,\ldots,L-1\right\}  $\ such that $t\equiv i\left(  \operatorname{mod}%
n\right)  $\ is at most $L/n$.\ \ For such a $t$, let $E_{t}^{\left(
v,i,k\right)  }$\ be the event that $\Delta\left(  x_{t},v,i\right)  \leq k$;
then $E_{t}^{\left(  v,i,k\right)  }$\ holds if and only if%
\[
x_{t}\left[  i\right]  =v\left[  i\right]  ,\ldots,x_{t}\left[  i-k+1\right]
=v\left[  i-k+1\right]
\]
(where we wrap around to $x_{t}\left[  n-1\right]  $\ after reaching
$x_{t}\left[  0\right]  $). \ This occurs with probability $2^{k}/2^{n}$\ over
$X$. \ Furthermore, by Proposition \ref{mixtime}, the $E_{t}^{\left(
v,i,k\right)  }$\ events for different $t$'s are independent. \ So let%
\[
\mu_{k}=\frac{L}{n}\cdot\frac{2^{k}}{2^{n}};
\]
then for fixed $v,i,k$, the expected number of $t$'s\ for which $E_{t}%
^{\left(  v,i,k\right)  }$\ holds is at most $\mu_{k}$. \ Thus by a Chernoff
bound, if $\mu_{k}\geq1$\ then%
\[
\Pr_{X}\left[  \left\vert \left\{  t:E_{t}^{\left(  v,i,k\right)  }\right\}
\right\vert >cn\cdot\mu_{k}\right]  <\left(  \frac{e^{cn-1}}{\left(
cn\right)  ^{cn}}\right)  ^{\mu_{k}}<\frac{1}{2^{2n}}%
\]
for sufficiently large $c$. \ Similarly, if $\mu_{k}<1$\ then%
\[
\Pr_{X}\left[  \left\vert \left\{  t:E_{t}^{\left(  v,i,k\right)  }\right\}
\right\vert >cn\right]  <\left(  \frac{e^{cn/\mu_{k}-1}}{\left(  cn/\mu
_{k}\right)  ^{cn/\mu_{k}}}\right)  ^{\mu_{k}}<\frac{1}{2^{2n}}%
\]
for sufficiently large $c$. \ By the union bound, then,%
\begin{align*}
\left\vert \left\{  t:E_{t}^{\left(  v,i,k\right)  }\right\}  \right\vert  &
\leq cn\cdot\left(  1+\mu_{k}\right) \\
&  =c\left(  n+\frac{L}{2^{n-k}}\right)
\end{align*}
for every $v,i,k$ triple \textit{simultaneously} with probability at least
$1-n^{2}2^{n}/2^{2n}=1-o\left(  1\right)  $. \ Summing over all $i$'s produces
the additional factor of $n$.
\end{proof}

\begin{lemma}
\label{sparsegood}If $X$ is sparse, then for every $v\in\left\{  0,1\right\}
^{n}$,%
\[
\Pr_{j,Y}\left[  v\in Y\left[  j\right]  \right]  =O\left(  \frac{n^{2}}%
{L}\right)  .
\]

\end{lemma}

\begin{proof}
By assumption, for every $k\in\left\{  0,\ldots,n\right\}  $,%
\begin{align*}
\Pr_{j}\left[  \Delta\left(  x_{j},v,j\operatorname{mod}n\right)  =k\right]
&  \leq\frac{\left\vert \left\{  t:\Delta\left(  x_{t},v,t\operatorname{mod}%
n\right)  =k\right\}  \right\vert }{L}\\
&  \leq\frac{cn}{L}\left(  n+\frac{L}{2^{n-k}}\right)  .
\end{align*}
Consider the probability that $v\in Y\left[  j\right]  $\ in the event that
$\Delta\left(  x_{j},v,j\operatorname{mod}n\right)  =k$.\ \ Clearly%
\[
\Pr_{Y}\left[  v\in\left\{  y_{j-n+1},\ldots,y_{j}\right\}  \right]  =\frac
{1}{2^{k}}.
\]
Also, Proposition \ref{mixtime}\ implies that for every $t\leq j-n$, the
probability that $y_{t}=v$\ is $2^{-n}$. \ So by the union bound,%
\[
\Pr_{Y}\left[  v\in\left\{  y_{0},\ldots,y_{j-n}\right\}  \right]  \leq
\frac{L}{2^{n}}.
\]
Then $\Pr_{j,Y}\left[  v\in Y\left[  j\right]  \right]  $\ equals%
\begin{align*}
\sum_{k=0}^{n}\left(
\begin{array}
[c]{c}%
\Pr_{j}\left[  \Delta\left(  x_{j},v,j\operatorname{mod}n\right)  =k\right]
\cdot\\
\Pr_{Y}\left[  v\in Y\left[  j\right]  ~|~\Delta\left(  x_{j}%
,v,j\operatorname{mod}n\right)  =k\right]
\end{array}
\right)   &  \leq\sum_{k=0}^{n}\frac{cn}{L}\left(  n+\frac{L}{2^{n-k}}\right)
\left(  \frac{1}{2^{k}}+\frac{L}{2^{n}}\right) \\
&  =O\left(  \frac{cn^{2}}{L}\right)
\end{align*}
as can be verified by breaking the sum into cases and doing some manipulations.
\end{proof}

The main result follows easily:

\begin{theorem}
\label{boolean}%
\[
\operatorname*{RLS}\left(  \left\{  0,1\right\}  ^{n}\right)  =\Omega\left(
\frac{2^{n/2}}{n^{2}}\right)  ,~~\operatorname*{QLS}\left(  \left\{
0,1\right\}  ^{n}\right)  =\Omega\left(  \frac{2^{n/4}}{n}\right)  .
\]

\end{theorem}

\begin{proof}
Take $\varepsilon=n^{2}/2^{n/2}$. \ Then by Theorem \ref{kappathm},\ it
suffices to show that a snake $X$ drawn from $\mathcal{D}_{h,L}$\ is $O\left(
\varepsilon\right)  $-good with probability at least\ $9/10$. \ First, since%
\[
\Pr_{X,j,Y\left[  j\right]  }\left[  X\cap Y=S_{X,Y}\right]  \geq0.9999
\]
by Lemma \ref{intersect}, Markov's inequality shows that%
\[
\Pr_{X}\left[  \Pr_{j,Y\left[  j\right]  }\left[  X\cap Y=S_{X,Y}\right]
\geq\frac{9}{10}\right]  \geq\frac{19}{20}.
\]
Second, by Lemma \ref{hammingball}, $X$ is sparse with probability $1-o\left(
1\right)  $, and by Lemma \ref{sparsegood}, if $X$ is sparse then%
\[
\Pr_{j,Y}\left[  v\in Y\left[  j\right]  \right]  =O\left(  \frac{n^{2}}%
{L}\right)  =O\left(  \varepsilon\right)
\]
for every $v$.\ \ So both requirements of Definition \ref{elgood}\ hold
simultaneously with probability at least $9/10$.
\end{proof}

\subsection{Constant-Dimensional Grid Graph\label{DDIM}}

In the Boolean hypercube case, $\mathcal{D}_{h,L}$\ was defined by a
`coordinate loop' instead of the usual random walk mainly for convenience.
\ When we move to the $d$-dimensional grid, though, the drawbacks of random
walks become more serious: first, the mixing time is too long, and second,
there are too many self-intersections, particularly if $d\leq4$. \ The snake
distribution will instead use straight lines of randomly chosen lengths
attached at the endpoints, as in Figure 7.2.%
%TCIMACRO{\FRAME{ftbpFU}{2.4267in}{2.0149in}{0pt}{\Qcb[The coordinate loop in
%$3$ dimensions]{In $d=3$ dimensions, a snake drawn from $\mathcal{D}_{h,L}%
%$\ moves a random distance left or right, then a random distance up or down,
%then a random distance inward or outward, etc.}}{\Qlb{linefig}}{linefig.eps}%
%width 2.4267in;  height 2.0149in;  depth 0pt;  original-width 10.3511in;
%original-height 7.7551in;  cropleft "0.4027";  croptop "0.9277";
%cropright "0.5958";  cropbottom "0.7143";
%filename 'linefig.eps';file-properties "XNPEU";}}}%
%BeginExpansion
\begin{figure}
[ptb]
\begin{center}
\includegraphics[
trim=4.168388in 5.539468in 4.183915in 0.560694in,
height=2.0149in,
width=2.4267in
]%
{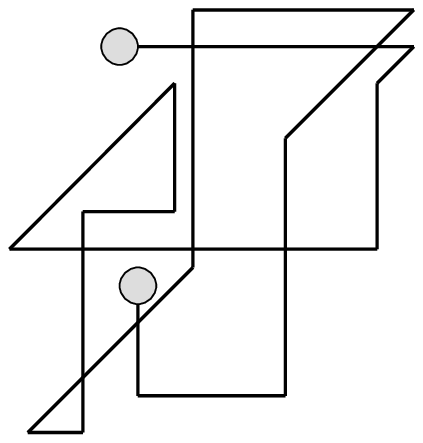}%
\caption[The coordinate loop in $3$ dimensions]{In $d=3$ dimensions, a snake
drawn from $\mathcal{D}_{h,L}$\ moves a random distance left or right, then a
random distance up or down, then a random distance inward or outward, etc.}%
\label{linefig}%
\end{center}
\end{figure}
%EndExpansion
Let $G_{d,N}$\ be a $d$-dimensional grid graph with $d\geq3$. \ That is,
$G_{d,N}$\ has $N$ vertices of the form $v=\left(  v\left[  0\right]
,\ldots,v\left[  d-1\right]  \right)  $, where each $v\left[  i\right]  $\ is
in $\left\{  1,\ldots,N^{1/d}\right\}  $ (assume for simplicity that $N$ is a
$d^{th}$\ power). \ Vertices $v$ and $w$ are adjacent if and only if
$\left\vert v\left[  i\right]  -w\left[  i\right]  \right\vert =1$\ for some
$i\in\left\{  0,\ldots,d-1\right\}  $, and $v\left[  j\right]  =w\left[
j\right]  $\ for all $j\neq i$ (so $G_{d,N}$\ does not wrap around at the boundaries).

Take $L=\sqrt{N}/100$, and define the snake distribution $\mathcal{D}_{h,L}%
$\ as follows. \ Starting from $x_{0}=h$, for each $T$\ take $x_{N^{1/d}%
\left(  T+1\right)  }$\ identical to $x_{N^{1/d}T}$, but with the $\left(
T\operatorname{mod}d\right)  ^{th}$\ coordinate $x_{N^{1/d}\left(  T+1\right)
}\left[  T\operatorname{mod}d\right]  $\ replaced by a uniform random value in
$\left\{  1,\ldots,N^{1/d}\right\}  $. \ Then take the vertices $x_{N^{1/d}%
T+1},\ldots,x_{N^{1/d}T+N^{1/d}-1}$\ to lie along the shortest path from
$x_{N^{1/d}T}$\ to $x_{N^{1/d}\left(  T+1\right)  }$,\ `stalling'\ at
$x_{N^{1/d}\left(  T+1\right)  }$\ once that vertex has been reached. \ Call%
\[
\Phi_{T}=\left(  x_{N^{1/d}T},\ldots,x_{N^{1/d}T+N^{1/d}-1}\right)
\]
a \textit{line} of vertices, whose \textit{direction} is $T\operatorname{mod}%
d$. \ As in the Boolean hypercube case, we have:

\begin{proposition}
\label{mixtime2}$\mathcal{D}_{h,L}$ mixes completely in $dN^{1/d}$ steps, in
the sense that if $T^{\ast}\geq T+d$, then $x_{N^{1/d}T^{\ast}}$\ is a uniform
random vertex\ conditioned on $x_{N^{1/d}T}$.
\end{proposition}

Lemma \ref{intersect}\ in Section \ref{BOOLEAN}\ goes through essentially
without change.

\begin{definition}
\label{sparse2}Letting $\Delta\left(  x,v,i\right)  $\ be as before, we say
$X$ is \textit{sparse} if there exists a constant $c$ (possibly dependent on
$d$) such that for all vertices $v$ and all $k$,%
\[
\left\vert \left\{  t:\Delta\left(  x_{t},v,\left\lfloor t/N^{1/d}%
\right\rfloor \operatorname{mod}d\right)  =k\right\}  \right\vert \leq\left(
c\log N\right)  \left(  N^{1/d}+\frac{L}{N^{1-k/d}}\right)  .
\]

\end{definition}

\begin{lemma}
\label{hammingball2}If $X$ is drawn from $\mathcal{D}_{h,L}$, then $X$ is
sparse with probability $1-o\left(  1\right)  $.
\end{lemma}

\begin{proof}
Similar to Lemma \ref{hammingball}. \ Let $\Phi_{T}$\ be a line of vertices
with direction $i=T\operatorname{mod}d$, and\ notice that $\Delta\left(
x_{t},v,i\right)  $\ is the same for every vertex $x_{t}$ in $\Phi_{T}$. \ Let
$E_{T}^{\left(  v,i,k\right)  }$\ denote the event that $\Delta\left(
x_{t},v,i\right)  \leq k$ for the $x_{t}$'s\ in $\Phi_{T}$. \ Then
$E_{T}^{\left(  v,i,k\right)  }$ occurs with probability $N^{\left(
k-1\right)  /d}/N$\ over $X$. \ Furthermore, if $\left\vert T-T^{\ast
}\right\vert \geq d$\ then $E_{T}^{\left(  v,i,k\right)  }$\ and $E_{T^{\ast}%
}^{\left(  v,i,k\right)  }$ are independent events. \ So let%
\[
\mu_{k}=L\cdot\frac{N^{\left(  k-1\right)  /d}}{N};
\]
then for fixed $v,i,k$,\ the expected number of lines for which $E_{T}%
^{\left(  v,i,k\right)  }$\ holds is at most $\mu_{k}$. \ Thus, by a Chernoff
bound, if $\mu_{k}\geq1$\ then%
\[
\Pr_{X}\left[  \left\vert \left\{  T:E_{T}^{\left(  v,i,k\right)  }\right\}
\right\vert >c\log N\cdot\mu_{k}\right]  <\left(  \frac{e^{c\log N-1}}{\left(
c\log N\right)  ^{c\log N}}\right)  ^{\mu_{k}}%
\]
which is at most $1/N^{2}$ for sufficiently large $c$. \ Similarly, if
$\mu_{k}<1$\ then letting $m=\left(  c\log N\right)  /\mu_{k}$,
\[
\Pr_{X}\left[  \left\vert \left\{  T:E_{T}^{\left(  v,i,k\right)  }\right\}
\right\vert >c\log N\right]  <\left(  \frac{e^{m-1}}{m^{m}}\right)  ^{\mu_{k}%
}<\frac{1}{N^{2}}%
\]
for sufficiently large $c$. \ So with probability $1-o\left(  1\right)  $\ it
holds that for all $v,k$, letting $i_{t}=\left\lfloor t/N^{1/d}\right\rfloor
\operatorname{mod}d$,%
\begin{align*}
\left\vert \left\{  t:\Delta\left(  x_{t},v,i_{t}\right)  =k\right\}
\right\vert  &  \leq c\log N\cdot\left(  1+\mu_{k}\right)  \cdot N^{1/d}\\
&  = \left(  c\log N\right)  \left(  N^{1/d}+\frac{L}{N^{1-k/d}}\right)  .
\end{align*}

\end{proof}

\begin{lemma}
\label{sparsegood2}If $X$ is sparse, then for every $v\in G_{d,N}$,%
\[
\Pr_{j,Y}\left[  v\in Y\left[  j\right]  \right]  =O\left(  \frac{N^{1/d}\log
N}{L}\right)  ,
\]
where the big-$O$ hides a constant dependent on $d$.
\end{lemma}

\begin{proof}
As in Lemma \ref{sparsegood}, setting $i_{j}=\left\lfloor j/N^{1/d}%
\right\rfloor \operatorname{mod}d$\ we obtain that $\Pr_{j,Y}\left[  v\in
Y\left[  j\right]  \right]  $ equals
\begin{align*}
& \sum_{k=1}^{d}\Pr_{j}\left[  \Delta\left(  x_{j},v,i_{j}\right)
=k\right] \Pr_{Y}\left[  v\in Y\left[  j\right]  ~|~\Delta\left(
x_{j},v,i_{j}\right) =k\right] \\ & \leq\sum_{k=1}^{d}\frac{c\log
N}{L}\left(  N^{1/d}+\frac {L}{N^{1-k/d}}\right)  \left(
\frac{1}{N^{\left(  k-1\right)  /d}}+\frac {L}{N}\right) \\ &
 =O\left(  \frac{N^{1/d}\log N}{L}\right)  .
\end{align*}

\end{proof}

By the same proof as for Theorem \ref{boolean}, taking $\varepsilon=\left(
\log N\right)  /N^{1/2-1/d}$\ yields the following:

\begin{theorem}
\label{grid}Neglecting a constant dependent on $d$, for all $d\geq3$%
\begin{align*}
\operatorname*{RLS}\left(  G_{d,N}\right)   &  =\Omega\left(  \frac
{N^{1/2-1/d}}{\log N}\right)  ,\\
\operatorname*{QLS}\left(  G_{d,N}\right)   &  =\Omega\left(  \sqrt
{\frac{N^{1/2-1/d}}{\log N}}\right)  .
\end{align*}

\end{theorem}

\chapter{Quantum Certificate Complexity\label{CER}}

This chapter studies the relationships between classical and quantum measures
of query complexity. \ Let $f:\mathcal{S}\rightarrow\left\{  0,1\right\}
$\ be a Boolean function with $\allowbreak\mathcal{S}\subseteq\left\{
0,1\right\}  ^{n}$, that takes input $Y=y_{1}\ldots y_{n}$. \ Then the
deterministic query complexity $\operatorname*{D}\left(  f\right)  $\ is the
minimum number of queries to the $y_{i}$'s need\-ed to evaluate $f$, if $Y$ is
chosen adversarially and if queries can be adaptive (that is, can depend on
the outcomes of previous queries). \ Also, the bounded-error randomized query
complexity, $\operatorname*{R}_{2}\left(  f\right)  $, is the minimum expected
number of queries needed by a randomized algorithm that, for each $Y$, outputs
$f\left(  Y\right)  $ with probability at least $2/3$. \ Here the `$2$' refers
to two-sided error; if instead we require $f\left(  Y\right)  $ to be output
with probability $1$ for every $Y$, we obtain $\operatorname*{R}_{0}\left(
f\right)  $, or zero-error randomized query complexity.

Analogously, $\operatorname*{Q}_{2}\left(  f\right)  $\ is the minimum number
of queries need\-ed by a quantum algorithm that outputs $f\left(  Y\right)  $
with probability at least $2/3$ for all $Y$. \ Also, for $k\in\left\{
0,1\right\}  $\ let $\operatorname*{Q}_{0}^{k}\left(  f\right)  $\ be the
minimum number of queries needed by a quantum algorithm that outputs $f\left(
Y\right)  $\ with probability $1$ if $f\left(  Y\right)  =k$, and with
probability at least $1/2$ if $f\left(  Y\right)  \neq k$.\ \ Then let
$\operatorname*{Q}_{0}\left(  f\right)  =\max\left\{  \operatorname*{Q}%
_{0}^{0}\left(  f\right)  ,\operatorname*{Q}_{0}^{1}\left(  f\right)
\right\}  $. \ If we require a single algorithm that succeeds with probability
$1$ for all $Y$, we obtain $\operatorname*{Q}_{E}\left(  f\right)  $, or exact
quantum query complexity. \ See Buhrman and de Wolf \cite{bw}\ for a more
detailed survey of these measures.

It is immediate that%
\[
\operatorname*{Q}{}_{2}\left(  f\right)  \leq\operatorname*{R}{}_{2}\left(
f\right)  \leq\operatorname*{R}{}_{0}\left(  f\right)  \leq\operatorname*{D}%
\left(  f\right)  \leq n,
\]
that $\operatorname*{Q}_{0}\left(  f\right)  \leq\operatorname*{R}_{0}\left(
f\right)  $, and that $\operatorname*{Q}_{E}\left(  f\right)  \leq
\operatorname*{D}\left(  f\right)  $. \ If $f$ is partial (i.e. $\mathcal{S}%
\neq\left\{  0,1\right\}  ^{n}$), then $\operatorname*{Q}_{2}\left(
f\right) $ can be superpolynomially smaller than
$\operatorname*{R}_{2}\left( f\right)  $; this is what makes Shor's
period-finding algorithm \cite{shor} possible. \ For total $f$, by
contrast, the largest known gap even between
$\operatorname*{D}\left(  f\right)  $\ and
$\operatorname*{Q}_{2}\left( f\right)  $\ is quadratic, and is
achieved by the $\operatorname*{OR}$ function on $n$ bits:
$\operatorname*{D}\left(  OR\right)  =n$\ (indeed
$\operatorname*{R}_{2}\left(  \operatorname*{OR}\right)
=\Omega\left( n\right)  $), whereas $\operatorname*{Q}_{2}\left(
\operatorname*{OR}\right) =\Theta\left(  \sqrt{n}\right)  $\ because
of Grover's search algorithm \cite{grover}. \ Furthermore, for total
$f$, Beals et al.\ \cite{bbcmw}\ showed
that $\operatorname*{D}\left(  f\right)  =O\left(  \operatorname*{Q}%
_{2}\left(  f\right)  ^{6}\right)  $, while de Wolf \cite{dewolf:thesis}%
\ showed that $\operatorname*{D}\left(  f\right)  =O\left(  \operatorname*{Q}%
_{2}\left(  f\right)  ^{2}\operatorname*{Q}_{0}\left(  f\right)  ^{2}\right)
$.

The result of Beals et al.\ \cite{bbcmw} relies on two intermediate
complexity measures, the \textit{certificate complexity}
$\operatorname*{C}\left( f\right)  $\ and \textit{block sensitivity}
$\operatorname*{bs}\left( f\right)  $, which are defined as follows.

\begin{definition}
A certificate for an input $X$ is a set $S\subseteq\left\{  1,\ldots
,n\right\}  $\ such that for all inputs $Y$ of $f$, if $y_{i}=x_{i}$\ for all
$i\in S$\ then $f\left(  Y\right)  =f\left(  X\right)  $. \ Then
$\operatorname*{C}^{X}\left(  f\right)  $\ is the minimum size of a
certificate for $X$, and $\operatorname*{C}\left(  f\right)  $\ is the maximum
of $\operatorname*{C}^{X}\left(  f\right)  $\ over all $X$.
\end{definition}

\begin{definition}
A sensitive block on input $X$ is a set $B\subseteq\left\{  1,\ldots
,n\right\}  $\ such that $f\left(  X^{\left(  B\right)  }\right)  \neq
f\left(  X\right)  $, where $X^{\left(  B\right)  }$\ is obtained from $X$ by
flipping $x_{i}$\ for each $i\in B$. \ Then $\operatorname*{bs}^{X}\left(
f\right)  $\ is the maximum number of disjoint sensitive blocks on $X$, and
$\operatorname*{bs}\left(  f\right)  $\ is the maximum of $\operatorname*{bs}%
^{X}\left(  f\right)  $\ over all $X$.
\end{definition}

Clearly $\operatorname*{bs}\left(  f\right)
\leq\operatorname*{C}\left( f\right)  \leq\operatorname*{D}\left(
f\right)  $. \ For total $f$, these measures are all polynomially
related: Nisan \cite{nisan}\ showed that $\operatorname*{C}\left(
f\right)  \leq\operatorname*{bs}\left(  f\right) ^{2}$, while Beals
et al.\ \cite{bbcmw}\ showed that $\operatorname*{D}\left( f\right)
\leq\operatorname*{C}\left(  f\right)  \operatorname*{bs}\left(
f\right)  $. \ Combining these results with
$\operatorname*{bs}\left( f\right)  =O\left(
\operatorname*{Q}_{2}\left(  f\right)  ^{2}\right)  $
(from the optimality of Grover's algorithm), one obtains $\operatorname*{D}%
\left(  f\right)  =O\left(  \operatorname*{Q}_{2}\left(  f\right)
^{6}\right)  $.

\section{Summary of Results\label{RESULTSCER}}

I investigate $\operatorname*{RC}\left(  f\right)  $ and $\operatorname*{QC}%
\left(  f\right)  $, the bounded-error randomized and quantum
generalizations of the certificate complexity
$\operatorname*{C}\left(  f\right)  $ (see Table 8.1). \ My
motivation is that, just as $\operatorname*{C}\left(
f\right)  $\ was used to show a polynomial relation between $\operatorname*{D}%
\left(  f\right)  $\ and $\operatorname*{Q}_{2}\left(  f\right)  $, so
$\operatorname*{RC}\left(  f\right)  $ and $\operatorname*{QC}\left(
f\right)  $\ can lead to new relations among fundamental query complexity measures.

\begin{table}[ptb]%
\begin{tabular}
[c]{l|ccc} & Deterministic & Randomized & Quantum \\\hline
Query complexity & $\operatorname*{D}\left(  f\right)  $ & $\operatorname*{R}%
_{2}\left(  f\right)  $ & $\operatorname*{Q}_{2}\left(  f\right)  $\\
Certificate complexity & $\operatorname*{C}\left(  f\right)  $ &
$\operatorname*{RC}\left(  f\right)  $ & $\operatorname*{QC}\left(  f\right)
$%
\end{tabular}
\caption[Query complexity and certificate complexity measures]{Query
complexity measures and their certificate complexity analogues.}%
\label{certable}%
\end{table}

What the certificate complexity $\operatorname*{C}\left(  f\right)
$\ measures is the number of \textit{queries} used to verify a certificate,
not the number of \textit{bits} used to communicate it. \ Thus, if we want to
generalize $\operatorname*{C}\left(  f\right)  $, we should assume the latter
is unbounded. \ A consequence is that without loss of generality, a
certificate is just a claimed value $X$\ for the input $Y$\footnote{Throughout
this chapter, I use $Y$ to denote the `actual' input being queried, and $X$ to
denote the `claimed' input.}---since any additional information that a prover
might provide, the verifier can compute for itself. \ The verifier's job is to
check that $f\left(  Y\right)  =f\left(  X\right)  $. \ With this in mind I
define $\operatorname*{RC}\left(  f\right)  $ as follows.

\begin{definition}
A randomized verifier for input $X$ is a randomized algorithm that, on input
$Y$ to $f$, (i) accepts with probability $1$ if $Y=X$, and (ii) rejects with
probability at least $1/2$ if $f\left(  Y\right)  \neq f\left(  X\right)  $.
\ (If $Y\neq X$ but $f\left(  Y\right)  =f\left(  X\right)  $,\ the acceptance
probability can be arbitrary.) $\ $Then $\operatorname*{RC}^{X}\left(
f\right)  $\ is the minimum expected number of queries used by a randomized
verifier for $X$, and $\operatorname*{RC}\left(  f\right)  $\ is the maximum
of $\operatorname*{RC}^{X}\left(  f\right)  $\ over all $X$.
\end{definition}

I define $\operatorname*{QC}\left(  f\right)  $\ analogously, with
quantum instead of randomized algorithms. \ The following justifies
the definition (the $\operatorname*{RC}\left(  f\right)  $ part was
originally shown by Raz et al.\ \cite{rtvv}).

\begin{proposition}
\label{onesided}Making the error probability two-sided rather than one-sided
changes $\operatorname*{RC}\left(  f\right)  $\ and $\operatorname*{QC}\left(
f\right)  $\ by at most a constant factor.
\end{proposition}

\begin{proof}
For $\operatorname*{RC}\left(  f\right)  $, let $r_{V}^{Y}$\ be the event that
verifier $V$ rejects on input $Y$, and let $d_{V}^{Y}$\ be the event that $V$
encounters a disagreement with $X$ on $Y$. \ We may assume $\Pr\left[
r_{V}^{Y}\,\,|\,\,d_{V}^{Y}\right]  =1$. \ Suppose that $\Pr\left[  r_{V}%
^{Y}\right]  \leq\varepsilon_{0}$ if $Y=X$ and $\Pr\left[  r_{V}^{Y}\right]
\geq1-\varepsilon_{1}$\ if $f\left(  Y\right)  \neq f\left(  X\right)  $. \ We
wish to lower-bound $\Pr\left[  d_{V}^{Y}\right]  $ for all $Y$ such that
$f\left(  Y\right)  \neq f\left(  X\right)  $. \ Observe that%
\begin{align*}
\Pr\left[  r_{V}^{Y}\,\wedge\,\urcorner d_{V}^{Y}\,\,|\,\,f\left(
Y\right) \neq f\left(  X\right)  \right]   &  \leq\Pr\left[
r_{V}^{X}\,\wedge \,\urcorner d_{V}^{X}\,\right] =\Pr\left[
r_{V}^{X}\,\right]  \leq\varepsilon_{0}.
\end{align*}
Hence for $f\left(  Y\right)  \neq f\left(  X\right)  $,%
\[
\Pr\left[  d_{V}^{Y}\right]  \geq\Pr\left[  r_{V}^{Y}\right]  -\Pr\left[
r_{V}^{Y}\,\wedge\,\urcorner d_{V}^{Y}\right]  \geq1-\varepsilon
_{1}-\varepsilon_{0}.
\]
Now let $V^{\ast}$\ be identical to $V$ except that, whenever $V$ rejects
despite having found no disagreement with $X$, $V^{\ast}$\ accepts. \ Clearly
$\Pr\left[  r_{V^{\ast}}^{X}\right]  =0$. \ Also, in the case $f\left(
Y\right)  \neq f\left(  X\right)  $,%
\[
\Pr\left[  r_{V^{\ast}}^{Y}\right]  =\Pr\left[  d_{V}^{Y}\right]
\geq1-\varepsilon_{1}-\varepsilon_{0}\text{.}%
\]
The result follows since $O\left(  1\right)  $\ repetitions suffice to boost
any constant error probability to any other constant error probability.

For $\operatorname*{QC}\left(  f\right)  $, suppose the verifier's final state
given input $Y$ is%
\[
\sum_{z}\alpha_{z}^{Y}\left\vert z\right\rangle \left(  \beta_{z}%
^{Y}\left\vert 0\right\rangle +\gamma_{z}^{Y}\left\vert 1\right\rangle
\right)
\]
where $\left\vert 0\right\rangle $\ is the reject state, $\left\vert
1\right\rangle $\ is the accept state, and $\left\vert \beta_{z}%
^{Y}\right\vert ^{2}+\left\vert \gamma_{z}^{Y}\right\vert ^{2}=1$\ for all
$z$. \ Suppose also that $A^{X}\geq1-\varepsilon_{0}$\ and that $A^{Y}%
\leq\varepsilon_{1}$\ whenever $f\left(  Y\right)  \neq f\left(  X\right)  $,
where $A^{Y}=\sum_{z}\left\vert \alpha_{z}^{Y}\gamma_{z}^{Y}\right\vert ^{2}$
is the probability of accepting. \ Then the verifier can make $A^{X}=1$\ by
performing the conditional rotation%
\[
\left(
\begin{array}
[c]{cc}%
\gamma_{z}^{X} & -\beta_{z}^{X}\\
\beta_{z}^{X} & \gamma_{z}^{X}%
\end{array}
\right)
\]
on the second register prior to measurement. \ In the case $f\left(  Y\right)
\neq f\left(  X\right)  $, this produces%
\begin{align*}
A^{Y}  &  =\sum_{z}\left\vert \alpha_{z}^{Y}\right\vert ^{2}\left\vert
\beta_{z}^{X}\beta_{z}^{Y}+\gamma_{z}^{X}\gamma_{z}^{Y}\right\vert ^{2}\\
&  \leq2\sum_{z}\left\vert \alpha_{z}^{Y}\right\vert ^{2}\left(  \left\vert
\beta_{z}^{X}\right\vert ^{2}+\left\vert \gamma_{z}^{Y}\right\vert ^{2}\right)
\\
&  \leq2\left(  \varepsilon_{0}+\varepsilon_{1}\right)  \text{.}%
\end{align*}

\end{proof}

It is immediate that $\operatorname*{QC}\left(  f\right)  \leq
\operatorname*{RC}\left(  f\right)  \leq\operatorname*{C}\left(  f\right)  $,
that $\operatorname*{QC}\left(  f\right)  =O\left(  \operatorname*{Q}%
_{2}\left(  f\right)  \right)  $, and that $\operatorname*{RC}\left(
f\right)  =O\left(  \operatorname*{R}_{2}\left(  f\right)  \right)  $. \ We
also have $\operatorname*{RC}\left(  f\right)  =\Omega\left(
\operatorname*{bs}\left(  f\right)  \right)  $, since a randomized verifier
for $X$ must query each sensitive block on $X$ with $1/2$ probability. \ This
suggests viewing $\operatorname*{RC}\left(  f\right)  $\ as an `alloy' of
block sensitivity and certificate complexity, an interpretation for which
Section \ref{GAP}\ gives some justification.

The results of this chapter are as follows. \ In Section \ref{CHARSEC}\ I show
that $\operatorname*{QC}\left(  f\right)  =\Theta\left(  \sqrt
{\operatorname*{RC}\left(  f\right)  }\right)  $ for all $f$ (partial or
total), precisely characterizing quantum certificate complexity in terms of
randomized certificate complexity. To do this, I first give a nonadaptive
characterization of $\operatorname*{RC}\left(  f\right)  $, and then apply the
adversary method of Ambainis \cite{ambainis}\ to lower-bound
$\operatorname*{QC}\left(  f\right)  $\ in terms of this characterization.
\ Then, in Section \ref{RRC}, I extend results on polynomials due to de Wolf
\cite{dewolf:thesis}\ and to Nisan and Smolensky (as described by Buhrman and
de Wolf \cite{bw}), to show that $\operatorname*{R}_{0}\left(  f\right)
=O\left(  \operatorname*{RC}\left(  f\right)  \operatorname*{ndeg}\left(
f\right)  \log n\right)  $\ for all total $f$, where $\operatorname*{ndeg}%
\left(  f\right)  $\ is the minimum degree of a polynomial $p$\ such that
$p\left(  X\right)  \neq0$\ if and only if $f\left(  X\right)  \neq0$.
\ Combining the results of Sections \ref{CHARSEC}\ and \ref{RRC} leads to a
new lower bound on quantum query complexity: that $\operatorname*{R}%
_{0}\left(  f\right)  =O\left(  \operatorname*{Q}_{2}\left(  f\right)
^{2}\operatorname*{Q}_{0}\left(  f\right)  \log n\right)  $ for all total $f$.
\ To my knowledge, this is the first quantum lower bound to use both the
adversary method and the polynomial method at different points in the argument.

Finally, in Section \ref{GAP}, I exhibit asymptotic gaps between
$\operatorname*{RC}\left(  f\right)  $\ and other query complexity measures,
including a total $f$ for which $\operatorname*{C}\left(  f\right)
=\Theta\left(  \operatorname*{QC}\left(  f\right)  ^{2.205}\right)  $, and a
symmetric partial $f$ for which $\operatorname*{QC}\left(  f\right)  =O\left(
1\right)  $\ yet $\operatorname*{Q}_{2}\left(  f\right)  =\Omega\left(  n/\log
n\right)  $. \ I conclude in Section \ref{OPENCER}\ with some open problems.

\section{Related Work\label{RELATEDCER}}

Raz et al.\ \cite{rtvv}\ studied a query complexity measure they
called $\operatorname*{ma}\left(  f\right)  $, for Merlin-Arthur. \
In my notation, $\operatorname*{ma}\left(  f\right)  $\ equals the
maximum of $\operatorname*{RC}^{X}\left(  f\right)  $\ over all $X$
with $f\left( X\right)  =1$. \ Raz et al.\ observed that
$\operatorname*{ma}\left(  f\right) =\operatorname*{ip}\left(
f\right)  $, where $\operatorname*{ip}\left( f\right)  $\ is the
number of queries needed given arbitrarily many rounds of
interaction with a prover. \ They also used error-correcting codes
to construct a total $f$ for which $\operatorname*{ma}\left(
f\right)  =O\left( 1\right)  $\ but $\operatorname*{C}\left(
f\right)  =\Omega\left(  n\right) $. \ This has similarities to the
construction, in Section \ref{SYM}, of a symmetric partial $f$ for
which $\operatorname*{QC}\left(  f\right)  =O\left( 1\right)  $\ but
$\operatorname*{Q}_{2}\left(  f\right)  =\Omega\left(  n/\log
n\right)  $. \ Aside from that and from Proposition \ref{onesided},
Raz et al.'s results do not overlap with the results here.

Watrous \cite{watrous}\ has investigated a different notion of `quantum
certificate complexity'---whether certificates that are quantum states can be
superpolynomially smaller than any classical certificate. \ Also, de Wolf
\cite{dewolf:ncc}\ has investigated `nondeterministic quantum query
complexity' in the alternate sense of algorithms that accept with zero
probability when $f\left(  Y\right)  =0$, and with positive probability when
$f\left(  Y\right)  =1$.

\section{\label{CHARSEC}Characterization of Quantum Certificate Complexity}

We wish to show that $\operatorname*{QC}\left(  f\right)  =\Theta\left(
\sqrt{\operatorname*{RC}\left(  f\right)  }\right)  $, precisely
characterizing quantum certificate complexity in terms of randomized
certificate complexity. \ The first step is to give a simpler characterization
of $\operatorname*{RC}\left(  f\right)  $.

\begin{lemma}
Call a randomized verifier for $X$ \textit{nonadaptive} if, on input $Y$, it
queries each $y_{i}$\ with independent probability $\lambda_{i}$, and rejects
if and only if it encounters a disagreement with $X$. \ (Thus, we identify
such a verifier with the vector $\left(  \lambda_{1},\ldots,\lambda
_{n}\right)  $.) \ Let $\operatorname*{RC}_{na}^{X}\left(  f\right)  $\ be the
minimum of $\lambda_{1}+\cdots+\lambda_{n}$\ over all nonadaptive verifiers
for $X$.\ \ Then $\operatorname*{RC}_{na}^{X}\left(  f\right)  =\Theta\left(
\operatorname*{RC}^{X}\left(  f\right)  \right)  $.
\end{lemma}

\begin{proof}
Clearly $\operatorname*{RC}_{na}^{X}\left(  f\right)  =\Omega\left(
\operatorname*{RC}^{X}\left(  f\right)  \right)  $. \ For the upper bound, we
can assume that a randomized verifier rejects immediately on finding a
disagreement with $X$, and accepts if it finds no disagreement. \ Let
$\mathcal{Y}=\left\{  Y:f\left(  Y\right)  \neq f\left(  X\right)  \right\}
$. \ Let $V$ be an optimal randomized verifier, and let $p_{t}\left(
Y\right)  $\ be the probability that $V$, when given input $Y\in\mathcal{Y}$,
finds a disagreement with $X$ on the $t^{th}$\ query. \ By Markov's
inequality, $V$ must have found a disagreement with probability at least $1/2$
after $T=\left\lceil 2\operatorname*{RC}^{X}\left(  f\right)  \right\rceil
$\ queries. \ So by the union bound%
\[
p_{1}\left(  Y\right)  +\cdots+p_{T}\left(  Y\right)  \geq\frac{1}{2}%
\]
for each $Y\in\mathcal{Y}$. \ Suppose we choose $t\in\left\{  1,\ldots
,T\right\}  $ uniformly at random and simulate the $t^{th}$\ query, pretending
that queries $1,\ldots,t-1$\ have already been made and have returned
agreement with $X$. \ Then we must find a disagreement with probability at
least $1/2T$. \ By repeating this procedure $4T$\ times, we can boost the
probability to $1-e^{-2}$. \ For $i\in\left\{  1,\ldots,n\right\}  $, let
$\lambda_{i}$\ be the probability that $y_{i}$\ is queried at least once.
\ Then $\lambda_{1}+\cdots+\lambda_{n}\leq4T$, whereas for each $Y\in
\mathcal{Y}$,%
\[
\sum_{i:y_{i}\neq x_{i}}\lambda_{i}\geq1-e^{-2}.
\]
It follows that, if each $y_{i}$\ is queried with independent probability
$\lambda_{i}$, then the probability that at least one $y_{i}$ disagrees with
$X$ is at least%
\[
1-\prod_{i:y_{i}\neq x_{i}}\left(  1-\lambda_{i}\right)  \geq1-\left(
1-\frac{1-e^{-2}}{n}\right)  ^{n}>0.57.
\]

\end{proof}

To obtain a lower bound on $\operatorname*{QC}\left(  f\right)  $, I
will use the following simple reformulation of Ambainis's adversary
method \cite{ambainis}.

\begin{theorem}
[Ambainis]\label{ambthm}Given a function $f:\mathcal{S}\rightarrow\left\{
0,1\right\}  $\ with $\mathcal{S}\subseteq\left\{  0,1\right\}  ^{n}$, let
$\beta$ be a function from $\mathcal{S}$\ to nonnegative reals, and let
$R:\mathcal{S}^{2}\rightarrow\left\{  0,1\right\}  $\ be a relation such that
$R\left(  X,Y\right)  =R\left(  Y,X\right)  $\ for all $X,Y$\ and $R\left(
X,Y\right)  =0$\ whenever $f\left(  X\right)  =f\left(  Y\right)  $. \ Let
$\delta_{0},\delta_{1}\in\left(  0,1\right]  $\ be such that for every
$X\in\mathcal{S}$\ and $i\in\left\{  1,\ldots,n\right\}  $,%
\begin{align*}
\sum_{Y\,:\,R\left(  X,Y\right)  =1}\beta\left(  Y\right)   &  \geq1,\\
\sum_{Y\,:\,R\left(  X,Y\right)  =1,x_{i}\neq y_{i}}\beta\left(  Y\right)   &
\leq\delta_{f\left(  X\right)  }.
\end{align*}
Then $\operatorname*{Q}_{2}\left(  f\right)  =\Omega\left(  \sqrt{\frac
{1}{\delta_{0}\delta_{1}}}\right)  $.
\end{theorem}

I now prove the main result of the section.

\begin{theorem}
\label{sqrt}For all $f$ (partial or total) and all $X$,%
\[
\operatorname*{QC}{}^{X}\left(  f\right)  =\Theta\left(  \sqrt
{\operatorname*{RC}{}^{X}\left(  f\right)  }\right)  .
\]

\end{theorem}

\begin{proof}
Let $\left(  \lambda_{1},\ldots,\lambda_{n}\right)  $ be an optimal
nonadaptive randomized verifier for $X$, and let%
\[
S=\lambda_{1}+\cdots+\lambda_{n}.
\]
First, $\operatorname*{QC}^{X}\left(  f\right)  =O\left(  \sqrt{S}\right)  $.
\ We can run a \textquotedblleft weighted Grover search,\textquotedblright\ in
which the proportion of basis states querying index $i$ is within a constant
factor of $\lambda_{i}/S$. \ (It suffices to use $n^{2}$\ basis states.) \ Let
$\mathcal{Y}=\left\{  Y:f\left(  Y\right)  \neq f\left(  X\right)  \right\}
$;\ then for any $Y\in\mathcal{Y}$, $O\left(  \sqrt{S}\right)  $\ iterations
suffice to find a disagreement with $X$ with probability $\Omega\left(
1\right)  $. Second, $\operatorname*{QC}^{X}\left(  f\right)  =\Omega\left(
\sqrt{S}\right)  $. \ Consider a matrix game in which Alice chooses an index
$i$\ to query and Bob chooses $Y\in\mathcal{Y}$; Alice wins if and only if
$y_{i}\neq x_{i}$. \ If both players are rational, then Alice wins with
probability $O\left(  1/S\right)  $, since otherwise Alice's strategy would
yield a verifier $\left(  \lambda_{1}^{\prime},\ldots,\lambda_{n}^{\prime
}\right)  $\ with%
\[
\lambda_{1}^{\prime}+\cdots+\lambda_{n}^{\prime}=o\left(  S\right)  .
\]
Hence by the minimax theorem, there exists a distribution $\mu$\ over
$\mathcal{Y}$\ such that for every $i$,%
\[
\Pr_{Y\in\mu}\left[  y_{i}\neq x_{i}\right]  =O\left(  \frac{1}{S}\right)  .
\]
Let $\beta\left(  X\right)  =1$\ and let $\beta\left(  Y\right)  =\mu\left(
Y\right)  $\ for each $Y\in\mathcal{Y}$. \ Also, let $R\left(  Y,Z\right)
=1$\ if and only if $Z=X$ for each $Y\in\mathcal{Y}$ and $Z\notin\mathcal{Y}$.
\ Then we can take $\delta_{f\left(  Y\right)  }=1$\ and $\delta_{f\left(
X\right)  }=O\left(  1/S\right)  $ in Theorem \ref{ambthm}. \ So the quantum
query complexity of distinguishing $X$ from an arbitrary $Y\in\mathcal{Y}$\ is
$\Omega\left(  \sqrt{S}\right)  $.
\end{proof}

\section{\label{RRC}Quantum Lower Bound for Total Functions}

The goal of this section is to show that%
\[
\operatorname*{R}\nolimits_{0}\left(  f\right)  =O\left(  \operatorname*{Q}%
\nolimits_{2}\left(  f\right)  ^{2}\operatorname*{Q}\nolimits_{0}\left(
f\right)  \log n\right)
\]
for all total $f$. \ Say that a real multilinear polynomial $p\left(
x_{1},\ldots,x_{n}\right)  $\ nondeterministically represents $f$ if for all
$X\in\left\{  0,1\right\}  ^{n}$, $p\left(  X\right)  \neq0$\ if and only if
$f\left(  X\right)  \neq0$.\ \ Let $\operatorname*{ndeg}\left(  f\right)  $ be
the minimum degree of a nondeterministic polynomial for $f$. \ Also, given
such a polynomial $p$, say that a monomial $M_{1}\in p$\ is \textit{covered}
by $M_{2}\in p$\ if $M_{2}$\ contains every variable in $M_{1}$.\ \ A monomial
$M$ is called a \textit{maxonomial} if it is not covered by any other monomial
of $p$. \ The following is a simple generalization of a lemma attributed in
\cite{bw} to Nisan and Smolensky.

\begin{lemma}
[Nisan-Smolensky]\label{maxo}Let $p$ nondeterministically represent $f$.
\ Then for every maxonomial $M$ of $p$ and $X\in f^{-1}\left(  0\right)  $,
there is a set $B$ of variables in $M$ such that $f\left(  X^{\left(
B\right)  }\right)  \neq f\left(  X\right)  $, where $X^{\left(  B\right)  }%
$\ is obtained from $X$\ by flipping the variables in $B$.
\end{lemma}

\begin{proof}
Obtain a restricted function $g$ from $f$, and a restricted polynomial $q$
from $p$, by setting each variable outside of $M$ to $x_{i}$. \ Then $g$
cannot be constant, since its representing polynomial $q$ contains $M$ as a
monomial. \ Thus there is a subset $B$ of variables in $M$ such that $g\left(
X^{\left(  B\right)  }\right)  =1$, and hence $f\left(  X^{\left(  B\right)
}\right)  =1$.
\end{proof}

Using Lemma \ref{maxo}, de Wolf \cite{dewolf:thesis}\ showed that
$\operatorname*{D}\left(  f\right)  \leq\operatorname*{C}\left(  f\right)
\operatorname*{ndeg}\left(  f\right)  $\ for all total $f$, slightly improving
the result $\operatorname*{D}\left(  f\right)  \leq\operatorname*{C}\left(
f\right)  \deg\left(  f\right)  $\ due to Buhrman and de Wolf \cite{bw}. \ In
Theorem \ref{degthm}, I will give an analogue of this result for
\textit{randomized} query and certificate complexities. \ However, I first
need a probabilistic lemma.

\begin{lemma}
\label{undom}Suppose we repeatedly apply the following procedure: first
identify the set $B$\ of maxonomials of $p$, then `shrink' each $M\in B$ with
(not necessarily independent) probability at least $1/2$. \ Shrinking $M$
means replacing it by an arbitrary monomial of degree $\deg\left(  M\right)
-1$. \ Then with high probability $p$ is a constant polynomial after $O\left(
\deg\left(  p\right)  \log n\right)  $\ iterations.
\end{lemma}

\begin{proof}
For any set $A$ of monomials, consider the weighting function%
\[
\omega\left(  A\right)  =\sum_{M\in A}\deg\left(  M\right)  !
\]
Let $S$ be the set of monomials of $p$. \ Initially $\omega\left(  S\right)
\leq n^{\deg\left(  p\right)  }\deg\left(  p\right)  !$, and we are done when
$\omega\left(  S\right)  =0$. \ The claim is that at every iteration,
$\omega\left(  B\right)  \geq\frac{1}{e}\omega\left(  S\right)  $. \ For every
$M^{\ast}\in S\setminus B$\ is covered by some $M\in B$, but a given $M\in
B$\ can cover at most $\tbinom{\deg\left(  M\right)  }{\ell}$\ distinct
$M^{\ast}$ with $\deg\left(  M^{\ast}\right)  =\ell$. \ Hence%
\begin{align*}
\omega\left(  S\setminus B\right)   &  \leq\sum_{M\in B}\sum_{\ell=0}%
^{\deg\left(  M\right)  -1}\tbinom{\deg\left(  M\right)  }{\ell}\ell!\\
&  \leq\sum_{M\in B}\deg\left(  M\right)  !\left(  \frac{1}{1!}+\frac{1}%
{2!}+\cdots\right) \\
&  \leq\left(  e-1\right)  \omega\left(  B\right)  .
\end{align*}

At every iteration, the contribution of each $M\in B$\ to $\omega\left(
A\right)  $\ has at least $1/2$ probability of shrinking from $\deg\left(
M\right)  !$\ to $\left(  \deg\left(  M\right)  -1\right)  !$\ (or to $0$ if
$\deg\left(  M\right)  =1$). \ When this occurs, the contribution of $M$ is at
least halved. \ Hence $\omega\left(  S\right)  $\ decreases by an expected
amount at least $\frac{1}{4e}\omega\left(  S\right)  $. \ Thus after%
\[
\log_{4e/\left(  4e-1\right)  }\left(  2n^{\deg\left(  p\right)  }\deg\left(
p\right)  !\right)  =O\left(  \deg\left(  p\right)  \log n\right)
\]
iterations, the expectation of $\omega\left(  S\right)  $\ is less than $1/2$,
so $S$ is empty with probability at least $1/2$.
\end{proof}

I can now prove the main result.\footnote{The proof of Theorem
\ref{degthm} that I gave previously \cite{aar:cer} makes a claim
that is both superfluous for proving the theorem and false. \ I am
grateful to Gatis Midrijanis for pointing this out to me.}

\begin{theorem}
\label{degthm}For total $f$,%
\[
\operatorname*{R}\nolimits_{0}\left(  f\right)  =O\left(  \operatorname*{RC}%
\left(  f\right)  \operatorname*{ndeg}\left(  f\right)  \log n\right)  .
\]

\end{theorem}

\begin{proof}
The algorithm is as follows.

\begin{quote}
\texttt{Repeat}

\qquad\texttt{Choose a }$\mathtt{0}$\texttt{-input
}$\mathtt{X}$\texttt{ compatible with all queries made so
far\footnote{Clearly, as long as $f$ is not a constant function,
there \textit{exists} a $0$-input $X$ compatible with all queries
made so far.}}

\qquad\texttt{Query a randomized }$\mathtt{0}$\texttt{-certificate
for }$\mathtt{X}$

\texttt{Until }$\mathtt{f}$\texttt{ has been restricted to a
constant function}
\end{quote}

Let $p$ be a polynomial that nondeterministically represents $f$. \
Then the key fact is that for every $0$-input $X$, when we query a
randomized $0$-certificate for $X$ we \textquotedblleft
hit\textquotedblright\ each maxonomial $M$ of $p$ with probability
at least $1/2$. \ Here hitting $M$ means querying a variable in $M$.
\ This is because, by Lemma \ref{maxo}, it is possible to change
$f\left(  X\right)  $\ from $0$ to $1$ just by flipping variables in
$M$. \ So a randomized certificate would be incorrect if it probed
those variables with probability less than $1/2$.

Therefore, each iteration of the algorithm shrinks each maxonomial
of $p$ with probability at least $1/2$.\ \ It follows from Lemma
\ref{undom} that the algorithm terminates after an expected number
of iterations $O\left( \deg\left(  p\right)  \log n\right)  $.
\end{proof}

Buhrman et al.\ \cite{bbcmw}\ showed that
$\operatorname*{ndeg}\left( f\right)
\leq2\operatorname*{Q}_{0}\left(  f\right)  $. \ Combining this with
Theorems \ref{sqrt}\ and \ref{degthm} yields a new relation between
classical and quantum query complexity.

\begin{corollary}
\label{rqqe}For all total $f$,%
\[
\operatorname*{R}\nolimits_{0}\left(  f\right)  =O\left(  \operatorname*{Q}%
\nolimits_{2}\left(  f\right)  ^{2}\operatorname*{Q}\nolimits_{0}\left(
f\right)  \log n\right)  .
\]

\end{corollary}

The best previous relation of this kind was $\operatorname*{R}\nolimits_{0}%
\left(  f\right)  =O\left(  \operatorname*{Q}\nolimits_{2}\left(  f\right)
^{2}\operatorname*{Q}\nolimits_{0}\left(  f\right)  ^{2}\right)  $, due to de
Wolf \cite{dewolf:thesis}. \ It is worth mentioning another corollary of
Theorems \ref{sqrt}\ and \ref{degthm}, this one purely classical:

\begin{corollary}
\label{r0r2}For all total $f$,
\[
\operatorname*{R}\nolimits_{0}\left(  f\right)  =O\left(  \operatorname*{R}%
\nolimits_{2}\left(  f\right)  \operatorname*{ndeg}\left(  f\right)  \log
n\right)
\]

\end{corollary}

Previously, no relation between $\operatorname*{R}_{0}$\ and
$\operatorname*{R}_{2}$\ better than $\operatorname*{R}_{0}\left(  f\right)
=O\left(  \operatorname*{R}_{2}\left(  f\right)  ^{3}\right)  $\ was known
(although no asymptotic gap between $\operatorname*{R}_{0}$\ and
$\operatorname*{R}_{2}$ is known either \cite{santha}).

\section{\label{GAP}Asymptotic Gaps}

Having related $\operatorname*{RC}\left(  f\right)  $\ and $\operatorname*{QC}%
\left(  f\right)  $\ to other query complexity measures in Section \ref{RRC},
in what follows I seek the largest possible asymptotic gaps among the
measures. \ In particular, I give a total $f$ for which $\operatorname*{RC}%
\left(  f\right)  =\Theta\left(  \operatorname*{C}\left(  f\right)
^{0.907}\right)  $\ and hence $\operatorname*{C}\left(  f\right)
=\Theta\left(  \operatorname*{QC}\left(  f\right)  ^{2.205}\right)  $, as well
as a total $f$ for which $\operatorname*{bs}\left(  f\right)  =\Theta\left(
\operatorname*{RC}\left(  f\right)  ^{0.922}\right)  $. \ Although these gaps
are the largest of which I know, Section \ref{LOCALSEC}\ shows that no `local'
technique can improve the relations $\operatorname*{C}\left(  f\right)
=O\left(  \operatorname*{RC}\left(  f\right)  ^{2}\right)  $\ and
$\operatorname*{RC}\left(  f\right)  =O\left(  \operatorname*{bs}\left(
f\right)  ^{2}\right)  $. \ Finally, Section \ref{SYM}\ uses combinatorial
designs to construct a symmetric partial $f$ for which $\operatorname*{RC}%
\left(  f\right)  $\ and $\operatorname*{QC}\left(  f\right)  $ are $O\left(
1\right)  $,\ yet $\operatorname*{Q}_{2}\left(  f\right)  =\Omega\left(
n/\log n\right)  $.

Wegener and Z\'{a}dori \cite{wz}\ exhibited total Boolean functions with
asymptotic gaps between $\operatorname*{C}\left(  f\right)  $\ and
$\operatorname*{bs}\left(  f\right)  $. \ In similar fashion, I give a
function family $\left\{  g_{t}\right\}  $ with an asymptotic gap between
$\operatorname*{C}\left(  g_{t}\right)  $\ and $\operatorname*{RC}\left(
g_{t}\right)  $. \ Let $g_{1}\left(  x_{1},\ldots,x_{29}\right)  $\ equal $1$
if and only if the Hamming weight of its input is $13$, $14$, $15$, or $16$.
\ (The parameter $29$ was found via computer search to produce a maximal
separation.) \ Then for $t>1$, let%
\[
g_{t}\left(  x_{1},\ldots,x_{29^{t}}\right)  =g_{0}\left[  g_{t-1}\left(
X_{1}\right)  ,\ldots,g_{t-1}\left(  X_{29}\right)  \right]
\]
where $X_{1}$\ is the first $29^{t-1}$\ input bits, $X_{2}$\ is the second
$29^{t-1}$, and so on. \ For $k\in\left\{  0,1\right\}  $, let%
\begin{align*}
\operatorname*{bs}{}^{k}\left(  f\right)   &  =\max_{f\left(  X\right)
=k}\operatorname*{bs}{}^{X}\left(  f\right)  ,\\
\operatorname*{C}{}^{k}\left(  f\right)   &  =\max_{f\left(  X\right)
=k}\operatorname*{C}{}^{X}\left(  f\right)  .
\end{align*}
\noindent\ Then since $\operatorname*{bs}^{0}\left(  g_{1}\right)
=\operatorname*{bs}^{1}\left(  g_{1}\right)  =17$, we have $\operatorname*{bs}%
\left(  g_{t}\right)  =17^{t}$. \ On the other hand, $\operatorname*{C}%
^{0}\left(  g_{1}\right)  =17$ but $\operatorname*{C}^{1}\left(  g_{1}\right)
=26$, so%
\begin{align*}
\operatorname*{C}{}^{1}\left(  g_{t}\right)   &  =13\operatorname*{C}{}%
^{1}\left(  g_{t-1}\right)  +13\operatorname*{C}{}^{0}\left(  g_{t-1}\right)
,\\
\operatorname*{C}{}^{0}\left(  g_{t}\right)   &  =17\max\left\{
\operatorname*{C}{}^{1}\left(  g_{t-1}\right)  ,\operatorname*{C}{}^{0}\left(
g_{t-1}\right)  \right\}  .
\end{align*}
Solving this recurrence yields $\operatorname*{C}\left(  g_{t}\right)
=\Theta\left(  22.725^{t}\right)  $. \ We can now show a gap between
$\operatorname*{C}$\ and $\operatorname*{RC}$.

\begin{proposition}
\label{gapprop}$\operatorname*{RC}\left(  g_{t}\right)  =\Theta\left(
\operatorname*{C}\left(  g_{t}\right)  ^{0.907}\right)  $.
\end{proposition}

\begin{proof}
Since $\operatorname*{bs}\left(  g_{t}\right)  =\Omega\left(
\operatorname*{C}\left(  g_{t}\right)  ^{0.907}\right)  $, it suffices to show
that $\operatorname*{RC}\left(  g_{t}\right)  =O\left(  \operatorname*{bs}%
\left(  g_{t}\right)  \right)  $. \ The randomized verifier $V$\ chooses an
input variable to query as follows. \ Let $X$\ be the claimed input, and let
$K=\sum_{i=1}^{29}g_{t-1}\left(  X_{i}\right)  $. \ Let $I_{0}=\left\{
i:g_{t-1}\left(  X_{i}\right)  =0\right\}  $\ and $I_{1}=\left\{
i:g_{t-1}\left(  X_{i}\right)  =1\right\}  $. \ With probability $p_{K}$, $V$
chooses an $i\in I_{1}$\ uniformly at random; otherwise $A$ chooses an $i\in
I_{0}$\ uniformly at random. \ Here $p_{K}$\ is as follows.%

\begin{tabular}
[c]{c|cccccc}%
$K$ & $\left[  0,12\right]  $ & $13$ & $14$ & $15$ & $16$ & $\left[
17,29\right]  $\\\hline
$p_{K}$ & $0$ & $\frac{13}{17}$ & $\frac{7}{12}$ & $\frac{5}{12}$ & $\frac
{4}{17}$ & $1$%
\end{tabular}

Once $i$ is chosen, $V$ repeats the procedure for $X_{i}$, and\ continues
recursively in this manner until reaching a variable $y_{j}$ to query. \ One
can check that if $g_{t}\left(  X\right)  \neq g_{t}\left(  Y\right)  $, then
$g_{t-1}\left(  X_{i}\right)  \neq g_{t-1}\left(  Y_{i}\right)  $\ with
probability at least $1/17$. \ Hence $x_{j}\neq y_{j}$\ with probability at
least $1/17^{t}$, and $\operatorname*{RC}\left(  g_{t}\right)  =O\left(
17^{t}\right)  $.
\end{proof}

By Theorem \ref{sqrt}, it follows that $\operatorname*{C}\left(  g_{t}\right)
=\Theta\left(  \operatorname*{QC}\left(  g_{t}\right)  ^{2.205}\right)  $.
\ This offers a surprising contrast with the query complexity setting, where
the best known gap between the deterministic and quantum measures is quadratic
($\operatorname*{D}\left(  f\right)  =\Theta\left(  \operatorname*{Q}%
_{2}\left(  f\right)  ^{2}\right)  $).

The family $\left\{  g_{t}\right\}  $\ happens \textit{not} to yield an
asymptotic gap between $\operatorname*{bs}\left(  f\right)  $\ and
$\operatorname*{RC}\left(  f\right)  $. \ The reason is that any input to
$g_{0}$\ can be covered perfectly by sensitive blocks of minimum size, with no
variables left over. In general, though, one can have $\operatorname*{bs}%
\left(  f\right)  =o\left(  \operatorname*{RC}\left(  f\right)
\right)  $. \ As reported by Bublitz et al.\ \cite{bsvw}, M.
Paterson found a total Boolean function $h_{1}\left(
x_{1},\ldots,x_{6}\right)  $\ such that
$\operatorname*{C}^{X}\left(  h_{1}\right)  =5$\ and $\operatorname*{bs}%
^{X}\left(  h_{1}\right)  =4$\ for all $X$. \ Composing $h_{1}$\ recursively
yields $\operatorname*{bs}\left(  h_{t}\right)  =\Theta\left(
\operatorname*{C}\left(  h_{t}\right)  ^{0.861}\right)  $\ and
$\operatorname*{bs}\left(  h_{t}\right)  =\Theta\left(  \operatorname*{RC}%
\left(  h_{t}\right)  ^{0.922}\right)  $, both of which are the largest such
gaps of which I know.

\subsection{\label{LOCALSEC}Local Separations}

It is a longstanding open question whether the relation $\allowbreak
\operatorname*{C}\left(  f\right)  \leq\operatorname*{bs}\left(  f\right)
^{2}$\ due to Nisan \cite{nisan}\ is tight. \ As a first step, one can ask
whether the relations $\operatorname*{C}\left(  f\right)  =O\left(
\operatorname*{RC}\left(  f\right)  ^{2}\right)  $\ and $\operatorname*{RC}%
\left(  f\right)  =O\left(  \operatorname*{bs}\left(  f\right)  ^{2}\right)
$\ are tight. \ In this section I introduce a notion of \textit{local proof}
in query complexity,\ and then show there is no local proof that
$\operatorname*{C}\left(  f\right)  =o\left(  \operatorname*{RC}\left(
f\right)  ^{2}\right)  $\ or that $\operatorname*{RC}\left(  f\right)
=o\left(  \operatorname*{bs}\left(  f\right)  ^{2}\right)  $. \ This implies
that proving either result would require techniques unlike those that are
currently known. \ My inspiration comes from computational complexity, where
researchers first formalized known methods of proof, including
\textit{relativizable proofs} \cite{bgs}\ and \textit{natural proofs}
\cite{rr}, and then argued that these methods were not powerful enough to
resolve the field's outstanding problems.

Let $G\left(  f\right)  $\ and $H\left(  f\right)  $\ be query complexity
measures obtained by maximizing over all inputs---that is,%
\begin{align*}
G\left(  f\right)   &  =\max_{X}G^{X}\left(  f\right)  ,\\
H\left(  f\right)   &  =\max_{X}H^{X}\left(  f\right)  .
\end{align*}
Call $B\subseteq\left\{  1,\ldots,n\right\}  $\ a \textit{minimal block} on
$X$ if $B$ is sensitive on $X$ (meaning $f\left(  X^{\left(  B\right)
}\right)  \neq f\left(  X\right)  $), and no sub-block $B^{\prime}\subset
B$\ is sensitive on $X$. \ Also, let $X$'s \textit{neighborhood}
$\mathcal{N}\left(  X\right)  $\ consist of $X$ together with $X^{\left(
B\right)  }$\ for every minimal block $B$ of $X$.\ \ Consider a proof that
$G\left(  f\right)  =O\left(  t\left(  H\left(  f\right)  \right)  \right)  $
for some nondecreasing $t$. \ I call the proof \textit{local} if it proceeds
by showing that for every input $X$,%
\[
G^{X}\left(  f\right)  =O\left(  \max_{Y\in\mathcal{N}\left(  X\right)
}\left\{  t\left(  H^{Y}\left(  f\right)  \right)  \right\}  \right)  .
\]
As a canonical example, Nisan's proof \cite{nisan}\ that $\operatorname*{C}%
\left(  f\right)  \leq\operatorname*{bs}\left(  f\right)  ^{2}$\ is local.
\ For each $X$, Nisan observes that (i) a maximal set of disjoint minimal
blocks is a certificate for $X$, (ii) such a set can contain at most
$\operatorname*{bs}^{X}\left(  f\right)  $\ blocks, and (iii) each block can
have size at most $\max_{Y\in\mathcal{N}\left(  X\right)  }\operatorname*{bs}%
^{Y}\left(  f\right)  $. \ Another example of a local proof is the
proof in Section \ref{CHARSEC}\ that $\operatorname*{RC}\left(
f\right) =O\left( \operatorname*{QC}\left(  f\right)  ^{2}\right) $.

\begin{proposition}
\label{localprop}There is no local proof showing that
$\operatorname*{C}\left( f\right)  =o\left( \operatorname*{RC}\left(
f\right)  ^{2}\right)  $\ or that $\operatorname*{RC}\left( f\right)
=o\left( \operatorname*{bs}\left( f\right)  ^{2}\right)  $ for all
total $f$.
\end{proposition}

\begin{proof}
The first part is easy:\ let $f\left(  X\right)  =1$\ if $\left|  X\right|
\geq\sqrt{n}$\ (where $\left|  X\right|  $\ denotes the Hamming weight of
$X$), and $f\left(  X\right)  =0$ otherwise. \ Consider the all-zero input
$0^{n}$. \ We have $\operatorname*{C}^{0^{n}}\left(  f\right)  =n-\left\lceil
\sqrt{n}\right\rceil +1$, but $\operatorname*{RC}^{0^{n}}\left(  f\right)
=O\left(  \sqrt{n}\right)  $, and indeed $\operatorname*{RC}^{Y}\left(
f\right)  =O\left(  \sqrt{n}\right)  $\ for all $Y\in\mathcal{N}\left(
0^{n}\right)  $. For the second part, arrange the input variables in a lattice
of size $\sqrt{n}\times\sqrt{n}$. \ Take $m=\Theta\left(  n^{1/3}\right)  $,
and let $g\left(  X\right)  $\ be the monotone Boolean function that outputs
$1$ if and only if $X$ contains a $1$\textit{-square }of size $m\times m$.
\ This is a square of $1$'s that can wrap around the edges of the lattice;
note that only the variables along the sides must be set to $1$, not those in
the interior. \ An example input, with a $1$-square of size $3\times3$, is
shown below.%
\[%
\begin{array}
[c]{ccccc}%
0 & 0 & 0 & 0 & 0\\
0 & 0 & 0 & 0 & 0\\
1 & 0 & 0 & 1 & 1\\
1 & 0 & 0 & 1 & 0\\
1 & 0 & 0 & 1 & 1
\end{array}
\]
Clearly $\operatorname*{bs}^{0^{n}}\left(  g\right)  =\Theta\left(
n^{1/3}\right)  $, since there can be at most $n/m^{2}$\ disjoint $1$-squares
of size $m\times m$. \ Also, $\operatorname*{bs}^{Y}\left(  g\right)
=\Theta\left(  n^{1/3}\right)  $\ for any $Y$\ that is $0$ except for a single
$1$-square. \ On the other hand, if we choose uniformly at random among all
such $Y$'s, then at any lattice site $i$, $\Pr_{Y}\left[  y_{i}=1\right]
=\Theta\left(  n^{-2/3}\right)  $. \ Hence $\operatorname*{RC}^{0^{n}}\left(
g\right)  =\Omega\left(  n^{2/3}\right)  $.
\end{proof}

\subsection{\label{SYM}Symmetric Partial Functions}

If $f$ is partial, then $\operatorname*{QC}\left(  f\right)  $\ can be much
smaller than $\operatorname*{Q}_{2}\left(  f\right)  $. \ This is strikingly
illustrated by the collision problem: let $\operatorname*{Col}\left(
Y\right)  =0$ if $Y=y_{1}\ldots y_{n}$\ is a one-to-one sequence and
$\operatorname*{Col}\left(  Y\right)  =1$ if $Y$ is a two-to-one sequence,
promised that one of these is the case. \ Then $\operatorname*{RC}\left(
\operatorname*{Col}\right)  =\operatorname*{QC}\left(  \operatorname*{Col}%
\right)  =O\left(  1\right)  $, since every one-to-one input differs from
every two-to-one input on at least $n/2$ of the $y_{i}$'s. \ On the other
hand, Chapter \ref{COL}\ showed that $\operatorname*{Q}_{2}\left(
\operatorname*{Col}\right)  =\Omega\left(  n^{1/5}\right)  $.

From the example of the collision problem, it is tempting to conjecture
that\ (say) $\operatorname*{Q}_{2}\left(  f\right)  =O\left(  n^{1/3}\right)
$\ whenever $\operatorname*{QC}\left(  f\right)  =O\left(  1\right)  $---that
is, `if every $0$-input is far from every $1$-input,\ then the quantum query
complexity is sublinear.' \ Here I disprove this conjecture, even for the
special case of symmetric functions such as $\operatorname*{Col}$. \ (Given a
finite set $\mathcal{H}$, a function $f:\mathcal{S}\rightarrow\left\{
0,1\right\}  $ where $\mathcal{S}\subseteq\mathcal{H}^{n}$\ is called
symmetric if $x_{1}\ldots x_{n}\in\mathcal{S}$\ implies $x_{\sigma\left(
1\right)  }\ldots x_{\sigma\left(  n\right)  }\in\mathcal{S}$\ and $f\left(
x_{1}\ldots x_{n}\right)  =f\left(  x_{\sigma\left(  1\right)  }\ldots
x_{\sigma\left(  n\right)  }\right)  $ for every permutation $\sigma$.)

The proof uses the following lemma, which can be found in Nisan and Wigderson
\cite{nw} for example.

\begin{lemma}
[Nisan-Wigderson]\label{nwlemma}For any $\gamma>1$, there exists a family of
sets%
\[
A_{1},\ldots,A_{m}\subseteq\left\{  1,\ldots,\left\lceil \gamma n\right\rceil
\right\}
\]
such that $m=\Omega\left(  2^{n/\gamma}\right)  $, $\left\vert A_{i}%
\right\vert =n$ for all $i$, and $\left\vert A_{i}\cap A_{j}\right\vert \leq
n/\gamma$\ for all $i\neq j$.
\end{lemma}

A lemma due to Ambainis \cite{ambainis:aa} is also useful. \ Let
$f:\mathcal{S}\rightarrow\left\{  0,1\right\}  $ where $\mathcal{S}%
\subseteq\left\{  0,1\right\}  ^{n}$ be a partial Boolean function, and let
$p:\left\{  0,1\right\}  ^{n}\rightarrow\mathbb{R}$ be a real-valued
multilinear polynomial. \ We say that $p$ approximates $f$ if (i) $p\left(
X\right)  \in\left[  0,1\right]  $\ for every input $X\in\left\{  0,1\right\}
^{n}$ (not merely those in $\mathcal{S}$), and (ii) $\left\vert p\left(
X\right)  -g\left(  X\right)  \right\vert \leq1/3$\ for every $X\in
\mathcal{S}$.

\begin{lemma}
[Ambainis]\label{amblemma}At most $2^{O\left(  \Delta\left(  n,d\right)
dn^{2}\right)  }$\ distinct Boolean functions (partial or total) can be
approximated by polynomials of degree $d$, where $\Delta\left(  n,d\right)
=\sum_{i=0}^{d}\tbinom{n}{i}$.
\end{lemma}

The result is an easy consequence of Lemmas \ref{nwlemma} and \ref{amblemma}.

\begin{theorem}
\label{symthm}There exists a symmetric partial $f$ for \allowbreak which
$\operatorname*{QC}\left(  f\right)  =O\left(  1\right)  $\ and
$\operatorname*{Q}_{2}\left(  f\right)  =\Omega\left(  n/\log n\right)  $.
\end{theorem}

\begin{proof}
Let $f:\mathcal{S}\rightarrow\left\{  0,1\right\}  $\ where $\mathcal{S}%
\subseteq\left\{  1,\ldots,3n\right\}  ^{n}$, and let $m=\Omega\left(
2^{n/3}\right)  $. \ Let $A_{1},\ldots,A_{m}\subseteq\left\{  1,\ldots
,3n\right\}  $\ be as in Lemma \ref{nwlemma}. \ We put $x_{1},\ldots,x_{n}%
$\ in $\mathcal{S}$\ if and only if $\left\{  x_{1},\ldots,x_{n}\right\}
=A_{j}$\ for some $j$. \ Clearly $\operatorname*{QC}\left(  f\right)
=O\left(  1\right)  $, since if $i\neq j$\ then every permutation of $A_{i}%
$\ differs from every permutation of $A_{j}$\ on at least $n/3$\ indices.
\ The number of symmetric $f$ with $\mathcal{S}$\ as above is $2^{m}%
=2^{\Omega\left(  2^{n/3}\right)  }$. \ We can convert any such $f$
to a Boolean function $g$ on $O\left(  n\log n\right)  $ variables.
\ But Beals et al.\ \cite{bbcmw}\ showed that, if
$\operatorname*{Q}_{2}\left(  g\right)  =T$, then $g$ is
approximated by a polynomial of degree at most $2T$. \ So by Lemma
\ref{amblemma}, if $\operatorname*{Q}_{2}\left(  g\right)  \leq T$\
for every
$g$ then%
\[
2T\cdot\Delta\left(  n\log n,2T\right)  \cdot\left(  n\log n\right)
^{2}=\Omega\left(  2^{n/3}\right)
\]
and we solve to obtain $T=\Omega\left(  n/\log n\right)  $.
\end{proof}

\section{\label{OPENCER}Open Problems}

Is $\widetilde{\deg}\left(  f\right)  =\Omega\left(  \sqrt{\operatorname*{RC}%
\left(  f\right)  }\right)  $, where $\widetilde{\deg}\left(
f\right)  $\ is the minimum degree of a polynomial approximating
$f$? \ In other words, can one lower-bound $\operatorname*{QC}\left(
f\right)  $\ using the polynomial method of Beals et al.\
\cite{bbcmw}, rather than the adversary method of Ambainis
\cite{ambainis}?

Also, is $\operatorname*{R}_{0}\left(  f\right)  =O\left(  \operatorname*{RC}%
\left(  f\right)  ^{2}\right)  $? \ If so we obtain the new relations
$\operatorname*{R}_{0}\left(  f\right)  =O\left(  \operatorname*{Q}_{2}\left(
f\right)  ^{4}\right)  $ and $\operatorname*{R}_{0}\left(  f\right)  =O\left(
\operatorname*{R}_{2}\left(  f\right)  ^{2}\right)  $.

\chapter{The Need to Uncompute\label{RFS}}

Like a classical algorithm, a quantum algorithm can solve problems
recursively by calling itself as a subroutine. \ When this is done,
though, the algorithm typically needs to call itself \textit{twice}
for each subproblem to be solved. \ The second call's purpose is to
uncompute `garbage' left over by the first call, and thereby enable
interference between different branches of the computation. \ Of
course, a factor of $2$ increase in running time hardly seems like a
big deal, when set against the speedups promised by quantum
computing. \ The problem is that these factors of $2$ multiply, with
each level of recursion producing an additional factor. \ Thus, one
might wonder whether the uncomputing\ step is really necessary, or
whether a cleverly designed algorithm might avoid it. \ This chapter
gives the first nontrivial example in which recursive uncomputation
is provably necessary.

The example concerns a long-neglected problem called \textit{Recursive Fourier
Sampling} (henceforth $\operatorname*{RFS}$),\ which was introduced by
Bernstein and Vazirani \cite{bv} in 1993 to prove the first oracle separation
between $\mathsf{BPP}$ and \thinspace$\mathsf{BQP}$. \ Many surveys on quantum
computing pass directly from the Deutsch-Jozsa algorithm \cite{dj}\ to the
dramatic results of Simon \cite{simon}\ and Shor \cite{shor}, without even
mentioning $\operatorname*{RFS}$. \ There are two likely reasons for this
neglect. \ First, the $\operatorname*{RFS}$\ problem seems artificial. \ It
was introduced for the sole purpose of proving an oracle result, and is unlike
all other problems for which a quantum speedup is known. \ (I will define
$\operatorname*{RFS}$ in Section \ref{PRELIMRFS}; but for now, it involves a
tree of depth $\log n$, where each vertex is labeled with a function to be
evaluated via a Fourier transform.) \ Second, the speedup for
$\operatorname*{RFS}$\ is only quasipolynomial ($n$ versus $n^{\log n}$),
rather than exponential as for the period-finding and hidden subgroup problems.

Nevertheless, I believe that $\operatorname*{RFS}$ merits renewed
attention---for it serves as an important link between quantum
computing and the ideas of classical complexity theory. \ One reason
is that, although other problems in $\mathsf{BQP}$---such as the
factoring, discrete logarithm, and `shifted Legendre symbol'
problems \cite{dhi}---are thought to be classically intractable,\
these problems are quite low-level by complexity-theoretic
standards. \ They, or their associated decision problems, are in
$\mathsf{NP}\cap\mathsf{coNP}$.\footnote{For the shifted Legendre
symbol problem, this is true assuming a number-theoretic conjecture
of Boneh
and Lipton \cite{bl}.} \ By contrast, Bernstein and Vazirani \cite{bv}%
\ showed that, as an oracle problem, $\operatorname*{RFS}$\ lies
outside $\mathsf{NP}$ and even $\mathsf{MA}$ (the latter result is
unpublished, though not difficult). \ Subsequently Watrous
\cite{watrous}\ gave an oracle $A$, based on an unrelated problem,
for which $\mathsf{BQP}^{A}\not \subset
\mathsf{MA}^{A}$.\footnote{Actually, to place $\mathsf{BQP}$ outside
$\mathsf{MA}$ relative to an oracle, it suffices to consider the
complement of Simon's problem (\textquotedblleft Does $f\left(
x\right) =f\left(  x\oplus s\right)  $ only when
$s=0$?\textquotedblright).} \ Also,
Green and Pruim \cite{gp}\ gave an oracle $B$\ for which $\mathsf{BQP}%
^{B}\not \subset \mathsf{P}^{\mathsf{NP}^{B}}$. \ However, Watrous' problem
was shown by Babai \cite{babai:am}\ to be in $\mathsf{AM}$, while Green and
Pruim's problem is in $\mathsf{BPP}$. \ Thus, neither problem can be used to
place $\mathsf{BQP}$\ outside higher levels of the polynomial hierarchy.

On the other hand, Umesh Vazirani and others have conjectured that
$\operatorname*{RFS}$\ is not in $\mathsf{PH}$, from which it would follow
that there exists an oracle $A$ relative to which $\mathsf{BQP}^{A}%
\not \subset \mathsf{PH}^{A}$. \ Proving this is, in my view, one of
the central open problems in quantum complexity theory. \ Its
solution seems likely to require novel techniques for constant-depth
circuit lower bounds.\footnote{For the $\operatorname*{RFS}$\
function can be represented by a low-degree real polynomial---this
follows from the existence of a polynomial-time quantum algorithm
for $\operatorname*{RFS}$, together with the result of Beals et al.\
\cite{bbcmw}\ relating quantum algorithms to low-degree polynomials.
\ As a result, the circuit lower bound technique of Razborov
\cite{razborov:ac0}\ and Smolensky \cite{smolensky}, which is based
on the nonexistence of low-degree polynomials, seems unlikely to
work. \ Even the random restriction method of Furst et al.\
\cite{fss}\ can be related to low-degree polynomials, as shown by
Linial et al.\ \cite{lmn}.}

In this chapter I examine the\ $\operatorname*{RFS}$\ problem from a different
angle. \ Could Bernstein and Vazirani's quantum algorithm for
$\operatorname*{RFS}$\ be improved even further, to give an
\textit{exponential} speedup over the classical algorithm? \ And could we use
$\operatorname*{RFS}$, not merely to place $\mathsf{BQP}$\ outside of
$\mathsf{PH}$ relative to an oracle, but to place it outside of $\mathsf{PH}%
$\ with (say) a logarithmic number of alternations?

My answer to both questions is a strong `no.' \ I study a large class of
variations on $\operatorname*{RFS}$, and show that all of them fall into one
of two classes:

\begin{enumerate}
\item[(1)] a trivial class, for which there exists a classical algorithm
making only one query, or

\item[(2)] a nontrivial class, for which any quantum algorithm needs
$2^{\Omega\left(  h\right)  }$\ queries, where $h$ is the height of the tree
to be evaluated. \ (By comparison, the Bernstein-Vazirani algorithm uses
$2^{h}$ queries, because of its need to uncompute garbage recursively at each
level of the tree.)
\end{enumerate}
Since $n^{h}$\ queries always suffice classically, this dichotomy theorem
implies that the speedup afforded by quantum computers is at most
quasipolynomial. \ It also implies that (nontrivial) $\operatorname*{RFS}$\ is
solvable in quantum polynomial time only when $h=O\left(  \log n\right)  $.

The plan is as follows. \ In Section \ref{PRELIMRFS} I define the
$\operatorname*{RFS}$ problem, and give Bernstein and\ Vazirani's
quantum algorithm for solving it. \ In Section \ref{LOWERRFS}, I use
the adversary method of Ambainis \cite{ambainis}\ to prove a lower
bound on the quantum query complexity of any $\operatorname*{RFS}$
variant. \ This bound, however, requires a parameter that I call the
\textquotedblleft nonparity coefficient\textquotedblright\ to be
large. \ Intuitively, given a Boolean function $g:\left\{
0,1\right\}  ^{n}\rightarrow\left\{  0,1\right\}  $, the nonparity
coefficient measures how far $g$ is from being the parity of some
subset of its input bits---not under the uniform distribution over
inputs (the standard assumption in Fourier analysis), but under an
adversarial distribution. \ The crux of the argument is that
\textit{either} the nonparity coefficient is zero (meaning the
$\operatorname*{RFS}$\ variant in question is trivial), or else it
is bounded below by a positive constant. \ This statement is proved
in Section \ref{LOWERRFS}, and seems like it might be of independent
interest. \ Section \ref{OPENRFS} concludes with some open problems.

\section{Preliminaries\label{PRELIMRFS}}

In ordinary Fourier sampling, we are given oracle access to a Boolean function
$A:\left\{  0,1\right\}  ^{n}\rightarrow\left\{  0,1\right\}  $,\ and are
promised that there exists a secret string $s\in\left\{  0,1\right\}  ^{n}%
$\ such that $A\left(  x\right)  =s\cdot x\left(
\operatorname{mod}2\right) $\ for all $x$. \ The problem is to find
$s$---or rather, since we need a problem with Boolean output, the
problem is to return $g\left(  s\right)  $, where $g:\left\{
0,1\right\}  ^{n}\rightarrow\left\{  0,1\right\}  $\ is some known
Boolean function. \ We can think of $g\left(  s\right)  $\ as the
\textquotedblleft hard-core bit\textquotedblright\ of $s$, and can
assume that $g$ itself is efficiently computable, or else that we
are given access to an oracle for $g$.

To obtain a height-$2$ recursive Fourier sampling tree, we simply compose this
problem. \ That is, we are no longer given direct access to $A\left(
x\right)  $, but instead are promised that $A\left(  x\right)  =g\left(
s_{x}\right)  $, where $s_{x}\in\left\{  0,1\right\}  ^{n}$ is the secret
string for another Fourier sampling problem. \ A query then takes the form
$\left(  x,y\right)  $, and produces as output $A_{x}\left(  y\right)
=s_{x}\cdot y\left(  \operatorname{mod}2\right)  $. \ As before, we are
promised that there exists an $s$ such that $A\left(  x\right)  =s\cdot
x\left(  \operatorname{mod}2\right)  $ for all $x$, meaning that the $s_{x}%
$\ strings must be chosen consistent with this promise. \ Again we must return
$g\left(  s\right)  $.

Continuing, we can define height-$h$ recursive Fourier sampling, or
$\operatorname*{RFS}_{h}$, recursively as follows. \ We are given oracle
access to a function $A\left(  x_{1},\ldots,x_{h}\right)  $ for all
$x_{1},\ldots,x_{h}\in\left\{  0,1\right\}  ^{n}$,\ and are promised that

\begin{enumerate}
\item[(1)] for each fixed $x_{1}^{\ast}$, $A\left(  x_{1}^{\ast},x_{2}%
,\ldots,x_{h}\right)  $\ is an instance of $\operatorname*{RFS}_{h-1}$ on
$x_{2},\ldots,x_{h}$, having answer bit $b\left(  x_{1}^{\ast}\right)
\in\left\{  0,1\right\}  $; and

\item[(2)] there exists a secret string $s\in\left\{  0,1\right\}  ^{n}$\ such
that $b\left(  x_{1}^{\ast}\right)  =s\cdot x_{1}^{\ast}\left(
\operatorname{mod}2\right)  $\ for each $x_{1}^{\ast}$.
\end{enumerate}

Again the answer bit to be returned is $g\left(  s\right)  $. \ Note that $g$
is assumed to be the same everywhere in the tree---though using the techniques
in this chapter, it would be straightforward to generalize to the case of
different $g$'s. \ As an example that will be used later, we could take
$g\left(  s\right)  =g_{\operatorname{mod}3}\left(  s\right)  $, where
$g_{\operatorname{mod}3}\left(  s\right)  =0$ if $\left\vert s\right\vert
\equiv0\left(  \operatorname{mod}3\right)  $ and $g_{\operatorname{mod}%
3}\left(  s\right)  =1$ otherwise, and $\left\vert s\right\vert $\ denotes the
Hamming weight of $s$. \ We do not want to take $g$ to be the parity of $s$,
for if we did then $g\left(  s\right)  $\ could be evaluated using a single
query. \ To see this, observe that if $x$\ is the all-$1$'s string, then
$s\cdot x\left(  \operatorname{mod}2\right)  $\ is the parity of $s$.

By an `input,' I will mean a complete assignment for the $\operatorname*{RFS}$
oracle (that is, $A\left(  x_{1},\ldots,x_{h}\right)  $\ for all $x_{1}%
,\ldots,x_{h}$). \ I will sometimes refer also to an `$\operatorname*{RFS}$
tree,' where each vertex at distance $\ell$ from the root has a label
$x_{1},\ldots,x_{\ell}$. \ If $\ell=h$\ then the vertex is a leaf; otherwise
it has $2^{n}$\ children, each with a label $x_{1},\ldots,x_{\ell},x_{\ell+1}$
for some $x_{\ell+1}$. \ The subtrees of the tree just correspond to the
sub-instances of $\operatorname*{RFS}$.

Bernstein and Vazirani \cite{bv}\ showed that $\operatorname*{RFS}_{\log n}$,
or $\operatorname*{RFS}$ with height $\log n$ (all logarithms are base $2$),
is solvable on a quantum computer in time polynomial in $n$. \ I include a
proof for completeness. \ Let $A=\left(  A_{n}\right)  _{n\geq0}$ be an oracle
that, for each $n$, encodes an instance of $\operatorname*{RFS}_{\log n}%
$\ whose answer is $\Psi_{n}$. \ Then let $L_{A}$ be the unary language
$\left\{  0^{n}:\Psi_{n}=1\right\}  $.

\begin{lemma}
\label{alg}$L_{A}\in\mathsf{EQP}^{A}\subseteq\mathsf{BQP}^{A}$ for any choice
of $A.$
\end{lemma}

\begin{proof}
$\operatorname*{RFS}_{1}$ can be solved exactly in four queries, with no
garbage bits left over. \ The algorithm is as follows: first prepare the state%
\[
2^{-n/2}\sum_{x\in\left\{  0,1\right\}  ^{n}}\left\vert x\right\rangle
\left\vert A\left(  x\right)  \right\rangle ,
\]
using one query to $A$. \ Then apply a phase flip conditioned on $A\left(
x\right)  =1$, and uncompute $A\left(  x\right)  $\ using a second query,
obtaining%
\[
2^{-n/2}\sum_{x\in\left\{  0,1\right\}  ^{n}}\left(  -1\right)  ^{A\left(
x\right)  }\left\vert x\right\rangle .
\]
Then apply a Hadamard gate to each bit of the $\left\vert x\right\rangle
$\ register. \ It can be checked that the resulting state is simply
$\left\vert s\right\rangle $. \ One can then compute $\left\vert
s\right\rangle \left\vert g\left(  s\right)  \right\rangle $ and uncompute
$\left\vert s\right\rangle $\ using two more queries to $A$, to obtain
$\left\vert g\left(  s\right)  \right\rangle $. \ To solve $RFS_{\log
n}\left(  n\right)  $, we simply apply the above algorithm recursively at each
level of the tree. \ The total number of queries used is $4^{\log n}=n^{2}$.

One can further reduce the number of queries to $2^{\log n}=n$\ by using the
\textquotedblleft one-call kickback trick,\textquotedblright\ described by
Cleve et al.\ \cite{cemm}. \ Here one prepares the state%
\[
2^{-n/2}\sum_{x\in\left\{  0,1\right\}  ^{n}}\left\vert x\right\rangle
\otimes\frac{\left\vert 1\right\rangle -\left\vert 0\right\rangle }{\sqrt{2}}%
\]
and then exclusive-$OR$'s $A\left(  x\right)  $\ into the second register.
\ This induces the desired phase $\left(  -1\right)  ^{A\left(  x\right)  }%
$\ without the need to uncompute $A\left(  x\right)  $. \ However, one still
needs to uncompute $\left\vert s\right\rangle $\ after computing $\left\vert
g\left(  s\right)  \right\rangle $.
\end{proof}

A remark on notation: to avoid confusion with subscripts, I denote the
$i^{th}$\ bit of string $x$ by $x\left[  i\right]  $.

\section{Quantum Lower Bound\label{LOWERRFS}}

In this section I prove a lower bound on the quantum query complexity of
$\operatorname*{RFS}$. \ Crucially, the bound should hold for any nontrivial
one-bit function of the secret strings, not just a specific function such as
$g_{\operatorname{mod}3}\left(  s\right)  $ defined in Section \ref{PRELIMRFS}%
. \ Let $\operatorname*{RFS}_{h}^{g}$\ be height-$h$ recursive Fourier
sampling in which the problem at each vertex is to return $g\left(  s\right)
$. \ The following notion turns out to be essential.

\begin{definition}
Given a Boolean function $g:\left\{  0,1\right\}
^{n}\rightarrow\left\{ 0,1\right\}  $ (partial or total), the
\textit{nonparity coefficient} $\mu\left(  g\right)  $ is the
largest $\mu^{\ast}$\ for which\ there exist distributions $D_{0}$\
over the $0$-inputs of $g$, and $D_{1}$\ over the $1$-inputs, such
that for all $z\in\left\{  0,1\right\}  ^{n}$, all $0$-inputs
$\widehat{s}_{0}$, and all $1$-inputs $\widehat{s}_{1}$, we have%
\[
\Pr_{s_{0}\in D_{0},s_{1}\in D_{1}}\left[  s_{0}\cdot z\equiv\widehat{s}%
_{1}\cdot z\left(  \operatorname{mod}2\right)  \,\,\,\vee\,\,\,s_{1}\cdot
z\equiv\widehat{s}_{0}\cdot z\left(  \operatorname{mod}2\right)  \right]
\geq\mu^{\ast}\text{.}%
\]
\end{definition}

Loosely speaking, the nonparity coefficient is high if there exist
distributions over $0$-inputs and $1$-inputs that make $g$ far from being a
parity function of a subset of input bits. \ The following proposition
develops some intuition about $\mu\left(  g\right)  $.

\begin{proposition}
\quad

\begin{enumerate}
\item[(i)] $\mu\left(  g\right)  \leq3/4$ for all nonconstant $g$.

\item[(ii)] $\mu\left(  g\right)  =0$\ if and only if $g$ can be written as
the parity (or the NOT of the parity) of a subset $B$ of input bits.
\end{enumerate}
\end{proposition}

\begin{proof}
\quad

\begin{enumerate}
\item[(i)] Given any $s_{0}\neq\widehat{s}_{1}$ and $s_{1}\neq\widehat{s}_{0}%
$, a uniform random $z$ will satisfy%
\[
\Pr_{z}\left[  s_{0}\cdot z\not \equiv \widehat{s}_{1}\cdot z\left(
\operatorname{mod}2\right)  \,\,\,\wedge\,\,\,s_{1}\cdot z\not \equiv
\widehat{s}_{0}\cdot z\left(  \operatorname{mod}2\right)  \right]  \geq
\frac{1}{4}\text{.}%
\]
(If $s_{0}\oplus\widehat{s}_{1}=s_{1}\oplus\widehat{s}_{0}$\ then this
probability will be $1/2$; otherwise it will be $1/4$.) \ So certainly there
is a fixed choice of $z$\ that works for random $s_{0}$\ and $s_{1}$.

\item[(ii)] For the `if' direction, take $z\left[  i\right]  =1$ if and only
if $i\in B$, and choose $\widehat{s}_{0}$\ and $\widehat{s}_{1}$\ arbitrarily.
\ This ensures that $\mu^{\ast}=0$. \ For the `only if' direction, if
$\mu\left(  g\right)  =0$, we can choose $D_{0}$ to have support on all
$0$-inputs, and $D_{1}$\ to have support on all $1$-inputs. \ Then there must
be a $z$\ such that $s_{0}\cdot z$\ is constant as we range over $0$-inputs,
and $s_{1}\cdot z$\ is constant as we range over $1$-inputs. \ Take $i\in
B$\ if and only if $z\left[  i\right]  =1$.
\end{enumerate}
\end{proof}

If $\mu\left(  g\right)  =0$, then $\operatorname*{RFS}_{h}^{g}$\ is easily
solvable using a single classical query. \ Theorem \ref{rfslb} will show that
for all $g$ (partial or total),%
\[
\operatorname*{Q}\nolimits_{2}\left(  \operatorname*{RFS}\nolimits_{h}%
^{g}\right)  =\Omega\left(  \left(  \frac{1}{1-\mu\left(  g\right)  }\right)
^{h/2}\right)  ,
\]
where $\operatorname*{Q}_{2}$\ is bounded-error quantum query complexity as
defined in Section \ref{BLACKBOX}. \ In other words, any $\operatorname*{RFS}$
problem with $\mu$\ bounded away from $0$ requires a number of queries
exponential in the tree height $h$.

However, there is an essential further part of the argument, which restricts
the values of $\mu\left(  g\right)  $\ itself. \ Suppose there existed a
family $\left\{  g_{n}\right\}  $\ of `pseudoparity' functions: that is,
$\mu\left(  g_{n}\right)  >0$ for all $n$,\ yet $\mu\left(  g_{n}\right)
=O(1/\log n)$. \ Then the best bound obtainable from Theorem \ref{rfslb}%
\ would be $\Omega\left(  \left(  1+1/\log n\right)  ^{h/2}\right)  $,
suggesting that $\operatorname*{RFS}_{\log^{2}n}^{g}$ might still be solvable
in quantum polynomial time. \ On the other hand, it would be unclear a priori
how to solve $\operatorname*{RFS}_{\log^{2}n}^{g}$\ classically with a
logarithmic number of alternations. \ Theorem \ref{egood} will rule out this
scenario by showing that pseudoparity\ functions do not exist: if $\mu\left(
g\right)  <0.146$ then $g$ is a parity function, and hence $\mu\left(
g\right)  =0$.

The theorem of Ambainis\ that we need is his \textquotedblleft most
general\textquotedblright\ lower bound from \cite{ambainis}, which he
introduced to show that the quantum query complexity of inverting a
permutation is $\Omega\left(  \sqrt{n}\right)  $, and which we used already in
Chapter \ref{PLS}. \ Let us restate the theorem in the present context.

\begin{theorem}
[Ambainis]\label{ambthmrfs}Let $X\subseteq f^{-1}\left(  0\right)  $\ and
$Y\subseteq f^{-1}\left(  1\right)  $\ be sets of inputs to function $f$.
\ Let $R\left(  x,y\right)  \geq0$ be a symmetric real-valued relation
function, and for $x\in X$, $y\in Y$, and index $i$, let%
\begin{align*}
\theta\left(  x,i\right)   &  =\frac{\sum_{y^{\ast}\in Y~:~x\left[  i\right]
\neq y^{\ast}\left[  i\right]  }R\left(  x,y^{\ast}\right)  }{\sum_{y^{\ast
}\in Y}R\left(  x,y^{\ast}\right)  },\\
\theta\left(  y,i\right)   &  =\frac{\sum_{x^{\ast}\in X~:~x^{\ast}\left[
i\right]  \neq y\left[  i\right]  }R\left(  x^{\ast},y\right)  }{\sum
_{y^{\ast}\in Y}R\left(  x^{\ast},y\right)  },
\end{align*}
where the denominators are all nonzero. \ Then $\operatorname*{Q}_{2}\left(
f\right)  =O\left(  1/\upsilon\right)  $\ where%
\[
\upsilon=\max_{x\in X,~y\in Y,~i~:~R\left(  x,y\right)  >0,~x\left[  i\right]
\neq y\left[  i\right]  }\sqrt{\theta\left(  x,i\right)  \theta\left(
y,i\right)  }.
\]

\end{theorem}

We are now ready to prove a lower bound for $\operatorname*{RFS}$.

\begin{theorem}
\label{rfslb}For all $g$ (partial or total), $\operatorname*{Q}_{2}\left(
\operatorname*{RFS}_{h}^{g}\right)  =\Omega\left(  \left(  1-\mu\left(
g\right)  \right)  ^{-h/2}\right)  $.
\end{theorem}

\begin{proof}
Let $X$ be the set of all $0$-inputs to $\operatorname*{RFS}_{h}^{g}$, and let
$Y$ be the set of all $1$-inputs. \ We will weight the inputs using the
distributions $D_{0},D_{1}$\ from the definition of the nonparity coefficient
$\mu\left(  g\right)  $. \ For all $x\in X$, let $p\left(  x\right)  $\ be the
product, over all vertices $v$ in the $\operatorname*{RFS}$\ tree for $x$, of
the probability of the secret string $s$ at $v$, if $s$ is drawn from
$D_{g\left(  s\right)  }$ (where we condition on $v$'s output bit, $g\left(
s\right)  $). \ Next, say that $x\in X$\ and $y\in Y$ \textit{differ
minimally} if, for all vertices $v$ of the $\operatorname*{RFS}$ tree, the
subtrees rooted at $v$ are identical in $x$ and in $y$\ whenever the answer
bit $g\left(  s\right)  $\ at $v$ is the same in $x$ and in $y$.\ \ If $x$ and
$y$ differ minimally, then we will set $R\left(  x,y\right)  =p\left(
x\right)  p\left(  y\right)  $; otherwise we will set $R\left(  x,y\right)
=0$. \ Clearly $R\left(  x,y\right)  =R\left(  y,x\right)  $\ for all $x\in
X,y\in Y$. \ Furthermore, we claim that $\theta\left(  x,i\right)
\theta\left(  y,i\right)  \leq\left(  1-\mu\left(  g\right)  \right)  ^{h}%
$\ for all $x,y$\ that differ minimally and all $i$ such that $x\left[
i\right]  \neq y\left[  i\right]  $. \ For suppose $y^{\ast}\in Y$\ is chosen
with probability proportional to $R\left(  x,y^{\ast}\right)  $, and $x^{\ast
}\in X$\ is chosen with probability proportional to $R\left(  x^{\ast
},y\right)  $. \ Then $\theta\left(  x,i\right)  \theta\left(  y,i\right)
$\ equals the probability that we would notice the switch from $x$\ to
$y^{\ast}$\ by monitoring $i$, times the probability that we would notice the
switch from $y$\ to $x^{\ast}$.

Let $v_{j}$\ be the $j^{th}$ vertex along the path in the $\operatorname*{RFS}%
$\ tree\ from the root to the leaf vertex $i$, for all $j\in\left\{
1,\ldots,h\right\}  $. \ Also, let $z_{j}\in\left\{  0,1\right\}  ^{n}$\ be
the label of the edge between $v_{j-1}$\ and $v_{j}$, and let $s_{x,j}$\ and
$s_{y,j}$\ be the secret strings at $v_{j}$\ in $x$\ and $y$\ respectively.
\ Then since $x$ and $y$ differ minimally, we must have $g\left(
s_{x,j}\right)  \neq g\left(  s_{y,j}\right)  $\ for all $j$---for otherwise
the subtrees rooted at $v_{j}$\ would be identical, which contradicts the
assumption $x\left[  i\right]  \neq y\left[  i\right]  $. \ So we can think of
the process of choosing $y^{\ast}$ as first choosing a random $s_{x,1}%
^{\prime}$\ from $D_{1}$ so that $1=g\left(  s_{x,1}^{\prime}\right)  \neq
g\left(  s_{x,1}\right)  =0$, then choosing a random $s_{x,2}^{\prime}$\ from
$D_{1-g\left(  s_{x,2}\right)  }$ so that $g\left(  s_{x,2}^{\prime}\right)
\neq g\left(  s_{x,2}\right)  $, and so on. \ Choosing $x^{\ast}$ is
analogous, except that whenever we used $D_{0}$\ in choosing $y^{\ast}$\ we
use $D_{1}$, and vice versa. \ Since the $2h$\ secret strings $s_{x,1}%
,\ldots,s_{x,h},s_{y,1},\ldots,s_{y,h}$\ to be updated are independent of one
another, it follows that%
\begin{align*}
\Pr\left[  y^{\ast}\left[  i\right]  \neq x\left[  i\right]  \right]
\Pr\left[  x^{\ast}\left[  i\right]  \neq y\left[  i\right]  \right]   &  =%
%TCIMACRO{\dprod \limits_{j=1}^{h}}%
%BeginExpansion
{\displaystyle\prod\limits_{j=1}^{h}}
%EndExpansion
\Pr_{s\in D_{0}}\left[  s\cdot z_{j}\not \equiv s_{x,j}\cdot z_{j}\right]
\Pr_{s\in D_{1}}\left[  s\cdot z_{j}\not \equiv s_{y,j}\cdot z_{j}\right] \\
&  \leq%
%TCIMACRO{\dprod \limits_{j=1}^{h}}%
%BeginExpansion
{\displaystyle\prod\limits_{j=1}^{h}}
%EndExpansion
\left(  1-\mu\left(  g\right)  \right) \\
&  =\left(  1-\mu\left(  g\right)  \right)  ^{h}%
\end{align*}
by the definition of $\mu\left(  g\right)  $. \ Therefore%
\[
\operatorname*{Q}\nolimits_{2}\left(  \operatorname*{RFS}\nolimits_{h}%
^{g}\right)  =\Omega\left(  \left(  1-\mu\left(  g\right)  \right)
^{-h/2}\right)
\]
by Theorem \ref{ambthmrfs}.
\end{proof}

Before continuing further, let me show that there is a natural, explicit
choice of $g$---the function $g_{\operatorname{mod}3}\left(  s\right)  $\ from
Section \ref{PRELIMRFS}---for which the nonparity coefficient is almost $3/4$.
\ Thus, for $g=g_{\operatorname{mod}3}$, the algorithm of Lemma \ref{alg}\ is
essentially optimal.

\begin{proposition}
\label{gns}$\mu\left(  g_{\operatorname{mod}3}\right)  =3/4-O\left(
1/n\right)  $.
\end{proposition}

\begin{proof}
Let $n\geq6$. \ Let $D_{0}$ be the uniform distribution over all $s$ with
$\left\vert s\right\vert =3\left\lfloor n/6\right\rfloor $ (so
$g_{\operatorname{mod}3}\left(  s\right)  =0$); likewise let $D_{1}$\ be the
uniform distribution over $s$ with $\left\vert s\right\vert =3\left\lfloor
n/6\right\rfloor +2$ ($g_{\operatorname{mod}3}\left(  s\right)  =1$). \ We
consider only the case of $s$ drawn from $D_{0}$; the $D_{1}$\ case is
analogous. \ We will show that for any $z$,
\[
\left\vert \Pr_{s\in D_{0}}\left[  s\cdot z\equiv0\right]  -\frac{1}%
{2}\right\vert =O\left(  \frac{1}{n}\right)
\]
(all congruences are $\operatorname{mod}2$). \ The theorem then follows, since
by the definition of the nonparity coefficient, given any $z$ the choices of
$s_{0}\in D_{0}$\ and $s_{1}\in D_{1}$\ are independent.

Assume without loss of generality that $1\leq\left\vert z\right\vert \leq n/2$
(if $\left\vert z\right\vert >n/2$, then replace $z$ by its complement). \ We
apply induction on $\left\vert z\right\vert $. \ If $\left\vert z\right\vert
=1$, then clearly%
\[
\Pr\left[  s\cdot z\equiv0\right]  =3\left\lfloor n/6\right\rfloor /n=\frac
{1}{2}\pm O\left(  \frac{1}{n}\right)  \text{.}%
\]
For $\left\vert z\right\vert \geq2$,\ let $z=z_{1}\oplus z_{2}$, where $z_{2}%
$\ contains only the rightmost $1$ of $z$ and $z_{1}$ contains all the other
$1$'s. \ Suppose the proposition holds for $\left\vert z\right\vert -1$.
\ Then%
\begin{align*}
\Pr\left[  s\cdot z\equiv0\right]  =  &  \Pr\left[  s\cdot z_{1}%
\equiv0\right]  \Pr\left[  s\cdot z_{2}\equiv0|s\cdot z_{1}\equiv0\right]  +\\
&  \Pr\left[  s\cdot z_{1}\equiv1\right]  \Pr\left[  s\cdot z_{2}%
\equiv1|s\cdot z_{1}\equiv1\right]  \text{,}%
\end{align*}
where%
\[
\Pr\left[  s\cdot z_{1}\equiv0\right]  =\frac{1}{2}+\alpha,\,\,\,\Pr\left[
s\cdot z_{1}\equiv1\right]  =\frac{1}{2}-\alpha
\]
for some $\left\vert \alpha\right\vert =O\left(  1/n\right)  $. \ Furthermore,
even conditioned on $s\cdot z_{1}$, the expected number of $1$'s in $s$
outside of $z_{1}$\ is $\left(  n-\left\vert z_{1}\right\vert \right)  /2\pm
O\left(  1\right)  $ and they are uniformly distributed. \ Therefore%
\[
\Pr\left[  s\cdot z_{2}\equiv b|s\cdot z_{1}\equiv b\right]  =\frac{1}%
{2}+\beta_{b}%
\]
for some $\left\vert \beta_{0}\right\vert ,\left\vert \beta_{1}\right\vert
=O\left(  1/n\right)  $. \ So%
\begin{align*}
\Pr\left[  s\cdot z\equiv0\right]   &  =\frac{1}{2}+\frac{\beta_{0}}{2}%
+\alpha\beta_{0}-\frac{\beta_{1}}{2}-\alpha\beta_{1}\\
&  =\frac{1}{2}\pm O\left(  \frac{1}{n}\right)  .
\end{align*}

\end{proof}

Finally it must be shown that pseudoparity functions do not exist.\ \ That is,
if $g$ is too close to a parity function for the bound of Theorem
\ref{rfslb}\ to apply, then $g$ actually \textit{is} a parity function, from
which it follows that $RFS_{h}^{g}$\ admits an efficient classical algorithm.

\begin{theorem}
\label{egood}Suppose $\mu\left(  g\right)  <0.146$. \ Then $g$ is a parity
function (equivalently, $\mu\left(  g\right)  =0$).
\end{theorem}

\begin{proof}
By linear programming duality, there exists a joint distribution $\mathcal{D}$
over $z\in\left\{  0,1\right\}  ^{n}$, $0$-inputs $\widehat{s}_{0}\in
g^{-1}\left(  0\right)  $, and $1$-inputs $\widehat{s}_{1}\in g^{-1}\left(
1\right)  $, such that for all $s_{0}\in g^{-1}\left(  0\right)  $ and
$s_{1}\in g^{-1}\left(  1\right)  $,%
\[
\Pr_{\left(  z,\widehat{s}_{0},\widehat{s}_{1}\right)  \in\mathcal{D}}\left[
s_{0}\cdot z\equiv\widehat{s}_{1}\cdot z\left(  \operatorname{mod}2\right)
\,\,\,\vee\,\,\,s_{1}\cdot z\equiv\widehat{s}_{0}\cdot z\left(
\operatorname{mod}2\right)  \right]  <\mu\left(  g\right)  \text{.}%
\]
Furthermore $\widehat{s}_{0}\cdot z\not \equiv \widehat{s}_{1}\cdot z\left(
\operatorname{mod}2\right)  $, since otherwise we could violate the hypothesis
by taking $s_{0}=\widehat{s}_{0}$\ or $s_{1}=\widehat{s}_{1}$. \ It follows
that there exists a joint distribution $\mathcal{D}^{\prime}$\ over
$z\in\left\{  0,1\right\}  ^{n}$ and $b\in\left\{  0,1\right\}  $ such that%
\[
\Pr_{\left(  z,b\right)  \in\mathcal{D}^{\prime}}\left[  s\cdot z\equiv
b\left(  \operatorname{mod}2\right)  \right]  >1-\mu\left(  g\right)
\]
for all $s\in g^{-1}\left(  0\right)  $, and%
\[
\Pr_{\left(  z,b\right)  \in\mathcal{D}^{\prime}}\left[  s\cdot z\not \equiv
b\left(  \operatorname{mod}2\right)  \right]  >1-\mu\left(  g\right)
\]
for all $s\in g^{-1}\left(  1\right)  $. \ But this implies that $g$ is a
bounded-error threshold function of parity functions. \ More precisely, there
exist probabilities $p_{z}$, summing to $1$, as well as $b_{z}\in\left\{
0,1\right\}  $\ such that for all $s\in\left\{  0,1\right\}  ^{n}$,%
\[
\Psi\left(  s\right)  =\sum_{z\in\left\{  0,1\right\}  ^{n}}p_{z}\left(
\left(  s\cdot z\right)  \oplus b_{z}\right)  \text{ is }\left\{
\begin{array}
[c]{ll}%
>1-\mu\left(  g\right)  & \text{if }g\left(  s\right)  =1\\
<\mu\left(  g\right)  & \text{if }g\left(  s\right)  =0.
\end{array}
\ \ \right.
\]
We will consider $\operatorname*{var}\left(  \Psi\right)  $, the variance of
the above quantity $\Psi\left(  s\right)  $\ if $s$ is drawn uniformly at
random from $\left\{  0,1\right\}  ^{n}$. \ First, if $p_{z}\geq1/2$\ for any
$z$, then $g\left(  s\right)  =\left(  s\cdot z\right)  \oplus b_{z}$\ is a
parity function and hence $\mu\left(  g\right)  =0$. \ So we can assume
without loss of generality that $p_{z}<1/2$\ for all $z$. \ Then since $s$ is
uniform, for each $z_{1}\neq z_{2}$\ we know that $\left(  s\cdot
z_{1}\right)  \oplus b_{z_{1}}$\ and $\left(  s\cdot z_{2}\right)  \oplus
b_{z_{2}}$\ are pairwise independent $\left\{  0,1\right\}  $ random
variables, both with expectation $1/2$. \ So%
\[
\operatorname*{var}\left(  \Psi\right)  =\frac{1}{4}%
%TCIMACRO{\tsum \nolimits_{z}}%
%BeginExpansion
{\textstyle\sum\nolimits_{z}}
%EndExpansion
p_{z}^{2}<\frac{1}{4}\left(  \left(  \frac{1}{2}\right)  ^{2}+\left(  \frac
{1}{2}\right)  ^{2}\right)  =\frac{1}{8}\text{.}%
\]
On the other hand, since $\Psi\left(  s\right)  $\ is always less than $\mu
$\ or greater than $1-\mu$,%
\[
\operatorname*{var}\left(  \Psi\right)  >\left(  \frac{1}{2}-\mu\right)
^{2}.
\]
Combining,%
\[
\mu>\frac{2-\sqrt{2}}{4}>0.146.
\]

\end{proof}

\section{Open Problems\label{OPENRFS}}

An intriguing open problem is whether Theorem \ref{rfslb}\ can be
proved using the polynomial method of Beals et al.\ \cite{bbcmw},
rather than the adversary method of Ambainis \cite{ambainis}. \ It
is known that one can lower-bound polynomial degree in terms of
block sensitivity, or the maximum number of disjoint changes to an
input that change the output value. \ The trouble is that the
$\operatorname*{RFS}$ function has block sensitivity $1$---the
\textquotedblleft sensitive blocks\textquotedblright\ of each input
tend to have small intersection, but are not disjoint. \ For this
reason, I implicitly used the quantum certificate complexity of
Chapter \ref{CER} rather than block sensitivity to prove a lower
bound.

I believe the constant of Theorem \ref{egood}\ can be improved. \ The smallest
nonzero $\mu\left(  g\right)  $\ value I know of is attained when $n=2$ and
$g=\operatorname*{OR}\left(  s\left[  1\right]  ,s\left[  2\right]  \right)  $:

\begin{proposition}
$\mu\left(  \operatorname*{OR}\right)  =1/3$.
\end{proposition}

\begin{proof}
First, $\mu\left(  \operatorname*{OR}\right)  \geq1/3$, since $D_{1}$\ can
choose $s\left[  1\right]  s\left[  2\right]  $\ to be $01$, $10$,\ or $11$
each with probability $1/3$; then for any $z\neq0$ and the unique $0$-input
$\widehat{s}_{0}=00$, we have $s_{1}\cdot z\not \equiv \widehat{s}_{0}\cdot z$
with probability at most $2/3$. \ Second, $\mu\left(  \operatorname*{OR}%
\right)  \leq1/3$, since applying linear programming duality, we can let the
pair $\left(  z,\widehat{s}_{1}\right)  $ equal $\left(  01,01\right)  $,
$\left(  10,10\right)  $, or $\left(  11,10\right)  $\ each with probability
$1/3$. \ Then $0\equiv s_{0}\cdot z\not \equiv \widehat{s}_{1}\cdot z\equiv
1$\ always, and for any $1$-input $s_{1}$, we have $s_{1}\cdot z\equiv
1\not \equiv \widehat{s}_{0}\cdot z$ with probability $2/3$.
\end{proof}

Finally, I conjecture that uncomputation is unavoidable not just for
$\operatorname*{RFS}$ but for many other recursive problems, such as game-tree
evaluation. \ Formally, the conjecture is that the quantum query complexity of
evaluating a game tree increases exponentially with depth as the number of
leaves is held constant, even if there is at most one winning move per vertex
(so that the tree can be evaluated with zero probability of error).

\chapter{Limitations of Quantum Advice\label{ADV}}

How many classical bits can \textquotedblleft really\textquotedblright\ be
encoded into $n$ qubits? \ Is it $n$, because of Holevo's Theorem
\cite{holevo}; $2n$, because of dense quantum coding \cite{bw}\ and quantum
teleportation \cite{bbcjpy}; exponentially many, because of quantum
fingerprinting \cite{bcww}; or infinitely many, because amplitudes are
continuous? \ The best general answer to this question is probably
\textit{mu}, the Zen word that \textquotedblleft unasks\textquotedblright\ a
question.\footnote{Another \textit{mu}-worthy question is, \textquotedblleft
Where does the power of quantum computing come from? \ Superposition?
\ Interference? \ The large size of Hilbert space?\textquotedblright}

To a computer scientist, however, it is natural to formalize the question in
terms of \textit{quantum one-way communication complexity}
\cite{bjk,bcww,klauck:cc,yao:fing}. \ The setting is as follows: Alice has an
$n$-bit string $x$, Bob has an $m$-bit string $y$, and together they wish to
evaluate $f\left(  x,y\right)  $\ where $f:\left\{  0,1\right\}  ^{n}%
\times\left\{  0,1\right\}  ^{m}\rightarrow\left\{  0,1\right\}  $\ is a
Boolean function. \ After examining her input $x=x_{1}\ldots x_{n}$, Alice can
send a single quantum message $\rho_{x}$\ to Bob, whereupon Bob, after
examining his input $y=y_{1}\ldots y_{m}$, can choose some basis in which to
measure $\rho_{x}$. \ He must then output a claimed value for $f\left(
x,y\right)  $. \ We are interested in how long Alice's message needs to be,
for Bob to succeed with high probability on any $x,y$ pair. \ Ideally the
length will be much smaller than if Alice had to send a classical message.

Communication complexity questions have been intensively studied in
theoretical computer science (see the book of Kushilevitz and Nisan
\cite{kn}\ for example). \ In both the classical and quantum cases, though,
most attention has focused on \textit{two-way }communication, meaning that
Alice and Bob get to send messages back and forth. \ I believe that the study
of one-way quantum communication presents two main advantages. \ First, many
open problems about two-way communication look gruesomely difficult---for
example, are the randomized and quantum communication complexities of every
total Boolean function polynomially related? \ We might gain insight into
these problems by tackling their one-way analogues first. \ And second,
because of its greater simplicity, the one-way model more directly addresses
our opening question: how much \textquotedblleft useful
stuff\textquotedblright\ can be packed into a quantum state? \ Thus, results
on one-way communication fall into the quantum information theory tradition
initiated by Holevo \cite{holevo}\ and others, as much as the communication
complexity tradition initiated by Yao \cite{yao:cc}.

Related to quantum one-way communication is the notion of \textit{quantum
advice}.\ \ As pointed out by Nielsen and Chuang \cite[p.203]{nc}, there is no
compelling physical reason to assume that the starting state of a quantum
computer is a computational basis state:\footnote{One might object that the
starting state is itself the outcome of some computational process, which
began no earlier than the Big Bang. \ However, (1) for all we know highly
entangled states were created in the Big Bang, and (2) $14$ billion years is a
long time.}

\begin{quote}
[W]e know that many systems in Nature `prefer' to sit in highly entangled
states of many systems; might it be possible to exploit this preference to
obtain extra computational power?\ \ It might be that having access to certain
states allows particular computations to be done much more easily than if we
are constrained to start in the computational basis.
\end{quote}

One way to interpret Nielsen and Chuang's provocative question is as follows.
\ Suppose we could request the \textit{best possible} starting state for a
quantum computer, knowing the language to be decided and the input length $n$
but not knowing the input itself.\footnote{If we knew the input, we would
simply request a starting state that contains the right answer!} \ Denote the
class of languages that we could then decide by $\mathsf{BQP/qpoly}$---meaning
quantum polynomial time, given an arbitrarily-entangled but polynomial-size
quantum advice state.\footnote{$\mathsf{BQP/qpoly}$ might remind readers of a
better-studied class called $\mathsf{QMA}$ (Quantum Merlin-Arthur). \ But
there are two key differences: first, advice can be trusted while proofs
cannot; second, proofs can be tailored to a particular input while advice
cannot.} \ How powerful is this class? \ If $\mathsf{BQP/qpoly}$ contained
(for example) the $\mathsf{NP}$-complete problems, then we would need to
rethink our most basic assumptions about the power of quantum computing. \ We
will see later that quantum advice is closely related to quantum one-way
communication, since we can think of an advice state as a one-way message sent
to an algorithm by a benevolent \textquotedblleft advisor.\textquotedblright

This chapter is about the \textit{limitations} of quantum advice and one-way
communication. \ It presents three contributions which are basically
independent of one another.

First, Section \ref{MESSAGES} shows that $\operatorname*{D}^{1}\left(
f\right)  =O\left(  mQ_{2}^{1}\left(  f\right)  \log\operatorname*{Q}_{2}%
^{1}\left(  f\right)  \right)  $\ for any Boolean function $f$, partial or
total. \ Here $\operatorname*{D}^{1}\left(  f\right)  $\ is deterministic
one-way communication complexity, $\operatorname*{Q}_{2}^{1}\left(  f\right)
$\ is bounded-error one-way quantum communication complexity, and $m$ is the
length of Bob's input. \ Intuitively, whenever the set of Bob's possible
inputs is not too large, Alice can send him a short classical message that
lets him learn the outcome of any measurement he would have wanted to make on
the quantum message $\rho_{x}$. \ It is interesting that a slightly tighter
bound for total functions---$\operatorname*{D}^{1}\left(  f\right)  =O\left(
mQ_{2}^{1}\left(  f\right)  \right)  $---follows easily from a result of
Klauck \cite{klauck:cc} together with a lemma of Sauer \cite{sauer} about
VC-dimension. \ However, the proof of the latter bound is highly
nonconstructive, and seems to fail for partial $f$.\nolinebreak

Using my communication complexity result, Section \ref{ADVSIM} shows
that
$\mathsf{BQP/qpoly}\subseteq\mathsf{PP/poly}$---in other words, $\mathsf{BQP}%
$\ with polynomial-size quantum advice can be simulated in $\mathsf{PP}$\ with
polynomial-size classical advice.\footnote{Given a complexity class
$\mathsf{C}$, the class $\mathsf{C/poly}$ consists of all languages decidable
by a $\mathsf{C}$\ machine, given a polynomial-size classical advice
string\ that depends only on the input length. \ See Chapter \ref{COMPLEXITY}
for more information about the complexity classes mentioned in this chapter.}
\ This resolves a question of Harry Buhrman (personal communication), who
asked whether quantum advice\ can be simulated in \textit{any} classical
complexity class with short classical advice. \ A corollary of this
containment is that we cannot hope to show an unrelativized separation between
quantum and classical advice (that is, that $\mathsf{BQP/poly}\neq
\mathsf{BQP/qpoly}$), without also showing that $\mathsf{PP}$\ does not have
polynomial-size circuits.

What makes this result surprising is that, in the minds of many computer
scientists, a quantum state is basically an exponentially long vector.
\ Indeed, this belief seems to fuel skepticism of quantum computing (see
Goldreich \cite{goldreich:qc} for example). \ But given an exponentially long
advice string, even a classical computer could decide any language whatsoever.
\ So one might\ imagine\ na\"{\i}vely\ that quantum advice would let us solve
problems that are not even recursively enumerable given classical advice of a
similar size! \ The failure of this na\"{\i}ve intuition supports the view
that a quantum superposition over $n$-bit strings is \textquotedblleft more
similar\textquotedblright\ to a probability distribution over $n$-bit strings
than to a $2^{n}$-bit string.

The second contribution of the chapter, in Section \ref{DPT}, is an
oracle relative to which $\mathsf{NP}$\ is not contained in\
$\mathsf{BQP/qpoly}$. \ Underlying this oracle separation is the
first correct proof of a \textit{direct product theorem} for quantum
search.\ \ Given an $N$-item database with $K$ marked items, the
direct product theorem says that if a quantum algorithm makes
$o\left(  \sqrt{N}\right)  $ queries, then the probability that the
algorithm finds all $K$ of the marked items decreases exponentially
in $K$. \ Notice that such a result does not follow from any
existing quantum lower bound. \ Earlier Klauck \cite{klauck:ts}\ had
claimed a weaker direct product theorem, based on the hybrid method
of Bennett et al.\ \cite{bbbv}, in a paper on quantum time-space
tradeoffs for sorting. \ Unfortunately, Klauck's proof is incorrect.
\ The proof uses the polynomial method of Beals et al.\
\cite{bbcmw},\ with the novel twist that we examine all
\textit{higher} derivatives of a polynomial (not just the first
derivative). \ The proof has already been improved by Klauck,
\v{S}palek, and de Wolf \cite{ksw}, who were able to recover and
even extend Klauck's original claims about quantum sorting.

The final contribution, in Section \ref{TRACEDIST}, is a new
\textit{trace distance method} for proving lower bounds on quantum
one-way communication complexity. \ Previously there was only one
basic lower bound technique: the VC-dimension method of Klauck
\cite{klauck:cc}, which relied on lower bounds for quantum random
access codes due to Ambainis et al.\ \cite{antv}\ and Nayak
\cite{nayak}. \ Using VC-dimension one can show, for example, that
$\operatorname*{Q}_{2}^{1}\left(  \operatorname*{DISJ}\right)
=\Omega\left(
n\right)  $, where the \textit{disjointness function}\ $\operatorname*{DISJ}%
:\left\{  0,1\right\}  ^{n}\times\left\{  0,1\right\}  ^{n}\rightarrow\left\{
0,1\right\}  $\ is defined by \ $\operatorname*{DISJ}\left(  x,y\right)  =1$
if and only if $x_{i}y_{i}=0$\ for all $i\in\left\{  1,\ldots,n\right\}  $.

For some problems, however, the VC-dimension method yields no nontrivial
quantum lower bound. \ Seeking to make this point vividly, Ambainis posed the
following problem. \ Alice is given two elements $x,y$\ of a finite field
$\mathbb{F}_{p}$ (where $p$ is prime); Bob is given another two elements
$a,b\in\mathbb{F}_{p}$. \ Bob's goal is to output $1$ if $y\equiv ax+b\left(
\operatorname{mod}p\right)  $ and $0$ otherwise. \ For this problem, the
VC-dimension method yields no randomized \textit{or} quantum lower bound
better than constant. \ On the other hand, the well-known fingerprinting
protocol for the equality function \cite{ry}\ seems to fail for Ambainis'
problem, because of the interplay between addition and multiplication. \ So it
is natural to conjecture that the randomized and even quantum one-way
complexities are $\Theta\left(  \log p\right)  $---that is, that no nontrivial
protocol exists for this problem.

Ambainis posed a second problem in the same spirit. \ Here Alice is given
$x\in\left\{  1,\ldots,N\right\}  $, Bob is given $y\in\left\{  1,\ldots
,N\right\}  $, and both players know a subset $S\subset\left\{  1,\ldots
,N\right\}  $. \ Bob's goal is to decide whether $x-y\in S$\ where subtraction
is modulo $N$. \ The conjecture is that if $S$ is chosen uniformly at random
with $\left\vert S\right\vert $ about $\sqrt{N}$, then with high probability
the randomized and quantum one-way complexities are both $\Theta\left(  \log
N\right)  $.

Using the trace distance method, I am able to show optimal quantum lower
bounds for both of Ambainis' problems. \ Previously, no nontrivial lower
bounds were known even for randomized protocols. \ The key idea is to consider
two probability distributions over Alice's quantum message $\rho_{x}$. \ The
first distribution corresponds to $x$ chosen uniformly at random; the second
corresponds to $x$ chosen uniformly conditioned on $f\left(  x,y\right)  =1$.
\ These distributions give rise to two mixed states $\rho$\ and $\rho_{y}$,
which Bob must be able to distinguish with non-negligible bias assuming he can
evaluate $f\left(  x,y\right)  $. \ I then show an upper bound on the trace
distance $\left\Vert \rho-\rho_{y}\right\Vert _{\operatorname*{tr}}$, which
implies that Bob cannot distinguish the distributions.

Theorem \ref{vardist}\ gives a very general condition under which
the trace distance method works;\ Corollaries \ref{coset}\ and
\ref{subset} then show that the condition is satisfied for Ambainis'
two problems. \ Besides showing a significant limitation of the
VC-dimension method, I hope the new method is a non-negligible step
towards proving that $\operatorname*{R}_{2}^{1}\left( f\right)
=O\left(  \operatorname*{Q}_{2}^{1}\left(  f\right)  \right)  $\ for
all total Boolean functions $f$, where
$\operatorname*{R}_{2}^{1}\left( f\right)  $\ is randomized one-way
complexity. \ I conclude in Section \ref{OPENADV}\ with some open
problems.

\section{Preliminaries\label{PRELIMADV}}

Following standard conventions, I denote by $\operatorname*{D}^{1}\left(
f\right)  $\ the deterministic one-way complexity of $f$, or the minimum
number of bits that Alice must send if her message is a function of $x$.
\ Also, $\operatorname*{R}_{2}^{1}\left(  f\right)  $, the bounded-error
randomized one-way complexity, is the minimum $k$ such that for every $x,y$,
if Alice sends Bob a $k$-bit message drawn from some distribution
$\mathcal{D}_{x}$, then Bob can output a bit $a$ such that $a=f\left(
x,y\right)  $\ with probability at least $2/3$. \ (The subscript $2$ means
that the error is two-sided.) \ The zero-error randomized complexity
$\operatorname*{R}_{0}^{1}\left(  f\right)  $\ is similar, except that Bob's
answer can never be wrong: he must output $f\left(  x,y\right)  $\ with
probability at least $1/2$ and otherwise declare failure.

The bounded-error quantum one-way complexity $\operatorname*{Q}_{2}^{1}\left(
f\right)  $\ is the minimum $k$ such that, if Alice sends Bob a mixed state
$\rho_{x}$\ of $k$ qubits, there exists a joint measurement of $\rho_{x}$\ and
$y$ enabling Bob to output an $a$ such that $a=f\left(  x,y\right)  $\ with
probability at least $2/3$. \ The zero-error and exact complexities
$\operatorname*{Q}_{0}^{1}\left(  f\right)  $\ and $\operatorname*{Q}_{E}%
^{1}\left(  f\right)  $\ are defined analogously. \ Requiring Alice's message
to be a pure state would increase these complexities by at most a factor of
$2$, since by Kraus' Theorem, every $k$-qubit mixed state can be realized as
half of a $2k$-qubit pure state. \ (Winter \cite{winter}\ has shown that this
factor of $2$ is tight.) \ See Klauck \cite{klauck:cc}\ for more detailed
definitions of quantum and classical one-way communication complexity measures.

It is immediate that $\operatorname*{D}^{1}\left(  f\right)  \geq
\operatorname*{R}_{0}^{1}\left(  f\right)  \geq\operatorname*{R}_{2}%
^{1}\left(  f\right)  \geq\operatorname*{Q}_{2}^{1}\left(  f\right)  $, that
$\operatorname*{R}_{0}^{1}\left(  f\right)  \geq\operatorname*{Q}_{0}%
^{1}\left(  f\right)  \geq\operatorname*{Q}_{2}^{1}\left(  f\right)  $, and
that $\operatorname*{D}^{1}\left(  f\right)  \geq\operatorname*{Q}_{E}%
^{1}\left(  f\right)  $. \ Also, for total $f$, Duri\v{s} et al.\
\cite{dhrs} showed that $\operatorname*{R}_{0}^{1}\left(  f\right)
=\Theta\left( \operatorname*{D}^{1}\left(  f\right)  \right)  $,
while Klauck \cite{klauck:cc}\ showed that
$\operatorname*{Q}_{E}^{1}\left(  f\right)
=\operatorname*{D}^{1}\left(  f\right)  $\ and that $\operatorname*{Q}_{0}%
^{1}\left(  f\right)  =\Theta\left(  \operatorname*{D}^{1}\left(  f\right)
\right)  $. \ In other words, randomized and quantum messages yield no
improvement for total functions if one is unwilling to tolerate a bounded
probability of error. \ This remains\ true even if Alice and Bob share
arbitrarily many EPR pairs \cite{klauck:cc}. \ As is often the case, the
situation is dramatically different for partial functions: there it is easy to
see that $\operatorname*{R}_{0}^{1}\left(  f\right)  $\ can be constant\ even
though $\operatorname*{D}^{1}\left(  f\right)  =\Omega\left(  n\right)  $: let
$f\left(  x,y\right)  =1$\ if $x_{1}y_{1}+\cdots+x_{n/2}y_{n/2}\geq
n/4$\ and\ $x_{n/2+1}y_{n/2+1}+\cdots+x_{n}y_{n}=0$ and $f\left(  x,y\right)
=0$\ if $x_{1}y_{1}+\cdots+x_{n/2}y_{n/2}=0$\ and\ $x_{n/2+1}y_{n/2+1}%
+\cdots+x_{n}y_{n}\geq n/4$, promised that one of these is the case.

Moreover, Bar-Yossef, Jayram, and Kerenidis \cite{bjk}\ have \textit{almost}
shown that $\operatorname*{Q}_{E}^{1}\left(  f\right)  $\ can be exponentially
smaller than $\operatorname*{R}_{2}^{1}\left(  f\right)  $. \ In particular,
they proved that separation for a \textit{relation}, meaning a problem for
which Bob has many possible valid outputs. \ For a partial function $f$ based
on their relation, they also showed that $\operatorname*{Q}_{E}^{1}\left(
f\right)  =\Theta\left(  \log n\right)  $\ whereas $\operatorname*{R}_{0}%
^{1}\left(  f\right)  =\Theta\left(  \sqrt{n}\right)  $; and they conjectured
(but did not prove) that $\operatorname*{R}_{2}^{1}\left(  f\right)
=\Theta\left(  \sqrt{n}\right)  $.

\subsection{Quantum Advice\label{ADVICE}}

Informally, $\mathsf{BQP/qpoly}$ is the class of languages decidable in
polynomial time on a quantum computer, given a polynomial-size quantum advice
state that depends only on the input length. \ I now make the definition more formal.

\begin{definition}
\label{qpolydef}A language $L$ is in $\mathsf{BQP/qpoly}$\ if there exists a
polynomial-size quantum circuit family $\left\{  C_{n}\right\}  _{n\geq1}$,
and a polynomial-size family of quantum states $\left\{  \left|  \psi
_{n}\right\rangle \right\}  _{n\geq1}$, such that for all $x\in\left\{
0,1\right\}  ^{n}$,

\begin{enumerate}
\item[(i)] If $x\in L\ $then $q\left(  x\right)  \geq2/3$, where $q\left(
x\right)  $\ is\ the probability that the first qubit is measured to be
$\left|  1\right\rangle $, after $C_{n}$\ is applied to the starting state
$\left|  x\right\rangle \otimes\left|  0\cdots0\right\rangle \otimes\left|
\psi_{n}\right\rangle $.

\item[(ii)] If $x\notin L$\ then$\ q\left(  x\right)  \leq1/3$.\footnote{If
the starting state is $\left\vert x\right\rangle \otimes\left\vert
0\cdots0\right\rangle \otimes\left\vert \varphi\right\rangle $\ for some
$\left\vert \varphi\right\rangle \neq\left\vert \psi_{n}\right\rangle $, then
the acceptance probability is not required to lie in $\left[  0,1/3\right]
\cup\left[  2/3,1\right]  $. \ Therefore, what I call $\mathsf{BQP/qpoly}%
$\ corresponds to what Nishimura and Yamakami \cite{ny}\ call $\mathsf{BQP/}%
^{\mathsf{\ast}}\mathsf{Qpoly}$. \ Also, it does not matter whether the
circuit family $\left\{  C_{n}\right\}  _{n\geq1}$\ is uniform, since we are
giving it advice anyway.}
\end{enumerate}
\end{definition}

The central open question about $\mathsf{BQP/qpoly}$\ is whether it equals
$\mathsf{BQP/poly}$, or $\mathsf{BQP}$ with polynomial-size \textit{classical}
advice. \ We do have a candidate for an oracle problem separating the two
classes: the \textit{group membership problem} of Watrous \cite{watrous},
which I will describe for completeness. \ Let $G_{n}$\ be a black box
group\footnote{In other words, we have a quantum oracle available that given
$x,y\in G_{n}$\ outputs $xy$ (i.e. exclusive-OR's $xy$ into an answer
register), and that given $x\in G_{n}$\ outputs $x^{-1}$.} whose elements are
uniquely labeled by $n$-bit strings,\ and let $H_{n}$\ be a subgroup of
$G_{n}$. \ Both $G_{n}$\ and $H_{n}$\ depend only on the input length\ $n$, so
we can assume that a nonuniform algorithm knows generating sets for both of
them. \ Given an element $x\in G_{n}$ as input, the problem is to decide
whether $x\in H_{n}$.

If $G_{n}$\ is \textquotedblleft sufficiently nonabelian\textquotedblright%
\ and $H_{n}$\ is exponentially large, we do not know how to solve this
problem in $\mathsf{BQP}$\ or even $\mathsf{BQP/poly}$. \ On the other hand,
we can solve it in $\mathsf{BQP/qpoly}$\ as follows. \ Let the quantum advice
state be an equal superposition over all elements of $H_{n}$:%
\[
\left\vert H_{n}\right\rangle =\frac{1}{\sqrt{\left\vert H_{n}\right\vert }%
}\sum_{y\in H_{n}}\left\vert y\right\rangle
\]
We can transform $\left\vert H_{n}\right\rangle $\ into%
\[
\left\vert xH_{n}\right\rangle =\frac{1}{\sqrt{\left\vert H_{n}\right\vert }%
}\sum_{y\in H_{n}}\left\vert xy\right\rangle
\]
by mapping $\left\vert y\right\rangle \left\vert 0\right\rangle $ to
$\left\vert y\right\rangle \left\vert xy\right\rangle $\ to $\left\vert
y\oplus x^{-1}xy\right\rangle \left\vert xy\right\rangle =\left\vert
0\right\rangle \left\vert xy\right\rangle $\ for each $y\in H_{n}$. Our
algorithm will first prepare the state $\left(  \left\vert 0\right\rangle
\left\vert H_{n}\right\rangle +\left\vert 1\right\rangle \left\vert
xH_{n}\right\rangle \right)  /\sqrt{2}$,\ then apply a Hadamard gate to the
first qubit, and finally measure the first qubit in the standard basis, in
order to distinguish the cases $\left\vert H_{n}\right\rangle =\left\vert
xH_{n}\right\rangle $\ and $\left\langle H_{n}|xH_{n}\right\rangle =0$\ with
constant bias. \ The first case occurs whenever $x\in H_{n}$, and the second
occurs whenever $x\notin H_{n}$.

Although the group membership problem provides intriguing evidence for the
power of quantum advice, we have no idea how to show that it is not also
solvable using classical advice. \ Indeed, apart from a result of Nishimura
and Yamakami \cite{ny}\ that $\mathsf{EESPACE}\not \subset \mathsf{BQP/qpoly}%
$, essentially nothing was known about the class $\mathsf{BQP/qpoly}$\ before
the work reported here.

\subsection{The Almost As Good As New Lemma\label{GOODASNEW}}

The following simple lemma, which was implicit in \cite{antv}, is used three
times in this chapter---in Theorems \ref{partialthm}, \ref{qpoly}, and
\ref{npqpoly}. \ It says that, if the outcome of measuring a quantum state
$\rho$\ could be predicted with near-certainty given knowledge of $\rho$, then
measuring $\rho$ will damage it only slightly. \ Recall that the trace
distance $\left\|  \rho-\sigma\right\|  _{\operatorname*{tr}}$\ between two
mixed states $\rho$\ and $\sigma$ equals $\frac{1}{2}\sum_{i}\left|
\lambda_{i}\right|  $, where $\lambda_{1},\ldots,\lambda_{N}$\ are the
eigenvalues of $\rho-\sigma$.

\begin{lemma}
\label{tracedist}Suppose a $2$-outcome measurement of a mixed state $\rho$
yields outcome $0$ with probability $1-\varepsilon$. \ Then after the
measurement, we can recover a state $\widetilde{\rho}$\ such that $\left\|
\widetilde{\rho}-\rho\right\|  _{\operatorname*{tr}}\leq\sqrt{\varepsilon}$.
\ This is true even if the measurement is a POVM (that is, involves
arbitrarily many ancilla qubits).
\end{lemma}

\begin{proof}
Let $\left|  \psi\right\rangle $\ be a purification of the entire system
($\rho$ plus ancilla). \ We can represent any measurement as a unitary $U$
applied to $\left|  \psi\right\rangle $, followed by a $1$-qubit measurement.
\ Let $\left|  \varphi_{0}\right\rangle $\ and $\left|  \varphi_{1}%
\right\rangle $\ be the two possible pure states after the measurement; then
$\left\langle \varphi_{0}|\varphi_{1}\right\rangle =0$\ and $U\left|
\psi\right\rangle =\alpha\left|  \varphi_{0}\right\rangle +\beta\left|
\varphi_{1}\right\rangle $ for some $\alpha,\beta$\ such that $\left|
\alpha\right|  ^{2}=1-\varepsilon$\ and $\left|  \beta\right|  ^{2}%
=\varepsilon$. \ Writing the measurement result as $\sigma=\left(
1-\varepsilon\right)  \left|  \varphi_{0}\right\rangle \left\langle
\varphi_{0}\right|  +\varepsilon\left|  \varphi_{1}\right\rangle \left\langle
\varphi_{1}\right|  $, it is easy to show that%
\[
\left\|  \sigma-U\left|  \psi\right\rangle \left\langle \psi\right|
U^{-1}\right\|  _{\operatorname*{tr}}=\sqrt{\varepsilon\left(  1-\varepsilon
\right)  }.
\]
So applying $U^{-1}$\ to $\sigma$,%
\[
\left\|  U^{-1}\sigma U-\left|  \psi\right\rangle \left\langle \psi\right|
\right\|  _{\operatorname*{tr}}=\sqrt{\varepsilon\left(  1-\varepsilon\right)
}.
\]
Let $\widetilde{\rho}$\ be the restriction of $U^{-1}\sigma U$\ to the
original qubits of $\rho$. \ Theorem 9.2 of Nielsen and Chuang \cite{nc} shows
that tracing out a subsystem never increases trace distance, so $\left\|
\widetilde{\rho}-\rho\right\|  _{\operatorname*{tr}}\leq\sqrt{\varepsilon
\left(  1-\varepsilon\right)  }\leq\sqrt{\varepsilon}$.
\end{proof}

\section{Simulating Quantum Messages\label{MESSAGES}}

Let $f:\left\{  0,1\right\}  ^{n}\times\left\{  0,1\right\}  ^{m}%
\rightarrow\left\{  0,1\right\}  $\ be a Boolean function. \ In this section I
first combine existing results to obtain the relation $\operatorname*{D}%
^{1}\left(  f\right)  =O\left(  mQ_{2}^{1}\left(  f\right)  \right)  $ for
total $f$, and then prove using a new method that $\operatorname*{D}%
^{1}\left(  f\right)  =O\left(  mQ_{2}^{1}\left(  f\right)  \log
\operatorname*{Q}_{2}^{1}\left(  f\right)  \right)  $\ for all $f$ (partial or total).

Define the \textit{communication matrix} $M_{f}$\ to be a $2^{n}\times2^{m}%
$\ matrix with $f\left(  x,y\right)  $\ in the $x^{th}$\ row and $y^{th}%
$\ column.\ \ Then letting $\operatorname*{rows}\left(  f\right)  $\ be the
number of distinct rows in $M_{f}$, the following is immediate.

\begin{proposition}
\label{nrows}For total $f$,%
\begin{align*}
\operatorname*{D}\nolimits^{1}\left(  f\right)   &  =\left\lceil \log
_{2}\operatorname*{rows}\left(  f\right)  \right\rceil ,\\
\operatorname*{Q}\nolimits_{2}^{1}\left(  f\right)   &  =\Omega\left(
\log\log\operatorname*{rows}\left(  f\right)  \right)  .
\end{align*}

\end{proposition}

Also, let the VC-dimension $\operatorname*{VC}\left(  f\right)  $\ equal the
maximum $k$ for which there exists a\ $2^{n}\times k$\ submatrix $M_{g}$\ of
$M_{f}$\ with $\operatorname*{rows}\left(  g\right)  =2^{k}$. \ Then Klauck
\cite{klauck:cc}\ observed the following, based on a lower bound for quantum
random access codes due to Nayak \cite{nayak}.

\begin{proposition}
[Klauck]\label{vc}$\operatorname*{Q}_{2}^{1}\left(  f\right)  =\Omega\left(
\operatorname*{VC}\left(  f\right)  \right)  $ for total $f$.
\end{proposition}

Now let $\operatorname*{cols}\left(  f\right)  $\ be the number of distinct
columns in $M_{f}$. \ Then Proposition \ref{vc}\ yields the following general
lower bound:

\begin{corollary}
\label{vccor}$\operatorname*{D}^{1}\left(  f\right)  =O\left(  mQ_{2}%
^{1}\left(  f\right)  \right)  $ for total $f$, where $m$ is the size of Bob's input.
\end{corollary}

\begin{proof}
It follows from a lemma of Sauer \cite{sauer}\ that%
\[
\operatorname*{rows}\left(  f\right)  \leq\sum_{i=0}^{\operatorname*{VC}%
\left(  f\right)  }\dbinom{\operatorname*{cols}\left(  f\right)  }{i}%
\leq\operatorname*{cols}\left(  f\right)  ^{\operatorname*{VC}\left(
f\right)  +1}.
\]
Hence $\operatorname*{VC}\left(  f\right)  \geq\log_{\operatorname*{cols}%
\left(  f\right)  }\operatorname*{rows}\left(  f\right)  -1$, so%
\begin{align*}
\operatorname*{Q}\nolimits_{2}^{1}\left(  f\right)  =\Omega\left(  \operatorname*{VC}%
\left(  f\right)  \right)   &  =\Omega\left(  \frac{\log\operatorname*{rows}%
\left(  f\right)  }{\log\operatorname*{cols}\left(  f\right)  }\right) \\
&  =\Omega\left(  \frac{\operatorname*{D}^{1}\left(  f\right)  }{m}\right)  .
\end{align*}

\end{proof}

In particular, $\operatorname*{D}^{1}\left(  f\right)  $\ and
$\operatorname*{Q}_{2}^{1}\left(  f\right)  $\ are polynomially related for
total $f$, whenever Bob's input is polynomially smaller than Alice's, and
Alice's input is not \textquotedblleft padded.\textquotedblright\ \ More
formally, $\operatorname*{D}^{1}\left(  f\right)  =O\left(  \operatorname*{Q}%
_{2}^{1}\left(  f\right)  ^{1/\left(  1-c\right)  }\right)  $ whenever
$m=O\left(  n^{c}\right)  $\ for some $c<1$ and $\operatorname*{rows}\left(
f\right)  =2^{n}$ (i.e. all rows of $M_{f}$\ are distinct). \ For then
$\operatorname*{D}^{1}\left(  f\right)  =n$ by Proposition \ref{nrows}, and
$\operatorname*{Q}_{2}^{1}\left(  f\right)  =\Omega\left(  \operatorname*{D}%
^{1}\left(  f\right)  /n^{c}\right)  =\Omega\left(  n^{1-c}\right)  $\ by
Corollary \ref{vccor}.

I now give a new method for replacing quantum messages by classical ones when
Bob's input is small. \ Although the best bound I know how to obtain with this
method---$\operatorname*{D}^{1}\left(  f\right)  =O\left(  mQ_{2}^{1}\left(
f\right)  \log\operatorname*{Q}_{2}^{1}\left(  f\right)  \right)  $---is
slightly weaker than the $\operatorname*{D}^{1}\left(  f\right)  =O\left(
mQ_{2}^{1}\left(  f\right)  \right)  $ of Corollary \ref{vccor}, our method
works for \textit{partial} Boolean functions as well as total ones.\ \ It also
yields a (relatively) efficient procedure by which Bob can reconstruct Alice's
quantum message, a fact I will exploit in Section \ref{ADVSIM}\ to show
$\mathsf{BQP/qpoly}\subseteq\mathsf{PP/poly}$. \ By contrast, the method based
on Sauer's Lemma seems to be nonconstructive.

\begin{theorem}
\label{partialthm}$\operatorname*{D}^{1}\left(  f\right)  =O\left(  mQ_{2}%
^{1}\left(  f\right)  \log\operatorname*{Q}_{2}^{1}\left(  f\right)  \right)
$ for all $f$ (partial or total).
\end{theorem}

\begin{proof}
Let $f:\mathcal{D}\rightarrow\left\{  0,1\right\}  $\ be a partial Boolean
function with $\mathcal{D}\subseteq\left\{  0,1\right\}  ^{n}\times\left\{
0,1\right\}  ^{m}$, and for all $x\in\left\{  0,1\right\}  ^{n}$, let
$\mathcal{D}_{x}=\left\{  y\in\left\{  0,1\right\}  ^{m}:\left(  x,y\right)
\in\mathcal{D}\right\}  $. \ Suppose Alice can send Bob a quantum state with
$\operatorname*{Q}_{2}^{1}\left(  f\right)  $\ qubits, that enables him to
compute $f\left(  x,y\right)  $\ for any $y\in\mathcal{D}_{x}$\ with error
probability at most $1/3$. \ Then she can also send him a boosted state $\rho$
with $K=O\left(  \operatorname*{Q}_{2}^{1}\left(  f\right)  \log
\operatorname*{Q}_{2}^{1}\left(  f\right)  \right)  $ qubits, such that for
all $y\in\mathcal{D}_{x}$,%
\[
\left\vert P_{y}\left(  \rho\right)  -f\left(  x,y\right)  \right\vert
\leq\frac{1}{\operatorname*{Q}_{2}^{1}\left(  f\right)  ^{10}},
\]
where $P_{y}\left(  \rho\right)  $\ is the probability that some measurement
$\Lambda\left[  y\right]  $\ yields a `$1$'\ outcome when applied to $\rho$.
\ We can assume for simplicity that $\rho$\ is a pure state $\left\vert
\psi\right\rangle \left\langle \psi\right\vert $; as discussed in Section
\ref{PRELIMADV}, this increases the message length by at most a factor of $2$.

Let $\mathcal{Y}$\ be any subset of $\mathcal{D}_{x}$\ satisfying $\left\vert
\mathcal{Y}\right\vert \leq\operatorname*{Q}_{2}^{1}\left(  f\right)  ^{2}$.
\ Then starting with $\rho$, Bob can measure $\Lambda\left[  y\right]  $\ for
each $y\in\mathcal{Y}$ in lexicographic order, reusing the same message
state\ again and again but uncomputing whatever garbage he generates while
measuring. \ Let $\rho_{t}$\ be the state after the $t^{th}$\ measurement;
thus $\rho_{0}=\rho=\left\vert \psi\right\rangle \left\langle \psi\right\vert
$. \ Since the probability that Bob outputs the wrong value of $f\left(
x,y\right)  $ on any given $y$ is at most $1/\operatorname*{Q}_{2}^{1}\left(
f\right)  ^{10}$,\ Lemma \ref{tracedist} implies that%
\[
\left\Vert \rho_{t}-\rho_{t-1}\right\Vert _{\operatorname*{tr}}\leq\sqrt
{\frac{1}{\operatorname*{Q}_{2}^{1}\left(  f\right)  ^{10}}}=\frac
{1}{\operatorname*{Q}_{2}^{1}\left(  f\right)  ^{5}}.
\]
Since trace distance satisfies the triangle inequality, this in turn implies
that%
\[
\left\Vert \rho_{t}-\rho\right\Vert _{\operatorname*{tr}}\leq\frac
{t}{\operatorname*{Q}_{2}^{1}\left(  f\right)  ^{5}}\leq\frac{1}%
{\operatorname*{Q}_{2}^{1}\left(  f\right)  ^{3}}.
\]
Now imagine an \textquotedblleft ideal scenario\textquotedblright\ in which
$\rho_{t}=\rho$ for every $t$; that is, the measurements do not damage $\rho
$\ at all. \ Then the maximum bias with which Bob could distinguish the actual
from the ideal scenario is%
\[
\left\Vert \rho_{0}\otimes\cdots\otimes\rho_{\left\vert \mathcal{Y}\right\vert
-1}-\rho^{\otimes\left\vert \mathcal{Y}\right\vert }\right\Vert
_{\operatorname*{tr}}\leq\frac{\left\vert \mathcal{Y}\right\vert
}{\operatorname*{Q}_{2}^{1}\left(  f\right)  ^{3}}\leq\frac{1}%
{\operatorname*{Q}_{2}^{1}\left(  f\right)  }.
\]
So by the union bound, Bob will output $f\left(  x,y\right)  $\ for every
$y\in\mathcal{Y}$\ simultaneously with probability at least%
\[
1-\frac{\left\vert \mathcal{Y}\right\vert }{\operatorname*{Q}_{2}^{1}\left(
f\right)  ^{10}}-\frac{1}{\operatorname*{Q}_{2}^{1}\left(  f\right)  }\geq0.9
\]
for sufficiently large $\operatorname*{Q}_{2}^{1}\left(  f\right)  $.

Now imagine that the communication channel is blocked, so Bob has to guess
what message Alice wants to send him. \ He does this by using the $K$-qubit
maximally mixed state $I$ in place of $\rho$.\ \ We can write $I$\ as%
\[
I=\frac{1}{2^{K}}\sum_{j=1}^{2^{K}}\left\vert \psi_{j}\right\rangle
\left\langle \psi_{j}\right\vert ,
\]
where $\left\vert \psi_{1}\right\rangle ,\ldots,\left\vert \psi_{2^{K}%
}\right\rangle $\ are orthonormal vectors such that $\left\vert \psi
_{1}\right\rangle =\left\vert \psi\right\rangle $.\ \ So if Bob uses the same
procedure as above except with $I$ instead of $\rho$, then for any
$\mathcal{Y}\subseteq\mathcal{D}_{x}$ with $\left\vert \mathcal{Y}\right\vert
\leq\operatorname*{Q}_{2}^{1}\left(  f\right)  ^{2}$, he will output $f\left(
x,y\right)  $ for every $y\in\mathcal{Y}$ simultaneously with probability at
least $0.9/2^{K}$.

The classical simulation of the quantum protocol is now as follows. \ Alice's
message to Bob consists of $T\leq K$\ inputs $y_{1},\ldots,y_{T}\in
\mathcal{D}_{x}$, together with $f\left(  x,y_{1}\right)  ,\ldots,f\left(
x,y_{T}\right)  $.\footnote{Strictly speaking, Bob will be able to compute
$f\left(  x,y_{1}\right)  ,\ldots,f\left(  x,y_{T}\right)  $\ for himself
given $y_{1},\ldots,y_{T}$; he does not need Alice to tell him the $f$
values.} \ Thus the message length is $mT+T=O\left(  mQ_{2}^{1}\left(
f\right)  \log\operatorname*{Q}_{2}^{1}\left(  f\right)  \right)  $. \ Here
are the semantics of Alice's message: \textit{\textquotedblleft Bob, suppose
you looped over all }$y\in\mathcal{D}_{x}$\textit{\ in lexicographic order;
and for each one, guessed that }$f\left(  x,y\right)  =\operatorname*{round}%
\left(  P_{y}\left(  I\right)  \right)  $\textit{, where }%
$\operatorname*{round}\left(  p\right)  $\textit{ is }$1$\textit{ if }%
$p\geq1/2$\textit{\ and }$0$\textit{\ if }$p<1/2$\textit{. \ Then }$y_{1}%
$\textit{\ is the first }$y$\textit{ for which you would guess the wrong value
of }$f\left(  x,y\right)  $\textit{. \ In general, let }$I_{t}$\textit{ be the
state obtained by starting from }$I$\textit{ and then measuring }%
$\Lambda\left[  y_{1}\right]  ,\ldots,\Lambda\left[  y_{t}\right]  $\textit{
in that order, given that the outcomes of the measurements are }$f\left(
x,y_{1}\right)  ,\ldots,f\left(  x,y_{t}\right)  $\textit{\ respectively.
\ (Note that }$I_{t}$\textit{\ is not changed by measurements of every }%
$y\in\mathcal{D}_{x}$\textit{\ up to }$y_{t}$\textit{, only by measurements of
}$y_{1},\ldots,y_{t}$\textit{.) \ If you looped over all }$y\in\mathcal{D}%
_{x}$\textit{\ in lexicographic order beginning from }$y_{t}$\textit{,\ then
}$y_{t+1}$\textit{ is the first }$y$\textit{ you would encounter for which
}$\operatorname*{round}\left(  P_{y}\left(  I_{t}\right)  \right)  \neq
f\left(  x,y\right)  $\textit{.\textquotedblright}

Given the sequence of $y_{t}$'s as defined above, it is obvious that Bob can
compute $f\left(  x,y\right)  $\ for any $y\in\mathcal{D}_{x}$. \ First, if
$y=y_{t}$\ for some $t$, then he simply outputs $f\left(  x,y_{t}\right)  $.
\ Otherwise, let $t^{\ast}$\ be the largest $t$ for which $y_{t}%
<y$\ lexicographically. \ Then Bob prepares a classical description of the
state $I_{t^{\ast}}$---which he can do since he knows $y_{1},\ldots
,y_{t^{\ast}}$\ and $f\left(  x,y_{1}\right)  ,\ldots,f\left(  x,y_{t^{\ast}%
}\right)  $---and then outputs $\operatorname*{round}\left(  P_{y}\left(
I_{t^{\ast}}\right)  \right)  $ as his claimed value of $f\left(  x,y\right)
$. \ Notice that, although Alice uses her knowledge of $\mathcal{D}_{x}$\ to
prepare her message, Bob does not need to know $\mathcal{D}_{x}$\ in order to
interpret the message. \ That is why the simulation works for partial as well
as total functions.

But why can we assume that the sequence of $y_{t}$'s stops at $y_{T}$\ for
some $T\leq K$? \ Suppose $T>K$; we will derive a contradiction. \ Let
$\mathcal{Y}=\left\{  y_{1},\ldots,y_{K+1}\right\}  $. \ Then $\left|
\mathcal{Y}\right|  =K+1\leq\operatorname*{Q}_{2}^{1}\left(  f\right)  ^{2}$,
so we know from previous reasoning that if Bob starts with $I$ and then
measures $\Lambda\left[  y_{1}\right]  ,\ldots,\Lambda\left[  y_{K+1}\right]
$\ in that order, he will observe $f\left(  x,y_{1}\right)  ,\ldots,f\left(
x,y_{K+1}\right)  $\ simultaneously with probability at least $0.9/2^{K}$.
\ But by the definition of $y_{t}$,\ the probability that $\Lambda\left[
y_{t}\right]  $\ yields the correct outcome is at most $1/2$,\ conditioned on
$\Lambda\left[  y_{1}\right]  ,\ldots,\Lambda\left[  y_{t-1}\right]  $\ having
yielded the correct outcomes. \ Therefore $f\left(  x,y_{1}\right)
,\ldots,f\left(  x,y_{K+1}\right)  $\ are observed simultaneously with
probability at most $1/2^{K+1}<0.9/2^{K}$, contradiction.
\end{proof}

\subsection{Simulating Quantum Advice\label{ADVSIM}}

I now apply the new simulation method to upper-bound the power of quantum advice.

\begin{theorem}
\label{qpoly}$\mathsf{BQP/qpoly}\subseteq\mathsf{PP/poly}$.
\end{theorem}

\begin{proof}
For notational convenience, let $L_{n}\left(  x\right)  =1$\ if input
$x\in\left\{  0,1\right\}  ^{n}$\ is in language $L$, and $L_{n}\left(
x\right)  =0$ otherwise. \ Suppose $L_{n}$\ is computed by a $\mathsf{BQP}$
machine using quantum advice of length $p\left(  n\right)  $. \ We will give a
$\mathsf{PP}$\ machine that computes $L_{n}$ using classical advice of length
$O\left(  np\left(  n\right)  \log p\left(  n\right)  \right)  $. \ Because of
the close connection between advice and one-way communication, the simulation
method will be essentially identical to that of Theorem \ref{partialthm}.

By using a boosted advice state on $K=O\left(  p\left(  n\right)  \log
p\left(  n\right)  \right)  $\ qubits,\ a polynomial-time quantum algorithm
$A$ can compute $L_{n}\left(  x\right)  $\ with error probability at most
$1/p\left(  n\right)  ^{10}$. \ Now the classical advice to the $\mathsf{PP}$
machine consists of $T\leq K$\ inputs $x_{1},\ldots,x_{T}\in\left\{
0,1\right\}  ^{n}$, together with $L_{n}\left(  x_{1}\right)  ,\ldots
,L_{n}\left(  x_{T}\right)  $. \ Let $I$ be the maximally mixed state on $K$
qubits. \ Also, let $P_{x}\left(  \rho\right)  $\ be the probability that $A$
outputs `$1$' on input $x$, given $\rho$\ as its advice state. \ Then $x_{1}%
$\ is the lexicographically first input $x$\ for which $\operatorname*{round}%
\left(  P_{x}\left(  I\right)  \right)  \neq L_{n}\left(  x\right)  $. \ In
general, let $I_{t}$\ be the state obtained by starting with $I$ as the advice
and then running $A$\ on $x_{1},\ldots,x_{t}$\ in that order (uncomputing
garbage along the way), if we postselect on $A$\ correctly outputting
$L_{n}\left(  x_{1}\right)  ,\ldots,L_{n}\left(  x_{t}\right)  $. \ Then
$x_{t+1}$\ is the lexicographically first $x>x_{t}$\ for which
$\operatorname*{round}\left(  P_{x}\left(  I_{t}\right)  \right)  \neq
L_{n}\left(  x\right)  $.

Given the classical advice, we can compute $L_{n}\left(  x\right)  $ as
follows: if $x\in\left\{  x_{1},\ldots,x_{T}\right\}  $\ then output
$L_{n}\left(  x_{t}\right)  $. \ Otherwise let $t^{\ast}$\ be the largest $t$
for which $x_{t}<x$\ lexicographically,\ and output $\operatorname*{round}%
\left(  P_{x}\left(  I_{t^{\ast}}\right)  \right)  $. \ The proof that this
algorithm works is the same as in Theorem \ref{partialthm}, and so is omitted
for brevity. \ All that needs to be shown is that the algorithm can be
implemented in $\mathsf{PP}$.

Adleman, DeMarrais, and Huang \cite{adh} (see also Fortnow and Rogers
\cite{fr}) showed that $\mathsf{BQP}\subseteq\mathsf{PP}$, by using what
physicists would call a \textquotedblleft Feynman
sum-over-histories.\textquotedblright\ \ Specifically, let $C$ be a
polynomial-size quantum circuit that starts in the all-$0$ state, and that
consists solely of Toffoli and Hadamard gates (Shi \cite{shi:gate}\ has shown
that this gate set is universal). \ Also, let $\alpha_{z}$\ be the amplitude
of basis state $\left\vert z\right\rangle $ after all gates in $C$ have been
applied. \ We can write $\alpha_{z}$ as a sum of exponentially many
contributions, $a_{1}+\cdots+a_{N}$, where each $a_{i}$\ is a rational real
number computable in classical polynomial time.\ \ So by evaluating the sum%
\[
\left\vert \alpha_{z}\right\vert ^{2}=\sum_{i,j=1}^{N}a_{i}a_{j},
\]
putting positive and negative terms on \textquotedblleft opposite sides of the
ledger,\textquotedblright\ a $\mathsf{PP}$\ machine can check whether
$\left\vert \alpha_{z}\right\vert ^{2}>\beta$ for any rational constant
$\beta$. \ It follows that a $\mathsf{PP}$\ machine can also check whether%
\[
\sum_{z~:~S_{1}\left(  z\right)  }\left\vert \alpha_{z}\right\vert ^{2}%
>\sum_{z~:~S_{0}\left(  z\right)  }\left\vert \alpha_{z}\right\vert ^{2}%
\]
(or equivalently, whether $\Pr\left[  S_{1}\right]  >\Pr\left[  S_{0}\right]
$) for any classical polynomial-time predicates $S_{1}$ and $S_{0}$.

Now suppose the circuit $C$ does the following, in the case $x\notin\left\{
x_{1},\ldots,x_{T}\right\}  $. \ It first prepares the $K$-qubit maximally
mixed state $I$ (as half of a $2K$-qubit pure state), and then runs $A$ on
$x_{1},\ldots,x_{t^{\ast}},x$\ in that order, using $I$\ as its advice state.
\ The claimed values of $L_{n}\left(  x_{1}\right)  ,\ldots,L_{n}\left(
x_{t^{\ast}}\right)  ,L_{n}\left(  x\right)  $\ are written to output
registers but not measured. \ For $i\in\left\{  0,1\right\}  $, let the
predicate $S_{i}\left(  z\right)  $\ hold if and only if basis state
$\left\vert z\right\rangle $\ contains the output sequence $L_{n}\left(
x_{1}\right)  ,\ldots,L_{n}\left(  x_{t^{\ast}}\right)  ,i$. \ Then it is not
hard to see that%
\[
P_{x}\left(  I_{t^{\ast}}\right)  =\frac{\Pr\left[  S_{1}\right]  }{\Pr\left[
S_{1}\right]  +\Pr\left[  S_{0}\right]  },
\]
so $P_{x}\left(  I_{t^{\ast}}\right)  >1/2$\ and hence $L_{n}\left(  x\right)
=1$\ if and only if $\Pr\left[  S_{1}\right]  >\Pr\left[  S_{0}\right]  $.
\ Since the case $x\in\left\{  x_{1},\ldots,x_{T}\right\}  $\ is trivial, this
shows that $L_{n}\left(  x\right)  $\ is computable in $\mathsf{PP/poly}$.
\end{proof}

Let me make five remarks about Theorem \ref{qpoly}. \ First, for the same
reason that Theorem \ref{partialthm}\ works for partial as well as total
functions, one actually obtains the stronger result that
$\mathsf{PromiseBQP/qpoly}\subseteq\mathsf{PromisePP/poly}$, where
$\mathsf{PromiseBQP}$\ and $\mathsf{PromisePP}$\ are the promise-problem
versions of $\mathsf{BQP}$\ and $\mathsf{PP}$\ respectively.

Second, as pointed out to me by Lance Fortnow, a corollary of Theorem
\ref{qpoly} is that we cannot hope to show an unrelativized separation between
$\mathsf{BQP/poly}$\ and $\mathsf{BQP/qpoly}$, without also showing that
$\mathsf{PP}$\ does not have polynomial-size circuits. \ For
$\mathsf{BQP/poly}\neq\mathsf{BQP/qpoly}$\ clearly implies
that\ $\mathsf{P/poly}\neq\mathsf{PP/poly}$. \ But the latter then implies
that $\mathsf{PP}\not \subset \mathsf{P/poly}$, since assuming $\mathsf{PP}%
\subset\mathsf{P/poly}$\ we could also obtain polynomial-size circuits for a
language $L\in\mathsf{PP/poly}$ by defining a new language $L^{\prime}%
\in\mathsf{PP}$, consisting of all $\left(  x,a\right)  $ pairs such that the
$\mathsf{PP}$ machine would accept $x$ given advice string $a$. \ The reason
this works is that $\mathsf{PP}$\ is a syntactically defined class.

Third, initially I showed that
$\mathsf{BQP/qpoly}\subseteq\mathsf{EXP/poly}$, by using a
simulation in which an $\mathsf{EXP}$\ machine keeps track of a
subspace $H$ of the advice Hilbert space to which the `true' advice
state must be close. \ In that simulation, the classical advice
specifies inputs $x_{1},\ldots,x_{T}$\ for which $\dim\left(
H\right)  $\ is at least halved; the observation that $\dim\left(
H\right)  $\ must be at least $1$ by the end then implies that
$T\leq K=O\left(  p\left(  n\right)  \log p\left(  n\right) \right)
$, meaning that the advice is of polynomial size. \ The huge
improvement from\ $\mathsf{EXP}$\ to $\mathsf{PP}$\ came solely from
working with \textit{measurement outcomes} and their
\textit{probabilities} instead of with \textit{subspaces} and their
\textit{dimensions}. \ We can compute the former using the same
\textquotedblleft Feynman sum-over-histories\textquotedblright\ that
Adleman et al.\ \cite{adh}\ used to show
$\mathsf{BQP}\subseteq\mathsf{PP}$, but I could not see any way to
compute the latter without explicitly storing and diagonalizing
exponentially large matrices.

Fourth, assuming $\mathsf{BQP/poly}\neq\mathsf{BQP/qpoly}$, Theorem
\ref{qpoly}\ is \textit{almost} the best result of its kind that one could
hope for, since the only classes known to lie between $\mathsf{BQP}$\ and
$\mathsf{PP}$\ and not known to equal either are obscure ones such as
$\mathsf{AWPP}$\ \cite{fr}. \ Initially the theorem\ seemed to me to prove
something stronger, namely that $\mathsf{BQP/qpoly}\subseteq
\mathsf{PostBQP/poly}$. \ Here $\mathsf{PostBQP}$\ is the class of languages
decidable by polynomial-size quantum circuits with \textit{postselection}%
---meaning the ability to measure a qubit that has a nonzero probability of
being $\left\vert 1\right\rangle $, and then \textit{assume} that the
measurement outcome will be $\left\vert 1\right\rangle $. \ Clearly
$\mathsf{PostBQP}$\ lies somewhere between $\mathsf{BQP}$\ and $\mathsf{PP}$;
one can think of it as a quantum analogue of the classical complexity class
$\mathsf{BPP}_{\mathsf{path}}$\ \cite{hht}. \ It turns out, however, that
$\mathsf{PostBQP}=\mathsf{PP}$ (see Chapter \ref{POST}).

Fifth, it is clear that Adleman et al.'s $\mathsf{BQP}\subseteq\mathsf{PP}%
$\ result \cite{adh}\ can be extended to show that $\mathsf{PQP}=\mathsf{PP}$.
\ Here $\mathsf{PQP}$\ is the quantum analogue of $\mathsf{PP}$---that is,
quantum polynomial time but where the probability of a correct answer need
only be bounded above $1/2$, rather than above $2/3$. \ It has been asked
whether Theorem \ref{qpoly} could similarly be extended to show that
$\mathsf{PQP/qpoly}=\mathsf{PP/poly}$. \ The answer is no---for indeed,
$\mathsf{PQP/qpoly}$\ contains every language whatsoever! \ To see this, given
any function $L_{n}:\left\{  0,1\right\}  ^{n}\rightarrow\left\{  0,1\right\}
$, let the quantum advice state be%
\[
\left\vert \psi_{n}\right\rangle =\frac{1}{2^{n/2}}\sum_{x\in\left\{
0,1\right\}  ^{n}}\left\vert x\right\rangle \left\vert L_{n}\left(  x\right)
\right\rangle .
\]
Then a $\mathsf{PQP}$\ algorithm to compute $L_{n}$ is as follows: given an
input $x\in\left\{  0,1\right\}  ^{n}$, first measure $\left\vert \psi
_{n}\right\rangle $ in the standard basis. \ If $\left\vert x\right\rangle
\left\vert L_{n}\left(  x\right)  \right\rangle $\ is observed, output
$L_{n}\left(  x\right)  $; otherwise output a uniform random bit.

\section{A Direct Product Theorem for Quantum Search\label{DPT}}

Can quantum computers solve $\mathsf{NP}$-complete problems\ in polynomial
time?\ \ In the early days of quantum computing, Bennett et al.\ \cite{bbbv}%
\ gave an oracle relative to which $\mathsf{NP}\not \subset \mathsf{BQP}$,
providing what is still the best evidence we have that the answer is no. \ It
is easy to extend Bennett et al.'s result to give an oracle relative to which
$\mathsf{NP}\not \subset \mathsf{BQP/poly}$; that is, $\mathsf{NP}$ is hard
even for nonuniform quantum algorithms. \ But when we try to show
$\mathsf{NP}\not \subset \mathsf{BQP/qpoly}$\ relative to an oracle, a new
difficulty arises: even if the oracle encodes $2^{n}$ exponentially hard
search problems for each input length $n$, the quantum advice, being an
\textquotedblleft exponentially large object\textquotedblright\ itself, might
somehow encode information about all $2^{n}$ problems. \ We need to argue that
even if so, only a miniscule fraction of that information can be extracted by
measuring the advice.

How does one prove such a statement? \ As it turns out, the task can be
reduced to proving a \textit{direct product theorem} for quantum search.
\ This is a theorem that in its weakest form says the following: given $N$
items, $K$ of which are marked, if we lack enough time to find even
\textit{one} marked item, then the probability of finding all $K$ items
decreases exponentially in $K$. \ For intuitively, suppose there were a
quantum advice state that let us efficiently find any one of $K$ marked items.
\ Then by \textquotedblleft guessing\textquotedblright\ the advice (i.e.
replacing it by a maximally mixed state), and then using the guessed advice
multiple times, we could efficiently find all $K$ of the items with a success
probability that our direct product theorem shows is impossible. \ This
reduction is formalized in Theorem \ref{npqpoly}.

But what about the direct product theorem itself? \ It seems like it
should be trivial to prove---for surely there are no devious
correlations by which success in finding one marked item leads to
success in finding all the others! \ So it is surprising that even a
weak direct product theorem eluded proof for years. \ In 2001,
Klauck \cite{klauck:ts}\ gave an attempted proof using the hybrid
method of Bennett et al.\ \cite{bbbv}. \ His motivation was to show
a limitation of space-bounded quantum sorting algorithms. \
Unfortunately, Klauck's proof is fallacious.\footnote{Specifically,
the last sentence in the proof of Lemma 5 in \cite{klauck:ts}\
(\textquotedblleft Clearly this probability is at least
$\operatorname*{Q}_{x}\left(  p_{x}-\alpha\right)
$\textquotedblright) is not justified by what precedes it.}

In this section I give the first correct proof of a direct product
theorem, based on the polynomial method of Beals et al.\
\cite{bbcmw}. \ Besides showing that $\mathsf{NP}\not \subset
\mathsf{BQP/qpoly}$\ relative to an oracle, my result can be used to
recover the conclusions in \cite{klauck:ts} about the hardness of
quantum sorting (see Klauck, \v{S}palek, and de Wolf \cite{ksw}\ for
details). \ I expect the result to have other applications as well.

I will need the following lemma of Beals et al.\ \cite{bbcmw}.

\begin{lemma}
[Beals et al.]\label{bbcmwlem}Suppose a quantum algorithm makes $T$ queries to
an oracle string $X\in\left\{  0,1\right\}  ^{N}$, and accepts with
probability $A\left(  X\right)  $. \ Then there exists a real polynomial $p$,
of degree at most $2T$, such that%
\[
p\left(  i\right)  =\operatorname*{EX}_{\left\vert X\right\vert =i}\left[
A\left(  X\right)  \right]
\]
for all integers $i\in\left\{  0,\ldots,N\right\}  $, where $\left\vert
X\right\vert $\ denotes the Hamming weight of $X$.
\end{lemma}

Lemma \ref{bbcmwlem}\ implies that, to lower-bound the number of queries $T$
made by a quantum algorithm, it suffices to lower-bound $\deg\left(  p\right)
$, where $p$ is a real polynomial representing the algorithm's expected
acceptance probability. \ As an example, any quantum algorithm that computes
the $\operatorname*{OR}$\ function on $N$ bits, with success probability at
least $2/3$, yields a polynomial $p$\ such that $p\left(  0\right)  \in\left[
0,1/3\right]  $\ and $p\left(  i\right)  \in\left[  2/3,1\right]  $\ for all
integers $i\in\left\{  1,\ldots,N\right\}  $. \ To lower-bound the degree of
such a polynomial, one can use an inequality proved by A. A. Markov in 1890
(\cite{aamarkov}; see also \cite{rivlin}):

\begin{theorem}
[A. A. Markov]\label{aamarkovthm}Given a real polynomial $p$ and constant
$N>0$, let $r^{\left(  0\right)  }=\max_{x\in\left[  0,N\right]  }\left\vert
p\left(  x\right)  \right\vert $\ and $r^{\left(  1\right)  }=\max
_{x\in\left[  0,N\right]  }\left\vert p^{\prime}\left(  x\right)  \right\vert
$. \ Then%
\[
\deg\left(  p\right)  \geq\sqrt{\frac{Nr^{\left(  1\right)  }}{2r^{\left(
0\right)  }}}.
\]

\end{theorem}

Theorem \ref{aamarkovthm} deals with the entire range $\left[  0,N\right]  $,
whereas in our setting $p\left(  x\right)  $\ is constrained only at the
integer points $x\in\left\{  0,\ldots,N\right\}  $. \ But as shown in
\cite{ez,ns,rc}, this is not a problem. \ For by elementary calculus,
$p\left(  0\right)  \leq1/3$\ and $p\left(  1\right)  \geq2/3$\ imply that
$p^{\prime}\left(  x\right)  \geq1/3$\ for some real $x\in\left[  0,1\right]
$, and therefore $r^{\left(  1\right)  }\geq1/3$. \ Furthermore, let $x^{\ast
}$\ be a point in $\left[  0,N\right]  $\ where $\left\vert p\left(  x^{\ast
}\right)  \right\vert =r^{\left(  0\right)  }$. \ Then $p\left(  \left\lfloor
x^{\ast}\right\rfloor \right)  \in\left[  0,1\right]  $\ and $p\left(
\left\lceil x^{\ast}\right\rceil \right)  \in\left[  0,1\right]  $\ imply that
$r^{\left(  1\right)  }\geq2\left(  r^{\left(  0\right)  }-1\right)  $. \ Thus%
\[
\deg\left(  p\right)  \geq\sqrt{\frac{Nr^{\left(  1\right)  }}{2r^{\left(
0\right)  }}}\geq\sqrt{\frac{N\max\left\{  1/3,2\left(  r^{\left(  0\right)
}-1\right)  \right\}  }{2r^{\left(  0\right)  }}}=\Omega\left(  \sqrt
{N}\right)  .
\]
This is the proof of\ Beals et al.\ \cite{bbcmw}\ that quantum
search requires $\Omega\left(  \sqrt{N}\right)  $\ queries.

When proving a direct product theorem, one can no longer apply Theorem
\ref{aamarkovthm} so straightforwardly. \ The reason is that the success
probabilities in question are extremely small, and therefore the maximum
derivative $r^{\left(  1\right)  }$\ could also be extremely small.
\ Fortunately, though, one can still prove a good lower bound on the degree of
the relevant polynomial $p$. \ The key is to look not just at the first
derivative of $p$, but at higher derivatives.

To start, we need a lemma about the behavior of functions under repeated differentiation.

\begin{lemma}
\label{derivative}Let $f:\mathbb{R}\rightarrow\mathbb{R}$\ be an infinitely
differentiable function such that for some positive integer $K$, we have
$f\left(  i\right)  =0$ for all $i\in\left\{  0,\ldots,K-1\right\}  $ and
$f\left(  K\right)  =\delta>0$. \ Also, let $r^{\left(  m\right)  }=\max
_{x\in\left[  0,N\right]  }\left\vert f^{\left(  m\right)  }\left(  x\right)
\right\vert $, where $f^{\left(  m\right)  }\left(  x\right)  $\ is the
$m^{th}$\ derivative of $f$ evaluated at $x$ (thus $f^{\left(  0\right)  }%
=f$). \ Then $r^{\left(  m\right)  }\geq\delta/m!$\ for all $m\in\left\{
0,\ldots,K\right\}  $.
\end{lemma}

\begin{proof}
We claim, by induction on $m$, that there exist $K-m+1$\ points $0\leq
x_{0}^{\left(  m\right)  }<\cdots<x_{K-m}^{\left(  m\right)  }\leq K$ such
that $f^{\left(  m\right)  }\left(  x_{i}^{\left(  m\right)  }\right)  =0$ for
all $i\leq K-m-1$\ and $f^{\left(  m\right)  }\left(  x_{K-m}^{\left(
m\right)  }\right)  \geq\delta/m!$. \ If we define $x_{i}^{\left(  0\right)
}=i$, then the base case $m=0$\ is immediate from the conditions of the lemma.
\ Suppose the claim is true for $m$; then by elementary calculus, for all
$i\leq K-m-2$\ there exists a point $x_{i}^{\left(  m+1\right)  }\in\left(
x_{i}^{\left(  m\right)  },x_{i+1}^{\left(  m\right)  }\right)  $\ such that
$f^{\left(  m+1\right)  }\left(  x_{i}^{\left(  m+1\right)  }\right)  =0$.
\ Notice that $x_{i}^{\left(  m+1\right)  }\geq x_{i}^{\left(  m\right)  }%
\geq\cdots\geq x_{i}^{\left(  0\right)  }=i$. \ So there is also a point
$x_{K-m-1}^{\left(  m+1\right)  }\in\left(  x_{K-m-1}^{\left(  m\right)
},x_{K-m}^{\left(  m\right)  }\right)  $ such that\
\begin{align*}
f^{\left(  m+1\right)  }\left(  x_{K-m-1}^{\left(  m+1\right)  }\right)   &
\geq\frac{f^{\left(  m\right)  }\left(  x_{K-m}^{\left(  m\right)  }\right)
-f^{\left(  m\right)  }\left(  x_{K-m-1}^{\left(  m\right)  }\right)
}{x_{K-m}^{\left(  m\right)  }-x_{K-m-1}^{\left(  m\right)  }}\\
&  \geq\frac{\delta/m!-0}{K-\left(  K-m-1\right)  }\\
&  =\frac{\delta}{\left(  m+1\right)  !}.
\end{align*}

\end{proof}

With the help of Lemma \ref{derivative}, one can sometimes lower-bound the
degree of a real polynomial even its first derivative is small throughout the
region of interest.\ To do so, I will use the following generalization of A.
A. Markov's inequality (Theorem \ref{aamarkovthm}), which was\ proved by A. A.
Markov's younger brother V. A. Markov\ in 1892 (\cite{vamarkov}; see also
\cite{rivlin}).

\begin{theorem}
[V. A. Markov]\label{vamarkovthm}Given a real polynomial $p$ of degree $d$ and
positive real number $N$, let $r^{\left(  m\right)  }=\max_{x\in\left[
0,N\right]  }\left\vert p^{\left(  m\right)  }\left(  x\right)  \right\vert $.
\ Then for all $m\in\left\{  1,\ldots,d\right\}  $,%
\begin{align*}
r^{\left(  m\right)  }  &  \leq\left(  \frac{2r^{\left(  0\right)  }}%
{N}\right)  ^{m}T_{d}^{\left(  m\right)  }\left(  1\right) \\
&  \leq\left(  \frac{2r^{\left(  0\right)  }}{N}\right)  ^{m}\frac
{d^{2}\left(  d^{2}-1^{2}\right)  \left(
d^{2}-2^{2}\right)  \cdot\cdots\cdot\left(  d%
^{2}-\left(  m-1\right)  ^{2}\right)  }{1\cdot3\cdot5\cdot\cdots\cdot\left(
2m-1\right)  }.
\end{align*}
Here $T_{d}\left(  x\right)  =\cos\left(  d\arccos x\right)  $\ is
the $d^{th}$\ Chebyshev polynomial of the first kind.
\end{theorem}

As demonstrated below, combining Theorem \ref{vamarkovthm}\ with Lemma
\ref{derivative} yields a lower bound on $\deg\left(  p\right)  $.

\begin{lemma}
\label{degreelb}Let $p$\ be a real polynomial such that

\begin{enumerate}
\item[(i)] $p\left(  x\right)  \in\left[  0,1\right]  $ at all integer points
$x\in\left\{  0,\ldots,N\right\}  $, and

\item[(ii)] for some positive integer $K\leq N$ and real $\delta>0$, we have
$p\left(  K\right)  =\delta$\ and $p\left(  i\right)  =0$\ for all
$i\in\left\{  0,\ldots,K-1\right\}  $.
\end{enumerate}

Then $\deg\left(  p\right)  =\Omega\left(  \sqrt{N\delta^{1/K}}\right)  $.
\end{lemma}

\begin{proof}
Let $p^{\left(  m\right)  }$\ and $r^{\left(  m\right)  }$ be as in Theorem
\ref{vamarkovthm}.\ \ Then for all $m\in\left\{  1,\ldots,\deg\left(
p\right)  \right\}  $, Theorem \ref{vamarkovthm}\ yields%
\[
r^{\left(  m\right)  }\leq\left(  \frac{2r^{\left(  0\right)  }}{N}\right)
^{m}\frac{\deg\left(  p\right)  ^{2m}}{1\cdot3\cdot5\cdot\cdots\cdot\left(
2m-1\right)  }%
\]
Rearranging,%
\[
\deg\left(  p\right)  \geq\sqrt{\frac{N}{2r^{\left(  0\right)  }}\left(
1\cdot3\cdot5\cdot\cdots\cdot\left(  2m-1\right)  \cdot r^{\left(  m\right)
}\right)  ^{1/m}}%
\]
for all $m\geq1$\ (if $m>\deg\left(  p\right)  $\ then $r^{\left(  m\right)
}=0$\ so the bound is trivial).

There are now two cases. \ First suppose $r^{\left(  0\right)  }\geq
2$.\ \ Then as discussed previously, condition (i) implies that $r^{\left(
1\right)  }\geq2\left(  r^{\left(  0\right)  }-1\right)  $, and hence that%
\[
\deg\left(  p\right)  \geq\sqrt{\frac{Nr^{\left(  1\right)  }}{2r^{\left(
0\right)  }}}\geq\sqrt{\frac{N\left(  r^{\left(  0\right)  }-1\right)
}{r^{\left(  0\right)  }}}=\Omega\left(  \sqrt{N}\right)
\]
by Theorem \ref{aamarkovthm}. \ Next suppose $r^{\left(  0\right)  }<2$.
\ Then $r^{\left(  m\right)  }\geq\delta/m!$\ for all $m\leq K$\ by Lemma
\ref{derivative}. \ So setting $m=K$\ yields%
\[
\deg\left(  p\right)  \geq\sqrt{\frac{N}{4}\left(  1\cdot3\cdot5\cdot
\cdots\cdot\left(  2K-1\right)  \cdot\frac{\delta}{K!}\right)  ^{1/K}}%
=\Omega\left(  \sqrt{N\delta^{1/K}}\right)  .
\]
Either way we are done.
\end{proof}

Strictly speaking, one does not need the full strength of Theorem
\ref{vamarkovthm}\ to prove a lower bound on $\deg\left(  p\right)  $ that
suffices for an oracle separation between $\mathsf{NP}$\ and
$\mathsf{BQP/qpoly}$. \ For one can show a \textquotedblleft
rough-and-ready\textquotedblright\ version of V. A. Markov's inequality by
applying A. A. Markov's inequality (Theorem \ref{aamarkovthm}) repeatedly, to
$p,p^{\left(  1\right)  },p^{\left(  2\right)  },$ and so on. \ This yields%
\[
r^{\left(  m\right)  }\leq\frac{2}{N}\deg\left(  p\right)  ^{2}r^{\left(
m-1\right)  }\leq\left(  \frac{2}{N}\deg\left(  p\right)  ^{2}\right)
^{m}r^{\left(  0\right)  }%
\]
for all $m$. \ If $\deg\left(  p\right)  $\ is small, then this upper bound on
$r^{\left(  m\right)  }$\ contradicts the lower bound of Lemma
\ref{derivative}. \ However, the lower bound on $\deg\left(  p\right)  $\ that
one gets from A. A. Markov's inequality is only $\Omega\left(  \sqrt
{N\delta^{1/K}/K}\right)  $, as opposed to $\Omega\left(  \sqrt{N\delta^{1/K}%
}\right)  $\ from Lemma \ref{degreelb}.\footnote{An earlier version of this
chapter claimed to prove $\deg\left(  p\right)  =\Omega\left(  \sqrt{NK}%
/\log^{3/2}\left(  1/\delta\right)  \right)  $, by applying
\textit{Bernstein's inequality} \cite{bernstein}\ rather than A. A.
Markov's to all derivatives $p^{\left(  m\right)  }$. \ I have since
discovered a flaw in that argument. \ In any case, the Bernstein\
lower bound is both unnecessary for an oracle separation, and
superseded by the later results of Klauck et al.\ \cite{ksw}.}

Shortly after seeing my proof of a weak direct product theorem, Klauck,
\v{S}palek, and de Wolf \cite{ksw}\ managed to improve the lower bound on
$\deg\left(  p\right)  $\ to the essentially tight $\Omega\left(
\sqrt{NK\delta^{1/K}}\right)  $. \ In particular, their bound implies that
$\delta$\ decreases exponentially in $K$ whenever $\deg\left(  p\right)
=o\left(  \sqrt{NK}\right)  $. \ They obtained this improvement by
\textit{factoring} $p$ instead of differentiating it as in Lemma
\ref{derivative}.

In any case, a direct product theorem follows trivially from what has already
been said.

\begin{theorem}
[Direct Product Theorem]\label{directprod}Suppose a quantum algorithm makes
$T$ queries to an oracle string $X\in\left\{  0,1\right\}  ^{N}$. \ Let
$\delta$\ be the minimum probability, over all $X$\ with Hamming weight
$\left\vert X\right\vert =K$, that the algorithm finds all $K$\ of the
`$1$'\ bits. \ Then $\delta\leq\left(  cT^{2}/N\right)  ^{K}$ for some
constant $c$.
\end{theorem}

\begin{proof}
Have the algorithm accept if it finds $K$ or more `$1$'\ bits and reject
otherwise. \ Let $p\left(  i\right)  $\ be the\ expected probability of
acceptance if $X$\ is drawn uniformly at random subject to $\left\vert
X\right\vert =i$. \ Then we know the following about $p$:

\begin{enumerate}
\item[(i)] $p\left(  i\right)  \in\left[  0,1\right]  $ at all integer points
$i\in\left\{  0,\ldots,N\right\}  $, since $p\left(  i\right)  $\ is a probability.

\item[(ii)] $p\left(  i\right)  =0$ for all $i\in\left\{  0,\ldots
,K-1\right\}  $, since there are not $K$ marked items to be found.

\item[(iii)] $p\left(  K\right)  \geq\delta$.
\end{enumerate}

Furthermore, Lemma \ref{bbcmwlem} implies that $p$\ is a polynomial in $i$
satisfying $\deg\left(  p\right)  \leq2T$. \ It follows from Lemma
\ref{degreelb} that $T=\Omega\left(  \sqrt{N\delta^{1/K}}\right)  $, or
rearranging, that $\delta\leq\left(  cT^{2}/N\right)  ^{K}$.
\end{proof}

The desired oracle separation can now be proven using standard complexity
theory tricks.

\begin{theorem}
\label{npqpoly}There exists an oracle relative to which $\mathsf{NP}%
\not \subset \mathsf{BQP/qpoly}$.
\end{theorem}

\begin{proof}
Given an oracle $A:\left\{  0,1\right\}  ^{\ast}\rightarrow\left\{
0,1\right\}  $, define the language $L_{A}$\ by $\left(  y,z\right)  \in
L_{A}$\ if and only if $y\leq z$\ lexicographically and there exists an $x$
such that $y\leq x\leq z$\ and $A\left(  x\right)  =1$. \ Clearly $L_{A}%
\in\mathsf{NP}^{A}$\ for all $A$. \ We argue that for some $A$, no
$\mathsf{BQP/qpoly}$\ machine $M$ with oracle access to $A$ can decide $L_{A}%
$. \ Without loss of generality we assume $M$ is fixed, so that only the
advice states $\left\{  \left\vert \psi_{n}\right\rangle \right\}  _{n\geq1}%
$\ depend on $A$. \ We also assume the advice is boosted, so that $M$'s error
probability on any input $\left(  y,z\right)  $ is $2^{-\Omega\left(
n^{2}\right)  }$.

Choose a set $S\subset\left\{  0,1\right\}  ^{n}$ subject to $\left\vert
S\right\vert =2^{n/10}$; then for all $x\in\left\{  0,1\right\}  ^{n}$, set
$A\left(  x\right)  =1$\ if and only if $x\in S$. \ We claim that by using
$M$, an algorithm could find all $2^{n/10}$\ elements of $S$\ with high
probability after only $2^{n/10}\operatorname*{poly}\left(  n\right)
$\ queries to $A$. \ Here is how: first use binary search (repeatedly halving
the distance between $y$\ and $z$) to find the lexicographically first element
of $S$.\ \ By Lemma \ref{tracedist}, the boosted advice state $\left\vert
\psi_{n}\right\rangle $\ is good for $2^{\Omega\left(  n^{2}\right)  }$\ uses,
so this takes only $\operatorname*{poly}\left(  n\right)  $\ queries. \ Then
use binary search to find the lexicographically second element, and so on
until all\ elements have been found.

Now replace $\left\vert \psi_{n}\right\rangle $\ by the maximally mixed state
as in Theorem \ref{partialthm}. This yields an algorithm that uses no advice,
makes $2^{n/10}\operatorname*{poly}\left(  n\right)  $ queries, and finds all
$2^{n/10}$\ elements of\ $S$\ with probability $2^{-O\left(
\operatorname*{poly}\left(  n\right)  \right)  }$. \ But taking $\delta
=2^{-O\left(  \operatorname*{poly}\left(  n\right)  \right)  }$,
$T=2^{n/10}\operatorname*{poly}\left(  n\right)  $, $N=2^{n}$, and
$K=2^{n/10}$, such an algorithm would satisfy $\delta\gg\left(  cT^{2}%
/N\right)  ^{K}$, which violates the bound of Theorem \ref{directprod}.
\end{proof}

Indeed one can show that $\mathsf{NP}\not \subset \mathsf{BQP/qpoly}%
$\ relative a random oracle with probability $1$.\footnote{First group the
oracle bits into polynomial-size blocks as Bennett and Gill \cite{bg} do, then
use the techniques of Chapter \ref{COL} to show that the acceptance
probability is a low-degree univariate polynomial in the number of all-$0$
blocks. \ The rest of the proof follows Theorem \ref{npqpoly}.}

\section{The Trace Distance Method\label{TRACEDIST}}

This section introduces a new method for proving lower bounds on quantum
one-way communication complexity. \ Unlike in Section \ref{MESSAGES}, here I
do not try to simulate quantum protocols using classical ones. \ Instead I
prove lower bounds for quantum protocols directly, by reasoning about the
trace distance between two possible distributions over Alice's quantum message
(that is, between two mixed states). \ The result is a method that works even
if Alice's and Bob's inputs are the same size.

I first state the method as a general theorem; then, in Section
\ref{APPLTRACE}, I apply the theorem to prove lower bounds for two problems of
Ambainis. \ Let $\left\Vert \mathcal{D}-\mathcal{E}\right\Vert $\ denote the
variation distance between probability distributions $\mathcal{D}$ and
$\mathcal{E}$.

\begin{theorem}
\label{vardist}Let $f:\left\{  0,1\right\}  ^{n}\times\left\{  0,1\right\}
^{m}\rightarrow\left\{  0,1\right\}  $\ be a total Boolean function. \ For
each $y\in\left\{  0,1\right\}  ^{m}$, let $\mathcal{A}_{y}$\ be a
distribution over $x\in\left\{  0,1\right\}  ^{n}$ such that $f\left(
x,y\right)  =1$. \ Let $\mathcal{B}$\ be a distribution over $y\in\left\{
0,1\right\}  ^{m}$, and let $\mathcal{D}_{k}$ be the distribution over
$\left(  \left\{  0,1\right\}  ^{n}\right)  ^{k}$\ formed by first choosing
$y\in\mathcal{B}$\ and then choosing $k$ samples independently from
$\mathcal{A}_{y}$. \ Suppose that $\Pr_{x\in\mathcal{D}_{1},y\in\mathcal{B}%
}\left[  f\left(  x,y\right)  =0\right]  =\Omega\left(  1\right)  $\ and that
$\left\Vert \mathcal{D}_{2}-\mathcal{D}_{1}^{2}\right\Vert \leq\delta.$ \ Then
$\operatorname*{Q}_{2}^{1}\left(  f\right)  =\Omega\left(  \log1/\delta
\right)  $.
\end{theorem}

\begin{proof}
Suppose that if Alice's input is $x$, then she sends Bob the $\ell$-qubit
mixed state $\rho_{x}$. \ Suppose also that for every $x\in\left\{
0,1\right\}  ^{n}$ and $y\in\left\{  0,1\right\}  ^{m}$, Bob outputs $f\left(
x,y\right)  $\ with probability at least $2/3$. \ Then by amplifying a
constant number of times, Bob's success probability can be made $1-\varepsilon
$\ for any constant $\varepsilon>0$. \ So with $L=O\left(  \ell\right)
$\ qubits of communication, Bob can distinguish the following two cases with
constant bias:

\textbf{Case I.} $\ y$ was drawn from $\mathcal{B}$\ and $x$ from
$\mathcal{D}_{1}$.

\textbf{Case II.} $y$ was drawn from $\mathcal{B}$\ and $x$ from
$\mathcal{A}_{y}$.

For in Case I, we assumed that $f\left(  x,y\right)  =0$\ with constant
probability, whereas in Case II, $f\left(  x,y\right)  =1$\ always. \ An
equivalent way to say this is that with constant probability over $y$, Bob can
distinguish the mixed states $\rho=\operatorname*{EX}_{x\in\mathcal{D}_{1}%
}\left[  \rho_{x}\right]  $\ and $\rho_{y}=\operatorname*{EX}_{x\in
\mathcal{A}_{y}}\left[  \rho_{x}\right]  $\ with constant bias. \ Therefore%
\[
\operatorname*{EX}_{y\in\mathcal{B}}\left[  \left\Vert \rho-\rho
_{y}\right\Vert _{\operatorname*{tr}}\right]  =\Omega\left(  1\right)  .
\]

We need an upper bound on the trace distance $\left\Vert \rho-\rho
_{y}\right\Vert _{\operatorname*{tr}}$\ that is more amenable to analysis.
\ Let $\lambda_{1},\ldots,\lambda_{2^{L}}$\ be the eigenvalues of $\rho
-\rho_{y}$. \ Then%
\begin{align*}
\left\Vert \rho-\rho_{y}\right\Vert _{\operatorname*{tr}}  &  =\frac{1}{2}%
\sum_{i=1}^{2^{L}}\left\vert \lambda_{i}\right\vert \\
&  \leq\frac{1}{2}\sqrt{2^{L}\sum_{i=1}^{2^{L}}\lambda_{i}^{2}}\\
&  =2^{L/2-1}\sqrt{\sum_{i,j=1}^{2^{L}}\left\vert \left(  \rho\right)
_{ij}-\left(  \rho_{y}\right)  _{ij}\right\vert ^{2}}%
\end{align*}
where $\left(  \rho\right)  _{ij}$\ is the $\left(  i,j\right)  $\ entry of
$\rho$. \ Here the second line uses the Cauchy-Schwarz inequality, and the
third line uses the unitary invariance of the Frobenius norm.

We claim that%
\[
\operatorname*{EX}_{y\in\mathcal{B}}\left[  \sum_{i,j=1}^{2^{L}}\left\vert
\left(  \rho\right)  _{ij}-\left(  \rho_{y}\right)  _{ij}\right\vert
^{2}\right]  \leq2\delta.
\]
From this claim it follows that%
\begin{align*}
\operatorname*{EX}_{y\in\mathcal{B}}\left[  \left\Vert \rho-\rho
_{y}\right\Vert _{\operatorname*{tr}}\right]   &  \leq2^{L/2-1}%
\operatorname*{EX}_{y\in\mathcal{B}}\left[  \sqrt{\sum_{i,j=1}^{2^{L}%
}\left\vert \left(  \rho\right)  _{ij}-\left(  \rho_{y}\right)  _{ij}%
\right\vert ^{2}}\right] \\
&  \leq2^{L/2-1}\sqrt{\operatorname*{EX}_{y\in\mathcal{B}}\left[  \sum
_{i,j=1}^{2^{L}}\left\vert \left(  \rho\right)  _{ij}-\left(  \rho_{y}\right)
_{ij}\right\vert ^{2}\right]  }\\
&  \leq\sqrt{2^{L-1}\delta}.
\end{align*}
Therefore the message length $L$ must be $\Omega\left(  \log1/\delta\right)
$\ to ensure that $\operatorname*{EX}_{y\in\mathcal{B}}\left[  \left\Vert
\rho-\rho_{y}\right\Vert _{\operatorname*{tr}}\right]  =\Omega\left(
1\right)  $.

Let us now prove the claim. \ We have%
\begin{align*}
\operatorname*{EX}_{y\in\mathcal{B}}\left[  \sum_{i,j=1}^{2^{L}}\left\vert
\left(  \rho\right)  _{ij}-\left(  \rho_{y}\right)  _{ij}\right\vert
^{2}\right]   &  =\sum_{i,j=1}^{2^{L}}\left(  \left\vert \left(  \rho\right)
_{ij}\right\vert ^{2}-2\operatorname{Re}\left(  \left(  \rho\right)
_{ij}^{\ast}\operatorname*{EX}_{y\in\mathcal{B}}\left[  \left(  \rho
_{y}\right)  _{ij}\right]  \right)  +\operatorname*{EX}_{y\in\mathcal{B}%
}\left[  \left\vert \left(  \rho_{y}\right)  _{ij}\right\vert ^{2}\right]
\right) \\
&  =\sum_{i,j=1}^{2^{L}}\left(  \operatorname*{EX}_{y\in\mathcal{B}}\left[
\left\vert \left(  \rho_{y}\right)  _{ij}\right\vert ^{2}\right]  -\left\vert
\left(  \rho\right)  _{ij}\right\vert ^{2}\right)  ,
\end{align*}
since $\operatorname*{EX}_{y\in\mathcal{B}}\left[  \left(  \rho_{y}\right)
_{ij}\right]  =\left(  \rho\right)  _{ij}$. \ For a given $\left(  i,j\right)
$\ pair,%
\begin{align*}
\operatorname*{EX}_{y\in\mathcal{B}}\left[  \left\vert \left(  \rho
_{y}\right)  _{ij}\right\vert ^{2}\right]  -\left\vert \left(  \rho\right)
_{ij}\right\vert ^{2}  &  =\operatorname*{EX}_{y\in\mathcal{B}}\left[
\left\vert \operatorname*{EX}_{x\in\mathcal{A}_{y}}\left[  \left(  \rho
_{x}\right)  _{ij}\right]  \right\vert ^{2}\right]  -\left\vert
\operatorname*{EX}_{x\in\mathcal{D}_{1}}\left[  \left(  \rho_{x}\right)
_{ij}\right]  \right\vert ^{2}\\
&  =\operatorname*{EX}_{y\in\mathcal{B},x,z\in\mathcal{A}_{y}}\left[  \left(
\rho_{x}\right)  _{ij}^{\ast}\left(  \rho_{z}\right)  _{ij}\right]
-\operatorname*{EX}_{x,z\in\mathcal{D}_{1}}\left[  \left(  \rho_{x}\right)
_{ij}^{\ast}\left(  \rho_{z}\right)  _{ij}\right] \\
&  =\sum_{x,z}\left(  \Pr_{\mathcal{D}_{2}}\left[  x,z\right]  -\Pr
_{\mathcal{D}_{1}^{2}}\left[  x,z\right]  \right)  \left(  \rho_{x}\right)
_{ij}^{\ast}\left(  \rho_{z}\right)  _{ij}.
\end{align*}
Now for all $x,z$,%
\[
\left\vert \sum_{i,j=1}^{2^{L}}\left(  \rho_{x}\right)  _{ij}^{\ast}\left(
\rho_{z}\right)  _{ij}\right\vert \leq\sum_{i,j=1}^{2^{L}}\left\vert \left(
\rho_{x}\right)  _{ij}\right\vert ^{2}\leq1.
\]
Hence%
\begin{align*}
\sum_{x,z}\left(  \Pr_{\mathcal{D}_{2}}\left[  x,z\right]  -\Pr_{\mathcal{D}%
_{1}^{2}}\left[  x,z\right]  \right)  \sum_{i,j=1}^{2^{L}}\left(  \rho
_{x}\right)  _{ij}^{\ast}\left(  \rho_{z}\right)  _{ij}  &  \leq\sum
_{x,z}\left(  \Pr_{\mathcal{D}_{2}}\left[  x,z\right]  -\Pr_{\mathcal{D}%
_{1}^{2}}\left[  x,z\right]  \right) \\
&  =2\left\Vert \mathcal{D}_{2}-\mathcal{D}_{1}^{2}\right\Vert \\
&  \leq2\delta,
\end{align*}
and we are done.
\end{proof}

The difficulty in extending Theorem \ref{vardist}\ to partial functions is
that the distribution $\mathcal{D}_{1}$\ might not make sense, since it might
assign a nonzero probability to some $x$ for which\ $f\left(  x,y\right)
$\ is undefined.

\subsection{Applications\label{APPLTRACE}}

In this subsection I apply Theorem \ref{vardist}\ to prove lower bounds for
two problems of Ambainis. \ To facilitate further research and to investigate
the scope of our method, I state the problems in a more general way than
Ambainis did. \ Given a group $G$, the \textit{coset problem}
$\operatorname*{Coset}\left(  G\right)  $\ is defined as follows. \ Alice is
given a left coset $C$\ of a subgroup in $G$, and Bob is given an element
$y\in G$. \ Bob must output $1$ if $y\in C$\ and $0$ otherwise. \ By
restricting the group $G$, we obtain many interesting and natural problems.
\ For example, if $p$ is prime then $\operatorname*{Coset}\left(
\mathbb{Z}_{p}\right)  $\ is just the equality problem, so the protocol of
Rabin and Yao \cite{ry}\ yields $\operatorname*{Q}_{2}^{1}\left(
\operatorname*{Coset}\left(  \mathbb{Z}_{p}\right)  \right)  =\Theta\left(
\log\log p\right)  $.

\begin{theorem}
\label{coset}$\operatorname*{Q}_{2}^{1}\left(  \operatorname*{Coset}\left(
\mathbb{Z}_{p}^{2}\right)  \right)  =\Theta\left(  \log p\right)  $.
\end{theorem}

\begin{proof}
The upper bound is obvious. \ For the lower bound, it suffices to consider a
function $f_{p}$\ defined as follows. \ Alice is given $\left\langle
x,y\right\rangle \in\mathbb{F}_{p}^{2}$\ and Bob is given $\left\langle
a,b\right\rangle \in\mathbb{F}_{p}^{2}$; then%
\[
f_{p}\left(  x,y,a,b\right)  =\left\{
\begin{array}
[c]{ll}%
1 & \text{if }y\equiv ax+b\left(  \operatorname{mod}p\right) \\
0 & \text{otherwise.}%
\end{array}
\right.
\]
Let $\mathcal{B}$ be the uniform distribution over $\left\langle
a,b\right\rangle \in\mathbb{F}_{p}^{2}$, and let $\mathcal{A}_{a,b}$\ be the
uniform distribution over $\left\langle x,y\right\rangle $\ such that $y\equiv
ax+b\left(  \operatorname{mod}p\right)  $. \ Thus $\mathcal{D}_{1}$\ is the
uniform distribution over $\left\langle x,y\right\rangle \in\mathbb{F}_{p}%
^{2}$; note that%
\[
\Pr_{\left\langle x,y\right\rangle \in\mathcal{D}_{1},\left\langle
a,b\right\rangle \in\mathcal{B}}\left[  f_{p}\left(  x,y,a,b\right)
=0\right]  =1-\frac{1}{p}.
\]
But what about the distribution $\mathcal{D}_{2}$, which is formed by first
drawing $\left\langle a,b\right\rangle \in\mathcal{B}$, and then drawing
$\left\langle x,y\right\rangle $ and $\left\langle z,w\right\rangle
$\ independently from $\mathcal{A}_{a,b}$? \ Given a pair $\left\langle
x,y\right\rangle ,\left\langle z,w\right\rangle \in\mathbb{F}_{p}^{2}$, there
are three cases regarding the probability of its being drawn from
$\mathcal{D}_{2}$:

\begin{enumerate}
\item[(1)] $\left\langle x,y\right\rangle =\left\langle z,w\right\rangle $
($p^{2}$ pairs). \ In this case%
\begin{align*}
\Pr_{\mathcal{D}_{2}}\left[  \left\langle x,y\right\rangle ,\left\langle
z,w\right\rangle \right]   &  =\sum_{\left\langle a,b\right\rangle
\in\mathbb{F}_{p}^{2}}\Pr\left[  \left\langle a,b\right\rangle \right]
\Pr\left[  \left\langle x,y\right\rangle ,\left\langle z,w\right\rangle
~|~\left\langle a,b\right\rangle \right] \\
&  =p\left(  \frac{1}{p^{2}}\cdot\frac{1}{p^{2}}\right)  =\frac{1}{p^{3}}.
\end{align*}

\item[(2)] $x\neq z$ ($p^{4}-p^{3}$ pairs). \ In this case there exists a
unique $\left\langle a^{\ast},b^{\ast}\right\rangle $\ such that $y\equiv
a^{\ast}x+b^{\ast}\left(  \operatorname{mod}p\right)  $\ and $w\equiv a^{\ast
}z+b^{\ast}\left(  \operatorname{mod}p\right)  $, so%
\begin{align*}
\Pr_{\mathcal{D}_{2}}\left[  \left\langle x,y\right\rangle ,\left\langle
z,w\right\rangle \right]   &  =\Pr\left[  \left\langle a^{\ast},b^{\ast
}\right\rangle \right]  \Pr\left[  \left\langle x,y\right\rangle ,\left\langle
z,w\right\rangle ~|~\left\langle a^{\ast},b^{\ast}\right\rangle \right] \\
&  =\frac{1}{p^{2}}\cdot\frac{1}{p^{2}}=\frac{1}{p^{4}}.
\end{align*}

\item[(3)] $x=z$ but $y\neq w$ ($p^{3}-p^{2}$ pairs). \ In this case
$\Pr_{\mathcal{D}_{2}}\left[  \left\langle x,y\right\rangle ,\left\langle
z,w\right\rangle \right]  =0$.
\end{enumerate}

Putting it all together,%
\begin{align*}
\left\Vert \mathcal{D}_{2}-\mathcal{D}_{1}^{2}\right\Vert  &  =\frac{1}%
{2}\left(  p^{2}\left\vert \frac{1}{p^{3}}-\frac{1}{p^{4}}\right\vert +\left(
p^{4}-p^{3}\right)  \left\vert \frac{1}{p^{4}}-\frac{1}{p^{4}}\right\vert
+\left(  p^{3}-p^{2}\right)  \left\vert 0-\frac{1}{p^{4}}\right\vert \right)
\\
&  =\frac{1}{p}-\frac{1}{p^{2}}.
\end{align*}
So taking $\delta=1/p-1/p^{2}$, we have $\operatorname*{Q}_{2}^{1}\left(
\operatorname*{Coset}\left(  \mathbb{Z}_{p}^{2}\right)  \right)
=\Omega\left(  \log\left(  1/\delta\right)  \right)  =\Omega\left(  \log
p\right)  $ by Theorem \ref{vardist}.
\end{proof}

I now consider Ambainis' second problem. \ Given a group $G$ and nonempty set
$S\subset G$\ with $\left\vert S\right\vert \leq\left\vert G\right\vert /2$,
the \textit{subset problem} $\operatorname*{Subset}\left(  G,S\right)  $\ is
defined as follows. \ Alice is given $x\in G$\ and Bob is given $y\in
G$;\ then Bob must output $1$ if $xy\in S$\ and $0$ otherwise.

Let $\mathcal{M}$\ be the distribution over $st^{-1}\in G$\ formed by drawing
$s$ and $t$ uniformly and independently from $S$. \ Then let $\Delta
=\left\Vert \mathcal{M}-\mathcal{D}_{1}\right\Vert $, where $\mathcal{D}_{1}%
$\ is the uniform distribution over $G$.

\begin{proposition}
\label{subset}For all $G,S$ such that $\left\vert S\right\vert \leq\left\vert
G\right\vert /2$,%
\[
\operatorname*{Q}\nolimits_{2}^{1}\left(  \operatorname*{Subset}\left(
G,S\right)  \right)  =\Omega\left(  \log1/\Delta\right)  .
\]

\end{proposition}

\begin{proof}
Let $\mathcal{B}$ be the uniform distribution over $y\in G$, and let
$\mathcal{A}_{y}$\ be the uniform distribution over $x$\ such that $xy\in S$.
\ Thus $\mathcal{D}_{1}$\ is the uniform distribution over $x\in G$; note that%
\[
\Pr_{x\in\mathcal{D}_{1},y\in\mathcal{B}}\left[  xy\notin S\right]
=1-\frac{\left\vert S\right\vert }{\left\vert G\right\vert }\geq\frac{1}{2}.
\]
We have%
\begin{align*}
\left\Vert \mathcal{D}_{2}-\mathcal{D}_{1}^{2}\right\Vert  &  =\frac{1}{2}%
\sum_{x,z\in G}\left\vert \frac{\left\vert \left\{  y\in G,s,t\in
S:xy=s,zy=t\right\}  \right\vert }{\left\vert G\right\vert \left\vert
S\right\vert ^{2}}-\frac{1}{\left\vert G\right\vert ^{2}}\right\vert \\
&  =\frac{1}{2}\sum_{x,z\in G}\left\vert \frac{\left\vert \left\{  s,t\in
S:xz^{-1}=st^{-1}\right\}  \right\vert }{\left\vert S\right\vert ^{2}}%
-\frac{1}{\left\vert G\right\vert ^{2}}\right\vert \\
&  =\frac{1}{2}\sum_{x\in G}\left\vert \frac{\left\vert \left\{  s,t\in
S:x=st^{-1}\right\}  \right\vert }{\left\vert S\right\vert ^{2}}-\frac
{1}{\left\vert G\right\vert }\right\vert \\
&  =\frac{1}{2}\sum_{x\in G}\left\vert \Pr_{\mathcal{M}}\left[  x\right]
-\frac{1}{\left\vert G\right\vert }\right\vert \\
&  =\left\Vert \mathcal{M}-\mathcal{D}_{1}\right\Vert \\
&  =\Delta.
\end{align*}
Therefore $\log\left(  1/\delta\right)  =\Omega\left(  \log1/\Delta\right)  $.
\end{proof}

Having lower-bounded $\operatorname*{Q}_{2}^{1}\left(  \operatorname*{Subset}%
\left(  G,S\right)  \right)  $ in terms of $1/\Delta$,\ it remains only to
upper-bound the variation distance $\Delta$. \ The following proposition
implies that for all constants $\varepsilon>0$, if $S$ is chosen uniformly at
random subject to $\left\vert S\right\vert =\left\vert G\right\vert
^{1/2+\varepsilon}$, then $\operatorname*{Q}_{2}^{1}\left(
\operatorname*{Subset}\left(  G,S\right)  \right)  =\Omega\left(  \log\left(
\left\vert G\right\vert \right)  \right)  $\ with constant probability\ over
$S$.

\begin{theorem}
\label{randset}For all groups $G$ and integers $K\in\left\{  1,\ldots
,\left\vert G\right\vert \right\}  $, if $S\subset G$ is chosen uniformly at
random subject to $\left\vert S\right\vert =K$, then $\Delta=O\left(
\sqrt{\left\vert G\right\vert }/K\right)  $ with $\Omega\left(  1\right)
$\ probability over $S$.
\end{theorem}

\begin{proof}
We have%
\[
\Delta=\frac{1}{2}\sum_{x\in G}\left\vert \Pr_{\mathcal{M}}\left[  x\right]
-\frac{1}{\left\vert G\right\vert }\right\vert \leq\frac{\sqrt{\left\vert
G\right\vert }}{2}\sqrt{\sum_{x\in G}\left(  \Pr_{\mathcal{M}}\left[
x\right]  -\frac{1}{\left\vert G\right\vert }\right)  ^{2}}%
\]
by the Cauchy-Schwarz inequality. \ We claim that%
\[
\operatorname*{EX}_{S}\left[  \sum_{x\in G}\left(  \Pr_{\mathcal{M}}\left[
x\right]  -\frac{1}{\left\vert G\right\vert }\right)  ^{2}\right]  \leq
\frac{c}{K^{2}}%
\]
for some constant $c$. \ From this it follows by Markov's inequality that%
\[
\Pr_{S}\left[  \sum_{x\in G}\left(  \Pr_{\mathcal{M}}\left[  x\right]
-\frac{1}{\left\vert G\right\vert }\right)  ^{2}\geq\frac{2c}{K^{2}}\right]
\leq\frac{1}{2}%
\]
and hence%
\[
\Delta\leq\frac{\sqrt{\left\vert G\right\vert }}{2}\sqrt{\frac{2c}{K^{2}}%
}=O\left(  \frac{\sqrt{\left\vert G\right\vert }}{K}\right)
\]
with probability at least $1/2$.

Let us now prove the claim. \ We have%
\[
\Pr_{\mathcal{M}}\left[  x\right]  =\Pr_{i,j}\left[  s_{i}s_{j}^{-1}=x\right]
=\Pr_{i,j}\left[  s_{i}=xs_{j}\right]  ,
\]
where $S=\left\{  s_{1},\ldots,s_{K}\right\}  $\ and $i,j$\ are drawn
uniformly and independently from $\left\{  1,\ldots,K\right\}  $. \ So by
linearity of expectation,%
\begin{align*}
\operatorname*{EX}_{S}\left[  \sum_{x\in G}\left(  \Pr_{\mathcal{M}}\left[
x\right]  -\frac{1}{\left\vert G\right\vert }\right)  ^{2}\right]   &
=\operatorname*{EX}_{S}\left[  \sum_{x\in G}\left(  \left(  \Pr_{i,j}\left[
s_{i}=xs_{j}\right]  \right)  ^{2}-\frac{2}{\left\vert G\right\vert }\Pr
_{i,j}\left[  s_{i}=xs_{j}\right]  +\frac{1}{\left\vert G\right\vert ^{2}%
}\right)  \right] \\
&  =\sum_{x\in G}\left(  \frac{1}{K^{4}}\sum_{i,j,k,l=1}^{K}p_{x,ijkl}\right)
-\frac{2}{\left\vert G\right\vert }\sum_{x\in G}\left(  \frac{1}{K^{2}}%
\sum_{i,j=1}^{K}p_{x,ij}\right)  +\frac{1}{\left\vert G\right\vert }%
\end{align*}
where%
\begin{align*}
p_{x,ij}  &  =\Pr_{S}\left[  s_{i}=xs_{j}\right]  ,\\
p_{x,ijkl}  &  =\Pr_{S}\left[  s_{i}=xs_{j}\wedge s_{k}=xs_{l}\right]  .
\end{align*}

First we analyze $p_{x,ij}$. \ Let $\operatorname*{ord}\left(  x\right)  $\ be
the order of $x$ in $G$. \ Of the $K^{2}$\ possible ordered\ pairs $\left(
i,j\right)  $, there are $K$ pairs with the \textquotedblleft
pattern\textquotedblright\ $ii$\ (meaning that $i=j$), and $K\left(
K-1\right)  $ pairs with the pattern $ij$\ (meaning that $i\neq j$). \ If
$\operatorname*{ord}\left(  x\right)  =1$ (that is, $x$\ is the identity),
then we have $p_{x,ij}=\Pr_{S}\left[  s_{i}=s_{j}\right]  $, so $p_{x,ij}%
=1$\ under the pattern $ii$, and $p_{x,ij}=0$ under the pattern $ij$.\ \ On
the other hand, if $\operatorname*{ord}\left(  x\right)  >1$, then
$p_{x,ij}=0$\ under the pattern $ii$, and $p_{x,ij}=\frac{1}{\left\vert
G\right\vert -1}$\ under the pattern $ij$. \ So%
\[
\frac{1}{K^{2}}\sum_{x\in G}\sum_{i,j=1}^{K}p_{x,ij}=\frac{1}{K^{2}}\left(
K+\left(  \left\vert G\right\vert -1\right)  \frac{K\left(  K-1\right)
}{\left\vert G\right\vert -1}\right)  =1.
\]

Though unnecessarily cumbersome, the above analysis was a warmup for
the more complicated case of $p_{x,ijkl}$. \ Table 10.1 lists the
expressions for $p_{x,ijkl}$, given $\operatorname*{ord}\left(
x\right)  $\ and the pattern of $\left( i,j,k,l\right)  $.

\begin{table}[ptb]
\begin{tabular} [c]{|ll|lll|}\hline Pattern & Number of
such $4$-tuples & $\operatorname*{ord}\left(  x\right) =1$ &
$\operatorname*{ord}\left(  x\right)  =2$ &
$\operatorname*{ord}\left( x\right)  >2$\\\hline iiii,iikk &
\thinspace$K^{2}$ & \multicolumn{1}{|c}{$1$} &
\multicolumn{1}{c}{$0$} & \multicolumn{1}{c|}{$0$}\\
ijij & $K\left(  K-1\right)  $ & \multicolumn{1}{|c}{$0$} &
\multicolumn{1}{c}{$\frac{1}{\left\vert G\right\vert -1}$} &
\multicolumn{1}{c|}{$\frac{1}{\left\vert G\right\vert -1}$}\\
ijji & $K\left(  K-1\right)  $ & \multicolumn{1}{|c}{$0$} &
\multicolumn{1}{c}{$\frac{1}{\left\vert G\right\vert -1}$} &
\multicolumn{1}{c|}{$0$}\\
iiil,iiki,ijii,ijjj & $4K\left(  K-1\right)  $ &
\multicolumn{1}{|c}{$0$} &
\multicolumn{1}{c}{$0$} & \multicolumn{1}{c|}{$0$}\\
ijki,ijjk & $2K\left(  K-1\right)  \left(  K-2\right)  $ &
\multicolumn{1}{|c}{$0$} & \multicolumn{1}{c}{$0$} &
\multicolumn{1}{c|}{$\frac{1}{\left(  \left\vert G\right\vert
-1\right)
\left(  \left\vert G\right\vert -2\right)  }$}\\
iikl,ijkk,ijik,ijkj & $4K\left(  K-1\right)  \left(  K-2\right)  $ &
\multicolumn{1}{|c}{$0$} & \multicolumn{1}{c}{$0$} & \multicolumn{1}{c|}{$0$%
}\\
ijkl & $K\left(  K-1\right)  \left(  K-2\right)  \left(  K-3\right)
$ & \multicolumn{1}{|c}{$0$} & \multicolumn{1}{c}{$\frac{1}{\left(
\left\vert G\right\vert -1\right)  \left(  \left\vert G\right\vert
-3\right)  }$} & \multicolumn{1}{c|}{$\frac{1}{\left(  \left\vert
G\right\vert -1\right) \left(  \left\vert G\right\vert -3\right)
}$}\\\hline
\end{tabular}
\label{pijkl} \caption[Expressions for $p_{x,ijkl}$]{Expressions for
$p_{x,ijkl}$}
\end{table}

Let $r$\ be the number of $x\in G$\ such that $\operatorname*{ord}\left(
x\right)  =2$, and let\ $r^{\prime}=\left\vert G\right\vert -r-1$\ be the
number such that $\operatorname*{ord}\left(  x\right)  >2$. \ Then%
\begin{align*}
\frac{1}{K^{4}}\sum_{x\in G}\sum_{i,j,k,l=1}^{K}p_{x,ijkl}  &  =\frac{1}%
{K^{4}}\left(
\begin{array}
[c]{c}%
K^{2}+\left(  2r+r^{\prime}\right)  \frac{K\left(  K-1\right)  }{\left\vert
G\right\vert -1}+2r^{\prime}\frac{K\left(  K-1\right)  \left(  K-2\right)
}{\left(  \left\vert G\right\vert -1\right)  \left(  \left\vert G\right\vert
-2\right)  }\\
+\left(  r+r^{\prime}\right)  \frac{K\left(  K-1\right)  \left(  K-2\right)
\left(  K-3\right)  }{\left(  \left\vert G\right\vert -1\right)  \left(
\left\vert G\right\vert -3\right)  }%
\end{array}
\right) \\
&  \leq\frac{1}{\left\vert G\right\vert -3}+O\left(  \frac{1}{K^{2}}\right)
\end{align*}
using the fact that $K\leq\left\vert G\right\vert $.

Putting it all together,%
\[
\operatorname*{EX}_{S}\left[  \sum_{x\in G}\left(  \Pr_{\mathcal{M}}\left[
x\right]  -\frac{1}{\left\vert G\right\vert }\right)  ^{2}\right]  \leq
\frac{1}{\left\vert G\right\vert -3}+O\left(  \frac{1}{K^{2}}\right)
-\frac{2}{\left\vert G\right\vert }+\frac{1}{\left\vert G\right\vert
}=O\left(  \frac{1}{K^{2}}\right)
\]
and we are done.
\end{proof}

From fingerprinting one also has the following upper bound. \ Let $q$ be the
periodicity of $S$, defined as the number of distinct sets $gS=\left\{
gs:s\in S\right\}  $ where $g\in G$.

\begin{proposition}
\label{fingerprint}$\operatorname*{R}_{2}^{1}\left(  \operatorname*{Subset}%
\left(  G,S\right)  \right)  =O\left(  \log\left\vert S\right\vert +\log\log
q\right)  $.
\end{proposition}

\begin{proof}
Assume for simplicity that $q=\left\vert G\right\vert $; otherwise we could
reduce to a subgroup $H\leq G$\ with $\left\vert H\right\vert =q$. \ The
protocol is as follows: Alice draws a uniform random prime $p$ from the range
$\left[  \left\vert S\right\vert ^{2}\log^{2}\left\vert G\right\vert
,2\left\vert S\right\vert ^{2}\log^{2}\left\vert G\right\vert \right]  $; she
then sends Bob the pair\ $\left(  p,x\operatorname{mod}p\right)  $\ where $x$
is interpreted as an integer. \ This takes $O\left(  \log\left\vert
S\right\vert +\log\log\left\vert G\right\vert \right)  $ bits. \ Bob outputs
$1$ if and only if there exists a $z\in G$\ such that $zy\in S$\ and $x\equiv
z\left(  \operatorname{mod}p\right)  $. \ To see the protocol's correctness,
observe that if\ $x\neq z$, then there at most $\log\left\vert G\right\vert
$\ primes $p$ such that $x-z\equiv0\left(  \operatorname{mod}p\right)  $,
whereas the relevant range contains $\Omega\left(  \frac{\left\vert
S\right\vert ^{2}\log^{2}\left\vert G\right\vert }{\log\left(  \left\vert
S\right\vert \log\left\vert G\right\vert \right)  }\right)  $\ primes.
\ Therefore, if $xy\notin S$, then by the union bound%
\[
\Pr_{p}\left[  \exists z:zy\in S,x\equiv z\left(  \operatorname{mod}p\right)
\right]  =O\left(  \left\vert S\right\vert \log\left\vert G\right\vert
\frac{\log\left(  \left\vert S\right\vert \log\left\vert G\right\vert \right)
}{\left\vert S\right\vert ^{2}\log^{2}\left\vert G\right\vert }\right)
=o\left(  1\right)  .
\]

\end{proof}

\section{Open Problems\label{OPENADV}}

Are $\operatorname*{R}_{2}^{1}\left(  f\right)  $ and $\operatorname*{Q}%
_{2}^{1}\left(  f\right)  $\ polynomially related for every total Boolean
function $f$? \ Also, can we exhibit \textit{any} asymptotic separation
between these measures? \ The best separation I know of is a factor of $2$:
for the equality function we have $\operatorname*{R}_{2}^{1}\left(
\operatorname*{EQ}\right)  \geq\left(  1-o\left(  1\right)  \right)  \log
_{2}n$, whereas Winter \cite{winter}\ has shown that $\operatorname*{Q}%
_{2}^{1}\left(  \operatorname*{EQ}\right)  \leq\left(  1/2+o\left(  1\right)
\right)  \log_{2}n$\ using a protocol involving mixed states.\footnote{If we
restrict ourselves to pure states, then $\left(  1-o\left(  1\right)  \right)
\log_{2}n$\ qubits are needed. \ Based on that fact, a previous version of
this chapter claimed incorrectly that $\operatorname*{Q}_{2}^{1}\left(
\operatorname*{EQ}\right)  \geq\left(  1-o\left(  1\right)  \right)  \log
_{2}n$.} \ This factor-$2$ savings is tight for equality: a simple counting
argument shows that $\operatorname*{Q}_{2}^{1}\left(  \operatorname*{EQ}%
\right)  \geq\left(  1/2-o\left(  1\right)  \right)  \log_{2}n$; and although
the usual randomized protocol for equality\ \cite{ry}\ uses $\left(
2+o\left(  1\right)  \right)  \log_{2}n$\ bits, there exist\ protocols based
on error-correcting codes that use only $\log_{2}\left(  cn\right)  =\log
_{2}n+O\left(  1\right)  $\ bits. \ All of this holds for any constant error
probability $0<\varepsilon<1/2$.

Can we lower-bound $\operatorname*{Q}_{2}^{1}\left(
\operatorname*{Coset}\left(  G\right) \right)  $\ for groups other
than $\mathbb{Z}_{p}^{2}$\ (such as $\mathbb{Z}_{2}^{n}$, or
nonabelian groups)? \ Also, can we characterize
$\operatorname*{Q}_{2}^{1}\left(  \operatorname*{Subset}\left(
G,S\right) \right)  $\ for all sets $S$, closing the gap between the
upper and lower bounds?

Is there an oracle relative to which $\mathsf{BQP/poly}\neq
\mathsf{BQP/qpoly}$?

Can we give oracles relative to which $\mathsf{NP}\cap\mathsf{coNP}%
$\ and $\mathsf{SZK}$\ are not contained in $\mathsf{BQP/qpoly}$? \
 Even more ambitiously, can we prove a direct product theorem for
quantum query complexity that applies to any partial or total
function (not just search)?

For all $f$ (partial or total), is $\operatorname*{R}_{2}^{1}\left(
f\right)  =O\left(  \sqrt{n}\right)  $\ whenever $\operatorname*{Q}_{2}%
^{1}\left(  f\right)  =O\left(  \log n\right)  $? \ In other words,
is the separation of Bar-Yossef et al.\ \cite{bjk} the best
possible?

Can the result $\operatorname*{D}^{1}\left(  f\right)  =O\left(
mQ_{2}^{1}\left(  f\right)  \log\operatorname*{Q}_{2}^{1}\left(
f\right) \right)  $\ for partial $f$ be improved to
$\operatorname*{D}^{1}\left( f\right)  =O\left(  mQ_{2}^{1}\left(
f\right)  \right)  $? \ I do not even know how to rule out
$\operatorname*{D}^{1}\left(  f\right)  =O\left(
m+\operatorname*{Q}_{2}^{1}\left(  f\right)  \right)  $.

In the Simultaneous Messages (SM) model, there is no direct
communication between Alice and Bob;\ instead, Alice and Bob both
send messages to a third party called the \textit{referee}, who then
outputs the function value. \ The complexity measure is the sum of
the two message lengths. \ Let $\operatorname*{R}_{2}^{||}\left(
f\right)  $\ and $\operatorname*{Q}_{2}^{||}\left(  f\right)  $ be
the randomized and quantum
bounded-error SM complexities of $f$ respectively, and let $\operatorname*{R}%
_{2}^{||,\operatorname*{pub}}\left(  f\right)  $\ be the randomized
SM complexity if Alice and Bob share an arbitrarily long random
string. \ Building on work by Buhrman et al.\ \cite{bcww}, Yao
\cite{yao:fing} showed that $\operatorname*{Q}_{2}^{||}\left(
f\right)  =O\left(  \log n\right) $\ whenever
$\operatorname*{R}_{2}^{||,\operatorname*{pub}}\left(  f\right)
=O\left(  1\right)  $. \ He then asked about the other direction:
for some $\varepsilon>0$, does
$\operatorname*{R}_{2}^{||,\operatorname*{pub}}\left(
f\right)  =O\left(  n^{1/2-\varepsilon}\right)  $\ whenever $\operatorname*{Q}%
_{2}^{||}\left(  f\right)  =O\left(  \log n\right)  $, and does
$\operatorname*{R}_{2}^{||}\left(  f\right)  =O\left(
n^{1-\varepsilon }\right)  $ whenever
$\operatorname*{Q}_{2}^{||}\left(  f\right)  =O\left( \log n\right)
$? \ In an earlier version of this chapter, I showed that
$\operatorname*{R}_{2}^{||}\left(  f\right)  =O\left( \sqrt{n}\left(
\operatorname*{R}_{2}^{||,\operatorname*{pub}}\left( f\right)  +\log
n\right)  \right)  $, which means that a positive answer to Yao's
first question would imply a positive answer to the second. \ Later
I learned that Yao independently proved the same result
\cite{yao:hw}. \ Here I ask a related question: can
$\operatorname*{Q}_{2}^{||}\left(
f\right)  $\ ever be exponentially smaller than $\operatorname*{R}%
_{2}^{||,\operatorname*{pub}}\left(  f\right)  $? \ (Buhrman et al.\
\cite{bcww}\ showed that $\operatorname*{Q}_{2}^{||}\left(  f\right)
$\ can be exponentially smaller than
$\operatorname*{R}_{2}^{||}\left(  f\right)  $.) \ Iordanis
Kerenidis has pointed out to me that, based on the hidden matching
problem of Bar-Yossef et al.\ \cite{bjk}\ discussed in Section \ref{PRELIMADV}%
,\ one can define a \textit{relation} for which $\operatorname*{Q}_{2}%
^{||}\left(  f\right)  $\ is exponentially smaller than $\operatorname*{R}%
_{2}^{||,\operatorname*{pub}}\left(  f\right)  $. \ However, as in the case of
$\operatorname*{Q}_{2}^{1}\left(  f\right)  $\ versus $\operatorname*{R}%
_{2}^{1}\left(  f\right)  $, it remains to extend that result to functions.

\chapter{Summary of Part \ref{LQC}\label{SUMLQC}}

From my unbiased perspective, quantum lower bounds are some of the deepest
results to have emerged from the study of quantum computing and information.
\ These results tell us that many problems we thought were intractable based
on classical intuition, really \textit{are} intractable according to our best
theory of the physical world. \ On the other hand, the \textit{reasons} for
intractability are much more subtle than in the classical case. \ In some
sense, this has to be true---for otherwise the reasons would apply even to
those problems for which dramatic quantum speedups exist.

We currently have two methods for proving lower bounds on quantum
query complexity: the polynomial method of Beals et al.\
\cite{bbcmw}, and the adversary method of Ambainis \cite{ambainis}.
\ The preceding chapters have illustrated what, borrowing from
Wigner \cite{wigner}, we might call the \textquotedblleft
unreasonable effectiveness\textquotedblright\ of these methods. \
Both continue to work far outside of their original design
specs---whether by proving \textit{classical} lower bounds, lower
bounds for exponentially small success probabilities (as in the
direct product theorem), or polynomial lower bounds for quantities
that have \textquotedblleft no right\textquotedblright\ to be
polynomials (as in the collision and set comparison problems). \ Yet
the two methods also have complementary limitations. \ The adversary
method is useless when the relevant probability gaps are small, or
when every $0$-input differs from every $1$-input in a constant
fraction of locations. \ Likewise, the polynomial method cannot be
applied to problems that lack permutation symmetry, at least using
the techniques we currently know. \ Thus, perhaps the most important
open problem in quantum lower bounds is to develop a new method that
overcomes the limitations of both the polynomial and the adversary
methods.\footnote{Along these lines, Barnum, Saks, and Szegedy
\cite{bss} have given what in some sense is a provably optimal
method, but their method (based on semidefinite programming)\ seems
too difficult to apply directly.}

In keeping with the theme of this thesis, I end Part \ref{LQC}\ by listing
some classical intuitions about computing, that a hypothetical being from
Conway's Game of Life could safely carry into the quantum universe.

\begin{itemize}
\item The collision problem is not that much easier than unordered search. For
despite being extremely far from any one-to-one function, a random two-to-one
function still \textit{looks} one-to-one unless we do an expensive search for collisions.

\item Finding a local minimum of a function is not that much easier than
finding a global minimum. This is because the paths leading to local minima
could be exponentially long.

\item If we want to distinguish an input $X$ from the set of all $Y$ such that
$f\left(  Y\right)  \neq f\left(  X\right)  $, then there is nothing much
better to do than to query nonadaptively according to the minimax strategy.

\item The difficulty of recursive Fourier sampling increases exponentially
with the height of the tree.

\item Given $n$ unrelated instances of a problem, but only enough time to
solve $o\left(  n\right)  $\ of them, the probability of succeeding on all $n$
instances decreases exponentially with $n$.

\item $\mathsf{NP}$-complete problems are probably hard, even with the help of
polynomial-size advice.
\end{itemize}

\part{Models and Reality\label{MAR}}

\begin{quote}
LS: So you believe quantum mechanics?

Me: Of course I do!

LS: So a thousand years from now, people will still be doing quantum mechanics?

Me: Well\ldots\ um\ldots\ I guess so\ldots

---Conversation between me and Lee Smolin
\end{quote}

\chapter{Skepticism of Quantum Computing\label{SKEP}}

\begin{quote}
\textquotedblleft QC of the sort that factors long numbers seems firmly rooted
in science fiction \ldots\ The present attitude would be analogous to, say,
Maxwell selling the Daemon of his famous thought experiment as a path to
cheaper electricity from heat.\textquotedblright

---Leonid Levin \cite{levin:qc}
\end{quote}

Quantum computing presents a dilemma:\ is it reasonable to study a type of
computer that has never been built, and might never be built in one's
lifetime? \ Some researchers strongly believe the answer is `no.' \ Their
objections generally fall into four categories:

\begin{enumerate}
\item[(A)] There is a fundamental physical reason why large quantum computers
can never be built.

\item[(B)] Even if (A) fails, large quantum computers will never be built in practice.

\item[(C)] Even if (A) and (B) fail, the speedup offered by quantum computers
is of limited theoretical interest.

\item[(D)] Even if (A), (B), and (C) fail, the speedup is of limited practical
value.\footnote{Because of the `even if' clauses, the objections seem to me
logically independent, so that there are $16$ possible positions regarding
them (or $15$ if one is against quantum computing). \ I ignore the possibility
that no speedup exists, in other words that $\mathsf{BPP}=\mathsf{BQP}$. \ By
`large quantum computer' I mean any computer much faster than its best
classical simulation, as a result of asymptotic complexity rather than the
speed of elementary operations. \ Such a computer need not be universal; it
might be specialized for (say) factoring.}
\end{enumerate}

The objections can be classified along two axes, as in Table 12.1.

\begin{table}[ptb]%
\begin{tabular}
[c]{lll}
& \textbf{Theoretical} & \textbf{Practical}\\
\textbf{Physical} & (A) & (B)\\
\textbf{Algorithmic} & (C) & (D)
\end{tabular}
\label{objtable}\caption[Four objections to quantum computing]{Four objections
to quantum computing.}%
\end{table}

This chapter focuses on objection (A), that quantum computing is impossible
for a fundamental physical reason. \ Among computer scientists, this objection
is most closely associated with Leonid Levin \cite{levin:qc}.\footnote{More
recently, Oded Goldreich \cite{goldreich:qc} has also put forward an argument
against quantum computing. Compared to Levin's arguments, Goldreich's is
easily understood: he believes that states arising in Shor's algorithm have
exponential \textquotedblleft non-degeneracy\textquotedblright\ and therefore
take exponential time to prepare, and that there is no burden on those who
hold this view to suggest a definition of non-degeneracy.} \ The following
passage captures much of the flavor of his critique:

\begin{quote}
The major problem [with quantum computing] is the requirement that basic
quantum equations hold to multi-hundredth if not millionth decimal positions
where the significant digits of the relevant quantum amplitudes reside. \ We
have never seen a physical law valid to over a dozen decimals. \ Typically,
every few new decimal places require major rethinking of most basic concepts.
\ Are quantum amplitudes still complex numbers to such accuracies or do they
become quaternions, colored graphs, or sick-humored gremlins?\ \cite{levin:qc}
\end{quote}

Among other things, Levin argues that quantum computing is analogous to the
unit-cost arithmetic model, and should be rejected for essentially the same
reasons; that claims to the contrary rest on a confusion between metric and
topological approximation; that quantum fault-tolerance theorems depend on
extravagant assumptions; and that even if a quantum computer failed, we could
not measure its state to prove a breakdown of quantum mechanics, and thus
would be unlikely to learn anything new.

A few responses to Levin's arguments can be offered immediately. \ First, even
classically, one can flip a coin a thousand times to produce probabilities of
order $2^{-1000}$. \ Should one dismiss such probabilities as unphysical? \ At
the very least, it is not obvious that amplitudes should behave differently
than probabilities with respect to error---since both evolve linearly, and
neither is directly observable.

Second, if Levin believes that quantum mechanics will fail, but is agnostic
about what will replace it, then his argument can be turned around. \ How do
we know that the successor to quantum mechanics will limit us to
$\mathsf{BPP}$, rather than letting us solve (say) $\mathsf{PSPACE}$-complete
problems? \ This is more than a logical point. \ Abrams and Lloyd
\cite{al}\ argue that a wide class of nonlinear variants of the
Schr\"{o}dinger equation would allow $\mathsf{NP}$-complete and even
$\mathsf{\#P}$-complete problems to be solved\ in polynomial time. \ And
Penrose \cite{penrose}, who proposed a model for `objective collapse' of the
wavefunction, believes that his proposal takes us outside the set of
computable functions entirely!

Third, to falsify quantum mechanics, it would suffice to show that a quantum
computer evolved to \textit{some} state far from the state that quantum
mechanics predicts. \ Measuring the exact state is unnecessary. \ Nobel prizes
have been awarded in the past `merely' for falsifying a previously held
theory, rather than replacing it by a new one. \ An example is the physics
Nobel awarded to Fitch \cite{fitch} and Cronin\ \cite{cronin}\ in 1980 for
discovering CP symmetry violation.

Perhaps the key to understanding Levin's unease about quantum computing lies
in his remark that \textquotedblleft we have never seen a physical law valid
to over a dozen decimals.\textquotedblright\ \ Here he touches on a serious
epistemological question:\ \textit{How far should we extrapolate from today's
experiments to where quantum mechanics has never been tested?} \ I will try to
address this question by reviewing the evidence for quantum mechanics. \ For
my purposes it will not suffice to declare the predictions of quantum
mechanics \textquotedblleft verified to one part in a
trillion,\textquotedblright\ because we have to distinguish at least three
different \textit{types} of prediction: \textit{interference},
\textit{entanglement}, and \textit{Schr\"{o}dinger cats}. \ Let us consider
these in turn.

\begin{enumerate}
\item[(1)] \textbf{Interference.} \ If the different paths that an electron
could take in its orbit around a nucleus did not interfere destructively,
canceling each other out, then electrons would not have quantized energy
levels. \ So being accelerating electric charges, they would lose energy and
spiral into their respective nuclei, and all matter would disintegrate. \ That
this has not happened---together with the results of (for example)
single-photon double-slit experiments---is compelling evidence for the reality
of quantum interference.

\item[(2)] \textbf{Entanglement.} \ One might accept that a single particle's
position is described by a wave in three-dimensional phase space,
but deny that two particles are described by a wave in
\textit{six}-dimensional phase space.\ \ However, the Bell
inequality experiments of Aspect et al.\ \cite{aspect}\ and
successors have convinced all but a few physicists that quantum
entanglement exists, can be maintained over large distances, and
cannot be explained by local hidden-variable theories.

\item[(3)] \textbf{Schr\"{o}dinger Cats.} \ Accepting two- and three-particle
entanglement is not the same as accepting that whole molecules,
cats, humans, and galaxies can be in coherent superposition states.
\ However, recently Arndt et al.\ \cite{arndt}\ have performed the
double-slit interference experiment using $C_{60}$\ molecules
(buckyballs) instead of photons; while Friedman et al.\
\cite{friedman}\ have found evidence that a superconducting current,
consisting of billions of electrons, can enter a coherent
superposition of flowing clockwise around a coil and flowing
counterclockwise (see Leggett \cite{leggett} for a survey of such
experiments). \ Though short of cats, these experiments at least
allow us to say the following: \textit{if we could build a
general-purpose quantum computer with as many components as have
already been placed into coherent superposition, then on certain
problems, that computer would outperform any computer in the world
today.}
\end{enumerate}

Having reviewed some of the evidence for quantum mechanics, we must now ask
what alternatives have been proposed that might also explain the evidence.
\ The simplest alternatives are those in which quantum states
\textquotedblleft spontaneously collapse\textquotedblright\ with some
probability, as in the GRW (Ghirardi-Rimini-Weber) theory \cite{grw}%
.\footnote{Penrose \cite{penrose}\ has proposed another such theory, but as
mentioned earlier, his theory suggests that the quantum computing model is
\textit{too} restrictive.}\ \ The drawbacks of the GRW theory include
violations of energy conservation, and parameters that must be fine-tuned to
avoid conflicting with experiments. \ More relevant for us, though, is that
the collapses postulated by the theory are only in the position basis, so that
quantum information stored in internal degrees of freedom (such as spin) is
unaffected. \ Furthermore, even if we extended the theory to collapse those
internal degrees, large quantum computers could still be built. \ For the
theory predicts roughly one collapse per particle per $10^{15}$\ seconds, with
a collapse affecting everything in a $10^{-7}$-meter vicinity. \ So even in
such a vicinity, one could perform a computation involving (say) $10^{10}%
$\ particles for $10^{5}$\ seconds. \ Finally, as pointed out to me by Rob
Spekkens, standard quantum error-correction techniques might be used to
overcome even GRW-type decoherence.

A second class of alternatives includes those of 't Hooft \cite{thooft}\ and
Wolfram \cite{wolfram}, in which something like a deterministic cellular
automaton underlies quantum mechanics. \ On the basis of his theory, 't Hooft
predicts that \textquotedblleft\lbrack i]t will never be possible to construct
a `quantum computer' that can factor a large number faster, and within a
smaller region of space, than a classical machine would do, if the latter
could be built out of parts at least as large and as slow as the Planckian
dimensions\textquotedblright\ \cite{thooft}. \ Similarly, Wolfram states\ that
``[i]ndeed within the usual formalism [of quantum mechanics] one can construct
quantum computers that may be able to solve at least a few specific problems
exponentially faster than ordinary Turing machines. \ But particularly after
my discoveries \ldots\ I strongly suspect that even if this is formally the
case, it will still not turn out to be a true representation of ultimate
physical reality, but will instead just be found to reflect various
idealizations made in the models used so far'' \cite[p.771]{wolfram}.

The obvious question then is how these theories account for Bell inequality
violations. \ I confess to being unable to understand 't Hooft's answer to
this question, except that he believes that the usual notions of causality and
locality might no longer apply in quantum gravity. \ As for Wolfram's theory,
which involves \textquotedblleft long-range threads\textquotedblright\ to
account for Bell inequality violations, I will show in Section \ref{WOLFRAM}
below that it fails Wolfram's own desiderata of causal and relativistic invariance.

\section{Bell Inequalities and Long-Range Threads\label{WOLFRAM}}

\begin{quote}
This section is excerpted from my review \cite{aar:rev}\ of Stephen
Wolfram's \textit{A New Kind of Science} \cite{wolfram}.
\end{quote}

The most interesting chapter of \textit{A New Kind of Science} is the ninth,
on `Fundamental Physics.' \ Here Wolfram confronts general relativity and
quantum mechanics, arguably the two most serious challenges to the
deterministic, cellular-automaton-based view of nature that he espouses.
\ Wolfram conjectures that spacetime is discrete at the Planck scale, of about
$10^{-33}$\ centimeters or $10^{-43}$\ seconds. \ This conjecture is not new,
and has received considerable attention recently in connection with the
holographic principle \cite{bousso}\ from black hole thermodynamics, which
Wolfram does not discuss. \ But are new ideas offered to substantiate the conjecture?

For Wolfram, spacetime is a \textit{causal network}, in which events
are vertices and edges specify the dependence relations between
events. \ Pages 486--496 and 508--515 discuss in detail how to
generate such a network from a simple set of rules. \ In particular,
we could start with a finite undirected `space graph' $G$. \ We then
posit a set of update rules, each of which replaces a subgraph by
another subgraph with the same number of outgoing edges. \ The new
subgraph must preserve any symmetries of the old one. \ Then each
event in the causal network corresponds to an application of an
update rule. \ If updating event $B$ becomes possible as a result of
event $A$, then we draw an edge from $A$ to $B$.

Properties of space are defined in terms of $G$. \ For example, if the number
of vertices in $G$ at distance at most $n$ from any given vertex grows as
$n^{D}$, then space can be said to have dimension $D$. \ (As for formalizing
this definition, Wolfram says only that there are \textquotedblleft some
subtleties. \ For example, to find a definite volume growth rate one does
still need to take some kind of limit---and one needs to avoid sampling too
many or too few\textquotedblright\ vertices \ (p. 1030).) \ Similarly, Wolfram
argues that the curvature information needed for general relativity, in
particular the Ricci tensor, can be read from the connectivity pattern of $G$.
\ Interestingly, to make the model as simple as possible, Wolfram does not
associate a bit to each vertex of $G$, representing (say) the presence or
absence of a particle. \ Instead particles are localized structures, or
`tangles,' in $G$.

An immediate problem is that one might obtain many nonequivalent causal
networks, depending on the order in which update rules are applied to $G$.
\ Wolfram calls a set of rules that allows such nondeterministic evolution a
`multiway system.' \ He recognizes, but rejects, a possible connection to
quantum mechanics:

\begin{quote}
The notion of `many-figured time' has been discussed since the 1950s in the
context of the many-worlds interpretation of quantum mechanics. \ There are
some similarities to the multiway systems that I\ consider here. \ But an
important difference is that while in the many-worlds approach, branchings are
associated with possible observation or measurement events, what I suggest
here is that they could be an intrinsic feature of even the very lowest-level
rules for the universe (p. 1035-6).
\end{quote}

It is unclear exactly what distinction is being drawn: is there any physical
event that is not associated with a \textit{possible} observation or
measurement? \ In any case, Wolfram opts instead for rule sets that are
`causal invariant': that is, that yield the same causal network regardless of
the order in which rules are applied. \ As noted by Wolfram, a sufficient
(though not necessary) condition for causal invariance is that no
`replaceable' subgraph overlaps itself or any other replaceable subgraph.

Wolfram points out an immediate analogy to special relativity, wherein
observers do not in general agree on the order in which spacelike separated
events occur, yet agree on any final outcome of the events. \ He is vague,
though, about how (say) the Lorentz transformations might be derived in a
causal network model:

\begin{quote}
There are many subtleties here, and indeed to explain the details of what is
going on will no doubt require quite a few new and rather abstract concepts.
But the general picture that I believe will emerge is that when particles move
faster they will appear to have more nodes associated with them (p. 529).
\end{quote}

Wolfram is \textquotedblleft certainly aware that many physicists will want to
know more details,\textquotedblright\ he says in the endnotes, about how a
discrete model of the sort he proposes can reproduce known features of
physics. \ But, although he chose to omit technical formalism from the
presentation, \textquotedblleft\lbrack g]iven my own personal background in
theoretical physics it will come as no surprise that I have often used such
formalism in the process of working out what I describe in these
sections\textquotedblright\ (p. 1043). \ The paradox is obvious: if technical
formalism would help convince physicists of his ideas, then what could Wolfram
lose by including it, say in the endnotes? \ If, on the other hand, such
formalism is irrelevant, then why does Wolfram even mention having used it?

Physicists' hunger for details will likely grow further when they read the
section on `Quantum Phenomena' (p. 537--545). \ Here Wolfram maintains that
quantum mechanics is only an approximation to an underlying classical (and
most likely deterministic) theory. \ Many physicists have sought such a
theory, from Einstein to (in modern times) 't Hooft \cite{thooft}. \ But a
series of results, beginning in the 1960's, has made it clear that such a
theory comes at a price. \ I will argue that, although Wolfram discusses these
results, he has not understood what they actually entail.

To begin, Wolfram is \textit{not} advocating a hidden-variable approach such
as Bohmian mechanics, in which the state vector is supplemented by an
`actual'\ eigenstate of a particular observable. \ Instead he thinks that, at
the lowest level, the state vector is not needed at all; it is merely a useful
construct for describing some (though presumably not all) higher-level
phenomena. \ Indeterminacy arises because of one's inability to know the exact
state of a system:

\begin{quote}
[I]f one knew all of the underlying details of the network that makes up our
universe, it should always be possible to work out the result of any
measurement. \ I strongly believe that the initial conditions for the universe
were quite simple. \ But like many of the processes we have seen in this book,
the evolution of the universe no doubt intrinsically generates apparent
randomness. \ And the result is that most aspects of the network that
represents the current state of our universe will seem essentially random (p. 543).
\end{quote}

Similarly, Wolfram explains as follows why an electron has wave properties:
``\ldots a network which represents our whole universe must also include us as
observers. \ And this means that there is no way that we can look at the
network from the outside and see the electron as a definite object'' (p. 538).
\ An obvious question then is how Wolfram accounts for the possibility of
quantum computing, assuming $\mathsf{BPP}\neq\mathsf{BQP}$. \ He gives an
answer in the final chapter:

\begin{quote}
Indeed within the usual formalism [of quantum mechanics] one can construct
quantum computers that may be able to solve at least a few specific problems
exponentially faster than ordinary Turing machines. \ But particularly after
my discoveries in Chapter 9 [`Fundamental Physics'], I strongly suspect that
even if this is formally the case, it will still not turn out to be a true
representation of ultimate physical reality, but will instead just be found to
reflect various idealizations made in the models used so far (p. 771).
\end{quote}

In the endnotes, though, where he explains quantum computing in more detail,
Wolfram seems to hedge about which idealizations he has in mind:

\begin{quote}
It does appear that only modest precision is needed for the initial
amplitudes. \ And it seems that perturbations from the environment can be
overcome using versions of error-correcting codes. \ But it remains unclear
just what might be needed actually to perform for example the final
measurements required (p. 1148).
\end{quote}

One might respond that, with or without quantum computing, Wolfram's proposals
can be ruled out on the simpler ground that they disallow Bell inequality
violations. \ However, Wolfram puts forward an imaginative hypothesis to
account for bipartite entanglement. \ When two particles (or `tangles' in the
graph $G$) collide, long-range `threads' may form between them, which remain
in place even if the particles are later separated:

\begin{quote}
The picture that emerges is then of a background containing a very large
number of connections that maintain an approximation to three-dimensional
space, together with a few threads that in effect go outside of that space to
make direct connections between particles (p. 544).
\end{quote}

The threads can produce Bell correlations, but are somehow too small (i.e.
contain too few edges) to transmit information in a way that violates causality.

There are several objections one could raise against this thread hypothesis.
\ What I will show is that, \textit{if} one accepts two of Wolfram's own
desiderata---determinism and causal invariance---then the hypothesis fails.
\ First, though, let me remark that Wolfram says little about what, to me, is
a more natural possibility than the thread hypothesis. \ This is an explicitly
\textit{quantum} cellular automaton or causal network, with a unitary
transition rule. \ The reason seems to be that he does not want continuity
anywhere in a model, not even in probabilities or amplitudes. \ In the notes,
he describes an experiment with a quantum cellular automaton as follows:

\begin{quote}
One might hope to be able to get an ordinary cellular automaton with a limited
set of possible values by choosing a suitable [phase rotation] $\theta$
[$\theta=\pi/4$ and $\theta=\pi/3$\ are given as examples in an illustration].
\ But in fact in non-trivial cases most of the cells generated at each step
end up having distinct values (p. 1060).
\end{quote}

This observation is unsurprising, given the quantum computing results
mentioned in Chapter \ref{QUANTUM}, to the effect that almost any nontrivial
gate set is universal (that is, can approximate any unitary matrix to any
desired precision, or any orthogonal matrix in case one is limited to reals).
\ Indeed, Shi \cite{shi:gate}\ has shown that a Toffoli gate, plus any gate
that does not preserve the computational basis, or a
controlled-$\operatorname*{NOT}$ gate plus any gate whose \textit{square} does
not preserve the computational basis, are both universal gate sets. \ In any
case, Wolfram does not address the fact that continuity in amplitudes seems
more `benign' than continuity in measurable quantities: the former, unlike the
latter, does not enable an infinite amount of computation to be performed in a
finite time. \ Also, as observed by Bernstein and Vazirani \cite{bv}, the
linearity of quantum mechanics implies that tiny errors in amplitudes will not
be magnified during a quantum computation.

I now proceed to the argument that Wolfram's thread hypothesis is inconsistent
with causal invariance and relativity. \ Let $\mathcal{R}$ be a set of graph
updating rules, which might be probabilistic. \ Then consider the following
four assertions (which, though not mathematically precise, will be clarified
by subsequent discussion).

\begin{enumerate}
\item[(1)] $\mathcal{R}$ satisfies causal invariance. \ That is, given any
initial graph (and choice of randomness if $\mathcal{R}$\ is probabilistic),
$\mathcal{R}$\ yields a unique causal network.

\item[(2)] $\mathcal{R}$ satisfies the relativity postulate. \ That is,
assuming the causal network approximates a flat Minkowski spacetime at a large
enough scale, there are no preferred inertial frames.

\item[(3)] $\mathcal{R}$ permits Bell inequality violations.

\item[(4)] Any updating rule in $\mathcal{R}$ is always considered to act on a
fixed graph, not on a distribution or superposition over graphs. \ This is
true even if parts of the initial graph are chosen at random, and even if
$\mathcal{R}$ is probabilistic.
\end{enumerate}

The goal is to show that, for any $\mathcal{R}$, at least one of these
assertions is false. \ Current physical theory would suggest that (1)-(3) are
true and that (4) is false. \ Wolfram, if I understand him correctly, starts
with (4) as a premise, and then introduces causal invariance to satisfy (1)
and (2), and long-range threads to satisfy (3). \ Of course, even to state the
two-party Bell inequalities requires some notion of randomness. \ And on pages
299--326, Wolfram discusses three mechanisms for introducing randomness into a
system: randomness in initial conditions, randomness from the environment
(i.e. probabilistic updating rules), and intrinsic randomness (i.e.
deterministic rules that produce pseudorandom output). \ However, all of these
mechanisms are compatible with (4), and so my argument will show that they are
inadequate assuming (1)-(3). \ The conclusion is that, in a model of the sort
Wolfram considers, randomness must play a more fundamental role than he allows.

In a standard Bell experiment, Alice and Bob are given input bits $x_{A}$\ and
$x_{B}$\ respectively, chosen uniformly and independently at random. \ Their
goal is, without communicating, to output bits $y_{A}$\ and $y_{B}$
respectively such that $y_{A}\oplus y_{B}=x_{A}\wedge x_{B}$. \ Under any
`local hidden variable' theory, Alice and Bob can succeed with probability at
most $3/4$; the optimal strategy is for them to ignore their inputs and output
(say) $y_{A}=0$\ and $y_{B}=0$. \ However, suppose Alice has a qubit $\rho
_{A}$\ and Bob a $\rho_{B}$, that are jointly in the Bell state $\left(
\left\vert 00\right\rangle +\left\vert 11\right\rangle \right)  /\sqrt{2}$.
\ Then there is a protocol\footnote{If $x_{A}=1$\ then Alice applies a $\pi
/8$\ phase rotation to $\rho_{A}$, and if $x_{B}=1$\ then Bob applies a
$-\pi/8$\ rotation to $\rho_{B}$. \ Both parties then measure in the standard
basis and output whatever they observe.} by which they can succeed with
probability $\left(  5+\sqrt{2}\right)  /8\approx0.802$.

To model this situation, let $A$ and $B$, corresponding to Alice and Bob, be
disjoint subgraphs of a graph $G$. \ Suppose that, at a large scale, $G$
approximates a Euclidean space of some dimension; and that any causal network
obtained by applying updates to $G$\ approximates a Minkowski spacetime. \ One
can think of $G$ as containing long-range threads from $A$ to $B$, though the
nature of the threads will not affect the conclusions. \ Encode Alice's input
$x_{A}$\ by (say) placing an edge between two specific vertices in $A$ if and
only if $x_{A}=1$. \ Encode $x_{B}$ similarly, and also supply Alice and Bob
with arbitrarily many correlated random bits. \ Finally, let us stipulate that
at the end of the protocol, there is an edge between two specific vertices in
$A$ if and only if $y_{A}=1$, and similarly for $y_{B}$. \ A technicality is
that we need to be able to identify which vertices correspond to $x_{A}$,
$y_{A}$, and so on, even as $G$ evolves over time. \ We could do this by
stipulating that (say) \textquotedblleft the $x_{A}$\ vertices are the ones
that are roots of complete binary trees of depth $3$,\textquotedblright%
\ and\ then choosing the rule set to guarantee that, throughout the protocol,
exactly two vertices have this property.

Call a variable `touched' after an update has been applied to a subgraph
containing any of the variable's vertices.\ \ Also, let $Z$ be an assignment
to all random variables: that is, $x_{A}$, $x_{B}$, the correlated random
bits, and the choice of randomness if $\mathcal{R}$\ is probabilistic. \ Then
for all $Z$ we need the following, based on what observers in different
inertial frames could perceive:

\begin{enumerate}
\item[(i)] There exists a sequence of updates under which $y_{A}$\ is output
before any of Bob's variables are touched.

\item[(ii)] There exists another sequence under which $y_{B}$\ is output
before any of Alice's variables are touched.
\end{enumerate}

Now it is easy to see that, if a Bell inequality violation occurs, then causal
invariance must be violated. \ Given $Z$, let $y_{A}^{\left(  1\right)
}\left(  Z\right)  $, $y_{B}^{\left(  1\right)  }\left(  Z\right)  $ be the
values of $y_{A},y_{B}$\ that are output under rule sequence (i), and let
$y_{A}^{\left(  2\right)  }\left(  Z\right)  $, $y_{B}^{\left(  2\right)
}\left(  Z\right)  $\ be the values output under sequence (ii). \ Then there
must exist some $Z$ for which either $y_{A}^{\left(  1\right)  }\left(
Z\right)  \neq y_{A}^{\left(  2\right)  }\left(  Z\right)  $ or $y_{B}%
^{\left(  1\right)  }\left(  Z\right)  \neq y_{B}^{\left(  2\right)  }\left(
Z\right)  $---for if not, then the entire protocol could be simulated under a
local hidden variable model. \ It follows that the outcome of the protocol can
depend on the order in which updates are applied.

To obtain a Bell inequality violation, something like the following seems to
be needed. \ We can encode `hidden variables'\ into $G$, representing the
outcomes of the possible measurements Bob could make on $\rho_{B}$. \ (We can
imagine, if we like, that the update rules are such that observing any one of
these variables destroys all the others. \ Also, we make no assumption of
contextuality.) \ Then, after Alice measures $\rho_{A}$, using the long-range
threads she updates Bob's hidden variables conditioned on her measurement
outcome. \ Similarly, Bob updates Alice's hidden variables conditioned on his
outcome. \ Since at least one party must access its hidden variables for there
to be Bell inequality violations, causal invariance is still violated. \ But a
sort of probabilistic causal invariance holds, in the sense that\ if we
marginalize out $A$ (the `Alice' part of $G$), then the \textit{distribution}
of values for each of Bob's hidden variables is the same before and after
Alice's update.\ \ The lesson is that, if we want both causal invariance and
Bell inequality violations, then we need to introduce probabilities at a
fundamental level---not merely to represent Alice and Bob's subjective
uncertainty about the state of $G$, but even to define whether a set of rules
is or is not causal invariant.

Note that I made no assumption about how the random bits were generated---i.e.
whether they were `truly random' or were the pseudorandom output of some
updating rule.\ \ The conclusion is also unaffected if we consider a
`deterministic' variant of Bell's theorem due to Greenberger, Horne, and
Zeilinger \cite{ghz}. \ There three parties, Alice, Bob, and Charlie, are
given input bits $x_{A}$, $x_{B}$, and $x_{C}$ respectively, satisfying the
promise that $x_{A}\oplus x_{B}\oplus x_{C}=0$. \ The goal is to output bits
$y_{A}$, $y_{B}$, and $y_{C}$\ such that $y_{A}\oplus y_{B}\oplus y_{C}%
=x_{A}\vee x_{B}\vee x_{C}$. \ Under a local hidden variable model, there is
no protocol that succeeds on all four possible inputs; but if the parties
share the GHZ state $\left(  \left\vert 011\right\rangle +\left\vert
101\right\rangle +\left\vert 110\right\rangle -\left\vert 000\right\rangle
\right)  /2$, then such a protocol exists. \ However, although the
\textit{output} is correct with certainty, assuming causal invariance one
cannot \textit{implement} the protocol without introducing randomness into the
underlying rules, exactly as in the two-party case.

After a version of the above argument was sent to Wolfram, Todd Rowland, an
employee of Wolfram, sent me email claiming that the argument fails for the
following reason. \ I assumed that there exist two sequences of updating
events, one in which Alice's measurement precedes Bob's and one in which Bob's
precedes Alice's. \ But I neglected the possibility that a \textit{single}
update, call it $E$, is applied to a subgraph that straddles the long-range
threads. \ The event $E$ would encompass both Alice and Bob's measurements, so
that neither would precede the other in any sequence of updates. \ We could
thereby obtain a rule set $\mathcal{R}$\ satisfying assertions (1), (3), and (4).

I argue that such an $\mathcal{R}$\ would nevertheless fail to satisfy (2).
\ For in effect we start with a flat Minkowski spacetime, and then take two
distinct events that are simultaneous in a particular inertial frame, and
identify them as being the \textit{same} event $E$. \ This can be visualized
as `pinching together' two horizontally separated points on a spacetime
diagram. \ (Actually a whole `V' of points must be pinched together, since
otherwise entanglement could not have been created.) \ However, what happens
in a different inertial frame? \ It would seem that $E$, a single event, is
perceived to occur at two separate times. \ That by itself might be thought
acceptable, but it implies that there exists a class of preferred inertial
frames: those in which $E$ is perceived to occur only once. \ Of course, even
in a flat spacetime, one could designate as `preferred' those frames in which
Alice and Bob's measurements are perceived to be simultaneous. \ A crucial
distinction, though, is that there one only obtains a class of preferred
frames after deciding which event at Alice's location, \textit{and} which at
Bob's location, should count as the `measurement.' \ Under Rowland's
hypothesis, by contrast, once one decides what counts as the measurement at
Alice's location, the decision at Bob's location is made automatically,
because of the identification of events that would otherwise be far apart.

\chapter{Complexity Theory of Quantum States\label{MLIN}}

In my view, the central weakness in the arguments of quantum
computing skeptics is their failure to suggest any answer the
following question: \textit{Exactly what property separates the
quantum states we are sure we can create, from the states that
suffice for Shor's factoring algorithm?}

I call such a property a \textquotedblleft Sure/Shor
separator.\textquotedblright\ \ The purpose of this chapter is to develop a
mathematical theory of Sure/Shor separators, and thereby illustrate what
I\ think a scientific discussion about the possibility of quantum computing
might look like. \ In particular, I will introduce \textit{tree states}, which
informally are those states $\left\vert \psi\right\rangle \in\mathcal{H}%
_{2}^{\otimes n}$\ expressible by a polynomial-size `tree' of addition and
tensor product gates. \ For example, $\alpha\left\vert 0\right\rangle
^{\otimes n}+\beta\left\vert 1\right\rangle ^{\otimes n}$\ and $\left(
\alpha\left\vert 0\right\rangle +\beta\left\vert 1\right\rangle \right)
^{\otimes n}$ are both tree states. \ Section \ref{SURESHOR}\ provides the
philosophical motivation for thinking of tree states as a possible Sure/Shor
separator; then Section \ref{CQS}\ formally defines tree states and many
related classes of quantum states. \ Next, Section \ref{BASICMLIN}%
\ investigates basic properties of tree states. \ Among other results, it
shows that any tree state is representable by a tree of polynomial size and
logarithmic depth; and that most states do not even have large inner product
with any tree state. \ Then Section \ref{RELMLIN}\ shows relationships among
tree size, circuit size, bounded-depth tree size, Vidal's $\chi$\ complexity
\cite{vidal}, and several other measures. \ It also relates questions about
quantum state classes to more traditional questions about computational
complexity classes.

But the main results of the chapter, proved in Section \ref{LOWERMLIN}, are
lower bounds on tree size for various natural families of quantum states. \ In
particular, Section \ref{ECC} analyzes \textquotedblleft subgroup
states,\textquotedblright\ which are uniform superpositions $\left\vert
S\right\rangle $\ over all elements of a subgroup $S\leq\mathbb{Z}_{2}^{n}$.
\ The importance of these states arises from their central role in stabilizer
codes, a type of quantum error-correcting code. \ I first show that if $S$ is
chosen uniformly at random, then with high probability $\left\vert
S\right\rangle $ cannot be represented by any tree of size $n^{o\left(  \log
n\right)  }$. \ This result has a corollary of independent
complexity-theoretic interest: the first superpolynomial gap between the
formula size and the multilinear formula size of a function $f:\left\{
0,1\right\}  ^{n}\rightarrow\mathbb{R}$. \ I then present two improvements of
the basic lower bound. \ First, I show that a random subgroup state cannot
even be \textit{approximated} well in trace distance by any tree of size
$n^{o\left(  \log n\right)  }$. \ Second, I \textquotedblleft
derandomize\textquotedblright\ the lower bound, by using Reed-Solomon codes to
construct an \textit{explicit} subgroup state\ with tree size $n^{\Omega
\left(  \log n\right)  }$.

Section \ref{DIVIS} analyzes the states that arise in Shor's factoring
algorithm---for example, a uniform superposition over all multiples of a fixed
positive integer $p$, written in binary.\ \ Originally, I had hoped to show a
superpolynomial tree size lower bound for these states as well. \ However, I
am only able to show such a bound assuming a number-theoretic conjecture.

The lower bounds use a sophisticated recent technique of Raz
\cite{raz,raz:nc2}, which was introduced to show that the permanent and
determinant of a matrix require superpolynomial-size multilinear formulas.
\ Currently, Raz's technique is only able to show\ lower bounds of the form
$n^{\Omega\left(  \log n\right)  }$, but I conjecture that $2^{\Omega\left(
n\right)  }$\ lower bounds hold in all of the cases discussed above.

One might wonder how superpolynomial tree size relates to more
\textit{physical} properties of a quantum state. \ Section \ref{PERSIST}%
\ addresses this question, by pointing out how Raz's lower bound technique is
connected to a notion that physicists call \textquotedblleft persistence of
entanglement\textquotedblright\ \cite{br,db}. \ On the other hand, I also give
examples showing that the connection is not exact.

Section \ref{MOTS}\ studies a weakening of tree size called \textquotedblleft
manifestly orthogonal tree size,\textquotedblright\ and shows that this
measure can sometimes be characterized \textit{exactly}, enabling us to prove
\textit{exponential} lower bounds. \ The techniques in Section \ref{MOTS}%
\ might be of independent interest to complexity theorists---one reason being
that they do not obviously \textquotedblleft naturalize\textquotedblright\ in
the sense of Razborov and Rudich \cite{rr}.

Section \ref{TREEBQP} addresses the following question. \ If the state of a
quantum computer at every time step is a tree state, then can the computer be
simulated classically? \ In other words, letting $\mathsf{TreeBQP}$\ be the
class of languages accepted by such a machine, does $\mathsf{TreeBQP}%
=\mathsf{BPP}$? \ A positive answer would make tree states more attractive as
a Sure/Shor separator. \ For once we admit any states incompatible with the
polynomial-time Church-Turing thesis, it seems like we might as well go all
the way, and admit \textit{all} states preparable by polynomial-size quantum
circuits! \ Although I leave this question open, I do show that
$\mathsf{TreeBQP}\subseteq\mathsf{\Sigma}_{3}^{\mathsf{P}}\cap\mathsf{\Pi}%
_{3}^{\mathsf{P}}$, where $\mathsf{\Sigma}_{3}^{\mathsf{P}}\cap\mathsf{\Pi
}_{3}^{\mathsf{P}}$\ is the third level of the polynomial hierarchy
$\mathsf{PH}$. \ By contrast, it is conjectured that $\mathsf{BQP}%
\not \subset \mathsf{PH}$, though admittedly not on strong evidence.

Section \ref{EXPER}\ discusses the implications of these results for
experimental physics. \ It advocates a dialectic between theory and
experiment, in which theorists would propose a class of quantum states that
encompasses everything seen so far, and then experimenters would try to
prepare states not in that class. \ It also asks whether states with
superpolynomial tree size have already been observed in condensed-matter
systems; and more broadly, what sort of evidence is needed to establish a
state's existence. \ Other issues addressed in Section \ref{EXPER}\ include
how to deal with mixed states and particle position and momentum states, and
the experimental relevance of asymptotic bounds. \ I conclude in Section
\ref{OPENMLIN}\ with some open problems.

\section{Sure/Shor Separators\label{SURESHOR}}

Given the discussion in Chapter \ref{SKEP}, I believe that the challenge for
quantum computing skeptics is clear. \ Ideally, come up with an alternative to
quantum mechanics---even an idealized toy theory---that can account for all
present-day experiments, yet would not allow large-scale quantum computation.
\ Failing that, \textit{at least say what you take quantum mechanics' domain
of validity to be}. \ One way to do this would be to propose a set $S$ of
quantum states that you believe corresponds to possible physical states of
affairs.\footnote{A skeptic might also specify what happens if a state
$\left\vert \psi\right\rangle \in S$\ is acted on by a unitary $U$ such that
$U\left\vert \psi\right\rangle \notin S$, but this will not be insisted upon.}
\ The set $S$ must contain all \textquotedblleft Sure states\textquotedblright%
\ (informally, the states that have already been demonstrated in the
lab), but no \textquotedblleft Shor states\textquotedblright\ (again
informally, the states that can be shown to suffice for factoring,
say, $500$-digit numbers). \ If $S$ satisfies both of these
constraints, then I call $S$ a \textit{Sure/Shor separator} (see
Figure 13.1).\begin{figure}[ptb]
\begin{center}
\includegraphics[
trim=0.9in 2.393715in 0.9in 2.5in,
height=2.4in,
width=2.6in
]{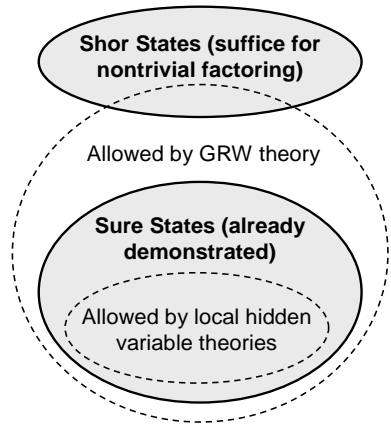}
\end{center}
\caption[Sure/Shor separators]{A Sure/Shor separator must contain all Sure
states but no Shor states. \ That is why neither local hidden variables nor
the GRW theory yields a Sure/Shor separator.}%
\label{sureshor}%
\end{figure}

Of course, an alternative theory need not involve a sharp cutoff between
possible and impossible states. \ So it is perfectly acceptable for a skeptic
to define a \textquotedblleft complexity measure\textquotedblright\ $C\left(
\left\vert \psi\right\rangle \right)  $\ for quantum states, and then say
something like the following: \ \textit{If }$\left\vert \psi_{n}\right\rangle
$\textit{\ is a state of }$n$\textit{ spins, and }$C\left(  \left\vert
\psi_{n}\right\rangle \right)  $\textit{\ is at most, say, }$n^{2}$\textit{,
then I predict that }$\left\vert \psi_{n}\right\rangle $\textit{\ can be
prepared using only \textquotedblleft polynomial effort.\textquotedblright%
\ \ Also, once prepared, }$\left\vert \psi_{n}\right\rangle $\textit{\ will be
governed by standard quantum mechanics to extremely high precision. \ All
states created to date have had small values of }$C\left(  \left\vert \psi
_{n}\right\rangle \right)  $\textit{. \ However, if }$C\left(  \left\vert
\psi_{n}\right\rangle \right)  $\textit{\ grows as, say, }$2^{n}$\textit{,
then I predict that }$\left\vert \psi_{n}\right\rangle $\textit{\ requires
\textquotedblleft exponential effort\textquotedblright\ to prepare, or else is
not even approximately governed by quantum mechanics, or else does not even
make sense in the context of an alternative theory. \ The states that arise in
Shor's factoring algorithm have exponential values of }$C\left(  \left\vert
\psi_{n}\right\rangle \right)  $\textit{. \ So as my Sure/Shor separator, I
propose the set of all infinite families of states }$\left\{  \left\vert
\psi_{n}\right\rangle \right\}  _{n\geq1}$\textit{, where }$\left\vert
\psi_{n}\right\rangle $\textit{\ has }$n$\textit{ qubits, such that }$C\left(
\left\vert \psi_{n}\right\rangle \right)  \leq p\left(  n\right)  $\textit{
for some polynomial }$p$\textit{.}

To understand the importance of Sure/Shor separators, it is helpful to think
through some examples. \ A major theme of Levin's arguments was that
exponentially small amplitudes are somehow unphysical. \ However, clearly we
cannot reject \textit{all} states with tiny amplitudes---for would anyone
dispute that the state $2^{-5000}\left(  \left\vert 0\right\rangle +\left\vert
1\right\rangle \right)  ^{\otimes10000}$\ is formed whenever $10,000$ photons
are each polarized at $45^{\circ}$?\ \ Indeed, once we accept $\left\vert
\psi\right\rangle $\ and $\left\vert \varphi\right\rangle $\ as Sure states,
we are almost \textit{forced} to accept $\left\vert \psi\right\rangle
\otimes\left\vert \varphi\right\rangle $\ as well---since we can imagine, if
we like, that $\left\vert \psi\right\rangle $\ and $\left\vert \varphi
\right\rangle $\ are prepared in two separate laboratories.\footnote{It might
be objected that in some theories, such as Chern-Simons theory, there is no
clear tensor product decomposition. \ However, the relevant question is
whether $\left\vert \psi\right\rangle \otimes\left\vert \varphi\right\rangle
$\ is a Sure state, \textit{given} that $\left\vert \psi\right\rangle $\ and
$\left\vert \varphi\right\rangle $\ are both Sure states that are
well-described in tensor product Hilbert spaces.} \ So considering a Shor
state such as%
\[
\left\vert \Phi\right\rangle =\frac{1}{2^{n/2}}\sum_{r=0}^{2^{n}-1}\left\vert
r\right\rangle \left\vert x^{r}\operatorname{mod}N\right\rangle ,
\]
what property of this state could quantum computing skeptics latch onto as
being physically extravagant? \ They might complain that $\left\vert
\Phi\right\rangle $\ involves entanglement across hundreds or thousands of
particles; but as mentioned in Chapter \ref{SKEP}, there are other states with
that same property, namely the \textquotedblleft Schr\"{o}dinger
cats\textquotedblright\ $\left(  \left\vert 0\right\rangle ^{\otimes
n}+\left\vert 1\right\rangle ^{\otimes n}\right)  /\sqrt{2}$, that should be
regarded as Sure states. \ Alternatively, the skeptics might object to the
\textit{combination} of exponentially small amplitudes with entanglement
across hundreds of particles. \ However, simply viewing a Schr\"{o}dinger cat
state in the Hadamard basis produces an equal superposition over all strings
of even parity, which has both properties. \ We seem to be on a slippery slope
leading to all of quantum mechanics! \ Is there any defensible place to draw a line?

The dilemma above is what led me to propose \textit{tree states} as
a possible Sure/Shor separator. \ The idea, which might seem more
natural to logicians than to physicists, is this. \ Once we accept
the linear combination and tensor product rules of quantum
mechanics---allowing $\alpha\left\vert \psi\right\rangle
+\beta\left\vert \varphi\right\rangle $\ and $\left\vert
\psi\right\rangle \otimes\left\vert \varphi\right\rangle $\ into our
set $S$ of possible states whenever $\left\vert \psi\right\rangle
,\left\vert \varphi\right\rangle \in S$---one of our few remaining
hopes for keeping $S$ a proper subset of the set of \textit{all}
states is to impose some restriction on how those two rules can be
iteratively applied. \ In particular, we could let $S$ be the
closure of $\left\{  \left\vert 0\right\rangle ,\left\vert
1\right\rangle \right\}  $\ under a \textit{polynomial number} of
linear combinations and tensor products. \ That is, $S$\ is the set
of all infinite families of states $\left\{  \left\vert
\psi_{n}\right\rangle \right\} _{n\geq1}$ with $\left\vert
\psi_{n}\right\rangle \in\mathcal{H}_{2}^{\otimes n}$, such that
$\left\vert \psi_{n}\right\rangle $\ can be expressed as a
\textquotedblleft tree\textquotedblright\ involving at most $p\left(
n\right)  $\ addition, tensor product, $\left\vert 0\right\rangle $,
and $\left\vert 1\right\rangle $\ gates for some polynomial $p$ (see
Figure 13.2).\begin{figure}[ptb]
\begin{center}
\includegraphics[
trim=2in 4.5in 2in -0.35in,
height=1.7in,
width=2.9464in
]{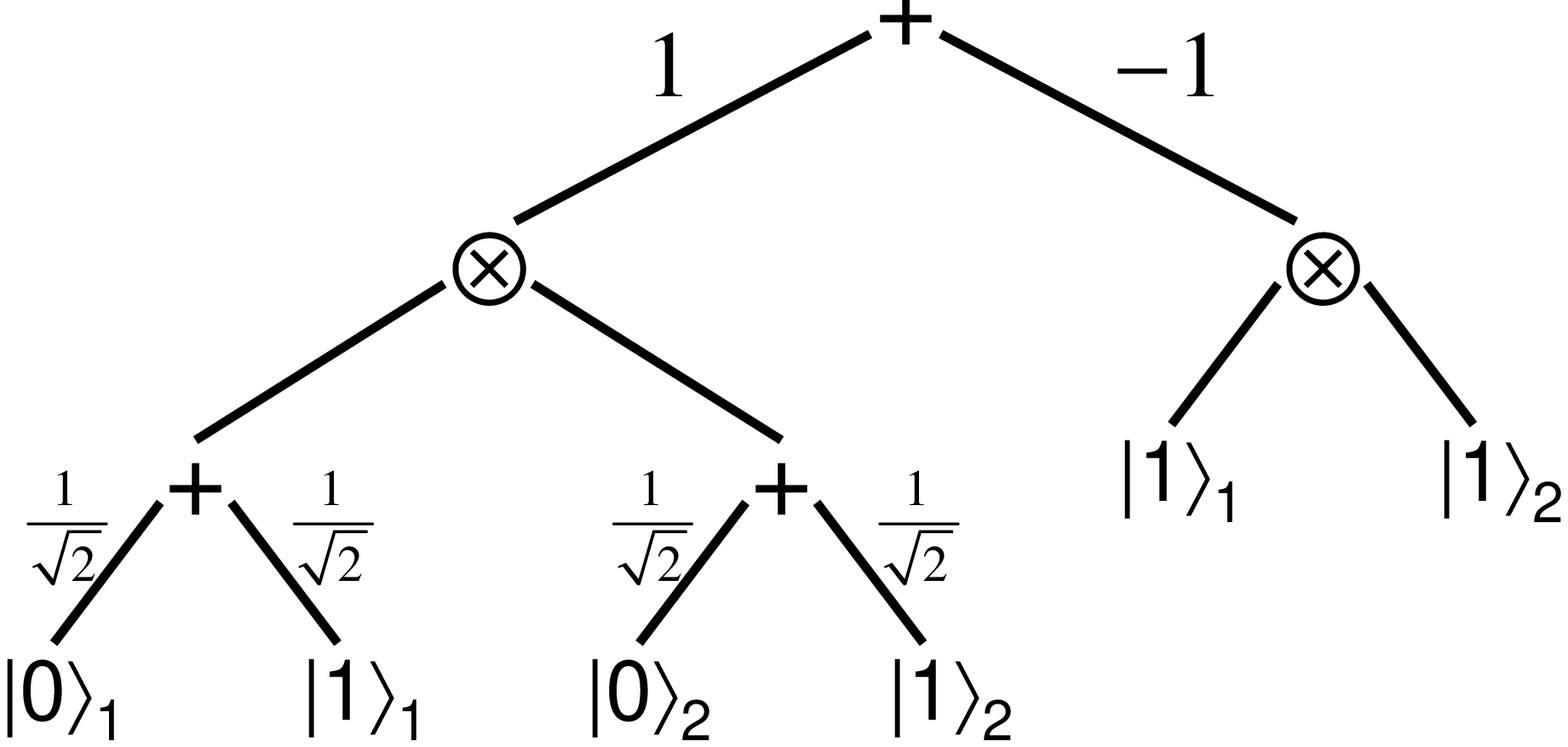}
\end{center}
\caption[Tree representing a quantum state]{Expressing $\left(  \left\vert
00\right\rangle +\left\vert 01\right\rangle +\left\vert 10\right\rangle
-\left\vert 11\right\rangle \right)  /2$\ by a tree of linear combination and
tensor product gates, with scalar multiplication along edges. \ Subscripts
denote the identity of a qubit.}%
\label{tree}%
\end{figure}

To be clear, I am \textit{not} advocating that \textquotedblleft all states in
Nature are tree states\textquotedblright\ as a serious physical
hypothesis.\ \ Indeed, even if I believed firmly in a breakdown of quantum
mechanics,\footnote{which I don't} there are other choices for the set $S$
that seem equally reasonable. \ For example, define \textit{orthogonal tree
states} similarly to tree states, except that we can only form the linear
combination $\alpha\left\vert \psi\right\rangle +\beta\left\vert
\varphi\right\rangle $\ if $\left\langle \psi|\varphi\right\rangle =0$.
\ Rather than choose among tree states, orthogonal tree states, and the other
candidate Sure/Shor separators that occurred to me, my approach will be to
prove everything I can about all of them. \ If I devote more space to tree
states than to others, that is simply because tree states are the subject of
the most interesting results. \ On the other hand, if one shows (for example)
that $\left\{  \left\vert \psi_{n}\right\rangle \right\}  $\ is not a tree
state, then one has also shown that $\left\{  \left\vert \psi_{n}\right\rangle
\right\}  $\ is not an orthogonal tree state. \ So many candidate separators
are related to each other; and indeed, their relationships will be a major
theme of the chapter.

In summary, to debate whether quantum computing is fundamentally impossible,
we need at least one proposal for how it \textit{could} be impossible. \ Since
even skeptics admit that quantum mechanics is valid within some
\textquotedblleft regime,\textquotedblright\ a key challenge for any such
proposal is to separate the regime of acknowledged validity from the quantum
computing regime. \ Though others will disagree, I do not see any choice but
to \textit{identify those two regimes with classes of quantum states}. \ For
gates and measurements that suffice for quantum computing have already been
demonstrated experimentally. \ Thus, if we tried to identify the two regimes
with classes of gates or measurements, then we could equally well talk about
the class of \textit{states} on which all $1$- and $2$-qubit operations behave
as expected. \ A similar argument would apply if we identified the two regimes
with classes of quantum circuits---since any \textquotedblleft
memory\textquotedblright\ that a quantum system retains of the previous gates
in a circuit, is part of the system's state by definition. \ So: states,
gates, measurements, circuits---what else is there?

I should stress that none of the above depends on the interpretation of
quantum mechanics. \ In particular, it is irrelevant whether we regard quantum
states as \textquotedblleft really out there\textquotedblright\ or as
representing subjective knowledge---since in either case, the question is
whether there can exist systems that we would \textit{describe} by $\left\vert
\psi\right\rangle $\ based on their observed behavior.

Once we agree to seek a Sure/Shor separator, we quickly find that the obvious
ideas---based on precision in amplitudes, or entanglement across of hundreds
of particles---are nonstarters. \ The only idea that seems plausible is to
limit the class of allowed quantum states to those with some kind of succinct
representation. \ That still leaves numerous possibilities; and for each one,
it might be a difficult problem to decide whether a given $\left\vert
\psi\right\rangle $\ is succinctly representable or not. \ Thus, constructing
a useful theory of Sure/Shor separators is a nontrivial task. \ This chapter
represents a first attempt.

\section{Classifying Quantum States\label{CQS}}

In both quantum and classical complexity theory, the objects studied
are usually sets of languages or Boolean functions. \ However, a
generic $n$-qubit quantum state requires exponentially many
classical bits to describe, and this suggests looking at \textit{the
complexity of quantum states themselves}. \ That is, which states
have polynomial-size classical descriptions of various kinds? \ This
question has been studied from several angles by Aharonov and
Ta-Shma \cite{at};\ Janzing, Wocjan, and Beth \cite{jwb};\ Vidal
\cite{vidal}; and Green et al.\ \cite{ghmp}. \ Here I propose a
general framework for the question. \ For simplicity, I limit myself
to pure states $\left\vert \psi _{n}\right\rangle
\in\mathcal{H}_{2}^{\otimes n}$ with the fixed orthogonal basis
$\left\{  \left\vert x\right\rangle :x\in\left\{  0,1\right\}
^{n}\right\}  $. \ Also, by `states' I mean infinite families of
states $\left\{  \left\vert \psi_{n}\right\rangle \right\}
_{n\geq1}$.

Like complexity classes, pure quantum states can be organized into a
hierarchy (see Figure 13.3). \ At the bottom are the classical basis
states, which have the form $\left\vert x\right\rangle $\ for some
$x\in\left\{  0,1\right\}  ^{n}$. \ We can generalize classical
states in two directions: to the class
$\mathsf{\otimes}_{\mathsf{1}}$ of separable states, which have the
form $\left(  \alpha_{1}\left\vert 0\right\rangle
+\beta_{1}\left\vert 1\right\rangle \right)
\otimes\cdots\otimes\left(  \alpha_{n}\left\vert 0\right\rangle
+\beta_{n}\left\vert 1\right\rangle \right)  $; and to the class
$\mathsf{\Sigma}_{\mathsf{1}}$, which consists of all states
$\left\vert \psi_{n}\right\rangle $\ that are superpositions of at
most $p\left( n\right)  $\ classical states, where $p$ is a
polynomial. \ At the next level, $\mathsf{\otimes}_{\mathsf{2}}$
contains the states that can be written as a tensor product of
$\mathsf{\Sigma}_{\mathsf{1}}$\ states, with qubits permuted
arbitrarily. \ Likewise, $\mathsf{\Sigma}${}$_{\mathsf{2}}$ contains
the states that can be written as a linear combination of a
polynomial number of $\mathsf{\otimes}_{\mathsf{1}}$ states. \ We
can continue indefinitely to $\mathsf{\Sigma}${}$_{\mathsf{3}}$,
$\mathsf{\otimes}_{\mathsf{3}}$, etc.
\ Containing the whole `tensor-sum hierarchy' $\mathsf{\cup}_{\mathsf{k}%
}\mathsf{\Sigma}${}$_{\mathsf{k}}=\mathsf{\cup}_{\mathsf{k}}\mathsf{\otimes
}_{\mathsf{k}}$\ is the class $\mathsf{Tree}$, of all states expressible by a
polynomial-size tree of additions and tensor products nested arbitrarily.
\ Formally, $\mathsf{Tree}$\ consists of all states $\left\vert \psi
_{n}\right\rangle $\ such that $\operatorname*{TS}\left(  \left\vert \psi
_{n}\right\rangle \right)  \leq p\left(  n\right)  $\ for some polynomial
$p$,\ where the \textit{tree size} $\operatorname*{TS}\left(  \left\vert
\psi_{n}\right\rangle \right)  $\ is defined as follows. \begin{figure}[ptb]
\begin{center}
\includegraphics[
trim=0.678569in 1.806069in 0.679377in 1.813617in,
height=2.8928in,
width=2.7164in
]{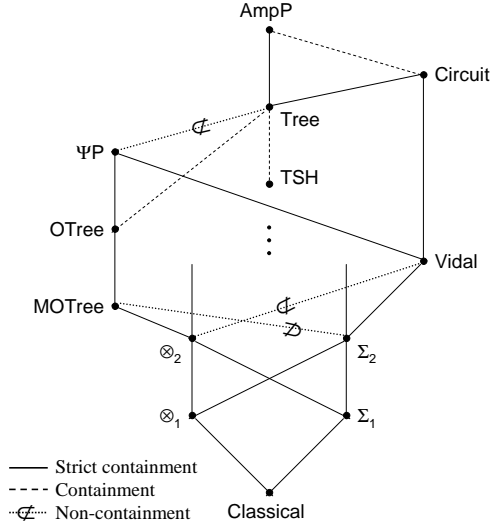}
\end{center}
\caption[Known relations among quantum state classes]{Known relations among
quantum state classes.}%
\label{classes}%
\end{figure}

\begin{definition}
A \textit{quantum state tree} over $\mathcal{H}_{2}^{\otimes n}$\ is a rooted
tree where each leaf vertex is labeled with $\alpha\left|  0\right\rangle
+\beta\left|  1\right\rangle $ for some $\alpha,\beta\in\mathsf{C}$, and each
non-leaf vertex (called a gate) is labeled with either $+$ or $\otimes$.
\ Each vertex $v$ is also labeled with a set $S\left(  v\right)
\subseteq\left\{  1,\ldots,n\right\}  $, such that

\begin{enumerate}
\item[(i)] If $v$ is a leaf then $\left|  S\left(  v\right)  \right|  =1$,

\item[(ii)] If $v$ is the root then $S\left(  v\right)  =\left\{
1,\ldots,n\right\}  $,

\item[(iii)] If $v$ is a $+$ gate and $w$ is a child of $v$, then $S\left(
w\right)  =S\left(  v\right)  $,

\item[(iv)] If $v$ is a $\otimes$\ gate and $w_{1},\ldots,w_{k}$\ are the
children of $v$, then $S\left(  w_{1}\right)  ,\ldots,S\left(  w_{k}\right)
$\ are pairwise disjoint and form a partition of $S\left(  v\right)  $.
\end{enumerate}

Finally, if $v$\ is a $+$ gate, then the outgoing edges\ of $v$ are labeled
with complex numbers. \ For each $v$, the subtree rooted at $v$ represents a
quantum state of the qubits in $S\left(  v\right)  $\ in the obvious way. \ We
require this state to be normalized for each $v$.\footnote{Requiring only the
\textit{whole} tree to represent a normalized state clearly yields no further
generality.}
\end{definition}

We say a tree is \textit{orthogonal} if it satisfies the further condition
that if $v$ is a $+$ gate, then any two children $w_{1},w_{2}$\ of $v$
represent $\left\vert \psi_{1}\right\rangle ,\left\vert \psi_{2}\right\rangle
$\ with $\left\langle \psi_{1}|\psi_{2}\right\rangle =0$. \ If the condition
$\left\langle \psi_{1}|\psi_{2}\right\rangle =0$ can be replaced by the
stronger condition that for all basis states $\left\vert x\right\rangle $,
either $\left\langle \psi_{1}|x\right\rangle =0$ or $\left\langle \psi
_{2}|x\right\rangle =0$, then we say the tree is \textit{manifestly
orthogonal}. \ Manifest orthogonality is an extremely unphysical definition; I
introduce it because it turns out to be interesting from a lower bounds perspective.

For reasons of convenience, let us define the \textit{size} $\left\vert
T\right\vert $\ of a tree $T$ to be the number of leaf vertices. \ Then given
a state $\left\vert \psi\right\rangle \in\mathcal{H}_{2}^{\otimes n}$, the
\textit{tree size} $\operatorname*{TS}\left(  \left\vert \psi\right\rangle
\right)  $\ is the minimum size of a tree that represents $\left\vert
\psi\right\rangle $. \ The \textit{orthogonal tree size} $\operatorname*{OTS}%
\left(  \left\vert \psi\right\rangle \right)  $\ and \textit{manifestly
orthogonal tree size} $\operatorname*{MOTS}\left(  \left\vert \psi
\right\rangle \right)  $\ are defined similarly. \ Then $\mathsf{OTree}$\ is
the class of $\left\vert \psi_{n}\right\rangle $\ such that
$\operatorname*{OTS}\left(  \left\vert \psi_{n}\right\rangle \right)  \leq
p\left(  n\right)  $ for some polynomial $p$, and $\mathsf{MOTree}$\ is the
class such that $\operatorname*{MOTS}\left(  \left\vert \psi_{n}\right\rangle
\right)  \leq p\left(  n\right)  $\ for some $p$.

It is easy to see that
\[
n\leq\operatorname*{TS}\left(  \left\vert \psi\right\rangle \right)
\leq\operatorname*{OTS}\left(  \left\vert \psi\right\rangle \right)
\leq\operatorname*{MOTS}\left(  \left\vert \psi\right\rangle \right)  \leq
n2^{n}%
\]
for every $\left\vert \psi\right\rangle $, and that the set of $\left\vert
\psi\right\rangle $\ such that $\operatorname*{TS}\left(  \left\vert
\psi\right\rangle \right)  <2^{n}$\ has measure $0$ in $\mathcal{H}%
_{2}^{\otimes n}$. \ Two other important properties of $\operatorname*{TS}$
and $\operatorname*{OTS}$ are as follows:

\begin{proposition}
\label{invariant}\quad

\begin{enumerate}
\item[(i)] $\operatorname*{TS}$ and $\operatorname*{OTS}$ are invariant under
local\footnote{Several people told me that a reasonable complexity measure
must be invariant under \textit{all} basis changes. Alas, this would imply
that all pure states have the same complexity!} basis changes, up to a
constant factor of $2$.

\item[(ii)] If $\left\vert \phi\right\rangle $ is obtained from $\left\vert
\psi\right\rangle $ by applying a $k$-qubit unitary, then $\operatorname*{TS}%
\left(  \left\vert \phi\right\rangle \right)  \leq k4^{k}\operatorname*{TS}%
\left(  \left\vert \psi\right\rangle \right)  $ and $\operatorname*{OTS}%
\left(  \left\vert \phi\right\rangle \right)  \leq k4^{k}\operatorname*{OTS}%
\left(  \left\vert \psi\right\rangle \right)  $.
\end{enumerate}
\end{proposition}

\begin{proof}
\quad

\begin{enumerate}
\item[(i)] Simply replace each occurrence of $\left\vert 0\right\rangle $\ in
the original tree by a tree for $\alpha\left\vert 0\right\rangle
+\beta\left\vert 1\right\rangle $, and each occurrence of $\left\vert
1\right\rangle $\ by a tree for $\gamma\left\vert 0\right\rangle
+\delta\left\vert 1\right\rangle $, as appropriate.

\item[(ii)] Suppose without loss of generality that the gate is applied to the
first $k$ qubits. \ Let $T$ be a tree representing $\left\vert \psi
\right\rangle $, and let $T_{y}$\ be the restriction of $T$\ obtained by
setting the first $k$ qubits to $y\in\left\{  0,1\right\}  ^{k}$. \ Clearly
$\left\vert T_{y}\right\vert \leq\left\vert T\right\vert $. \ Furthermore, we
can express $\left\vert \phi\right\rangle $\ in the form $\sum_{y\in\left\{
0,1\right\}  ^{k}}S_{y}T_{y}$, where each $S_{y}$\ represents a $k$-qubit
state and hence is expressible by a tree of size $k2^{k}$.
\end{enumerate}
\end{proof}

One can also define the $\varepsilon$\textit{-approximate tree size}
$\operatorname*{TS}_{\varepsilon}\left(  \left|  \psi\right\rangle \right)
$\ to be the minimum size of a tree representing a state $\left|
\varphi\right\rangle $\ such that $\left|  \left\langle \psi|\varphi
\right\rangle \right|  ^{2}\geq1-\varepsilon$, and define $\operatorname*{OTS}%
_{\varepsilon}\left(  \left|  \psi\right\rangle \right)  $\ and
$\operatorname*{MOTS}_{\varepsilon}\left(  \left|  \psi\right\rangle \right)
$\ similarly.

\begin{definition}
An arithmetic formula (over the ring $\mathbb{C}$ and $n$ variables) is a
rooted binary tree where each leaf vertex is labeled with either a complex
number or a variable in $\left\{  x_{1},\ldots,x_{n}\right\}  $, and each
non-leaf vertex is labeled with either $+$\ or $\times$. \ Such a tree
represents a polynomial $p\left(  x_{1},\ldots,x_{n}\right)  $\ in the obvious
way. \ We call a polynomial \textit{multilinear} if no variable appears raised
to a higher power than $1$, and an arithmetic formula multilinear if the
polynomials computed by each of its subtrees are multilinear.
\end{definition}

The \textit{size} $\left\vert \Phi\right\vert $\ of a multilinear formula
$\Phi$\ is the number of leaf vertices. \ Given a multilinear polynomial $p$,
the multilinear formula size $\operatorname*{MFS}\left(  p\right)  $\ is the
minimum size of a multilinear formula that represents $p$. \ Then given a
function $f:\left\{  0,1\right\}  ^{n}\rightarrow\mathbb{C}$, we define
\[
\operatorname*{MFS}\left(  f\right)  =\min_{p~:~p\left(  x\right)  =f\left(
x\right)  ~\forall x\in\left\{  0,1\right\}  ^{n}}\operatorname*{MFS}\left(
p\right)  .
\]
(Actually $p$ turns out to be unique \cite{ns}.) \ We can also define the
$\varepsilon$-approximate\ multilinear formula size of $f$,
\[
\operatorname*{MFS}\nolimits_{\varepsilon}\left(  f\right)  =\min
_{p~:~\left\Vert p-f\right\Vert _{2}^{2}\leq\varepsilon}\operatorname*{MFS}%
\left(  p\right)
\]
where $\left\Vert p-f\right\Vert _{2}^{2}=\sum_{x\in\left\{  0,1\right\}
^{n}}\left\vert p\left(  x\right)  -f\left(  x\right)  \right\vert ^{2}$.
\ (This metric is closely related to the inner product $\sum_{x}p\left(
x\right)  ^{\ast}f\left(  x\right)  $, but is often more convenient to work
with.) \ Now given a state $\left\vert \psi\right\rangle =\sum_{x\in\left\{
0,1\right\}  ^{n}}\alpha_{x}\left\vert x\right\rangle $ in $\mathcal{H}%
_{2}^{\otimes n}$, let $f_{\psi}$\ be the function from $\left\{  0,1\right\}
^{n}$\ to $\mathbb{C}$\ defined by $f_{\psi}\left(  x\right)  =\alpha_{x}$.

\begin{theorem}
\label{iff}For all $\left\vert \psi\right\rangle $,

\begin{enumerate}
\item[(i)] $\operatorname*{MFS}\left(  f_{\psi}\right)  =O\left(
\operatorname*{TS}\left(  \left\vert \psi\right\rangle \right)  \right)  $.

\item[(ii)] $\operatorname*{TS}\left(  \left|  \psi\right\rangle \right)
=O\left(  \operatorname*{MFS}\left(  f_{\psi}\right)  +n\right)  $.

\item[(iii)] $\operatorname*{MFS}_{\delta}\left(  f_{\psi}\right)  =O\left(
\operatorname*{TS}_{\varepsilon}\left(  \left\vert \psi\right\rangle \right)
\right)  $ where $\delta=2-2\sqrt{1-\varepsilon}$.

\item[(iv)] $\operatorname*{TS}_{2\varepsilon}\left(  \left\vert
\psi\right\rangle \right)  =O\left(  \operatorname*{MFS}_{\varepsilon}\left(
f_{\psi}\right)  +n\right)  $.
\end{enumerate}
\end{theorem}

\begin{proof}

\begin{enumerate}
\item[(i)] Given a tree representing $\left\vert \psi\right\rangle $, replace
every unbounded fan-in gate by a collection of binary gates, every $\otimes
$\ by $\times$, every $\left\vert 1\right\rangle _{i}$\ vertex by $x_{i}$, and
every $\left\vert 0\right\rangle _{i}$\ vertex by a formula for $1-x_{i}$.
\ Push all multiplications by constants at the edges down to $\times$\ gates
at the leaves.

\item[(ii)] Given a multilinear formula $\Phi$ for $f_{\psi}$, let $p\left(
v\right)  $\ be the polynomial computed at vertex $v$ of $\Phi$, and let
$S\left(  v\right)  $\ be the set of variables that appears in $p\left(
v\right)  $. \ First, call $\Phi$ \textit{syntactic} if at every $\times$ gate
with children $v$ and $w$, $S\left(  v\right)  \cap S\left(  w\right)
=\varnothing$. \ A lemma of Raz \cite{raz}\ states that we can always make
$\Phi$ syntactic without increasing its size.

Second, at every $+$\ gate $u$ with children $v$ and $w$, enlarge both
$S\left(  v\right)  $\ and $S\left(  w\right)  $\ to $S\left(  v\right)  \cup
S\left(  w\right)  $, by multiplying $p\left(  v\right)  $\ by $x_{i}+\left(
1-x_{i}\right)  $\ for every $x_{i}\in S\left(  w\right)  \setminus S\left(
v\right)  $, and multiplying $p\left(  w\right)  $\ by $x_{i}+\left(
1-x_{i}\right)  $\ for every $x_{i}\in S\left(  v\right)  \setminus S\left(
w\right)  $. \ Doing this does not invalidate any $\times$\ gate that is an
ancestor of $u$, since by the assumption that $\Phi$ is syntactic, $p\left(
u\right)  $\ is never multiplied by any polynomial containing variables in
$S\left(  v\right)  \cup S\left(  w\right)  $. \ Similarly, enlarge $S\left(
r\right)  $\ to $\left\{  x_{1},\ldots,x_{n}\right\}  $\ where $r$ is the root
of $\Phi$.

Third, call $v$ \textit{max-linear} if $\left\vert S\left(  v\right)
\right\vert =1$\ but $\left\vert S\left(  w\right)  \right\vert >1$ where $w$
is the parent of $v$. \ If $v$ is max-linear and $p\left(  v\right)
=a+bx_{i}$, then replace the tree rooted at $v$ by a tree computing
$a\left\vert 0\right\rangle _{i}+\left(  a+b\right)  \left\vert 1\right\rangle
_{i}$. \ Also, replace all multiplications by constants higher in $\Phi$ by
multiplications at the edges. \ (Because of the second step, there are no
additions by constants higher in $\Phi$.) \ Replacing every $\times$\ by
$\otimes$\ then gives a tree representing $\left\vert \psi\right\rangle $,
whose size is easily seen to be $O\left(  \left\vert \Phi\right\vert
+n\right)  $\ .

\item[(iii)] Apply the reduction from part (i). \ Let the resulting
multilinear formula compute polynomial $p$; then%
\[
\sum_{x\in\left\{  0,1\right\}  ^{n}}\left\vert p\left(  x\right)  -f_{\psi
}\left(  x\right)  \right\vert ^{2}=2-2\sum_{x\in\left\{  0,1\right\}  ^{n}%
}p\left(  x\right)  \overline{f_{\psi}\left(  x\right)  }\leq2-2\sqrt
{1-\varepsilon}=\delta.
\]

\item[(iv)] Apply the reduction from part (ii). \ Let $\left(  \beta
_{x}\right)  _{x\in\left\{  0,1\right\}  ^{n}}$\ be the resulting amplitude
vector; since this vector might not be normalized, divide each $\beta_{x}$ by
$\sum_{x}\left\vert \beta_{x}\right\vert ^{2}$ to produce $\beta_{x}^{\prime}%
$. \ Then%
\begin{align*}
\left\vert \sum_{x\in\left\{  0,1\right\}  ^{n}}\beta_{x}^{\prime}%
\overline{\alpha_{x}}\right\vert ^{2}  &  =1-\frac{1}{2}\sum_{x\in\left\{
0,1\right\}  ^{n}}\left\vert \beta_{x}^{\prime}-\alpha_{x}\right\vert ^{2}\\
&  \geq1-\frac{1}{2}\left(  \sqrt{\sum_{x\in\left\{  0,1\right\}  ^{n}%
}\left\vert \beta_{x}^{\prime}-\beta_{x}\right\vert ^{2}}+\sqrt{\sum
_{x\in\left\{  0,1\right\}  ^{n}}\left\vert \beta_{x}-\alpha_{x}\right\vert
^{2}}\right)  ^{2}\\
&  \geq1-\frac{1}{2}\left(  2\sqrt{\varepsilon}\right)  ^{2}=1-2\varepsilon.
\end{align*}

\end{enumerate}
\end{proof}

Besides $\mathsf{Tree}$, $\mathsf{OTree}$, and $\mathsf{MOTree}$, four other
classes of quantum states deserve mention:

$\mathsf{Circuit}$, a circuit analog of $\mathsf{Tree}$, contains the states
$\left\vert \psi_{n}\right\rangle =\sum_{x}\alpha_{x}\left\vert x\right\rangle
$\ such that for all $n$, there exists a multilinear arithmetic circuit of
size $p\left(  n\right)  $\ over the complex numbers that outputs $\alpha_{x}%
$\ given $x$ as input, for some polynomial $p$. \ (Multilinear circuits are
the same as multilinear trees, except that they allow unbounded fanout---that
is, polynomials computed at intermediate points can be reused arbitrarily many times.)

$\mathsf{AmpP}$ contains the states $\left|  \psi_{n}\right\rangle =\sum
_{x}\alpha_{x}\left|  x\right\rangle $\ such that for all $n,b$, there exists
a classical circuit of size $p\left(  n+b\right)  $\ that outputs $\alpha_{x}$
to $b$ bits of precision given $x$ as input, for some polynomial $p$.

$\mathsf{Vidal}$ contains the states that are `polynomially entangled' in the
sense of Vidal \cite{vidal}. \ Given a partition of $\left\{  1,\ldots
,n\right\}  $\ into $A$ and $B$, let $\chi_{A}\left(  \left\vert \psi
_{n}\right\rangle \right)  $\ be the minimum $k$ for which $\left\vert
\psi_{n}\right\rangle $\ can be written as $\sum_{i=1}^{k}\alpha_{i}\left\vert
\varphi_{i}^{A}\right\rangle \otimes\left\vert \varphi_{i}^{B}\right\rangle $,
where $\left\vert \varphi_{i}^{A}\right\rangle $\ and $\left\vert \varphi
_{i}^{B}\right\rangle $\ are states of qubits in $A$ and $B$
respectively.\ \ \ ($\chi_{A}\left(  \left\vert \psi_{n}\right\rangle \right)
$\ is known as the \textit{Schmidt rank}; see \cite{nc}\ for more
information.) \ Let $\chi\left(  \left\vert \psi_{n}\right\rangle \right)
=\max_{A}\chi_{A}\left(  \left\vert \psi_{n}\right\rangle \right)  $. \ Then
$\left\vert \psi_{n}\right\rangle \in\mathsf{Vidal}$\ if and only if
$\chi\left(  \left\vert \psi_{n}\right\rangle \right)  \leq p\left(  n\right)
$\ for some polynomial $p$.

$\mathsf{\Psi P}$ contains the states $\left\vert \psi_{n}\right\rangle $ such
that for all $n$ and $\varepsilon>0$, there exists a quantum circuit of size
$p\left(  n+\log\left(  1/\varepsilon\right)  \right)  $\ that maps the
all-$0$ state to a state some part of which has trace distance at most
$1-\varepsilon$ from $\left\vert \psi_{n}\right\rangle $, for some polynomial
$p$. \ Because of the Solovay-Kitaev Theorem \cite{kitaev:ec,nc},
$\mathsf{\Psi P}$\ is invariant under the choice of universal gate set.

\section{Basic Results\label{BASICMLIN}}

Before studying the tree size of specific quantum states, it would be nice to
know in general how tree size behaves as a complexity measure. \ In this
section I prove three rather nice properties of tree size.

\begin{theorem}
\label{logdepth}For all $\varepsilon>0$, there exists a tree representing
$\left\vert \psi\right\rangle $\ of size $O\left(  \operatorname*{TS}\left(
\left\vert \psi\right\rangle \right)  ^{1+\varepsilon}\right)  $\ and depth
$O\left(  \log\operatorname*{TS}\left(  \left\vert \psi\right\rangle \right)
\right)  $, as well as a manifestly orthogonal tree of size $O\left(
\operatorname*{MOTS}\left(  \left\vert \psi\right\rangle \right)
^{1+\varepsilon}\right)  $\ and depth $O\left(  \log\operatorname*{MOTS}%
\left(  \left\vert \psi\right\rangle \right)  \right)  $.
\end{theorem}

\begin{proof}
A classical theorem of Brent \cite{brent} says that given an arithmetic
formula $\Phi$, there exists an equivalent formula of depth $O\left(
\log\left\vert \Phi\right\vert \right)  $\ and size $O\left(  \left\vert
\Phi\right\vert ^{c}\right)  $, where $c$ is a constant. \ Bshouty, Cleve, and
Eberly \cite{bce}\ (see also Bonet and Buss \cite{bb:formulae}) improved
Brent's theorem to show that $c$\ can be taken to be $1+\varepsilon$\ for any
$\varepsilon>0$. \ So it suffices to show that, for `division-free' formulas,
these theorems preserve multilinearity (and in the $\operatorname*{MOTS}%
$\ case, preserve manifest orthogonality).

Brent's theorem is proven by induction on $\left\vert \Phi\right\vert $.
\ Here is a sketch: choose a subformula $I$ of $\Phi$ size between $\left\vert
\Phi\right\vert /3$\ and $2\left\vert \Phi\right\vert /3$\ (which one can show
always exists). \ Then identifying a subformula with the polynomial computed
at its root, $\Phi\left(  x\right)  $\ can be written as $G\left(  x\right)
+H\left(  x\right)  I\left(  x\right)  $\ for some formulas $G$ and $H$.
\ Furthermore, $G$ and $H$\ are both obtainable from $\Phi$ by removing $I$
and then applying further restrictions. \ So $\left\vert G\right\vert $ and
$\left\vert H\right\vert $ are both at most $\left\vert \Phi\right\vert
-\left\vert I\right\vert +O\left(  1\right)  $. \ Let $\widehat{\Phi}$\ be a
formula equivalent to $\Phi$ that evaluates $G$, $H$, and $I$ separately, and
then returns $G\left(  x\right)  +H\left(  x\right)  I\left(  x\right)  $.
\ Then $\left\vert \widehat{\Phi}\right\vert $\ is larger than $\left\vert
\Phi\right\vert $\ by at most a constant factor, while by the induction
hypothesis,\ we can assume the formulas for $G$, $H$, and $I$ have logarithmic
depth. \ Since the number of induction steps is $O\left(  \log\left\vert
\Phi\right\vert \right)  $, the total depth is logarithmic and the total
blowup in formula size is polynomial in $\left\vert \Phi\right\vert $.
\ Bshouty, Cleve, and Eberly's improvement uses a more careful decomposition
of $\Phi$, but the basic idea is the same.

Now, if $\Phi$ is syntactic multilinear, then clearly $G$, $H$, and $I$ are
also syntactic multilinear. \ Furthermore, $H$ cannot share variables with
$I$, since otherwise a subformula of $\Phi$ containing $I$ would have been
multiplied by a subformula containing variables from $I$. \ Thus
multilinearity is preserved. \ To see that manifest orthogonality is
preserved, suppose we are evaluating $G$ and $H$ `bottom up,' and let $G_{v}%
$\ and $H_{v}$\ be the polynomials computed at vertex $v$ of $\Phi$. \ Let
$v_{0}=\operatorname*{root}\left(  I\right)  $, let $v_{1}$\ be the parent of
$v_{0}$, let $v_{2}$\ be the parent of $v_{1}$, and so on until $v_{k}%
=\operatorname*{root}\left(  \Phi\right)  $. \ It is clear that, for every
$x$, either $G_{v_{0}}\left(  x\right)  =0$\ or $H_{v_{0}}\left(  x\right)
=0$. \ Furthermore, suppose that property holds for $G_{v_{i-1}},H_{v_{i-1}}$;
then by induction it holds for $G_{v_{i}},H_{v_{i}}$. \ If $v_{i}$\ is a
$\times$\ gate, then this follows from multilinearity (if $\left\vert
\psi\right\rangle $\ and $\left\vert \varphi\right\rangle $\ are manifestly
orthogonal, then $\left\vert 0\right\rangle \otimes\left\vert \psi
\right\rangle $\ and $\left\vert 0\right\rangle \otimes\left\vert
\varphi\right\rangle $\ are also manifestly orthogonal). \ If $v_{i}$\ is a
$+$ gate, then letting $\operatorname*{supp}\left(  p\right)  $\ be the set of
$x$ such that $p\left(  x\right)  \neq0$, any polynomial $p$ added to
$G_{v_{i-1}}$\ or $H_{v_{i-1}}$\ must have%
\[
\operatorname*{supp}\left(  p\right)  \cap\left(  \operatorname*{supp}\left(
G_{v_{i-1}}\right)  \cup\operatorname*{supp}\left(  H_{v_{i-1}}\right)
\right)  =\emptyset,
\]
and manifest orthogonality follows.
\end{proof}

\begin{theorem}
\label{prepare}Any $\left\vert \psi\right\rangle $\ can be prepared by a
quantum circuit of size polynomial in $\operatorname*{OTS}\left(  \left\vert
\psi\right\rangle \right)  $. \ Thus $\mathsf{OTree}\subseteq\mathsf{\Psi P}$.
\end{theorem}

\begin{proof}
Let $\Gamma\left(  \left\vert \psi\right\rangle \right)  $\ be the minimum
size of a circuit needed to prepare $\left\vert \psi\right\rangle
\in\mathcal{H}_{2}^{\otimes n}$\ starting from $\left\vert 0\right\rangle
^{\otimes n}$.\ \ The claim, by induction on $\Gamma\left(  \left\vert
\psi\right\rangle \right)  $, is that $\Gamma\left(  \left\vert \psi
\right\rangle \right)  \leq q\left(  \operatorname*{OTS}\left(  \left\vert
\psi\right\rangle \right)  \right)  $ for some polynomial $q$. \ The base case
$\operatorname*{OTS}\left(  \left\vert \psi\right\rangle \right)  =1$\ is
clear. \ Let $T$ be an orthogonal state tree for $\left\vert \psi\right\rangle
$, and assume without loss of generality that every gate has fan-in $2$ (this
increases $\left\vert T\right\vert $ by at most a constant factor). \ Let
$T_{1}$\ and $T_{2}$\ be the subtrees of $\operatorname*{root}\left(
T\right)  $, representing states $\left\vert \psi_{1}\right\rangle $\ and
$\left\vert \psi_{2}\right\rangle $\ respectively; note that $\left\vert
T\right\vert =\left\vert T_{1}\right\vert +\left\vert T_{2}\right\vert $.
\ First suppose $\operatorname*{root}\left(  T\right)  $\ is a $\otimes
$\ gate; then clearly $\Gamma\left(  \left\vert \psi\right\rangle \right)
\leq\Gamma\left(  \left\vert \psi_{1}\right\rangle \right)  +\Gamma\left(
\left\vert \psi_{2}\right\rangle \right)  $.

Second, suppose $\operatorname*{root}\left(  T\right)  $\ is a $+$\ gate, with
$\left\vert \psi\right\rangle =\alpha\left\vert \psi_{1}\right\rangle
+\beta\left\vert \psi_{2}\right\rangle $\ and $\left\langle \psi_{1}|\psi
_{2}\right\rangle =0$. \ Let $U$ be a quantum circuit that prepares
$\left\vert \psi_{1}\right\rangle $, and $V$\ be a circuit\ that prepares
$\left\vert \psi_{2}\right\rangle $.\ \ Then we can prepare $\alpha\left\vert
0\right\rangle \left\vert 0\right\rangle ^{\otimes n}+\beta\left\vert
1\right\rangle U^{-1}V\left\vert 0\right\rangle ^{\otimes n}$. \ Observe that
$U^{-1}V\left\vert 0\right\rangle ^{\otimes n}$\ is orthogonal to $\left\vert
0\right\rangle ^{\otimes n}$, since $\left\vert \psi_{1}\right\rangle
=U\left\vert 0\right\rangle ^{\otimes n}$\ is orthogonal to $\left\vert
\psi_{2}\right\rangle =V\left\vert 0\right\rangle ^{\otimes n}$. \ So applying
a $\operatorname*{NOT}$\ to the first register, conditioned on the
$\operatorname*{OR}$\ of the bits in the second register, yields $\left\vert
0\right\rangle \otimes\left(  \alpha\left\vert 0\right\rangle ^{\otimes
n}+\beta U^{-1}V\left\vert 0\right\rangle ^{\otimes n}\right)  $, from which
we obtain $\alpha\left\vert \psi_{1}\right\rangle +\beta\left\vert \psi
_{2}\right\rangle $\ by applying $U$ to the second register. \ The size of the
circuit used is $O\left(  \left\vert U\right\vert +\left\vert V\right\vert
+n\right)  $, with a possible constant-factor blowup arising from the need to
condition on the first register. \ If we are more careful, however, we can
combine the `conditioning' steps across multiple levels of the recursion,
producing a circuit of size $\left\vert V\right\vert +O\left(  \left\vert
U\right\vert +n\right)  $. \ By symmetry, we can also reverse the roles of $U$
and $V$ to obtain a circuit of size $\left\vert U\right\vert +O\left(
\left\vert V\right\vert +n\right)  $. \ Therefore%
\[
\Gamma\left(  \left\vert \psi\right\rangle \right)  \leq\min\left\{
\Gamma\left(  \left\vert \psi_{1}\right\rangle \right)  +c\Gamma\left(
\left\vert \psi_{2}\right\rangle \right)  +cn,\,\,c\Gamma\left(  \left\vert
\psi_{2}\right\rangle \right)  +\Gamma\left(  \left\vert \psi_{1}\right\rangle
\right)  +cn\right\}
\]
for some constant $c\geq2$. \ Solving this recurrence we find that
$\Gamma\left(  \left\vert \psi\right\rangle \right)  $\ is polynomial in
$\operatorname*{OTS}\left(  \left\vert \psi\right\rangle \right)  $.
\end{proof}

\begin{theorem}
\label{nonapx}If $\left\vert \psi\right\rangle \in\mathcal{H}_{2}^{\otimes n}%
$\ is chosen uniformly at random under the Haar measure, then
$\operatorname*{TS}_{1/16}\left(  \left\vert \psi\right\rangle \right)
=2^{\Omega\left(  n\right)  }$\ with probability $1-o\left(  1\right)  $.
\end{theorem}

\begin{proof}
To generate a uniform random state $\left\vert \psi\right\rangle =\sum
_{x\in\left\{  0,1\right\}  ^{n}}\alpha_{x}\left\vert x\right\rangle $, we can
choose $\widehat{\alpha}_{x},\widehat{\beta}_{x}\in\mathbb{R}$\ for each $x$
independently from a Gaussian distribution with mean $0$ and variance $1$,
then let $\alpha_{x}=\left(  \widehat{\alpha}_{x}+i\widehat{\beta}_{x}\right)
/\sqrt{R}$\ where $R=\sum_{x\in\left\{  0,1\right\}  ^{n}}\left(
\widehat{\alpha}_{x}^{2}+\widehat{\beta}_{x}^{2}\right)  $. \ Let%
\[
\Lambda_{\psi}=\left\{  x:\left(  \operatorname{Re}\alpha_{x}\right)
^{2}<\frac{1}{4\cdot2^{n}}\right\}  ,
\]
and let $\mathcal{G}$\ be the set of $\left\vert \psi\right\rangle $\ for
which $\left\vert \Lambda_{\psi}\right\vert <2^{n}/5$. \ The claim is that
$\Pr_{\left\vert \psi\right\rangle }\left[  \left\vert \psi\right\rangle
\in\mathcal{G}\right]  =1-o\left(  1\right)  $. \ First, $\operatorname*{EX}%
\left[  R\right]  =2^{n+1}$, so by a standard Hoeffding-type bound,
$\Pr\left[  R<2^{n}\right]  $\ is doubly-exponentially small in $n$. \ Second,
assuming $R\geq2^{n}$, for each $x$%
\[
\Pr\left[  x\in\Lambda_{\psi}\right]  \leq\Pr\left[  \widehat{\alpha}_{x}%
^{2}<\frac{1}{4}\right]  =\operatorname{erf}\left(  \frac{1}{4\sqrt{2}%
}\right)  <0.198,
\]
and the claim follows by a Chernoff bound.

For $g:\left\{  0,1\right\}  ^{n}\rightarrow\mathbb{R}$, let $A_{g}=\left\{
x:\operatorname*{sgn}\left(  g\left(  x\right)  \right)  \neq
\operatorname*{sgn}\left(  \operatorname{Re}\alpha_{x}\right)  \right\}  $,
where $\operatorname*{sgn}\left(  y\right)  $\ is $1$ if $y\geq0$\ and
$-1$\ otherwise.\ \ Then if $\left\vert \psi\right\rangle \in\mathcal{G}$,
clearly%
\[
\sum_{x\in\left\{  0,1\right\}  ^{n}}\left\vert g\left(  x\right)  -f_{\psi
}\left(  x\right)  \right\vert ^{2}\geq\frac{\left\vert A_{g}\right\vert
-\left\vert \Lambda_{\psi}\right\vert }{4\cdot2^{n}}%
\]
where $f_{\psi}\left(  x\right)  =\operatorname{Re}\alpha_{x}$, and thus%
\[
\left\vert A_{g}\right\vert \leq\left(  4\left\Vert g-f_{\psi}\right\Vert
_{2}^{2}+\frac{1}{5}\right)  2^{n}.
\]
Therefore to show that $\operatorname*{MFS}_{1/15}\left(  f_{\psi}\right)
=2^{\Omega\left(  n\right)  }$\ with probability $1-o\left(  1\right)  $, we
need only show that for almost all Boolean functions $f:\left\{  0,1\right\}
^{n}\rightarrow\left\{  -1,1\right\}  $, there is no arithmetic formula $\Phi$
of size $2^{o\left(  n\right)  }$\ such that
\[
\left\vert \left\{  x:\operatorname*{sgn}\left(  \Phi\left(  x\right)
\right)  \neq f\left(  x\right)  \right\}  \right\vert \leq0.49\cdot2^{n}.
\]
Here an arithmetic formula is real-valued, and can include addition,
subtraction, and multiplication gates of fan-in $2$ as well as constants. \ We
do not need to assume multilinearity, and it is easy to see that the
assumption of bounded fan-in is without loss of generality. \ Let $W$ be the
set of Boolean functions \textit{sign-represented} by an arithmetic formula
$\Phi$ of size $2^{o\left(  n\right)  }$, in the sense that
$\operatorname*{sgn}\left(  \Phi\left(  x\right)  \right)  =f\left(  x\right)
$ for all $x$. \ Then it suffices to show that $\left\vert W\right\vert
=2^{2^{o\left(  n\right)  }}$, since the number of functions sign-represented
on an $0.51$\ fraction of inputs is at most $\left\vert W\right\vert
\cdot2^{2^{n}H\left(  0.51\right)  }$.\ \ (Here $H$ denotes the binary entropy function.)

Let $\Phi$ be an arithmetic formula that takes as input the binary string
$x=\left(  x_{1},\ldots,x_{n}\right)  $\ as well as constants $c_{1}%
,c_{2},\ldots$. \ Let $\Phi_{c}$\ denote $\Phi$ under a particular assignment
$c$ to $c_{1},c_{2},\ldots$.\ \ Then a result of Gashkov \cite{gashkov} (see
also Tur\'{a}n and Vatan \cite{tv}), which follows from Warren's Theorem
\cite{warren}\ in real algebraic geometry, shows that as we range over all
$c$, $\Phi_{c}$ sign-represents at most $\left(  2^{n+4}\left\vert
\Phi\right\vert \right)  ^{\left\vert \Phi\right\vert }$\ distinct Boolean
functions, where $\left\vert \Phi\right\vert $\ is the size of $\Phi$.
\ Furthermore, excluding constants, the number of distinct arithmetic formulas
of size $\left\vert \Phi\right\vert $\ is at most $\left(  3\left\vert
\Phi\right\vert ^{2}\right)  ^{\left\vert \Phi\right\vert }$. \ When
$\left\vert \Phi\right\vert =2^{o\left(  n\right)  }$, this gives $\left(
3\left\vert \Phi\right\vert ^{2}\right)  ^{\left\vert \Phi\right\vert }%
\cdot\left(  2^{n+4}\left\vert \Phi\right\vert \right)  ^{\left\vert
\Phi\right\vert }=2^{2^{o\left(  n\right)  }}$. \ Therefore
$\operatorname*{MFS}_{1/15}\left(  f_{\psi}\right)  =2^{\Omega\left(
n\right)  }$; by Theorem \ref{iff}, part (iii), this implies that
$\operatorname*{TS}_{1/16}\left(  \left\vert \psi\right\rangle \right)
=2^{\Omega\left(  n\right)  }$.
\end{proof}

A corollary of Theorem \ref{nonapx} is the following `nonamplification'
property: there exist states that can be approximated to within, say,
$1\%$\ by trees of polynomial size, but that require exponentially large trees
to approximate to within a smaller margin (say $0.01\%$).

\begin{corollary}
\label{nonamp}For all $\delta\in\left(  0,1\right]  $, there exists a state
$\left\vert \psi\right\rangle $\ such that $\operatorname*{TS}_{\delta}\left(
\left\vert \psi\right\rangle \right)  =n$\ but $\operatorname*{TS}%
_{\varepsilon}\left(  \left\vert \psi\right\rangle \right)  =2^{\Omega\left(
n\right)  }$ where $\varepsilon=\delta/32-\delta^{2}/4096$.
\end{corollary}

\begin{proof}
It is clear from Theorem \ref{nonapx}\ that there exists a state $\left\vert
\varphi\right\rangle =\sum_{x\in\left\{  0,1\right\}  ^{n}}\alpha
_{x}\left\vert x\right\rangle $\ such that $\operatorname*{TS}_{1/16}\left(
\left\vert \varphi\right\rangle \right)  =2^{\Omega\left(  n\right)  }$\ and
$\alpha_{0^{n}}=0$. \ Take $\left\vert \psi\right\rangle =\sqrt{1-\delta
}\left\vert 0\right\rangle ^{\otimes n}+\sqrt{\delta}\left\vert \varphi
\right\rangle $. \ Since $\left\vert \left\langle \psi|0\right\rangle
^{\otimes n}\right\vert ^{2}=1-\delta$, we have $\operatorname*{MOTS}_{\delta
}\left(  \left\vert \psi\right\rangle \right)  =n$. \ On the other hand,
suppose some $\left\vert \phi\right\rangle =\sum_{x\in\left\{  0,1\right\}
^{n}}\beta_{x}\left\vert x\right\rangle $\ with $\operatorname*{TS}\left(
\left\vert \phi\right\rangle \right)  =2^{o\left(  n\right)  }$ satisfies
$\left\vert \left\langle \phi|\psi\right\rangle \right\vert ^{2}%
\geq1-\varepsilon$. \ Then%
\[
\sum_{x\neq0^{n}}\left(  \sqrt{\delta}\alpha_{x}-\beta_{x}\right)  ^{2}%
\leq2-2\sqrt{1-\varepsilon}.
\]
Thus, letting $f_{\varphi}\left(  x\right)  =\alpha_{x}$, we have
$\operatorname*{MFS}\nolimits_{c}\left(  f_{\varphi}\right)  =O\left(
\operatorname*{TS}\left(  \left\vert \phi\right\rangle \right)  \right)
$\ where $c=\left(  2-2\sqrt{1-\varepsilon}\right)  /\delta$. \ By Theorem
\ref{iff}, part (iv), this implies that $\operatorname*{TS}_{2c}\left(
\left\vert \varphi\right\rangle \right)  =O\left(  \operatorname*{TS}\left(
\left\vert \phi\right\rangle \right)  \right)  $. \ But $2c=1/16$\ when
$\varepsilon=\delta/32-\delta^{2}/4096$, contradiction.
\end{proof}

\section{Relations Among Quantum State Classes\label{RELMLIN}}

This section presents some results about the quantum state hierarchy
introduced in Section \ref{CQS}. \ Theorem \ref{trivrelate}\ shows simple
inclusions and separations, while Theorem \ref{psipampp} shows that
separations higher in the hierarchy would imply major complexity class
separations (and vice versa).

\begin{theorem}
\label{trivrelate}\quad

\begin{enumerate}
\item[(i)] $\mathsf{Tree}\cup\mathsf{Vidal}\subseteq\mathsf{Circuit}%
\subseteq\mathsf{AmpP}$.

\item[(ii)] All states in $\mathsf{Vidal}$ have tree size $n^{O\left(  \log
n\right)  }$.

\item[(iii)] $\mathsf{\Sigma}_{\mathsf{2}}\subseteq\mathsf{Vidal}$ but
$\mathsf{\otimes}_{\mathsf{2}}\not \subset \mathsf{Vidal}$.

\item[(iv)] $\mathsf{\otimes}_{\mathsf{2}}\subsetneq\mathsf{MOTree}$.

\item[(v)] $\mathsf{\Sigma}_{\mathsf{1}}$, $\mathsf{\Sigma}_{\mathsf{2}}$,
$\mathsf{\Sigma}_{\mathsf{3}}$, $\mathsf{\otimes}_{\mathsf{1}}$,
$\mathsf{\otimes}_{\mathsf{2}}$, and $\mathsf{\otimes}_{\mathsf{3}}$ are all
distinct. \ Also, $\mathsf{\otimes}_{\mathsf{3}}\neq\mathsf{\Sigma
}_{\mathsf{4}}\cap\mathsf{\otimes}_{\mathsf{4}}$.
\end{enumerate}
\end{theorem}

\begin{proof}

\begin{enumerate}
\item[(i)] $\mathsf{Tree}\subseteq\mathsf{Circuit}$ since any multilinear tree
is also a multilinear circuit. \ $\mathsf{Circuit}\subseteq\mathsf{AmpP}%
$\ since the circuit yields a polynomial-time algorithm for computing the
amplitudes. \ For\ $\mathsf{Vidal}\subseteq\mathsf{Circuit}$, we use an idea
of Vidal \cite{vidal}: given $\left\vert \psi_{n}\right\rangle \in
\mathsf{Vidal}$, for all $j\in\left\{  1,\ldots,n\right\}  $\ we can express
$\left\vert \psi_{n}\right\rangle $\ as%
\[
\sum_{i=1}^{\chi\left(  \left\vert \psi\right\rangle \right)  }\alpha
_{ij}\left\vert \phi_{i}^{\left[  1\ldots j\right]  }\right\rangle
\otimes\left\vert \phi_{i}^{\left[  j+1\ldots n\right]  }\right\rangle
\]
where $\chi\left(  \left\vert \psi_{n}\right\rangle \right)  $\ is
polynomially bounded. \ Furthermore, Vidal showed that each $\left\vert
\phi_{i}^{\left[  1\ldots j\right]  }\right\rangle $\ can be written as a
linear combination of states of the form $\left\vert \phi_{i}^{\left[  1\ldots
j-1\right]  }\right\rangle \otimes\left\vert 0\right\rangle $\ and $\left\vert
\phi_{i}^{\left[  1\ldots j-1\right]  }\right\rangle \otimes\left\vert
1\right\rangle $---the point being that the set of $\left\vert \phi
_{i}^{\left[  1\ldots j-1\right]  }\right\rangle $\ states is the same,
independently of $\left\vert \phi_{i}^{\left[  1\ldots j\right]
}\right\rangle $. \ This immediately yields a polynomial-size multilinear
circuit for $\left\vert \psi_{n}\right\rangle $.

\item[(ii)] Given $\left\vert \psi_{n}\right\rangle \in\mathsf{Vidal}$, we can
decompose $\left\vert \psi_{n}\right\rangle $\ as%
\[
\sum_{i=1}^{\chi\left(  \left\vert \psi\right\rangle \right)  }\alpha
_{i}\left\vert \phi_{i}^{\left[  1\ldots n/2\right]  }\right\rangle
\otimes\left\vert \phi_{i}^{\left[  n/2+1\ldots n\right]  }\right\rangle .
\]
Then $\chi\left(  \left\vert \phi_{i}^{\left[  1\ldots n/2\right]
}\right\rangle \right)  \leq\chi\left(  \left\vert \psi_{n}\right\rangle
\right)  $\ and $\chi\left(  \left\vert \phi_{i}^{\left[  n/2+1\ldots
n\right]  }\right\rangle \right)  \leq\chi\left(  \left\vert \psi
_{n}\right\rangle \right)  $ for all $i$, so we can recursively decompose
these states in the same manner. \ It follows that $\operatorname*{TS}\left(
\left\vert \psi_{n}\right\rangle \right)  \leq2\chi\left(  \left\vert
\psi\right\rangle \right)  \operatorname*{TS}\left(  \left\vert \psi
_{n/2}\right\rangle \right)  $; solving this recurrence relation yields
$\operatorname*{TS}\left(  \left\vert \psi_{n}\right\rangle \right)
\leq\left(  2\chi\left(  \left\vert \psi\right\rangle \right)  \right)  ^{\log
n}=n^{O\left(  \log n\right)  }$.

\item[(iii)] $\mathsf{\Sigma}_{\mathsf{2}}\subseteq\mathsf{Vidal}$ follows
since a sum of $t$ separable states has $\chi\leq t$,\ while $\mathsf{\otimes
}_{\mathsf{2}}\not \subset \mathsf{Vidal}$\ follows from the example of $n/2$
Bell pairs: $2^{-n/4}\left(  \left\vert 00\right\rangle +\left\vert
11\right\rangle \right)  ^{\otimes n/2}$.

\item[(iv)] $\mathsf{\otimes}_{\mathsf{2}}\subseteq\mathsf{MOTree}$ is
obvious, while\ $\mathsf{MOTree}\not \subset \mathsf{\otimes}_{\mathsf{2}}%
$\ follows from the example of $\left\vert P_{n}^{i}\right\rangle $, an equal
superposition over all $n$-bit strings of parity $i$. \ The following
recursive formulas imply that $\operatorname*{MOTS}\left(  \left\vert
P_{n}^{i}\right\rangle \right)  \leq4\operatorname*{MOTS}\left(  \left\vert
P_{n/2}^{i}\right\rangle \right)  =O\left(  n^{2}\right)  $:%
\begin{align*}
\left\vert P_{n}^{0}\right\rangle  &  =\frac{1}{\sqrt{2}}\left(  \left\vert
P_{n/2}^{0}\right\rangle \left\vert P_{n/2}^{0}\right\rangle +\left\vert
P_{n/2}^{1}\right\rangle \left\vert P_{n/2}^{1}\right\rangle \right)  ,\\
\left\vert P_{n}^{1}\right\rangle  &  =\frac{1}{\sqrt{2}}\left(  \left\vert
P_{n/2}^{0}\right\rangle \left\vert P_{n/2}^{1}\right\rangle +\left\vert
P_{n/2}^{1}\right\rangle \left\vert P_{n/2}^{0}\right\rangle \right)  .
\end{align*}
On the other hand, $\left\vert P_{n}\right\rangle \notin\mathsf{\otimes
}_{\mathsf{2}}$\ follows from $\left\vert P_{n}\right\rangle \notin
\mathsf{\Sigma}_{\mathsf{1}}$\ together with the fact that $\left\vert
P_{n}\right\rangle $\ has no nontrivial tensor product decomposition.

\item[(v)] $\mathsf{\otimes}_{\mathsf{1}}\not \subset \mathsf{\Sigma
}_{\mathsf{1}}$ and $\mathsf{\Sigma}_{\mathsf{1}}\not \subset \mathsf{\otimes
}_{\mathsf{1}}\ $are obvious.$\ \ \mathsf{\otimes}_{\mathsf{2}}\not \subset
\mathsf{\Sigma}_{\mathsf{2}}$ (and hence $\mathsf{\otimes}_{\mathsf{1}}%
\neq\mathsf{\otimes}_{\mathsf{2}}$) follows from part (iii). \ $\mathsf{\Sigma
}_{\mathsf{2}}\not \subset \mathsf{\otimes}_{\mathsf{2}}$\ (and hence
$\mathsf{\Sigma}_{\mathsf{1}}\neq\mathsf{\Sigma}_{\mathsf{2}}$) follows from
part (iv), together with the fact that $\left\vert P_{n}\right\rangle $ has a
$\mathsf{\Sigma}_{\mathsf{2}}$\ formula based on the Fourier transform:%
\[
\left\vert P_{n}\right\rangle =\frac{1}{\sqrt{2}}\left(  \left(
\frac{\left\vert 0\right\rangle +\left\vert 1\right\rangle }{\sqrt{2}}\right)
^{\otimes n}+\left(  \frac{\left\vert 0\right\rangle -\left\vert
1\right\rangle }{\sqrt{2}}\right)  ^{\otimes n}\right)  .
\]
$\mathsf{\Sigma}_{\mathsf{2}}\neq\mathsf{\Sigma}_{\mathsf{3}}$ follows from
$\mathsf{\otimes}_{\mathsf{2}}\not \subset \mathsf{\Sigma}_{\mathsf{2}}$\ and
$\mathsf{\otimes}_{\mathsf{2}}\subseteq\mathsf{\Sigma}_{\mathsf{3}}$. \ Also,
$\mathsf{\Sigma}_{\mathsf{3}}\not \subset \mathsf{\otimes}_{\mathsf{3}}$
follows from $\mathsf{\Sigma}_{\mathsf{2}}\neq\mathsf{\Sigma}_{\mathsf{3}}$,
together with the fact that we can easily construct states in $\mathsf{\Sigma
}_{\mathsf{3}}\setminus\mathsf{\Sigma}_{\mathsf{2}}$\ that have no nontrivial
tensor product decomposition---for example,%
\[
\frac{1}{\sqrt{2}}\left(  \left\vert 0\right\rangle ^{\otimes n}+\left(
\frac{\left\vert 01\right\rangle +\left\vert 10\right\rangle }{\sqrt{2}%
}\right)  ^{\otimes n/2}\right)  .
\]
$\mathsf{\otimes}_{\mathsf{2}}\neq\mathsf{\otimes}_{\mathsf{3}}\ $follows from
$\mathsf{\Sigma}_{\mathsf{2}}\not \subset \mathsf{\otimes}_{\mathsf{2}}$\ and
$\mathsf{\Sigma}_{\mathsf{2}}\subseteq\mathsf{\otimes}_{\mathsf{3}}$.
\ Finally, $\mathsf{\otimes}_{\mathsf{3}}\neq\mathsf{\Sigma}_{\mathsf{4}}%
\cap\mathsf{\otimes}_{\mathsf{4}}$\ follows from $\mathsf{\Sigma}_{\mathsf{3}%
}\not \subset \mathsf{\otimes}_{\mathsf{3}}$\ and $\mathsf{\Sigma}%
_{\mathsf{3}}\subseteq\mathsf{\Sigma}_{\mathsf{4}}\cap\mathsf{\otimes
}_{\mathsf{4}}$.
\end{enumerate}
\end{proof}

\begin{theorem}
\label{psipampp}\quad

\begin{enumerate}
\item[(i)] $\mathsf{BQP}=\mathsf{P}^{\mathsf{\#P}}$ implies $\mathsf{AmpP}%
\subseteq\mathsf{\Psi P}$.

\item[(ii)] $\mathsf{AmpP}\subseteq\mathsf{\Psi P}$ implies $\mathsf{NP}%
\subseteq\mathsf{BQP/poly}.$

\item[(iii)] $\mathsf{P}=\mathsf{P}^{\mathsf{\#P}}$ implies $\mathsf{\Psi
P}\subseteq\mathsf{AmpP}$.

\item[(iv)] $\mathsf{\Psi P}\subseteq\mathsf{AmpP}$ implies $\mathsf{BQP}%
\subseteq\mathsf{P/poly}$.
\end{enumerate}
\end{theorem}

\begin{proof}

\begin{enumerate}
\item[(i)] First, $\mathsf{BQP}=\mathsf{P}^{\mathsf{\#P}}$\ implies
$\mathsf{BQP/poly}=\mathsf{P}^{\mathsf{\#P}}\mathsf{/poly}$, since given a
$\mathsf{P}^{\mathsf{\#P}}\mathsf{/poly}$\ machine $M$, the language
consisting of all $\left(  x,a\right)  $\ such that $M$ accepts on input $x$
and advice $a$ is clearly in $\mathsf{BQP}$. \ So assume $\mathsf{BQP/poly}%
=\mathsf{P}^{\mathsf{\#P}}\mathsf{/poly}$, and consider a state $\left\vert
\psi\right\rangle =\sum_{x\in\left\{  0,1\right\}  ^{n}}\alpha_{x}\left\vert
x\right\rangle $\ with\ $\left\vert \psi\right\rangle \in\mathsf{AmpP}$. \ By
the result of Bernstein and Vazirani \cite{bv}\ that $\mathsf{BQP}%
\subseteq\mathsf{P}^{\mathsf{\#P}}$, for all $b$ there exists a quantum
circuit of size polynomial in $n$ and $b$ that approximates $p_{0}=\sum
_{y\in\left\{  0,1\right\}  ^{n-1}}\left\vert \alpha_{0y}\right\vert ^{2}$, or
the probability that the first qubit is measured to be $0$, to $b$ bits of
precision. \ So by uncomputing garbage, we can prepare a state close to
$\sqrt{p_{0}}\left\vert 0\right\rangle +\sqrt{1-p_{0}}\left\vert
1\right\rangle $. \ Similarly, given a superposition over length-$k$ prefixes
of $x$, we can prepare a superposition over length-$\left(  k+1\right)  $
prefixes of $x$ by approximating the conditional measurement probabilities.
\ We thus obtain a state close to $\sum_{x}\left\vert \alpha_{x}\right\vert
\left\vert x\right\rangle $. \ The last step is to approximate the phase of
each $\left\vert x\right\rangle $, apply that phase, and uncompute to obtain a
state close to $\sum_{x}\alpha_{x}\left\vert x\right\rangle $.

\item[(ii)] Given a $SAT$ instance $\varphi$, first use Valiant-Vazirani
\cite{vv} to produce a formula $\varphi^{\prime}$\ that (with non-negligible
probability) has one satisfying assignment if $\varphi$\ is satisfiable and
zero otherwise. \ Then let $\alpha_{x}=1$\ if $x$ is a satisfying assignment
for $\varphi^{\prime}$\ and $\alpha_{x}=0$\ otherwise; clearly $\left\vert
\psi\right\rangle =\sum_{x}\alpha_{x}\left\vert x\right\rangle $ is in
$\mathsf{AmpP}$. \ By the assumption $\mathsf{AmpP}\subseteq\mathsf{\Psi P}$,
there exists a polynomial-size quantum circuit that approximates $\left\vert
\psi\right\rangle $, and thereby finds the unique satisfying assignment for
$\varphi^{\prime}$\ if it exists.

\item[(iii)] As in part (i), $\mathsf{P}=\mathsf{P}^{\mathsf{\#P}}$\ implies
$\mathsf{P/poly}=\mathsf{P}^{\mathsf{\#P}}\mathsf{/poly}$. \ The containment
$\mathsf{\Psi P}\subseteq\mathsf{AmpP}$\ follows since we can approximate
amplitudes to polynomially many bits of precision in $\mathsf{\#P}$.

\item[(iv)] As is well known \cite{bv}, any quantum computation can be made
`clean' in the sense that it accepts if and only if a particular basis state
(say $\left\vert 0\right\rangle ^{\otimes n}$) is measured. \ The implication
follows easily.
\end{enumerate}
\end{proof}

\section{Lower Bounds\label{LOWERMLIN}}

We want to show that certain quantum states of interest to us are not
represented by trees of polynomial size. \ At first this seems like a hopeless
task. \ Proving superpolynomial formula-size lower bounds for `explicit'
functions is a notoriously hard open problem, as it would imply complexity
class separations such as $\mathsf{NC}^{1}\neq\mathsf{P}$.

Here, though, we are only concerned with \textit{multilinear} formulas.
\ Could this make it easier to prove a lower bound? \ The answer is not
obvious, but very recently, for reasons unrelated to quantum computing, Raz
\cite{raz,raz:nc2}\ showed the first superpolynomial lower bounds on
multilinear formula size. \ In particular, he showed that multilinear formulas
computing the permanent or determinant of an $n\times n$\ matrix over any
field have size $n^{\Omega\left(  \log n\right)  }$.

Raz's technique is a beautiful combination of the Furst-Saxe-Sipser method of
random restrictions \cite{fss}, with matrix rank arguments as used in
communication complexity. \ I now outline the method. \ Given a function
$f:\left\{  0,1\right\}  ^{n}\rightarrow\mathbb{C}$, let $P$ be a partition of
the input variables $x_{1},\ldots,x_{n}$\ into two collections $y=\left(
y_{1},\ldots,y_{n/2}\right)  $\ and $z=\left(  z_{1},\ldots,z_{n/2}\right)  $.
\ This yields a function $f_{P}\left(  y,z\right)  :\left\{  0,1\right\}
^{n/2}\times\left\{  0,1\right\}  ^{n/2}\rightarrow\mathbb{C}$. \ Then let
$M_{f|P}$\ be a $2^{n/2}\times2^{n/2}$\ matrix whose rows are labeled by
assignments $y\in\left\{  0,1\right\}  ^{n/2}$, and whose columns are labeled
by assignments $z\in\left\{  0,1\right\}  ^{n/2}$. \ The $\left(  y,z\right)
$\ entry of $M_{f|P}$\ is $f_{P}\left(  y,z\right)  $. \ Let
$\operatorname*{rank}\left(  M_{f|P}\right)  $\ be the rank of $M_{f|P}$ over
the complex numbers. \ Finally, let $\mathcal{P}$\ be the uniform distribution
over all partitions $P$.

The following, Corollary 3.6 in \cite{raz:nc2}, is one statement of Raz's main
theorem; recall that $\operatorname*{MFS}\left(  f\right)  $\ is the minimum
size of a multilinear formula for $f$.

\begin{theorem}
[\cite{raz:nc2}]\label{razthm}Suppose that%
\[
\Pr_{P\in\mathcal{P}}\left[  \operatorname*{rank}\left(  M_{f|P}\right)
\geq2^{n/2-\left(  n/2\right)  ^{1/8}/2}\right]  =n^{-o\left(  \log n\right)
}.
\]
Then $\operatorname*{MFS}\left(  f\right)  =n^{\Omega\left(  \log n\right)  }$.
\end{theorem}

An immediate corollary yields lower bounds on \textit{approximate} multilinear
formula size. \ Given an $N\times N$ matrix $M=\left(  m_{ij}\right)  $, let
$\operatorname*{rank}\nolimits_{\varepsilon}\left(  M\right)  =\min
_{L~:~\left\Vert L-M\right\Vert _{2}^{2}\leq\varepsilon}\operatorname*{rank}%
\left(  L\right)  $\ where $\left\Vert L-M\right\Vert _{2}^{2}=\sum
_{i,j=1}^{N}\left\vert \ell_{ij}-m_{ij}\right\vert ^{2}$.

\begin{corollary}
\label{razcor}Suppose that%
\[
\Pr_{P\in\mathcal{P}}\left[  \operatorname*{rank}\nolimits_{\varepsilon
}\left(  M_{f|P}\right)  \geq2^{n/2-\left(  n/2\right)  ^{1/8}/2}\right]
=n^{-o\left(  \log n\right)  }.
\]
Then $\operatorname*{MFS}_{\varepsilon}\left(  f\right)  =n^{\Omega\left(
\log n\right)  }$.
\end{corollary}

\begin{proof}
Suppose $\operatorname*{MFS}_{\varepsilon}\left(  f\right)  =n^{o\left(  \log
n\right)  }$. \ Then for all $g$ such that $\left\Vert f-g\right\Vert _{2}%
^{2}\leq\varepsilon$, we would have $\operatorname*{MFS}\left(  g\right)
=n^{o\left(  \log n\right)  }$, and therefore%
\[
\Pr_{P\in\mathcal{P}}\left[  \operatorname*{rank}\left(  M_{g|P}\right)
\geq2^{n/2-\left(  n/2\right)  ^{1/8}/2}\right]  =n^{-\Omega\left(  \log
n\right)  }.
\]
by Theorem \ref{razthm}. \ But $\operatorname*{rank}\nolimits_{\varepsilon
}\left(  M_{f|P}\right)  \leq\operatorname*{rank}\left(  M_{g|P}\right)  $,
and hence%
\[
\Pr_{P\in\mathcal{P}}\left[  \operatorname*{rank}\nolimits_{\varepsilon
}\left(  M_{f|P}\right)  \geq2^{n/2-\left(  n/2\right)  ^{1/8}/2}\right]
=n^{-\Omega\left(  \log n\right)  },
\]
contradiction.
\end{proof}

Another simple corollary gives lower bounds in terms of \textit{restrictions}
of $f$. \ Let $\mathcal{R}_{\ell}$ be the following distribution over
restrictions $R$: choose $2\ell$\ variables of $f$ uniformly at random, and
rename them $y=\left(  y_{1},\ldots,y_{\ell}\right)  $\ and $z=\left(
z_{1},\ldots,z_{\ell}\right)  $. \ Set each of the remaining $n-2\ell
$\ variables to $0$\ or $1$ uniformly and independently at random. \ This
yields a restricted function $f_{R}\left(  y,z\right)  $. \ Let $M_{f|R}$\ be
a $2^{\ell}\times2^{\ell}$\ matrix whose $\left(  y,z\right)  $\ entry is
$f_{R}\left(  y,z\right)  $.

\begin{corollary}
\label{razcor2}Suppose that%
\[
\Pr_{R\in\mathcal{R}_{\ell}}\left[  \operatorname*{rank}\left(  M_{f|R}%
\right)  \geq2^{\ell-\ell^{1/8}/2}\right]  =n^{-o\left(  \log n\right)  }%
\]
where $\ell=n^{\delta}$\ for some constant $\delta\in\left(  0,1\right]  $.
\ Then $\operatorname*{MFS}\left(  f\right)  =n^{\Omega\left(  \log n\right)
}$.
\end{corollary}

\begin{proof}
Under the hypothesis, clearly there exists a \textit{fixed} restriction
$g:\left\{  0,1\right\}  ^{2\ell}\rightarrow\mathbb{C}$\ of $f$, which leaves
$2\ell$\ variables unrestricted, such that%
\[
\Pr_{P\in\mathcal{P}}\left[  \operatorname*{rank}\left(  M_{g|P}\right)
\geq2^{\ell-\ell^{1/8}/2}\right]  =n^{-o\left(  \log n\right)  }%
=\ell^{-o\left(  \log\ell\right)  }.
\]
Then by Theorem \ref{razthm},%
\[
\operatorname*{MFS}\left(  f\right)  \geq\operatorname*{MFS}\left(  g\right)
=\ell^{\Omega\left(  \log\ell\right)  }=n^{\Omega\left(  \log n\right)  }.
\]

\end{proof}

The following sections apply Raz's theorem to obtain $n^{\Omega\left(  \log
n\right)  }$\ tree size\ lower bounds\ for two classes of quantum states:
states arising in quantum error-correction in Section \ref{ECC}, and (assuming
a number-theoretic conjecture) states arising in Shor's factoring algorithm in
Section \ref{DIVIS}.

\subsection{Subgroup States\label{ECC}}

Let the elements of $\mathbb{Z}_{2}^{n}$\ be labeled by $n$-bit strings.
\ Given a subgroup $S\leq\mathbb{Z}_{2}^{n}$, we define the \textit{subgroup
state} $\left\vert S\right\rangle $\ as follows:%
\[
\left\vert S\right\rangle =\frac{1}{\sqrt{\left\vert S\right\vert }}\sum_{x\in
S}\left\vert x\right\rangle .
\]
Coset states arise as codewords in the class of quantum error-correcting codes
known as stabilizer codes \cite{cs,gottesman:heis,steane}. \ Our interest in
these states, however, arises from their large tree size rather than their
error-correcting properties.

Let $\mathcal{E}$\ be the following distribution over subgroups $S$. \ Choose
an $n/2\times n$\ matrix $A$ by setting each entry to $0$ or $1$ uniformly and
independently. \ Then let $S=\left\{  x~|~Ax\equiv0\left(  \operatorname{mod}%
2\right)  \right\}  $. \ By Theorem \ref{iff}, part (i), it suffices to
lower-bound the multilinear formula size of the function $f_{S}\left(
x\right)  $, which is $1$ if $x\in S$\ and $0$\ otherwise.

\begin{theorem}
\label{ecclb}If $S$ is drawn from $\mathcal{E}$, then $\operatorname*{MFS}%
\left(  f_{S}\right)  =n^{\Omega\left(  \log n\right)  }$ (and hence
$\operatorname*{TS}\left(  \left\vert S\right\rangle \right)  =n^{\Omega
\left(  \log n\right)  }$), with probability $\Omega\left(  1\right)  $\ over
$S$.
\end{theorem}

\begin{proof}
Let $P$ be a uniform random partition of the inputs $x_{1},\ldots,x_{n}$\ of
$f_{S}$\ into two sets $y=\left(  y_{1},\ldots,y_{n/2}\right)  $\ and
$z=\left(  z_{1},\ldots,z_{n/2}\right)  $. \ Let $M_{S|P}$\ be the
$2^{n/2}\times2^{n/2}$\ matrix whose $\left(  y,z\right)  $\ entry is
$f_{S|P}\left(  y,z\right)  $; then we need to show that $\operatorname*{rank}%
\left(  M_{S|P}\right)  $\ is large with high probability. \ Let $A_{y}$\ be
the $n/2\times n/2$\ submatrix of the $n/2\times n$\ matrix $A$ consisting of
all rows that correspond to $y_{i}$\ for some $i\in\left\{  1,\ldots
,n/2\right\}  $,\ and similarly let $A_{z}$\ be the $n/2\times n/2$%
\ submatrix\ corresponding to $z$. \ Then it is easy to see that, so long as
$A_{y}$\ and $A_{z}$\ are both invertible, for all $2^{n/2}$\ settings of $y$
there exists a \textit{unique} setting of $z$ for which $f_{S|P}\left(
y,z\right)  =1$. \ This then implies that $M_{S|P}$\ is a permutation of the
identity matrix, and hence that $\operatorname*{rank}\left(  M_{S|P}\right)
=2^{n/2}$. \ Now, the probability that a random $n/2\times n/2$\ matrix over
$\mathbb{Z}_{2}$\ is invertible is%
\[
\frac{1}{2}\cdot\frac{3}{4}\cdot\cdots\cdot\frac{2^{n/2}-1}{2^{n/2}}>0.288.
\]
So the probability that $A_{y}$ and $A_{z}$ are both invertible is at least
$0.288^{2}$. \ By Markov's inequality, it follows that for at least an $0.04$
fraction of $S$'s, $\operatorname*{rank}\left(  M_{S|P}\right)  =2^{n/2}$\ for
at least an $0.04$\ fraction of $P$'s. \ Theorem \ref{razthm}\ then yields the
desired result.
\end{proof}

Aaronson and Gottesman \cite{ag}\ showed how to prepare any
$n$-qubit subgroup state using a quantum circuit of size $O\left(
n^{2}/\log n\right)  $. \ So a corollary of Theorem \ref{ecclb} is
that $\mathsf{\Psi P}\not \subset \mathsf{Tree}$. \ Since $f_{S}$
clearly has a (non-multilinear) arithmetic formula of size $O\left(
nk\right)  $, a second corollary is the following.

\begin{corollary}
\label{mlinsep}There exists a family of functions $f_{n}:\left\{  0,1\right\}
^{n}\rightarrow\mathbb{R}$\ that has polynomial-size arithmetic formulas, but
no polynomial-size multilinear formulas.
\end{corollary}

The reason Corollary \ref{mlinsep}\ does not follow from Raz's results is that
polynomial-size formulas for the permanent and determinant are not known; the
smallest known formulas for the determinant have size $n^{O\left(  \log
n\right)  }$\ (see \cite{bcs}).

We have shown that not all subgroup states are tree states, but it is still
conceivable that all subgroup states are extremely well \textit{approximated}
by tree states. \ Let us now rule out the latter possibility. \ We first need
a lemma about matrix rank, which follows from the Hoffman-Wielandt inequality.

\begin{lemma}
\label{hw}Let $M$ be an $N\times N$\ complex matrix, and let $I_{N}$\ be the
$N\times N$\ identity matrix. \ Then $\left\Vert M-I_{N}\right\Vert _{2}%
^{2}\geq N-\operatorname*{rank}\left(  M\right)  $.
\end{lemma}

\begin{proof}
The Hoffman-Wielandt inequality \cite{hw}\ (see also \cite{astvw}) states that
for any two $N\times N$\ matrices $M,P$,%
\[
\sum_{i=1}^{N}\left(  \sigma_{i}\left(  M\right)  -\sigma_{i}\left(  P\right)
\right)  ^{2}\leq\left\Vert M-P\right\Vert _{2}^{2},
\]
where $\sigma_{i}\left(  M\right)  $\ is the $i^{th}$\ singular value of
$M$\ (that is, $\sigma_{i}\left(  M\right)  =\sqrt{\lambda_{i}\left(
M\right)  }$, where $\lambda_{1}\left(  M\right)  \geq\cdots\geq\lambda
_{N}\left(  M\right)  \geq0$\ are the eigenvalues of $MM^{\ast}$, and
$M^{\ast}$\ is the conjugate transpose of $M$). \ Clearly $\sigma_{i}\left(
I_{N}\right)  =1$\ for all $i$. \ On the other hand, $M$ has only
$\operatorname*{rank}\left(  M\right)  $\ nonzero singular values, so%
\[
\sum_{i=1}^{N}\left(  \sigma_{i}\left(  M\right)  -\sigma_{i}\left(
I_{N}\right)  \right)  ^{2}\geq N-\operatorname*{rank}\left(  M\right)  .
\]

\end{proof}

Let $\widehat{f}_{S}\left(  x\right)  =f_{S}\left(  x\right)  /\sqrt
{\left\vert S\right\vert }$ be $f_{S}\left(  x\right)  $ normalized to have
$\left\Vert \widehat{f}_{S}\right\Vert _{2}^{2}=1$.

\begin{theorem}
\label{ecclbapprox}For all constants $\varepsilon\in\left[  0,1\right)  $, if
$S$ is drawn from $\mathcal{E}$, then $\operatorname*{MFS}_{\varepsilon
}\left(  \widehat{f}_{S}\right)  =n^{\Omega\left(  \log n\right)  }$ with
probability $\Omega\left(  1\right)  $\ over $S$.
\end{theorem}

\begin{proof}
As in Theorem \ref{ecclb}, we look at the matrix $M_{S|P}$\ induced by a
random partition $P=\left(  y,z\right)  $. \ We already know that for at least
an $0.04$\ fraction of $S$'s, the $y$ and $z$ variables are in one-to-one
correspondence for at least an $0.04$\ fraction of $P$'s. \ In that case
$\left\vert S\right\vert =2^{n/2}$, and therefore $M_{S|P}$\ is a permutation
of $I/\sqrt{\left\vert S\right\vert }=I/2^{n/4}$\ where $I$ is the identity.
\ It follows from Lemma \ref{hw}\ that for all matrices $M$ such that
$\left\Vert M-M_{S|P}\right\Vert _{2}^{2}\leq\varepsilon$,%
\[
\operatorname*{rank}\left(  M\right)  \geq2^{n/2}-\left\Vert \sqrt{\left\vert
S\right\vert }\left(  M-M_{S|P}\right)  \right\Vert _{2}^{2}\geq\left(
1-\varepsilon\right)  2^{n/2}%
\]
and therefore $\operatorname*{rank}\nolimits_{\varepsilon}\left(
M_{S|P}\right)  \geq\left(  1-\varepsilon\right)  2^{n/2}$. \ Hence%
\[
\Pr_{P\in\mathcal{P}}\left[  \operatorname*{rank}\nolimits_{\varepsilon
}\left(  M_{f|P}\right)  \geq2^{n/2-\left(  n/2\right)  ^{1/8}/2}\right]
\geq0.04,
\]
and the result follows from Corollary \ref{razcor}.
\end{proof}

A corollary of Theorem \ref{ecclbapprox}\ and of Theorem \ref{iff}, part
(iii), is that $\operatorname*{TS}_{\varepsilon}\left(  \left\vert
S\right\rangle \right)  =n^{\Omega\left(  \log n\right)  }$ with probability
$\Omega\left(  1\right)  $\ over $S$, for all $\varepsilon<1$.

Finally, let me show how to derandomize the lower bound for subgroup states,
using ideas pointed out to me by Andrej Bogdanov. \ In the proof of Theorem
\ref{ecclb}, all we needed about the matrix $A$ was that a random $k\times
k$\ submatrix has full rank with $\Omega\left(  1\right)  $ probability, where
$k=n/2$. \ If we switch from the field $\mathbb{F}_{2}$\ to $\mathbb{F}%
_{2^{d}}$ for some $d\geq\log_{2}n$, then it is easy to construct explicit
$k\times n$\ matrices with this same property. \ For example, let%
\[
V=\left(
\begin{array}
[c]{cccc}%
1^{0} & 1^{1} & \cdots & 1^{k-1}\\
2^{0} & 2^{1} & \cdots & 2^{k-1}\\
\vdots & \vdots &  & \vdots\\
n^{0} & n^{1} & \cdots & n^{k-1}%
\end{array}
\right)
\]
be the $n\times k$ Vandermonde matrix, where $1,\ldots,n$\ are labels of
elements in $\mathbb{F}_{2^{d}}$. \ \textit{Any} $k\times k$\ submatrix of $V$
has full rank, because the Reed-Solomon (RS) code that $V$ represents is a
perfect erasure code.\footnote{In other words, because a degree-$\left(
k-1\right)  $ polynomial is determined by its values at any $k$ points.}
\ Hence, there exists an explicit state of $n$ \textquotedblleft
qupits\textquotedblright\ with $p=2^{d}$\ that has tree size $n^{\Omega\left(
\log n\right)  }$---namely the uniform superposition over all elements of the
set $\left\{  x~|~V^{T}x=0\right\}  $, where $V^{T}$\ is the transpose of $V$.

To replace qupits by qubits, we concatenate the RS and Hadamard codes to
obtain a \textit{binary} linear erasure code with parameters almost as good as
those of the original RS code. \ More explicitly, interpret $\mathbb{F}%
_{2^{d}}$\ as the field of polynomials over $\mathbb{F}_{2}$, modulo some
irreducible of degree $d$. \ Then let $m\left(  a\right)  $\ be the $d\times
d$\ Boolean matrix that maps $q\in\mathbb{F}_{2^{d}}$\ to $aq\in
\mathbb{F}_{2^{d}}$, where $q$\ and $aq$\ are encoded by their $d\times
1$\ vectors of coefficients. \ Let $H$ map a length-$d$ vector to its
length-$2^{d}$ Hadamard encoding. \ Then $Hm\left(  a\right)  $\ is a
$2^{d}\times d$\ Boolean matrix that maps $q\in\mathbb{F}_{2^{d}}$\ to the
Hadamard encoding of $aq$. \ We can now define an $n2^{d}\times kd$%
\ \textquotedblleft binary Vandermonde matrix\textquotedblright\ as follows:%
\[
V_{\operatorname*{bin}}=\left(
\begin{array}
[c]{cccc}%
Hm\left(  1^{0}\right)  & Hm\left(  1^{1}\right)  & \cdots & Hm\left(
1^{k-1}\right) \\
Hm\left(  2^{0}\right)  & Hm\left(  2^{1}\right)  & \cdots & Hm\left(
2^{k-1}\right) \\
\vdots & \vdots & ~ & \vdots\\
Hm\left(  n^{0}\right)  & Hm\left(  n^{1}\right)  & \cdots & Hm\left(
n^{k-1}\right)
\end{array}
\right)  .
\]
For the remainder of the section, fix $k=n^{\delta}$\ for some $\delta
<1/2$\ and $d=O\left(  \log n\right)  $.

\begin{lemma}
\label{kd}A $\left(  kd+c\right)  \times kd$\ submatrix of
$V_{\operatorname*{bin}}$\ chosen uniformly at random has rank $kd$\ (that is,
full rank) with probability at least $2/3$, for $c$ a sufficiently large constant.
\end{lemma}

\begin{proof}
The first claim is that $\left\vert V_{\operatorname*{bin}}u\right\vert
\geq\left(  n-k\right)  2^{d-1}$\ for all nonzero vectors $u\in\mathbb{F}%
_{2}^{kd}$, where $\left\vert ~~\right\vert $\ represents the number of
`$1$'\ bits. \ To see this, observe that for all nonzero $u$, the
\textquotedblleft codeword vector\textquotedblright\ $Vu\in\mathbb{F}_{2^{d}%
}^{n}$\ must have at least $n-k$ nonzero entries by the Fundamental Theorem of
Algebra, where here $u$ is interpreted as an element of $\mathbb{F}_{2^{d}%
}^{k}$. \ Furthermore, the Hadamard code maps any nonzero entry in $Vu$\ to
$2^{d-1}$\ nonzero bits in $V_{\operatorname*{bin}}u\in\mathbb{F}_{2}^{n2^{d}%
}$.

Now let $W$\ be a uniformly random $\left(  kd+c\right)  \times kd$\ submatrix
of $V_{\operatorname*{bin}}$. \ By the above claim, for any fixed nonzero
vector $u\in\mathbb{F}_{2}^{kd}$,%
\[
\Pr_{W}\left[  Wu=0\right]  \leq\left(  1-\frac{\left(  n-k\right)  2^{d-1}%
}{n2^{d}}\right)  ^{kd+c}=\left(  \frac{1}{2}+\frac{k}{2n}\right)  ^{kd+c}.
\]
So by the union bound, $Wu$\ is nonzero for all nonzero $u$ (and hence $W$ is
full rank) with probability at least%
\[
1-2^{kd}\left(  \frac{1}{2}+\frac{k}{2n}\right)  ^{kd+c}=1-\left(  1+\frac
{k}{n}\right)  ^{kd}\left(  \frac{1}{2}+\frac{k}{2n}\right)  ^{c}.
\]
Since $k=n^{1/2-\Omega\left(  1\right)  }$\ and $d=O\left(  \log n\right)  $,
the above quantity is at least $2/3$\ for sufficiently large $c$.
\end{proof}

Given an $n2^{d}\times1$\ Boolean vector $x$, let $f\left(  x\right)  =1$\ if
$V_{\operatorname*{bin}}^{T}x=0$\ and $f\left(  x\right)  =0$\ otherwise. \ Then:

\begin{theorem}
\label{explicit}$\operatorname*{MFS}\left(  f\right)  =n^{\Omega\left(  \log
n\right)  }$.
\end{theorem}

\begin{proof}
Let $V_{y}$\ and $V_{z}$ be two disjoint $kd\times\left(  kd+c\right)
$\ submatrices of $V_{\operatorname*{bin}}^{T}$ chosen uniformly at random.
\ Then by Lemma \ref{kd}\ together with the union bound, $V_{y}$\ and $V_{z}%
$\ both have full rank with probability at least $1/3$. \ Letting $\ell=kd+c$,
it follows that%
\[
\Pr_{R\in\mathcal{R}_{\ell}}\left[  \operatorname*{rank}\left(  M_{f|R}%
\right)  \geq2^{\ell-c}\right]  \geq\frac{1}{3}=n^{-o\left(  \log n\right)  }%
\]
by the same reasoning as in Theorem \ref{ecclb}. \ Therefore
$\operatorname*{MFS}\left(  f\right)  =n^{\Omega\left(  \log n\right)  }$\ by
Corollary \ref{razcor2}.
\end{proof}

Let $\left\vert S\right\rangle $\ be a uniform superposition over all $x$ such
that $f\left(  x\right)  =1$; then a corollary of Theorem \ref{explicit}\ is
that $\operatorname*{TS}\left(  \left\vert S\right\rangle \right)
=n^{\Omega\left(  \log n\right)  }$. \ Naturally, using the ideas of
Theorem\ \ref{ecclbapprox} one can also show that $\operatorname*{TS}%
_{\varepsilon}\left(  \left\vert S\right\rangle \right)  =n^{\Omega\left(
\log n\right)  }$\ for all $\varepsilon<1$.

\subsection{Shor States\label{DIVIS}}

Since the motivation for this work was to study possible Sure/Shor separators,
an obvious question is, \textit{do states arising in Shor's algorithm have
superpolynomial tree size?} \ Unfortunately, I am only able to answer this
question assuming a number-theoretic conjecture. \ To formalize the
question,\ let%
\[
\frac{1}{2^{n/2}}\sum_{r=0}^{2^{n}-1}\left\vert r\right\rangle \left\vert
x^{r}\operatorname{mod}N\right\rangle
\]
be a Shor state. \ It will be convenient to measure the second register, so
that the state of the first register has the form%
\[
\left\vert a+p\mathbb{Z}\right\rangle =\frac{1}{\sqrt{I}}\sum_{i=0}%
^{I}\left\vert a+pi\right\rangle
\]
for some integers $a<p$\ and $I=\left\lfloor \left(  2^{n}-a-1\right)
/p\right\rfloor $. \ Here $a+pi$\ is written out in binary using $n$ bits.
\ Clearly a lower bound on $\operatorname*{TS}\left(  \left\vert
a+p\mathbb{Z}\right\rangle \right)  $\ would imply an equivalent lower bound
for the joint state of the two registers.

To avoid some technicalities, assume $p$ is prime (since the goal is to prove
a \textit{lower} bound, this assumption is without loss of generality).
\ Given an $n$-bit string $x=x_{n-1}\ldots x_{0}$, let $f_{n,p,a}\left(
x\right)  =1$\ if $x\equiv a\left(  \operatorname{mod}p\right)  $\ and
$f_{n,p,a}\left(  x\right)  =0$\ otherwise. \ Then $\operatorname*{TS}\left(
\left\vert a+p\mathbb{Z}\right\rangle \right)  =\Theta\left(
\operatorname*{MFS}\left(  f_{n,p,a}\right)  \right)  $\ by Theorem \ref{iff},
so from now on we will focus attention on $f_{n,p,a}$.

\begin{proposition}
\label{trivshor}\quad

\begin{enumerate}
\item[(i)] Let $f_{n,p}=f_{n,p,0}$. \ Then$\ \operatorname*{MFS}\left(
f_{n,p,a}\right)  \leq\operatorname*{MFS}\left(  f_{n+\log p,p}\right)  $,
meaning that we can set $a=0$ without loss of generality.

\item[(ii)] $\operatorname*{MFS}\left(  f_{n,p}\right)  =O\left(
\operatorname*{min}\left\{  n2^{n}/p,np\right\}  \right)  $.
\end{enumerate}
\end{proposition}

\begin{proof}
\quad

\begin{enumerate}
\item[(i)] Take the formula for $f_{n+\log p,p}$, and restrict the most
significant $\log p$\ bits to sum to a number congruent to
$-a\operatorname{mod}p$ (this is always possible since $x\rightarrow2^{n}%
x$\ is an isomorphism of $\mathbb{Z}_{p}$).

\item[(ii)] For $\operatorname*{MFS}\left(  f_{n,p}\right)  =O\left(
n2^{n}/p\right)  $, write out the $x$'s for which $f_{n,p}\left(  x\right)
=1$\ explicitly. \ For $\operatorname*{MFS}\left(  f_{n,p}\right)  =O\left(
np\right)  $, use the Fourier transform, similarly to Theorem \ref{trivrelate}%
, part (v):%
\[
f_{n,p}\left(  x\right)  =\frac{1}{p}\sum_{h=0}^{p-1}%
%TCIMACRO{\dprod \limits_{j=0}^{n-1}}%
%BeginExpansion
{\displaystyle\prod\limits_{j=0}^{n-1}}
%EndExpansion
\exp\left(  \frac{2\pi ih}{p}\cdot2^{j}x_{j}\right)  .
\]
This immediately yields a sum-of-products formula of size $O\left(  np\right)
$.
\end{enumerate}
\end{proof}

I now state the number-theoretic conjecture.

\begin{conjecture}
\label{primes}There exist constants $\gamma,\delta\in\left(  0,1\right)
$\ and a prime $p=\Omega\left(  2^{n^{\delta}}\right)  $ for which the
following holds. \ Let the set $A$ consist of $n^{\delta}$\ elements of
$\left\{  2^{0},\ldots,2^{n-1}\right\}  $ chosen uniformly at random. \ Let
$S$ consist of all $2^{n^{\delta}}$ sums of subsets of $A$,\ and let
$S\operatorname{mod}p=\left\{  x\operatorname{mod}p:x\in S\right\}  $. \ Then%
\[
\Pr_{A}\left[  \left\vert S\operatorname{mod}p\right\vert \geq\left(
1+\gamma\right)  \frac{p}{2}\right]  =n^{-o\left(  \log n\right)  }.
\]

\end{conjecture}

\begin{theorem}
\label{conjimp}Conjecture \ref{primes}\ implies that\ $\operatorname*{MFS}%
\left(  f_{n,p}\right)  =n^{\Omega\left(  \log n\right)  }$\ and hence
$\operatorname*{TS}\left(  \left\vert p\mathbb{Z}\right\rangle \right)
=n^{\Omega\left(  \log n\right)  }$.
\end{theorem}

\begin{proof}
Let $f=f_{n,p}$ and $\ell=n^{\delta}$. \ Let $R$\ be a restriction of $f$ that
renames $2\ell$\ variables $y_{1},\ldots,y_{\ell},z_{1},\ldots,z_{\ell}$, and
sets each of the remaining $n-2\ell$\ variables to $0$\ or $1$. \ This leads
to a new function, $f_{R}\left(  y,z\right)  $, which is $1$ if $y+z+c\equiv
0\left(  \operatorname{mod}p\right)  $\ and $0$ otherwise for some constant
$c$. \ Here we are defining $y=2^{a_{1}}y_{1}+\cdots+2^{a_{\ell}}y_{\ell}%
$\ and $z=2^{b_{1}}z_{1}+\cdots+2^{b_{\ell}}z_{\ell}$\ where $a_{1}%
,\ldots,a_{\ell},b_{1},\ldots,b_{\ell}$\ are the appropriate place values.
\ Now suppose $y\operatorname{mod}p$\ and $z\operatorname{mod}p$\ both assume
at least $\left(  1+\gamma\right)  p/2$\ distinct values as we range over all
$x\in\left\{  0,1\right\}  ^{n}$. \ Then by the pigeonhole principle, for at
least $\gamma p$\ possible values of $y\operatorname{mod}p$, there exists a
unique possible value of $z\operatorname{mod}p$\ for which $y+z+c\equiv
0\left(  \operatorname{mod}p\right)  $\ and hence $f_{R}\left(  y,z\right)
=1$. \ So $\operatorname*{rank}\left(  M_{f|R}\right)  \geq\gamma p$, where
$M_{f|R}$\ is the $2^{\ell}\times2^{\ell}$\ matrix whose $\left(  y,z\right)
$\ entry is $f_{R}\left(  y,z\right)  $. \ It follows that assuming Conjecture
\ref{primes},%
\[
\Pr_{R\in\mathcal{R}_{\ell}}\left[  \operatorname*{rank}\left(  M_{f|R}%
\right)  \geq\gamma p\right]  =n^{-o\left(  \log n\right)  }.
\]
Furthermore, $\gamma p\geq2^{\ell-\ell^{1/8}/2}$\ for sufficiently large $n$
since $p=\Omega\left(  2^{n^{\delta}}\right)  $. \ Therefore
$\operatorname*{MFS}\left(  f\right)  =n^{\Omega\left(  \log n\right)  }$\ by
Corollary \ref{razcor2}.
\end{proof}

Using the ideas of Theorem \ref{ecclbapprox},\ one can show that under the
same conjecture, $\operatorname*{MFS}_{\varepsilon}\left(  f_{n,p}\right)
=n^{\Omega\left(  \log n\right)  }$\ and $\operatorname*{TS}_{\varepsilon
}\left(  \left\vert p\mathbb{Z}\right\rangle \right)  =n^{\Omega\left(  \log
n\right)  }$ for all $\varepsilon<1$---in other words, there exist Shor states
that cannot be approximated by polynomial-size trees.

Originally, I had stated Conjecture \ref{primes}\ without any restriction on
how the set $S$ is formed. \ The resulting conjecture was far more general
than I needed, and indeed was falsified by Carl Pomerance (personal communication).

\subsection{Tree Size and Persistence of Entanglement\label{PERSIST}}

In this section I pursue a deeper understanding of the tree size lower bounds,
by discussing a \textit{physical} property of quantum states that is related
to error-correction as well as to superpolynomial tree size. \ D\"{u}r and
Briegel \cite{db} call a quantum state \textquotedblleft persistently
entangled,\textquotedblright\ if (roughly speaking)\ it remains highly
entangled even after a limited amount of interaction with its environment.
\ As an illustration, the Schr\"{o}dinger cat state $\left(  \left\vert
0\right\rangle ^{\otimes n}+\left\vert 1\right\rangle ^{\otimes n}\right)
/\sqrt{2}$\ is in some sense highly entangled, but it is \textit{not}
persistently entangled, since measuring a single qubit in the standard basis
destroys all entanglement.

By contrast, consider the \textquotedblleft cluster states\textquotedblright%
\ defined by Briegel and Raussendorf \cite{br}. \ These states have attracted
a great deal of attention because of their application to quantum computing
via $1$-qubit measurements only \cite{rbb}. \ For our purposes, a
two-dimensional cluster state is an equal superposition over all settings of a
$\sqrt{n}\times\sqrt{n}$\ array of bits,\ with each basis state having a phase
of $\left(  -1\right)  ^{r}$, where $r$ is the number of horizontally or
vertically adjacent pairs of bits that are both `$1$'. \ D\"{u}r and Briegel
\cite{db} showed that such states are persistently entangled in a precise
sense:\ one can distill $n$-partite entanglement from them even after each
qubit has interacted with a heat bath for an amount of time independent of $n$.

Persistence of entanglement seems related to how one shows tree size lower
bounds using Raz's technique. \ For to apply Corollary \ref{razcor2}, one
basically \textquotedblleft measures\textquotedblright\ most of a state's
qubits, then partitions the unmeasured qubits into two subsystems of equal
size, and argues that with high probability those two subsystems are still
almost maximally entangled. \ The connection is not perfect, though. \ For one
thing, setting most of the qubits to $0$ or $1$ uniformly at random is not the
same as measuring them. \ For another, Theorem \ref{razthm}\ yields
$n^{\Omega\left(  \log n\right)  }$\ tree size lower bounds without the need
to trace out a subset of qubits. \ It suffices for the \textit{original} state
to be almost maximally entangled, no matter how one partitions it into two
subsystems of equal size.

But what about $2$-D cluster states---do \textit{they} have tree size
$n^{\Omega\left(  \log n\right)  }$? \ I strongly conjecture that the answer
is `yes.' \ However, proving this conjecture will almost certainly require
going beyond Theorem \ref{razthm}. \ One will want to use random restrictions
that respect the $2$-D neighborhood structure of cluster states---similar to
the restrictions used by Raz \cite{raz}\ to show that the permanent and
determinant have multilinear formula size $n^{\Omega\left(  \log n\right)  }$.

I end this section by showing that there exist states that are persistently
entangled in the sense of D\"{u}r and Briegel \cite{db}, but that have
polynomial tree size. \ In particular, D\"{u}r and Briegel showed that even
\textit{one}-dimensional cluster states are persistently entangled. \ On the
other hand:

\begin{proposition}
\label{1d}Let%
\[
\left\vert \psi\right\rangle =\frac{1}{2^{n/2}}\sum_{x\in\left\{  0,1\right\}
^{n}}\left(  -1\right)  ^{x_{1}x_{2}+x_{2}x_{3}+\cdots+x_{n-1}x_{n}}\left\vert
x\right\rangle .
\]
Then $\operatorname*{TS}\left(  \left\vert \psi\right\rangle \right)
=O\left(  n^{4}\right)  $.
\end{proposition}

\begin{proof}
Given bits $i,j,k$, let $\left\vert P_{n}^{ijk}\right\rangle $\ be an equal
superposition over all $n$-bit strings $x_{1}\ldots x_{n}$\ such that
$x_{1}=i$, $x_{n}=k$, and $x_{1}x_{2}+\cdots+x_{n-1}x_{n}\equiv j\left(
\operatorname{mod}2\right)  $. \ Then%
\begin{align*}
\left\vert P_{n}^{i0k}\right\rangle  &  =\frac{1}{\sqrt{8}}\left(
\begin{array}
[c]{c}%
\left\vert P_{n/2}^{i00}\right\rangle \left\vert P_{n/2}^{00k}\right\rangle
+\left\vert P_{n/2}^{i10}\right\rangle \left\vert P_{n/2}^{01k}\right\rangle
+\left\vert P_{n/2}^{i00}\right\rangle \left\vert P_{n/2}^{10k}\right\rangle
+\left\vert P_{n/2}^{i10}\right\rangle \left\vert P_{n/2}^{11k}\right\rangle
+\\
\left\vert P_{n/2}^{i01}\right\rangle \left\vert P_{n/2}^{00k}\right\rangle
+\left\vert P_{n/2}^{i11}\right\rangle \left\vert P_{n/2}^{01k}\right\rangle
+\left\vert P_{n/2}^{i01}\right\rangle \left\vert P_{n/2}^{11k}\right\rangle
+\left\vert P_{n/2}^{i11}\right\rangle \left\vert P_{n/2}^{10k}\right\rangle
\end{array}
\right)  ,\\
\left\vert P_{n}^{i1k}\right\rangle  &  =\frac{1}{\sqrt{8}}\left(
\begin{array}
[c]{c}%
\left\vert P_{n/2}^{i00}\right\rangle \left\vert P_{n/2}^{01k}\right\rangle
+\left\vert P_{n/2}^{i10}\right\rangle \left\vert P_{n/2}^{00k}\right\rangle
+\left\vert P_{n/2}^{i00}\right\rangle \left\vert P_{n/2}^{11k}\right\rangle
+\left\vert P_{n/2}^{i10}\right\rangle \left\vert P_{n/2}^{10k}\right\rangle
+\\
\left\vert P_{n/2}^{i01}\right\rangle \left\vert P_{n/2}^{01k}\right\rangle
+\left\vert P_{n/2}^{i11}\right\rangle \left\vert P_{n/2}^{00k}\right\rangle
+\left\vert P_{n/2}^{i01}\right\rangle \left\vert P_{n/2}^{10k}\right\rangle
+\left\vert P_{n/2}^{i11}\right\rangle \left\vert P_{n/2}^{11k}\right\rangle
\end{array}
\right)  .
\end{align*}
Therefore $\operatorname*{TS}\left(  \left\vert P_{n}^{ijk}\right\rangle
\right)  \leq16\operatorname*{TS}\left(  \left\vert P_{n/2}^{ijk}\right\rangle
\right)  $, and solving this recurrence relation yields
\[\operatorname*{TS}%
\left(  \left\vert P_{n}^{ijk}\right\rangle \right)  =O\left(
n^{4}\right).
\]
Finally observe that%
\[
\left\vert \psi\right\rangle =\left(  \frac{\left\vert 0\right\rangle
+\left\vert 1\right\rangle }{\sqrt{2}}\right)  ^{\otimes n}-\frac{\left\vert
P_{n}^{010}\right\rangle +\left\vert P_{n}^{011}\right\rangle +\left\vert
P_{n}^{110}\right\rangle +\left\vert P_{n}^{111}\right\rangle }{\sqrt{2}}.
\]

\end{proof}

\section{Manifestly Orthogonal Tree Size\label{MOTS}}

This section studies the manifestly orthogonal tree size of coset
states:\footnote{All results apply equally well to the subgroup states of
Section \ref{ECC}; the greater generality of coset states is just for
convenience.} states having the form%
\[
\left\vert C\right\rangle =\frac{1}{\sqrt{\left\vert C\right\vert }}\sum_{x\in
C}\left\vert x\right\rangle
\]
where $C=\left\{  x~|~Ax\equiv b\right\}  $\ is a coset in $\mathbb{Z}_{2}%
^{n}$. \ In particular, I present a \textit{tight} characterization of
$\operatorname*{MOTS}\left(  \left\vert C\right\rangle \right)  $, which
enables me to prove \textit{exponential} lower bounds on it, in contrast to
the $n^{\Omega\left(  \log n\right)  }$ lower bounds for ordinary tree size.
\ This characterization also yields a separation between orthogonal and
manifestly orthogonal tree size; and an algorithm for computing
$\operatorname*{MOTS}\left(  \left\vert C\right\rangle \right)  $\ whose
complexity is only singly exponential in $n$. \ My proof technique is
independent of Raz's, and is highly tailored to take advantage of manifest
orthogonality. \ However, even if this technique finds no other application,
it has two features that I\ hope will make it of independent interest to
complexity theorists. \ First, it yields tight lower bounds, and second, it
does not obviously \textquotedblleft naturalize\textquotedblright\ in the
sense of Razborov and Rudich \cite{rr}. \ Rather, it takes advantage of
certain structural properties of coset states that do not seem to hold for
random states.

Given a state $\left\vert \psi\right\rangle $, recall that the manifestly
orthogonal tree size $\operatorname*{MOTS}\left(  \left\vert \psi\right\rangle
\right)  $ is the minimum size of a tree representing $\left\vert
\psi\right\rangle $, in which all additions are of two states $\left\vert
\psi_{1}\right\rangle ,\left\vert \psi_{2}\right\rangle $\ with
\textquotedblleft disjoint supports\textquotedblright---that
is,\ either\ $\left\langle \psi_{1}|x\right\rangle =0$ or $\left\langle
\psi_{2}|x\right\rangle =0$ for every basis state $\left\vert x\right\rangle
$. \ Here the size $\left\vert T\right\vert $\ of $T$ is the number of leaf
vertices. \ We can assume without loss of generality that every $+$\ or
$\otimes$\ vertex has at least one child, and that every child of a
$+$\ vertex is a $\otimes$\ vertex and vice versa. \ Also, given a set
$S\subseteq\left\{  0,1\right\}  ^{n}$, let%
\[
\left\vert S\right\rangle =\frac{1}{\sqrt{\left\vert S\right\vert }}\sum_{x\in
S}\left\vert x\right\rangle
\]
be a uniform superposition over the elements of $S$, and let
$M\left( S\right) := \operatorname*{MOTS}\left(  \left\vert
S\right\rangle \right)  $.

Let $C=\left\{  x:Ax\equiv b\right\}  $ be a subgroup in $\mathbb{Z}_{2}^{n}$,
for some $A\in\mathbb{Z}_{2}^{k\times n}$\ and $b\in\mathbb{Z}_{2}^{k}%
$.\ \ Let $\left[  n\right]  =\left\{  1,\ldots,n\right\}  $, and let $\left(
I,J\right)  $\ be a nontrivial partition of $\left[  n\right]  $ (one where
$I$\ and $J$ are both nonempty). \ Then clearly there exist distinct cosets
$C_{I}^{\left(  1\right)  },\ldots,C_{I}^{\left(  H\right)  }$\ in the $I$
subsystem, and distinct cosets\ $C_{J}^{\left(  1\right)  },\ldots
,C_{J}^{\left(  H\right)  }$\ in the $J$\ subsystem, such that%
\[
C=%
%TCIMACRO{\dbigcup \limits_{h\in\left[  H\right]  }}%
%BeginExpansion
{\displaystyle\bigcup\limits_{h\in\left[  H\right]  }}
%EndExpansion
C_{I}^{\left(  h\right)  }\otimes C_{J}^{\left(  h\right)  }.
\]
The $C_{I}^{\left(  h\right)  }$'s\ and $C_{J}^{\left(  h\right)  }$'s\ are
unique up to ordering.\ \ Furthermore, the quantities $\left\vert
C_{I}^{\left(  h\right)  }\right\vert $, $\left\vert C_{J}^{\left(  h\right)
}\right\vert $, $M\left(  C_{I}^{\left(  h\right)  }\right)  $, and $M\left(
C_{J}^{\left(  h\right)  }\right)  $\ remain unchanged as we range over
$h\in\left[  H\right]  $. \ For this reason I suppress the dependence on $h$
when mentioning them.

For various sets $S$, the strategy will be to analyze $M\left(  S\right)
/\left\vert S\right\vert $, the ratio of tree size to cardinality. \ We can
think of this ratio as the \textquotedblleft price per pound\textquotedblright%
\ of $S$: the number of vertices that we have to pay per basis state that we
cover. \ The following lemma says that, under that cost measure, a coset is
\textquotedblleft as good a deal\textquotedblright\ as any of its subsets:

\begin{lemma}
\label{induclem}For all cosets $C$,%
\[
\frac{M\left(  C\right)  }{\left\vert C\right\vert }=\min\left(
\frac{M\left(  S\right)  }{\left\vert S\right\vert }\right)
\]
where the minimum is over nonempty $S\subseteq C$.
\end{lemma}

\begin{proof}
By induction on $n$. \ The base case $n=1$\ is obvious, so\ assume the lemma
true for $n-1$. \ Choose $S^{\ast}\subseteq C$ to minimize $M\left(  S^{\ast
}\right)  /\left\vert S^{\ast}\right\vert $. \ Let $T$\ be a manifestly
orthogonal tree for $\left\vert S^{\ast}\right\rangle $ of minimum size, and
let $v$ be the root of $T$. \ We can assume without loss of generality that
$v$ is a $\otimes$ vertex, since otherwise $v$ has some $\otimes$\ child
representing a set $R\subset S^{\ast}$\ such that $M\left(  R\right)
/\left\vert R\right\vert \leq M\left(  S^{\ast}\right)  /\left\vert S^{\ast
}\right\vert $. \ Therefore for some nontrivial partition $\left(  I,J\right)
$\ of $\left[  n\right]  $, and some $S_{I}^{\ast}\subseteq\left\{
0,1\right\}  ^{\left\vert I\right\vert }$ and $S_{J}^{\ast}\subseteq\left\{
0,1\right\}  ^{\left\vert J\right\vert }$, we have%
\begin{align*}
\left\vert S^{\ast}\right\rangle  &  =\left\vert S_{I}^{\ast}\right\rangle
\otimes\left\vert S_{J}^{\ast}\right\rangle ,\\
\left\vert S^{\ast}\right\vert  &  =\left\vert S_{I}^{\ast}\right\vert
\left\vert S_{J}^{\ast}\right\vert ,\\
M\left(  S^{\ast}\right)   &  =M\left(  S_{I}^{\ast}\right)  +M\left(
S_{J}^{\ast}\right)  ,
\end{align*}
where the last equality holds because if $M\left(  S^{\ast}\right)  <M\left(
S_{I}^{\ast}\right)  +M\left(  S_{J}^{\ast}\right)  $, then $T$ was not a
minimal tree for $\left\vert S^{\ast}\right\rangle $. \ Then%
\[
\frac{M\left(  S^{\ast}\right)  }{\left\vert S^{\ast}\right\vert }%
=\frac{M\left(  S_{I}^{\ast}\right)  +M\left(  S_{J}^{\ast}\right)
}{\left\vert S_{I}^{\ast}\right\vert \left\vert S_{J}^{\ast}\right\vert }%
=\min\left(  \frac{M\left(  S_{I}\right)  +M\left(  S_{J}\right)  }{\left\vert
S_{I}\right\vert \left\vert S_{J}\right\vert }\right)
\]
where the minimum is over nonempty $S_{I}\subseteq\left\{  0,1\right\}
^{\left\vert I\right\vert }$\ and $S_{J}\subseteq\left\{  0,1\right\}
^{\left\vert J\right\vert }$\ such that $S_{I}\otimes S_{J}\subseteq C$. \ Now
there must be an $h$ such that $S_{I}^{\ast}\subseteq C_{I}^{\left(  h\right)
}$\ and $S_{J}^{\ast}\subseteq C_{J}^{\left(  h\right)  }$, since otherwise
some $x\notin C$\ would be assigned nonzero amplitude. \ By the induction
hypothesis,%
\[
\frac{M\left(  C_{I}\right)  }{\left\vert C_{I}\right\vert }=\min\left(
\frac{M\left(  S_{I}\right)  }{\left\vert S_{I}\right\vert }\right)
,~~~~~~~~~~\frac{M\left(  C_{J}\right)  }{\left\vert C_{J}\right\vert }%
=\min\left(  \frac{M\left(  S_{J}\right)  }{\left\vert S_{J}\right\vert
}\right)  ,
\]
where the minima are over nonempty $S_{I}\subseteq C_{I}^{\left(  h\right)  }$
and $S_{J}\subseteq C_{J}^{\left(  h\right)  }$\ respectively. \ Define
$\beta=\left\vert S_{I}\right\vert \cdot\left\vert S_{J}\right\vert /M\left(
S_{J}\right)  $\ and $\gamma=\left\vert S_{J}\right\vert \cdot\left\vert
S_{I}\right\vert /M\left(  S_{I}\right)  $. \ Then since setting $S_{I}%
:=C_{I}^{\left(  h\right)  }$\ and $S_{J}:=C_{J}^{\left(  h\right)  }%
$\ maximizes the four quantities $\left\vert S_{I}\right\vert $,\ $\left\vert
S_{J}\right\vert $, $\left\vert S_{I}\right\vert /M\left(  S_{I}\right)  $,
and $\left\vert S_{J}\right\vert /M\left(  S_{J}\right)  $\ simultaneously,
this choice also maximizes $\beta$\ and $\gamma$ simultaneously. \ Therefore
it maximizes their harmonic mean,%
\[
\frac{\beta\gamma}{\beta+\gamma}=\frac{\left\vert S_{I}\right\vert \left\vert
S_{J}\right\vert }{M\left(  S_{I}\right)  +M\left(  S_{J}\right)  }%
=\frac{\left\vert S\right\vert }{M\left(  S\right)  }.
\]
We have proved that setting $S:=C_{I}^{\left(  h\right)  }\otimes
C_{J}^{\left(  h\right)  }$\ maximizes $\left\vert S\right\vert /M\left(
S\right)  $, or equivalently minimizes $M\left(  S\right)  /\left\vert
S\right\vert $. \ The one remaining observation is that taking the disjoint
sum of $C_{I}^{\left(  h\right)  }\otimes C_{J}^{\left(  h\right)  }$\ over
all $h\in\left[  H\right]  $\ leaves the ratio $M\left(  S\right)  /\left\vert
S\right\vert $ unchanged. \ So setting $S:=C$\ also minimizes $M\left(
S\right)  /\left\vert S\right\vert $, and we are done.
\end{proof}

I can now give a recursive characterization of $M\left(  C\right)  $.

\begin{theorem}
\label{motstight}If $n\geq2$, then%
\[
M\left(  C\right)  =\left\vert C\right\vert \min\left(  \frac{M\left(
C_{I}\right)  +M\left(  C_{J}\right)  }{\left\vert C_{I}\right\vert \left\vert
C_{J}\right\vert }\right)
\]
where the minimum is over nontrivial partitions $\left(  I,J\right)  $ of
$\left[  n\right]  $.
\end{theorem}

\begin{proof}
The upper bound is obvious; let us prove the lower bound. \ Let $T$ be a
manifestly orthogonal tree for $\left\vert C\right\rangle $ of minimum size,
and let $v^{\left(  1\right)  },\ldots,v^{\left(  L\right)  }$\ be the topmost
$\otimes$\ vertices in $T$. \ Then there exists a partition $\left(
S^{\left(  1\right)  },\ldots,S^{\left(  L\right)  }\right)  $\ of $C$ such
that the subtree rooted at $v^{\left(  i\right)  }$\ represents $\left\vert
S^{\left(  i\right)  }\right\rangle $. \ We have%
\[
\left\vert T\right\vert =M\left(  S^{\left(  1\right)  }\right)
+\cdots+M\left(  S^{\left(  L\right)  }\right)  =\left\vert S^{\left(
1\right)  }\right\vert \frac{M\left(  S^{\left(  1\right)  }\right)
}{\left\vert S^{\left(  1\right)  }\right\vert }+\cdots+\left\vert S^{\left(
L\right)  }\right\vert \frac{M\left(  S^{\left(  L\right)  }\right)
}{\left\vert S^{\left(  L\right)  }\right\vert }.
\]
Now let\ $\eta=\min_{i}\left(  M\left(  S^{\left(  i\right)  }\right)
/\left\vert S^{\left(  i\right)  }\right\vert \right)  $. \ We will construct
a partition $\left(  R^{\left(  1\right)  },\ldots,R^{\left(  H\right)
}\right)  $\ of $C$ such that $M\left(  R^{\left(  h\right)  }\right)
/\left\vert R^{\left(  h\right)  }\right\vert =\eta$\ for all $h\in\left[
H\right]  $, which will imply a new tree $T^{\prime}$ with $\left\vert
T^{\prime}\right\vert \leq\left\vert T\right\vert $. \ Choose $j\in\left[
L\right]  $\ such that $M\left(  S^{\left(  j\right)  }\right)  /\left\vert
S^{\left(  j\right)  }\right\vert =\eta$, and suppose vertex $v^{\left(
j\right)  }$\ of $T$ expresses $\left\vert S^{\left(  j\right)  }\right\rangle
$ as $\left\vert S_{I}\right\rangle \otimes\left\vert S_{J}\right\rangle
$\ for some nontrivial partition $\left(  I,J\right)  $. \ Then%
\[
\eta=\frac{M\left(  S^{\left(  j\right)  }\right)  }{\left\vert S^{\left(
j\right)  }\right\vert }=\frac{M\left(  S_{I}\right)  +M\left(  S_{J}\right)
}{\left\vert S_{I}\right\vert \left\vert S_{J}\right\vert }%
\]
where $M\left(  S^{\left(  j\right)  }\right)  =M\left(  S_{I}\right)
+M\left(  S_{J}\right)  $ follows from the minimality of $T$. \ As in Lemma
\ref{induclem}, there must be an $h$ such that $S_{I}\subseteq C_{I}^{\left(
h\right)  }$\ and $S_{J}\subseteq C_{J}^{\left(  h\right)  }$. \ But Lemma
\ref{induclem}\ then implies that $M\left(  C_{I}\right)  /\left\vert
C_{I}\right\vert \leq M\left(  S_{I}\right)  /\left\vert S_{I}\right\vert
$\ and that $M\left(  C_{J}\right)  /\left\vert C_{J}\right\vert \leq M\left(
S_{J}\right)  /\left\vert S_{J}\right\vert $. \ Combining these bounds with
$\left\vert C_{I}\right\vert \geq\left\vert S_{I}\right\vert $\ and
$\left\vert C_{J}\right\vert \geq\left\vert S_{J}\right\vert $, we obtain by a
harmonic mean inequality that%
\[
\frac{M\left(  C_{I}\otimes C_{J}\right)  }{\left\vert C_{I}\otimes
C_{J}\right\vert }\leq\frac{M\left(  C_{I}\right)  +M\left(  C_{J}\right)
}{\left\vert C_{I}\right\vert \left\vert C_{J}\right\vert }\leq\frac{M\left(
S_{I}^{\ast}\right)  +M\left(  S_{J}^{\ast}\right)  }{\left\vert S_{I}^{\ast
}\right\vert \left\vert S_{J}^{\ast}\right\vert }=\eta.
\]
So setting $R^{\left(  h\right)  }:=C_{I}^{\left(  h\right)  }\otimes
C_{J}^{\left(  h\right)  }$\ for all $h\in\left[  H\right]  $\ yields a new
tree $T^{\prime}$\ no larger than $T$. \ Hence by the minimality of $T$,%
\[
M\left(  C\right)  =\left\vert T\right\vert =\left\vert T^{\prime}\right\vert
=H\cdot M\left(  C_{I}\otimes C_{J}\right)  =\frac{\left\vert C\right\vert
}{\left\vert C_{I}\right\vert \left\vert C_{J}\right\vert }\cdot\left(
M\left(  C_{I}\right)  +M\left(  C_{J}\right)  \right)  .
\]

\end{proof}

One can express Theorem\ \ref{motstight}\ directly in terms of the matrix $A$
as follows. \ Let $M\left(  A\right)  =M\left(  C\right)
=\operatorname*{MOTS}\left(  \left\vert C\right\rangle \right)  $\ where
$C=\left\{  x:Ax\equiv b\right\}  $\ (the vector $b$ is irrelevant, so long as
$Ax\equiv b$\ is solvable). \ Then%
\begin{equation}
M\left(  A\right)  =\min\left(  2^{\operatorname*{rank}\left(  A_{I}\right)
+\operatorname*{rank}\left(  A_{J}\right)  -\operatorname*{rank}\left(
A\right)  }\left(  M\left(  A_{I}\right)  +M\left(  A_{J}\right)  \right)
\right)  \tag{*}\label{marank}%
\end{equation}
where the minimum is over all nontrivial partitions $\left(  A_{I}%
,A_{J}\right)  $\ of the columns of $A$. \ As a base case, if $A$ has only one
column, then $M\left(  A\right)  =2$ if $A=0$\ and $M\left(  A\right)
=1$\ otherwise. \ This immediately implies the following.

\begin{corollary}
\label{algcor}There exists a deterministic $O\left(  n3^{n}\right)  $-time
algorithm that computes $M\left(  A\right)  $, given $A$ as input.
\end{corollary}

\begin{proof}
First compute $\operatorname*{rank}\left(  A^{\ast}\right)  $\ for all
$2^{n-1}$\ matrices $A^{\ast}$\ that are formed by choosing a subset of the
columns of $A$. \ This takes time $O\left(  n^{3}2^{n}\right)  $.\ \ Then
compute $M\left(  A^{\ast}\right)  $\ for all $A^{\ast}$\ with one column,
then for all $A^{\ast}$\ with two columns, and so on, applying the formula
(\ref{marank})\ recursively. \ This takes time%
\[
\sum_{t=1}^{n}\dbinom{n}{t}t2^{t}=O\left(  n3^{n}\right)  .
\]

\end{proof}

Another easy consequence of Theorem \ref{motstight} is that the language
$\left\{  A:M\left(  A\right)  \leq s\right\}  $\ is in $\mathsf{NP}$. \ I do
not know whether this language is $\mathsf{NP}$-complete but suspect it is.

As mentioned above, my characterization makes it possible to prove exponential
lower bounds on the manifestly orthogonal tree size of coset states.

\begin{theorem}
\label{explb}Suppose the entries of $A\in\mathbb{Z}_{2}^{k\times n}$ are drawn
uniformly and independently at random, where $k\in\left[  4\log_{2}n,\frac
{1}{2}\sqrt{n\ln2}\right]  $. \ Then $M\left(  A\right)  $\ $=\left(
n/k^{2}\right)  ^{\Omega\left(  k\right)  }$\ with probability $\Omega\left(
1\right)  $ over $A$.
\end{theorem}

\begin{proof}
Let us upper-bound the probability that certain \textquotedblleft bad
events\textquotedblright\ occur when $A$ is drawn. \ The first bad event is
that $A$ contains an all-zero column. \ This occurs with probability at most
$2^{-k}n=o\left(  1\right)  $. \ The second bad event is that there exists a
$k\times d$\ submatrix of $A$ with $d\geq12k$\ that has rank at most $2k/3$.
\ This also occurs with probability $o\left(  1\right)  $. \ For we claim
that, if $A^{\ast}$\ is drawn uniformly at random from $\mathbb{Z}%
_{2}^{k\times d}$, then%
\[
\Pr_{A_{I}}\left[  \operatorname*{rank}\left(  A^{\ast}\right)  \leq r\right]
\leq\dbinom{d}{r}\left(  \frac{2^{r}}{2^{k}}\right)  ^{d-r}.
\]
To see this, imagine choosing the columns of $A^{\ast}$\ one by one. \ For
$\operatorname*{rank}\left(  A^{\ast}\right)  $ to be at most $r$, there must
be at least $d-r$\ columns that are linearly dependent on the previous
columns. \ But each column is dependent on the previous ones with probability
at most $2^{r}/2^{k}$. \ The claim then follows from the union bound. \ So the
probability that \textit{any} $k\times d$\ submatrix of $A$ has rank at most
$r$ is at most%
\[
\dbinom{n}{d}\dbinom{d}{r}\left(  \frac{2^{r}}{2^{k}}\right)  ^{d-r}\leq
n^{d}d^{r}\left(  \frac{2^{r}}{2^{k}}\right)  ^{d-r}.
\]
Set $r=2k/3$ and $d=12k$; then the above is at most%
\[
\exp\left\{  12k\log n+\frac{2k}{3}\log\left(  12k\right)  -\left(
12k-\frac{2k}{3}\right)  \frac{k}{3}\right\}  =o\left(  1\right)
\]
where we have used the fact that $k\geq4\log n$.

Assume that neither bad event occurs, and let $\left(  A_{I}^{\left(
0\right)  },A_{J}^{\left(  0\right)  }\right)  $ be a partition of the columns
of $A$ that minimizes the expression (\ref{marank}). \ Let $A^{\left(
1\right)  }=A_{I}^{\left(  0\right)  }$\ if $\left\vert A_{I}^{\left(
0\right)  }\right\vert \geq\left\vert A_{J}^{\left(  0\right)  }\right\vert
$\ and $A^{\left(  1\right)  }=A_{J}^{\left(  0\right)  }$\ otherwise, where
$\left\vert A_{I}^{\left(  0\right)  }\right\vert $\ and $\left\vert
A_{J}^{\left(  0\right)  }\right\vert $\ are the numbers of columns in
$A_{I}^{\left(  0\right)  }$\ and $A_{J}^{\left(  0\right)  }$\ respectively
(so that $\left\vert A_{I}^{\left(  0\right)  }\right\vert +\left\vert
A_{J}^{\left(  0\right)  }\right\vert =n$). \ Likewise, let $\left(
A_{I}^{\left(  1\right)  },A_{J}^{\left(  1\right)  }\right)  $\ be an optimal
partition of the columns of $A^{\left(  1\right)  }$, and let $A^{\left(
2\right)  }=A_{I}^{\left(  1\right)  }$\ if $\left\vert A_{I}^{\left(
1\right)  }\right\vert \geq\left\vert A_{J}^{\left(  1\right)  }\right\vert
$\ and $A^{\left(  2\right)  }=A_{J}^{\left(  1\right)  }$\ otherwise.
\ Continue in this way until an $A^{\left(  t\right)  }$\ is reached such that
$\left\vert A^{\left(  t\right)  }\right\vert =1$. \ Then an immediate
consequence of (\ref{marank})\ is that $M\left(  A\right)  \geq Z^{\left(
0\right)  }\cdot\cdots\cdot Z^{\left(  t-1\right)  }$\ where%
\[
Z^{\left(  \ell\right)  }=2^{\operatorname*{rank}\left(  A_{I}^{\left(
\ell\right)  }\right)  +\operatorname*{rank}\left(  A_{J}^{\left(
\ell\right)  }\right)  -\operatorname*{rank}\left(  A^{\left(  \ell\right)
}\right)  }%
\]
and $A^{\left(  0\right)  }=A$.

Call $\ell$ a \textquotedblleft balanced cut\textquotedblright\ if
$\min\left\{  \left\vert A_{I}^{\left(  \ell\right)  }\right\vert ,\left\vert
A_{J}^{\left(  \ell\right)  }\right\vert \right\}  \geq12k$, and an
\textquotedblleft unbalanced cut\textquotedblright\ otherwise. \ If $\ell$ is
a balanced cut, then $\operatorname*{rank}\left(  A_{I}^{\left(  \ell\right)
}\right)  \geq2k/3$\ and $\operatorname*{rank}\left(  A_{J}^{\left(
\ell\right)  }\right)  \geq2k/3$, so $Z^{\left(  \ell\right)  }\geq2^{k/3}$.
\ If $\ell$ is an unbalanced cut, then call $\ell$ a \textquotedblleft
freebie\textquotedblright\ if $\operatorname*{rank}\left(  A_{I}^{\left(
\ell\right)  }\right)  +\operatorname*{rank}\left(  A_{J}^{\left(
\ell\right)  }\right)  =\operatorname*{rank}\left(  A^{\left(  \ell\right)
}\right)  $. \ There can be at most $k$ freebies, since for each one,
$\operatorname*{rank}\left(  A^{\left(  \ell+1\right)  }\right)
<\operatorname*{rank}\left(  A^{\left(  \ell\right)  }\right)  $\ by the
assumption that all columns of $A$\ are nonzero. \ For the other unbalanced
cuts, $Z^{\left(  \ell\right)  }\geq2$.

Assume $\left\vert A^{\left(  \ell+1\right)  }\right\vert =\left\vert
A^{\left(  \ell\right)  }\right\vert /2$\ for each balanced cut and
$\left\vert A^{\left(  \ell+1\right)  }\right\vert =\left\vert A^{\left(
\ell\right)  }\right\vert -12k$\ for each unbalanced cut. \ Then if the goal
is to minimize $Z^{\left(  0\right)  }\cdot\cdots\cdot Z^{\left(  t-1\right)
}$, clearly the best strategy is to perform balanced cuts first, then
unbalanced cuts until $\left\vert A^{\left(  \ell\right)  }\right\vert
=12k^{2}$, at which point we can use the $k$ freebies. \ Let $B$ be the number
of balanced cuts; then%
\[
Z^{\left(  0\right)  }\cdot\cdots\cdot Z^{\left(  t-1\right)  }=\left(
2^{k/3}\right)  ^{B}2^{\left(  n/2^{B}-12k^{2}\right)  /12k}.
\]
This is minimized by taking $B=\log_{2}\left(  \frac{n\ln2}{4k^{2}}\right)  $,
in which case $Z^{\left(  0\right)  }\cdot\cdots\cdot Z^{\left(  t-1\right)
}=\left(  n/k^{2}\right)  ^{\Omega\left(  k\right)  }$.
\end{proof}

A final application of my characterization is to separate orthogonal from
manifestly orthogonal tree size.

\begin{corollary}
\label{otreesep}There exist states with polynomially-bounded orthogonal tree
size, but manifestly orthogonal tree size $n^{\Omega\left(  \log n\right)  }$.
\ Thus $\mathsf{OTree}\neq\mathsf{MOTree}$.
\end{corollary}

\begin{proof}
Set $k=4\log_{2}n$, and let $C=\left\{  x:Ax\equiv0\right\}  $\ where $A$\ is
drawn uniformly at random from $\mathbb{Z}_{2}^{k\times n}$. \ Then by Theorem
\ref{explb},%
\[
\operatorname*{MOTS}\left(  \left\vert C\right\rangle \right)  =\left(
n/k^{2}\right)  ^{\Omega\left(  k\right)  }=n^{\Omega\left(  \log n\right)  }%
\]
with probability $\Omega\left(  1\right)  $ over $A$. \ On the other hand, if
we view $\left\vert C\right\rangle $\ in the Fourier basis (that is, apply a
Hadamard to every qubit), then the resulting state has only $2^{k}=n^{16}%
$\ basis states with nonzero amplitude, and hence has orthogonal tree size at
most $n^{17}$. \ So by Proposition \ref{invariant}, part (i),
$\operatorname*{OTS}\left(  \left\vert C\right\rangle \right)  \leq2n^{17}$ as well.
\end{proof}

Indeed, the orthogonal tree states of Corollary \ref{otreesep}\ are
superpositions over polynomially many separable states, so it also follows
that $\mathsf{\Sigma}_{\mathsf{2}}\not \subset \mathsf{MOTree}$.

\section{Computing With Tree States\label{TREEBQP}}

Suppose a quantum computer is restricted to being in a tree state at all
times. \ (We can imagine that if the tree size ever exceeds some polynomial
bound, the quantum computer explodes, destroying our laboratory.) \ Does the
computer then have an efficient classical simulation? \ In other words,
letting $\mathsf{TreeBQP}$\ be the class of languages accepted by such a
machine, does $\mathsf{TreeBQP=BPP}$? \ A positive answer would make tree
states more attractive as a Sure/Shor separator. \ For once we admit any
states incompatible with the polynomial-time Church-Turing thesis, it seems
like we might as well go all the way, and admit \textit{all} states preparable
by polynomial-size quantum circuits! \ The $\mathsf{TreeBQP}$\ versus
$\mathsf{BPP}$\ problem is closely related to the problem of finding an
efficient (classical) algorithm to \textit{learn} multilinear formulas. \ In
light of Raz's lower bound, and of the connection between lower bounds and
learning noticed by Linial, Mansour, and Nisan \cite{lmn}, the latter problem
might be less hopeless than it looks. \ In this section I show a weaker
result: that $\mathsf{TreeBQP}$\ is contained in $\mathsf{\Sigma}%
_{3}^{\mathsf{P}}\cap\mathsf{\Pi}_{3}^{\mathsf{P}}$, the third level of the
polynomial hierarchy. \ Since $\mathsf{BQP}$\ is not known to lie in
$\mathsf{PH}$, this result could be taken as weak evidence that
$\mathsf{TreeBQP\neq BQP}$. \ (On the other hand, we do not yet have oracle
evidence even for\ $\mathsf{BQP}\not \subset \mathsf{AM}$, though not for lack
of trying \cite{aar:rfs}.)

\begin{definition}
\label{formulabqp}$\mathsf{TreeBQP}$\ is the class of languages accepted by a
$\mathsf{BQP}$\ machine subject to the constraint that at every time step $t$,
the machine's state $\left\vert \psi^{\left(  t\right)  }\right\rangle $\ is
exponentially close to a tree state. \ More formally, the initial state is
$\left\vert \psi^{\left(  0\right)  }\right\rangle =\left\vert 0\right\rangle
^{\otimes\left(  p\left(  n\right)  -n\right)  }\otimes\left\vert
x\right\rangle $ (for an input $x\in\left\{  0,1\right\}  ^{n}$\ and
polynomial bound $p$), and a uniform classical polynomial-time algorithm
generates a sequence of gates $g^{\left(  1\right)  },\ldots,g^{\left(
p\left(  n\right)  \right)  }$. \ Each $g^{\left(  t\right)  }$\ can be either
be selected from some finite universal basis of unitary gates (as will be
shown in Theorem \ref{nice}, part (i), the choice of gate set does not
matter), or can be a $1$-qubit measurement. \ When we perform a measurement,
the state evolves to one of two possible pure states, with the usual
probabilities, rather than to a mixed state. \ We require that the final gate
$g^{\left(  p\left(  n\right)  \right)  }$\ is a measurement of the first
qubit. \ If at least one intermediate state $\left\vert \psi^{\left(
t\right)  }\right\rangle $\ had\ $\operatorname*{TS}_{1/2^{\Omega\left(
n\right)  }}\left(  \left\vert \psi^{\left(  t\right)  }\right\rangle \right)
>p\left(  n\right)  $,\ then the outcome of the final measurement is chosen
adversarially; otherwise it is given by the usual Born probabilities. \ The
measurement must return $1$\ with probability at least $2/3$\ if the input is
in the language, and with probability at most $1/3$\ otherwise.
\end{definition}

Some comments on the definition: I allow $\left\vert \psi^{\left(  t\right)
}\right\rangle $\ to deviate from a tree state by an exponentially small
amount, in order to make the model independent of the choice of gate set. \ I
allow intermediate measurements because otherwise it is unclear even how to
simulate $\mathsf{BPP}$.\footnote{If we try to simulate $\mathsf{BPP}$ in the
standard way, we might produce complicated entanglement between the
computation register and the register containing the random bits, and no
longer have a tree state.} \ The rule for measurements follows the
\textquotedblleft Copenhagen interpretation,\textquotedblright\ in the sense
that if a qubit is measured to be $1$, then subsequent computation is not
affected by what would have happened were the qubit measured to be $0$. \ In
particular, if measuring $0$ would have led to states of tree size greater
than $p\left(  n\right)  $, that does not invalidate the results of the path
where $1$ is measured.

The following theorem shows that $\mathsf{TreeBQP}$\ has many of the
properties one would want it to have.

\begin{theorem}
\label{nice}\quad

\begin{enumerate}
\item[(i)] The definition of $\mathsf{TreeBQP}$\ is invariant under the choice
of gate set.

\item[(ii)] The probabilities $\left(  1/3,2/3\right)  $\ can be replaced by
any $\left(  p,1-p\right)  $ with $2^{-2^{\sqrt{\log n}}}<p<1/2$.

\item[(iii)] $\mathsf{BPP}\subseteq\mathsf{TreeBQP}\subseteq\mathsf{BQP}$.
\end{enumerate}
\end{theorem}

\begin{proof}
\quad

\begin{enumerate}
\item[(i)] The Solovay-Kitaev Theorem \cite{kitaev:ec,nc} shows that given a
universal gate set, one can approximate any $k$-qubit unitary to accuracy
$1/\varepsilon$\ using $k$\ qubits and a circuit of size $O\left(
\operatorname*{polylog}\left(  1/\varepsilon\right)  \right)  $. \ So let
$\left\vert \psi^{\left(  0\right)  }\right\rangle ,\ldots,\left\vert
\psi^{\left(  p\left(  n\right)  \right)  }\right\rangle \in\mathcal{H}%
_{2}^{\otimes p\left(  n\right)  }$ be a sequence of states, with $\left\vert
\psi^{\left(  t\right)  }\right\rangle $\ produced from $\left\vert
\psi^{\left(  t-1\right)  }\right\rangle $\ by applying a $k$-qubit unitary
$g^{\left(  t\right)  }$\ (where $k=O\left(  1\right)  $).\ \ Then using a
polynomial-size circuit, one can approximate each $\left\vert \psi^{\left(
t\right)  }\right\rangle $ to accuracy $1/2^{\Omega\left(  n\right)  }$, as in
the definition of $\mathsf{TreeBQP}$. \ Furthermore, since the approximation
circuit for $g^{\left(  t\right)  }$ acts only on $k$\ qubits, any
intermediate state $\left\vert \varphi\right\rangle $\ it produces satisfies
$\operatorname*{TS}\nolimits_{1/2^{\Omega\left(  n\right)  }}\left(
\left\vert \varphi\right\rangle \right)  \leq k4^{k}\operatorname*{TS}%
\nolimits_{1/2^{\Omega\left(  n\right)  }}\left(  \left\vert \psi^{\left(
t-1\right)  }\right\rangle \right)  $\ by Proposition \ref{invariant}.

\item[(ii)] To amplify to a constant probability, run $k$ copies of the
computation in tensor product, then output the majority answer. \ By part (i),
outputting the majority can increase the tree size by a factor of at
most\ $2^{k+1}$. \ To amplify to $2^{-2^{\sqrt{\log n}}}$, observe that the
Boolean majority function on $k$ bits has a multilinear formula of size
$k^{O\left(  \log k\right)  }$. \ For let $T_{k}^{h}\left(  x_{1},\ldots
,x_{k}\right)  $\ equal $1$ if $x_{1}+\cdots+x_{k}\geq h$\ and $0$ otherwise;
then%
\[
T_{k}^{h}\left(  x_{1},\ldots,x_{k}\right)  =1-\prod_{i=0}^{h}\left(
1-T_{\left\lfloor k/2\right\rfloor }^{i}\left(  x_{1},\ldots,x_{\left\lfloor
k/2\right\rfloor }\right)  T_{\left\lceil k/2\right\rceil }^{h-i}\left(
x_{\left\lfloor k/2\right\rfloor +1},\ldots,x_{k}\right)  \right)  ,
\]
so $\operatorname*{MFS}\left(  T_{k}^{h}\right)  \leq2h\max_{i}%
\operatorname*{MFS}\left(  T_{\left\lceil k/2\right\rceil }^{h}\right)
+O\left(  1\right)  $, and solving this recurrence yields $\operatorname*{MFS}%
\left(  T_{k}^{k/2}\right)  =k^{O\left(  \log k\right)  }$. \ Substituting
$k=2^{\sqrt{\log n}}$\ into $k^{O\left(  \log k\right)  }$\ yields
$n^{O\left(  1\right)  }$, meaning the tree size increases by at most a
polynomial factor.

\item[(iii)] To simulate $\mathsf{BPP}$, just perform a classical reversible
computation, applying a Hadamard followed by a measurement to some qubit
whenever we need a random bit. \ Since the number of basis states with nonzero
amplitude is at most $2$, the simulation is clearly in $\mathsf{TreeBQP}$.
\ The other containment is obvious.
\end{enumerate}
\end{proof}

\begin{theorem}
\label{inph}$\mathsf{TreeBQP}\subseteq\mathsf{\Sigma}_{3}^{\mathsf{P}}%
\cap\mathsf{\Pi}_{3}^{\mathsf{P}}$.
\end{theorem}

\begin{proof}
Since $\mathsf{TreeBQP}$\ is closed under complement, it suffices to show that
$\mathsf{TreeBQP}\subseteq\mathsf{\Pi}_{3}^{\mathsf{P}}$. \ Our proof will
combine approximate counting with a predicate to verify the correctness of a
$\mathsf{TreeBQP}$\ computation. \ Let $C$ be a uniformly-generated quantum
circuit, and let $M=\left(  m^{\left(  1\right)  },\ldots,m^{\left(  p\left(
n\right)  \right)  }\right)  $\ be a sequence of binary measurement outcomes.
\ We adopt the convention that after making a measurement, the state vector is
\textit{not} rescaled to have norm $1$. \ That way the probabilities across
all `measurement branches' continue to sum to $1$. \ Let $\left\vert
\psi_{M,x}^{\left(  0\right)  }\right\rangle ,\ldots,\left\vert \psi
_{M,x}^{\left(  p\left(  n\right)  \right)  }\right\rangle $\ be the sequence
of unnormalized pure states under measurement outcome sequence $M$ and input
$x$, where $\left\vert \psi_{M,x}^{\left(  t\right)  }\right\rangle
=\sum_{y\in\left\{  0,1\right\}  ^{p\left(  n\right)  }}\alpha_{y,M,x}%
^{\left(  t\right)  }\left\vert y\right\rangle $. \ Also, let $\Lambda\left(
M,x\right)  $\ express that $\operatorname*{TS}_{1/2^{\Omega\left(  n\right)
}}\left(  \left\vert \psi_{M,x}^{\left(  t\right)  }\right\rangle \right)
\leq p\left(  n\right)  $ for every $t$. \ Then $C$ accepts if%
\[
W_{x}=\sum_{M\,:\,\Lambda\left(  M,x\right)  }\sum_{y\in\left\{  0,1\right\}
^{p\left(  n\right)  -1}}\left\vert \alpha_{1y,M,x}^{\left(  p\left(
n\right)  \right)  }\right\vert ^{2}\geq\frac{2}{3},
\]
while $C$ rejects if $W_{x}\leq1/3$. \ If we could compute each $\left\vert
\alpha_{1y,M,x}^{\left(  p\left(  n\right)  \right)  }\right\vert $
efficiently (as well as $\Lambda\left(  M,x\right)  $), we would then have a
$\mathsf{\Pi}_{2}^{\mathsf{P}}$\ predicate expressing that $W_{x}\geq2/3$.
\ This follows since we can do approximate counting via hashing in
$\mathsf{AM}\subseteq\mathsf{\Pi}_{2}^{\mathsf{P}}$\ \cite{gs}, and thereby
verify that an exponentially large sum of nonnegative terms is at least $2/3$,
rather than at most $1/3$. \ The one further fact we need is that in our
$\mathsf{\Pi}_{2}^{\mathsf{P}}$\ ($\forall\exists$) predicate, we can take the
existential quantifier to range over tuples of `candidate solutions'---that
is, $\left(  M,y\right)  $\ pairs together with lower bounds $\beta$ on
$\left\vert \alpha_{1y,M,x}^{\left(  p\left(  n\right)  \right)  }\right\vert
$.

It remains only to show how we verify that $\Lambda\left(  M,x\right)
$\ holds and that $\left|  \alpha_{1y,M,x}^{\left(  p\left(  n\right)
\right)  }\right|  =\beta$. \ First, we extend the existential quantifier so
that it guesses not only $M$ and $y$, but also a sequence of trees $T^{\left(
0\right)  },\ldots,T^{\left(  p\left(  n\right)  \right)  }$, representing
$\left|  \psi_{M,x}^{\left(  0\right)  }\right\rangle ,\ldots,\left|
\psi_{M,x}^{\left(  p\left(  n\right)  \right)  }\right\rangle $ respectively.
\ Second, using the last universal quantifier to range over $\widehat{y}%
\in\left\{  0,1\right\}  ^{p\left(  n\right)  }$, we verify the following:

\begin{enumerate}
\item[(1)] $T^{\left(  0\right)  }$ is a fixed tree representing $\left|
0\right\rangle ^{\otimes\left(  p\left(  n\right)  -n\right)  }\otimes\left|
x\right\rangle $.

\item[(2)] $\left|  \alpha_{1y,M,x}^{\left(  p\left(  n\right)  \right)
}\right|  $ equals its claimed value to $\Omega\left(  n\right)  $\ bits of precision.

\item[(3)] Let $g^{\left(  1\right)  },\ldots,g^{\left(  p\left(  n\right)
\right)  }$\ be the gates applied by $C$. \ Then for all $t$ and $\widehat{y}%
$, if $g^{\left(  t\right)  }$\ is unitary then $\alpha_{\widehat{y}%
,M,x}^{\left(  t\right)  }=\left\langle \widehat{y}\right\vert \cdot
g^{\left(  t\right)  }\left\vert \psi_{M,x}^{\left(  t-1\right)
}\right\rangle $ to $\Omega\left(  n\right)  $\ bits of precision. \ Here the
right-hand side is a sum of $2^{k}$\ terms ($k$ being the number of qubits
acted on by $g^{\left(  t\right)  }$), each term efficiently computable given
$T^{\left(  t-1\right)  }$. \ Similarly, if $g^{\left(  t\right)  }$\ is a
measurement of the $i^{th}$\ qubit, then $\alpha_{\widehat{y},M,x}^{\left(
t\right)  }=\alpha_{\widehat{y},M,x}^{\left(  t-1\right)  }$\ if the $i^{th}%
$\ bit of $\widehat{y}$\ equals $m^{\left(  t\right)  }$, while $\alpha
_{\widehat{y},M,x}^{\left(  t\right)  }=0$\ otherwise.
\end{enumerate}
\end{proof}

In the proof of Theorem \ref{inph}, the only fact about tree states I needed
was that $\mathsf{Tree}\subseteq\mathsf{AmpP}$; that is, there is a
polynomial-time classical algorithm that computes the amplitude $\alpha_{x}%
$\ of any basis state $\left\vert x\right\rangle $. \ So if we define
$\mathsf{AmpP}$-$\mathsf{BQP}$\ analogously to $\mathsf{TreeBQP}$ except that
any states in $\mathsf{AmpP}$\ are allowed, then $\mathsf{AmpP}$%
-$\mathsf{BQP}\subseteq\mathsf{\Sigma}_{3}^{\mathsf{P}}\cap\mathsf{\Pi}%
_{3}^{\mathsf{P}}$\ as well.

\section{The Experimental Situation\label{EXPER}}

The results of this chapter suggest an obvious challenge for
experimenters: \textit{prepare non-tree states in the lab}. \ For
were this challenge met, it would rule out one way in which quantum
mechanics could fail, just as the Bell inequality experiments of
Aspect et al.\ \cite{aspect}\ did twenty years ago. \ If they
wished, quantum computing skeptics could then propose a new
candidate Sure/Shor separator, and experimenters could try to rule
out \textit{that} one, and so on. \ The result would be to divide
the question of whether quantum computing is possible into a series
of smaller questions about which states can be prepared. \ In my
view, this would aid progress in two ways: by helping experimenters
set clear goals, and by forcing theorists to state clear
conjectures.

However, my experimental challenge raises some immediate questions.\ \ In
particular, what would it \textit{mean} to prepare a non-tree state? \ How
would we know if we succeeded? \ Also, have non-tree states already been
prepared (or observed)? \ The purpose of this section is to set out my
thoughts about these questions.

First of all, when discussing experiments, it goes without saying that we must
convert asymptotic statements into statements about specific values of $n$.
\ The central tenet of computational complexity theory is that this is
possible. \ Thus, instead of asking whether $n$-qubit states with tree size
$2^{\Omega\left(  n\right)  }$\ can be prepared, we ask whether $200$-qubit
states with tree size at least (say) $2^{80}$\ can be prepared. \ Even though
the second question does not logically imply anything about the first, the
second is closer to what we ultimately care about anyway. \ Admittedly,
knowing that $\operatorname*{TS}\left(  \left\vert \psi_{n}\right\rangle
\right)  =n^{\Omega\left(  \log n\right)  }$\ tells us little about
$\operatorname*{TS}\left(  \left\vert \psi_{100}\right\rangle \right)  $\ or
$\operatorname*{TS}\left(  \left\vert \psi_{200}\right\rangle \right)  $,
especially since in Raz's paper \cite{raz}, the constant in the exponent
$\Omega\left(  \log n\right)  $\ is taken to be $10^{-6}$\ (though this can
certainly be improved). \ Thus, proving tight lower bounds for small $n$ is
one of the most important problems left open by this chapter. \ In Section
\ref{MOTS} I show how to solve this problem for the case of manifestly
orthogonal tree size.

A second objection is that my formalism applies only to pure states, but in
reality all states are mixed. \ However, there are several natural ways to
extend the formalism to mixed states. \ \ Given a mixed state $\rho$, we could
minimize tree size over all purifications of $\rho$, or minimize the expected
tree size $\sum_{i}\left\vert \alpha_{i}\right\vert ^{2}\operatorname*{TS}%
\left(  \left\vert \psi_{i}\right\rangle \right)  $, or maximum $\max
_{i}\operatorname*{TS}\left(  \left\vert \psi_{i}\right\rangle \right)  $,
over all decompositions $\rho=\sum_{i}\alpha_{i}\left\vert \psi_{i}%
\right\rangle \left\langle \psi_{i}\right\vert $.

A third objection is a real quantum state might be a \textquotedblleft
soup\textquotedblright\ of free-wandering fermions and bosons, with no
localized subsystems corresponding to qubits. \ How can one determine the tree
size of such a state? \ The answer is that one cannot. \ Any complexity
measure for particle position and momentum states would have to be quite
different from the measures considered in this chapter. \ On the other hand,
the states of interest for quantum computing usually \textit{do} involve
localized qubits. \ Indeed, even if quantum information is stored in particle
positions, one might force each particle into two sites (corresponding to
$\left\vert 0\right\rangle $\ and $\left\vert 1\right\rangle $), neither of
which can be occupied by any other particle. \ In that case it again becomes
meaningful to discuss tree size.

But how do we verify that a state with large tree size was prepared? \ Of
course, if $\left\vert \psi\right\rangle $\ is preparable by a polynomial-size
quantum circuit, then \textit{assuming quantum mechanics is valid} (and
assuming our gates behave as specified), we can always test whether a given
state $\left\vert \varphi\right\rangle $\ is close to $\left\vert
\psi\right\rangle $\ or not. \ Let $U$\ map $\left\vert 0\right\rangle
^{\otimes n}$\ to $\left\vert \psi\right\rangle $; then it suffices to test
whether $U^{-1}\left\vert \varphi\right\rangle $\ is close to $\left\vert
0\right\rangle ^{\otimes n}$. \ However, in the experiments under discussion,
the validity of quantum mechanics is the very point in question. \ And once we
allow Nature to behave in arbitrary ways, a skeptic could explain \textit{any}
experimental result without having to invoke states with large tree size.

The above fact has often been urged against me, but as it stands, it is no
different from the fact that one could explain any astronomical observation
without abandoning the Ptolemaic system. \ The issue here is not one of proof,
but of accumulating observations that are consistent with\ the hypothesis of
large tree size, and inconsistent with alternative hypotheses if we disallow
special pleading. \ So for example, to test whether the subgroup state%
\[
\left\vert S\right\rangle =\frac{1}{\sqrt{\left\vert S\right\vert }}\sum_{x\in
S}\left\vert x\right\rangle
\]
was prepared, we might use CNOT gates to map $\left\vert x\right\rangle $\ to
$\left\vert x\right\rangle \left\vert v^{T}x\right\rangle $ for some vector
$v\in\mathbb{Z}_{2}^{n}$. \ Based on our knowledge of $S$, we could then
predict whether the qubit $\left\vert v^{T}x\right\rangle $\ should be
$\left\vert 0\right\rangle $, $\left\vert 1\right\rangle $, or an equal
mixture of $\left\vert 0\right\rangle $ and $\left\vert 1\right\rangle $ when
measured. \ Or we could apply Hadamard gates to all $n$ qubits of $\left\vert
S\right\rangle $, then perform the same test for the subgroup dual to $S$.
\ In saying that a system is in state $\left\vert S\right\rangle $, it is not
clear if we \textit{mean} anything more than that it responds to all such
tests in expected ways. \ Similar remarks apply to Shor states and cluster states.

In my view, tests of the sort described above are certainly
\textit{sufficient}, so the interesting question is whether they are
\textit{necessary}, or whether weaker and more indirect tests would
also suffice. \ This question rears its head when we ask whether
non-tree states have already been observed. \ For as pointed out to
me by Anthony Leggett, there exist systems studied in
condensed-matter physics that are strong candidates for having
superpolynomial tree size. \ An example is the magnetic salt
LiHo$_{x}$Y$_{1-x}$F$_{4}$ studied by Ghosh et al.\ \cite{grac},
which, like the cluster states of Briegel and Raussendorf \cite{br},
basically consists of a lattice of spins subject to pairwise
nearest-neighbor Hamiltonians. \ The main differences are that the
salt lattice is 3-D instead of 2-D, is tetragonal instead of cubic,
and is irregular in that not every site is occupied by a spin. \
Also, there are weak interactions even between spins that are not
nearest neighbors. \ But none of these differences seem likely to
change a superpolynomial tree size into a polynomial one.

For me, the main issues are (1) how precisely can we
characterize\footnote{By \textquotedblleft
characterize,\textquotedblright\ I mean give an explicit formula for
the amplitudes at a particular time $t$, in some standard basis. \
If a state is characterized as the ground state of a Hamiltonian,
then we first need to solve for the amplitudes before we can prove
tree size lower bounds using Raz's method.} the quantum state of the
magnetic salt, and (2) how strong the evidence is that that
\textit{is} the state. \ What Ghosh et al.\ \cite{grac}\ did was to
calculate bulk properties of the salt, such as its magnetic
susceptibility and specific heat, with and without taking into
account the quantum entanglement generated by the nearest-neighbor
Hamiltonians. \ They found that including entanglement yielded a
better fit to the experimentally measured values. \ However, this is
clearly a far cry from preparing a system in a state of one's
choosing by applying a known pulse sequence, and then applying any
of a vast catalog of tests to verify that the state was prepared. \
So it would be valuable to have more direct evidence that states
qualitatively like cluster states can exist in Nature.

In summary, the ideas of this chapter underscore the importance of
current experimental work on large, persistently entangled quantum
states; but they also suggest a new motivation and perspective for
this work. \ They suggest that we reexamine known condensed-matter
systems with a new goal in mind: understanding the complexity of
their associated quantum states. \ They also suggest that 2-D
cluster states and random subgroup states are interesting in a way
that 1-D spin chains and Schr\"{o}dinger cat states are not. \ Yet
when experimenters try to prepare states of the former type, they
often see it as merely a stepping stone towards demonstrating
error-correction or another quantum computing benchmark. \ Thus,
Knill et al.\ \cite{klmn} prepared\footnote{Admittedly, what they
really prepared is the `pseudo-pure' state
$\rho=\varepsilon\left\vert \psi\right\rangle \left\langle
\psi\right\vert +\left(  1-\varepsilon\right)  I$, where $I$ is the
maximally mixed state and $\varepsilon\approx10^{-5}$. \ Braunstein
et al.\ \cite{bcjlps}\ have shown that, if the number of qubits $n$
is less than about $14$, then such states cannot be entangled. \
That is, there exists a representation of $\rho$\ as a mixture of
pure states, each of which is separable and therefore has tree size
$O\left(  n\right)  $. \ This is a well-known limitation of the
liquid NMR technology used by Knill et al.\ \ Thus, a key challenge
is to replicate the successes of liquid NMR using
colder qubits.} the $5$-qubit state%
\[
\left\vert \psi\right\rangle =\frac{1}{4}\left(
\begin{array}
[c]{c}%
\left\vert 00000\right\rangle +\left\vert 10010\right\rangle
+\left\vert 01001\right\rangle +\left\vert 10100\right\rangle \\
+\left\vert 01010\right\rangle -\left\vert 11011\right\rangle
-\left\vert 00110\right\rangle -\left\vert 11000\right\rangle \\
-\left\vert 11101\right\rangle -\left\vert 00011\right\rangle
-\left\vert 11110\right\rangle -\left\vert 01111\right\rangle \\
-\left\vert 10001\right\rangle -\left\vert 01100\right\rangle
-\left\vert 10111\right\rangle +\left\vert 00101\right\rangle
\end{array}
\right)  ,
\]
for which $\operatorname*{MOTS}\left(  \left\vert \psi\right\rangle \right)
=40$\ from the decomposition%
\[
\left\vert \psi\right\rangle =\frac{1}{4}\left(
\begin{array}
[c]{c}%
\left(  \left\vert 01\right\rangle +\left\vert 10\right\rangle \right)
\otimes\left(  \left\vert 010\right\rangle -\left\vert 111\right\rangle
\right)  +\left(  \left\vert 01\right\rangle -\left\vert 10\right\rangle
\right)  \otimes\left(  \left\vert 001\right\rangle -\left\vert
100\right\rangle \right) \\
-\left(  \left\vert 00\right\rangle +\left\vert 11\right\rangle \right)
\otimes\left(  \left\vert 011\right\rangle +\left\vert 110\right\rangle
\right)  +\left(  \left\vert 00\right\rangle -\left\vert 11\right\rangle
\right)  \otimes\left(  \left\vert 000\right\rangle +\left\vert
101\right\rangle \right)
\end{array}
\right)  ,
\]
and for which I conjecture $\operatorname*{TS}\left(  \left\vert
\psi\right\rangle \right)  =40$ as well. \ However, the sole motivation of the
experiment was to demonstrate a $5$-qubit quantum error-correcting code. \ In
my opinion, whether states with large tree size can be prepared is a
fundamental question in its own right. \ Were that question studied directly,
perhaps we could address it for larger numbers of qubits.

Let me end by stressing that, in the perspective I am advocating, there is
nothing sacrosanct about tree size as opposed to other complexity measures.
\ This chapter concentrated on tree size because it is the subject of our main
results, and because it is better to be specific than vague. \ On the other
hand, Sections \ref{BASICMLIN}, \ \ref{RELMLIN}, and \ref{MOTS}\ contain
numerous results about orthogonal tree size, manifestly orthogonal tree size,
Vidal's $\chi$\ complexity, and other measures. \ Readers dissatisfied with
\textit{all} of these measures are urged to propose new ones, perhaps
motivated directly by experiments. \ I see nothing wrong with having multiple
ways to quantify the complexity of quantum states, and much wrong with having
no ways.

\section{Conclusion and Open Problems\label{OPENMLIN}}

A crucial step in quantum computing was to separate the question of whether
quantum computers can be built from the question of what one could do with
them. \ This separation allowed computer scientists to make great advances on
the latter question, despite knowing nothing about the former. \ I have
argued, however, that the tools of computational complexity theory are
relevant to both questions. \ The claim that large-scale quantum computing is
possible in principle is really a claim that certain \textit{states} can
exist---that quantum mechanics will not break down if we try to prepare those
states. \ Furthermore, what distinguishes these states from states we have
seen must be more than precision in amplitudes, or the number of qubits
maintained coherently. \ The distinguishing property should instead be some
sort of \textit{complexity}. \ That is, Sure states should have succinct
representations of a type that Shor states do not.

I have tried to show that, by adopting this viewpoint, we make the debate
about whether quantum computing is possible less ideological and more
scientific. \ By studying particular examples of Sure/Shor separators, quantum
computing skeptics would strengthen their case---for they would then have a
plausible research program aimed at identifying what, exactly, the barriers to
quantum computation are. \ I hope, however, that the `complexity theory of
quantum states' initiated here will be taken up by quantum computing
proponents as well. \ This theory offers a new perspective on the transition
from classical to quantum computing, and a new connection between quantum
computing and the powerful circuit lower bound techniques of classical
complexity theory.

I end with some open problems.

\begin{enumerate}
\item[(1)] Can Raz's technique be improved to show exponential tree size lower bounds?

\item[(2)] Can we prove Conjecture \ref{primes}, implying an $n^{\Omega\left(
\log n\right)  }$\ tree size lower bound for Shor states?

\item[(3)] Let $\left\vert \varphi\right\rangle $\ be a uniform superposition
over all $n$-bit strings of Hamming weight $n/2$. \ It is easy to show by
divide-and-conquer that $\operatorname*{TS}\left(  \left\vert \varphi
\right\rangle \right)  =n^{O\left(  \log n\right)  }$. \ Is this upper bound
tight? \ More generally, can we show a superpolynomial tree size lower bound
for any state with permutation symmetry?

\item[(4)] Is $\mathsf{Tree}=\mathsf{OTree}$? \ That is, are there tree states
that are not orthogonal tree states?

\item[(5)] Is the tensor-sum hierarchy of Section \ref{CQS}\ infinite? \ That
is, do we have $\mathsf{\Sigma}${}$_{\mathsf{k}}\neq\mathsf{\Sigma}$%
{}$_{\mathsf{k+1}}$\ for all $k$?

\item[(6)] Is $\mathsf{TreeBQP}=\mathsf{BPP}$? \ That is, can a quantum
computer that is always in a tree state be simulated classically? \ The key
question seems to be whether the concept class of multilinear formulas is
efficiently learnable.

\item[(7)] Is there a practical method to compute the tree size of, say,
$10$-qubit states? \ Such a method would have great value in interpreting
experimental results.
\end{enumerate}

\chapter{Quantum Search of Spatial Regions\label{GG}}

\begin{quote}
\textit{This chapter represents joint work with Andris Ambainis.}
\end{quote}

The goal of Grover's quantum search algorithm \cite{grover}\ is to search an
`unsorted database' of size $n$ in a number of queries proportional to
$\sqrt{n}$. \ Classically, of course, order $n$ queries are needed. \ It is
sometimes asserted that, although the speedup of Grover's algorithm\ is only
quadratic, this speedup is \textit{provable}, in contrast to the exponential
speedup of Shor's factoring algorithm \cite{shor}. \ But is that really true?
\ Grover's algorithm is typically imagined as speeding up combinatorial
search---and we do not know whether every problem in $\mathsf{NP}$ can be
classically solved quadratically faster than the \textquotedblleft
obvious\textquotedblright\ way, any more than we know whether factoring is in
$\mathsf{BPP}$.

\begin{figure}
[ptb]
\begin{center}
\includegraphics[
height=113.5pt,
width=243.0625pt
]%
{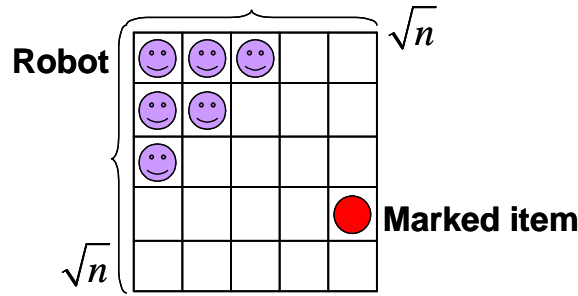}%
\caption[Quantum robot searching a 2D grid]{A quantum robot, in a
superposition over locations, searching for a
marked item on a 2D grid of size $\sqrt{n}\times\sqrt{n}$.}%
\label{gridfig}%
\end{center}
\end{figure}

But could Grover's algorithm speed up search of a \textit{physical region}?
\ Here the basic problem, it seems to us, is the time needed for signals to
travel across the region. \ For if we are interested in the fundamental limits
imposed by physics, then we should acknowledge that the speed of light is
finite, and that a bounded region of space can store only a finite amount of
information, according to the holographic principle\ \cite{bousso}. \ We
discuss the latter constraint in detail in Section \ref{PHYS}; for now, we say
only that it suggests a model in which a `quantum robot' occupies a
superposition over finitely many locations, and moving the robot from one
location to an adjacent one takes unit time. \ In such a model, the time
needed to search a region could depend critically on its spatial layout. \ For
example,\ if the $n$ entries are arranged on a line, then even to move the
robot from one end to the other takes $n-1$ steps. \ But what if the entries
are arranged on, say, a $2$-dimensional square grid (Figure \ref{gridfig})?%
%TCIMACRO{\FRAME{ftbpFU}{243.0625pt}{113.5pt}{0pt}{\Qcb{A quantum robot, in a
%superposition over locations, searching for a marked item on a 2D grid of size
%$\sqrt{n}\times\sqrt{n}$.}}{\Qlb{gridfig}}{Figure}%
%maintain-aspect-ratio TRUE;  display "USEDEF";  valid_file "T";
%width 243.0625pt;  height 113.5pt;  depth 0pt;  original-width 344.3125pt;
%original-height 159.375pt;  cropleft "0";  croptop "1";  cropright "1";
%cropbottom "0";  tempfilename 'gridfig.eps';tempfile-properties "XNPR";}}}%
%BeginExpansion
%EndExpansion

\section{Summary of Results\label{SUMGG}}

This chapter gives the first systematic treatment of quantum search of spatial
regions, with `regions' modeled as connected graphs. \ Our main result is
positive: we show that a quantum robot can search a $d$-dimensional hypercube
with $n$ vertices for a unique marked vertex in time $O\left(  \sqrt{n}%
\log^{3/2}n\right)  $\ when $d=2$, or $O\left(  \sqrt{n}\right)  $
when $d\geq3$. \ This matches (or in the case of $2$ dimensions,
nearly matches) the $\Omega\left(  \sqrt{n}\right)  $\ lower bound
for quantum search, and supports the view that Grover search of a
physical region presents no problem of principle.\ Our basic
technique is divide-and-conquer; indeed, once the idea is pointed
out, an upper bound of $O\left(  n^{1/2+\varepsilon}\right) $\
follows readily. \ However, to obtain the tighter bounds is more
difficult; for that we use the amplitude-amplification framework of
Brassard et al.\ \cite{bhmt}.

Section \ref{GRID} presents the main results;\ Section
\ref{MULTIPLE}\ shows further that, when there are $k$ or more
marked vertices, the search time becomes $O\left(
\sqrt{n}\log^{5/2}n\right)  $\ when $d=2$, or $\Theta\left(
\sqrt{n}/k^{1/2-1/d}\right)  $\ when $d\geq3$.\ \ Also, Section
\ref{IRREG} generalizes our algorithm to arbitrary graphs that have
`hypercube-like' expansion properties. \ Here the best bounds we can
achieve are $\sqrt {n}2^{O\left(  \sqrt{\log n}\right)  }$\ when
$d=2$, or $O\left(  \sqrt {n}\operatorname*{polylog}n\right)  $ when
$d>2$\ (note that $d$ need not be an integer). \ Table 14.1
summarizes the results.

\begin{table}[ptb]
\label{sumresults}
\begin{tabular}
[c]{c|cc}
& $d=2$ & $d>2$\\\hline
\multicolumn{1}{r|}{Hypercube, $1$ marked item} &
\multicolumn{1}{|l}{$O\left(  \sqrt{n}\log^{3/2}n\right)  $} &
\multicolumn{1}{l}{$\Theta\left(  \sqrt{n}\right)  $}\\
\multicolumn{1}{r|}{Hypercube, $k$ or more marked items} &
\multicolumn{1}{|l}{$O\left(  \sqrt{n}\log^{5/2}n\right)  $} &
\multicolumn{1}{l}{$\Theta\left(  \frac{\sqrt{n}}{k^{1/2-1/d}}\right)  $}\\
\multicolumn{1}{r|}{Arbitrary graph, $k$ or more marked items} &
\multicolumn{1}{|l}{$\sqrt{n}2^{O\left(  \sqrt{\log n}\right)  }$} &
\multicolumn{1}{l}{$\widetilde{\Theta}\left(  \frac{\sqrt{n}}{k^{1/2-1/d}%
}\right)  $}%
\end{tabular}
\caption[Summary of bounds for spatial search]{Upper and lower
bounds for quantum search on a $d$-dimensional graph given in this
chapter. \ The symbol $\widetilde{\Theta}$\ means that the upper
bound includes a polylogarithmic term. \ Note that, if $d=2$, then
$\Omega\left( \sqrt{n}\right)  $\ is always a lower bound, for any
number of
marked items.}%
\end{table}

Section \ref{APPL}\ shows, as an unexpected application of our search
algorithm, that the quantum communication complexity of the well-known
\textit{disjointness problem} is $O\left(  \sqrt{n}\right)  $. \ This improves
an $O\left(  \sqrt{n}c^{\log^{\ast}n}\right)  $\ upper bound of H\o yer and de
Wolf \cite{hoyerdewolf},\ and matches the $\Omega\left(  \sqrt{n}\right)
$\ lower bound of Razborov \cite{razborov:cc}.

The rest of the chapter is about the formal model that underlies our
results. \ Section \ref{PHYS} sets the stage for this model, by
exploring the ultimate limits on information storage imposed by
properties of space and time. \ This discussion\ serves only to
motivate our results; thus, it can be safely skipped by readers
unconcerned with the physical universe. \ In Section \ref{MODEL}\ we
define \textit{quantum query algorithms on graphs}, a model similar
to quantum query algorithms as defined in Section \ref{BLACKBOX},
but with the added requirement that unitary operations be `local'
with respect to some graph. \ In Section \ref{LOCALGG}\ we address
the difficult question, which also arises in work on quantum random
walks \cite{aakv}\ and quantum cellular automata \cite{watrous:ca},
of what `local' means. Section \ref{GENERALGG}\ proves general facts
about our model, including an upper bound of $O\left(
\sqrt{n\delta}\right)  $\ for the time needed to search any graph
with diameter $\delta$, and a proof (using the hybrid argument of
Bennett et al.\ \cite{bbbv}) that this upper bound is tight for
certain graphs. \ We conclude in Section \ref{OPENGG}\ with some
open problems.

\section{Related Work\label{PREVGG}}

In a paper on `Space searches with a quantum robot,' Benioff
\cite{benioff:robot}\ asked whether Grover's algorithm\ can speed up search of
a physical region, as opposed to a combinatorial search space. \ His answer
was discouraging: for a $2$-D grid of size $\sqrt{n}\times\sqrt{n}$, Grover's
algorithm is no faster than classical search. \ The reason is that, during
each of the $\Theta\left(  \sqrt{n}\right)  $ Grover iterations, the algorithm
must use order $\sqrt{n}$\ steps just to travel across the grid and return to
its starting point for the diffusion step. \ On the other hand, Benioff noted,
Grover's algorithm does yield some speedup for grids of dimension $3$ or
higher, since those grids have diameter less than $\sqrt{n}$.

Our results show that Benioff's claim is mistaken: by using Grover's algorithm
more carefully, one can search a $2$-D grid for a single marked vertex in
$O\left(  \sqrt{n}\log^{3/2}n\right)  $ time. \ To us this illustrates why one
should not assume an algorithm is optimal on heuristic grounds. \ Painful
experience---for example, the \textquotedblleft obviously
optimal\textquotedblright\ $O\left(  n^{3}\right)  $\ matrix multiplication
algorithm \cite{strassen}---is what taught computer scientists to see the
proving of lower bounds as more than a formality.

Our setting is related to that of quantum random walks on graphs
\cite{aakv,ccdfgs,cfg,skw}. \ In an earlier version of this chapter,
we asked whether quantum walks might yield an alternative spatial
search algorithm, possibly even one that outperforms our
divide-and-conquer algorithm. \ Motivated by this question,\ Childs
and Goldstone \cite{cg} managed to show that in the continuous-time
setting, a quantum walk can search a $d$-dimensional hypercube for a
single marked vertex in time $O\left( \sqrt{n}\log n\right)  $ when
$d=4$, or $O\left(  \sqrt{n}\right)  $ when $d\geq5$. \ Our
algorithm was still faster in $3$ or fewer dimensions\ (see Table
14.2). \ Subsequently, however, Ambainis, Kempe, and Rivosh
\cite{akr}\ gave an algorithm based on a discrete-time quantum walk,
which was as fast as ours in $3$ or more dimensions, and faster in
$2$ dimensions. \ In particular, when $d=2$\ their algorithm used
only $O\left(  \sqrt{n}\log n\right)  $\ time to find a unique
marked vertex. \ Childs and Goldstone \cite{cg2} then gave a
continuous-time quantum walk algorithm with the same performance,
and related this algorithm to properties of the Dirac equation. \ It
is still open whether $O\left(  \sqrt{n}\right)  $\ time is
achievable in $2$ dimensions.

\begin{table}[ptb]
\label{bounds}
\begin{tabular}
[c]{l|cccc}
& $d=2$ & $d=3$ & $d=4$ & $d\geq5$\\\hline
This chapter & \multicolumn{1}{|l}{$O\left(  \sqrt{n}\log^{3/2}n\right)  $} &
\multicolumn{1}{l}{$O\left(  \sqrt{n}\right)  $} &
\multicolumn{1}{l}{$O\left(  \sqrt{n}\right)  $} &
\multicolumn{1}{l}{$O\left(  \sqrt{n}\right)  $}\\
\cite{cg} & \multicolumn{1}{|l}{$O\left(  n\right)  $} &
\multicolumn{1}{l}{$O\left(  n^{5/6}\right)  $} & \multicolumn{1}{l}{$O\left(
\sqrt{n}\log n\right)  $} & \multicolumn{1}{l}{$O\left(  \sqrt{n}\right)  $}\\
\cite{akr,cg2} & \multicolumn{1}{|l}{$O\left(  \sqrt{n}\log n\right)  $} &
\multicolumn{1}{l}{$O\left(  \sqrt{n}\right)  $} &
\multicolumn{1}{l}{$O\left(  \sqrt{n}\right)  $} &
\multicolumn{1}{l}{$O\left(  \sqrt{n}\right)  $}%
\end{tabular}
\caption[Divide-and-conquer versus quantum walks]{Time needed to find a unique
marked item in a $d$-dimensional hypercube, using the divide-and-conquer
algorithms of this chapter, the original quantum walk algorithm of Childs and
Goldstone \cite{cg}, and the improved walk algorithms of Ambainis, Kempe, and
Rivosh \cite{akr}\ and Childs and Goldstone \cite{cg2}.}%
\label{divwalk}%
\end{table}

Currently, the main drawback of the quantum walk approach is that all analyses
have relied heavily on symmetries in the underlying graph. \ If even minor
`defects' are introduced, it is no longer known how to upper-bound the running
time. \ By contrast, the analysis of our divide-and-conquer algorithm is
elementary, and does not depend on eigenvalue bounds. \ We can therefore show
that the algorithm works for any graphs with sufficiently good expansion properties.

Childs and Goldstone \cite{cg}\ argued that the quantum walk approach has the
advantage of requiring fewer auxiliary qubits than the divide-and-conquer
approach. \ However, the need for many qubits\ was an artifact of how we
implemented the algorithm in a previous version of the chapter. \ The current
version uses only \textit{one} qubit.

\section{The Physics of Databases\label{PHYS}}

Theoretical computer science generally deals with the limit as some resource
(such as time or memory) increases to infinity. \ What is not always
appreciated is that, as the resource bound increases, physical constraints may
come into play that were negligible at `sub-asymptotic'\ scales.\ \ We believe
theoretical computer scientists ought to know something about such
constraints, and to account for them when possible. \ For if the constraints
are ignored on the ground that they \textquotedblleft never matter in
practice,\textquotedblright\ then the obvious question arises: why use
asymptotic analysis in the first place, rather than restricting attention to
those instance sizes that occur in practice?

A constraint of particular interest for us is the \textit{holographic
principle} \cite{bousso},\ which arose from black-hole thermodynamics. \ The
principle states that the information content of any spatial region is
upper-bounded by its \textit{surface area} (not volume), at a rate of one bit
per Planck area, or about $1.4\times10^{69}$\ bits per square meter.
\ Intuitively, if one tried to build a spherical hard disk with mass density
$\upsilon$, one could not keep expanding it forever. \ For as soon as the
radius reached the Schwarzschild bound of $r=\sqrt{3/\left(  8\pi
\upsilon\right)  }$ (in Planck units, $c=G=\hbar=k=1$), the hard disk would
collapse to form a black hole, and thus its contents would be irretrievable.

Actually the situation is worse than that: even a \textit{planar} hard disk of
constant mass density would collapse to form a black hole once its radius
became sufficiently large, $r=\Theta\left(  1/\upsilon\right)  $. \ (We assume
here that the hard disk is disc-shaped. \ A linear or $1$-D hard disk could
expand indefinitely without collapse.) \ It is possible, though, that a hard
disk's information content could asymptotically exceed its mass. \ For
example, a black hole's mass is proportional to the radius of its event
horizon, but the entropy is proportional to the \textit{square} of the radius
(that is, to the surface area). \ Admittedly, inherent difficulties with
storage and retrieval make a black hole horizon less than ideal as a hard
disk. \ However, even a weakly-gravitating system could store information at a
rate asymptotically exceeding its mass-energy. \ For instance, Bousso
\cite{bousso}\ shows that an enclosed ball of radiation with radius $r$ can
store\ $n=\Theta\left(  r^{3/2}\right)  $ bits, even though its energy grows
only as $r$. \ Our results in Section\ \ref{SCATTERED}\ will imply that a
quantum robot could (in principle!) search such a `radiation disk' for a
marked item in time $O\left(  r^{5/4}\right)  =O\left(  n^{5/6}\right)  $.
\ This is some improvement over the trivial $O\left(  n\right)  $ upper bound
for a $1$-D hard disk, though it falls short of the desired $O\left(  \sqrt
{n}\right)  $.

In general, if $n=r^{c}$ bits are scattered throughout a $3$-D ball of radius
$r$ (where $c\leq3$\ and the bits' locations are known), we will show in
Theorem \ref{scatterthm}\ that the time needed to search for a `$1$'\ bit
grows as $n^{1/c+1/6}=r^{1+c/6}$ (omitting logarithmic factors). \ In
particular, if $n=\Theta\left(  r^{2}\right)  $\ (saturating the holographic
bound), then the time grows as $n^{2/3}$ or $r^{4/3}$. \ To achieve a search
time of $O\left(  \sqrt{n}\operatorname*{polylog}n\right)  $, the bits would
need to be concentrated on a $2$-D surface.

Because of the holographic principle, we see that it is not only quantum
mechanics that yields a $\Omega\left(  \sqrt{n}\right)  $\ lower bound on the
number of steps needed for unordered search. \ If the items to be searched are
laid out spatially, then general relativity in $3+1$\ dimensions independently
yields the same bound, $\Omega\left(  \sqrt{n}\right)  $, up to a constant
factor.\footnote{Admittedly, the holographic principle is part of quantum
gravity and not general relativity \textit{per se}. \ All that matters for us,
though, is that the principle seems logically independent of
quantum-mechanical linearity, which is what produces the \textquotedblleft
other\textquotedblright\ $\Omega\left(  \sqrt{n}\right)  $\ bound.}
\ Interestingly, in $d+1$\ dimensions the relativity bound would be
$\Omega\left(  n^{1/\left(  d-1\right)  }\right)  $, which\ for $d>3$\ is
weaker than the quantum mechanics bound. \ Given that our two fundamental
theories yield the same lower bound, it is natural to ask whether that bound
is tight. \ The answer seems to be that it is \textit{not} tight, since (i)
the entropy on a black hole horizon is not efficiently accessible\footnote{In
the case of a black hole horizon, waiting for the bits to be emitted as
Hawking radiation---as recent evidence suggests that they are \cite{sv}%
---takes time proportional to $r^{3}$,\ which is much too long.}, and (ii)
weakly-gravitating systems are subject to the \textit{Bekenstein bound}
\cite{bekenstein}, an even stronger entropy constraint than the holographic bound.

Yet it is still of basic interest to know whether $n$ bits in a radius-$r$
ball can be searched in time $o\left(  \min\left\{  n,r\sqrt{n}\right\}
\right)  $---that is, whether it is possible to do \textit{anything} better
than either brute-force quantum search (with the drawback pointed out by
Benioff \cite{benioff:robot}), or classical search. \ Our results show that it
is possible.

From a physical point of view, several questions naturally arise: (1) whether
our complexity measure is realistic; (2) how to account for time dilation; and
(3) whether given the number of bits we are imagining, cosmological bounds are
also relevant. \ Let us address these questions in turn.

(1) One could argue that to maintain a `quantum database' of size
$n$ requires $n$ computing elements (\cite{zalka}, though see also
\cite{rg}). \ So why not just exploit those elements to search the
database in \textit{parallel}? \ Then it becomes trivial to show
that the search time is limited only by the radius of the database,
so the algorithms of this chapter are unnecessary. \ Our response is
that, while there might be $n$ `passive' computing elements (capable
of storing data), there might be many fewer `active'\ elements,
which we consequently wish to place in a superposition over
locations. \ This assumption seems physically unobjectionable. \ For
a particle (and indeed any object) really does have an indeterminate
location, not merely an indeterminate internal state (such as spin)
\textit{at} some location. \ We leave as an open problem, however,
whether our assumption is valid for specific quantum computer
architectures such as ion traps.

(2) So long as we invoke general relativity, should we not also
consider the effects of time dilation? \ Those effects are indeed
pronounced near a black hole horizon. \ Again, though, for our upper
bounds we will have in mind systems\ far from the Schwarzschild
limit, for which any time dilation is by at most a constant factor
independent of $n$.

(3) How do cosmological considerations affect our analysis? \ Bousso
\cite{bousso:vac}\ argues that, in a spacetime with positive
cosmological constant $\Lambda>0$, the total number of bits
accessible to any one experiment is at most $3\pi/\left(
\Lambda\ln2\right)  $, or roughly $10^{122}$ given current
experimental bounds \cite{perlmutter}\ on $\Lambda $.\footnote{Also,
Lloyd \cite{lloyd}\ argues that the total number of bits accessible
\textit{up till now} is at most the square of the number of Planck
times elapsed so far, or about $\left(  10^{61}\right)  ^{2}=10^{122}%
$.\ \ Lloyd's bound, unlike Bousso's, does not depend on $\Lambda$ being
positive. The numerical coincidence between the two bounds reflects the
experimental finding \cite{perlmutter,ryden}\ that we live in a transitional
era, when both $\Lambda$\ and \textquotedblleft dust\textquotedblright%
\ contribute significantly to the universe's net energy balance ($\Omega
_{\Lambda}\approx0.7$, $\Omega_{\operatorname*{dust}}\approx0.3$). \ In
earlier times dust (and before that radiation)\ dominated, and Lloyd's bound
was tighter. \ In later times $\Lambda$\ will dominate, and Bousso's bound
will be tighter. \ \textit{Why} we should live in such a transitional era is
unknown.} \ Intuitively, even if the universe is spatially infinite, most of
it recedes too quickly from any one observer to be harnessed as computer memory.

One response to this result is to assume an idealization in which $\Lambda
$\ vanishes, although Planck's constant $\hbar$ does not vanish. \ As
justification, one could argue that without the idealization $\Lambda=0$,
\textit{all} asymptotic bounds in computer science are basically fictions.
\ But perhaps a better response is to accept the $3\pi/\left(  \Lambda
\ln2\right)  $ bound, and then ask how close one can come to
\textit{saturating} it in different scenarios. \ Classically, the maximum
number of bits that can be searched is, in a crude
model\footnote{Specifically, neglecting gravity and other forces that could
counteract the effect of $\Lambda$.}, actually proportional to $1/\sqrt
{\Lambda}\approx10^{61}$ rather than $1/\Lambda$. \ The reason is that if a
region had much more than $1/\sqrt{\Lambda}$\ bits, then after $1/\sqrt
{\Lambda}$\ Planck times---that is, about $10^{10}$\ years, or roughly the
current age of the universe---most of the region would have receded beyond
one's cosmological horizon. \ What our results suggest is that, using a
quantum robot, one could come closer to saturating the cosmological
bound---since, for example, a $2$-D region of size $1/\Lambda$ can be searched
in time $O\left(  \frac{1}{\sqrt{\Lambda}}\operatorname*{polylog}\frac
{1}{\sqrt{\Lambda}}\right)  $. \ How anyone could \textit{prepare} (say) a
database of size much greater than $1/\sqrt{\Lambda}$\ remains unclear, but if
such a database existed, it could be searched!

\section{The Model\label{MODELGG}}

As discussed in Part \ref{LQC}, much of what is known about the power of
quantum computing comes from the \textit{black-box} or \textit{query}
model---in which one counts only the number of queries to an oracle, not the
number of computational steps. \ We will take this model as the starting point
for a formal definition of quantum robots. \ Doing so will focus attention on
our main concern: how much harder is it to evaluate a function when its inputs
are spatially separated? \ As it turns out, all of our algorithms
\textit{will} be efficient as measured by the number of gates and auxiliary
qubits needed to implement them.

For simplicity, we assume that a robot's goal is to evaluate a Boolean
function $f:\left\{  0,1\right\}  ^{n}\rightarrow\left\{  0,1\right\}  $,
which could be partial or total. \ A `region of space' is a connected
undirected graph $G=\left(  V,E\right)  $\ with vertices $V=\left\{
v_{1},\ldots,v_{n}\right\}  $. \ Let $X=x_{1}\ldots x_{n}\in\left\{
0,1\right\}  ^{n}$\ be an input to $f$; then each bit\ $x_{i}$\ is available
only at vertex $v_{i}$. \ We assume the robot knows $G$ and the vertex labels
in advance, and so is ignorant only of the $x_{i}$ bits. \ We thus sidestep a
major difficulty for quantum walks \cite{aakv}, which is how to ensure that a
process on an unknown graph is unitary.

At any time, the robot's state has the form%
\[
\sum\alpha_{i,z}\left\vert v_{i},z\right\rangle \text{.}%
\]
Here $v_{i}\in V$\ is a vertex, representing the robot's location; and $z$ is
a bit string (which can be arbitrarily long), representing the robot's
internal configuration. \ The state evolves via an alternating sequence of
$T$\ algorithm\ steps and $T$ oracle steps:%
\[
U^{\left(  1\right)  }\rightarrow O^{\left(  1\right)  }\rightarrow U^{\left(
1\right)  }\rightarrow\cdots\rightarrow U^{\left(  T\right)  }\rightarrow
O^{\left(  T\right)  }\text{.}%
\]
An oracle step $O^{\left(  t\right)  }$ maps each basis state $\left\vert
v_{i},z\right\rangle $\ to $\left\vert v_{i},z\oplus x_{i}\right\rangle $,
where $x_{i}$\ is exclusive-OR'ed into the first bit of $z$. \ An algorithm
step $U^{\left(  t\right)  }$ can be any unitary matrix that (1) does not
depend on $X$, and (2) acts `locally' on $G$. \ How to make the second
condition precise is the subject of Section \ref{LOCALGG}.

The initial state of the algorithm is $\left\vert v_{1},0\right\rangle $.
\ Let $\alpha_{i,z}^{\left(  t\right)  }\left(  X\right)  $\ be the amplitude
of $\left\vert v_{i},z\right\rangle $\ immediately after the $t^{th}$\ oracle
step; then the algorithm succeeds with probability $1-\varepsilon$\ if%
\[
\sum_{\left\vert v_{i},z\right\rangle \,:\,z_{OUT}=f\left(  X\right)
}\left\vert \alpha_{i,z}^{\left(  T\right)  }\left(  X\right)  \right\vert
^{2}\geq1-\varepsilon
\]
for all inputs $X$, where $z_{OUT}$\ is a bit of $z$ representing the output.

\subsection{Locality Criteria\label{LOCALGG}}

Classically, it is easy to decide whether a stochastic matrix acts
\textit{locally} with respect to a graph $G$:\ it does if it moves probability
only along the edges of $G$. \ In the quantum case, however, interference
makes the question much more subtle. \ In this section we propose three
criteria for whether a unitary matrix $U$ is local. \ Our algorithms can be
implemented using the most restrictive of these criteria, whereas our lower
bounds apply to all three of them.

The first criterion we call \textit{Z-locality} (for zero): $U$ is Z-local if,
given any pair of non-neighboring vertices $v_{1},v_{2}$ in $G$, $U$ ``sends
no amplitude''\ from $v_{1}$\ to $v_{2}$; that is, the corresponding entries
in $U$ are all $0$. \ The second criterion, \textit{C-locality} (for
composability), says that this is not enough: not only must $U$ send amplitude
only between neighboring vertices, but it must be composed of a product of
commuting unitaries, each of which acts on a single edge. \ The third
criterion is perhaps the most natural one to a physicist: $U$ is
\textit{H-local} (for Hamiltonian) if it can be obtained by applying a
locally-acting, low-energy Hamiltonian for some fixed amount of time.\ \ More
formally, let $U_{i,z\rightarrow i^{\ast},z^{\ast}}$\ be the entry in the
$\left|  v_{i},z\right\rangle $\ column and $\left|  v_{i^{\ast}},z^{\ast
}\right\rangle $\ row of $U$.

\begin{definition}
$U$ is Z-local if $U_{i,z\rightarrow i^{\ast},z^{\ast}}=0$\ whenever $i\neq
i^{\ast}$\ and $\left(  v_{i},v_{i^{\ast}}\right)  $\ is not an edge of $G$.
\end{definition}

\begin{definition}
$U$ is C-local if the basis states can be partitioned into subsets
$P_{1},\ldots,P_{q}$\ such that

\begin{enumerate}
\item[(i)] $U_{i,z\rightarrow i^{\ast},z^{\ast}}=0$\ whenever $\left\vert
v_{i},z\right\rangle $\ and $\left\vert v_{i^{\ast}},z^{\ast}\right\rangle
$\ belong to distinct $P_{j}$'s, and

\item[(ii)] for each $j$, all basis states in $P_{j}$\ are either from the
same vertex or from two adjacent vertices.
\end{enumerate}
\end{definition}

\begin{definition}
$U$ is H-local if $U=e^{iH}$\ for some Hermitian $H$ with eigenvalues of
absolute value at most $\pi$, such that $H_{i,z\rightarrow i^{\ast},z^{\ast}%
}=0$\ whenever $i\neq i^{\ast}$\ and $\left(  v_{i},v_{i^{\ast}}\right)  $\ is
not an edge in $E$.
\end{definition}

If a unitary matrix is C-local, then it is also Z-local and H-local. \ For the
latter implication, note that any unitary $U$ can be written as $e^{iH}$\ for
some $H$ with eigenvalues of absolute value at most $\pi$. \ So we can write
the unitary $U_{j}$\ acting on each $P_{j}$ as $e^{iH_{j}}$; then since the
$U_{j}$'s\ commute,%
\[
\prod U_{j}=e^{i\sum H_{j}}\text{.}%
\]
Beyond that, though, how are the locality criteria related? \ Are they
approximately equivalent? \ If not, then does a problem's complexity in our
model ever depend on which criterion is chosen? \ Let us emphasize that these
questions are \textit{not} answered by, for example, the Solovay-Kitaev
theorem (see \cite{nc}), that an $n\times n$\ unitary matrix can be
approximated using a number of gates polynomial in $n$. \ For recall that the
definition of C-locality requires the edgewise operations to commute---indeed,
without that requirement, one could produce any unitary matrix at all. \ So
the relevant question, which we leave open, is whether any Z-local or H-local
unitary can be approximated by a product of, say, $O\left(  \log n\right)
$\ C-local unitaries. \ (A product of $O\left(  n\right)  $ such unitaries
trivially suffices, but that is far too many.) \ Again, the algorithms in this
chapter will use C-local unitaries, whereas the lower bounds will apply even
to Z-local and H-local unitaries.

\section{General Bounds\label{GENERALGG}}

Given a Boolean function $f:\left\{  0,1\right\}
^{n}\rightarrow\left\{ 0,1\right\}  $, the quantum query complexity
$Q\left(  f\right)  $ is the minimum $T$ for which there exists a
$T$-query quantum algorithm that evaluates $f$ with probability at
least $2/3$ on all inputs. \ (We will always be interested in the
\textit{two-sided, bounded-error} complexity, denoted $Q_{2}\left(
f\right) $ elsewhere in this thesis.) \ Similarly, given a graph $G$
with $n$ vertices labeled $1,\ldots,n$, we let $Q\left(  f,G\right)
$ be the minimum $T$ for which there exists a $T$-query quantum
robot on $G$ that evaluates $f$\ with probability $2/3$. \ Here the
algorithm steps must be C-local; we use $Q^{Z}\left(  f,G\right)  $\
and $Q^{H}\left( f,G\right)  $\ to denote the corresponding measure
with Z-local and H-local steps respectively. \ Clearly $Q\left(
f,G\right)  \geq Q^{Z}\left(  f,G\right)  $\ and $Q\left( f,G\right)
\geq Q^{H}\left( f,G\right)  $; we do not know whether all three
measures are asymptotically equivalent.

Let $\delta_{G}$\ be the diameter of $G$, and\ call $f$ \textit{nondegenerate}
if it depends on all $n$ input bits.

\begin{proposition}
\label{immed}For all $f,G$,

\begin{enumerate}
\item[(i)] $Q\left(  f,G\right)  \leq2n-3$.

\item[(ii)] $Q\left(  f,G\right)  \leq\left(  2\delta_{G}+1\right)  Q\left(
f\right)  $.

\item[(iii)] $Q\left(  f,G\right)  \geq Q\left(  f\right)  $.

\item[(iv)] $Q\left(  f,G\right)  \geq\delta_{G}/2$ if $f$ is nondegenerate.
\end{enumerate}
\end{proposition}

\begin{proof}
\quad

\begin{enumerate}
\item[(i)] Starting from the root, a spanning tree for $G$ can be traversed in
$2\left(  n-1\right)  -1$\ steps (there is no need to return to the root).

\item[(ii)] We can simulate a query in $2\delta_{G}$\ steps, by fanning out
from the start vertex $v_{1}$ and then returning. \ Applying a unitary at
$v_{1}$\ takes $1$ step.

\item[(iii)] Obvious.

\item[(iv)] There exists a vertex $v_{i}$ whose distance to $v_{1}$\ is at
least $\delta_{G}/2$, and $f$\ could depend on $x_{i}$.
\end{enumerate}
\end{proof}

We now show that the model is robust.

\begin{proposition}
\label{robustgg}For nondegenerate $f$, the following change $Q\left(
f,G\right)  $\ by at most a constant factor.

\begin{enumerate}
\item[(i)] Replacing the initial state $\left\vert v_{1},0\right\rangle $\ by
an arbitrary (known) $\left\vert \psi\right\rangle $.

\item[(ii)] Requiring the final state to be localized at some vertex $v_{i}%
$\ with probability at least $1-\varepsilon$, for a constant $\varepsilon>0$.

\item[(iii)] Allowing multiple algorithm steps between each oracle step (and
measuring the complexity by the number of algorithm steps).
\end{enumerate}
\end{proposition}

\begin{proof}

\begin{enumerate}
\item[(i)] We can transform $\left\vert v_{1},0\right\rangle $\ to $\left\vert
\psi\right\rangle $\ (and hence $\left\vert \psi\right\rangle $\ to
$\left\vert v_{1},0\right\rangle $) in $\delta_{G}=O\left(  Q\left(
f,G\right)  \right)  $ steps, by fanning out from $v_{1}$\ along the edges of
a minimum-height spanning tree.

\item[(ii)] Assume without loss of generality that $z_{OUT}$\ is accessed only
once, to write the output. \ Then after $z_{OUT}$\ is accessed, uncompute
(that is, run the algorithm backwards) to localize the final state at $v_{1}$.
\ The state can then be localized at any $v_{i}$\ in $\delta_{G}=O\left(
Q\left(  f,G\right)  \right)  $ steps. \ We can succeed with any constant
probability by repeating this procedure a constant number of times.

\item[(iii)] The oracle step $O$ is its own inverse, so we can implement a
sequence $U_{1},U_{2},\ldots$\ of algorithm steps as follows (where $I$ is the
identity):%
\[
U_{1}\rightarrow O\rightarrow I\rightarrow O\rightarrow U_{2}\rightarrow\cdots
\]

\end{enumerate}
\end{proof}

A function of particular interest is $f=\operatorname*{OR}\left(  x_{1}%
,\ldots,x_{n}\right)  $, which outputs $1$ if and only if $x_{i}=1$\ for some
$i$. \ We first give a general upper bound on $Q\left(  \operatorname*{OR}%
,G\right)  $\ in terms of the diameter of $G$. \ (Throughout the chapter, we
sometimes omit floor and ceiling signs if they clearly have no effect on the asymptotics.)

\begin{proposition}
\label{upperdg}%
\[
Q\left(  \operatorname*{OR},G\right)  =O\left(  \sqrt{n\delta_{G}}\right)  .
\]

\end{proposition}

\begin{proof}
Let $\tau$ be a minimum-height spanning tree for $G$, rooted at $v_{1}$. \ A
depth-first search on $\tau$ uses $2n-2$\ steps. \ Let $S_{1}$\ be the set of
vertices visited by depth-first search in steps $1$ to $\delta_{G}$, $S_{2}%
$\ be those visited in steps $\delta_{G}+1$\ to $2\delta_{G}$, and so on.
\ Then%
\[
S_{1}\cup\cdots\cup S_{2n/\delta_{G}}=V\text{.}%
\]
Furthermore, for each $S_{j}$ there is a classical algorithm $A_{j}$, using at
most $3\delta_{G}$ steps, that starts at $v_{1}$, ends at $v_{1}$, and outputs
`$1$' if and only if $x_{i}=1$\ for some $v_{i}\in S_{j}$. \ Then we simply
perform Grover search at $v_{1}$\ over all $A_{j}$; since each iteration takes
$O\left(  \delta_{G}\right)  $\ steps and there are $O\left(  \sqrt
{2n/\delta_{G}}\right)  $ iterations, the number of steps is $O\left(
\sqrt{n\delta_{G}}\right)  $.
\end{proof}

The bound of Proposition \ref{upperdg}\ is tight:

\begin{theorem}
\label{lowerdg}For all $\delta$,\ there exists a graph $G$ with diameter
$\delta_{G}=\delta$\ such that%
\[
Q\left(  \operatorname*{OR},G\right)  =\Omega\left(  \sqrt{n\delta}\right)  .
\]
Indeed, $Q^{Z}\left(  f,G\right)  $\ and $Q^{H}\left(  f,G\right)  $\ are also
$\Omega\left(  \sqrt{n\delta}\right)  $.
\end{theorem}

\begin{proof}
For simplicity, we first consider the C-local and Z-local cases, and then
discuss what changes in the H-local\ case. \ Let $G$ be a `starfish' with
central vertex $v_{1}$\ and $M=2\left(  n-1\right)  /\delta$ legs
$L_{1},\ldots,L_{M}$, each of length $\delta/2$ (see Figure \ref{starfish}).%
%TCIMACRO{\FRAME{ftbpFU}{125.375pt}{111.375pt}{0pt}{\Qcb{The `starfish' graph
%$G$. \ The marked item is at one of the tip vertices.}}{\Qlb{starfish}%
%}{Figure}{\special{ language "Scientific Word";  type "GRAPHIC";
%maintain-aspect-ratio TRUE;  display "USEDEF";  valid_file "T";
%width 125.375pt;  height 111.375pt;  depth 0pt;  original-width 176.4375pt;
%original-height 156.1875pt;  cropleft "0";  croptop "1";  cropright "1";
%cropbottom "0";  tempfilename 'starfish.eps';tempfile-properties "XNPR";}}}%
%BeginExpansion
\begin{figure}
[ptb]
\begin{center}
\includegraphics[
height=111.375pt,
width=125.375pt
]%
{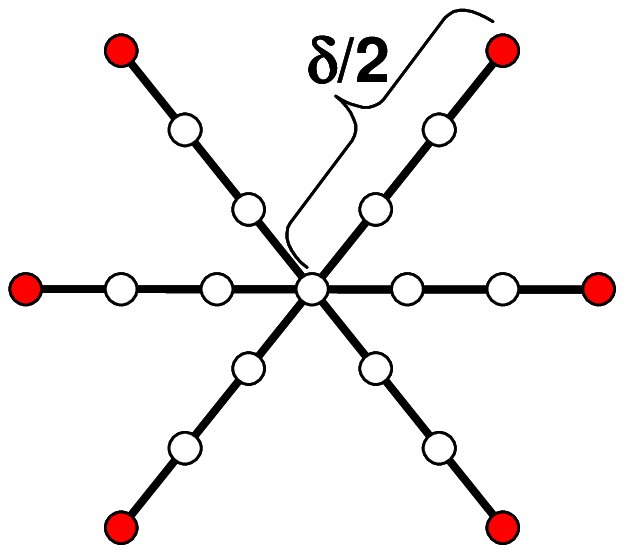}%
\caption[The `starfish' graph]{The `starfish' graph $G$. \ The
marked item is at one of the tip
vertices.}%
\label{starfish}%
\end{center}
\end{figure}
%EndExpansion
We use the hybrid argument of Bennett et al.\ \cite{bbbv}. \ Suppose
we run the algorithm on the all-zero input $X_{0}$. \ Then define
the \textit{query magnitude} $\Gamma_{j}^{\left(  t\right)  }$\ to
be the probability of finding
the robot in leg\ $L_{j}$ immediately after the $t^{th}$\ query:%
\[
\Gamma_{j}^{\left(  t\right)  }=\sum_{v_{i}\in L_{j}\,}\sum_{z\,}\left\vert
\alpha_{i,z}^{\left(  t\right)  }\left(  X_{0}\right)  \right\vert
^{2}\text{.}%
\]
Let $T$ be the total number of queries, and let $w=T/\left(  c\delta\right)  $
for some constant $0<c<1/2$. \ Clearly%
\[
\sum_{q=0}^{w-1}\sum_{j=1}^{M}\Gamma_{j}^{\left(  T-qc\delta\right)  }\leq
\sum_{q=0}^{w-1}1=w\text{.}%
\]
Hence there must exist a leg $L_{j^{\ast}}$\ such that
\[
\sum_{q=0}^{w-1}\Gamma_{j^{\ast}}^{\left(  T-qc\delta\right)  }\leq\frac{w}%
{M}=\frac{w\delta}{2\left(  n-1\right)  }.
\]
Let $v_{i^{\ast}}$ be the tip vertex of $L_{j^{\ast}}$, and let $Y$ be the
input which is $1$\ at $v_{i^{\ast}}$\ and $0$ elsewhere. \ Then let $X_{q}%
$\ be a hybrid input, which is $X_{0}$\ during queries $1$ to $T-qc\delta$,
but $Y$ during queries $T-qc\delta+1$ to $T$. \ Also, let%
\[
\left\vert \psi^{\left(  t\right)  }\left(  X_{q}\right)  \right\rangle
=\sum_{i,z}\alpha_{i,z}^{\left(  t\right)  }\left(  X_{q}\right)  \left\vert
v_{i},z\right\rangle
\]
be the algorithm's state after $t$ queries when run on $X_{q}$, and let%
\begin{align*}
D\left(  q,r\right)   &  =\left\Vert \left\vert \psi^{\left(  T\right)
}\left(  X_{q}\right)  \right\rangle -\left\vert \psi^{\left(  T\right)
}\left(  X_{r}\right)  \right\rangle \right\Vert _{2}^{2}\\
&  =\sum_{v_{i}\in G\,}\sum_{z\,}\left\vert \alpha_{i,z}^{\left(  T\right)
}\left(  X_{q}\right)  -\alpha_{i,z}^{\left(  T\right)  }\left(  X_{r}\right)
\right\vert ^{2}\text{.}%
\end{align*}
Then for all $q\geq1$,\ we claim that $D\left(  q-1,q\right)  \leq
4\Gamma_{j^{\ast}}^{\left(  T-qc\delta\right)  }$. \ For by unitarity, the
Euclidean distance between $\left\vert \psi^{\left(  t\right)  }\left(
X_{q-1}\right)  \right\rangle $\ and $\left\vert \psi^{\left(  t\right)
}\left(  X_{q}\right)  \right\rangle $\ can only increase as a result of
queries $T-qc\delta+1$\ through $T-\left(  q-1\right)  c\delta$. \ But no
amplitude from outside $L_{j^{\ast}}$\ can reach $v_{i^{\ast}}$\ during that
interval, since the distance is $\delta/2$\ and there are only $c\delta
<\delta/2$\ time steps. \ Therefore, switching from $X_{q-1}$\ to $X_{q}$\ can
only affect amplitude that is in $L_{j^{\ast}}$\ immediately after query
$T-qc\delta$:%
\begin{align*}
D\left(  q-1,q\right)   &  \leq\sum_{v_{i}\in L_{j^{\ast}}\,}\sum
_{z\,}\left\vert \alpha_{i,z}^{\left(  T-qc\delta\right)  }\left(
X_{q}\right)  -\left(  -\alpha_{i,z}^{\left(  T-qc\delta\right)  }\left(
X_{q}\right)  \right)  \right\vert ^{2}\\
&  =4\sum_{v_{i}\in L_{j^{\ast}}\,}\sum_{z\,}\left\vert \alpha_{i,z}^{\left(
T-qc\delta\right)  }\left(  X_{0}\right)  \right\vert ^{2}=4\Gamma_{j^{\ast}%
}^{\left(  T-qc\delta\right)  }.
\end{align*}
It follows that%
\[
\sqrt{D\left(  0,w\right)  }\leq\sum_{q=1}^{w}\sqrt{D\left(  q-1,q\right)
}\leq2\sum_{q=1}^{w}\sqrt{\Gamma_{j^{\ast}}^{\left(  T-qc\delta\right)  }}%
\leq2w\sqrt{\frac{\delta}{2\left(  n-1\right)  }}=\frac{T}{c}\sqrt{\frac
{2}{\delta\left(  n-1\right)  }}.
\]
Here the first inequality uses the triangle inequality, and the third uses the
Cauchy-Schwarz inequality. \ Now assuming the algorithm is correct we need
$D\left(  0,w\right)  =\Omega\left(  1\right)  $, which implies that
$T=\Omega\left(  \sqrt{n\delta}\right)  $.

In the H-local case, it is no longer true that no amplitude from outside
$L_{j^{\ast}}$\ can reach $v_{i^{\ast}}$\ in $c\delta$\ time steps. \ But if
$c$ is a small enough constant, then the amount of amplitude that can reach
$v_{i^{\ast}}$\ decreases exponentially in $\delta$. \ To see this,\ assume
without loss of generality that all amplitude not in $L_{j^{\ast}}$\ starts in
the state $\left\vert v_{0},\psi\right\rangle $, where $\left\vert
\psi\right\rangle $ is some superposition over auxiliary qubits. \ Let $H$ be
the local Hamiltonian that acts between the $t^{th}$\ and $\left(  t+1\right)
^{st}$\ queries,\ all of whose eigenvalues have absolute value at most $\pi$.
\ Since $H$\ is Hermitian, we can decompose it as $V\Lambda V^{-1}$\ where
$V$\ is unitary and $\Lambda$ is diagonal. \ So by Taylor series expansion,%
\[
e^{iH}=\sum_{j\geq0}\frac{i^{j}}{j!}V\Lambda^{j}V^{-1}\text{.}%
\]
Now let $S$\ be the set of basis states $\left\vert v_{b},z_{b}\right\rangle
$\ such that\ the distance from $v_{0}$\ to $v_{b}$\ is $\ell$, for some
$\ell>4\pi$. \ Notice that for all $j<\ell$ and $\left\vert v_{b}%
,z_{b}\right\rangle \in S$, we have%
\[
\left\langle v_{b},z_{b}\right\vert H^{j}\left\vert v_{0},\psi\right\rangle
=\left\langle v_{b},z_{b}\right\vert V\Lambda^{j}V^{-1}\left\vert v_{0}%
,\psi\right\rangle =0
\]
by the locality of $H$.\ \ Therefore%
\begin{align*}
\sum_{\left\vert v_{b},z_{b}\right\rangle \in S}\left\vert \left\langle
v_{b},z_{b}\right\vert e^{iH}\left\vert v_{0},\psi\right\rangle \right\vert
^{2}  &  =\sum_{\left\vert v_{b},z_{b}\right\rangle \in S}\left\vert
\sum_{j\geq\ell}\frac{i^{j}}{j!}\left\langle v_{b},z_{b}\right\vert
V\Lambda^{j}V^{-1}\left\vert v_{0},\psi\right\rangle \right\vert ^{2}\\
&  \leq\left(  \sum_{j\geq\ell}\sqrt{\sum_{\left\vert v_{b},z_{b}\right\rangle
\in S}\left\vert \frac{i^{j}}{j!}\left\langle v_{b},z_{b}\right\vert
V\Lambda^{j}V^{-1}\left\vert v_{0},\psi\right\rangle \right\vert ^{2}}\right)
^{2}\\
&  \leq\left(  \sum_{j\geq\ell}\sqrt{\frac{\pi^{j}}{j!}}\right)  ^{2}\\
&  \leq\frac{4\pi^{\ell}}{\ell!}.
\end{align*}
Here the second line uses the triangle inequality, the third line uses the
fact that $V\Lambda^{j}V^{-1}$\ has maximum eigenvalue at most $\pi^{j}$\ (and
therefore $\left(  i^{j}/j!\right)  V\Lambda^{j}V^{-1}$\ has maximum
eigenvalue at most $\pi^{j}/j!$), and the fourth line uses the fact that
$\ell>4\pi$. \ Intuitively, the probability that $H$ sends the robot\ a
distance $\ell$\ from $v_{0}$\ is at most $4\pi^{\ell}/\ell!$, which decreases
exponentially in $\ell$. \ One can now use a Chernoff-Hoeffding bound to
upper-bound the probability that $c\delta$\ local Hamiltonians, applied in
succession, ever move the robot a distance $\delta/2$\ from $v_{0}$. \ It is
clear that the resulting upper bound is $2^{-\Omega\left(  \delta\right)  }%
$\ for small enough $c$. \ Therefore%
\[
D\left(  q-1,q\right)  \leq4\Gamma_{j^{\ast}}^{\left(  T-qc\delta\right)
}+2^{-\Omega\left(  \delta\right)  }%
\]
and the remainder of the proof goes through as before.
\end{proof}

\section{Search on Grids\label{GRID}}

Let $\mathcal{L}_{d}\left(  n\right)  $\ be a $d$-dimensional grid
graph of size $n^{1/d}\times\cdots\times n^{1/d}$. \ That is, each
vertex is specified by $d$ coordinates\
$i_{1},\ldots,i_{d}\in\left\{  1,\ldots,n^{1/d}\right\} $, and is
connected to the at most $2d$ vertices obtainable by adding or
subtracting $1$ from a single coordinate (boundary vertices have
fewer than $2d$\ neighbors). \ We write simply $\mathcal{L}_{d}$\
when $n$ is clear from context. \ In this section we present our
main positive results: that $Q\left(
\operatorname*{OR},\mathcal{L}_{d}\right)  =\Theta\left(  \sqrt
{n}\right)  $ for $d\geq3$, and $Q\left(  \operatorname*{OR},\mathcal{L}%
_{2}\right)  =O\left(  \sqrt{n}\operatorname*{polylog}n\right)  $\
for $d=2$.

Before proving these claims, let us develop some intuition by
showing weaker bounds, taking the case $d=2$ for illustration. \
Clearly $Q\left( \operatorname*{OR},\mathcal{L}_{2}\right)  =O\left(
n^{3/4}\right)  $: we
simply partition $\mathcal{L}_{2}\left(  n\right)  $\ into $\sqrt{n}%
$\ subsquares, each a copy of $\mathcal{L}_{2}\left(
\sqrt{n}\right)  $. \ In $5\sqrt{n}$\ steps, the robot can travel
from the start vertex to any subsquare $C$, search $C$\ classically
for a marked vertex, and then return to the start vertex. \ Thus, by
searching all $\sqrt{n}$\ of the $C$'s in superposition and applying
Grover's algorithm, the robot can search the grid in time $O\left(
n^{1/4}\right)  \times5\sqrt{n}=O\left(  n^{3/4}\right)  $.

Once we know that, we might as well partition $\mathcal{L}_{2}\left(
n\right)  $\ into $n^{1/3}$\ subsquares, each a copy of $\mathcal{L}%
_{2}\left(  n^{2/3}\right)  $. \ Searching any one of these
subsquares by the previous algorithm takes time $O\left(  \left(
n^{2/3}\right)  ^{3/4}\right) =O\left(  \sqrt{n}\right)  $, an
amount of time that also suffices to travel to the subsquare and
back from the start vertex. \ So using Grover's algorithm, the robot
can search $\mathcal{L}_{2}\left(  n\right)  $\ in time $O\left(
\sqrt{n^{1/3}}\cdot\sqrt{n}\right)  =O\left(  n^{2/3}\right)  $. \
We can continue recursively in this manner to make the running time
approach $O\left(  \sqrt{n}\right)  $. \ The trouble is that, with
each additional layer of recursion, the robot needs to repeat the
search more often to upper-bound the error probability. \ Using this
approach, the best bounds we could obtain are roughly $O\left(
\sqrt{n}\operatorname*{polylog}n\right) $\ for $d\geq3$, or
$\sqrt{n}2^{O\left(  \sqrt{\log n}\right)  }$\ for $d=2$. \ In what
follows, we use the amplitude amplification approach of Brassard et
al.\ \cite{bhmt}\ to improve these bounds, in the case of a single
marked vertex, to $O\left(  \sqrt{n}\right)  $\ for $d\geq3$
(Section \ref{D3}) and $O\left(  \sqrt{n}\log^{3/2}n\right)  $\ for
$d=2$\ (Section \ref{D2}). \ Section \ref{MULTIPLE} generalizes
these results to the case of multiple marked vertices.

Intuitively, the reason the case $d=2$\ is special is that there,
the diameter of the grid is $\Theta\left(  \sqrt{n}\right)  $,\
which matches exactly the time needed for Grover search. \ For
$d\geq3$, by contrast, the robot can travel across the grid in much
less time than is needed to search it.

\subsection{Amplitude Amplification\label{AASEC}}

We start by describing amplitude amplification \cite{bhmt}, a
generalization of Grover search. \ Let $A$ be a quantum algorithm
that, with probability $\epsilon$, outputs a correct answer together
with a witness that proves the answer correct. (For example, in the
case of search, the algorithm outputs a vertex label $i$ such that
$x_{i}=1$.) \ Amplification generates a new algorithm that calls $A$
order $1/\sqrt{\epsilon}$ times, and that produces both a correct
answer and a witness with probability $\Omega\left(  1\right) $. \
In particular, assume $A$ starts in basis state $\left\vert
s\right\rangle $, and let $m$\ be a positive integer. \ Then the
amplification procedure works as follows:

\begin{enumerate}
\item[(1)] Set $\left\vert \psi_{0}\right\rangle =A\left\vert s\right\rangle $.

\item[(2)] For $i=1$ to $m$ set $\left\vert \psi_{i+1}\right\rangle
=ASA^{-1}W\left\vert \psi_{i}\right\rangle $, where

\begin{itemize}
\item $W$ flips the phase of basis state $\left\vert y\right\rangle $\ if and
only if $\left\vert y\right\rangle $\ contains a description of a
correct witness, and

\item $S$ flips the phase of basis state $\left\vert y\right\rangle $\ if and
only if $\left\vert y\right\rangle =\left\vert s\right\rangle $.
\end{itemize}
\end{enumerate}

We can decompose $\left\vert \psi_{0}\right\rangle $ as
$\sin\alpha\left\vert \Psi_{\operatorname*{succ}}\right\rangle
+\cos\alpha\left\vert \Psi _{\operatorname*{fail}}\right\rangle $,
where $\left\vert \Psi _{\operatorname*{succ}}\right\rangle $ is a
superposition over basis states
containing a correct witness and $\left\vert \Psi_{\operatorname*{fail}%
}\right\rangle $ is a superposition over all other basis states. \
Brassard et al. \cite{bhmt}\ showed the following:

\begin{lemma}
[\cite{bhmt}]\label{bhmtlem}$|\psi_{i}\rangle=\sin\left[  \left(
2i+1\right) \alpha\right]  \left\vert
\Psi_{\operatorname*{succ}}\right\rangle +\cos\left[  \left(
2i+1\right)  \alpha\right]  \left\vert \Psi
_{\operatorname*{fail}}\right\rangle $.
\end{lemma}

If measuring $\left\vert \psi_{0}\right\rangle $ gives a correct
witness with probability $\epsilon$, then $\left\vert
\sin\alpha\right\vert ^{2}=\epsilon$ and $\left\vert
\alpha\right\vert \geq1/\sqrt{\epsilon}$. \ So taking
$m=O(1/\sqrt{\epsilon})$ yields $\sin\left[  \left(  2m+1\right)
\alpha\right]  \approx1$. \ For our algorithms, though, the
multiplicative constant under the big-O also matters. \ To
upper-bound this constant, we prove the following lemma.

\begin{lemma}
\label{Ampl}Suppose a quantum algorithm $A$ outputs a correct answer
and witness with probability exactly $\epsilon$. \ Then by using
$2m+1$ calls to
$A$\ or $A^{-1}$, where%
\[
m\leq\frac{\pi}{4\arcsin\sqrt{\epsilon}}-\frac{1}{2},
\]
we can output a correct answer and witness with probability at least%
\[
\left(  1-\frac{\left(  2m+1\right)  ^{2}}{3}\epsilon\right)  \left(
2m+1\right)  ^{2}\epsilon.
\]

\end{lemma}

\begin{proof}
We perform $m$ steps of amplitude amplification, which requires
$2m+1$\ calls
$A$ or $A^{-1}$. \ By Lemma \ref{bhmtlem}, this yields the final state%
\[
\sin\left[  \left(  2m+1\right)  \alpha\right]  \left\vert \Psi
_{\operatorname*{succ}}\right\rangle +\cos\left[  \left(
2m+1\right) \alpha\right]  \left\vert
\Psi_{\operatorname*{fail}}\right\rangle .
\]
where $\alpha=\arcsin\sqrt{\epsilon}$. \ Therefore the success
probability is
\begin{align*}
\sin^{2}\left[  \left(  2m+1\right)  \arcsin\sqrt{\epsilon}\right]
&
\geq\sin^{2}\left[  \left(  2m+1\right)  \sqrt{\epsilon}\right] \\
&  \geq\left(  \left(  2m+1\right)  \sqrt{\epsilon}-\frac{\left(
2m+1\right)
^{3}}{6}\epsilon^{3/2}\right)  ^{2}\\
&  \geq\left(  2m+1\right)  ^{2}\epsilon-\frac{\left(  2m+1\right)  ^{4}}%
{3}\epsilon^{2}.
\end{align*}
Here the first line uses the monotonicity of $\sin^{2}x$\ in the
interval $\left[  0,\pi/2\right]  $, and the second line uses the
fact that $\sin x\geq x-x^{3}/6$ for all $x\geq0$ by Taylor series
expansion.
\end{proof}

Note that there is no need to uncompute any garbage left by $A$,
beyond the
uncomputation that happens \textquotedblleft automatically\textquotedblright%
\ within the amplification procedure.

\subsection{Dimension At Least 3\label{D3}}

Our goal is the following:

\begin{theorem}
\label{sqrtsrchfull}If $d\geq3$, then $Q\left(  \operatorname*{OR}%
,\mathcal{L}_{d}\right)  =\Theta\left(  \sqrt{n}\right)  $.
\end{theorem}

In this section, we prove Theorem \ref{sqrtsrchfull}\ for the
special case of a unique marked vertex; then, in Sections
\ref{MULTIPLE} and \ref{UNKNOWN}, we
will generalize to multiple marked vertices. \ Let $\operatorname*{OR}%
^{\left(  k\right)  }$\ be the problem of deciding whether there are
no marked vertices or exactly $k$ of them, given that one of these
is true. \ Then:

\begin{theorem}
\label{sqrtsrch}If $d\geq3$, then $Q\left(
\operatorname*{OR}^{\left( 1\right)  },\mathcal{L}_{d}\right)
=\Theta\left(  \sqrt{n}\right)  $.
\end{theorem}

Choose constants $\beta\in\left(  2/3,1\right)  $\ and $\mu\in\left(
1/3,1/2\right)  $\ such that $\beta\mu>1/3$ (for example,
$\beta=4/5$\ and $\mu=5/11$ will work). \ Let $\ell_{0}$\ be a large
positive integer; then for all positive integers $R$, let
$\ell_{R}=\ell_{R-1}\left\lceil \ell _{R-1}^{1/\beta-1}\right\rceil
$. \ Also let $n_{R}=\ell_{R}^{d}$.
%, and let
%$N_{R}=n_{0}^{\left(  1/\beta\right)  ^{R}}$\ be the \textquotedblleft
%non-rounded\textquotedblright\ version of $n_{R}$. \
Assume for simplicity that $n=n_{R}$\ for some $R$; in other words,
that the hypercube $\mathcal{L}_{d}\left(  n_{R}\right)  $\ to be
searched has sides of length $\ell_{R}$. \ Later we will remove this
assumption.

Consider the following recursive algorithm $\mathcal{A}$. \ If
$n=n_{0}$, then search $\mathcal{L}_{d}\left(  n_{0}\right)  $
classically, returning $1$ if a
marked vertex is found and $0$ otherwise. \ Otherwise partition $\mathcal{L}%
_{d}\left(  n_{R}\right)  $\ into $n_{R}/n_{R-1}$\ subcubes, each
one a copy of $\mathcal{L}_{d}\left(  n_{R-1}\right)  $. \ Take the
algorithm that consists of picking a subcube $C$ uniformly at
random, and then running $\mathcal{A}$ recursively on $C$. \ Amplify
this algorithm $\left( n_{R}/n_{R-1}\right)  ^{\mu}$ times.

The intuition behind the exponents is that $n_{R-1}\approx
n_{R}^{\beta}$, so searching $\mathcal{L}_{d}\left(  n_{R-1}\right)
$\ should take about $n_{R}^{\beta/2}$\ steps, which dominates the
$n_{R}^{1/d}$ steps needed to travel across the hypercube when
$d\geq3$. \ Also, at level $R$ we want to amplify a number of times
that is less than $\left(  n_{R}/n_{R-1}\right) ^{1/2}$ by some
polynomial amount, since full amplification would be inefficient. \
The reason for the constraint $\beta\mu>1/3$\ will appear in the
analysis.

We now provide a more explicit description of $\mathcal{A}$, which
shows that $\mathcal{A}$ can be implemented using C-local unitaries
and only a single bit of workspace. \ At any time, the quantum
robot's state will have the form $\sum_{i,z}\alpha_{i,z}\left\vert
v_{i},z\right\rangle $, where $v_{i}$\ is a vertex of
$\mathcal{L}_{d}\left(  n_{R}\right)  $ and $z$ is a single bit that
records whether or not a marked vertex has been found. \ Given a
subcube $C$,
let $v\left(  C\right)  $ be the \textquotedblleft corner\textquotedblright%
\ vertex of $C$; that is, the vertex that is minimal in all $d$
coordinates. \ Then the initial state when searching $C$\ will be
$\left\vert v\left( C\right)  ,0\right\rangle $. \ Beware, however,
that \textquotedblleft initial state\textquotedblright\ in this
context\ just means the state $\left\vert s\right\rangle $\ from
Section \ref{AASEC}. \ Because of the way amplitude amplification
works, $\mathcal{A}$\ will often be invoked on $C$ with other
initial states, and even run in reverse.

Below we give pseudocode for $\mathcal{A}$. \ Our procedure calls
the three unitaries $A$, $W$, and $S$\ from Section \ref{AASEC} as
subroutines. \ For convenience, we write
$\mathcal{A}_{R},A_{R},W_{R},S_{R}$\ to denote the level of
recursion that is currently active.

\begin{algorithm}
[$\mathcal{A}_{R}$]\label{algw}Searches a subcube $C$ of size
$n_{R}$ for the marked vertex, and amplifies the result to have
larger probability. \ Default initial state: $\left\vert v\left(
C\right)  ,0\right\rangle $.

\textbf{If }$R=0$\textbf{\ then:}

\begin{enumerate}
\item[(1)] Use classical C-local operations to visit all $n_{0}$\ vertices of
$C$ in any order. \ At each $v_{i}\in C$, use a query transformation
to map the state $\left\vert v_{i},z\right\rangle $\ to $\left\vert
v_{i},z\oplus x_{i}\right\rangle $.

\item[(2)] Return to $v\left(  C\right)  $.
\end{enumerate}

\textbf{If }$R\geq1$\textbf{\ then:}

\begin{enumerate}
\item[(1)] Let $m_{R}$ be the smallest integer such that $2m_{R}+1\geq\left(
n_{R}/n_{R-1}\right)  ^{\mu}$.

\item[(2)] Call $A_{R}$.

\item[(3)] For $i=1$ to $m_{R}$, call $W_{R}$, then $A_{R}^{-1}$, then $S_{R}%
$, then $A_{R}$. \
\end{enumerate}
\end{algorithm}

Suppose $\mathcal{A}_{R}$\ is run on the initial state $\left\vert
v\left( C\right)  ,0\right\rangle $, and let
$C_{1},\ldots,C_{n_{R}/n_{0}}$\ be the \textit{minimal subcubes} in
$C$---meaning those of size $n_{0}$. \ Then the
final state after $\mathcal{A}_{R}$\ terminates should be%
\[
\frac{1}{\sqrt{n_{R}/n_{0}}}\sum_{i=1}^{n_{R}/n_{0}}\left\vert
v\left( C_{i}\right)  ,0\right\rangle
\]
if $C$ does not contain the marked vertex. \ Otherwise the final
state should have non-negligible overlap with $\left\vert v\left(
C_{i^{\ast}}\right) ,1\right\rangle $, where $C_{i^{\ast}}$\ is the
minimal subcube in $C$ that contains the marked vertex. \ In
particular, if $R=0$, then the final state should be $\left\vert
v\left(  C\right)  ,1\right\rangle $\ if $C$ contains the marked
vertex, and $\left\vert v\left(  C\right)  ,0\right\rangle $\
otherwise.

The two phase-flip subroutines, $W_{R}$\ and $S_{R}$, are both
trivial to
implement. \ To apply $W_{R}$, map each basis state $\left\vert v_{i}%
,z\right\rangle $\ to $\left(  -1\right)  ^{z}\left\vert
v_{i},z\right\rangle $. \ To apply $S_{R}$, map each basis state
$\left\vert v_{i},z\right\rangle $\ to $-\left\vert
v_{i},z\right\rangle $\ if $v_{i}=v\left(  C\right)  $\ for some
subcube $C$ of size $n_{R}$,\ and to $\left\vert
v_{i},z\right\rangle $\ otherwise. \ Below we give pseudocode for
$A_{R}$.

\begin{algorithm}
[$A_{R}$]\label{alga}Searches a subcube $C$ of size $n_{R}$\ for the
marked vertex. \ Default initial state: $\left\vert v\left(
C\right) ,0\right\rangle $.

\begin{enumerate}
\item[(1)] Partition $C$ into $n_{R}/n_{R-1}$\ smaller subcubes $C_{1}%
,\ldots,C_{n_{R}/n_{R-1}}$, each of size $n_{R-1}$.

\item[(2)] For all $j\in\left\{  1,\ldots,d\right\}  $, let $V_{j}$\ be the
set of corner vertices $v\left(  C_{i}\right)  $\ that differ from
$v\left( C\right)  $\ only in the first $j$ coordinates. \ Thus
$V_{0}=\left\{ v\left(  C\right)  \right\}  $, and in general
$\left\vert V_{j}\right\vert =\ell_{R}^{j}$. \ For $j=1$\ to $d$,
let $\left\vert V_{j}\right\rangle $ be
the state%
\[
\left\vert V_{j}\right\rangle =\frac{1}{\ell_{R}^{j/2}}\sum_{v\left(
C_{i}\right)  \in V_{j}}\left\vert v\left(  C_{i}\right)
,0\right\rangle
\]
Apply a sequence of transformations $U_{1}$, $U_{2}$, $\ldots$,
$U_{d}$ where $U_{j}$ is a unitary that maps $\left\vert
V_{j-1}\right\rangle $\ to $\left\vert V_{j}\right\rangle $\ by
applying C-local\ unitaries that move amplitude only along the
$j^{th}$\ coordinate.

\item[(3)] Call $\mathcal{A}_{R-1}$\ recursively, to search $C_{1}%
,\ldots,C_{n_{R}/n_{R-1}}$\ in superposition and amplify the
results.
\end{enumerate}
\end{algorithm}

If $A_{R}$\ is run on the initial state $\left\vert v\left(
C\right)
,0\right\rangle $, then the final state should be%
\[
\frac{1}{\sqrt{n_{R}/n_{R-1}}}\sum_{i=1}^{n_{R}/n_{0}}\left\vert
\phi _{i}\right\rangle ,
\]
where $\left\vert \phi_{i}\right\rangle $\ is the correct final
state when $\mathcal{A}_{R-1}$\ is run on subcube $C_{i}$ with
initial state $\left\vert v\left(  C_{i}\right)  ,0\right\rangle $.
\ A key point is that there is no need for $A_{R}$\ to call
$\mathcal{A}_{R-1}$\ twice, once to compute and once to
uncompute---for the uncomputation is already built in to
$\mathcal{A}$. \ This is what will enable us to prove an upper bound
of $O\left(  \sqrt {n}\right)  $\ instead of $O\left(
\sqrt{n}2^{R}\right)  =O\left(  \sqrt
{n}\operatorname*{polylog}n\right)  $.

We now analyze the running time of $\mathcal{A}$.

\begin{lemma}
\label{d3lemma}$\mathcal{A}_{R}$ uses $O\left(  n_{R}^{\mu}\right)
$\ steps.
\end{lemma}

\begin{proof}
Let $T_{\mathcal{A}}\left(  R\right)  $\ and $T_{A}\left(  R\right)
$\ be the total numbers of steps used by $\mathcal{A}_{R}$ and
$A_{R}$ respectively in searching $\mathcal{L}_{d}\left(
n_{R}\right)  $. \ Then we have
$T_{\mathcal{A}}\left(  0\right)  =O\left(  1\right)  $, and%
\begin{align*}
T_{\mathcal{A}}\left(  R\right)   &  \leq\left(  2m_{R}+1\right)
T_{A}\left(
R\right)  +2m_{R},\\
T_{A}\left(  R\right)   &  \leq dn_{R}^{1/d}+T_{\mathcal{A}}\left(
R-1\right)
\end{align*}
for all $R\geq1$. \ For $W_{R}$\ and $S_{R}$\ can both be
implemented in a single step, while $A_{R}$\ uses
$d\ell_{R}=dn_{R}^{1/d}$\ steps to move the
robot across the hypercube. \ Combining,%
\begin{align*}
T_{\mathcal{A}}\left(  R\right)   &  \leq\left(  2m_{R}+1\right)
\left(
dn_{R}^{1/d}+T_{\mathcal{A}}\left(  R-1\right)  \right)  +2m_{R}\\
&  \leq\left(  \left(  n_{R}/n_{R-1}\right)  ^{\mu}+2\right)  \left(
dn_{R}^{1/d}+T_{\mathcal{A}}\left(  R-1\right)  \right)  +\left(
n_{R}/n_{R-1}\right)  ^{\mu}+1\\
&  =O\left(  \left(  n_{R}/n_{R-1}\right)  ^{\mu}n_{R}^{1/d}\right)
+\left( \left(  n_{R}/n_{R-1}\right)  ^{\mu}+2\right)
T_{\mathcal{A}}\left(
R-1\right) \\
&  =O\left(  \left(  n_{R}/n_{R-1}\right)  ^{\mu}n_{R}^{1/d}\right)
+\left(
n_{R}/n_{R-1}\right)  ^{\mu}T_{\mathcal{A}}\left(  R-1\right) \\
&  =O\left(  \left(  n_{R}/n_{R-1}\right)  ^{\mu}n_{R}^{1/d}+\left(
n_{R}/n_{R-2}\right)  ^{\mu}n_{R-1}^{1/d}+\cdots+\left(
n_{R}/n_{0}\right)
^{\mu}n_{1}^{1/d}\right) \\
&  =n_{R}^{\mu}\cdot O\left(
\frac{n_{R}^{1/d}}{n_{R-1}^{\mu}}+\frac
{n_{R-1}^{1/d}}{n_{R-2}^{\mu}}+\cdots+\frac{n_{1}^{1/d}}{n_{0}^{\mu}}\right)
\\
&  =n_{R}^{\mu}\cdot O\left(
n_{R}^{1/d-\beta\mu}+\cdots+n_{2}^{1/d-\beta\mu
}+n_{1}^{1/d-\beta\mu}\right) \\
&  =n_{R}^{\mu}\cdot O\left(  n_{R}^{1/d-\beta\mu}+\left(
n_{R}^{1/d-\beta \mu}\right)  ^{1/\beta}+\cdots+\left(
n_{R}^{1/d-\beta\mu}\right)
^{1/\beta^{R-1}}\right) \\
&  =O\left(  n_{R}^{\mu}\right)  .
\end{align*}
Here the second line follows because $2m_{R}+1\leq\left(  n_{R}/n_{R-1}%
\right)  ^{\mu}+2$, the fourth because the $\left(
n_{R}/n_{R-1}\right)  ^{\mu}$\ terms increase doubly exponentially,
so adding $2$ to each will not affect the asymptotics; the seventh
because $n^{\mu}_i =
\Omega\left(\left(n^{\mu}_{i+1}\right)^{\beta}\right)$, the eighth
because $n_{R-1}\leq n_{R}^{\beta}$; and the last because
$\beta\mu>1/3\geq 1/d$, hence $n_{1}^{1/d-\beta\mu}<1$.
\end{proof}

Next we need to lower-bound the success probability. \ Say that $\mathcal{A}%
$\ or $A$\ \textquotedblleft succeeds\textquotedblright\ if a
measurement in the standard basis yields the result $\left\vert
v\left(  C_{i^{\ast}}\right) ,1\right\rangle $, where
$C_{i^{\ast}}$\ is the minimal subcube that contains the marked
vertex. \ Of course, the marked vertex itself can then be found in
$n_{0}=O\left(  1\right)  $\ steps.

\begin{lemma}
\label{d3lemma2}Assuming there is a unique marked vertex,
$\mathcal{A}_{R}$ succeeds with probability $\Omega\left(
1/n_{R}^{1-2\mu}\right)  $.
\end{lemma}

\begin{proof}
Let $P_{\mathcal{A}}\left(  R\right)  $\ and $P_{A}\left(  R\right)
$\ be the success probabilities of $\mathcal{A}_{R}$ and $A_{R}$
respectively when searching $\mathcal{L}_{d}\left(  n_{R}\right)  $.
\ Then clearly $P_{\mathcal{A}}\left(  0\right)  =1$, and
$P_{A}\left(  R\right)  =\left( n_{R-1}/n_{R}\right)
P_{\mathcal{A}}\left(  R-1\right)  $\ for all $R\geq1$.
\ So by Lemma \ref{Ampl},%
\begin{align*}
P_{\mathcal{A}}\left(  R\right)   &  \geq\left(  1-\frac{1}{3}\left(
2m_{R}+1\right)  ^{2}P_{A}\left(  R\right)  \right)  \left(
2m_{R}+1\right)
^{2}P_{A}\left(  R\right) \\
&  =\left(  1-\frac{1}{3}\left(  2m_{R}+1\right)  ^{2}\frac{n_{R-1}}{n_{R}%
}P_{\mathcal{A}}\left(  R-1\right)  \right)  \left(  2m_{R}+1\right)
^{2}\frac{n_{R-1}}{n_{R}}P_{\mathcal{A}}\left(  R-1\right) \\
&  \geq\left(  1-\frac{1}{3}\left(  n_{R}/n_{R-1}\right)
^{2\mu}\frac {n_{R-1}}{n_{R}}P_{\mathcal{A}}\left(  R-1\right)
\right)  \left( n_{R}/n_{R-1}\right)
^{2\mu}\frac{n_{R-1}}{n_{R}}P_{\mathcal{A}}\left(
R-1\right) \\
&  \geq\left(  1-\frac{1}{3}\left(  n_{R-1}/n_{R}\right)
^{1-2\mu}\right)
\left(  n_{R-1}/n_{R}\right)  ^{1-2\mu}P_{\mathcal{A}}\left(  R-1\right) \\
&  \geq\left(  n_{0}/n_{R}\right)  ^{1-2\mu}%
%TCIMACRO{\dprod \limits_{r=1}^{R}}%
%BeginExpansion
{\displaystyle\prod\limits_{r=1}^{R}}
%EndExpansion
\left(  1-\frac{1}{3}\left(  n_{R-1}/n_{R}\right)  ^{1-2\mu}\right) \\
&  \geq \left(  n_{0}/n_{R}\right)  ^{1-2\mu}%
%TCIMACRO{\dprod \limits_{r=1}^{R}}%
%BeginExpansion
{\displaystyle\prod\limits_{r=1}^{R}}
%EndExpansion
\left(  1-\frac{1}{3 n_{R}^{\left(  1-\beta\right)  \left(  1-2\mu\right)  }%
}\right) \\
&  \geq\left(  n_{0}/n_{R}\right)  ^{1-2\mu}\left(
1-\sum_{r=1}^{R}\frac
{1}{3N_{R}^{\left(  1-\beta\right)  \left(  1-2\mu\right)  }}\right) \\
&  =\Omega\left(  1/n_{R}^{1-2\mu}\right)  .
\end{align*}
Here the third line follows because $2m_{R}+1\geq n_{R-1}/n_{R}$ and
the function $x-\frac{1}{3}x^{2}$\ is nondecreasing in the interval
$\left[  0,1\right]  $;\ the fourth because
$P_{\mathcal{A}}\left(  R-1\right)  \leq1$; the sixth because $n_{R-1}%
\leq n_{R}^{\beta}$; and the last because $\beta<1$\ and $\mu<1/2$,
the $n_{R}$'s increase doubly exponentially, and $n_{0}$\ is
sufficiently large.
\end{proof}

Finally, take $\mathcal{A}_{R}$ itself and amplify it to success
probability $\Omega\left(  1\right)  $ by running it
$O(n_{R}^{1/2-\mu})$ times. \ This yields an algorithm for searching
$\mathcal{L}_{d}\left(  n_{R}\right) $\ with overall running time
$O\left(n_{R}^{1/2}\right)  $, which implies that $Q\left(  \operatorname*{OR}%
^{\left(  1\right)  },\mathcal{L}_{d}\left(  n_{R}\right)  \right)
=O\left( n_{R}^{1/2}\right)  $.

All that remains is to handle values of $n$ that do not equal
$n_{R}$\ for any
$R$. \ The solution is simple: first find the largest $R$ such that\ $n_{R}%
<n$. \ Then set $n^{\prime}=n_{R}\left\lceil
n^{1/d}/\ell_{R}\right\rceil ^{d}$, and embed $\mathcal{L}_{d}\left(
n\right)  $\ into the larger hypercube $\mathcal{L}_{d}\left(
n^{\prime}\right)  $. \ Clearly $Q\left( \operatorname*{OR}^{\left(
1\right)  },\mathcal{L}_{d}\left(  n\right)
\right)  \leq Q\left(  \operatorname*{OR}^{\left(  1\right)  },\mathcal{L}%
_{d}\left(  n^{\prime}\right)  \right)  $. \ Also notice that
$n^{\prime }=O\left(  n\right)  $\ and that $n^{\prime}=O\left(
n_{R}^{1/\beta}\right) =O\left(  n_{R}^{3/2}\right)  $. \ Next
partition $\mathcal{L}_{d}\left( n^{\prime}\right)  $\ into
$n^{\prime}/n_{R}$\ subcubes, each a copy of $\mathcal{L}_{d}\left(
n_{R}\right)  $. \ The algorithm will now have one
additional level of recursion, which chooses a subcube of $\mathcal{L}%
_{d}\left(  n^{\prime}\right)  $\ uniformly at random, runs $\mathcal{A}_{R}%
$\ on that subcube, and then amplifies the resulting procedure
$\Theta\left(
\sqrt{n^{\prime}/n_{R}}\right)  $\ times. \ The total time is now%
\[
O\left(  \sqrt{\frac{n^{\prime}}{n_{R}}}\left(  \left(
n^{\prime}\right)
^{1/d}+n_{R}^{1/2}\right)  \right)  =O\left(  \sqrt{\frac{n^{\prime}}{n_{R}}%
}n_{R}^{1/2}\right)  =O\left(  \sqrt{n}\right)  ,
\]
while the success probability is $\Omega\left(  1\right)  $.\ \ This
completes Theorem \ref{sqrtsrch}.

\subsection{Dimension 2\label{D2}}

In the $d=2$\ case, the best we can achieve is the following:

\begin{theorem}
\label{sqrtsrch2full}$Q\left(
\operatorname*{OR},\mathcal{L}_{2}\right) =O\left(
\sqrt{n}\log^{5/2}n\right)  $.
\end{theorem}

Again, we start with the single marked vertex case and postpone the
general case to Sections \ref{MULTIPLE} and \ref{UNKNOWN}.

\begin{theorem}
\label{sqrtsrch2}$Q\left(  \operatorname*{OR}^{\left(  1\right)  }%
,\mathcal{L}_{2}\right)  =O\left(  \sqrt{n}\log^{3/2}n\right)  $.
\end{theorem}

For $d\geq3$, we performed amplification on large (greater than
$O\left( 1/n^{1-2\mu}\right)  $) probabilities only once, at the
end. \ For $d=2$, on the other hand, any algorithm that we construct
with any nonzero success probability will have running time
$\Omega\left(  \sqrt{n}\right)  $, simply because that is the
diameter of the grid. \ If we want to keep the running time
$O\left(\sqrt{n}\right)$, then we can only perform $O\left( 1\right)
$ amplification steps at the end. \ Therefore we need to keep the
success probability relatively high throughout the recursion,\
meaning that we suffer an increase in the running time, since
amplification to high probabilities is less efficient.

The procedures $\mathcal{A}_{R}$,\ $A_{R}$, $W_{R}$, and $S_{R}$ are
identical to those in Section \ref{D3}; all that changes are the
parameter settings. \ For all integers $R\geq0$, we now let
$n_{R}=\ell_{0}^{2R}$, for some odd integer $\ell_{0}\geq3$\ to be
set later. \ Thus, $\mathcal{A}_{R}$\ and $A_{R}$\ search the square
grid $\mathcal{L}_{2}\left(  n_{R}\right)  $ of size
$\ell_{0}^{R}\times\ell_{0}^{R}$.\ \ Also, let $m=\left(  \ell
_{0}-1\right)  /2$; then $\mathcal{A}_{R}$\ applies $m$ steps of
amplitude amplification to $A_{R}$.

We now prove the counterparts of Lemmas \ref{d3lemma}\ and
\ref{d3lemma2} for the two-dimensional case.

\begin{lemma}
\label{d2lemma}$\mathcal{A}_{R}$ uses $O\left(
R\ell_{0}^{R+1}\right)  $\ steps.
\end{lemma}

\begin{proof}
Let $T_{\mathcal{A}}\left(  R\right)  $\ and $T_{A}\left(  R\right)
$\ be the time used by $\mathcal{A}_{R}$ and $A_{R}$ respectively in
searching $\mathcal{L}_{2}\left(  n_{R}\right)  $. \ Then
$T_{\mathcal{A}}\left(
0\right)  =1$, and for all $R\geq1$,%
\begin{align*}
T_{\mathcal{A}}\left(  R\right)   &  \leq\left(  2m+1\right)
T_{A}\left(
R\right)  +2m,\\
T_{A}\left(  R\right)   &  \leq2n_{R}^{1/2}+T_{\mathcal{A}}\left(
R-1\right) .
\end{align*}
Combining,%
\begin{align*}
T_{\mathcal{A}}\left(  R\right)   &  \leq\left(  2m+1\right)  \left(
2n_{R}^{1/2}+T_{\mathcal{A}}\left(  R-1\right)  \right)  +2m\\
&  =\ell_{0}\left(  2\ell_{0}^{R}+T_{\mathcal{A}}\left(  R-1\right)
\right)
+\ell_{0}-1\\
&  =O\left(  \ell_{0}^{R+1}+\ell_{0}T_{\mathcal{A}}\left(
R-1\right)  \right)
\\
&  =O\left(  R\ell_{0}^{R+1}{}\right)  .
\end{align*}

\end{proof}

\begin{lemma}
\label{d2lemma2}$\mathcal{A}_{R}$ succeeds with probability
$\Omega\left( 1/R\right)  $.
\end{lemma}

\begin{proof}
Let $P_{\mathcal{A}}\left(  R\right)  $\ and $P_{A}\left(  R\right)
$\ be the success probabilities of $\mathcal{A}_{R}$ and $A_{R}$
respectively when searching $\mathcal{L}_{2}\left(  n_{R}\right)  $.
\ Then $P_{A}\left( R\right)  =P_{\mathcal{A}}\left(  R-1\right)
/\ell_{0}^{2}$\ for all $R\geq
1$. \ So by Lemma \ref{Ampl}, and using the fact that $2m+1=\ell_{0}$,%
\begin{align*}
P_{\mathcal{A}}\left(  R\right)   &  \geq\left(  1-\frac{\left(
2m+1\right) ^{2}}{3}P_{A}\left(  R\right)  \right)  \left(
2m+1\right)  ^{2}P_{A}\left(
R\right) \\
&  =\left(  1-\frac{\ell_{0}^{2}}{3}\frac{P_{\mathcal{A}}\left(
R-1\right) }{\ell_{0}^{2}}\right)
\ell_{0}^{2}\frac{P_{\mathcal{A}}\left(  R-1\right)
}{\ell_{0}^{2}}\\
&  =P_{\mathcal{A}}\left(  R-1\right)
-\frac{1}{3}P_{\mathcal{A}}^{2}\left(
R-1\right) \\
&  =\Omega\left(  1/R\right)  .
\end{align*}
This is because $\Omega\left(  R\right)  $\ iterations of the map
$x_{R}:=x_{R-1}-\frac{1}{3}x_{R-1}^{2}$\ are needed to drop from
(say) $2/R$\ to $1/R$, and $x_{0}=P_{\mathcal{A}}\left(  0\right)
=1$ is greater than $2/R$.
\end{proof}

We can amplify $\mathcal{A}_{R}$\ to success probability
$\Omega\left( 1\right)  $\ by repeating it $O\left(  \sqrt{R}\right)
$\ times. \ This yields an algorithm for searching
$\mathcal{L}_{2}\left(  n_{R}\right)
$\ that uses $O\left(  R^{3/2}\ell_{0}^{R+1}\right)  =O\left(  \sqrt{n_{R}%
}R^{3/2}\ell_{0}\right)  $ steps in total. \ We can minimize this
expression subject to $\ell_{0}^{2R}=n_{R}$ by taking $\ell_{0}$\ to
be constant and $R$\ to be $\Theta\left(  \log n_{R}\right)  $,
which yields $Q\left( \operatorname*{OR}^{\left(  1\right)
},\mathcal{L}_{2}\left(  n_{R}\right) \right)  =O\left(
\sqrt{n_{R}}\log n_{R}^{3/2}\right)  $. \ If $n$\ is not of the form
$\ell_{0}^{2R}$,\ then we simply find the smallest integer $R$ such
that $n<\ell_{0}^{2R}$, and embed $\mathcal{L}_{2}\left(  n\right)
$\ in the larger grid $\mathcal{L}_{2}\left(  \ell_{0}^{2R}\right)
$. \ Since $\ell _{0}$ is a constant, this increases the running
time by at most a constant factor. \ We have now proved Theorem
\ref{sqrtsrch2}.

\subsection{Multiple Marked Items\label{MULTIPLE}}

What about the case in which there are multiple $i$'s\ with
$x_{i}=1$? \ If there are $k$ marked items (where $k$ need not be
known in advance), then Grover's algorithm can find a marked item
with high probability in $O\left( \sqrt{n/k}\right)  $\ queries, as
shown by Boyer\ et al. \cite{bbht}. \ In our setting, however, this
is too much to hope for---since even if there are many marked
vertices, they might all be in a faraway part of the hypercube. \
Then
$\Omega\left(  n^{1/d}\right)  $ steps are needed, even if $\sqrt{n/k}%
<n^{1/d}$. \ Indeed, we can show a stronger lower bound. \ Recall
that $\operatorname*{OR}^{\left(  k\right)  }$\ is the problem of
deciding whether there are no marked vertices or exactly $k$ of
them.

\begin{theorem}
\label{mmi}For all constants $d\geq2$,
\[
Q\left(  \operatorname*{OR}\nolimits^{\left(  k\right)  },\mathcal{L}%
_{d}\right)  =\Omega\left(  \frac{\sqrt{n}}{k^{1/2-1/d}}\right)  .
\]

\end{theorem}

\begin{proof}
For simplicity, we assume that both $k^{1/d}$ and $\left(
n/3^{d}k\right) ^{1/d}$ are integers. \ (In the general case, we can
just replace $k$ by $\left\lceil k^{1/d}\right\rceil ^{d}$ and $n$
by the largest number of the form $\left(  3m\left\lceil
k^{1/d}\right\rceil \right)  ^{d}$ which is less than $n$. \ This
only changes the lower bound by a lower order term.)

We use a hybrid argument almost identical to that of Theorem
\ref{lowerdg}. \ Divide $\mathcal{L}_{d}$ into $n/k$ subcubes, each
having $k$ vertices and side length $k^{1/d}$. \ Let $S$ be a
regularly-spaced set of $M=n/\left( 3^{d}k\right)  $\ of these
subcubes, so that any two subcubes in $S$ have distance at least
$2k^{1/d}$\ from one another. \ Then choose a subcube $C_{j}\in S$\
uniformly at random and mark all $k$ vertices in $C_{j}$. \ This
enables us to consider each $C_{j}\in S$\ itself as a
\textit{single} vertex (out of $M$ in total), having distance at
least $2k^{1/d}$\ to every other vertex.

More formally, given a subcube $C_{j}\in S$, let
$\widetilde{C}_{j}$\ be the set of vertices\ consisting of $C_{j}$\
and the $3^{d}-1$\ subcubes surrounding it. \ (Thus,
$\widetilde{C}_{j}$ is a subcube of side length $3k^{1/d}$.) \ Then
the query magnitude of $\widetilde{C}_{j}$\ after the
$t^{th}$\ query is%
\[
\Gamma_{j}^{\left(  t\right)
}=\sum_{v_{i}\in\widetilde{C}_{j}\,}\sum _{z\,}\left\vert
\alpha_{i,z}^{\left(  t\right)  }\left(  X_{0}\right) \right\vert
^{2},
\]
where $X_{0}$\ is the all-zero input. \ Let $T$ be the number of
queries, and let $w=T/\left(  ck^{1/d}\right)  $ for some constant
$c>0$. \ Then as in
Theorem \ref{lowerdg}, there must exist a subcube $\widetilde{C}_{j^{\ast}}%
$\ such that%
\[
\sum_{q=0}^{w-1}\Gamma_{j^{\ast}}^{\left(  T-qck^{1/d}\right)
}\leq\frac {w}{M}=\frac{3^{d}kw}{n}.
\]
Let \thinspace$Y$ be the input which is $1$ in $C_{j^{\ast}}$\ and
$0$ elsewhere; then let $X_{q}$\ be a hybrid input which is $X_{0}$\
during queries $1$ to $T-qck^{1/d}$, but $Y$ during queries
$T-qck^{1/d}+1$ to $T$.
\ Next let%
\[
D\left(  q,r\right)  =\sum_{v_{i}\in G\,}\sum_{z\,}\left\vert \alpha
_{i,z}^{\left(  T\right)  }\left(  X_{q}\right)
-\alpha_{i,z}^{\left(
T\right)  }\left(  X_{r}\right)  \right\vert ^{2}\text{.}%
\]
Then as in Theorem \ref{lowerdg}, for all $c<1$\ we have $D\left(
q-1,q\right)  \leq4\Gamma_{j^{\ast}}^{\left(  T-qck^{1/d}\right)
}$.\ For in the $ck^{1/d}$\ queries from $T-qck^{1/d}+1$\ through
$T-\left(  q-1\right) ck^{1/d}$, no amplitude originating outside
$\widetilde{C}_{j^{\ast}}$\ can travel a distance $k^{1/d}$\ and
thereby reach $C_{j^{\ast}}$. \ Therefore switching from $X_{q-1}$\
to $X_{q}$\ can only affect amplitude that is in
$\widetilde{C}_{j^{\ast}}$\ immediately after query $T-qck^{1/d}$. \
It
follows that%
\[
\sqrt{D\left(  0,w\right)  }\leq\sum_{q=1}^{w}\sqrt{D\left(
q-1,q\right)
}\leq2\sum_{q=1}^{w}\sqrt{\Gamma_{j^{\ast}}^{\left(  T-qck^{1/d}\right)  }%
}\leq2w\sqrt{\frac{3^{d}k}{n}}=\frac{2\sqrt{3^{d}}k^{1/2-1/d}T}{c\sqrt{n}}.
\]
Hence $T=\Omega\left(  \sqrt{n}/k^{1/2-1/d}\right)  $\ for constant
$d$, since assuming the algorithm is correct we need $D\left(
0,w\right)  =\Omega\left( 1\right)  $.
\end{proof}

Notice that if $k\approx n$, then the bound of Theorem \ref{mmi}
becomes
$\Omega\left(  n^{1/d}\right)  $ which is just the diameter of $\mathcal{L}%
_{d}$. \ Also, if $d=2$, then $1/2-1/d=0$ and the bound is simply
$\Omega\left(  \sqrt{n}\right)  $ independent of $k$. \ The bound of
Theorem \ref{mmi}\ can be achieved (up to a constant factor that
depends on $d$) for $d\geq3$, and nearly achieved for $d=2$. \ We
first construct an algorithm for the case when $k$ is known.

\begin{theorem}
\label{mmiub}\quad

\begin{enumerate}
\item[(i)] For $d\geq3$,%
\[
Q\left(  \operatorname*{OR}\nolimits^{\left(  k\right)  },\mathcal{L}%
_{d}\right)  =O\left(  \frac{\sqrt{n}}{k^{1/2-1/d}}\right)  .
\]

\item[(ii)] For $d=2$,%
\[
Q\left(  \operatorname*{OR}\nolimits^{\left(  k\right)  },\mathcal{L}%
_{2}\right)  =O\left(  \sqrt{n}\log^{3/2}n\right)  .
\]

\end{enumerate}
\end{theorem}

To prove Theorem \ref{mmiub}, we first divide $\mathcal{L}_{d}\left(
n\right)  $ into $n/\gamma$ subcubes, each of size
$\gamma^{1/d}\times \cdots\times\gamma^{1/d}$ (where $\gamma$ will
be fixed later). \ Then in each subcube, we choose one vertex
uniformly at random.

\begin{lemma}
\label{rm}If $\gamma\geq k$, then the probability that exactly one
marked vertex is chosen is at least $k/\gamma-\left(
k/\gamma\right)  ^{2}$.
\end{lemma}

\begin{proof}
Let $x$ be a marked vertex. \ The probability that $x$ is chosen is
$1/\gamma $. \ Given that $x$ is chosen, the probability that one of
the other marked vertices, $y$, is chosen is $0$ if $x$ and $y$
belong to the same subcube, or $1/\gamma$\ if they belong to
different subcubes. \ Therefore, the probability
that $x$ alone is chosen is at least%
\[
\frac{1}{\gamma}\left(  1-\frac{k-1}{\gamma}\right)
\geq\frac{1}{\gamma }\left(  1-\frac{k}{\gamma}\right)  .
\]
Since the events \textquotedblleft$x$ alone is
chosen\textquotedblright\ are mutually disjoint, we conclude that
the probability that exactly one marked vertex is chosen is at least
$k/\gamma-\left(  k/\gamma\right)  ^{2}$.
\end{proof}

In particular, fix $\gamma$ so that $\gamma/3<k<2\gamma/3$; then
Lemma \ref{rm}\ implies that the probability of choosing exactly one
marked vertex is at least $2/9$. \ The algorithm is now as follows.
\ As in the lemma, subdivide $\mathcal{L}_{d}\left(  n\right)  $
into $n/\gamma$ subcubes and choose one location at random from
each. \ Then run the algorithm for the unique-solution case (Theorem
\ref{sqrtsrch} or \ref{sqrtsrch2}) on the chosen locations only, as
if they were vertices of $\mathcal{L}_{d}\left( n/\gamma\right)  $.

The running time in the unique case was $O\left(
\sqrt{n/\gamma}\right)  $ for $d\geq3$ or
\[
O\left(  \sqrt{\frac{n}{\gamma}}\log^{3/2}\left(  n/\gamma\right)
\right) =O\left(  \sqrt{\frac{n}{\gamma}}\log^{3/2}n\right)
\]
for $d=2$. \ However, each local unitary in the original algorithm
now becomes
a unitary affecting two vertices $v$\ and $w$ in neighboring subcubes $C_{v}%
$\ and $C_{w}$. \ When placed side by side, $C_{v}$\ and $C_{w}$
form a rectangular box of size
$2\gamma^{1/d}\times\gamma^{1/d}\times\cdots \times\gamma^{1/d}$. \
Therefore the distance between $v$ and $w$ is at most $\left(
d+1\right)  \gamma^{1/d}$. \ It follows that each local unitary in
the original algorithm takes $O\left(  d\gamma^{1/d}\right)  $ time
in the new
algorithm. \ For $d\geq3$, this results in an overall running time of%
\[
O\left(  \sqrt{\frac{n}{\gamma}}d\gamma^{1/d}\right)  =O\left(
d\frac
{\sqrt{n}}{\gamma^{1/2-1/d}}\right)  =O\left(  \frac{\sqrt{n}}{k^{1/2-1/d}%
}\right)  .
\]
For $d=2$\ we obtain%
\[
O\left(  \sqrt{\frac{n}{\gamma}}\gamma^{1/2}\log^{3/2}n\right)
=O\left( \sqrt{n}\log^{3/2}n\right)  .
\]

\subsection{Unknown Number of Marked Items\label{UNKNOWN}}

We now show how to deal with an unknown $k$. \ Let $\operatorname*{OR}%
\nolimits^{\left(  \geq k\right)  }$\ be the problem of deciding
whether there are no marked vertices or \textit{at least} $k$ of
them, given that one of these is true.

\begin{theorem}
\label{mmiub1}\quad

\begin{enumerate}
\item[(i)] For $d\geq3$,%
\[
Q\left(  \operatorname*{OR}\nolimits^{\left(  \geq k\right)  },\mathcal{L}%
_{d}\right)  =O\left(  \frac{\sqrt{n}}{k^{1/2-1/d}}\right)  .
\]

\item[(ii)] For $d=2$,%
\[
Q\left(  \operatorname*{OR}\nolimits^{\left(  \geq k\right)  },\mathcal{L}%
_{2}\right)  =O\left(  \sqrt{n}\log^{5/2}n\right)  .
\]

\end{enumerate}
\end{theorem}

\begin{proof}
We use the straightforward `doubling' approach of Boyer et al.
\cite{bbht}:

\begin{enumerate}
\item[(1)] For $j=0$\ to $\log_{2}\left(  n/k\right)  $

\begin{itemize}
\item Run the algorithm of Theorem \ref{mmiub} with subcubes of size
$\gamma_{j}=2^{j}k$.

\item If a marked vertex is found, then output $1$ and halt.
\end{itemize}

\item[(2)] Query a random vertex $v$, and output $1$ if $v$ is a marked vertex
and $0$ otherwise.
\end{enumerate}

Let $k^{\ast}\geq k$ be the number of marked vertices. $\ $If
$k^{\ast}\leq n/3$, then there exists a $j\leq\log_{2}\left(
n/k\right)  $ such that $\gamma_{j}/3\leq
k^{\ast}\leq2\gamma_{j}/3$. \ So Lemma \ref{rm} implies that the
$j^{th}$\ iteration of step (1) finds a marked vertex with
probability at least $2/9$. \ On the other hand, if $k^{\ast}\geq
n/3$, then step (2) finds a marked vertex with probability at least
$1/3$. \ For $d\geq3$, the time used
in step (1) is at most%
\[
\sum_{j=0}^{\log_{2}\left(  n/k\right)  }\frac{\sqrt{n}}{\gamma_{j}^{1/2-1/d}%
}=\frac{\sqrt{n}}{k^{1/2-1/d}}\left[  \sum_{j=0}^{\log_{2}\left(
n/k\right)
}\frac{1}{2^{j\left(  1/2-1/d\right)  }}\right]  =O\left(  \frac{\sqrt{n}%
}{k^{1/2-1/d}}\right)  ,
\]
the sum in brackets being a decreasing geometric series. \ For
$d=2$, the time is $O\left(  \sqrt{n}\log^{5/2}n\right)  $, since
each iteration takes $O\left(  \sqrt{n}\log^{3/2}n\right)  $ time
and there are at most $\log n$ iterations. \ In neither case does
step (2) affect the bound, since $k\leq n$\ implies that
$n^{1/d}\leq\sqrt{n}/k^{1/2-1/d}$.
\end{proof}

Taking $k=1$ gives algorithms for unconstrained $\operatorname*{OR}$
with running times $O(\sqrt{n})$ for $d\geq3$ and
$O(\sqrt{n}\log^{5/2}n)$ for $d=2$, thereby establishing Theorems
\ref{sqrtsrchfull} and \ref{sqrtsrch2full}.

\section{Search on Irregular Graphs\label{IRREG}}

In Section \ref{PREVGG}, we claimed that our divide-and-conquer
approach has the advantage of being \textit{robust}: it works not
only for highly symmetric graphs such as hypercubes, but for any
graphs having comparable expansion properties. \ Let us now
substantiate this claim.

Say a family of connected graphs $\left\{  G_{n}=\left(
V_{n},E_{n}\right) \right\}  $ is $d$\textit{-dimensional} if there
exists a $\kappa>0$ such that
for all $n,\ell$ and $v\in V_{n}$,%
\[
\left\vert B\left(  v,\ell\right)  \right\vert \geq\min\left(
\kappa\ell ^{d},n\right)  ,
\]
where $B\left(  v,\ell\right)  $\ is the set of vertices having
distance at most $\ell$ from $v$ in $G_{n}$. \ Intuitively, $G_{n}$
is $d$-dimensional (for $d\geq2$ an integer) if its expansion
properties are at least as good as those of the hypercube
$\mathcal{L}_{d}\left(  n\right)  $.\footnote{In general, it makes
sense to consider non-integer $d$ as well.} \ It is immediate that
the diameter of $G_{n}$ is at most $\left(  n/\kappa\right) ^{1/d}$.
\ Note, though, that $G_{n}$ might not be an expander graph in the
usual sense, since we have not required that every sufficiently
small \textit{set} of vertices has many neighbors.

Our goal is to show the following.

\begin{theorem}
\label{irregthm}If $G$ is $d$-dimensional, then

\begin{enumerate}
\item[(i)] For a constant $d>2$,%
\[
Q\left(  \operatorname*{OR},G\right)  =O\left(  \sqrt{n}%
\operatorname*{polylog}n\right)  .
\]

\item[(ii)] For $d=2$,%
\[
Q\left(  \operatorname*{OR},G\right)  =\sqrt{n}2^{O\left(
\sqrt{\log n}\right)  }.
\]

\end{enumerate}
\end{theorem}

In proving part (i),\ the intuition is simple: we want to decompose
$G$ recursively into subgraphs (called \textit{clusters}), which
will serve the same role as subcubes did in the hypercube case. \
The procedure is as follows. \ For some constant $n_{1}>1$, first
choose $\left\lceil n/n_{1}\right\rceil $ vertices uniformly at
random to be designated as $1$-\textit{pegs}. \ Then form
$1$\textit{-clusters} by assigning each vertex in $G$ to its closest
$1$-peg, as in a Voronoi diagram. \ (Ties are broken randomly.) \
Let $v\left(  C\right)  $ be the peg of cluster $C$.\ \ Next, split
up any $1$-cluster $C$ with more than $n_{1}$\ vertices into
$\left\lceil \left\vert C\right\vert /n_{1}\right\rceil $
arbitrarily-chosen $1$-clusters, each with size at most $n_{1}$ and
with $v\left(  C\right)
$\ as its $1$-peg. \ Observe that%
\[
\sum_{i=1}^{\left\lceil n/n_{1}\right\rceil }\left\lceil
\frac{\left\vert
C_{i}\right\vert }{n_{1}}\right\rceil \leq2\left\lceil \frac{n}{n_{1}%
}\right\rceil ,
\]
where $n=\left\vert C_{1}\right\vert +\cdots+\left\vert
C_{\left\lceil n/n_{1}\right\rceil }\right\vert $. \ Therefore, the
splitting-up step can at most double the number of clusters.

In the next iteration, set $n_{2}=n_{1}^{1/\beta}$, for some
constant $\beta\in\left(  2/d,1\right)  $. \ Choose $2\left\lceil
n/n_{2}\right\rceil $\ vertices uniformly at random as $2$-pegs. \
Then form $2$-clusters by assigning each $1$-cluster $C$\ to the
$2$-peg\ that is closest to the $1$-peg $v\left(  C\right)  $. \
Given a $2$-cluster $C^{\prime}$, let $\left\vert
C^{\prime}\right\vert $\ be the number of $1$-clusters in
$C^{\prime}$. \ Then as before, split up any $C^{\prime}$\ with
$\left\vert C^{\prime}\right\vert
>n_{2}/n_{1}$ into $\left\lceil \left\vert C^{\prime}\right\vert /\left(
n_{2}/n_{1}\right)  \right\rceil $\ arbitrarily-chosen $2$-clusters,
each with size at most $n_{2}/n_{1}$\ and with $v\left(
C^{\prime}\right)  $\ as its
$2$-peg. \ Continue recursively in this manner, setting $n_{R}=n_{R-1}%
^{1/\beta}$\ and choosing $2^{R-1}\left\lceil n/n_{R}\right\rceil $\
vertices as $R$-pegs for each $R$. \ Stop at the maximum $R$ such
that $n_{R}\leq n$. \ For technical convenience, set $n_{0}=1$, and
consider each vertex\ $v$ to be the $0$-peg of the $0$-cluster
$\left\{  v\right\}  $.

At the end we have a tree of clusters, which can be searched
recursively just as in the hypercube case. \ In more detail, basis
states now have the form $\left\vert v,z,C\right\rangle $,\ where
$v$ is a vertex, $z$ is an answer bit, and $C$\ is the (label of
the) cluster currently being searched. \ (Unfortunately, because
multiple $R$-clusters can have the same peg, a single auxiliary
qubit no longer suffices.) \ Also, let $K^{\prime}\left( C\right)
$\ be the number of $\left(  R-1\right)  $-clusters in $R$-cluster
$C$; then $K^{\prime}\left(  C\right)  \leq K\left(  R\right)  $\
where $K\left(  R\right)  =2\left\lceil n_{R}/n_{R-1}\right\rceil $.
\ If $K^{\prime}\left(  C\right)  <K\left(  R\right)  $, then place
$K\left( R\right)  -K^{\prime}\left(  C\right)  $\ \textquotedblleft
dummy\textquotedblright\ $\left(  R-1\right)  $-clusters in $C$,
each of which has $\left(  R-1\right)  $-peg\ $v\left(  C\right)  $.

The algorithm $\mathcal{A}_{R}$\ from Section \ref{D3}\ now does the
following, when invoked on the initial state $\left\vert v\left(
C\right) ,0,C\right\rangle $, where $C$\ is an $R$-cluster. \ If
$R=0$, then $\mathcal{A}_{R}$ uses a query transformation to prepare
the state $\left\vert v\left(  C\right)  ,1,C\right\rangle $\ if
$v\left(  C\right)  $\ is the marked vertex and $\left\vert v\left(
C\right)  ,0,C\right\rangle $\ otherwise. \ If $R\geq1$ and $C$ is
not a dummy cluster, then $\mathcal{A}_{R}$\ performs $m_{R}$\
steps of amplitude amplification on $A_{R}$, where $m_{R}$ is the
largest integer such that $2m_{R}+1\leq
\sqrt{n_{R}/n_{R-1}}$.\footnote{In the hypercube case, we performed
fewer
amplifications in order to lower the running time from $\sqrt{n}%
\operatorname*{polylog}n$\ to $\sqrt{n}$. \ Here, though, the
splitting-up step produces a $\operatorname*{polylog}n$\ factor
anyway.} \ If $C$ is a dummy cluster, then $\mathcal{A}_{R}$ does
nothing for an appropriate number of steps, and then returns that no
marked item was found.

We now describe the subroutine $A_{R}$, for $R\geq1$. \ When invoked
with $\left\vert v\left(  C\right)  ,0,C\right\rangle $ as its
initial state,
$A_{R}$\ first prepares a uniform superposition%
\[
\frac{1}{\sqrt{K\left(  R\right)  }}\sum_{i=1}^{K\left(  R\right)
}\left\vert v\left(  C_{i}\right)  ,0,C_{i}\right\rangle .
\]
It then calls $\mathcal{A}_{R-1}$\ recursively, to search $C_{1}%
,\ldots,C_{K\left(  R\right)  }$\ in superposition and amplify the
results.

For $R\geq1$, define the \textit{radius} of an $R$-cluster $C$ to be
the maximum, over all $\left(  R-1\right)  $-clusters\ $C^{\prime}$\
in $C$, of the distance from $v\left(  C\right)  $\ to $v\left(
C^{\prime}\right)  $. \ Also, call an $R$-cluster \textit{good} if
it has radius at most $\ell_{R}$, where $\ell_{R}=\left(
\frac{2}{\kappa}n_{R}\ln n\right)  ^{1/d}$.

\begin{lemma}
\label{radiuslem}With probability $1-o\left(  1\right)  $ over the
choice of clusters, all clusters are good.
\end{lemma}

\begin{proof}
Let $v$\ be the $\left(  R-1\right)  $-peg of an $\left(  R-1\right)
$-cluster. \ Then $\left\vert B\left(  v,\ell\right)  \right\vert
\geq \kappa\ell^{d}$, where $B\left(  v,\ell\right)  $\ is the ball
of radius $\ell$\ about $v$. \ So the probability that $v$\ has
distance greater than
$\ell_{R}$\ to the nearest $R$-peg\ is at most%
\[
\left(  1-\frac{\kappa\ell_{R}^{d}}{n}\right)  ^{\left\lceil n/n_{R}%
\right\rceil }\leq\left(  1-\frac{2\ln n}{n/n_{R}}\right)
^{n/n_{R}}<\frac {1}{n^{2}}.
\]
Furthermore, the total number of pegs is easily seen to be $O\left(
n\right) $. \ It follows by the union bound that \textit{every}
$\left(  R-1\right) $-peg\ for \textit{every} $R$ has distance at
most $\ell_{R}$\ to the nearest $R$-peg, with probability $1-O\left(
1/n\right)  =1-o\left(  1\right)  $ over the choice of clusters.
\end{proof}

We now analyze the running time and success probability of
$\mathcal{A}_{R}$.

\begin{lemma}
\label{irreglem}$\mathcal{A}_{R}$ uses $O\left(  \sqrt{n_{R}}\log
^{1/d}n\right)  $ steps, assuming that all clusters are good.
\end{lemma}

\begin{proof}
Let $T_{\mathcal{A}}\left(  R\right)  $\ and $T_{A}\left(  R\right)
$\ be the time used by $\mathcal{A}_{R}$ and $A_{R}$ respectively in
searching an
$R$-cluster. \ Then we have%
\begin{align*}
T_{\mathcal{A}}\left(  R\right)   &
\leq\sqrt{n_{R}/n_{R-1}}T_{A}\left(
R\right)  ,\\
T_{A}\left(  R\right)   &  \leq\ell_{R}+T_{\mathcal{A}}\left(
R-1\right)
\end{align*}
with the base case $T_{\mathcal{A}}\left(  0\right)  =1$. \ Combining,%
\begin{align*}
T_{\mathcal{A}}\left(  R\right)   &  \leq\sqrt{n_{R}/n_{R-1}}\left(
\ell
_{R}+T_{\mathcal{A}}\left(  R-1\right)  \right)  \\
&  \leq\sqrt{n_{R}/n_{R-1}}\ell_{R}+\sqrt{n_{R}/n_{R-2}}\ell_{R-1}%
+\cdots+\sqrt{n_{R}/n_{0}}\ell_{1}\\
&  =\sqrt{n_{R}}\cdot O\left(  \frac{\left(  n_{R}\ln n\right)  ^{1/d}}%
{\sqrt{n_{R-1}}}+\cdots+\frac{\left(  n_{1}\ln n\right)  ^{1/d}}{\sqrt{n_{0}}%
}\right)  \\
&  =\sqrt{n_{R}}\left(  \ln^{1/d}n\right)  \cdot O\left(
n_{R}^{1/d-\beta
/2}+\cdots+n_{1}^{1/d-\beta/2}\right)  \\
&  =\sqrt{n_{R}}\left(  \ln^{1/d}n\right)  \cdot O\left(
n_{1}^{1/d-\beta /2}+\left(  n_{1}^{1/d-\beta/2}\right)
^{1/\beta}+\cdots+\left(
n_{1}^{1/d-\beta/2}\right)  ^{\left(  1/\beta\right)  ^{R-1}}\right)  \\
&  =O\left(  \sqrt{n_{R}}\log^{1/d}n\right)  ,
\end{align*}
where the last line holds because $\beta>2/d$ and therefore $n_{1}%
^{1/d-\beta/2}<1$.
\end{proof}

\begin{lemma}
\label{irreglem2}$\mathcal{A}_{R}$ succeeds with probability
$\Omega\left( 1/\operatorname*{polylog}n_{R}\right)  $ in searching
a graph of size $n=n_{R}$, assuming there is a unique marked vertex.
\end{lemma}

\begin{proof}
For all $R\geq0$, let $C_{R}$\ be the $R$-cluster that contains the
marked vertex, and\ let $P_{\mathcal{A}}\left(  R\right)  $\ and
$P_{A}\left( R\right)  $\ be the success probabilities of
$\mathcal{A}_{R}$ and $A_{R}$ respectively when searching $C_{R}$. \
Then for all $R\geq1$, we have $P_{A}\left(  R\right)
=P_{\mathcal{A}}\left(  R-1\right)  /\left(  2K\left(
R\right)  \right)  $, and therefore%
\begin{align*}
P_{\mathcal{A}}\left(  R\right)    & \geq\left(  1-\frac{\left(
2m_{R}+1\right)  ^{2}}{3}P_{A}\left(  R\right)  \right)  \left(
2m_{R}+1\right)  ^{2}P_{A}\left(  R\right)  \\
& =\left(  1-\frac{\left(  2m_{R}+1\right)  ^{2}}{3}\cdot\frac{P_{\mathcal{A}%
}\left(  R-1\right)  }{2K\left(  R\right)  }\right)  \left(
2m_{R}+1\right)
^{2}\frac{P_{\mathcal{A}}\left(  R-1\right)  }{2K\left(  R\right)  }\\
& =\Omega\left(  P_{\mathcal{A}}\left(  R-1\right)  \right)  \\
& =\Omega\left(  1/\operatorname*{polylog}n_{R}\right)  .
\end{align*}
Here the third line holds because $\left(  2m_{R}+1\right)
^{2}\approx n_{R}/n_{R-1}\approx K\left(  R\right)  /2$, and the
last line because $R=\Theta\left(  \log\log n_{R}\right)  $.
\end{proof}

Finally, we repeat $\mathcal{A}_{R}$\ itself
$O(\operatorname*{polylog}n_{R})$ times, to achieve success
probability $\Omega\left(  1\right)  $\ using $O\left(
\sqrt{n_{R}}\operatorname*{polylog}n_{R}\right)  $\ steps in total.
\ Again, if $n$\ is not equal to $n_{R}$\ for any $R$, then we
simply find the largest $R$ such that $n_{R}<n$, and then add one
more level of recursion that searches a random $R$-cluster and
amplifies the result $\Theta\left( \sqrt{n/n_{R}}\right)  $\ times.
\ The resulting algorithm uses $O\left(
\sqrt{n}\operatorname*{polylog}n\right)  $\ steps, thereby
establishing part (i) of Theorem \ref{irregthm}\ for the case of a
unique marked vertex. \ The generalization to multiple marked
vertices is straightforward.

\begin{corollary}
\label{irregcor}If $G$ is $d$-dimensional for a constant $d>2$, then%
\[
Q\left(  \operatorname*{OR}\nolimits^{\left(  \geq k\right)
},G\right)
=O\left(  \frac{\sqrt{n}\operatorname*{polylog}\frac{n}{k}}{k^{1/2-1/d}%
}\right)  .
\]

\end{corollary}

\begin{proof}
Assume without loss of generality that $k=o\left(  n\right)  $,
since otherwise a marked item is trivially found in $O\left(
n^{1/d}\right)
$\ steps. \ As in Theorem \ref{mmiub1}, we give an algorithm $\mathcal{B}%
$\ consisting of $\log_{2}\left(  n/k\right)  +1$\ iterations. \ In
iteration $j=0$, choose $\left\lceil n/k\right\rceil $\ vertices
$w_{1},\ldots ,w_{\left\lceil n/k\right\rceil }$\ uniformly at
random. \ Then run the algorithm for the unique marked vertex case,
but instead of taking all vertices in $G$ as $0$-pegs, take only
$w_{1},\ldots,w_{\left\lceil n/k\right\rceil }$. \ On the other
hand, still choose the $1$-pegs, $2$-pegs, and so on uniformly at
random from among all vertices in $G$. \ For all $R$, the number of
$R$-pegs\ should be\ $\left\lceil \left(  n/k\right)
/n_{R}\right\rceil $. \ In general, in iteration $j$\ of
$\mathcal{B}$, choose\ $\left\lceil n/\left(  2^{j}k\right)
\right\rceil $\ vertices
$w_{1},\ldots,w_{\left\lceil n/\left(  2^{j}k\right)  \right\rceil }%
$\ uniformly at random, and then run the algorithm for a\ unique
marked vertex
as if $w_{1},\ldots,w_{\left\lceil n/\left(  2^{j}k\right)  \right\rceil }%
$\ were the only vertices in the graph.

It is easy to see that, assuming there are $k$ or more marked
vertices, with\ probability $\Omega\left(  1\right)  $ there exists
an iteration $j$ such that exactly one of
$w_{1},\ldots,w_{\left\lceil n/\left(  2^{j}k\right) \right\rceil
}$\ is marked. \ Hence $\mathcal{B}$ succeeds with probability
$\Omega\left(  1\right)  $. \ It remains only to upper-bound
$\mathcal{B}$'s running time.

In\ iteration $j$, notice that Lemma \ref{radiuslem}\ goes through
if we use $\ell_{R}^{\left(  j\right)  }:=\left(
\frac{2}{\kappa}2^{j}kn_{R}\ln\frac {n}{k}\right)  ^{1/d}$ instead
of $\ell_{R}$. \ That is, with probability $1-O\left(  k/n\right)
=1-o\left(  1\right)  $\ over the choice of clusters, every
$R$-cluster\ has radius at most $\ell_{R}^{\left(  j\right)  }$. \
So letting $T_{\mathcal{A}}\left(  R\right)  $\ be the running time
of $\mathcal{A}_{R}$\ on an $R$-cluster, the recurrence in Lemma
\ref{irreglem}
becomes%
\[
T_{\mathcal{A}}\left(  R\right)  \leq\sqrt{n_{R}/n_{R-1}}\left(
\ell _{R}^{\left(  j\right)  }+T_{\mathcal{A}}\left(  R-1\right)
\right) =O\left(  \sqrt{n_{R}}\left(  2^{j}k\log\left(  n/k\right)
\right) ^{1/d}\right)  ,
\]
which is%
\[
O\left(  \frac{\sqrt{n}\log^{1/d}\frac{n}{k}}{\left(  2^{j}k\right)
^{1/2-1/d}}\right)
\]
if $n_{R}=\Theta\left(  n/\left(  2^{j}k\right)  \right)  $. \ As
usual, the case where there is no $R$ such that $n_{R}=\Theta\left(
n/\left( 2^{j}k\right)  \right)  $\ is trivially handled by adding
one more level of
recursion. \ If we factor in the $O\left(  1/\operatorname*{polylog}%
n_{R}\right)  $\ repetitions of $\mathcal{A}_{R}$\ needed to boost
the\ success probability to $\Omega\left(  1\right)  $, then the
total running
time of iteration $j$\ is%
\[
O\left(  \frac{\sqrt{n}\operatorname*{polylog}\frac{n}{k}}{\left(
2^{j}k\right)  ^{1/2-1/d}}\right)  .
\]
Therefore $\mathcal{B}$'s running time is%
\[
O\left(  \sum_{j=0}^{\log_{2}\left(  n/k\right)  }\frac{\sqrt{n}%
\operatorname*{polylog}n}{\left(  2^{j}k\right)  ^{1/2-1/d}}\right)
=O\left( \frac{\sqrt{n}\operatorname*{polylog}n}{k^{1/2-1/d}}\right)
.
\]

\end{proof}

For the $d=2$\ case, the best upper bound we can show is
$\sqrt{n}2^{O\left( \sqrt{\log n}\right)  }$. \ This is obtained by
simply modifying $\mathcal{A}_{R}$ to have a deeper recursion tree.
\ Instead of taking
$n_{R}=n_{R-1}^{1/\mu}$ for some $\mu$, we take $n_{R}=2^{\sqrt{\log n}%
}n_{R-1}=2^{R\sqrt{\log n}}$, so that the total number of levels is
$\left\lceil \sqrt{\log n}\right\rceil $. \ Lemma \ref{radiuslem}\
goes through without modification, while\ the recurrence for the
running time
becomes%
\begin{align*}
T_{\mathcal{A}}\left(  R\right)   &  \leq\sqrt{n_{R}/n_{R-1}}\left(
\ell
_{R}+T_{\mathcal{A}}\left(  R-1\right)  \right)  \\
&  \leq\sqrt{n_{R}/n_{R-1}}\ell_{R}+\sqrt{n_{R}/n_{R-2}}\ell_{R-1}%
+\cdots+\sqrt{n_{R}/n_{0}}\ell_{1}\\
&  =O\left(  2^{\sqrt{\log n}\left(  R/2\right)  }\sqrt{\ln
n}+\cdots
+2^{\sqrt{\log n}\left(  R/2\right)  }\sqrt{\ln n}\right)  \\
&  =\sqrt{n}2^{O\left(  \sqrt{\log n}\right)  }.
\end{align*}
Also, since the success probability decreases by at most a constant
factor at each level, we have that $P_{\mathcal{A}}\left(  R\right)
=2^{-O\left(
\sqrt{\log n}\right)  }$, and hence $2^{O\left(  \sqrt{\log n}\right)  }%
$\ amplification steps suffice to boost the success probability to
$\Omega\left(  1\right)  $. \ Handling multiple marked items adds an
additional factor of $\log n$, which is absorbed into $2^{O\left(
\sqrt{\log n}\right)  }$. \ This completes Theorem \ref{irregthm}.

\subsection{Bits Scattered on a Graph\label{SCATTERED}}

In Section \ref{PHYS}, we discussed several ways to pack a given
amount of entropy into a spatial region of given dimensions. \
However, we said nothing about how the entropy is
\textit{distributed} within the region. \ It might be uniform, or
concentrated on the boundary, or distributed in some other way. \ So
we need to answer the following: suppose that in some graph, $h$\
out of the $n$ vertices \textit{might} be marked, and we know which
$h$ those are. \ Then how much time is needed to determine whether
any of the $h$ \textit{is} marked? \ If the graph is the hypercube
$\mathcal{L}_{d}$\ for $d\geq2$\ or is $d$-dimensional for $d>2$,
then the results of the previous sections imply that $O\left(
\sqrt{n}\operatorname*{polylog}n\right)  $\ steps suffice. \
However, we wish to use fewer steps, taking advantage of the fact
that $h$ might be much smaller than $n$. \ Formally, suppose we are
given a graph $G$ with $n$ vertices, of which \thinspace$h$ are
potentially marked. \ Let $\operatorname*{OR}\nolimits^{\left(
h,\geq k\right)  }$ be the problem of deciding whether $G$ has no
marked vertices or at least $k$ of them, given that one of these is
the case.

\begin{proposition}
\label{scatterlb}For all integer constants $d\geq2$, there exists a
$d$-dimensional graph $G$ such that%
\[
Q\left(  \operatorname*{OR}\nolimits^{\left(  h,\geq k\right)
},G\right) =\Omega\left(  n^{1/d}\left(  \frac{h}{k}\right)
^{1/2-1/d}\right)  .
\]

\end{proposition}

\begin{proof}
[Proof]Let $G$ be the $d$-dimensional hypercube
$\mathcal{L}_{d}\left( n\right)  $. \ Create $h/k$ subcubes of
potentially marked vertices, each having $k$ vertices and side
length $k^{1/d}$. \ Space these subcubes out in
$\mathcal{L}_{d}\left(  n\right)  $\ so that the distance between
any pair of them is $\Omega\left(  \left(  nk/h\right)
^{1/d}\right)  $. \ Then choose a subcube $C$\ uniformly at random
and mark all $k$ vertices in $C$. \ This enables us to consider each
subcube as a single vertex, having distance $\Omega\left(  \left(
nk/h\right)  ^{1/d}\right)  $ to every other vertex. \ The lower
bound now follows by a hybrid argument essentially identical to that
of Theorem \ref{mmi}.
\end{proof}

In particular, if $d=2$\ then $\Omega\left(  \sqrt{n}\right)  $ time
is always needed, since the potentially marked vertices might all be
far from the start vertex. \ The lower bound of Proposition
\ref{scatterlb}\ can be achieved up to a polylogarithmic factor.

\begin{proposition}
\label{scatterthm}If $G$ is $d$-dimensional\ for a constant $d>2$, then%
\[
Q\left(  \operatorname*{OR}\nolimits^{\left(  h,\geq k\right)
},G\right)
=O\left(  n^{1/d}\left(  \frac{h}{k}\right)  ^{1/2-1/d}\operatorname*{polylog}%
\frac{h}{k}\right)  .
\]

\end{proposition}

\begin{proof}
Assume without loss of generality that $k=o\left(  h\right)  $,
since
otherwise a marked item is trivially found.\ \ Use algorithm $\mathcal{B}%
$\ from Corollary \ref{irregcor}, with the following simple change.
\ In iteration $j$, choose\ $\left\lceil h/\left(  2^{j}k\right)
\right\rceil $\ potentially marked vertices
$w_{1},\ldots,w_{\left\lceil h/\left( 2^{j}k\right)  \right\rceil
}$\ uniformly at random, and then run the algorithm for a\ unique
marked vertex as if $w_{1},\ldots,w_{\left\lceil h/\left(
2^{j}k\right)  \right\rceil }$\ were the only vertices in the graph.
\ That is, take $w_{1},\ldots,w_{\left\lceil h/\left(  2^{j}k\right)
\right\rceil }$\ as $0$-pegs; then for all $R\geq1$, choose
$\left\lceil h/\left(  2^{j}kn_{R}\right)  \right\rceil $\ vertices
of $G$ uniformly at random as $R$-pegs. \ Lemma \ref{radiuslem}\
goes through if we use
$\widehat{\ell}_{R}^{\left(  j\right)  }:=\left(  \frac{2}{\kappa}\frac{n}%
{h}2^{j}kn_{R}\ln\frac{h}{k}\right)  ^{1/d}$ instead of $\ell_{R}$.\
\ So
following Corollary \ref{irregcor}, the running time of iteration $j$\ is now%
\[
O\left(  \sqrt{n_{R}}\left(  \frac{n}{h}2^{j}k\right)  ^{1/d}%
\operatorname*{polylog}\frac{h}{k}\right)  =O\left(  n^{1/d}\left(
\frac {h}{2^{j}k}\right)
^{1/2-1/d}\operatorname*{polylog}\frac{h}{k}\right)
\]
if $n_{R}=\Theta\left(  h/\left(  2^{j}k\right)  \right)  $. \
Therefore the
total running time is%
\[
O\left(  \sum_{j=0}^{\log_{2}\left(  h/k\right)  }n^{1/d}\left(
\frac {h}{2^{j}k}\right)
^{1/2-1/d}\operatorname*{polylog}\frac{h}{k}\right)
=O\left(  n^{1/d}\left(  \frac{h}{k}\right)  ^{1/2-1/d}\operatorname*{polylog}%
\frac{h}{k}\right)  .
\]

\end{proof}

Intuitively, Proposition \ref{scatterthm}\ says that the worst case
for search occurs when the $h$ potential marked vertices are
scattered evenly throughout the graph.

\section{Application to Disjointness\label{APPL}}

In this section we show how our results can be used to strengthen a seemingly
unrelated result in quantum computing. \ Suppose Alice has a string
$X=x_{1}\ldots x_{n}\in\left\{  0,1\right\}  ^{n}$, and Bob has a string
$Y=y_{1}\ldots y_{n}\in\left\{  0,1\right\}  ^{n}$. \ In the
\textit{disjointness problem}, Alice and Bob must decide with high probability
whether there exists an $i$ such that $x_{i}=y_{i}=1$, using as few bits of
communication as possible. \ Buhrman, Cleve, and Wigderson \cite{bcw}%
\ observed that in the quantum setting, Alice and Bob can solve this problem
using only $O\left(  \sqrt{n}\log n\right)  $\ qubits of communication. \ This
was subsequently improved by H\o yer and de Wolf \cite{hoyerdewolf}\ to
$O\left(  \sqrt{n}c^{\log^{\ast}n}\right)  $, where $c$\ is a constant and
$\log^{\ast}n$\ is the iterated logarithm function. \ Using the search
algorithm of Theorem \ref{sqrtsrchfull}, we can improve this to $O\left(
\sqrt{n}\right)  $, which matches the celebrated $\Omega\left(  \sqrt
{n}\right)  $\ lower bound of Razborov \cite{razborov:cc}.

\begin{theorem}
The bounded-error quantum communication complexity of the
disjointness problem is $O\left( \sqrt{n}\right)  $.
\end{theorem}

\begin{proof}
The protocol is as follows. \ Alice and Bob both store their inputs in a $3$-D
cube $\mathcal{L}_{3}\left(  n\right)  $ (Figure \ref{alicebob}); that is,
they let $x_{jkl}=x_{i}$\ and $y_{jkl}=y_{i}$, where $i=n^{2/3}j+n^{1/3}%
k+l+1$\ and $j,k,l\in\left\{  0,\ldots,n^{1/3}-1\right\}  $.%
%TCIMACRO{\FRAME{ftbpFU}{3.6073in}{1.0652in}{0pt}{\Qcb{Alice and Bob
%`synchronize' locations on their respective cubes.}}{\Qlb{alicebob}}%
%{Figure}{\special{ language "Scientific Word";  type "GRAPHIC";
%display "USEDEF";  valid_file "T";  width 3.6073in;  height 1.0652in;
%depth 0pt;  original-width 5.1128in;  original-height 1.4806in;
%cropleft "0";  croptop "1";  cropright "1";  cropbottom "0";
%tempfilename 'alicebob.eps';tempfile-properties "XNPR";}}}%
%BeginExpansion
\begin{figure}
[ptb]
\begin{center}
\includegraphics[
height=1.0652in,
width=3.6073in
]%
{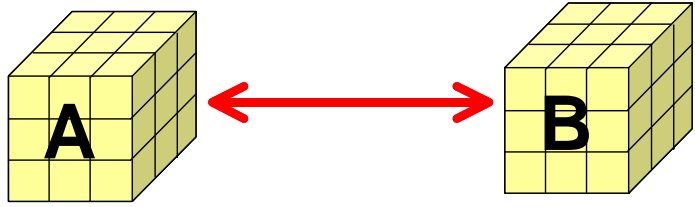}%
\caption[Disjointness in $O\left(\sqrt n\right)$
communication]{Alice and Bob `synchronize'
locations on their respective cubes.}%
\label{alicebob}%
\end{center}
\end{figure}
%EndExpansion
Throughout, they maintain a joint state of the form%
\begin{equation}
\sum\alpha_{j,k,l,z_{A},z_{B},c}\left\vert v_{jkl},z_{A}\right\rangle
\otimes\left\vert c\right\rangle \otimes\left\vert v_{jkl},z_{B}\right\rangle
, \label{theform}%
\end{equation}
where $c$\ is used for communication between the players, and $z_{A}$ and
$z_{B}$\ store the answers to queries. \ Thus, whenever Alice is at location
$\left(  j,k,l\right)  $\ of her cube, Bob is at location $\left(
j,k,l\right)  $\ of his cube. \ To decide whether there exists a $\left(
j,k,l\right)  $\ with $x_{jkl}=y_{jkl}=1$, Alice simply runs our search
algorithm for an unknown number of marked items, but with two changes.
\ First, after each query, Alice inverts her phase if and only if
$x_{jkl}=y_{jkl}=1$;\ this requires $2$ qubits of communication from Bob, to
send $y_{jkl}$\ to Alice and then to erase\ it. \ Second, before each movement
step, Alice tells Bob in which of the six possible directions she is going to
move. \ That way, Bob can synchronize his location with Alice's, and thereby
maintain the state in the form (\ref{theform}). \ This requires $6$ qubits of
communication from Alice, to send the direction to Bob and then to erase it.
\ Notice that no further communication is necessary, since there are no
auxiliary registers in our algorithm that need to be communicated. \ Since the
algorithm uses $O\left(  \sqrt{n}\right)  $\ steps, the number of qubits
communicated in the disjointness protocol is therefore also $O\left(  \sqrt
{n}\right)  $.
\end{proof}

\section{Open Problems\label{OPENGG}}

As discussed in Section \ref{LOCALGG}, a salient open problem raised by this
work is to prove relationships among Z-local, C-local, and H-local unitary
matrices. \ In particular, can any Z-local or H-local unitary be approximated
by a product of a small number of C-local unitaries? \ Also, is it true that
$Q\left(  f,G\right)  =\Theta\left(  Q^{Z}\left(  f,G\right)  \right)
=\Theta\left(  Q^{H}\left(  f,G\right)  \right)  $\ for all $f,G$?

A second problem is to obtain interesting lower bounds in our model. \ For
example, let $G$\ be a $\sqrt{n}\times\sqrt{n}$\ grid, and suppose $f\left(
X\right)  =1$\ if and only if every row of $G$ contains a vertex $v_{i}$\ with
$x_{i}=1$. \ Clearly $Q\left(  f,G\right)  =O\left(  n^{3/4}\right)  $, and we
conjecture that this is optimal. \ However, we were unable to show any lower
bound better than $\Omega\left(  \sqrt{n}\right)  $.

Finally, what is the complexity of finding a unique marked vertex on
a 2-D square grid? \ As mentioned in Section \ref{PREVGG}, Ambainis,
Kempe, and Rivosh \cite{akr}\ showed that $Q\left(
\operatorname*{OR}^{\left(  1\right) },\mathcal{L}_{2}\right)
=O\left(  \sqrt{n}\log n\right)  $. \ Can the remaining factor of
$\log n$\ be removed?

\chapter{Quantum Computing and Postselection\label{POST}}

\begin{quote}
\textquotedblleft Gill, in his seminal paper on probabilistic complexity
classes, defined the class $\mathsf{PP}$ and asked whether the class was
closed under intersection. \ In 1990, Fenner and Kurtz and later myself,
decided to try a new approach to the question: Consider a class defined like
$\mathsf{PP}$ but with additional restrictions, show that this class is closed
under intersection and then show the class was really the same as
$\mathsf{PP}$.\textquotedblright

---Lance Fortnow, \textit{My Computational Complexity Web Log} \cite{fortnow:blog}

 (the approach didn't succeed, though as this chapter will show, all
 it was missing was quantum mechanics)
\end{quote}

\textit{Postselection} is the power of discarding all runs of a
computation in which a given event does not occur. \ Clearly, such
an ability would let us solve $\mathsf{NP}$-complete problems in
polynomial time, since we could guess a random solution, and then
postselect on its being correct. \ But would postselection let us do
\textit{more} than $\mathsf{NP}$? \ Using a classical computer, the
class of problems we could efficiently solve coincides with a class
called $\mathsf{BPP}_{\mathsf{path}}$, which was defined by Han,
Hemaspaandra, and Thierauf \cite{hht} and which sits somewhere
between $\mathsf{MA}$ and $\mathsf{PP}$.

This chapter studies the power of postselection when combined with quantum
computing. \ In Section \ref{POSTBQP}, I define a new complexity class called
$\mathsf{PostBQP}$, which is similar to $\mathsf{BQP}$\ except that we can
measure a qubit that has some nonzero probability of being $\left\vert
1\right\rangle $, and \textit{assume} the outcome will be $\left\vert
1\right\rangle $. \ The main result is that $\mathsf{PostBQP}$\ equals the
classical complexity class $\mathsf{PP}$.

My original motivation, which I explain in Section \ref{FANTASY}, was to
analyze the computational power of \textquotedblleft fantasy\textquotedblright%
\ versions of quantum mechanics, and thereby gain insight into why quantum
mechanics is the way it is. \ For example, I show in Section \ref{FANTASY}
that if we changed the measurement probability rule from $\left\vert
\psi\right\vert ^{2}$\ to $\left\vert \psi\right\vert ^{p}$\ for some $p\neq
2$, or allowed linear but nonunitary gates, then we could simulate
postselection, and hence solve $\mathsf{PP}$-complete problems in
polynomial\ time. \ I was also motivated by a concept that I call
\textit{anthropic computing}: arranging things so that you're more likely to
exist if a computer produces a desired output than if it doesn't. \ As a
simple example, under the many-worlds interpretation of quantum mechanics, you
might kill yourself in all universes where a computer's output is incorrect.
\ My result implies that, using this \textquotedblleft
technique,\textquotedblright\ the class of problems that you could efficiently
solve is exactly $\mathsf{PP}$.

However, it recently dawned on me that the $\mathsf{PostBQP}=\mathsf{PP}%
$\ result is also interesting for purely classical reasons. \ In particular,
it yields an almost-trivial, quantum computing based proof that $\mathsf{PP}%
$\ is closed under intersection. \ This proof does not use rational
approximations, threshold polynomials, or any of the other
techniques pioneered by Beigel, Reingold, and Spielman \cite{brs} in
their justly-celebrated original proof. \ Another immediate
corollary of my new characterization of $\mathsf{PP}$\ is a result
originally due to Fortnow and Reingold \cite{fr:pp}: that
$\mathsf{PP}$\ is closed under polynomial-time truth-table
reductions. \ Indeed, I can show that $\mathsf{PP}$\ is closed under
$\mathsf{BQP}$\ truth-table reductions, which is a new result as far
as I know. \ I conclude in Section \ref{OPENPOST}\ with some open
problems.

\section{The Class $\mathsf{PostBQP}$\label{POSTBQP}}

I hereby define a complexity class:

\begin{definition}
\label{postbqpdef}$\mathsf{PostBQP}$ (Postselected Bounded-Error Quantum
Polynomial-Time) is the class of languages $L$ for which there exists a
uniform family of polynomial-size quantum circuits such that for all inputs
$x$,

\begin{enumerate}
\item[(i)] At the end of the computation, the first qubit has a nonzero
probability of being measured to be $\left\vert 1\right\rangle $.

\item[(ii)] If $x\in L$, then conditioned on the first qubit being $\left\vert
1\right\rangle $, the second qubit is $\left\vert 1\right\rangle $\ with
probability at least $2/3$.

\item[(iii)] If $x\notin L$, then conditioned on the first qubit being
$\left\vert 1\right\rangle $, the second qubit is $\left\vert 1\right\rangle
$\ with probability at most $1/3$.
\end{enumerate}
\end{definition}

We can think of $\mathsf{PostBQP}$\ as the \textquotedblleft
nondeterministic\textquotedblright\ version of $\mathsf{BQP}$. \ Admittedly,
there are already three other contenders for that title: $\mathsf{QMA}$,
defined by Watrous \cite{watrous}; $\mathsf{QCMA}$, defined by Aharonov and
Naveh \cite{an}; and $\mathsf{NQP}$, defined by Adleman, DeMarrais, and Huang
\cite{adh} (which turns out to equal $\mathsf{coC}_{\mathsf{=}}\mathsf{P}$
\cite{fghp}). \ As we will see, $\mathsf{PostBQP}$\ contains all of these as subclasses.

It is immediate that $\mathsf{NP}\subseteq\mathsf{PostBQP}\subseteq
\mathsf{PP}$. \ For the latter inclusion, we can use the same techniques used
by Adleman, DeMarrais, and Huang \cite{adh}\ to show that $\mathsf{BQP}%
\subseteq\mathsf{PP}$, but sum only over paths where the first qubit is
$\left\vert 1\right\rangle $\ at the end. \ This is made explicit in Theorem
\ref{qpoly}\ of Chapter \ref{ADV}.

How robust is $\mathsf{PostBQP}$? \ Just as Bernstein and Vazirani
\cite{bv}\ showed that intermediate measurements don't increase the power of
ordinary quantum computers, so it's easy to show that intermediate
postselection steps don't increase the power of $\mathsf{PostBQP}$. \ Whenever
we want to postselect on a qubit being $\left\vert 1\right\rangle $, we simply
CNOT that qubit into a fresh ancilla qubit that is initialized to $\left\vert
0\right\rangle $ and that will never be written to again. \ Then, at the end,
we compute the AND of all the ancilla qubits, and swap the result into the
first qubit. \ It follows that we can repeat a $\mathsf{PostBQP}$\ computation
a polynomial number of times, and thereby amplify the probability gap from
$\left(  1/3,2/3\right)  $\ to $\left(  2^{-p\left(  n\right)  }%
,1-2^{-p\left(  n\right)  }\right)  $\ for any polynomial $p$.

A corollary of the above observations is that $\mathsf{PostBQP}$\ has strong
closure properties.

\begin{proposition}
\label{closure}$\mathsf{PostBQP}$ is closed under union, intersection, and
complement. \ Indeed, it is closed under $\mathsf{BQP}$\ truth table
reductions, meaning that $\mathsf{PostBQP}=\mathsf{BQP}_{\mathsf{\Vert
,}\operatorname*{classical}}^{\mathsf{PostBQP}}$,\ where $\mathsf{BQP}%
_{\mathsf{\Vert,}\operatorname*{classical}}^{\mathsf{PostBQP}}$\ is the class
of problems solvable by a $\mathsf{BQP}$\ machine\ that can make a polynomial
number of nonadaptive classical queries to a $\mathsf{PostBQP}$ oracle.
\end{proposition}

\begin{proof}
Clearly $\mathsf{PostBQP}$ is closed under complement. \ To show closure under
intersection,\ let $L_{1},L_{2}\in\mathsf{PostBQP}$. \ Then to decide whether
$x\in L_{1}\cap L_{2}$, run amplified\ computations (with error probability at
most $1/6$) to decide if $x\in L_{1}$\ and if $x\in L_{2}$, postselect on both
computations succeeding, and accept if and only if both accept. \ It follows
that $\mathsf{PostBQP}$\ is closed under union as well.

In general, suppose a $\mathsf{BQP}_{\mathsf{\Vert,}\operatorname*{classical}%
}^{\mathsf{PostBQP}}$\ machine $M$\ submits queries $q_{1},\ldots,q_{p\left(
n\right)  }$\ to the $\mathsf{PostBQP}$\ oracle. \ Then run amplified
computations (with error probability at most, say, $\frac{1}{10p\left(
n\right)  }$) to decide the answers to these queries, and postselect on all
$p\left(  n\right)  $ of them succeeding. \ By the union bound, if $M$ had
error probability $\varepsilon$\ with a perfect $\mathsf{PostBQP}$\ oracle,
then its new error probability is at most $\varepsilon+1/10$, which can easily
be reduced through amplification.
\end{proof}

One might wonder why Proposition \ref{closure}\ doesn't go through with
\textit{adaptive} queries. \ The reason is subtle: suppose we have two
$\mathsf{PostBQP}$\ computations, the second of which relies on the output of
the first. \ Then even if the first computation is amplified a polynomial
number of times, it still has an exponentially small probability of error.
\ But since the second computation uses postselection, \textit{any} nonzero
error probability could be magnified arbitrarily, and is therefore too large.

I now prove the main result.

\begin{theorem}
\label{postbqppp}$\mathsf{PostBQP}=\mathsf{PP}$.
\end{theorem}

\begin{proof}
We have already observed that $\mathsf{PostBQP}\subseteq\mathsf{PP}$. \ For
the other direction, let $f:\left\{  0,1\right\}  ^{n}\rightarrow\left\{
0,1\right\}  $\ be an efficiently computable Boolean function and let
$s=\left\vert \left\{  x:f\left(  x\right)  =1\right\}  \right\vert $. \ Then
we need to decide in $\mathsf{PostBQP}$\ whether $s<2^{n-1}$ or $s\geq2^{n-1}%
$. \ (As a technicality, we can guarantee using padding that $s>0$.)

The algorithm is as follows: first prepare the state $2^{-n/2}\sum
_{x\in\left\{  0,1\right\}  ^{n}}\left\vert x\right\rangle \left\vert f\left(
x\right)  \right\rangle $. \ Then following Abrams and Lloyd \cite{al}, apply
Hadamard gates to all $n$ qubits in the first register and
postselect\footnote{Postselection is actually overkill here, since the first
register has at least $1/4$ probability of being $\left\vert 0\right\rangle
^{\otimes n}$.} on that register being $\left\vert 0\right\rangle ^{\otimes
n}$,\ to obtain $\left\vert 0\right\rangle ^{\otimes n}\left\vert
\psi\right\rangle $\ where%
\[
\left\vert \psi\right\rangle =\frac{\left(  2^{n}-s\right)  \left\vert
0\right\rangle +s\left\vert 1\right\rangle }{\sqrt{\left(  2^{n}-s\right)
^{2}+s^{2}}}.
\]
Next, for some positive real numbers $\alpha,\beta$\ to be specified later,
prepare $\alpha\left\vert 0\right\rangle \left\vert \psi\right\rangle
+\beta\left\vert 1\right\rangle H\left\vert \psi\right\rangle $ where%
\[
H\left\vert \psi\right\rangle
=\frac{\sqrt{1/2}\left(2^{n}\right)\left\vert 0\right\rangle
+\sqrt{1/2}\left(  2^{n}-2s\right)  \left\vert 1\right\rangle
}{\sqrt{\left(
2^{n}-s\right)  ^{2}+s^{2}}}%
\]
is the result of applying a Hadamard gate to $\left\vert \psi\right\rangle $.
\ Then postselect on the second qubit being $\left\vert 1\right\rangle
$.\ This yields the reduced state%
\[
\left\vert \varphi_{\beta/\alpha}\right\rangle =\frac{\alpha s\left\vert
0\right\rangle +\beta\sqrt{1/2}\left(  2^{n}-2s\right)  \left\vert
1\right\rangle }{\sqrt{\alpha^{2}s^{2}+\left(  \beta^{2}/2\right)  \left(
2^{n}-2s\right)  ^{2}}}%
\]
in the first qubit.

Suppose $s<2^{n-1}$, so that $s$ and $\sqrt{1/2}\left(  2^{n}-2s\right)
$\ are both at least $1$. \ Then we claim there exists an integer $i\in\left[
-n,n\right]  $\ such that, if we set $\beta/\alpha=2^{i}$, then $\left\vert
\varphi_{2^{i}}\right\rangle $\ is close to the state $\left\vert
+\right\rangle =\left(  \left\vert 0\right\rangle +\left\vert 1\right\rangle
\right)  /\sqrt{2}$:%
\[
\left\vert \left\langle +|\varphi_{2^{i}}\right\rangle \right\vert \geq
\frac{1+\sqrt{2}}{\sqrt{6}}>0.985.
\]
For since $\sqrt{1/2}\left(  2^{n}-2s\right)  /s$ lies between $2^{-n}$\ and
$2^{n}$, there must be an integer $i\in\left[  -n,n-1\right]  $ such that
$\left\vert \varphi_{2^{i}}\right\rangle $\ and $\left\vert \varphi_{2^{i+1}%
}\right\rangle $\ fall on opposite sides of $\left\vert +\right\rangle $\ in
the first quadrant (see Figure \ref{ppfig}).\ \ So the worst case is that
$\left\langle +|\varphi_{2^{i}}\right\rangle =\left\langle +|\varphi_{2^{i+1}%
}\right\rangle $, which occurs when $\left\vert \varphi_{2^{i}}\right\rangle
=\sqrt{2/3}\left\vert 0\right\rangle +\sqrt{1/3}\left\vert 0\right\rangle
$\ and $\left\vert \varphi_{2^{i+1}}\right\rangle =\sqrt{1/3}\left\vert
0\right\rangle +\sqrt{2/3}\left\vert 0\right\rangle $. \ On the other hand,
suppose $s\geq2^{n-1}$, so\ that $\sqrt{1/2}\left(  2^{n}-2s\right)  \leq0$.
\ Then $\left\vert \varphi_{2^{i}}\right\rangle $\ never lies in the first or
third quadrants, and therefore $\left\vert \left\langle +|\varphi_{2^{i}%
}\right\rangle \right\vert \leq1/\sqrt{2}<0.985$.%
%TCIMACRO{\FRAME{ftbpFU}{2.3694in}{2.3869in}{0pt}{\Qcb{If $s$\ and $2^{n}%
%-2s$\ are both positive, then as we vary the ratio of $\beta$\ to $\alpha$, we
%eventually get close to $\left\vert +\right\rangle =\left(  \left\vert
%0\right\rangle +\left\vert 1\right\rangle \right)  /\sqrt{2}$ (dashed lines).
%\ On the other hand, if $2^{n}-2s$ is not positive (dotted line), then we
%never even get into the first quadrant.}}{}{ppfig.eps}%
%width 2.3694in;  height 2.3869in;  depth 0pt;  original-width 10.3511in;
%original-height 7.7551in;  cropleft "0.2690";  croptop "0.6428";
%cropright "0.7721";  cropbottom "0.2632";
%filename 'ppfig.eps';file-properties "XNPEU";}}}%
%BeginExpansion
\begin{figure}
[ptb]
\begin{center}
\includegraphics[
trim=1.7in 3in 1.7in 2.770122in, height=2.3869in, width=2.3694in
]%
{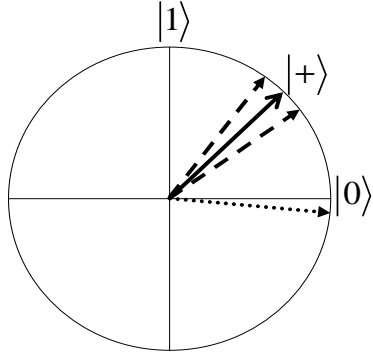}%
\caption[Simulating $\mathsf{PP}$ using postselection]{If $s$\ and
$2^{n}-2s$\ are both positive, then as we vary the ratio of $\beta$\
to $\alpha$, we eventually get close to $\left\vert +\right\rangle
=\left( \left\vert 0\right\rangle +\left\vert 1\right\rangle \right)
/\sqrt{2}$ (dashed lines). \ On the other hand, if $2^{n}-2s$ is not
positive
(dotted line), then we never even get into the first quadrant.}%
\label{ppfig}%
\end{center}
\end{figure}
%EndExpansion

It follows that, by repeating the whole algorithm $n\left(  2n+1\right)
$\ times (as in Proposition \ref{closure}), with $n$\ invocations for each
integer $i\in\left[  -n,n\right]  $, we can learn whether $s<2^{n-1}$ or
$s\geq2^{n-1}$\ with exponentially small probability of error.
\end{proof}

Combining Proposition \ref{closure} with Theorem \ref{postbqppp} immediately
yields that $\mathsf{PP}$\ is closed under intersection, as well as under
$\mathsf{BQP}$\ truth-table reductions.

\section{Fantasy Quantum Mechanics\label{FANTASY}}

\begin{quote}
\textquotedblleft It is striking that it has so far not been possible to find
a logically consistent theory that is close to quantum mechanics, other than
quantum mechanics itself.\textquotedblright

---Steven Weinberg, \textit{Dreams of a Final Theory} \cite{weinberg:dreams}
\end{quote}

Is quantum mechanics an island in theoryspace? \ By \textquotedblleft
theoryspace,\textquotedblright\ I mean the space of logically conceivable
physical theories, with two theories close to each other if they differ in few
respects. \ An \textquotedblleft island\textquotedblright\ in theoryspace is
then a natural and interesting theory, whose neighbors are all somehow
perverse or degenerate. \ The Standard Model is not an island, because we do
not know of any compelling (non-anthropic) reason why the masses and coupling
constants should have the values they do. \ Likewise, general relativity is
probably not an island, because of alternatives such as the Brans-Dicke theory.

To many physicists, however, quantum mechanics \textit{does} seem like an
island: change any one aspect, and the whole theory becomes inconsistent or
nonsensical. \ There are many mathematical results supporting this opinion:
for example, Gleason's Theorem \cite{gleason}\ and other \textquotedblleft
derivations\textquotedblright\ of the $\left\vert \psi\right\vert ^{2}%
$\ probability rule \cite{deutsch:dec,zurek}; arguments for why amplitudes
have to be complex numbers, rather than real numbers or quaternions
\cite{cfs,hardy}; and \textquotedblleft absurd\textquotedblright\ consequences
of allowing nonlinear transformations between states
\cite{al,gisin,polchinski}. \ The point of these results is to provide some
sort of explanation for why quantum mechanics has the properties it does.

In 1998, Abrams and Lloyd \cite{al}\ suggested that computational complexity
could also be pressed into such an explanatory role. \ In particular, they
showed that under almost any nonlinear variant of quantum mechanics, one could
build a \textquotedblleft nonlinear quantum computer\textquotedblright\ able
to solve $\mathsf{NP}$-complete and even $\mathsf{\#P}$-complete problems in
polynomial time.\footnote{A caveat is that it remains an open problem whether
this can be done fault-tolerantly. \ The answer might depend on the allowed
types of nonlinear gate. \ On the other hand, if arbitrary $1$-qubit nonlinear
gates can be implemented without error, then even $\mathsf{PSPACE}$-complete
problems can be solved in polynomial time. \ This is tight, since nonlinear
quantum computers can also be simulated in $\mathsf{PSPACE}$. \ I will give
more details in a forthcoming survey paper \cite{aar:np}.} \ One
interpretation of their result is that we should look very hard for
nonlinearities in experiments! \ But a different interpretation, the one I
prefer, is that their result provides independent evidence that quantum
mechanics is linear.\footnote{Note that I would \textit{not} advocate this
interpretation if it was merely (say) graph isomorphism that was efficiently
solvable in nonlinear quantum mechanics, just as I do not take Shor's
factoring algorithm as evidence for the falsehood of ordinary quantum
mechanics. \ I will explain in \cite{aar:np} why I think this distinction,
between $\mathsf{NP}$-complete\ problems and specific $\mathsf{NP}%
$-intermediate problems, is a justified one.}

In this section I build on Theorem \ref{postbqppp}\ to offer similar
\textquotedblleft evidence\textquotedblright\ that quantum mechanics is
unitary, and that the measurement rule is $\left\vert \psi\right\vert ^{2}$.

Let $\mathsf{BQP}_{\text{\textsf{nu}}}$\ be the class of problems solvable by
a uniform family of polynomial-size, bounded-error quantum circuits, where the
circuits can consist of arbitrary $1$- and $2$-qubit \textit{invertible}
linear transformations, rather than just unitary transformations.
\ Immediately before a measurement, the amplitude $\alpha_{x}$\ of each basis
state $\left\vert x\right\rangle $\ is divided by $\sqrt{\sum_{y}\left\vert
\alpha_{y}\right\vert ^{2}}$ to normalize it.

\begin{proposition}
\label{nuglobal}$\mathsf{BQP}_{\text{\textsf{nu}}}=\mathsf{PP}$.
\end{proposition}

\begin{proof}
The inclusion $\mathsf{BQP}_{\text{\textsf{nu}}}\subseteq\mathsf{PP}$\ follows
easily from Adleman, DeMarrais, and Huang's proof that $\mathsf{BQP}%
\subseteq\mathsf{PP}$ \cite{adh}, which does not depend on unitarity. \ For
the other direction, by Theorem \ref{postbqppp}\ it suffices to show that
$\mathsf{PostBQP}\subseteq\mathsf{BQP}_{\text{\textsf{nu}}}$. \ To postselect
on a qubit being $\left\vert 1\right\rangle $, we simply apply the $1$-qubit
nonunitary operation%
\[
\left(
\begin{array}
[c]{cc}%
2^{-q\left(  n\right)  } & 0\\
0 & 1
\end{array}
\right)
\]
for some sufficiently large polynomial $q$.
\end{proof}

Next, for any nonnegative real number $p$, define $\mathsf{BQP}_{p}%
$\ similarly to $\mathsf{BQP}$, except that when we measure, the probability
of obtaining a basis state $\left\vert x\right\rangle $ equals $\left\vert
\alpha_{x}\right\vert ^{p}/\sum_{y}\left\vert \alpha_{y}\right\vert ^{p}$
rather than $\left\vert \alpha_{x}\right\vert ^{2}$. \ Thus $\mathsf{BQP}%
_{2}=\mathsf{BQP}$. \ Assume that all gates are unitary and that there are no
intermediate measurements, just a single standard-basis measurement at the end.

\begin{theorem}
\label{bqpp}$\mathsf{PP}\subseteq\mathsf{BQP}_{p}\subseteq\mathsf{PSPACE}$ for all
constants $p\neq2$, with $\mathsf{BQP}_{p}%
=\mathsf{PP}$\ when $p\in\left\{  4,6,8,\ldots\right\}  $.
\end{theorem}

\begin{proof}
To simulate $\mathsf{PP}$ in $\mathsf{BQP}_{p}$, run the algorithm of Theorem
\ref{postbqppp},\ having initialized $O\left(  \frac{1}{\left\vert
p-2\right\vert }\operatorname*{poly}\left(  n\right)  \right)  $ ancilla
qubits to $\left\vert 0\right\rangle $. \ Suppose the algorithm's state at
some point is $\sum_{x}\alpha_{x}\left\vert x\right\rangle $, and we want to
postselect on the event $\left\vert x\right\rangle \in S$, where $S$ is a
subset of basis states. \ Here is how: if $p<2$, then\ for some sufficiently
large polynomial $q$, apply Hadamard gates to $c=2q\left(  n\right)  /\left(
2-p\right)  $\ fresh ancilla qubits conditioned on $\left\vert x\right\rangle
\in S$. \ The result is to increase the \textquotedblleft probability
mass\textquotedblright\ of each $\left\vert x\right\rangle \in S$\ from
$\left\vert \alpha_{x}\right\vert ^{p}$\ to%
\[
2^{c}\cdot\left\vert 2^{-c/2}\alpha_{x}\right\vert ^{p}=2^{\left(  2-p\right)
c/2}\left\vert \alpha_{x}\right\vert ^{p}=2^{q\left(  n\right)  }\left\vert
\alpha_{x}\right\vert ^{p},
\]
while the probability mass of each $\left\vert x\right\rangle \notin
S$\ remains unchanged. \ Similarly, if $p>2$, then apply Hadamard gates to
$c=2q\left(  n\right)  /\left(  p-2\right)  $\ fresh ancilla qubits
conditioned on $\left\vert x\right\rangle \notin S$. \ This decreases the
probability mass of each $\left\vert x\right\rangle \notin S$\ from
$\left\vert \alpha_{x}\right\vert ^{p}$\ to $2^{c}\cdot\left\vert
2^{-c/2}\alpha_{x}\right\vert ^{p}=2^{-q\left(  n\right)  }\left\vert
\alpha_{x}\right\vert ^{p}$,\ while the probability mass of each $\left\vert
x\right\rangle \in S$\ remains unchanged. \ The final observation is that
Theorem \ref{postbqppp}\ still goes through if $p\neq2$. \ For it suffices to
distinguish the case$\ \left\vert \left\langle +|\varphi_{2^{i}}\right\rangle
\right\vert >0.985$ from $\left\vert \left\langle +|\varphi_{2^{i}%
}\right\rangle \right\vert \leq1/\sqrt{2}$\ with exponentially small
probability of error, using polynomially many copies of the state $\left\vert
\varphi_{2^{i}}\right\rangle $. \ But we can do this for any $p$,\ since all
$\left\vert \psi\right\vert ^{p}$\ rules behave well under tensor products (in
the sense that $\left\vert \alpha\beta\right\vert ^{p}=\left\vert
\alpha\right\vert ^{p}\left\vert \beta\right\vert ^{p}$).

The inclusion $\mathsf{BQP}_{p}\subseteq\mathsf{PSPACE}$\ follows
easily from the techniques used by Bernstein and Vazirani \cite{bv}\
to show $\mathsf{BQP}\subseteq\mathsf{PSPACE}$. \ Let $S$ be the set
of accepting states; then simply compute $\sum_{x\in S}\left\vert
\alpha _{x}\right\vert ^{p}$\ and $\sum_{x\notin S}\left\vert
\alpha_{x}\right\vert ^{p}$\ and see which is greater.

To simulate $\mathsf{BQP}_{p}$ in $\mathsf{PP}$ when $p\in\left\{
4,6,8,\ldots\right\}  $, we generalize the technique of Adleman, DeMarrais,
and Huang \cite{adh}, which handled the case $p=2$. \ As in Theorem
\ref{qpoly} in Chapter \ref{ADV}, assume that all gates are Hadamard or
Toffoli gates; then we can write each amplitude $\alpha_{x}$ as a sum of
exponentially many contributions, $a_{x,1}+\cdots+a_{x,N}$, where each
$a_{x,i}$\ is a rational real number computable in classical polynomial
time.\ \ Then letting $S$ be the set of accepting states, it suffices to test
whether%
\begin{align*}
\sum_{x\in S}\left\vert \alpha_{x}\right\vert ^{p}  &  =\sum_{x\in S}%
\alpha_{x}^{p}\\
&  =\sum_{x\in S}\left(  \sum_{i\in\left\{  1,\ldots,N\right\}  }%
a_{x,i}\right)  ^{p}\\
&  =\sum_{x\in S}\sum_{B\subseteq\left\{  1,\ldots,N\right\}  ,\left\vert
B\right\vert =p}%
%TCIMACRO{\dprod \limits_{i\in B}}%
%BeginExpansion
{\displaystyle\prod\limits_{i\in B}}
%EndExpansion
a_{x,i}%
\end{align*}
is greater than $\sum_{x\notin S}\left\vert \alpha_{x}\right\vert ^{p}$, which
we can do in $\mathsf{PP}$.
\end{proof}

\section{Open Problems\label{OPENPOST}}

The new proof that $\mathsf{PP}$\ is closed under intersection came as a total
surprise to me. \ But on reflection, it goes a long way toward convincing me
of a thesis expressed in Chapter \ref{PROLOGUE}: that quantum computing offers
a new perspective from which to revisit the central questions of classical
complexity theory. \ What other classical complexity classes can we
characterize in quantum terms, and what other questions can we answer by that means?

A first step might be to prove even stronger closure properties for
$\mathsf{PP}$. \ Recall from Proposition \ref{closure} that $\mathsf{PostBQP}%
$\ is closed under polynomial-time truth-table reductions. \ Presumably this
can't be generalized to closure under Turing reductions, since if it could
then we would have $\mathsf{PP}=\mathsf{P}^{\mathsf{PP}}$, which is considered
unlikely.\footnote{Indeed, Beigel \cite{beigel}\ gave an oracle relative to
which $\mathsf{P}^{\mathsf{NP}}\not \subset \mathsf{PP}$.} \ But can we show
closure under nonadaptive \textit{quantum} reductions? \ More formally, let
$\mathsf{BQP}_{\mathsf{\Vert}}^{\mathsf{PostBQP}}$\ be the class of problems
solvable by a $\mathsf{BQP}$\ machine\ that can make a single quantum query,
which consists of a list of polynomially many questions for a
$\mathsf{PostBQP}$ oracle. \ Then does $\mathsf{BQP}_{\mathsf{\Vert}%
}^{\mathsf{PostBQP}}$ equal $\mathsf{PostBQP}$? \ The difficulty in showing
this seems to be uncomputing garbage after the $\mathsf{PostBQP}$\ oracle is simulated.

As for fantasy quantum mechanics, an interesting open question is whether
$\mathsf{BQP}_{p}=\mathsf{PP}$ for all nonnegative real numbers $p\neq2$. \ An
obvious idea for simulating $\mathsf{BQP}_{p}$\ in $\mathsf{PP}$ would be to
use a Taylor series expansion for the probability masses $\left\vert
\alpha_{x}\right\vert ^{p}$. \ Unfortunately, I have no idea how to get fast
enough convergence.

\chapter{The Power of History\label{QCHV}}

Quantum mechanics lets us calculate the probability that (say) an electron
will be found in an excited state if measured at a particular time. \ But it
is silent about \textit{multiple-time} or \textit{transition} probabilities:
that is, what is the probability that the electron will be in an excited state
at time $t_{1}$, given that it was in its ground state at an earlier time
$t_{0}$?\ \ The usual response is that this question is meaningless, unless of
course the electron was \textit{measured} (or otherwise known with probability
$1$) to be in its ground state at $t_{0}$. \ A different response---pursued by
Schr\"{o}dinger \cite{schrodinger}, Bohm \cite{bohm}, Bell \cite{bell}, Nelson
\cite{nelson}, Dieks \cite{dieks}, and others---treats the question as
provisionally meaningful, and then investigates how one might answer it
mathematically. \ Specific attempts at answers are called \textquotedblleft
hidden-variable theories.\textquotedblright

The appeal of hidden-variable theories is that they provide one possible
solution to the measurement problem. \ For they allow us to apply unitary
quantum mechanics to the entire universe (including ourselves), yet still
discuss the probability of a future observation conditioned on our current
observations. \ Furthermore, they let us do so without making any assumptions
about decoherence or the nature of observers. \ For example, even if an
observer were placed in coherent superposition, that observer would still have
a sequence of definite experiences, and the probability of any such sequence
could be calculated.

This chapter initiates the study of hidden variables from a quantum computing
perspective. \ I restrict attention to the simplest possible setting: that of
discrete time, a finite-dimensional Hilbert space, and a fixed orthogonal
basis. \ Within this setting, I reformulate known hidden-variable theories due
to Dieks \cite{dieks} and Schr\"{o}dinger \cite{schrodinger}, and also
introduce a new theory based on network flows. \ However, a more important
contribution is the \textit{axiomatic approach} that I use. \ I propose five
axioms for hidden-variable theories, and then compare theories against each
other based on which of the axioms they satisfy. \ A central question in this
approach is which subsets of axioms can be satisfied simultaneously.

In a second part of the chapter, I make the connection to quantum computing
explicit, by studying the computational complexity of simulating
hidden-variable theories. \ Below I describe the computational results.

\section{The Complexity of Sampling Histories\label{SAMPHIST}}

It is often stressed that hidden-variable theories yield exactly the same
predictions as ordinary quantum mechanics. \ On the other hand, these theories
describe a different picture of physical reality, with an additional layer of
dynamics beyond that of a state vector evolving unitarily. \ I address a
question that, to my knowledge, had never been raised before: \textit{what is
the computational complexity of simulating that additional dynamics?} \ In
other words, if we could examine a hidden variable's entire history, then
could we solve problems in polynomial time that are intractable even for
quantum computers?

I present strong evidence that the answer is yes. \ The Graph Isomorphism
problem asks whether two graphs $G$ and $H$ are isomorphic; while given a
basis for a lattice $\mathcal{L}\in\mathbb{R}^{n}$, the Approximate Shortest
Vector problem asks for a nonzero vector in $\mathcal{L}$ within a $\sqrt{n}%
$\ factor of the shortest one. \ I show that both problems are
efficiently solvable by sampling a hidden variable's history,
provided the hidden-variable theory satisfies a reasonable axiom.\ \
By contrast, despite a decade of effort, neither problem is known to
lie in $\mathsf{BQP}$. \ Thus, if we let $\mathsf{DQP}$\ (Dynamical
Quantum Polynomial-Time) be the class of problems solvable in the
new model, then this already provides circumstantial evidence that
$\mathsf{BQP}$\ is strictly contained in $\mathsf{DQP}$.

However, the evidence is stronger than this. \ For I actually show that
$\mathsf{DQP}$\ contains the entire class Statistical Zero Knowledge, or
$\mathsf{SZK}$. \ Furthermore, Chapter \ref{COL} showed that relative to an
oracle, $\mathsf{SZK}$\ is not contained in $\mathsf{BQP}$. \ Combining the
result that $\mathsf{SZK}\subseteq\mathsf{DQP}$ with the oracle separation of
Chapter \ref{COL}, one obtains that $\mathsf{BQP}\neq\mathsf{DQP}$\ relative
to an oracle as well.

Besides solving $\mathsf{SZK}$\ problems, I also show that by sampling
histories, one could search an unordered database of\ $N$\ items for a single
\textquotedblleft marked item\textquotedblright\ using only $O\left(
N^{1/3}\right)  $\ database queries. \ By comparison, Grover's quantum search
algorithm \cite{grover}\ requires $\Theta\left(  N^{1/2}\right)  $ queries,
while classical algorithms require $\Theta\left(  N\right)  $ queries. \ On
the other hand, I also show that the $N^{1/3}$\ upper bound is the best
possible---so even in the histories model, one cannot search an $N$-item
database in $\left(  \log N\right)  ^{c}$\ steps for some fixed power $c$.
\ This implies that $\mathsf{NP}\not \subset \mathsf{DQP}$ relative to an
oracle, which in turn suggests that $\mathsf{DQP}$ is \textit{still} not
powerful enough to solve $\mathsf{NP}$-complete problems in polynomial time.

At this point I should address a concern that many readers will have. \ Once
we extend quantum mechanics by positing the \textquotedblleft
unphysical\textquotedblright\ ability to sample histories, isn't it completely
unsurprising if we can then solve problems that were previously intractable?
\ I believe the answer is no, for three reasons.

First, almost every change that makes the quantum computing model more
powerful, seems to make it \textit{so much} more powerful that $\mathsf{NP}%
$-complete\ and even harder problems become solvable efficiently. \ To give
some examples, $\mathsf{NP}$-complete\ problems can be solved in polynomial
time using a nonlinear Schr\"{o}dinger equation, as shown by Abrams and Lloyd
\cite{al}; using closed timelike curves, as shown by Brun \cite{brun}\ and
Bacon \cite{bacon} (and conjectured by Deutsch \cite{deutsch:tt}); or using a
measurement rule of the form $\left\vert \psi\right\vert ^{p}$\ for any
$p\neq2$,\ as shown in Chapter \ref{POST}. \ It is also easy to see that we
could solve $\mathsf{NP}$-complete\ problems\ if, given a quantum state
$\left\vert \psi\right\rangle $, we could request a classical description of
$\left\vert \psi\right\rangle $, such as a list of amplitudes or a preparation
procedure.\footnote{For as Abrams and Lloyd \cite{al}\ observed, we can so
arrange things that $\left\vert \psi\right\rangle =\left\vert 0\right\rangle
$\ if an $\mathsf{NP}$-complete instance of interest to us has no solution,
but\ $\left\vert \psi\right\rangle =\sqrt{1-\varepsilon}\left\vert
0\right\rangle +\sqrt{\varepsilon}\left\vert 1\right\rangle $\ for some tiny
$\varepsilon$\ if it has a solution.} \ By contrast, the $\mathsf{DQP}$ model
is the first independently motivated model I know of that seems more powerful
than quantum computing, but only \textit{slightly} so.\footnote{One can define
other, less motivated, models with the same property by allowing
\textquotedblleft non-collapsing measurements\textquotedblright\ of quantum
states,\ but these models are very closely related to $\mathsf{DQP}$.
\ Indeed, a key ingredient in the results of this chapter will be to show that
certain kinds of non-collapsing measurements can be \textit{simulated} using
histories.} \ Moreover, the striking fact that unordered search takes about
$N^{1/3}$\ steps in the $\mathsf{DQP}$ model, as compared to $N$ steps
classically and $N^{1/2}$\ quantum-mechanically, suggests that $\mathsf{DQP}$
somehow \textquotedblleft continues a sequence\textquotedblright\ that begins
with $\mathsf{P}$\ and $\mathsf{BQP}$. \ It would be interesting to find a
model in which search takes $N^{1/4}$\ or $N^{1/5}$\ steps.

The second reason the results are surprising is that, given a hidden variable,
the distribution over its possible values at any \textit{single} time is
governed by standard quantum mechanics, and is therefore efficiently samplable
on a quantum computer. \ So if examining the variable's history confers any
extra computational power, then it can only be because of
\textit{correlations} between the variable's values at different times.

The third reason is the criterion for success. \ I am not saying merely that
one can solve Graph Isomorphism under \textit{some} hidden-variable theory; or
even that, under any theory satisfying the indifference axiom, there exists an
algorithm to solve it; but rather that there exists a \textit{single}
algorithm that solves Graph Isomorphism under any theory satisfying
indifference. \ Thus, we must consider even theories that are specifically
designed to thwart such an algorithm.

But what is the motivation for these results? \ The first motivation is that,
within the community of physicists who study hidden-variable theories such as
Bohmian mechanics, there is great interest in actually \textit{calculating}
the hidden-variable trajectories for specific physical systems \cite{pdh,gm}.
\ My results show that, when many interacting particles are involved, this
task might be fundamentally intractable, even if a quantum computer were
available. \ The second motivation is that, in classical computer science,
studying \textquotedblleft unrealistic\textquotedblright\ models of
computation has often led to new insights into realistic ones;\ and likewise I
expect that the $\mathsf{DQP}$ model\ could lead to new results
about\ standard quantum computation. \ Indeed, in a sense this has already
happened---for the collision lower bound of Chapter \ref{COL} grew out of work
on the $\mathsf{BQP}$ versus $\mathsf{DQP}$ question.

\section{Outline of Chapter\label{OUTLINE}}

Sections \ref{HV} through \ref{SCHROD} develop the axiomatic approach to
hidden variables; then Sections \ref{MODEL}\ through \ref{SEARCH}\ study the
computational complexity of sampling hidden-variable histories.

Section \ref{HV} formally defines hidden-variable theories in my sense; then
Section \ref{BOHM} contrasts these theories with related ideas such as Bohmian
mechanics and modal interpretations. \ Section \ref{OBJECTIONS} addresses the
most common objections to my approach: for example, that the implicit
dependence on a fixed basis is unacceptable.

In Section \ref{AXIOMS}, I introduce five possible axioms for hidden-variable
theories. \ These are indifference to the identity operation; robustness to
small perturbations; commutativity with respect to spacelike-separated
unitaries; commutativity for the special case of product states; and
invariance under decomposition of mixed states into pure states. \ Ideally, a
theory would satisfy all of these axioms. \ However, I show in Section
\ref{IMPOS}\ that no theory satisfies both indifference and commutativity; no
theory satisfies both indifference and a stronger version of robustness; no
theory satisfies indifference, robustness, and decomposition invariance; and
no theory satisfies a stronger version of decomposition invariance.

In Section \ref{SPECIFIC}\ I shift from negative to positive results.
\ Section \ref{FLOW}\ presents a hidden-variable theory called the
\textit{flow theory} or $\mathcal{FT}$, which is based on the Max-Flow-Min-Cut
theorem from combinatorial optimization. \ The idea is to define a network of
\textquotedblleft pipes\textquotedblright\ from basis states at an initial
time to basis states at a final time, and then route as much probability mass
as possible through these pipes. \ The capacity of each pipe depends on the
corresponding entry of the unitary acting from the initial to final time. \ To
find the probability of transitioning from basis state $\left\vert
i\right\rangle $ to basis state $\left\vert j\right\rangle $, we then
determine how much of the flow originating at $\left\vert i\right\rangle $\ is
routed along the pipe to $\left\vert j\right\rangle $. \ The main results are
that $\mathcal{FT}$ is well-defined and that it is robust to small
perturbations. \ Since $\mathcal{FT}$ trivially satisfies the indifference
axiom, this implies that the indifference and robustness axioms can be
satisfied simultaneously, which was not at all obvious\ \textit{a priori}.

Section \ref{SCHROD}\ presents a second theory that I call the
\textit{Schr\"{o}dinger theory} or $\mathcal{ST}$, since it is based on a pair
of integral equations introduced in a 1931 paper of Schr\"{o}dinger
\cite{schrodinger}. \ Schr\"{o}dinger conjectured, but was unable to prove,
the existence and uniqueness of a solution to these equations; the problem was
not settled until the work of Nagasawa \cite{nagasawa}\ in the 1980's. \ In
the discrete setting the problem is simpler, and I give a self-contained proof
of existence using a matrix scaling technique due to Sinkhorn \cite{sinkhorn}.
\ The idea is as follows: we want to convert a unitary matrix that maps one
quantum state to another, into a nonnegative matrix whose $i^{th}$\ column
sums to the initial probability of basis state $\left\vert i\right\rangle $,
and whose $j^{th}$\ row sums to the final probability of basis state
$\left\vert j\right\rangle $. \ To do so, we first replace each entry of the
unitary matrix by its absolute value, then normalize each column to sum to the
desired initial probability, then normalize each row to sum to the desired
final probability. \ But then the columns are no longer normalized correctly,
so we normalize them \textit{again}, then normalize the rows again, and so on.
\ I show that this iterative process converges, from which it follows that
$\mathcal{ST}$ is well-defined. \ I also observe that $\mathcal{ST}%
$\ satisfies the indifference and product commutativity axioms, and violates
the decomposition invariance axiom. \ I conjecture that $\mathcal{ST}$
satisfies the robustness axiom; proving that conjecture is one of the main
open problems of the chapter.

In Section \ref{MODEL} I shift attention to the complexity of sampling
histories. \ I formally define $\mathsf{DQP}$\ as the class of problems
solvable by a classical polynomial-time algorithm with access to a
\textquotedblleft history oracle.\textquotedblright\ \ Given a sequence of
quantum circuits as input, this oracle returns a sample from a corresponding
distribution over histories of a hidden variable, according to some
hidden-variable theory $\mathcal{T}$. \ The oracle can choose $\mathcal{T}%
$\ \textquotedblleft adversarially,\textquotedblright\ subject to the
constraint that $\mathcal{T}$\ satisfies the indifference and robustness
axioms. \ Thus, a key result from Section \ref{MODEL}\ that I rely on is that
there \textit{exists} a hidden-variable theory satisfying indifference and robustness.

Section \ref{RESULTS}\ establishes the most basic facts about $\mathsf{DQP}$:
for example, that $\mathsf{BQP}\subseteq\mathsf{DQP}$, and that $\mathsf{DQP}%
$\ is independent of the choice of gate set. \ Then Section \ref{JUGGLE}%
\ presents the \textquotedblleft juggle subroutine,\textquotedblright\ a
crucial ingredient in both of the main hidden-variable algorithms. \ Given a
state of the form $\left(  \left\vert a\right\rangle +\left\vert
b\right\rangle \right)  /\sqrt{2}$ or $\left(  \left\vert a\right\rangle
-\left\vert b\right\rangle \right)  /\sqrt{2}$, the goal of this subroutine is
to \textquotedblleft juggle\textquotedblright\ a hidden variable between
$\left\vert a\right\rangle $\ and $\left\vert b\right\rangle $, so that when
we inspect the hidden variable's history, both $\left\vert a\right\rangle
$\ and $\left\vert b\right\rangle $ are observed with high probability. \ The
difficulty is that this needs to work under \textit{any} indifferent
hidden-variable theory.

Next, Section \ref{SZK}\ combines the juggle subroutine with a technique of
Valiant and Vazirani \cite{vv}\ to prove that $\mathsf{SZK}\subseteq
\mathsf{DQP}$, from which it follows in particular that Graph Isomorphism\ and
Approximate Shortest Vector\ are in\ $\mathsf{DQP}$. \ Then Section
\ref{SEARCH}\ applies the juggle subroutine to search an $N$-item database in
$O\left(  N^{1/3}\right)  $\ queries, and also proves that this $N^{1/3}%
$\ bound is optimal.

I conclude in Section \ref{DISC}\ with some directions for further research.

\section{Hidden-Variable Theories\label{HV}}

Suppose we have an $N\times N$\ unitary matrix $U$, acting on a state%
\[
\left\vert \psi\right\rangle =\alpha_{1}\left\vert 1\right\rangle
+\cdots+\alpha_{N}\left\vert N\right\rangle ,
\]
where $\left\vert 1\right\rangle ,\ldots,\left\vert N\right\rangle $\ is a
standard orthogonal basis. \ Let%
\[
U\left\vert \psi\right\rangle =\beta_{1}\left\vert 1\right\rangle
+\cdots+\beta_{N}\left\vert N\right\rangle .
\]
Then can we construct a stochastic matrix $S$, which maps the vector of
probabilities%
\[
\overrightarrow{p}=\left[
\begin{array}
[c]{c}%
\left\vert \alpha_{1}\right\vert ^{2}\\
\vdots\\
\left\vert \alpha_{N}\right\vert ^{2}%
\end{array}
\right]
\]
induced by measuring $\left\vert \psi\right\rangle $, to the vector%
\[
\overrightarrow{q}=\left[
\begin{array}
[c]{c}%
\left\vert \beta_{1}\right\vert ^{2}\\
\vdots\\
\left\vert \beta_{N}\right\vert ^{2}%
\end{array}
\right]
\]
induced by measuring $U\left\vert \psi\right\rangle $? \ Trivially yes. \ The
following matrix maps \textit{any} vector of probabilities to $\overrightarrow
{q}$, ignoring the input vector $\overrightarrow{p}$\ entirely:
\[
S_{\mathcal{PT}}=\left[
\begin{array}
[c]{ccc}%
\left\vert \beta_{1}\right\vert ^{2} & \cdots & \left\vert \beta
_{1}\right\vert ^{2}\\
\vdots &  & \vdots\\
\left\vert \beta_{N}\right\vert ^{2} & \cdots & \left\vert \beta
_{N}\right\vert ^{2}%
\end{array}
\right]  .
\]
Here $\mathcal{PT}$\ stands for \textit{product theory}. \ The product theory
corresponds to a strange picture of physical reality, in which memories and
records are completely unreliable, there being no causal connection between
states of affairs at earlier and later times.

So we would like $S$ to depend on $U$\ itself somehow, not just on $\left\vert
\psi\right\rangle $\ and $U\left\vert \psi\right\rangle $. \ Indeed, ideally
$S$\ would be a function \textit{only} of $U$, and not of $\left\vert
\psi\right\rangle $. \ But this is impossible, as the following example shows.
\ Let $U$ be a $\pi/4$\ rotation, and let $\left\vert +\right\rangle =\left(
\left\vert 0\right\rangle +\left\vert 1\right\rangle \right)  /\sqrt{2}$ and
$\left\vert -\right\rangle =\left(  \left\vert 0\right\rangle -\left\vert
1\right\rangle \right)  /\sqrt{2}$. \ Then $U\left\vert +\right\rangle
=\left\vert 1\right\rangle $\ implies that%
\[
S\left(  \left\vert +\right\rangle ,U\right)  =\left[
\begin{array}
[c]{cc}%
0 & 0\\
1 & 1
\end{array}
\right]  ,
\]
whereas $U\left\vert -\right\rangle =\left\vert 0\right\rangle $\ implies that%
\[
S\left(  \left\vert -\right\rangle ,U\right)  =\left[
\begin{array}
[c]{cc}%
1 & 1\\
0 & 0
\end{array}
\right]  .
\]

On the other hand, it is easy to see that, if $S$ can depend on $\left\vert
\psi\right\rangle $ as well as $U$, then there are infinitely many choices for
the function $S\left(  \left\vert \psi\right\rangle ,U\right)  $. \ Every
choice reproduces the predictions of quantum mechanics perfectly when
restricted to single-time probabilities. \ So how can we possibly choose among
them? \ My approach in Sections \ref{AXIOMS}\ and \ref{SPECIFIC} will be to
write down axioms that we would like $S$ to satisfy, and then investigate
which of the axioms can be satisfied simultaneously.

Formally, a \textit{hidden-variable theory} is a family of functions $\left\{
S_{N}\right\}  _{N\geq1}$, where each $S_{N}$\ maps an $N$-dimensional mixed
state $\rho$\ and an $N\times N$\ \ unitary matrix $U$\ onto a singly
stochastic matrix $S_{N}\left(  \rho,U\right)  $. \ I will often suppress the
dependence on $N$, $\rho$, and $U$, and occasionally use subscripts such as
$\mathcal{PT}$\ or $\mathcal{FT}$\ to indicate the theory in question. \ Also,
if\ $\rho=\left\vert \psi\right\rangle \left\langle \psi\right\vert $\ is a
pure state I may write $S\left(  \left\vert \psi\right\rangle ,U\right)
$\ instead of $S\left(  \left\vert \psi\right\rangle \left\langle
\psi\right\vert ,U\right)  $.

Let $\left(  M\right)  _{ij}$\ denote the entry in the $i^{th}$\ column and
$j^{th}$\ row of matrix $M$. \ Then $\left(  S\right)  _{ij}$ is the
probability that the hidden variable takes value $\left\vert j\right\rangle $
after $U$\ is applied, conditioned on it taking value $\left\vert
i\right\rangle $\ before $U$ is applied. \ At a minimum, any theory must
satisfy the following marginalization axiom: for all $j\in\left\{
1,\ldots,N\right\}  $,\vspace{0pt}%
\[
\sum_{i}\left(  S\right)  _{ij}\left(  \rho\right)  _{ii}=\left(  U\rho
U^{-1}\right)  _{jj}\text{.}%
\]
This says that after $U$\ is applied, the hidden variable takes value
$\left\vert j\right\rangle $\ with probability $\left(  U\rho U^{-1}\right)
_{jj}$, which is the usual Born probability.

Often it will be convenient to refer, not to $S$ itself, but to the matrix
$P\left(  \rho,U\right)  $ of joint probabilities\ whose $\left(  i,j\right)
$\ entry is $\left(  P\right)  _{ij}=\left(  S\right)  _{ij}\left(
\rho\right)  _{ii}$. \ The $i^{th}$\ column of $P$\ must sum to $\left(
\rho\right)  _{ii}$, and the $j^{th}$\ row must sum to $\left(  U\rho
U^{-1}\right)  _{jj}$. \ Indeed, I will define the theories $\mathcal{FT}%
$\ and $\mathcal{ST}$\ by first specifying the matrix $P$, and then setting
$\left(  S\right)  _{ij}:=\left(  P\right)  _{ij}/\left(  \rho\right)  _{ii}$.
\ This approach has the drawback that if $\left(  \rho\right)  _{ii}=0$, then
the $i^{th}$\ column of $S$ is undefined. \ To get around this, I adopt the
convention that%
\[
S\left(  \rho,U\right)  :=\lim_{\varepsilon\rightarrow0^{+}}S\left(
\rho_{\varepsilon},U\right)
\]
where $\rho_{\varepsilon}=\left(  1-\varepsilon\right)  \rho+\varepsilon
I$\ and $I$ is the $N\times N$\ maximally mixed state. \ Technically, the
limits%
\[
\lim_{\varepsilon\rightarrow0^{+}}\frac{\left(  P\left(  \rho_{\varepsilon
},U\right)  \right)  _{ij}}{\left(  \rho_{\varepsilon}\right)  _{ii}}%
\]
might not exist, but in the cases of interest it will be obvious that they do.

\subsection{Comparison with Previous Work\label{BOHM}}

Before going further, I should contrast my approach with previous approaches
to hidden variables, the most famous of which is Bohmian mechanics
\cite{bohm}. \ My main difficulty with Bohmian mechanics is that it commits
itself to a Hilbert space of particle positions and momenta. \ Furthermore, it
is crucial that the positions and momenta be \textit{continuous}, in order for
particles to evolve deterministically. \ To see this, let $\left\vert
L\right\rangle $\ and $\left\vert R\right\rangle $\ be discrete positions, and
suppose a particle is in state $\left\vert L\right\rangle $\ at time $t_{0}$,
and state $\left(  \left\vert L\right\rangle +\left\vert R\right\rangle
\right)  /\sqrt{2}$ at a later time $t_{1}$. \ Then a hidden variable
representing the position would have entropy $0$ at $t_{1}$, since it is
always $\left\vert L\right\rangle $ then; but entropy $1$\ at $t_{1}$, since
it is $\left\vert L\right\rangle $\ or $\left\vert R\right\rangle $\ both with
$1/2$ probability. \ Therefore the earlier value cannot determine the later
one.\footnote{Put differently, Bohm's conservation of probability result
breaks down because the \textquotedblleft wavefunctions\textquotedblright\ at
$t_{0}$\ and $t_{1}$ are degenerate, with all amplitude concentrated on
finitely many points. \ But in a discrete Hilbert space, \textit{every}
wavefunction is degenerate in this sense!} \ It follows that Bohmian mechanics
is incompatible with the belief that\ all physical observables are discrete.
\ But in my view, there are strong reasons to hold that belief, which include
black hole entropy bounds; the existence of a natural minimum length scale
($10^{-33}$ cm); results on area quantization in quantum gravity \cite{rs};
the fact that many physical quantities once thought to be continuous have
turned out to be discrete; the infinities of quantum field theory;\ the
implausibility of analog \textquotedblleft hypercomputers\textquotedblright;
and conceptual problems raised by the independence of the continuum hypothesis.

Of course there exist stochastic analogues of Bohmian mechanics, among them
Nelsonian mechanics \cite{nelson}\ and Bohm and Hiley's \textquotedblleft
stochastic interpretation\textquotedblright\ \cite{bh}. \ But it is not
obvious why we should prefer these to other stochastic hidden-variable
theories. \ From a quantum-information perspective, it is much more natural to
take an abstract approach---one that allows arbitrary finite-dimensional
Hilbert spaces, and that does not rule out any transition rule \textit{a
priori}.

Stochastic hidden variables have also been considered in the context of modal
interpretations; see Dickson \cite{dickson}, Bacciagaluppi and Dickson
\cite{bd},\ and Dieks \cite{dieks}\ for example. \ However, the central
assumptions in that work are extremely different from mine. \ In modal
interpretations, a pure state evolving unitarily poses no problems at all: one
simply rotates the hidden-variable basis along with the state, so that the
state always represents a \textquotedblleft possessed
property\textquotedblright\ of the system in the current basis. \ Difficulties
arise only for mixed states; and there, the goal is to track a whole set of
possessed properties. \ By contrast, my approach is to fix an orthogonal
basis, then track a single hidden variable that is an element of that basis.
\ The issues raised by pure states and mixed states are essentially the same.

Finally I should mention the consistent-histories interpretation of Griffiths
\cite{griffiths} and Gell-Mann and Hartle \cite{gh}. \ This interpretation
assigns probabilities to various histories through a quantum system, so long
as the \textquotedblleft interference\textquotedblright\ between those
histories is negligible. \ Loosely speaking, then, the situations where
consistent histories make sense are precisely the ones where the question of
transition probabilities can be avoided.

\subsection{Objections\label{OBJECTIONS}}

Hidden-variable theories, as I define them, are open to several technical
objections. \ For example, I required transition probabilities for only one
orthogonal observable. \ What about other observables? \ The problem is that,
according to the Kochen-Specker theorem,\ we cannot assign consistent values
to all observables at any \textit{single} time, let alone give transition
probabilities for those values. \ This is an issue in any setting, not just
mine. \ The solution I prefer is to postulate a fixed orthogonal basis of
\textquotedblleft distinguishable experiences,\textquotedblright\ and to
interpret a measurement in any other basis as a unitary followed by a
measurement in the fixed basis. \ As mentioned in Section \ref{BOHM}, modal
interpretations opt for a different solution, which involves sets of bases
that change over time with the state itself.

Another objection is that the probability of transitioning from basis state
$\left\vert i\right\rangle $\ at time $t_{1}$\ to basis state $\left\vert
j\right\rangle $\ at time $t_{2}$\ might depend on how finely\ we divide the
time interval between $t_{1}$\ and $t_{2}$. \ In other words, for some state
$\left\vert \psi\right\rangle $\ and unitaries $V,W$, we might have%
\[
S\left(  \left\vert \psi\right\rangle ,WV\right)  \neq S\left(  V\left\vert
\psi\right\rangle ,W\right)  S\left(  \left\vert \psi\right\rangle ,V\right)
\]
(a similar point was made by Gillespie \cite{gillespie}). \ Indeed, this is
true for any hidden-variable theory other than the product theory
$\mathcal{PT}$. \ To see this, observe that for all unitaries $U$ and states
$\left\vert \psi\right\rangle $, there exist unitaries $V,W$\ such that
$U=WV$\ and $V\left\vert \psi\right\rangle =\left\vert 1\right\rangle $.
\ Then applying $V$\ destroys all information in the hidden variable (that is,
decreases its entropy to $0$); so if we then apply $W$, then the variable's
final value must be uncorrelated with the initial value. \ In other words,
$S\left(  V\left\vert \psi\right\rangle ,W\right)  S\left(  \left\vert
\psi\right\rangle ,V\right)  $\ must equal $S_{\mathcal{PT}}\left(  \left\vert
\psi\right\rangle ,U\right)  $.\ \ It follows that to any hidden-variable
theory we must associate a time scale, or some other rule for deciding when
the transitions take place.

In response, it should be noted that exactly the same problem arises in
\textit{continuous}-time stochastic hidden-variable theories. \ For if a state
$\left\vert \psi\right\rangle $\ is governed by the Schr\"{o}dinger equation
$d\left\vert \psi\right\rangle /dt=iH_{t}\left\vert \psi\right\rangle $, and a
hidden variable's probability distribution $\overrightarrow{p}$ is governed by
the stochastic equation $d\overrightarrow{p}/d\tau=A_{\tau}\overrightarrow{p}%
$, then there is still an arbitrary parameter $d\tau/dt$ on which the dynamics depend.

Finally, it will be objected that I have ignored special relativity. \ In
Section \ref{AXIOMS}\ I will define a \textit{commutativity axiom}, which
informally requires that the stochastic matrix $S$ not depend on the temporal
order of spacelike separated events. \ Unfortunately, we will see that when
entangled states are involved, commutativity is irreconcilable with another
axiom that seems even more basic. \ The resulting nonlocality has the same
character as the nonlocality of Bohmian mechanics---that is, one cannot use it
to send superluminal signals in the usual sense, but it is unsettling nonetheless.

\section{Axioms for Hidden-Variable Theories\label{AXIOMS}}

I now state five axioms that we might like hidden-variable theories to satisfy.

\textbf{Indifference.} \ The indifference axiom says that if $U$\ is
block-diagonal, then $S$ should also be block-diagonal with the same block
structure or some refinement thereof. Formally, let a \textit{block} be a
subset $B\subseteq\left\{  1,\ldots,N\right\}  $\ such that $\left(  U\right)
_{ij}=0$ for all $i\in B,j\notin B$ and $i\notin B,j\in B$. \ Then for all
blocks $B$, we should have $\left(  S\right)  _{ij}=0$ for all $i\in B,j\notin
B$ and $i\notin B,j\in B$. \ In particular, indifference implies that given
any state $\rho$\ in a tensor product space $\mathcal{H}_{A}\otimes
\mathcal{H}_{B}$, and any unitary $U$ that acts only on $\mathcal{H}_{A}%
$\ (that is, never maps a basis state $\left\vert i_{A}\right\rangle
\otimes\left\vert i_{B}\right\rangle $\ to $\left\vert j_{A}\right\rangle
\otimes\left\vert j_{B}\right\rangle $\ where $i_{B}\neq j_{B}$), the
stochastic matrix $S\left(  \rho,U\right)  $\ acts only on $\mathcal{H}_{A}%
$\ as well.

\textbf{Robustness.} \ A theory is robust if it is insensitive to small errors
in a state or unitary (which, in particular, implies continuity). \ Suppose we
obtain $\widetilde{\rho}$ and $\widetilde{U}$\ by perturbing $\rho$\ and
$U$\ respectively.\ \ Then for all polynomials $p$, there should exist a
polynomial $q$\ such that for all $N$,%
\[
\vspace{0pt}\left\Vert P\left(  \widetilde{\rho},\widetilde{U}\right)
-P\left(  \rho,U\right)  \right\Vert _{\infty}\leq\frac{1}{p\left(  N\right)
}%
\]
where $\left\Vert M\right\Vert _{\infty}=\max_{ij}\left\vert \left(  M\right)
_{ij}\right\vert $, whenever $\left\Vert \widetilde{\rho}-\rho\right\Vert
_{\infty}\leq1/q\left(  N\right)  $ and $\left\Vert \widetilde{U}-U\right\Vert
_{\infty}\leq1/q\left(  N\right)  $.\ \ Robustness has an important advantage
for quantum computing: if a hidden-variable theory is robust then the set of
gates used to define the unitaries $U_{1},\ldots,U_{T}$\ is irrelevant, since
by the Solovay-Kitaev Theorem (see \cite{kitaev:ec,nc}), any universal quantum
gate set can simulate any other to a precision $\varepsilon$\ with $O\left(
\log^{c}1/\varepsilon\right)  $\ overhead.

\textbf{Commutativity.} \ Let $\rho_{AB}$\ be a bipartite state, and let
$U_{A}$ and $U_{B}$\ act only on subsystems $A$ and $B$ respectively. \ Then
commutativity means that the order in which $U_{A}$ and $U_{B}$ are applied is
irrelevant:%
\[
S\left(  U_{A}\rho_{AB}U_{A}^{-1},U_{B}\right)  S\left(  \rho_{AB}%
,U_{A}\right)  \vspace{0pt}=S\left(  U_{B}\rho_{AB}U_{B}^{-1},U_{A}\right)
S\left(  \rho_{AB},U_{B}\right)  \text{.}%
\]

\textbf{Product Commutativity.} \ A theory is product commutative if it
satisfies commutativity for all separable pure states $\left\vert
\psi\right\rangle =\left\vert \psi_{A}\right\rangle \otimes\left\vert \psi
_{B}\right\rangle $.

\textbf{Decomposition Invariance.} \ A theory is decomposition invariant if%

\[
S\left(  \rho,U\right)  =\sum_{i=1}^{N}p_{i}S\left(  \left\vert \psi
_{i}\right\rangle \left\langle \psi_{i}\right\vert ,U\right)
\]
for every decomposition%
\[
\rho=\sum_{i=1}^{N}p_{i}\left\vert \psi_{i}\right\rangle \left\langle \psi
_{i}\right\vert
\]
of $\rho$\ into pure states. \ Theorem \ref{decomp}, part (ii) will show that
the analogous axiom for $P\left(  \rho,U\right)  $\ is unsatisfiable.

\subsection{Comparing Theories\label{COMPARE}}

To fix ideas, let us compare some hidden-variable theories with respect to the
above axioms. \ We have already seen the product theory $\mathcal{PT}$\ in
Section \ref{HV}. \ It is easy to show that $\mathcal{PT}$\ satisfies
robustness, commutativity, and decomposition invariance. \ However, I consider
$\mathcal{PT}$ unsatisfactory because it violates indifference: even if a
unitary $U$ acts only on the first of two qubits, $S_{\mathcal{PT}}\left(
\rho,U\right)  $\ will readily produce transitions involving the second qubit.

Recognizing this problem, Dieks \cite{dieks}\ proposed an alternative theory
that amounts to the following.\footnote{Dieks (personal communication) says he
would no longer defend this theory.} \ First partition the set of basis states
into minimal blocks $B_{1},\ldots,B_{m}$ between which $U$ never sends
amplitude. \ Then apply the product theory separately to each block; that is,
if $i$ and $j$ belong to the same block $B_{k}$\ then set%
\[
\left(  S\right)  _{ij}=\frac{\left(  U\rho U^{-1}\right)  _{jj}}%
{\sum_{\widehat{j}\in B_{k}}\left(  U\rho U^{-1}\right)  _{\widehat{j}%
\widehat{j}}},
\]
and otherwise set $\left(  S\right)  _{ij}=0$. \ The resulting \textit{Dieks
theory}, $\mathcal{DT}$,\ satisfies indifference by construction. \ However,
it does not satisfy robustness (or even continuity), since the set of blocks
can change if we replace `$0$' entries in $U$ by arbitrarily small nonzero entries.

In Section \ref{SPECIFIC}\ I will introduce two other
hidden-variable theories, the flow theory $\mathcal{FT}$\ and the
Schr\"{o}dinger theory $\mathcal{ST}$. \ Table 16.1 lists which
axioms the four theories
satisfy.%

\begin{table}
\begin{tabular}
[c]{lllll}
& $\mathcal{PT}$ (Product) & $\mathcal{DT}$ (Dieks) & $\mathcal{FT}$ (Flow) &
$\mathcal{ST}$ (Schr\"{o}dinger)\\
Indifference & No & Yes & Yes & Yes\\
Robustness & Yes & No & Yes & ?\\
Commutativity & Yes & No & No & No\\
Product Commutativity & Yes & Yes & No & Yes\\
Decomposition Invariance & Yes & Yes & No & No
\end{tabular}
\label{axiomtable} \caption[Four hidden-variable theories and the
axioms they satisfy]{Four hidden-variable theories and the axioms
they satisfy}
\end{table}

If we could prove that $\mathcal{ST}$\ satisfies robustness, then
Table 1 together with the impossibility results of Section
\ref{IMPOS}\ would completely characterize which of the axioms can
be satisfied simultaneously.

\section{Impossibility Results\label{IMPOS}}

This section shows that certain sets of axioms cannot be satisfied by any
hidden-variable theory. \ I first show that the failure of $\mathcal{DT}%
$,\ $\mathcal{FT}$, and $\mathcal{ST}$\ to satisfy commutativity is inherent,
and not a fixable technical problem.

\begin{theorem}
\label{nogo}No hidden-variable theory satisfies both indifference and commutativity.
\end{theorem}

\begin{proof}
Assume indifference holds, and let our initial state be $\left\vert
\psi\right\rangle = \frac {\left\vert 00\right\rangle +\left\vert
11\right\rangle }{\sqrt{2}}$. \ Suppose $U_{A}$\ applies a $\pi/8$
rotation to the first qubit, and $U_{B}$\ applies a\ $-\pi/8$\
rotation to the
second qubit. \ Then%
\begin{align*}
U_{A}\left\vert \psi\right\rangle  &  =U_{B}\left\vert \psi\right\rangle
=\frac{1}{\sqrt{2}}\left(  \cos\frac{\pi}{8}\left\vert 00\right\rangle
-\sin\frac{\pi}{8}\left\vert 01\right\rangle +\sin\frac{\pi}{8}\left\vert
10\right\rangle +\cos\frac{\pi}{8}\left\vert 11\right\rangle \right)  ,\\
U_{A}U_{B}\left\vert \psi\right\rangle  &  =U_{B}U_{A}\left\vert
\psi\right\rangle =\frac{1}{2}\left(  \left\vert 00\right\rangle -\left\vert
01\right\rangle +\left\vert 10\right\rangle +\left\vert 11\right\rangle
\right)  .
\end{align*}
Let $v_{t}$\ be the value of the hidden variable after $t$ unitaries have been
applied. \ Let $E$ be the event that $v_{0}=\left\vert 00\right\rangle
$\ initially, and $v_{2}=\left\vert 10\right\rangle $\ at the end.\ \ If
$U_{A}$\ is applied before $U_{B}$,\ then the unique `path' from $v_{0}$\ to
$v_{2}$\ consistent with indifference sets $v_{1}=\left\vert 10\right\rangle
$. \ So%
\[
\Pr\left[  E\right]  \leq\Pr\left[  v_{1}=\left\vert 10\right\rangle \right]
=\frac{1}{2}\sin^{2}\frac{\pi}{8}.
\]
But if $U_{B}$\ is applied before $U_{A}$, then the probability that
$v_{0}=\left\vert 11\right\rangle $\ and $v_{2}=\left\vert 10\right\rangle
$\ is at most $\frac{1}{2}\sin^{2}\frac{\pi}{8}$, by the same reasoning.
\ Thus, since $v_{2}$\ must equal $\left\vert 10\right\rangle $\ with
probability $1/4$, and since the only possibilities for $v_{0}$\ are
$\left\vert 00\right\rangle $\ and $\left\vert 11\right\rangle $,%
\[
\Pr\left[  E\right]  \geq\frac{1}{4}-\frac{1}{2}\sin^{2}\frac{\pi}{8}>\frac
{1}{2}\sin^{2}\frac{\pi}{8}.
\]
We conclude that commutativity is violated.
\end{proof}

Let me remark on the relationship between Theorem \ref{nogo}\ and Bell's
Theorem. \ Any hidden-variable theory that is \textquotedblleft
local\textquotedblright\ in Bell's sense would immediately satisfy both
indifference and commutativity. \ However, the converse is not obvious, since
there might be nonlocal information in the states $U_{A}\left\vert
\psi\right\rangle $ or $U_{B}\left\vert \psi\right\rangle $,\ which an
indifferent commutative theory could exploit but a local one could not.
\ Theorem \ref{nogo} rules out this possibility, and in that sense is a
strengthening of Bell's Theorem.

The next result places limits on decomposition invariance.

\begin{theorem}
\label{decomp}\quad

\begin{enumerate}
\item[(i)] No theory satisfies indifference, robustness, and decomposition invariance.

\item[(ii)] No theory has the property that%
\[
P\left(  \rho,U\right)  =\sum_{i=1}^{N}p_{i}P\left(  \left\vert \psi
_{i}\right\rangle \left\langle \psi_{i}\right\vert ,U\right)
\]
for every decomposition $\sum_{i=1}^{N}p_{i}\left\vert \psi_{i}\right\rangle
\left\langle \psi_{i}\right\vert $\ of $\rho$.
\end{enumerate}
\end{theorem}

\begin{proof}
\quad

\begin{enumerate}
\item[(i)] Suppose the contrary. \ Let%
\begin{align*}
R_{\theta}  &  =\left[
\begin{array}
[c]{cc}%
\cos\theta & -\sin\theta\\
\sin\theta & \cos\theta
\end{array}
\right]  ,\\
\left\vert \varphi_{\theta}\right\rangle  &  =\cos\theta\left\vert
0\right\rangle +\sin\theta\left\vert 1\right\rangle .
\end{align*}
Then for every $\theta$ not a multiple of $\pi/2$, we must have%
\begin{align*}
S\left(  \left\vert \varphi_{-\theta}\right\rangle ,R_{\theta}\right)   &
=\left[
\begin{array}
[c]{cc}%
1 & 1\\
0 & 0
\end{array}
\right]  ,\\
S\left(  \left\vert \varphi_{\pi/2-\theta}\right\rangle ,R_{\theta}\right)
&  =\left[
\begin{array}
[c]{cc}%
0 & 0\\
1 & 1
\end{array}
\right]  .
\end{align*}
So by decomposition invariance, letting $I=\left(  \left\vert 0\right\rangle
\left\langle 0\right\vert +\left\vert 1\right\rangle \left\langle 1\right\vert
\right)  /2$\ denote the maximally mixed state,%
\[
S\left(  I,R_{\theta}\right)  =S\left(  \frac{\left\vert \varphi_{-\theta
}\right\rangle \left\langle \varphi_{-\theta}\right\vert +\left\vert
\varphi_{\pi/2-\theta}\right\rangle \left\langle \varphi_{\pi/2-\theta
}\right\vert }{2},R_{\theta}\right)  =\left[
\begin{array}
[c]{cc}%
\frac{1}{2} & \frac{1}{2}\\
\frac{1}{2} & \frac{1}{2}%
\end{array}
\right]
\]
and therefore%
\[
P\left(  I,R_{\theta}\right)  =\left[
\begin{array}
[c]{cc}%
\frac{\left(  \rho\right)  _{00}}{2} & \frac{\left(  \rho\right)  _{11}}{2}\\
\frac{\left(  \rho\right)  _{00}}{2} & \frac{\left(  \rho\right)  _{11}}{2}%
\end{array}
\right]  =\left[
\begin{array}
[c]{cc}%
\frac{1}{4} & \frac{1}{4}\\
\frac{1}{4} & \frac{1}{4}%
\end{array}
\right]  .
\]
By robustness, this holds for $\theta=0$ as well. \ \ But this is a
contradiction, since by indifference $P\left(  I,R_{0}\right)  $\ must be half
the identity.

\item[(ii)] Suppose the contrary; then%
\[
P\left(  I,R_{\pi/8}\right)  =\frac{P\left(  \left\vert 0\right\rangle
,R_{\pi/8}\right)  +P\left(  \left\vert 1\right\rangle ,R_{\pi/8}\right)  }%
{2}.
\]
So considering transitions from $\left\vert 0\right\rangle $\ to $\left\vert
1\right\rangle $,%
\[
\left(  P\left(  I,R_{\pi/8}\right)  \right)  _{01}=\frac{\left(  P\left(
\left\vert 0\right\rangle ,R_{\pi/8}\right)  \right)  _{11}+0}{2}=\frac{1}%
{2}\sin^{2}\frac{\pi}{8}.
\]
But%
\[
P\left(  I,R_{\pi/8}\right)  =\frac{P\left(  \left\vert \varphi_{\pi
/8}\right\rangle ,R_{\pi/8}\right)  +P\left(  \left\vert \varphi_{5\pi
/8}\right\rangle ,R_{\pi/8}\right)  }{2}%
\]
also. \ Since $R_{\pi/8}\left\vert \varphi_{\pi/8}\right\rangle =\left\vert
\varphi_{\pi/4}\right\rangle $, we have%
\begin{align*}
\left(  P\left(  I,R_{\pi/8}\right)  \right)  _{01}  &  \geq\frac{1}{2}\left(
P\left(  \left\vert \varphi_{\pi/8}\right\rangle ,R_{\pi/8}\right)  \right)
_{01}\\
&  \geq\frac{1}{2}\left(  \frac{1}{2}-\left(  P\left(  \left\vert \varphi
_{\pi/8}\right\rangle ,R_{\pi/8}\right)  \right)  _{11}\right) \\
&  \geq\frac{1}{2}\left(  \frac{1}{2}-\sin^{2}\frac{\pi}{8}\right) \\
&  >\frac{1}{2}\sin^{2}\frac{\pi}{8}%
\end{align*}
which is a contradiction.
\end{enumerate}
\end{proof}

Notice that all three conditions in Theorem \ref{decomp}, part (i) were
essential---for $\mathcal{PT}$\ satisfies robustness and decomposition
invariance, $\mathcal{DT}$\ satisfies indifference and decomposition
invariance, and $\mathcal{FT}$\ satisfies indifference and robustness.

The last impossibility result says that no hidden-variable theory satisfies
both indifference and \textquotedblleft strong continuity,\textquotedblright%
\ in the sense that for all $\varepsilon>0$ there exists $\delta>0$\ such that
$\left\Vert \widetilde{\rho}-\rho\right\Vert \leq\delta$ implies $\left\Vert
S\left(  \widetilde{\rho},U\right)  -S\left(  \rho,U\right)  \right\Vert
\leq\varepsilon$. \ To see this, let%
\begin{align*}
U  &  =\left[
\begin{array}
[c]{ccc}%
1 & 0 & 0\\
0 & \frac{1}{\sqrt{2}} & -\frac{1}{\sqrt{2}}\\
0 & \frac{1}{\sqrt{2}} & \frac{1}{\sqrt{2}}%
\end{array}
\right]  ,\\
\rho &  =\sqrt{1-2\delta^{2}}\left\vert 0\right\rangle +\delta\left\vert
1\right\rangle +\delta\left\vert 2\right\rangle ,\\
\widetilde{\rho}  &  =\sqrt{1-2\delta^{2}}\left\vert 0\right\rangle
+\delta\left\vert 1\right\rangle -\delta\left\vert 2\right\rangle .
\end{align*}
Then by indifference,%
\[
S\left(  \rho,U\right)  =\left[
\begin{array}
[c]{ccc}%
1 & 0 & 0\\
0 & 0 & 0\\
0 & 1 & 1
\end{array}
\right]  ,~~~~~~~~S\left(  \widetilde{\rho},U\right)  =\left[
\begin{array}
[c]{ccc}%
1 & 0 & 0\\
0 & 1 & 1\\
0 & 0 & 0
\end{array}
\right]  .
\]
This is the reason why I defined robustness in terms of the joint
probabilities matrix $P$ rather than the stochastic matrix $S$. \ On the other
hand, note that by giving up indifference, one \textit{can} satisfy strong
continuity, as is shown by $\mathcal{PT}$.

\section{Specific Theories\label{SPECIFIC}}

This section presents two nontrivial examples of hidden-variable theories: the
flow theory in Section \ref{FLOW}, and the Schr\"{o}dinger\ theory in Section
\ref{SCHROD}.

\subsection{Flow Theory\label{FLOW}}

\begin{figure}
[ptb]
\begin{center}
\includegraphics[
trim=1.414995in 3.647223in 2.031921in 0.608775in,
height=1.7783in,
width=3.4803in
]%
{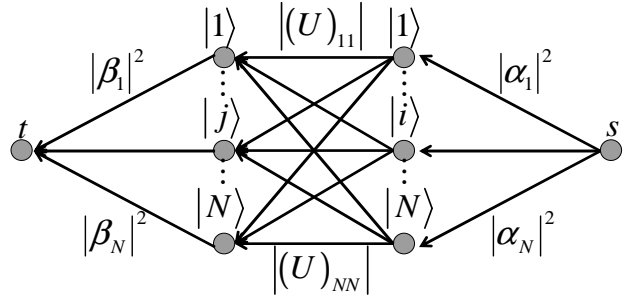}%
\caption[Flow network corresponding to a unitary matrix]{A network
(weighted directed graph with source and sink)
corresponding to the unitary $U$ and state $\left\vert \psi\right\rangle $}%
\label{flowfig}%
\end{center}
\end{figure}

The idea of the flow theory is to convert a unitary matrix into a weighted
directed graph, and then route probability mass through that graph like oil
through pipes. \ Given a unitary $U$, let
\[
\left[
\begin{array}
[c]{c}%
\beta_{1}\\
\vdots\\
\beta_{N}%
\end{array}
\right]  =\left[
\begin{array}
[c]{ccc}%
\left(  U\right)  _{11} & \cdots & \left(  U\right)  _{N1}\\
\vdots &  & \vdots\\
\left(  U\right)  _{1N} & \cdots & \left(  U\right)  _{NN}%
\end{array}
\right]  \left[
\begin{array}
[c]{c}%
\alpha_{1}\\
\vdots\\
\alpha_{N}%
\end{array}
\right]  ,
\]
where for the time being%
\begin{align*}
\left\vert \psi\right\rangle  &  =\alpha_{1}\left\vert 1\right\rangle
+\cdots+\alpha_{N}\left\vert N\right\rangle ,\\
U\left\vert \psi\right\rangle  &  =\beta_{1}\left\vert 1\right\rangle
+\cdots+\beta_{N}\left\vert N\right\rangle
\end{align*}
are pure states. \ Then consider the network $G$\ shown in Figure
\ref{flowfig}.%
%TCIMACRO{\FRAME{ftbpFU}{3.4803in}{1.7783in}{0pt}{\Qcb{A network (weighted
%directed graph with source and sink) corresponding to the unitary $U$ and
%state $\left\vert \psi\right\rangle $}}{\Qlb{flowfig}}{flow.eps}%
%width 3.4803in;  height 1.7783in;  depth 0pt;  original-width 10.3511in;
%original-height 7.7551in;  cropleft "0.1367";  croptop "0.9215";
%cropright "0.8037";  cropbottom "0.4703";
%filename 'flow.eps';file-properties "XNPEU";}}}%
%BeginExpansion
%EndExpansion
We have a source vertex $s$, a sink vertex $t$, and $N$ input and $N$ output
vertices labeled by basis states $\left\vert 1\right\rangle ,\ldots,\left\vert
N\right\rangle $. \ Each edge of the form $\left(  s,\left\vert i\right\rangle
\right)  $\ has capacity $\left\vert \alpha_{i}\right\vert ^{2}$, each edge
$\left(  \left\vert i\right\rangle ,\left\vert j\right\rangle \right)  $ has
capacity $\left\vert \left(  U\right)  _{ij}\right\vert $, and each edge
$\left(  \left\vert j\right\rangle ,t\right)  $ has capacity $\left\vert
\beta_{j}\right\vert ^{2}$. \ A natural question is how much probability mass
can flow from $s$ to $t$ without violating the capacity constraints.\ \ Rather
surprisingly, I will show that one unit of mass (that is, all of it) can.
\ Interestingly, this result would be false if edge $\left(  \left\vert
i\right\rangle ,\left\vert j\right\rangle \right)  $\ had capacity $\left\vert
\left(  U\right)  _{ij}\right\vert ^{2}$\ (or even $\left\vert \left(
U\right)  _{ij}\right\vert ^{1+\varepsilon}$)\ instead of $\left\vert \left(
U\right)  _{ij}\right\vert $. \ I will also show that there exists a mapping
from networks to maximal flows in those networks, that is \textit{robust} in
the sense that a small change in edge capacities produces only a small change
in the amount of flow through any edge.

The proofs of these theorems use classical results from the theory of network
flows (see \cite{clrs} for an introduction). \ In particular, let a
\textit{cut} be a set of edges that separates $s$ from $t$; the \textit{value}
of a cut is the sum of the capacities of its edges. \ Then a fundamental
result called the \textit{Max-Flow-Min-Cut Theorem}\ \cite{ff}\ says that the
maximum possible amount of flow from $s$ to $t$ equals the minimum value of
any cut. \ Using that result I can show the following.

\begin{theorem}
\label{flow}One unit of flow can be routed from $s$ to $t$ in $G$.
\end{theorem}

\begin{proof}
By the above, it suffices to show that any cut $C$ in $G$ has value at least
$1$. \ Let $A$ be the set of $i\in\left\{  1,\ldots,N\right\}  $ such
that$\ \left(  s,\left\vert i\right\rangle \right)  \notin C$, and let $B$ be
the set of $j$ such that $\left(  \left\vert j\right\rangle ,t\right)  \notin
C$. \ Then $C$\ must contain every edge $\left(  \left\vert i\right\rangle
,\left\vert j\right\rangle \right)  $\ such that $i\in A$\ and $j\in B$, and
we can assume without loss of generality that $C$ contains no other edges.
\ So the value of $C$ is%
\[
\sum_{i\notin A}\left\vert \alpha_{i}\right\vert ^{2}+\sum_{j\notin
B}\left\vert \beta_{j}\right\vert ^{2}+\sum_{i\in A,~j\in B}\left\vert \left(
U\right)  _{ij}\right\vert .
\]
Therefore we need to prove the matrix inequality%
\[
\left(  1-\sum_{i\in A}\left\vert \alpha_{i}\right\vert ^{2}\right)  +\left(
1-\sum_{j\in B}\left\vert \beta_{j}\right\vert ^{2}\right)  +\sum_{i\in
A,~j\in B}\left\vert \left(  U\right)  _{ij}\right\vert \geq1,
\]
or%
\begin{equation}
1+\sum_{i\in A,~j\in B}\left\vert \left(  U\right)  _{ij}\right\vert \geq
\sum_{i\in A}\left\vert \alpha_{i}\right\vert ^{2}+\sum_{j\in B}\left\vert
\beta_{j}\right\vert ^{2}. \label{maxeq}%
\end{equation}
Let $U$ be fixed, and consider the maximum of the right-hand side of equation
(\ref{maxeq}) over all $\left\vert \psi\right\rangle $. \ Since%
\[
\beta_{j}=\sum_{i}\left(  U\right)  _{ij}\alpha_{i},
\]
this maximum is equal to the largest eigenvalue $\lambda$\ of the positive
semidefinite matrix%
\[
\sum_{i\in A}\left\vert i\right\rangle \left\langle i\right\vert +\sum_{j\in
B}\left\vert u_{j}\right\rangle \left\langle u_{j}\right\vert
\]
where for each $j$,%
\[
\left\vert u_{j}\right\rangle =\left(  U\right)  _{1j}\left\vert
1\right\rangle +\cdots+\left(  U\right)  _{Nj}\left\vert N\right\rangle .
\]
Let $H_{A}$\ be the subspace of states spanned by $\left\{  \left\vert
i\right\rangle :i\in A\right\}  $, and let $H_{B}$\ be the subspace spanned by
$\left\{  \left\vert u_{j}\right\rangle :j\in B\right\}  $. \ Also, let
$L_{A}\left(  \left\vert \psi\right\rangle \right)  $\ be the length of the
projection of $\left\vert \psi\right\rangle $\ onto $H_{A}$, and let
$L_{B}\left(  \left\vert \psi\right\rangle \right)  $\ be the length of the
projection of $\left\vert \psi\right\rangle $ onto $H_{B}$. \ Then since the
$\left\vert i\right\rangle $'s and $\left\vert u_{j}\right\rangle $'s form
orthogonal bases for $H_{A}$\ and $H_{B}$\ respectively, we have%
\begin{align*}
\lambda &  =\max_{\left\vert \psi\right\rangle }\left(  \sum_{i\in
A}\left\vert \left\langle i|\psi\right\rangle \right\vert ^{2}+\sum_{j\in
B}\left\vert \left\langle u_{j}|\psi\right\rangle \right\vert ^{2}\right) \\
&  =\max_{\left\vert \psi\right\rangle }\left(  L_{A}\left(  \left\vert
\psi\right\rangle \right)  ^{2}+L_{B}\left(  \left\vert \psi\right\rangle
\right)  ^{2}\right)  .
\end{align*}
So letting $\theta$\ be the angle between $H_{A}$\ and $H_{B}$,%
\begin{align*}
\lambda &  =2\cos^{2}\frac{\theta}{2}\\
&  =1+\cos\theta\\
&  \leq1+\max_{\left\vert a\right\rangle \in H_{A},~\left\vert b\right\rangle
\in H_{B}}\left\vert \left\langle a|b\right\rangle \right\vert \\
&  =1+\max_{\substack{\left\vert \gamma_{1}\right\vert ^{2}+\cdots+\left\vert
\gamma_{N}\right\vert ^{2}=1\\\left\vert \delta_{1}\right\vert ^{2}%
+\cdots+\left\vert \delta_{N}\right\vert ^{2}=1}}\left\vert \left(  \sum_{i\in
A}\gamma_{i}\left\langle i\right\vert \right)  \left(  \sum_{j\in B}\delta
_{j}\left\vert u_{j}\right\rangle \right)  \right\vert \\
&  \leq1+\sum_{i\in A,~j\in B}\left\vert \left(  U\right)  _{ij}\right\vert
\end{align*}
which completes the theorem.
\end{proof}

Observe that Theorem \ref{flow}\ still holds if $U$ acts on a mixed state
$\rho$, since we can write $\rho$\ as a convex combination of pure states
$\left\vert \psi\right\rangle \left\langle \psi\right\vert $, construct a flow
for each $\left\vert \psi\right\rangle $\ separately, and then take a convex
combination of the flows.

Using Theorem \ref{flow}, I now define the flow theory $\mathcal{FT}$. \ Let
$F\left(  \rho,U\right)  $ be the set of maximal flows for $\rho
,U$---representable by $N\times N$ arrays of real numbers $f_{ij}$\ such that
$0\leq f_{ij}\leq\left\vert \left(  U\right)  _{ij}\right\vert $\ for all
$i,j$, and also%
\[
\sum_{j}f_{ij}=\left(  \rho\right)  _{ii},~~~\sum_{i}f_{ij}=\left(  U\rho
U^{-1}\right)  _{jj}.
\]
Clearly $F\left(  \rho,U\right)  $ is a convex polytope, which Theorem
\ref{flow}\ asserts is nonempty. \ Form a maximal flow $f^{\ast}\left(
\rho,U\right)  \in F\left(  \rho,U\right)  $ as follows: first let
$f_{11}^{\ast}$\ be the maximum of $f_{11}$\ over all $f\in F\left(
\rho,U\right)  $. \ Then let $f_{12}^{\ast}$\ be the maximum of $f_{12}$\ over
all $f\in F\left(  \rho,U\right)  $\ such that $f_{11}=f_{11}^{\ast}$.
\ Continue to loop through all $i,j$\ pairs in lexicographic order, setting
each $f_{ij}^{\ast}$\ to its maximum possible value consistent with the
$\left(  i-1\right)  N+j-1$\ previous values. \ Finally, let $\left(
P\right)  _{ij}=f_{ij}^{\ast}$ for all $i,j$. \ As discussed in Section
\ref{HV}, given $P$\ we can easily obtain the stochastic matrix $S$ by
dividing the $i^{th}$\ column by $\left(  \rho\right)  _{ii}$, or taking a
limit in case $\left(  \rho\right)  _{ii}=0$.

It is easy to check that $\mathcal{FT}$\ so defined satisfies the indifference
axiom. \ Showing that $\mathcal{FT}$\ satisfies robustness is harder. \ Our
proof is based on the Ford-Fulkerson algorithm \cite{ff}, a classic algorithm
for computing maximal flows that works by finding a sequence of
\textquotedblleft augmenting paths,\textquotedblright\ each of which increases
the flow from $s$ to $t$ by some positive amount.

\begin{theorem}
\label{fdrobust}$\mathcal{FT}$ satisfies robustness.
\end{theorem}

\begin{proof}
Let $G$ be an arbitrary flow network with source $s$, sink $t$, and directed
edges $e_{1},\ldots,e_{m}$, where each $e_{i}$\ has capacity $c_{i}$\ and
leads from $v_{i}$\ to $w_{i}$. \ It will be convenient to introduce a
fictitious edge $e_{0}$\ from $t$ to $s$ with unlimited capacity; then
maximizing the flow through $G$ is equivalent to maximizing the flow through
$e_{0}$. \ Suppose we produce a new network $\widetilde{G}$ by increasing a
single capacity $c_{i^{\ast}}$ by some $\varepsilon>0$. \ Let $f^{\ast}$\ be
the optimal flow for $G$, obtained by first maximizing the flow $f_{0}%
$\ through $e_{0}$, then maximizing the flow $f_{1}$\ through $e_{1}$ holding
$f_{0}$ fixed, and so on up to $f_{m}$. \ Let $\widetilde{f}^{\ast}$\ be the
maximal flow for $\widetilde{G}$\ produced in the same way. \ We claim that
for all $i\in\left\{  0,\ldots,m\right\}  $,%
\[
\left\vert \widetilde{f}_{i}^{\ast}-f_{i}^{\ast}\right\vert \leq\varepsilon.
\]
To see that the theorem follows from this claim: first, if $f^{\ast}$\ is
robust under adding $\varepsilon$\ to $c_{i^{\ast}}$, then it must also be
robust under subtracting $\varepsilon$\ from $c_{i^{\ast}}$.\ \ Second, if we
change $\rho,U$\ to $\widetilde{\rho},\widetilde{U}$\ such that $\left\Vert
\widetilde{\rho}-\rho\right\Vert _{\infty}\leq1/q\left(  N\right)  $ and
$\left\Vert \widetilde{U}-U\right\Vert _{\infty}\leq1/q\left(  N\right)  $,
then we can imagine the $N^{2}+2N$ edge capacities are changed one by one,
so\ that%
\begin{align*}
\left\Vert f^{\ast}\left(  \widetilde{\rho},\widetilde{U}\right)  -f^{\ast
}\left(  \rho,U\right)  \right\Vert _{\infty}  &  \leq\sum_{ij}\left\vert
\left\vert \left(  \widetilde{U}\right)  _{ij}\right\vert -\left\vert \left(
U\right)  _{ij}\right\vert \right\vert +\sum_{i}\left\vert \left(
\widetilde{\rho}\right)  _{ii}-\left(  \rho\right)  _{ii}\right\vert \\
&  ~~~~~~~~~~~~+\sum_{j}\left\vert \left(  \widetilde{U}\widetilde{\rho
}\widetilde{U}^{-1}\right)  _{jj}-\left(  U\rho U^{-1}\right)  _{jj}%
\right\vert \\
&  \leq\frac{4N^{2}}{q\left(  N\right)  }.
\end{align*}
(Here we have made no attempt to optimize the bound.)

We now prove the claim. \ To do so we describe an iterative algorithm for
computing $f^{\ast}$. \ First maximize the flow $f_{0}$\ through $e_{0}$, by
using the Ford-Fulkerson algorithm\ to find a maximal flow from $s$ to $t$.
\ Let $f^{\left(  0\right)  }$ be the resulting flow, and let $G^{\left(
1\right)  }$\ be the residual network that corresponds to $f^{\left(
0\right)  }$. \ For each $i$, that is, $G^{\left(  1\right)  }$\ has an edge
$e_{i}=\left(  v_{i},w_{i}\right)  $\ of capacity $c_{i}^{\left(  1\right)
}=c_{i}-f_{i}^{\left(  0\right)  }$, and an edge $\overline{e}_{i}=\left(
w_{i},v_{i}\right)  $\ of capacity $\overline{c}_{i}^{\left(  1\right)
}=f_{i}^{\left(  0\right)  }$.\ \ Next maximize $f_{1}$\ subject to $f_{0}$ by
using the Ford-Fulkerson algorithm to find \textquotedblleft augmenting
cycles\textquotedblright\ from $w_{1}$\ to $v_{1}$\ and back to $w_{1}$ in
$G^{\left(  1\right)  }\setminus\left\{  e_{0},\overline{e}_{0}\right\}  $.
\ Continue in this manner until each of $f_{1},\ldots,f_{m}$\ has been
maximized subject to the previous $f_{i}$'s. \ Finally set $f^{\ast
}=f^{\left(  m\right)  }$.

Now, one way to compute $\widetilde{f}^{\ast}$ is to start with $f^{\ast}$,
then repeatedly \textquotedblleft correct\textquotedblright\ it by applying
the same iterative algorithm to maximize $\widetilde{f}_{0}$, then
$\widetilde{f}_{1}$, and so on. \ Let $\varepsilon_{i}=\left\vert
\widetilde{f}_{i}^{\ast}-f_{i}^{\ast}\right\vert $;\ then we need to show that
$\varepsilon_{i}\leq\varepsilon$\ for all $i\in\left\{  0,\ldots,m\right\}  $.
\ The proof is by induction on $i$. \ Clearly $\varepsilon_{0}\leq\varepsilon
$, since increasing $c_{i^{\ast}}$ by $\varepsilon$\ can increase the value of
the minimum cut from $s$ to $t$ by at most $\varepsilon$. \ Likewise, after we
maximize $\widetilde{f}_{0}$,\ the value of the minimum cut from $w_{1}$\ to
$v_{1}$\ can increase by at most $\varepsilon-\varepsilon_{0}+\varepsilon
_{0}=\varepsilon$. \ For of the at most $\varepsilon$\ new units of flow from
$w_{1}$\ to $v_{1}$\ that increasing $c_{i^{\ast}}$\ made available,
$\varepsilon_{0}$\ of them were \textquotedblleft taken up\textquotedblright%
\ in maximizing $\widetilde{f}_{0}$, but the process of maximizing
$\widetilde{f}_{0}$\ could have again increased the minimum cut from $w_{1}%
$\ to $v_{1}$ by up to $\varepsilon_{0}$. \ Continuing in this way,%
\[
\varepsilon_{2}\leq\varepsilon-\varepsilon_{0}+\varepsilon_{0}-\varepsilon
_{1}+\varepsilon_{1}=\varepsilon,
\]
and so on up to $\varepsilon_{m}$. \ This completes the proof.
\end{proof}

That $\mathcal{FT}$\ violates decomposition invariance now follows from
Theorem \ref{decomp}, part (i). \ One can also show that $\mathcal{FT}%
$\ violates product commutativity, by considering the following example:\ let
$\left\vert \psi\right\rangle =\left\vert \varphi_{\pi/4}\right\rangle
\otimes\left\vert \varphi_{-\pi/8}\right\rangle $ be a $2$-qubit initial
state, and let $R_{\pi/4}^{A}$\ and $R_{\pi/4}^{B}$\ be $\pi/4$\ rotations
applied to the first and second qubits respectively. \ Then%
\[
S\left(  R_{\pi/4}^{A}\left\vert \psi\right\rangle ,R_{\pi/4}^{B}\right)
S\left(  \left\vert \psi\right\rangle ,R_{\pi/4}^{A}\right)  \vspace{0pt}\neq
S\left(  R_{\pi/4}^{B}\left\vert \psi\right\rangle ,R_{\pi/4}^{A}\right)
S\left(  \left\vert \psi\right\rangle ,R_{\pi/4}^{B}\right)  .
\]
We omit a proof for brevity.

\subsection{Schr\"{o}dinger Theory\label{SCHROD}}

The final hidden-variable theory, which I call the \textit{Schr\"{o}dinger
theory} or $\mathcal{ST}$, is the most interesting one mathematically. \ The
idea---to make a matrix into a stochastic matrix via row and column
rescaling---is natural enough that we came upon it independently, only later
learning that it originated in a 1931 paper of
Schr\"{o}dinger\ \cite{schrodinger}. \ The idea was subsequently developed by
Fortet \cite{fortet}, Beurling \cite{beurling}, Nagasawa \cite{nagasawa}, and
others. \ My goal is to give what (to my knowledge) is the first
self-contained, reasonably accessible presentation of the main result in this
area; and to interpret that result in what I think is the correct way: as
providing one example of a hidden-variable theory, whose strengths and
weaknesses should be directly compared to those of other theories.

Most of the technical difficulties in
\cite{beurling,fortet,nagasawa,schrodinger}\ arise because the stochastic
process being constructed involves continuous time and particle positions.
\ Here I eliminate those difficulties by restricting attention to discrete
time and finite-dimensional Hilbert spaces. \ I thereby obtain a generalized
version\footnote{In $\left(  r,c\right)  $-scaling, we are given an invertible
real matrix, and the goal is to rescale all rows and columns to sum to $1$.
\ The generalized version is to rescale the rows and columns to given values
(not necessarily $1$).} of a problem that computer scientists know as $\left(
r,c\right)  $\textit{-scaling of matrices} \cite{sinkhorn,fl,lsw}.

As in the case of the flow theory, given a unitary $U$ acting on a state
$\rho$, the first step is to replace each entry of $U$ by its absolute value,
obtaining a nonnegative matrix $U^{\left(  0\right)  }$\ defined by $\left(
U^{\left(  0\right)  }\right)  _{ij}:=\left\vert \left(  U\right)
_{ij}\right\vert $. \ We then wish to find nonnegative column multipliers
$\alpha_{1},\ldots,\alpha_{N}$ and row multipliers $\beta_{1},\ldots,\beta
_{N}$ such that for all $i,j$,%
\begin{align}
\alpha_{i}\beta_{1}\left(  U^{\left(  0\right)  }\right)  _{i1}+\cdots
+\alpha_{i}\beta_{N}\left(  U^{\left(  0\right)  }\right)  _{iN}  &  =\left(
\rho\right)  _{ii},\label{coleq}\\
\alpha_{1}\beta_{j}\left(  U^{\left(  0\right)  }\right)  _{1j}+\cdots
+\alpha_{N}\beta_{j}\left(  U^{\left(  0\right)  }\right)  _{Nj}  &  =\left(
U\rho U^{-1}\right)  _{jj}. \label{roweq}%
\end{align}
If we like, we can interpret the $\alpha_{i}$'s\ and $\beta_{j}$'s as
dynamical variables that reach equilibrium precisely when equations
(\ref{coleq})\ and (\ref{roweq}) are satisfied. \ Admittedly, it might be
thought physically implausible that such a complicated dynamical process
should take place at every instant of time. \ On the other hand, it is hard to
imagine a more \textquotedblleft benign\textquotedblright\ way to convert
$U^{\left(  0\right)  }$ into a joint probabilities matrix, than by simply
rescaling its rows and columns.

I will show that multipliers satisfying (\ref{coleq})\ and (\ref{roweq}%
)\ always exist. \ The intuition of a dynamical process reaching equilibrium
turns out to be key to the proof. \ For all $t\geq0$, let%
\begin{align*}
\left(  U^{\left(  2t+1\right)  }\right)  _{ij}  &  =\frac{\left(
\rho\right)  _{ii}}{\sum_{k}\left(  U^{\left(  2t\right)  }\right)  _{ik}%
}\left(  U^{\left(  2t\right)  }\right)  _{ij},\\
\left(  U^{\left(  2t+2\right)  }\right)  _{ij}  &  =\frac{\left(  U\rho
U^{-1}\right)  _{jj}}{\sum_{k}\left(  U^{\left(  2t+1\right)  }\right)  _{kj}%
}\left(  U^{\left(  2t+1\right)  }\right)  _{ij}.
\end{align*}
In words, we obtain\ $U^{\left(  2t+1\right)  }$\ by normalizing each column
$i$ of $U^{\left(  2t\right)  }$\ to sum to $\left(  \rho\right)  _{ii}$;
likewise we obtain $U^{\left(  2t+2\right)  }$\ by normalizing each row $j$ of
$U^{\left(  2t+1\right)  }$\ to sum to $\left(  U\rho U^{-1}\right)  _{jj}$.
\ The crucial fact is that the above process always converges to some
$P\left(  \rho,U\right)  =\lim_{t\rightarrow\infty}U^{\left(  t\right)  }$.
\ We can therefore take%
\begin{align*}
\alpha_{i}  &  =\prod_{t=0}^{\infty}\frac{\left(  \rho\right)  _{ii}}{\sum
_{k}\left(  U^{\left(  2t\right)  }\right)  _{ik}},\\
\beta_{j}  &  =\prod_{t=0}^{\infty}\frac{\left(  U\rho U^{-1}\right)  _{jj}%
}{\sum_{k}\left(  U^{\left(  2t+1\right)  }\right)  _{kj}}%
\end{align*}
for all $i,j$. \ Although I will not prove it here, it turns out that this
yields the \textit{unique} solution to equations (\ref{coleq})\ and
(\ref{roweq}),\ up to a global rescaling of the form $\alpha_{i}%
\rightarrow\alpha_{i}c$\ for all $i$ and $\beta_{j}\rightarrow\beta_{j}%
/c$\ for all $j$\ \cite{nagasawa}.

The convergence proof will reuse a result about network flows from Section
\ref{FLOW}, in order to define a nondecreasing \textquotedblleft progress
measure\textquotedblright\ based on Kullback-Leibler distance.

\begin{theorem}
\label{robust}The limit $P\left(  \rho,U\right)  =\lim_{t\rightarrow\infty
}U^{\left(  t\right)  }$ exists.
\end{theorem}

\begin{proof}
A consequence of Theorem \ref{flow} is that for every $\rho,U$, there exists
an $N\times N$ array of nonnegative real numbers $f_{ij}$ such that

\begin{enumerate}
\item[(1)] $f_{ij}=0$ whenever $\left\vert \left(  U\right)  _{ij}\right\vert
=0$,

\item[(2)] $f_{i1}+\cdots+f_{iN}=\left(  \rho\right)  _{ii}$ for all $i$, and

\item[(3)] $f_{1j}+\cdots+f_{Nj}=\left(  U\rho U^{-1}\right)  _{jj}$\ for all
$j$.
\end{enumerate}

Given any such array, define a progress measure%
\[
Z^{\left(  t\right)  }=%
%TCIMACRO{\dprod \limits_{ij}}%
%BeginExpansion
{\displaystyle\prod\limits_{ij}}
%EndExpansion
\left(  U^{\left(  t\right)  }\right)  _{ij}^{f_{ij}},
\]
where we adopt the convention $0^{0}=1$. \ We claim that $Z^{\left(
t+1\right)  }\geq Z^{\left(  t\right)  }$\ for all $t\geq1$. \ To see this,
assume without loss of generality that we are on an odd step $2t+1$, and let
$C_{i}^{\left(  2t\right)  }=\sum_{j}\left(  U^{\left(  2t\right)  }\right)
_{ij}$ be the $i^{th}$\ column sum before we normalize it. \ Then%
\begin{align*}
Z^{\left(  2t+1\right)  }  &  =%
%TCIMACRO{\dprod \limits_{ij}}%
%BeginExpansion
{\displaystyle\prod\limits_{ij}}
%EndExpansion
\left(  U^{\left(  2t+1\right)  }\right)  _{ij}^{f_{ij}}\\
&  =%
%TCIMACRO{\dprod \limits_{ij}}%
%BeginExpansion
{\displaystyle\prod\limits_{ij}}
%EndExpansion
\left(  \frac{\left(  \rho\right)  _{ii}}{C_{i}^{\left(  2t\right)  }}\left(
U^{\left(  2t\right)  }\right)  _{ij}\right)  ^{f_{ij}}\\
&  =\left(
%TCIMACRO{\dprod \limits_{ij}}%
%BeginExpansion
{\displaystyle\prod\limits_{ij}}
%EndExpansion
\left(  U^{\left(  2t\right)  }\right)  _{ij}^{f_{ij}}\right)  \left(
%TCIMACRO{\dprod \limits_{i}}%
%BeginExpansion
{\displaystyle\prod\limits_{i}}
%EndExpansion
\left(  \frac{\left(  \rho\right)  _{ii}}{C_{i}^{\left(  2t\right)  }}\right)
^{f_{i1}+\cdots+f_{iN}}\right) \\
&  =Z^{\left(  2t\right)  }\cdot%
%TCIMACRO{\dprod \limits_{i}}%
%BeginExpansion
{\displaystyle\prod\limits_{i}}
%EndExpansion
\left(  \frac{\left(  \rho\right)  _{ii}}{C_{i}^{\left(  2t\right)  }}\right)
^{\left(  \rho\right)  _{ii}}.
\end{align*}
As a result of the $2t^{th}$\ normalization step, we had $\sum_{i}%
C_{i}^{\left(  2t\right)  }=1$. \ Subject to that constraint, the maximum of%
\[%
%TCIMACRO{\dprod \limits_{i}}%
%BeginExpansion
{\displaystyle\prod\limits_{i}}
%EndExpansion
\left(  C_{i}^{\left(  2t\right)  }\right)  ^{\left(  \rho\right)  _{ii}}%
\]
over the $C_{i}^{\left(  2t\right)  }$'s\ occurs when $C_{i}^{\left(
2t\right)  }=\left(  \rho\right)  _{ii}$\ for all $i$---a simple calculus fact
that follows from the nonnegativity of Kullback-Leibler distance. \ This
implies that $Z^{\left(  2t+1\right)  }\geq Z^{\left(  2t\right)  }$.
\ Similarly, normalizing rows leads to $Z^{\left(  2t+2\right)  }\geq
Z^{\left(  2t+1\right)  }$. \

It follows that the limit $P\left(  \rho,U\right)  =\lim_{t\rightarrow\infty
}U^{\left(  t\right)  }$\ exists. \ For suppose not; then some $C_{i}^{\left(
t\right)  }$\ is bounded away from $\left(  \rho\right)  _{ii}$, so there
exists an $\varepsilon>0$\ such that $Z^{\left(  t+1\right)  }\geq\left(
1+\varepsilon\right)  Z^{\left(  t\right)  }$\ for all even $t$.\ \ But this
is a contradiction, since $Z^{\left(  0\right)  }>0$\ and $Z^{\left(
t\right)  }\leq1$ for all $t$.
\end{proof}

Besides showing that $P\left(  \rho,U\right)  $\ is well-defined, Theorem
\ref{robust}\ also yields a procedure to \textit{calculate} $P\left(
\rho,U\right)  $\ (as well as the $\alpha_{i}$'s and $\beta_{j}$'s). \ It can
be shown that this procedure converges to within entrywise error $\varepsilon$
after a number steps polynomial in $N$ and $1/\varepsilon$. \ Also, once we
have $P\left(  \rho,U\right)  $,\ the stochastic matrix $S\left(
\rho,U\right)  $ is readily obtained by normalizing each column of $P\left(
\rho,U\right)  $\ to sum to $1$. \ This completes the definition of the
Schr\"{o}dinger theory\ $\mathcal{ST}$.

It is immediate that $\mathcal{ST}$\ satisfies indifference. \ Also:

\begin{proposition}
\label{sdprod}$\mathcal{ST}$ satisfies product commutativity.
\end{proposition}

\begin{proof}
Given a state $\left\vert \psi\right\rangle =\left\vert \psi_{A}\right\rangle
\otimes\left\vert \psi_{B}\right\rangle $,\ let $U_{A}\otimes I$\ act only on
$\left\vert \psi_{A}\right\rangle $\ and let $I\otimes U_{B}$\ act only on
$\left\vert \psi_{B}\right\rangle $. \ Then we claim that%
\[
S\left(  \left\vert \psi\right\rangle ,U_{A}\otimes I\right)  =S\left(
\left\vert \psi_{A}\right\rangle ,U_{A}\right)  \otimes I.
\]
The reason is simply that multiplying all amplitudes in $\left\vert \psi
_{A}\right\rangle $ and $U_{A}\left\vert \psi_{A}\right\rangle $ by a constant
factor $\alpha_{x}$, as we do for each basis state $\left\vert x\right\rangle
$ of $\left\vert \psi_{B}\right\rangle $, has no effect on the scaling
procedure that produces $S\left(  \left\vert \psi_{A}\right\rangle
,U_{A}\right)  $. \ Similarly%
\[
S\left(  \left\vert \psi\right\rangle ,I\otimes U_{B}\right)  =I\otimes
S\left(  \left\vert \psi_{B}\right\rangle ,U_{B}\right)  .
\]
It follows that%
\begin{align*}
S\left(  \left\vert \psi_{A}\right\rangle ,U_{A}\right)  \otimes S\left(
\left\vert \psi_{B}\right\rangle ,U_{B}\right)   &  =S\left(  U_{A}\left\vert
\psi_{A}\right\rangle \otimes\left\vert \psi_{B}\right\rangle ,I\otimes
U_{B}\right)  S\left(  \left\vert \psi\right\rangle ,U_{A}\otimes I\right)
\vspace{0pt}\\
&  =S\left(  \left\vert \psi_{A}\right\rangle \otimes U_{B}\left\vert \psi
_{B}\right\rangle ,U_{A}\otimes I\right)  S\left(  \left\vert \psi
\right\rangle ,I\otimes U_{B}\right)  .
\end{align*}

\end{proof}

On the other hand, numerical simulations readily show that $\mathcal{ST}%
$\ violates decomposition invariance, even when $N=2$ (I omit a concrete
example for brevity).

\section{The Computational Model\label{MODEL}}

I now explain the histories model of computation, building up to the
complexity class $\mathsf{DQP}$. \ From now on, the states $\rho$\ that we
consider will always be pure states of $\ell=\log_{2}N$ qubits. \ That is,
$\rho=\left\vert \psi\right\rangle \left\langle \psi\right\vert $\ where%
\[
\left\vert \psi\right\rangle =\sum_{x\in\left\{  0,1\right\}  ^{\ell}}%
\alpha_{x}\left\vert x\right\rangle .
\]

The algorithms of this chapter will work under \textit{any} hidden-variable
theory that satisfies the indifference axiom. \ On the other hand, if we take
into account that even in theory (let alone in practice), a generic unitary
cannot be represented exactly with a finite universal gate set, only
approximated arbitrarily well, then we also need the robustness axiom. \ Thus,
it is reassuring that there \textit{exists} a hidden-variable theory\ (namely
$\mathcal{FT}$) that satisfies both indifference and robustness.

Let a quantum computer have the initial state $\left\vert 0\right\rangle
^{\otimes\ell}$, and suppose we apply a sequence $\mathcal{U}=\left(
U_{1},\ldots,U_{T}\right)  $\ of unitary operations, each of which is
implemented by a polynomial-size quantum circuit. \ Then a \textit{history} of
a hidden variable through the computation\ is a sequence $H=\left(
v_{0},\ldots,v_{T}\right)  $ of basis states, where $v_{t}$\ is the variable's
value immediately after $U_{t}$\ is applied (thus $v_{0}=\left\vert
0\right\rangle ^{\otimes\ell}$). \ Given any hidden-variable theory
$\mathcal{T}$, we can obtain a probability distribution $\Omega\left(
\mathcal{U},\mathcal{T}\right)  $\ over histories by just applying
$\mathcal{T}$\ repeatedly, once for each $U_{t}$, to obtain the stochastic
matrices%
\[
S\left(  \left\vert 0\right\rangle ^{\otimes\ell},U_{1}\right)  ,~~S\left(
U_{1}\left\vert 0\right\rangle ^{\otimes\ell},U_{2}\right)  ,~~\ldots
~~S\left(  U_{T-1}\cdots U_{1}\left\vert 0\right\rangle ^{\otimes\ell}%
,U_{T}\right)  .
\]
Note that $\Omega\left(  \mathcal{U},\mathcal{T}\right)  $\ is a Markov
distribution; that is, each $v_{t}$\ is independent of the other $v_{i}$'s
conditioned on $v_{t-1}$\ and $v_{t+1}$. \ Admittedly, $\Omega\left(
\mathcal{U},\mathcal{T}\right)  $\ could depend on the precise way in which
the combined circuit $U_{T}\cdots U_{1}$\ is \textquotedblleft
sliced\textquotedblright\ into component circuits $U_{1},\ldots,U_{T}$. \ But
as noted in Section \ref{OBJECTIONS}, such dependence on the granularity of
unitaries is unavoidable in any hidden-variable theory other than
$\mathcal{PT}$.

Given a hidden-variable theory $\mathcal{T}$, let $\mathcal{O}\left(
\mathcal{T}\right)  $\ be an oracle that takes as input a positive integer
$\ell$, and a sequence of quantum circuits $\mathcal{U}=\left(  U_{1}%
,\ldots,U_{T}\right)  $\ that act on $\ell$ qubits. \ Here each $U_{t}$\ is
specified by a sequence $\left(  g_{t,1},\ldots,g_{t,m\left(  t\right)
}\right)  $\ of gates chosen from some finite universal gate set $\mathcal{G}%
$. \ The oracle $\mathcal{O}\left(  \mathcal{T}\right)  $\ returns as output a
sample $\left(  v_{0},\ldots,v_{T}\right)  $\ from the history distribution
$\Omega\left(  \mathcal{U},\mathcal{T}\right)  $\ defined previously. \ Now
let $A$ be a deterministic classical Turing machine that is given oracle
access to $\mathcal{O}\left(  \mathcal{T}\right)  $. \ The machine $A$
receives an input $x$, makes a single oracle query to $\mathcal{O}\left(
\mathcal{T}\right)  $, then produces an output based on the response. \ We say
a set of strings $L$ is in $\mathsf{DQP}$\ if there exists an $A$ such that
for all sufficiently large $n$ and inputs $x\in\left\{  0,1\right\}  ^{n}$,
and all theories $\mathcal{T}$ satisfying the indifference and robustness
axioms, $A$ correctly decides whether $x\in L$\ with probability at least
$2/3$, in time polynomial in $n$.

Let me make some remarks about the above definition. \ There is no real
significance in the requirement that $A$ be deterministic and classical, and
that it be allowed only one query to $\mathcal{O}\left(  \mathcal{T}\right)
$. \ I made this choice only because it suffices for the upper bounds; it
might be interesting to consider the effects of other choices. \ However,
other aspects of the definition are not arbitrary. \ The order of quantifiers
matters; we want a single $A$ that works for \textit{any} hidden-variable
theory satisfying indifference and robustness. \ Also, we require $A$ to
succeed only for sufficiently large $n$ since by choosing a large enough
polynomial $q\left(  N\right)  $\ in the statement of the robustness axiom, an
adversary might easily make $A$ incorrect on a finite number of instances.

\subsection{Basic Results\label{RESULTS}}

Having defined the complexity class $\mathsf{DQP}$, in this subsection I
establish its most basic properties. \ First of all, it is immediate that
$\mathsf{BQP}\subseteq\mathsf{DQP}$; that is, sampling histories is at least
as powerful as standard quantum computation. \ For $v_{1}$, the first
hidden-variable value returned by $\mathcal{O}\left(  \mathcal{T}\right)  $,
can be seen as simply the result of applying a polynomial-size quantum circuit
$U_{1}$ to the initial state $\left\vert 0\right\rangle ^{\otimes\ell}$ and
then measuring in the standard basis. \ A key further observation is the following.

\begin{theorem}
\label{anyuniv}Any universal gate set\ yields the same complexity class
$\mathsf{DQP}$. \ By universal, we mean that any unitary matrix (real or
complex) can be approximated, without the need for ancilla qubits.
\end{theorem}

\begin{proof}
Let $\mathcal{G}$\ and $\mathcal{G}^{\prime}$\ be universal gate sets. \ Also,
let $\mathcal{U}=\left(  U_{1},\ldots,U_{T}\right)  $ be a sequence of $\ell
$-qubit unitaries, each specified by a polynomial-size quantum circuit over
$\mathcal{G}$. \ We have $T,\ell=O\left(  \operatorname*{poly}\left(
n\right)  \right)  $\ where $n$ is the input length. \ We can also assume
without loss of generality that $\ell\geq n$, since otherwise we simply insert
$n-\ell$ dummy qubits that are never acted on (by the indifference axiom, this
will not affect the results). \ We want to approximate $\mathcal{U}$\ by
another sequence of $\ell$-qubit unitaries, $\mathcal{U}^{\prime}=\left(
U_{1}^{\prime},\ldots,U_{T}^{\prime}\right)  $, where each $U_{t}^{\prime}%
$\ is specified by a quantum circuit over $\mathcal{G}^{\prime}$. \ In
particular, for all $t$\ we want $\left\Vert U_{t}^{\prime}-U_{t}\right\Vert
_{\infty}\leq2^{-\ell^{2}T}$. \ By the Solovay-Kitaev Theorem
\cite{kitaev:ec,nc}, we can achieve this\ using $\operatorname*{poly}\left(
n,\ell^{2}T\right)  =\operatorname*{poly}\left(  n\right)  $ gates from
$\mathcal{G}^{\prime}$; moreover, the circuit for $U_{t}^{\prime}$\ can be
constructed in polynomial time given the circuit for $U_{t}$.

Let $\left\vert \psi_{t}\right\rangle =U_{t}\cdots U_{1}\left\vert
0\right\rangle ^{\otimes\ell}$\ and $\left\vert \psi_{t}^{\prime}\right\rangle
=U_{t}^{\prime}\cdots U_{1}^{\prime}\left\vert 0\right\rangle ^{\otimes\ell}$.
\ Notice that for all $t\in\left\{  1,\ldots,T\right\}  $,%
\begin{align*}
\left\Vert \left\vert \psi_{t}^{\prime}\right\rangle -\left\vert \psi
_{t}\right\rangle \right\Vert _{\infty}  &  \leq2^{\ell}\left(  \left\Vert
\left\vert \psi_{t-1}^{\prime}\right\rangle -\left\vert \psi_{t-1}%
\right\rangle \right\Vert _{\infty}+2^{-\ell^{2}T}\right) \\
&  \leq T2^{\ell T}\left(  2^{-\ell^{2}T}\right)  =T2^{-\ell\left(
\ell-1\right)  T},
\end{align*}
since $\left\Vert \left\vert \psi_{0}^{\prime}\right\rangle -\left\vert
\psi_{0}\right\rangle \right\Vert _{\infty}=0$. \ Here $\left\Vert
~\right\Vert _{\infty}$\ denotes\ the maximum entrywise difference between two
vectors in $\mathbb{C}^{2^{\ell}}$. \ Also, given a theory $\mathcal{T}$, let
$P_{t}$\ and $P_{t}^{\prime}$\ be the joint probabilities matrices
corresponding to $U_{t}$\ and $U_{t}^{\prime}$\ respectively. \ Then by the
robustness axiom, there exists a polynomial $q$\ such that if $\left\Vert
U_{t}^{\prime}-U_{t}\right\Vert _{\infty}\leq1/q\left(  2^{\ell}\right)  $ and
$\left\Vert \left\vert \psi_{t-1}^{\prime}\right\rangle -\left\vert \psi
_{t-1}\right\rangle \right\Vert _{\infty}\leq1/q\left(  2^{\ell}\right)  $,
then $\left\Vert P_{t}-P_{t}^{\prime}\right\Vert _{\infty}\leq2^{-3\ell}$.
\ For all such polynomials $q$, we have $2^{-\ell^{2}T}\leq1/q\left(  2^{\ell
}\right)  $\ and $T2^{-\ell\left(  \ell-1\right)  T}\leq1/q\left(  2^{\ell
}\right)  $\ for sufficiently large $n\leq\ell$. \ Therefore $\left\Vert
P_{t}-P_{t}^{\prime}\right\Vert _{\infty}\leq2^{-3\ell}$\ for all $t$ and
sufficiently large $n$.

Now assume $n$ is sufficiently large, and\ consider the distributions
$\Omega\left(  \mathcal{U},\mathcal{T}\right)  $\ and $\Omega\left(
\mathcal{U}^{\prime},\mathcal{T}\right)  $\ over classical histories
$H=\left(  v_{0},\ldots,v_{T}\right)  $. \ For all $t\in\left\{
1,\ldots,T\right\}  $ and $x\in\left\{  0,1\right\}  ^{\ell}$, we have%
\[
\left\vert \Pr_{\Omega\left(  \mathcal{U},\mathcal{T}\right)  }\left[
v_{t}=\left\vert x\right\rangle \right]  -\Pr_{\Omega\left(  \mathcal{U}%
^{\prime},\mathcal{T}\right)  }\left[  v_{t}=\left\vert x\right\rangle
\right]  \right\vert \leq2^{\ell}\left(  2^{-3\ell}\right)  =2^{-2\ell}.
\]
It follows by the union bound that the variation distance $\left\Vert
\Omega\left(  \mathcal{U}^{\prime},\mathcal{T}\right)  -\Omega\left(
\mathcal{U},\mathcal{T}\right)  \right\Vert $\ is at most%
\[
T2^{\ell}\left(  2^{-2\ell}\right)  =\frac{T}{2^{\ell}}\leq\frac{T}{2^{n}}.
\]
In other words, $\Omega\left(  \mathcal{U}^{\prime},\mathcal{T}\right)  $\ can
be distinguished from $\Omega\left(  \mathcal{U},\mathcal{T}\right)  $\ with
bias at most $T/2^{n}$, which is exponentially small. \ So any classical
postprocessing algorithm that succeeds with high probability given $H\in
\Omega\left(  \mathcal{U},\mathcal{T}\right)  $, also succeeds with high
probability given $H\in\Omega\left(  \mathcal{U}^{\prime},\mathcal{T}\right)
$. \ This completes the theorem.
\end{proof}

Unfortunately, the best upper bound on $\mathsf{DQP}$\ I have been
able to show is $\mathsf{DQP}\subseteq\mathsf{EXP}$; that is, any
problem in $\mathsf{DQP}$\ is solvable in deterministic exponential
time. \ The proof is trivial:\ let $\mathcal{T}$\ be the flow theory
$\mathcal{FT}$. \ Then by using the Ford-Fulkerson algorithm, we can
clearly construct the requisite maximum flows in time polynomial in
$2^{\ell}$\ (hence exponential in $n$), and thereby calculate the
probability of each possible history $\left(  v_{1},\ldots
,v_{T}\right)  $\ to suitable precision.

\section{The Juggle Subroutine\label{JUGGLE}}

This section presents a crucial subroutine that will be used in both main
algorithms: the algorithm for simulating statistical zero knowledge in Section
\ref{SZK}, and the algorithm for search in $N^{1/3}$\ queries in Section
\ref{SEARCH}. \ Given an $\ell$-qubit state $\left(  \left\vert a\right\rangle
+\left\vert b\right\rangle \right)  /\sqrt{2}$, where $\left\vert
a\right\rangle $\ and $\left\vert b\right\rangle $\ are unknown basis states,
the goal of the juggle subroutine\ is to learn both $a$ and $b$. \ The name
arises because the strategy will be to \textquotedblleft
juggle\textquotedblright\ a hidden variable, so that if it starts out at
$\left\vert a\right\rangle $\ then with non-negligible probability it
transitions to $\left\vert b\right\rangle $, and vice versa. \ Inspecting the
entire history of the hidden variable will then reveal both $a$ and $b$, as desired.

To produce this behavior, we will exploit a basic feature of quantum
mechanics: that observable information in one basis can become unobservable
phase information in a different basis. \ We will apply a sequence of
unitaries that hide all information about $a$ and $b$ in phases, thereby
forcing the hidden variable to \textquotedblleft forget\textquotedblright%
\ whether it started at $\left\vert a\right\rangle $\ or $\left\vert
b\right\rangle $. \ We will then invert those unitaries to return the state to
$\left(  \left\vert a\right\rangle +\left\vert b\right\rangle \right)
/\sqrt{2}$, at which point the hidden variable, having \textquotedblleft
forgotten\textquotedblright\ its initial value, must be unequal to that value
with probability $1/2$.

I now give the subroutine. \ Let $\left\vert \psi\right\rangle =\left(
\left\vert a\right\rangle +\left\vert b\right\rangle \right)  /\sqrt{2}$\ be
the initial state. \ The first unitary, $U_{1}$, consists of Hadamard gates on
$\ell-1$\ qubits chosen uniformly at random, and the identity operation on the
remaining qubit, $i$. \ Next $U_{2}$ consists of a Hadamard gate on qubit $i$.
\ Finally $U_{3}$\ consists of Hadamard gates on all $\ell$ qubits. \ Let
$a=a_{1}\ldots a_{\ell}$\ and $b=b_{1}\ldots b_{\ell}$. \ Then since $a\neq
b$, we have $a_{i}\neq b_{i}$\ with probability at least $1/\ell$. \ Assuming
that occurs, the state%
\[
U_{1}\left\vert \psi\right\rangle =\frac{1}{2^{\ell/2}}\left(  \sum
_{z\in\left\{  0,1\right\}  ^{\ell}~:~z_{i}=a_{i}}\left(  -1\right)  ^{a\cdot
z-a_{i}z_{i}}\left\vert z\right\rangle +\sum_{z\in\left\{  0,1\right\}
^{\ell}~:~z_{i}=b_{i}}\left(  -1\right)  ^{b\cdot z-b_{i}z_{i}}\left\vert
z\right\rangle \right)
\]
assigns nonzero amplitude to all $2^{\ell}$\ basis states. \ Then $U_{2}%
U_{1}\left\vert \psi\right\rangle $\ assigns nonzero amplitude to $2^{\ell-1}%
$\ basis states $\left\vert z\right\rangle $, namely those for which $a\cdot
z\equiv b\cdot z\left(  \operatorname{mod}2\right)  $. \ Finally $U_{3}%
U_{2}U_{1}\left\vert \psi\right\rangle =\left\vert \psi\right\rangle $.

Let $v_{t}$ be the value of the hidden variable after $U_{t}$\ is applied.
\ Then assuming $a_{i}\neq b_{i}$, I claim that $v_{3}$\ is independent of
$v_{0}$. \ So in particular, if $v_{0}=\left\vert a\right\rangle $\ then
$v_{3}=\left\vert b\right\rangle $\ with $1/2$ probability, and if
$v_{0}=\left\vert b\right\rangle $\ then $v_{3}=\left\vert a\right\rangle
$\ with $1/2$ probability. \ To see this, observe that when $U_{1}$\ is
applied, there is no interference between basis states $\left\vert
z\right\rangle $ such that $z_{i}=a_{i}$,\ and those such that $z_{i}=b_{i}$.
\ So by the indifference axiom, the probability mass at $\left\vert
a\right\rangle $\ must spread out evenly among all $2^{\ell-1}$\ basis states
that agree with $a$\ on the $i^{th}$\ bit, and similarly for the probability
mass at $\left\vert b\right\rangle $. \ Then after $U_{2}$\ is applied,
$v_{2}$\ can differ from $v_{1}$\ only on the $i^{th}$\ bit, again by the
indifference axiom. \ So each basis state\ of $U_{2}U_{1}\left\vert
\psi\right\rangle $\ must receive an equal contribution from probability mass
originating at $\left\vert a\right\rangle $, and probability mass originating
at $\left\vert b\right\rangle $. \ Therefore $v_{2}$\ is independent of
$v_{0}$, from which it follows that $v_{3}$\ is independent of $v_{0}$\ as well.

Unfortunately, the juggle subroutine only works with probability $1/\left(
2\ell\right)  $---for it requires that $a_{i}\neq b_{i}$,\ and even then,
inspecting the history $\left(  v_{0},v_{1},\ldots\right)  $\ only reveals
both $\left\vert a\right\rangle $\ and $\left\vert b\right\rangle $\ with
probability $1/2$. \ Furthermore, the definition of $\mathsf{DQP}$ does not
allow more than one call to the history oracle. \ However, all we need to do
is pack multiple subroutine calls into a single oracle call. \ That is, choose
$U_{4}$\ similarly to $U_{1}$\ (except with a different value of $i$), and set
$U_{5}=U_{2}$\ and $U_{6}=U_{3}$. \ Do the same with $U_{7}$, $U_{8}$, and
$U_{9}$, and so on. \ Since $U_{3},U_{6},U_{9},\ldots$\ all return the quantum
state to $\left\vert \psi\right\rangle $, the effect is that of multiple
independent juggle attempts. \ With $2\ell^{2}$\ attempts, we can make the
failure probability at most $\left(  1-1/\left(  2\ell\right)  \right)
^{2\ell^{2}}<e^{-\ell}$.

As a final remark, it is easy to see that the juggle subroutine works equally
well with states of the form $\left\vert \psi\right\rangle =\left(  \left\vert
a\right\rangle -\left\vert b\right\rangle \right)  /\sqrt{2}$. \ This will
prove useful in Section \ref{SEARCH}.

\section{Simulating $\mathsf{SZK}$\label{SZK}}

This section shows that $\mathsf{SZK}\subseteq\mathsf{DQP}$. \ Here
$\mathsf{SZK}$, or Statistical Zero Knowledge, was originally defined as the
class of problems that possess a certain kind of \textquotedblleft
zero-knowledge proof protocol\textquotedblright---that is, a protocol between
an omniscient prover and a verifier, by which the verifier becomes convinced
of the answer to a problem, yet without learning anything else about the
problem. \ However, for present purposes this cryptographic definition of
$\mathsf{SZK}$\ is irrelevant. \ For Sahai and Vadhan \cite{sv}\ have given an
alternate and much simpler characterization: a problem is in $\mathsf{SZK}$ if
and only if it can be reduced to a problem called Statistical Difference,
which involves deciding whether two probability distributions are close or far.

More formally, let $P_{0}$\ and $P_{1}$\ be functions that map
$n$-bit strings to $q\left(n\right)$-bit strings for some polynomial
$q$, and that are specified by classical polynomial-time algorithms.
\ Let $\Lambda_{0}$ and $\Lambda_{1}$\ be the probability
distributions over $P_{0}\left( x\right)  $\ and $P_{1}\left(
x\right) $\ respectively, if $x\in\left\{  0,1\right\} ^{n}$ is
chosen uniformly at random. \ Then the problem is to decide whether
$\left\Vert \Lambda _{0}-\Lambda_{1}\right\Vert $\ is less than
$1/3$\ or greater than $2/3$,
given that one of these is the case. \ Here%
\[
\left\Vert \Lambda_{0}-\Lambda_{1}\right\Vert
=\frac{1}{2}\sum_{y\in\left\{ 0,1\right\}
^{q\left(n\right)}}\left\vert \Pr_{x\in\left\{  0,1\right\}
^{n}}\left[ P_{0}\left(  x\right) =y\right]  -\Pr_{x\in\left\{
0,1\right\}  ^{n}}\left[ P_{1}\left( x\right)  =y\right] \right\vert
\]
is the variation distance between $\Lambda_{0}$ and $\Lambda_{1}$.

To illustrate, let us see why Graph Isomorphism is in $\mathsf{SZK}$. \ Given
two graphs $G_{0}$\ and $G_{1}$, take $\Lambda_{0}$\ to be the uniform
distribution over all permutations of $G_{0}$, and $\Lambda_{1}$\ to be
uniform over all permutations of $G_{1}$. \ This way, if $G_{0}$\ and $G_{1}%
$\ are isomorphic, then $\Lambda_{0}$ and $\Lambda_{1}$\ will be identical, so
$\left\Vert \Lambda_{0}-\Lambda_{1}\right\Vert =0$. \ On the other hand, if
$G_{0}$\ and $G_{1}$\ are non-isomorphic, then $\Lambda_{0}$ and $\Lambda_{1}%
$\ will be perfectly distinguishable, so $\left\Vert \Lambda_{0}-\Lambda
_{1}\right\Vert =1$. \ Since $\Lambda_{0}$\ and $\Lambda_{1}$ are clearly
samplable by polynomial-time algorithms, it follows that any instance of Graph
Isomorphism\ can be expressed as an instance of Statistical Difference. \ For
a proof that Approximate Shortest Vector is in $\mathsf{SZK}$, the reader is
referred to Goldreich and Goldwasser \cite{gg} (see also Aharonov and Ta-Shma
\cite{at}).

The proof will use the following \textquotedblleft amplification
lemma\textquotedblright\ from \cite{sv}:\footnote{Note that in this lemma, the
constants $1/3$\ and $2/3$ are not arbitrary; it is important for technical
reasons that $\left(  2/3\right)  ^{2}>1/3$.}

\begin{lemma}
[Sahai and Vadhan]\label{amp}Given efficiently-samplable distributions
$\Lambda_{0}$\ and $\Lambda_{1}$, we can construct new efficiently-samplable
distributions $\Lambda_{0}^{\prime}$\ and $\Lambda_{1}^{\prime}$, such that if
$\left\Vert \Lambda_{0}-\Lambda_{1}\right\Vert \leq1/3$ then $\left\Vert
\Lambda_{0}^{\prime}-\Lambda_{1}^{\prime}\right\Vert \leq2^{-n}$, while if
$\left\Vert \Lambda_{0}-\Lambda_{1}\right\Vert \geq2/3$\ then $\left\Vert
\Lambda_{0}^{\prime}-\Lambda_{1}^{\prime}\right\Vert \geq1-2^{-n}$.
\end{lemma}

In particular, Lemma \ref{amp} means we can assume without loss of generality
that either $\left\Vert \Lambda_{0}-\Lambda_{1}\right\Vert \leq2^{-n^{c}}$\ or
$\left\Vert \Lambda_{0}-\Lambda_{1}\right\Vert \geq1-2^{-n^{c}}$\ for some
constant $c>0$.

Having covered the necessary facts about $\mathsf{SZK}$, we can now proceed to
the main result.

\begin{theorem}
\label{szk}$\mathsf{SZK}\subseteq\mathsf{DQP}$.
\end{theorem}

\begin{proof}
We show how to solve Statistical Difference\ by using a history
oracle. \ For simplicity, we start with the special case where
$P_{0}$\ and $P_{1}$\ are
both one-to-one functions. \ In this case, the circuit sequence $\mathcal{U}%
$\ given to the history oracle does the following: it first prepares the state%
\[
\frac{1}{2^{\left(  n+1\right)  /2}}\sum_{b\in\left\{  0,1\right\}
,x\in\left\{  0,1\right\}  ^{n}}\left\vert b\right\rangle \left\vert
x\right\rangle \left\vert P_{b}\left(  x\right)  \right\rangle .
\]
It then applies the juggle subroutine to the joint state of the
$\left\vert b\right\rangle $ and $\left\vert x\right\rangle $
registers, taking $\ell =n+1$. \ Notice that by the indifference
axiom, the hidden variable will never transition from one value of
$P_{b}\left(  x\right)  $\ to another---exactly as if we had\
\textit{measured} the third register in the standard basis. \ All
that matters is the reduced state $\left\vert \psi\right\rangle $ of
the first two registers, which has the form $\left(  \left\vert
0\right\rangle \left\vert x_{0}\right\rangle +\left\vert
1\right\rangle \left\vert x_{1}\right\rangle \right)  /\sqrt{2}$\
for some $x_{0},x_{1}$\ if $\left\Vert
\Lambda_{0}-\Lambda_{1}\right\Vert =0$, and $\left\vert
b\right\rangle \left\vert x\right\rangle $\ for some $b,x$\ if
$\left\Vert \Lambda _{0}-\Lambda_{1}\right\Vert =1$. \ We have
already seen that the juggle subroutine can distinguish these two
cases: when the hidden-variable history is inspected, it will
contain two values of the $\left\vert b\right\rangle $ register in
the former case, and only one value in the latter case. \ Also,
clearly the case $\left\Vert \Lambda_{0}-\Lambda_{1}\right\Vert \leq2^{-n^{c}%
}$\ is statistically indistinguishable from $\left\Vert \Lambda_{0}%
-\Lambda_{1}\right\Vert =0$\ with respect to the subroutine, and
likewise $\left\Vert \Lambda_{0}-\Lambda_{1}\right\Vert
\geq1-2^{-n^{c}}$\ is indistinguishable from $\left\Vert
\Lambda_{0}-\Lambda_{1}\right\Vert =1$.

We now consider the general case, where $P_{0}$\ and $P_{1}$\ need
not be one-to-one. \ Our strategy is to reduce to the one-to-one
case, by using a well-known hashing technique of Valiant and
Vazirani \cite{vv}.\ \ Let $\mathcal{D}_{n,k}$\ be the uniform
distribution over all affine functions mapping $\left\{  0,1\right\}
^{n}$\ to $\left\{  0,1\right\}  ^{k}$, where we identify those sets
with the finite fields $\mathbb{F}_{2}^{n}$\ and
$\mathbb{F}_{2}^{k}$ respectively. \ What Valiant and Vazirani
showed is that,
for all subsets $A\subseteq\left\{  0,1\right\}  ^{n}$\ such that $2^{k-2}%
\leq\left\vert A\right\vert \leq2^{k-1}$, and all $s\in\left\{
0,1\right\}
^{k}$,%
\[
\Pr_{h\in\mathcal{D}_{n,k}}\left[  \left\vert A\cap h^{-1}\left(
s\right) \right\vert =1\right]  \geq\frac{1}{8}.
\]
As a corollary, the expectation over $h\in\mathcal{D}_{n,k}$\ of%
\[
\left\vert \left\{  s\in\left\{  0,1\right\}  ^{k}:\left\vert A\cap
h^{-1}\left(  s\right)  \right\vert =1\right\}  \right\vert
\]
is at least $2^{k}/8$. \ It follows that, if $x$\ is drawn uniformly
at random
from $A$, then%
\[
\Pr_{h,x}\left[  \left\vert A\cap h^{-1}\left(  h\left(  x\right)
\right) \right\vert =1\right]  \geq\frac{2^{k}/8}{\left\vert
A\right\vert }\geq \frac{1}{4}.
\]
This immediately suggests the following algorithm for the
many-to-one case. \ Draw $k$\ uniformly at random from $\left\{
2,\ldots,n+1\right\}  $; then draw
$h_{0},h_{1}\in\mathcal{D}_{n,k}$. \ Have $\mathcal{U}$\ prepare the
state%
\[
\frac{1}{2^{\left(  n+1\right)  /2}}\sum_{b\in\left\{  0,1\right\}
,x\in\left\{  0,1\right\}  ^{n}}\left\vert b\right\rangle \left\vert
x\right\rangle \left\vert P_{b}\left(  x\right)  \right\rangle
\left\vert h_{b}\left(  x\right)  \right\rangle ,
\]
and then apply the juggle subroutine to the joint state of the
$\left\vert b\right\rangle $\ and $\left\vert x\right\rangle $\
registers, ignoring the $\left\vert P_{b}\left(  x\right)
\right\rangle $\ and $\left\vert h_{b}\left(  x\right)
\right\rangle $ registers as before.

Suppose $\left\Vert \Lambda_{0}-\Lambda_{1}\right\Vert =0$. \ Also,
given $x \in \left\{0,1\right\}^{n}$ and $i \in \left\{0,1\right\}$,
let $A_{i}=P_{i}^{-1}\left( P_{i}\left(x\right)\right)
$\ and $H_{i}=h_{i}^{-1}\left(  h_{i}\left(x\right)\right)  $, and suppose $2^{k-2}%
\leq\left\vert A_{0}\right\vert =\left\vert A_{1}\right\vert
\leq2^{k-1}$.
\ Then%
\[
\Pr_{s,h_{0},h_{1}}\left[  \left\vert A_{0}\cap H_{0} \right\vert
=1\wedge\left\vert A_{1}\cap H_{1} \right\vert =1\right] \geq\left(
\frac{1}{4}\right) ^{2},
\]
since the events $\left\vert A_{0}\cap H_{0} \right\vert =1$\ and
$\left\vert A_{1}\cap H_{1} \right\vert =1$\ are independent of each
other conditioned on $x$. \ Assuming both events occur, as before
the juggle subroutine will reveal both $\left\vert 0\right\rangle
\left\vert x_{0}\right\rangle $\ and $\left\vert 1\right\rangle
\left\vert x_{1}\right\rangle $\ with high probability, where
$x_{0}$\ and $x_{1}$\ are the unique elements of $A_{0}\cap H_{0}$\
and $A_{1}\cap H_{1}$\ respectively. \ By contrast, if $\left\Vert
\Lambda_{0}-\Lambda_{1}\right\Vert =1$\ then only one value of the
$\left\vert b\right\rangle $\ register will ever be observed. \
Again, replacing $\left\Vert \Lambda_{0}-\Lambda_{1}\right\Vert =0$\
by $\left\Vert \Lambda_{0}-\Lambda_{1}\right\Vert \leq2^{-n^{c}}$,
and $\left\Vert \Lambda_{0}-\Lambda_{1}\right\Vert =1$\ by
$\left\Vert \Lambda_{0}-\Lambda _{1}\right\Vert \geq1-2^{-n^{c}}$,
can have only a negligible effect on the history distribution.

Of course, the probability that the correct value of $k$ is chosen,
and that $A_{0}\cap H_{0}$\ and $A_{1}\cap H_{1}$\ both have a
unique element, could be as low as $1/\left( 16n\right) $. \ To deal
with this, we simply increase the number of calls to the juggle
subroutine by an $O\left(  n\right)  $ factor, drawing new values of
$k,h_{0},h_{1}$ for each call. \ We pack multiple subroutine calls
into a single oracle call as described in Section \ref{JUGGLE},
except that now we uncompute the entire state (returning it to
$\left\vert 0\cdots0\right\rangle $) and then recompute it between
subroutine calls. \ A final remark: since the algorithm that calls
the history oracle is deterministic, we \textquotedblleft
draw\textquotedblright\ new values of $k,h_{0},h_{1}$ by having $\mathcal{U}%
$\ prepare a uniform superposition over all possible values. \ The
indifference axiom justifies this procedure, by guaranteeing that
within each\ call to the juggle subroutine, the hidden-variable
values of $k$, $h_{0}$, and $h_{1}$\ remain constant.
\end{proof}

Recall from Chapter \ref{COL} that there exists an oracle $A$ relative to
which $\mathsf{SZK}^{A}\not \in \mathsf{BQP}^{A}$. \ By contrast, since
Theorem \ref{szk} is easily seen to relativize, we have $\mathsf{SZK}^{A}%
\in\mathsf{DQP}^{A}$ for all oracles $A$. \ It follows that there exists an
oracle $A$ relative to which $\mathsf{BQP}^{A}\neq\mathsf{DQP}^{A}$.

\section{Search in $N^{1/3}$ Queries\label{SEARCH}}

Given a Boolean function $f:\left\{  0,1\right\}  ^{n}\rightarrow\left\{
0,1\right\}  $, the database search problem is simply to find a string
$x$\ such that $f\left(  x\right)  =1$. \ We can assume without loss of
generality that this \textquotedblleft marked item\textquotedblright\ $x$ is
unique.\footnote{For if there are multiple marked items, then we can reduce to
the unique marked item case by using the Valiant-Vazirani hashing technique
described in Theorem \ref{szk}.} \ We want to find it using as few queries to
$f$ as possible, where a query returns $f\left(  y\right)  $\ given $y$.

Let $N=2^{n}$. \ Then classically, of course, $\Theta\left(
N\right) $\ queries are necessary and sufficient. \ By querying $f$
in superposition, Grover's algorithm \cite{grover}\ finds $x$ using
$O\left(  N^{1/2}\right) $\ queries, together with
$\widetilde{O}\left(  N^{1/2}\right)  $\ auxiliary computation steps
(here the $\widetilde{O}$\ hides a factor of the form $\left(  \log
N\right)  ^{c}$). \ Bennett et al.\ \cite{bbbv}\ showed that any
quantum algorithm needs $\Omega\left(  N^{1/2}\right)  $\ queries.

In this section, I show how to find the marked item by sampling histories,
using only $O\left(  N^{1/3}\right)  $\ queries and $\widetilde{O}\left(
N^{1/3}\right)  $\ computation steps. \ Formally, the model is as
follows.\ \ Each of the quantum circuits $U_{1},\ldots,U_{T}$\ that algorithm
$A$ gives to the history oracle $\mathcal{O}\left(  \mathcal{T}\right)  $ is
now able to query $f$. \ Suppose $U_{t}$ makes $q_{t}$\ queries to $f$; then
the total number of queries made by $A$ is defined to be $Q=q_{1}+\cdots
+q_{T}$. \ The total number of \textit{computation} steps is at least the
number of steps required to write down $U_{1},\ldots,U_{T}$, but could be greater.

\begin{theorem}
\label{searchthm}In the $\mathsf{DQP}$ model, we can search a database of $N$
items for a unique marked item using $O\left(  N^{1/3}\right)  $ queries and
$\widetilde{O}\left(  N^{1/3}\right)  $\ computation steps.
\end{theorem}

\begin{proof}
Assume without loss of generality that $N=2^{n}$ with $n|3$, and that each
database item is labeled by an $n$-bit string. \ Let $x\in\left\{
0,1\right\}  ^{n}$ be the label of the unique marked item. \ Then the sequence
of quantum circuits $\mathcal{U}$ does the following: it first runs $O\left(
2^{n/3}\right)  $\ iterations of Grover's algorithm, in order to produce the
$n$-qubit\ state $\alpha\left\vert x\right\rangle +\beta\sum_{y\in\left\{
0,1\right\}  ^{n}}\left\vert y\right\rangle $, where%
\begin{align*}
\alpha &  =\sqrt{\frac{1}{2^{n/3}+2^{-n/3+1}+1}},\\
\beta &  =2^{-n/3}\alpha
\end{align*}
(one can check that this state is normalized). \ Next $\mathcal{U}$ applies
Hadamard gates to the first $n/3$\ qubits. \ This yields the state%
\[
2^{-n/6}\alpha\sum_{y\in\left\{  0,1\right\}  ^{n/3}}\left(  -1\right)
^{x_{A}\cdot y}\left\vert y\right\rangle \left\vert x_{B}\right\rangle
+2^{n/6}\beta\sum_{z\in\left\{  0,1\right\}  ^{2n/3}}\left\vert 0\right\rangle
^{\otimes n/3}\left\vert z\right\rangle ,
\]
where $x_{A}$\ consists of the first $n/3$ bits of $x$, and $x_{B}$\ consists
of the remaining $2n/3$ bits. \ Let $Y$\ be the set of $2^{n/3}$ basis states
of the form $\left\vert y\right\rangle \left\vert x_{B}\right\rangle $, and
$Z$\ be the set of $2^{2n/3}$\ basis states of the form $\left\vert
0\right\rangle ^{\otimes n/3}\left\vert z\right\rangle $.

Notice that $2^{-n/6}\alpha=2^{n/6}\beta$. \ So with the sole exception of
$\left\vert 0\right\rangle ^{\otimes n/3}\left\vert x_{B}\right\rangle $
(which belongs to both $Y$ and $Z$),\ the \textquotedblleft
marked\textquotedblright\ basis states in $Y$\ have the same amplitude as the
\textquotedblleft unmarked\textquotedblright\ basis states in $Z$. \ This is
what we wanted. \ Notice also that, if we manage to find any $\left\vert
y\right\rangle \left\vert x_{B}\right\rangle \in Y$, then we can find $x$
itself using $2^{n/3}$ further classical queries: simply test all\ possible
strings that end in $x_{B}$. \ Thus, the goal of our algorithm will be to
cause the hidden variable to visit an element of $Y$, so that inspecting the
variable's history reveals that element.

As in Theorem \ref{szk}, the tools that we need are the juggle
subroutine, and a way of reducing many basis states to two. \ Let
$s$ be drawn uniformly at random from $\left\{  0,1\right\} ^{n/3}$.
\ Then $\mathcal{U}$\ appends a third register to the state, and
sets it equal to $\left\vert z\right\rangle $\ if the first two
registers have the form $\left\vert 0\right\rangle ^{\otimes
n/3}\left\vert z\right\rangle $, or to $\left\vert s,y\right\rangle
$\ if they have the form $\left\vert y\right\rangle \left\vert
x_{B}\right\rangle $. \ Disregarding the basis state $\left\vert
0\right\rangle ^{\otimes n/3}\left\vert x_{B}\right\rangle $ for
convenience, the result is%
\[
2^{-n/6}\alpha\left(  \sum_{y\in\left\{  0,1\right\}  ^{n/3}}\left(
-1\right)  ^{x_{A}\cdot y}\left\vert y\right\rangle \left\vert x_{B}%
\right\rangle \left\vert s,y\right\rangle +\sum_{z\in\left\{  0,1\right\}
^{2n/3}}\left\vert 0\right\rangle ^{\otimes n/3}\left\vert z\right\rangle
\left\vert z\right\rangle \right)  .
\]
Next $\mathcal{U}$\ applies the juggle subroutine to the joint state of the
first two registers. \ Suppose the hidden-variable value has the form
$\left\vert 0\right\rangle ^{\otimes n/3}\left\vert z\right\rangle \left\vert
z\right\rangle $ (that is, lies outside $Y$). \ Then with probability
$2^{-n/3}$\ over $s$, the first $n/3$\ bits of $z$ are equal to $s$. \ Suppose
this event occurs. \ Then conditioned on the third register being $\left\vert
z\right\rangle $, the reduced state of the first two registers is%
\[
\frac{\left(  -1\right)  ^{x_{A}\cdot z_{B}}\left\vert z_{B}\right\rangle
\left\vert x_{B}\right\rangle +\left\vert 0\right\rangle ^{\otimes
n/3}\left\vert z\right\rangle }{\sqrt{2}},
\]
where $z_{B}$\ consists of the last $n/3$\ bits of $z$. \ So it follows from
Section \ref{JUGGLE}\ that with\ probability $\Omega\left(  1/n\right)  $, the
juggle subroutine will cause the hidden variable to transition from
$\left\vert 0\right\rangle ^{\otimes n/3}\left\vert z\right\rangle $\ to
$\left\vert z_{B}\right\rangle \left\vert x_{B}\right\rangle $,\ and hence
from $Z$ to $Y$.

The algorithm calls the juggle subroutine $\Theta\left(  2^{n/3}n\right)
=\Theta\left(  N^{1/3}\log N\right)  $\ times, drawing a new value of $s$ and
recomputing the third register after each call. \ Each call moves the hidden
variable from $Z$ to $Y$ with independent\ probability $\Omega\left(
2^{-n/3}/n\right)  $; therefore with high probability \textit{some} call does
so. \ Note that this juggling phase does not involve any database queries.
\ Also, as in Theorem \ref{szk}, \textquotedblleft drawing\textquotedblright%
\ $s$ really means preparing a uniform superposition over all possible $s$.
\ Finally, the probability that the hidden variable ever visits the basis
state $\left\vert 0\right\rangle ^{\otimes n/3}\left\vert x_{B}\right\rangle
$\ is exponentially small (by the union bound), which justifies our having
disregarded it.
\end{proof}

A curious feature of Theorem \ref{searchthm}\ is the tradeoff between queries
and computation steps. \ Suppose we had run $Q$ iterations of Grover's
algorithm, or in other words made $Q$ queries to $f$. \ Then provided
$Q\leq\sqrt{N}$, the marked state $\left\vert x\right\rangle $\ would have
occurred with probability $\Omega\left(  Q^{2}/N\right)  $, meaning that
$\widetilde{O}\left(  N/Q^{2}\right)  $\ calls to the juggle subroutine would
have been sufficient to find $x$. \ Of course, the choice of $Q$ that
minimizes $\max\left\{  Q,N/Q^{2}\right\}  $ is $Q=N^{1/3}$. \ On the other
hand, had we been willing to spend $\widetilde{O}\left(  N\right)
$\ computation steps, we could have found $x$\ with only a \textit{single}
query!\footnote{One should not make too much of this fact; one way to
interpret it is simply that the \textquotedblleft number of
queries\textquotedblright\ should be redefined as $Q+T$\ rather than $Q$.}
\ Thus, one might wonder whether some other algorithm could push the number of
queries below $N^{1/3}$, without simultaneously increasing the number of
computation steps. \ The following theorem rules out that possibility.

\begin{theorem}
\label{lowerbound}In the $\mathsf{DQP}$ model,\ $\Omega\left(  N^{1/3}\right)
$\ computation steps are needed to search an $N$-item database for a unique
marked item. \ As a consequence, there exists an oracle relative to which
$\mathsf{NP}\not \subset \mathsf{DQP}$; that is, $\mathsf{NP}$-complete
problems are not efficiently solvable by sampling histories.
\end{theorem}

\begin{proof}
Let $N=2^{n}$ and $f:\left\{  0,1\right\}  ^{n}\rightarrow\left\{
0,1\right\}  $. \ Given a sequence of quantum circuits
$\mathcal{U}=\left( U_{1},\ldots,U_{T}\right)  $ that query $f$, and
assuming that $x\in\left\{ 0,1\right\}  ^{n}$\ is the unique string
such that $f\left(  x\right)  =1$, let $\left\vert \psi_{t}\left(
x\right)  \right\rangle $\ be the quantum state after $U_{t}$\ is
applied but before $U_{t+1}$\ is. \ Then the \textquotedblleft
hybrid argument\textquotedblright\ of Bennett et al.\ \cite{bbbv}\
implies that, by simply changing the location of the marked item
from $x$ to $x^{\ast}$, we can ensure that%
\[
\left\Vert \left\vert \psi_{t}\left(  x\right)  \right\rangle -\left\vert
\psi_{t}\left(  x^{\ast}\right)  \right\rangle \right\Vert =O\left(
\frac{Q_{t}^{2}}{N}\right)
\]
where $\left\Vert ~~\right\Vert $\ represents trace distance, and\ $Q_{t}$ is
the total number of queries made to $f$ by $U_{1},\ldots,U_{t}$. \ Therefore
$O\left(  Q_{t}^{2}/N\right)  $\ provides an upper bound on the probability of
noticing the $x\rightarrow x^{\ast}$ change\ by monitoring $v_{t}$, the value
of the hidden variable after $U_{t}$\ is applied. \ So by the union bound, the
probability of noticing the change by monitoring the entire history $\left(
v_{1},\ldots,v_{T}\right)  $\ is at most of order%
\[
\sum_{t=1}^{T}\frac{Q_{t}^{2}}{N}\leq\frac{TQ_{T}^{2}}{N}.
\]
This cannot be $\Omega\left(  1\right)  $ unless $T=\Omega\left(
N^{1/3}\right)  $\ or $Q_{T}=\Omega\left(  N^{1/3}\right)  $, either of which
implies an $\Omega\left(  N^{1/3}\right)  $\ lower bound on the total number
of steps.

To obtain an oracle relative to which $\mathsf{NP}\not \subset \mathsf{DQP}$,
we can now use a standard and well-known \textquotedblleft diagonalization
method\textquotedblright\ due to Baker, Gill, and Solovay \cite{bgs} to
construct an infinite sequence of exponentially hard search problems, such
that any $\mathsf{DQP}$\ machine fails on at least one of the problems,
whereas there exists an $\mathsf{NP}$\ machine that succeeds on all of them.
\ Details are omitted.
\end{proof}

\section{Conclusions and Open Problems\label{DISC}}

The idea that certain observables in quantum mechanics might have trajectories
governed by dynamical laws has reappeared many times: in Schr\"{o}dinger's
1931 stochastic approach \cite{schrodinger}, Bohmian mechanics \cite{bohm},
modal interpretations \cite{bd,dickson,dieks}, and elsewhere. \ Yet because
all of these proposals yield the same predictions for single-time
probabilities, if we are to decide between them it must be on the basis of
internal mathematical considerations. \ One message of this chapter has been
that such considerations can actually get us quite far.

To focus attention on the core issues, I restricted attention to the simplest
possible setting: discrete time, a finite-dimensional Hilbert space, and a
single orthogonal basis. \ Within this setting, I proposed what seem like
reasonable axioms that any hidden-variable theory should satisfy: for example,
indifference to the identity operation, robustness to small perturbations, and
independence of the temporal order of spacelike-separated events. \ I then
showed that not all of these axioms can be satisfied simultaneously. \ But
perhaps more surprisingly, I also showed that certain subsets of axioms
\textit{can} be satisfied for quite nontrivial reasons. \ In showing that the
indifference and robustness axioms can be simultaneously satisfied, Section
\ref{SPECIFIC}\ revealed an unexpected connection between unitary matrices and
the classical theory of network flows.

As mentioned previously, an important open problem is to show that the
Schr\"{o}dinger theory\ satisfies robustness. \ Currently, I can only show
that the matrix $P_{\mathcal{ST}}\left(  \rho,U\right)  $\ is robust to
\textit{exponentially} small perturbations, not polynomially small ones. \ The
problem is that if any row or column sum in the $U^{\left(  t\right)  }%
$\ matrix is extremely small, then the $\left(  r,c\right)  $-scaling process
will magnify tiny errors in the entries. \ Intuitively, though, this effect
should be washed out by later scaling steps.

A second open problem is whether there exists a theory that satisfies
indifference, as well as commutativity for all separable \textit{mixed} states
(not just separable pure states). \ A third problem is to investigate other
notions of robustness---for example, robustness to small
\textit{multiplicative} rather than additive errors.

On the complexity side, perhaps the most interesting problem left open by this
chapter is the computational complexity of simulating Bohmian mechanics. \ I
strongly conjecture that this problem, like the hidden-variable problems we
have seen, is strictly harder than simulating an ordinary quantum computer.
\ The trouble is that Bohmian mechanics does not quite fit in the framework of
this chapter: as discussed in Section \ref{OBJECTIONS}, we cannot have
deterministic hidden-variable trajectories for discrete degrees of freedom
such as qubits. \ Even worse, Bohmian mechanics violates the continuous
analogue of the indifference axiom. \ On the other hand, this means that by
trying to implement (say) the juggle subroutine with Bohmian trajectories, one
might learn not only about Bohmian mechanics and its relation to quantum
computation, but also about how essential the indifference axiom really is for
our implementation.

Another key open problem is to show better upper bounds on $\mathsf{DQP}$.
\ Recall that I was only able to show $\mathsf{DQP}\subseteq\mathsf{EXP}$, by
giving a classical exponential-time algorithm to simulate the flow theory
$\mathcal{FT}$. \ Can we improve this to (say) $\mathsf{DQP}\subseteq
\mathsf{PSPACE}$? \ Clearly it would suffice to give a $\mathsf{PSPACE}$
algorithm that computes the transition probabilities for some theory
$\mathcal{T}$\ satisfying the indifference and robustness axioms. \ On the
other hand, this might not be \textit{necessary}---that is, there might be an
indirect simulation method that does not work by computing (or even sampling
from) the distribution over histories. \ It would also be nice to pin down the
complexities of simulating specific hidden-variable theories, such as
$\mathcal{FT}$\ and $\mathcal{ST}$.

\chapter{Summary of Part \ref{MAR}\label{SUMMAR}}

Recall our hypothetical visitor from Conway's Game of Life, on a
complexity safari of the physical universe. \ Based on the results
in Part \ref{MAR}, the following are some intuitions about efficient
computation that I would advise our visitor to toss in the garbage.

\begin{itemize}
\item We can be fairly confident that the class of functions efficiently
computable in the physical world coincides with $\mathsf{P}$ (or
$\mathsf{BPP}$, which is presumably equal).

\item Although there are models of efficient computation more powerful than
$\mathsf{P}$, involving the manipulation of arbitrary real or complex numbers,
these models will inevitably blow up small errors in the numbers nonlinearly,
and must be therefore be unphysical.

\item A robot, moving at unit speed, would need order $n$ steps to search a
spatial region of size $n$ for a marked item.

\item The ability to see one's entire \textquotedblleft
history\textquotedblright\ in a single time step cannot yield any
complexity-theoretic advantage, since one could always just record the history
as one went along, at the cost of a polynomial increase in memory.
\end{itemize}

On the other hand, just as in Part \ref{LQC}, we have seen that many
of the intuitions in our visitor's suitcase are good to go. \ For
example:

\begin{itemize}
\item If the items in a database have distance $d$ from one another, then the
time needed to search the database is about $d$ times what it would be if the
items had unit distance from one another.

\item It is possible to choose a probability distribution over histories, in
such a way that state $i$ is never followed in a history by state $j$ if the
corresponding transition probability is zero, and such that a small change to
the transition matrices produces only a small change in the history distribution.

\item If, at the moment of your death, your entire life's history flashed
before you in an instant, then you could probably still not solve
$\mathsf{NP}$-complete problems in polynomial time.
\end{itemize}

\bibliographystyle{plain}
\bibliography{thesis}

\end{document}